\documentclass[11pt]{report}

\usepackage[T1]{fontenc}
\usepackage{lmodern}
\usepackage{amsmath,amssymb,amsthm,mathtools}
\usepackage[margin=1.1in]{geometry}
\usepackage{graphicx}
\usepackage{tikz}
\usetikzlibrary{arrows.meta}
\usepackage{array}
\usepackage{booktabs}
\usepackage{tabularx}
\usepackage{xltabular}
\usepackage{xcolor}
\definecolor{linkblue}{RGB}{20,60,150}
\usepackage[colorlinks=true,linkcolor=linkblue,citecolor=linkblue,urlcolor=linkblue]{hyperref}
\usepackage{bookmark}

\theoremstyle{plain}
\newtheorem{theorem}{Theorem}[chapter]
\newtheorem{proposition}[theorem]{Proposition}
\newtheorem{lemma}[theorem]{Lemma}
\newtheorem{corollary}[theorem]{Corollary}
\theoremstyle{definition}
\newtheorem{definition}[theorem]{Definition}
\newtheorem{construction}[theorem]{Construction}
\newtheorem{example}[theorem]{Example}

\newcommand{\lettersections}[1]{%
  \setcounter{section}{0}%
  \renewcommand{\thesection}{#1.\arabic{section}}%
  \renewcommand{\theHsection}{#1.\arabic{section}}}
\newcommand{\chaptersections}{%
  \renewcommand{\thesection}{\thechapter.\arabic{section}}%
  \renewcommand{\theHsection}{\theHchapter.\arabic{section}}}

\newcolumntype{L}{>{\raggedright\arraybackslash}X}

\let\keybibitem\bibitem
\renewcommand{\bibitem}[1]{\keybibitem[#1]{#1}}
\let\keythebibliography\thebibliography
\renewcommand{\thebibliography}[1]{%
  \keythebibliography{Shulman08}\addcontentsline{toc}{chapter}{\bibname}}
\renewcommand{\bibname}{References}

\title{\textbf{Foundations of Algebraic Architecture Theory}\\[0.6em]
\large\itshape A Rising Sea of Geometry, Transport, Comparison, and Reconstruction}
\author{Hiroyuki Nakahata\\
\normalsize Independent Researcher\\
\normalsize ORCID: \href{https://orcid.org/0009-0008-5928-0234}{0009-0008-5928-0234}}
\date{\normalsize September 2026}

\begin{document}

\nocite{Stacks,FGMPS04,Spivak12,SAGA,GB92,AG97,Goguen92,Gibson26,AB11,AMB12,Young26,CC77,GRS00,SSVW17,Shulman08,BS81,JRW12,Lawvere63,BW85,Kelly82,BMW12,BIAN9,ReS,McLarty03}

\maketitle

\begin{abstract}
AI-generated software changes make it increasingly important to determine what
a change preserves, where local consistency fails to extend globally, and which
alternatives remain. This paper develops the foundations of Algebraic
Architecture Theory (AAT) from Atoms, typed primitive facts, and Laws, equations
that objects must satisfy. We make explicit the choice of what counts as
structure and which operations and laws to preserve, calling this choice a
\emph{reading}. Objects and structure-preserving morphisms are defined relative
to a reading. From finite Atom families we construct cores closed under
operations and geometries equipped with sites and coefficients.

The theory addresses five kinds of decision: gluing, diagnosis, transport,
classification of changes, and reconstruction. Its culmination is a
reconstruction theorem: the category of full geometries and all their
structure-preserving morphisms is equivalent to a category of local models
given by primitive data and independently defined compatibility conditions.
Objects are recovered up to isomorphism, and morphisms between fixed endpoints
are recovered uniquely. This allows the overall structure and its changes to
be assembled from local descriptions written by separate parties.

To support these decisions, we construct a degree-one \v{C}ech obstruction to
gluing local states. Under torsor and sheaf conditions, its vanishing is
equivalent to the existence of a global state; under the conditions required
for comparison with repair semantics, it also corresponds to the existence of
a global repair. We compare diagnoses along evaluation-preserving changes of
resolution, cover, and coefficients, and give conditions for preserving first
cohomology. Invariance for all finite Law families for which both resolutions are
adequate is decidable when computable finite data are given.

Transport along exact changes is characterized by a universal property and
commutes with base change through an invertible comparison on exact pointed
pullback squares. Comparisons of two routes generated from the same square,
finite comparison diagram, and geometry factor into an invertible comparison
and an idempotent normalization. We characterize when observations alone
determine comparison preservation and classify all compatible lifts.

We encode independently defined lens (model synchronization) and protocol
semantics in this framework, proving equivalence of law satisfaction and a
one-to-one correspondence between semantics-preserving and typed morphisms.
Through these correspondences, common theorems classify and count operation-preserving
changes and yield unique extensions of morphisms from finite tables. The
corresponding Lean declarations are listed in the appendix.

\end{abstract}

\tableofcontents

\chapter*{Introduction}
\addcontentsline{toc}{chapter}{Introduction}

Software designs keep changing. We replace the representation of data, factor
processing out into shared components, and split one system into several
services. In every case the question is what is preserved across the change.
A change of data representation may leave the displayed shipping address
unchanged while altering what happens when that address is updated. To compare
the two representations, we must specify both what is read and which operations
must be preserved. This distinction will recur throughout the paper.

AI-generated changes make these questions increasingly important. As more
candidate changes become available, we need grounds for deciding which ones
preserve the required structure, where local consistency fails to extend to
the whole, and how much freedom remains among the permitted changes. This
paper develops mathematical objects and theorems for making these decisions.
We call the resulting foundations Algebraic Architecture Theory (AAT).

The starting point is an explicit choice of what counts as structure and what
a change must preserve. We call this choice a \emph{reading}. For the order
example, a reading specifies whether the comparison concerns only the displayed
address or also the update and the payment data it must retain. AAT defines
architecture objects and their structure-preserving maps relative to such
choices. The following five questions explain what we want to do with these
objects before we describe how they are constructed.

\section*{Five questions}

\textbf{When do local states fit together?}
Suppose that three services record the time of the same event using their own
clocks. For each connection, we specify a fixed offset for converting a time
from one clock to the other. If conversion from the first clock through the
second and third and back to the first does not recover the original time,
no assignment of offsets to the three clocks can satisfy all these conversion
rules. Each service's records may be internally consistent, yet following the
connections around the cycle reveals a contradiction. Can we correct each
service's timestamps to obtain records satisfying every specified conversion
rule? The question is whether a global state exists that simultaneously
satisfies the relations required on overlaps (Chapter~\ref{chap:2}).

\textbf{When does a change of reading preserve a diagnosis?}
Suppose that we extract a list of shipping addresses from order data, omitting
internal payment data. Conditions involving only addresses can still be
checked from this list. But checking conditions on individual entries and
detecting discrepancies between services are different tasks. In the clock
example, inspecting each service separately does not reveal the discrepancy
around the cycle. If we group several services into a single region for
inspection, we must also retain the connection and conversion data needed
for that diagnosis. Which information can we omit, and how can we repartition
the regions we inspect, while still detecting the same discrepancies that
corrections cannot remove (Chapter~\ref{chap:3})?

\textbf{When do different routes of change agree?}
When migrating an order-processing API, we want a direct migration from the
old version to the new one and a migration through an intermediate version to
yield the same order and the same behavior when it is read or updated.
Being able to undo each conversion does not by itself ensure agreement
between the two routes.
The order of operations matters too. We may split order processing and
inventory management into services and then extract the operations involved
in order cancellation, or extract those operations first and apply the same
splitting policy. Do the resulting designs agree on completion of the
cancellation and release of the inventory reservation, preserving the condition
that a completed cancellation has no outstanding reservation? What is needed
to put the data and operations obtained by the two routes in correspondence
and ensure that they satisfy the same conditions
(Chapters~\ref{chap:4}--\ref{chap:5})?

\textbf{Which changes preserve a comparison, and how can we distinguish them?}
Suppose that we extract the shipping-address update into a shared component.
The displayed address may remain the same even though an update mishandles
the payment data. We want moving old data to the new format and then updating
the address to yield the same order state, including payment data, as updating
first and then moving. There may be more than one invertible change of the internal representations
that preserves this correspondence between old and new.
Can the display and the pattern of calls alone tell us whether a candidate
change is compatible, or must we also inspect the internal values passed
through operations? We seek to identify the information needed, construct all
compatible changes, and describe the freedom in choosing them
(Chapters~\ref{chap:6}--\ref{chap:7}).

\textbf{When do local descriptions determine the whole?}
Suppose that the modernization of an intrabank transfer system is divided
among the account, payment, and accounting teams. Each team describes the
types of data it handles, its operations and required conditions, and the
correspondence between the old and new designs. For a given transfer, the
teams' descriptions must refer to the same transaction and amount, with
compatible completion conditions, if they are to combine into a single
transfer process. Their old-to-new correspondences must agree too: a
transaction migrated by the payment team and its record migrated by the
accounting team must still refer to the same transaction. What must be
described, and what must agree across teams, to assemble an overall design
and a single migration preserving its operations and conditions? When can
we also rule out distinct migrations that differ only where the descriptions
say nothing (Chapter~\ref{chap:8})?

These questions share several ingredients: objects, operations, local readouts
and their overlaps, and equations to be satisfied. We now describe the objects
that bring these ingredients together.

\section*{Architecture relative to a reading}

What facts are to be retained? An \emph{Atom} is a typed primitive fact, such
as the existence of a component or a named operation. A finite family of
Atoms together with relations is a \emph{configuration}. Adding structure
data, such as state sets and interpretations of operations, gives an
\emph{architecture object}. The equations that objects must satisfy are
specified as \emph{Laws}. These specify, for example, the required relationship
between reading and updating a state.

Which objects can arise under the permitted operations? From the extracted
Atom family we form a base object and generate a \emph{core}: it carries
the smallest family of objects closed under those operations. The core keeps
the choices of object formation, operations, and equations together. This
lets us ask what a change must preserve beyond the extracted facts alone.

To return to the three-service example, we also need to read information on
parts and compare it on overlaps. We choose local contexts, covers that
specify how the parts cover a whole, and coefficients that measure differences.
A core equipped with the covers, overlaps, coefficients, and local data is a
\emph{full geometry}. These are the objects on which the later chapters carry
out comparison and reconstruction.

The choices of vocabulary, extraction rules, object formation, equations,
operations, and locality together constitute a reading. Making the reading
explicit fixes the meaning of each preservation question. In the order
example, the choice to retain updates as well as displayed values changes
which maps are admitted. Chapter~\ref{chap:1} defines these choices and the
maps that compare them.

The \emph{Rising Sea} of the title comes from a metaphor with which
Grothendieck described his way of doing mathematics. A problem is treated not
as a hard shell to be cracked with a chisel, but as something that opens by
itself when immersed in a sea whose level rises quietly
(\cite[\href{https://webusers.imj-prg.fr/~leila.schneps/grothendieckcircle/RetSbis.pdf}{Part~III, note~122, ``La mer qui monte\ldots''}]{ReS};
the English phrase \emph{rising sea} follows the translation in
\cite{McLarty03}). Here the common objects allow different software questions
to be addressed by the same mathematical constructions. To explain how this
works, we next describe the connection to two independently defined semantics.

\section*{Correspondence with the semantics of CS}

In model synchronization, a \emph{lens} consists of states, views, reads, and
updates satisfying specified laws. In a protocol, named operations connect
states at control points, and observations record what can be read from those
states. We define these semantics independently in \S\S\ref{sec:1.8}--\ref{sec:1.9}.
For lenses, we construct Atoms and operations together with Laws that hold
if and only if the original laws hold (Proposition~\ref{prop:1.41}). Typed
morphisms recover the original semantics-preserving morphisms
(Proposition~\ref{prop:1.43}). For protocols, finite operation tables satisfying the declared path relations
and observation conditions determine realizations, and vertex maps satisfying
preservation conditions recover
semantics-preserving morphisms (Propositions~\ref{prop:1.37}
and~\ref{prop:1.38}).

These correspondences give a route from a software question to an AAT theorem
and back: construct the AAT input, prove the correspondence, apply the common
theorem, and interpret its conclusion in the original semantics. Each
correspondence states whether it gives an equivalence, an implication, or
preservation in one direction.

For example, consider the product lens used for the shipping address and
payment data. Of the four candidate changes that preserve the displayed
address and the unset payment value, only two also preserve updates. In the
worker protocol example, sixteen candidates preserve the correspondence of
control points, but only four preserve the handoff of contexts
(\S\ref{sec:7.8}). In both examples, an operation keeps an internal state
unchanged. Preserving that operation forces the changes of internal state at
its two ends to agree. The common classification theorem expresses this as
one permutation per connected component of the operation graph
(Theorem~\ref{thm:7.24}). Thus the same constraint explains both reductions in
the number of choices.

The correspondence also identifies, by a group isomorphism, the changes that
preserve a comparison in the original semantics and in the typed construction
(Theorem~\ref{thm:4.41}). Restricting the views of a lens also restricts its reads, updates, and lens
isomorphisms to the corresponding states (Proposition~\ref{prop:5.34}).
For protocols, realizations and semantics-preserving adapters restrict to
selected vertices, operations, and relations (Proposition~\ref{prop:5.36}). Section~\ref{sec:6.8}
examines which operations remain after normalization. At the end of the paper,
the same reconstruction principle yields unique extensions of morphisms from
compatible finite tables for lenses and protocols (Propositions~\ref{prop:8.26}
and~\ref{prop:8.28}).

\section*{Main constructions and their consequences}

The four parts organize the results as follows. The table gives a reading
guide; the paragraphs below state the principal conditions and conclusions.

\begin{xltabular}{\linewidth}{@{}>{\hsize=.7\hsize}L>{\hsize=1.15\hsize}L>{\hsize=1.15\hsize}L@{}}
\toprule
Part and chapters & Question & Results developed \\
\midrule
\endhead
I: Geometry and Local Consistency (Chapters~\ref{chap:1}--\ref{chap:2}) & What structure is compared, and when do local states fit together? & Construction of objects and geometry; obstructions to gluing and their relation to repair \\
\addlinespace[3pt]
II: Diagnosis and Transport (Chapters~\ref{chap:3}--\ref{chap:4}) & When does a diagnosis survive a change of reading, and when can structure be carried along a change? & Diagnostic invariance; universal transport and coherence of comparisons \\
\addlinespace[3pt]
III: Comparison and Normalization (Chapters~\ref{chap:5}--\ref{chap:6}) & When do two construction routes agree, and what does normalization retain? & Base-change comparisons; factorization and classification of invertibility \\
\addlinespace[3pt]
IV: Classification and Reconstruction (Chapters~\ref{chap:7}--\ref{chap:8}) & Which changes preserve a comparison, and when do local descriptions determine the whole? & Classification of compatible changes; recovery of objects and morphisms \\
\bottomrule
\end{xltabular}

\textbf{Geometry and local consistency (Part~I).}
Chapter~\ref{chap:1} constructs the core generated by a finite Atom family
and a reading. Its objects are exactly those reachable by finite sequences
of the permitted operations (Theorem~\ref{thm:1.18}). Contexts and covers
define a site; compatible restriction maps give a presheaf of rings and its
sheafification (\S\S\ref{sec:1.5}--\ref{sec:1.6}). The categories of
extraction, cores, and geometries are connected by projections that retain
the preceding levels of structure (Theorem~\ref{thm:1.31}).

Chapter~\ref{chap:2} expresses the satisfaction of Laws geometrically.
Under an algebraic presentation of the evaluation data and a localization
condition, simultaneous vanishing of the residuals is equivalent to
factorization through the space defined by the ideal of the equations
(Theorem~\ref{thm:2.9}). Differences of local states on overlaps determine a
degree-one \v{C}ech obstruction class. When local corrections give torsor
structures compatible with restriction and the states form a sheaf, vanishing
of this class is equivalent to the existence of a global state
(Theorem~\ref{thm:2.22}). Under the correspondence and completeness conditions
relating repair semantics to equations, a coefficient isomorphism identifies
their obstruction classes and connects vanishing with global repair
(Theorem~\ref{thm:2.49}, Corollary~\ref{cor:2.50}). An explicit
integer-coefficient example supplies an obstruction for the diagnostic
comparisons of the next chapter (\S\ref{sec:2.8}).

\textbf{Diagnosis and transport (Part~II).}
Chapter~\ref{chap:3} asks whether a diagnosis is preserved when the reading
changes. It constructs the coarsest resolution retaining all chosen Law
values (Theorem~\ref{thm:3.4}) and comparison maps between diagnostic
complexes. Under Condition~C on covers and coefficients, the comparison
induces an isomorphism on first cohomology (Theorem~\ref{thm:3.17}). Uniform
invariance, over all finite Law families for which both resolutions are
adequate, reduces to vanishing of kernels and cokernels indexed by nonempty
subsets of values. It is decidable when the required finite enumerations,
value tables, supports, and incidence data are computable
(Theorem~\ref{thm:3.22}, Proposition~\ref{prop:3.23}). The chapter also maps
the integer obstruction to the diagnosis and, under conditions on relation
components and common representatives, reflects vanishing
(Theorems~\ref{thm:3.37} and~\ref{thm:3.40}).

Chapter~\ref{chap:4} addresses a different question: carrying the structure
itself along a change. An exact change gives a bijective reindexing of
extracted Atoms. The transported core is characterized by unique
factorization of further changes (Theorem~\ref{thm:4.5}); transport to
geometry and its coherence with composition and projections are then
established (Theorems~\ref{thm:4.11}, \ref{thm:4.15}, and~\ref{thm:4.16}).
Specified comparisons may still differ from the canonical ones. For finite
comparison diagrams, the obstruction to removing these discrepancies
vanishes exactly when one can reselect edges to make all faces commute
(Theorem~\ref{thm:4.23}). Diagnosis and transport thus answer distinct
preservation questions about the structures constructed in Part~I.

\textbf{Comparison and normalization (Part~III).}
Chapter~\ref{chap:5} compares transporting structure with selecting an input
scope by pullback. It constructs the corresponding original input as a fiber
product and pulls back objects, operations, and Laws. For exact pointed
pullback squares, transport and pullback are related by an invertible
Beck--Chevalley comparison (Theorem~\ref{thm:5.11}). This is the precise
sense in which the two orders agree. For refinements that enlarge the
extracted content, backward transport instead requires reflection of
extraction on the realized locus (Theorem~\ref{thm:5.25}).

The chapter then builds two routes of full geometries from the same square,
finite comparison diagram, and geometry. Their generated comparison
$\beta=E\alpha$ combines an invertible comparison $\alpha$ with a
normalization $E$ chosen according to the diagnosis (\S\ref{sec:5.10}).
Chapter~\ref{chap:6} proves that, under the preservation conditions on
operations, residuals, invariants, and signatures, normalization defines an
idempotent morphism of full geometries (Theorem~\ref{thm:6.14}). The generated
comparison is invertible exactly when its normalization factor is the
identity (Theorem~\ref{thm:6.16}). Otherwise, the same comparison still gives
an isomorphism between the retained images in the idempotent completion
(Theorem~\ref{thm:6.12}). This distinguishes agreement after normalization
from recovery of the original structure.

\textbf{Classification and reconstruction (Part~IV).}
Chapter~\ref{chap:7} studies changes of the two endpoints that preserve a
chosen comparison. It separates two tasks: deciding whether a given change
is compatible, and constructing a compatible change from one specified after
normalization. Observations suffice for the first task exactly when the
kernel of the observation homomorphism is contained in the comparison-preserving
group (Theorem~\ref{thm:7.9}). For an invertible comparison and canonical
normalizations with a common presentation of the accompanying data, the
restriction homomorphism has a section. The compatible lifts of each fixed
normalized change form a torsor under its kernel
(Theorems~\ref{thm:7.19} and~\ref{thm:7.21}). Such lifts can exist even when
the normalized information does not determine compatibility of a given
change (\S\S\ref{sec:7.4}--\ref{sec:7.5}). For systems whose operations keep
hidden states, the classification by connected components also gives a split
short exact sequence onto the visible group (Theorem~\ref{thm:7.25}).

Chapter~\ref{chap:8} asks how to recover the objects and maps themselves
from local descriptions. The reconstruction principle has three requirements:
local readouts distinguish morphisms, compatible local morphisms can be
assembled, and local objects can be assembled (Theorem~\ref{thm:8.12}). The
main theorem proves these requirements for the primitive data of full
geometries, including types, operations, Laws, local geometry, and realization.
It gives an equivalence between the category of full geometries with all
their structure-preserving morphisms and the category of local models given
by primitive data and independently defined compatibility conditions
(Theorem~\ref{thm:8.18}). Objects are recovered up to isomorphism; morphisms
between fixed endpoints are recovered uniquely, with evaluation, composition,
and isomorphisms preserved (Corollary~\ref{cor:8.19}).

Here ``local'' refers to fragments of the data describing the geometry and
its maps, including the contexts and covers themselves. In Chapter~\ref{chap:2},
by contrast, the geometry and cover were fixed and the states were glued.
The reconstruction theorem applies to design descriptions divided among
teams when the required primitive data and compatibility conditions are
specified (\S\ref{sec:8.6}). In general the reconstruction uses compatible
data over all finite fragments. The special cases for lenses and protocols
explain when a finite table suffices (\S\ref{sec:8.7}); for finite models,
we also distinguish sufficient information from the checking procedure and
its computational cost (\S\ref{sec:8.8}).

\section*{Organization and reading guide}

The four parts comprise eight chapters. We assume basic familiarity with
sets, groups, rings, modules, categories, functors, and natural transformations.
The Preliminaries fix the notation; Chapter~\ref{chap:1} introduces AAT.
The chapters using sites, sheaves, cohomology, and idempotent completion
provide the needed definitions and references.

Begin with the objects and maps of Chapter~\ref{chap:1}, especially the core,
local geometry, and three-level projections (\S\S\ref{sec:1.4}--\ref{sec:1.7}).
For software applications, also read the lens and protocol semantics and
their encoding (\S\S\ref{sec:1.8}--\ref{sec:1.10}). Then choose a route:

\begin{itemize}
\item For local consistency and diagnosis, continue through
Chapters~\ref{chap:2}--\ref{chap:3}.
\item For transport and comparison of construction routes, concentrate on
Chapters~\ref{chap:4}--\ref{chap:6}, using the obstruction constructions of
Chapter~\ref{chap:2} and the diagnostic comparisons of Chapter~\ref{chap:3}
where they are invoked.
\item For classification and reconstruction, follow the construction of the
generated comparison in \S\ref{sec:5.10}, then read
Chapters~\ref{chap:6}--\ref{chap:8}, returning to the earlier transport and
base-change constructions as needed.
\end{itemize}

Related Work compares the objects, assumptions, admissible morphisms, and
conclusions with those of the closely related literature. Appendix~\ref{app:A}
lists the Lean declarations corresponding to the definitions, constructions,
and results, together with their assumptions and fixed source versions.
Appendix~\ref{app:B} gives the source locations and build instructions for
checking the proofs.

\lettersections{P}
\chapter*{Preliminaries and Notation}
\addcontentsline{toc}{chapter}{Preliminaries and Notation}

We assume the basic facts about sets, groups, rings, and modules, and about
categories, functors, and natural transformations.

\section{Sizes of sets and categories}\label{sec:P.1}

The natural numbers are $\mathbb{N}=\{0,1,2,\ldots\}$, and we write
$\mathbb{Z}$ and $\mathbb{Q}$ for the ring of integers and the field of
rational numbers, respectively. A family indexed by a set $I$ is written
$(X_i)_{i\in I}$. A finite family is a family whose index set $I$ is finite.
Finiteness of each $X_i$ is a condition separate from finiteness of the index
set.

Sizes of sets are handled in classical set theory with the axiom of choice,
together with Grothendieck universes. We fix universes
$\mathbb{U}\in\mathbb{V}$ as needed, and call a set that belongs to
$\mathbb{U}$ a $\mathbb{U}$-small set. We write
${\mathbf{Set}}_{\mathbb{U}}$ for the category of small sets and maps between
them, and treat this category itself in the larger universe $\mathbb{V}$. Where the
universe in use is clear, we omit the subscript and write ${\mathbf{Set}}$.

A category $\mathcal{C}$ is $\mathbb{U}$-small if its objects and its
morphisms form $\mathbb{U}$-small sets. The category $\mathcal{C}$ is locally $\mathbb{U}$-small if the set of
morphisms between each pair of objects $X,Y$ is $\mathbb{U}$-small, and
essentially $\mathbb{U}$-small if it is equivalent to a $\mathbb{U}$-small
category. Constructions that involve functor
categories use a universe large enough to contain their objects and
morphisms.

\section{Maps, morphisms, and composition}\label{sec:P.2}

The composite of maps $f:X\to Y$ and $g:Y\to Z$ is written
\[
gf=g\circ f:X\longrightarrow Z,
\qquad (gf)(x)=g(f(x)).
\]
The map on the right acts first. We use the same order of composition for
morphisms of categories and for functors. The identity morphism of an object
$X$ is written $\mathrm{id}_X$, and the identity functor of a category
$\mathcal{C}$ is written $\mathrm{Id}_{\mathcal{C}}$.

For a map $f:X\to Y$ and subsets $S\subseteq X$ and $T\subseteq Y$, we write
$f(S)$ for the image, $f^{-1}(T)$ for the preimage, and $f|_S:S\to Y$ for the
restriction. The fiber over $y\in Y$ is
\[
X_y=f^{-1}(\{y\})=\{x\in X\mid f(x)=y\}.
\]
A bijection is written $f:X\xrightarrow{\sim}Y$, and its inverse map is
written $f^{-1}:Y\to X$.

For a category $\mathcal{C}$, we write $\mathrm{Ob}(\mathcal{C})$ for its
collection of objects and $\mathrm{Hom}_{\mathcal{C}}(X,Y)$ for the set of
morphisms from $X$ to $Y$. When the category is clear, the latter is
abbreviated to $\mathrm{Hom}(X,Y)$. A morphism $f:X\to Y$ is an isomorphism if
there is a morphism $g:Y\to X$ with $gf=\mathrm{id}_X$ and
$fg=\mathrm{id}_Y$. Such a $g$ is unique, and we denote it by $f^{-1}$. We write
$X\cong Y$ for isomorphic objects, and $\mathrm{Aut}_{\mathcal{C}}(X)$ for the
automorphism group of $X$.

Commutativity of a diagram means that the composites along any two paths with
the same domain and codomain are equal. For example, the commutativity
condition for the square formed by $f:X\to Y$, $f':X'\to Y'$,
$a:X\to X'$, and $b:Y\to Y'$ is $bf=f'a$. When a comparison is made through a natural isomorphism, that natural
isomorphism is specified as part of the data of the diagram.

\section[Functors, natural transformations, and equivalences of
categories]{Functors, natural transformations,\texorpdfstring{\\}{ }and
equivalences of categories}\label{sec:P.3}

The actions of a functor $F:\mathcal{C}\to\mathcal{D}$ on objects and on
morphisms are written $F(X)$ and $F(f)$, respectively. The opposite category is written
$\mathcal{C}^{\mathrm{op}}$, and a contravariant functor is regarded as a
functor whose domain is $\mathcal{C}^{\mathrm{op}}$.

A natural transformation $\alpha:F\Rightarrow G$ between functors
$F,G:\mathcal{C}\to\mathcal{D}$ is a family that assigns to each object $X$ a
morphism $\alpha_X:F(X)\to G(X)$ such that, for each morphism $f:X\to Y$,
\[
G(f)\alpha_X=\alpha_YF(f).
\]
If every $\alpha_X$ is an isomorphism, $\alpha$ is called a natural
isomorphism and is written $\alpha:F\xRightarrow{\sim}G$. The composite of
natural transformations $\alpha:F\Rightarrow G$ and $\beta:G\Rightarrow H$ is
given by $(\beta\alpha)_X=\beta_X\alpha_X$.

A functor $F:\mathcal{C}\to\mathcal{D}$ is fully faithful if, for every pair
of objects $X,Y$, the map
\[
F_{X,Y}:\mathrm{Hom}_{\mathcal{C}}(X,Y)
\longrightarrow\mathrm{Hom}_{\mathcal{D}}(F(X),F(Y)),
\qquad f\longmapsto F(f)
\]
is a bijection. If every $D\in\mathrm{Ob}(\mathcal{D})$ is isomorphic to some
$F(X)$, $F$ is called essentially surjective.

An equivalence of categories $\mathcal{C}\simeq\mathcal{D}$ is given by
functors $F:\mathcal{C}\to\mathcal{D}$ and $G:\mathcal{D}\to\mathcal{C}$
together with natural isomorphisms
\[
\eta:\mathrm{Id}_{\mathcal{C}}\xRightarrow{\sim}GF,
\qquad
\varepsilon:FG\xRightarrow{\sim}\mathrm{Id}_{\mathcal{D}}.
\]
We call $G$ a quasi-inverse of $F$. For the standard definitions of
categories, functors, natural transformations, and equivalences of categories
we follow
\cite[\href{https://stacks.math.columbia.edu/tag/0013}{Tag~0013}]{Stacks}.

\section{Notation for coefficients and algebra}\label{sec:P.4}

A coefficient ring $k$ is a commutative ring with identity, and ring
homomorphisms preserve the identity. We write ${\mathbf{CommRing}}$ for the
category of commutative rings and ${\mathbf{Ab}}$ for the category of abelian
groups. Modules over $k$ are unital, $1m=m$, and we write $\mathrm{Mod}_k$ for
their category. A commutative $k$-algebra is a pair consisting of a commutative ring $A$
and a structure homomorphism $k\to A$. We allow the zero ring. These
conventions for commutative rings with identity and unital modules agree with
\cite[\href{https://stacks.math.columbia.edu/tag/0006}{Tag~0006},
\href{https://stacks.math.columbia.edu/tag/00AQ}{Tag~00AQ}]{Stacks}.

We write $A/I$ for the quotient ring of a ring $A$ by an ideal $I$, and $M/N$
for the quotient module of a module $M$ by a submodule $N$. The set of linear
maps between modules is written $\mathrm{Hom}_k(M,N)$, and the kernel and the
image of a homomorphism $u$ are written $\ker u$ and $\mathrm{im}\,u$. Abelian
groups and modules are written additively, and the zero element and the zero
map are both written $0$, according to context. General groups and
automorphism groups are written multiplicatively, with identity element $1$.

The tensor product over a fixed coefficient ring is written $M\otimes_k N$. A
change of coefficient ring is specified by a ring homomorphism
$\varphi:k\to k'$, and the extension of scalars of a module along it is
written $k'\otimes_k M$.

\section{Numbering and references}\label{sec:P.5}

Definitions, constructions, theorems, lemmas, propositions, corollaries, and
examples are numbered in one common sequence within each chapter of the main
body. For example, ``Definition~2.1'', ``Proposition~2.2'', and
``Definition~2.3'' all refer to items of Chapter~\ref{chap:2}. Sections, equations,
figures, and tables are each numbered within each chapter, and are referred to
as ``\S2.1'', ``(2.1)'', ``Figure~2.1'', and ``Table~2.1''. In the
Preliminaries the letter P, and in the appendices the letters A and B, are
used in place of the chapter number.

A citation gives a key such as \cite{Stacks}
together with the section or theorem number cited. Citations of the Stacks
project also give, in addition to section numbers, its permanent identifiers
called Tags. The bibliographic data are collected in the References.

\chaptersections

\part[Geometry and Local Consistency]{Geometry and Local Consistency\\[0.8em]{\large Chapters 1--2}}
\chapter{Construction of Relative Architecture}\label{chap:1}

\section*{Overview of the Chapter}

In this chapter we construct the objects and maps used to compare the structure of
software. The starting point is to make explicit the choice of what to read as
structure and which operations and laws to preserve.

For example, consider a state consisting of a displayed value and an internal value.
A read returns the displayed value, and an update keeps the internal value and
replaces the displayed value. If a change of the state representation leaves the
displayed value unchanged, reads are preserved. However, applying the change and
then updating need not give the same final state as updating and then applying the
change. To preserve both reads and updates, the two orders must also give the same
result.

Such a comparison requires both the structure of objects and the way operations act
on that structure. In AAT, a typed primitive fact is called an Atom, a family of
Atoms together with relations is called a configuration, and an object that also
keeps state sets, the interpretation of operations, and the like is called an architecture
object. The equations that objects must satisfy are specified as Laws, and we
provide the local structure for reading information part by part and gluing it
together. The choice of vocabulary, extraction rules, object formation, equations,
operations, and locality used in these constructions is called a reading.
Relativizing architecture to a reading means making this choice explicit together
with the objects.

The structure of the chapter and its main results are as follows.

\begin{xltabular}{\linewidth}{@{}LLL@{}}
\toprule
Question & Construction & Outcome \\
\midrule
\endhead
What to take as objects, and which operations to permit & Equip an Atom family with relations, structure, and operations & A core with a family of objects closed under operations (\S\S\ref{sec:1.1}--\ref{sec:1.4}, Theorem~\ref{thm:1.18}) \\
\addlinespace[3pt]
Which information to read locally and glue & Choose contexts, covers, and coefficients & A geometry with a site and a ring for each context (\S\S\ref{sec:1.5}--\ref{sec:1.6}) \\
\addlinespace[3pt]
What to preserve when the reading changes & Define morphisms comparing the data at each stage & The tower of projections connecting extraction, core, and geometry (\S\ref{sec:1.7}, Theorem~\ref{thm:1.31}) \\
\addlinespace[3pt]
How to express existing semantics of CS & Construct objects, operations, and Laws from lenses and protocols & Equivalence of the validity of laws, and recovery of semantics-preserving morphisms (\S\S\ref{sec:1.8}--\ref{sec:1.10}, Propositions~\ref{prop:1.41} and~\ref{prop:1.43}) \\
\bottomrule
\end{xltabular}

\section{Atoms and relative extraction}\label{sec:1.1}

To identify a single fact, one must specify, in addition to its content, what the
fact is about and from which viewpoint it is stated. An Atom carries this
information in five coordinates.

\begin{definition}[Vocabulary of Atoms]\label{def:1.1}

Let $K$, $X$, $S$, $P$, and $T$ be the sets of kinds, axes, subjects, predicates,
and payloads, respectively. The set of Atoms $\mathrm{At}$ is a nonempty set
equipped with maps
\[
\begin{aligned}
\mathrm{kind}&:\mathrm{At}\to K,&
\mathrm{axis}&:\mathrm{At}\to X,&
\mathrm{subject}&:\mathrm{At}\to S,\\
\mathrm{predicate}&:\mathrm{At}\to P,&
\mathrm{payload}&:\mathrm{At}\to T
\end{aligned}
\]
The combined map
$\mathrm{At}\to K\times X\times S\times P\times T$ is required to be injective.
Hence two Atoms are equal if and only if all five coordinates are equal.
The kind expresses the type of the fact, the axis the structure under attention,
and the subject what the fact is about. The predicate and the payload give what is
stated about that subject. For example, the existence of a component, a named
operation, a state quantity of an object, and a relation between two objects can
be expressed by Atoms of different kinds. The payload holds names, types, values,
and the like. Facts treated together are collected into a family of Atoms, and
their relations into a configuration.

\end{definition}

\begin{definition}[Atom families]\label{def:1.2}

An Atom family is a subset $F\subseteq\mathrm{At}$. Its support and its
restriction to an axis $x\in X$ are defined by
\[
\mathrm{supp}(F)=\{\mathrm{subject}(a)\mid a\in F\},
\qquad
F|_x=\{a\in F\mid \mathrm{axis}(a)=x\}
\]
Finiteness of an Atom family means finiteness of $F$.
An extracted family may in general be infinite. The composition reading of
\S\ref{sec:1.2} takes finite families as input.

The data from which extraction is performed are called a source. Even for the same
source, the facts extracted may vary with the choice of vocabulary and resolution.
An extraction doctrine collects this choice together with the conditions for
adopting an Atom.

\end{definition}

\begin{definition}[Extraction doctrines]\label{def:1.3}

An extraction doctrine $D$ consists of a set of sources $\mathrm{Src}_D$,
parameters $v_D,\gamma_D,\rho_D$ for vocabulary, meaning, and resolution, a
normalization map $N_D:\mathrm{Src}_D\to\mathrm{Src}_D$ of sources, and the
following four predicates.

\begin{xltabular}{\linewidth}{@{}LL@{}}
\toprule
Predicate & Meaning \\
\midrule
\endhead
$V_D(v,a)$ & The vocabulary $v$ admits the Atom $a$ \\
\addlinespace[3pt]
$M_D(\gamma,s,a)$ & The meaning reading $\gamma$ admits $a$ on the source $s$ \\
\addlinespace[3pt]
$R_D(\rho,s,a)$ & The resolution $\rho$ admits $a$ on $s$ \\
\addlinespace[3pt]
$E_D(s,a)$ & $a$ holds in the semantics of the source \\
\bottomrule
\end{xltabular}

After normalizing the source, we adopt the Atoms that satisfy all four conditions.
The extraction predicate and the extracted family are defined by
\[
\begin{aligned}
\mathrm{Extracts}_D(s,a)
&\Longleftrightarrow
V_D(v_D,a)\land M_D(\gamma_D,N_D(s),a)\\
&\hspace{2em}\land R_D(\rho_D,N_D(s),a)\land E_D(N_D(s),a),\\
\mathrm{Atomize}_D(s)
&=\{a\in\mathrm{At}\mid \mathrm{Extracts}_D(s,a)\}
\end{aligned}
\]
At this stage we use the structure of $N_D$ as a map; constructions that need
idempotence state it explicitly as an additional condition.

\end{definition}

\begin{proposition}[Existence and uniqueness of the extracted family]\label{prop:1.4}

Fix $D$ and $s\in\mathrm{Src}_D$. Then an Atom family $F$ satisfying
\[
\forall a\in\mathrm{At},\qquad
a\in F\ \Longleftrightarrow\ \mathrm{Extracts}_D(s,a)
\]
exists and is unique.

\end{proposition}

\begin{proof}

The family $\mathrm{Atomize}_D(s)$ of Definition~\ref{def:1.3} satisfies this
condition. If two families $F,G$ satisfy the condition, then
$a\in F\Longleftrightarrow a\in G$ for every Atom, and hence $F=G$ by
extensionality of sets.

\end{proof}

What this uniqueness fixes is the family of facts for a given source and doctrine.
A readout of that family is given by a separate map $o:F\to Y$.
If distinct Atoms satisfy $o(a)=o(b)$, this readout does not distinguish them.
The construction of the extracted family and the amount of information read out of
it can therefore be described separately.

\section{Configurations, objects, and operations}\label{sec:1.2}

An Atom family is the collection of adopted facts. To treat it as a single object,
we add the relations between the facts, together with the state sets and the
interpretation of operations. The relations are kept in a configuration, and the concrete structure
data in an architecture object.

\begin{definition}[Configuration]\label{def:1.5}

A configuration is a triple $C=(F_C,R_C,I_C)$, where $F_C\subseteq\mathrm{At}$
and $R_C,I_C\subseteq\mathrm{At}\times\mathrm{At}$.
Here $R_C$ expresses relations and $I_C$ specified identifications.
When the identifications are used as an equivalence relation, we require its
reflexivity, symmetry, and transitivity.
When $R_C,I_C\subseteq F_C\times F_C$, we say that the relations and
identifications are supported by the family.

A composition reading is a rule that assigns to each finite family $F$ a
configuration
\[
\mathrm{Comp}(F)=(F,R_F,I_F),
\qquad R_F,I_F\subseteq F\times F
\]
A configuration obtained by restricting the relations and identifications to a
finite subfamily is called a molecule. It is the finite unit for handling several
Atoms together with their relations when comparing local structure.

\end{definition}

\begin{definition}[Architecture object]\label{def:1.6}

An architecture object is a triple $A=(C_A,S_A,Q_A)$, where $C_A$ is a
configuration, $S_A$ is structure data of a chosen type, and $Q_A$ is a quantity
of a chosen type; the types themselves are included in the data of the object.
The structure data may hold state sets and transition maps, graphs, diagrams,
algebras, and the like. The quantity may hold specified values or results of
evaluation. The totality of such objects, in a universe of sufficient size, is
written $\mathrm{ArchObj}(\mathrm{At})$.

An object reading is a rule that sends a configuration $C$ to an object
$\mathrm{Form}(C)$ with $C_{\mathrm{Form}(C)}=C$.
Fixing this rule, from a finite family we obtain the construction
\[
F\longmapsto \mathrm{Comp}(F)\longmapsto
\mathrm{Form}(\mathrm{Comp}(F))
\]
One may also change the state sets or the interpretation of operations while
keeping the configuration the same, in which case the result is a different
architecture object.

\end{definition}

\begin{definition}[Morphisms of configurations and operations]\label{def:1.7}

An operation between objects is accompanied by an action on Atoms, relations, and
identifications. A morphism of configurations expresses this action.

A morphism of configurations $h:C\to D$ is a map
$h_{\mathrm{At}}:\mathrm{At}\to\mathrm{At}$ such that
\[
\begin{aligned}
a\in F_C&\Longrightarrow h_{\mathrm{At}}(a)\in F_D,\\
(a,b)\in R_C&\Longrightarrow
(h_{\mathrm{At}}(a),h_{\mathrm{At}}(b))\in R_D,\\
(a,b)\in I_C&\Longrightarrow
(h_{\mathrm{At}}(a),h_{\mathrm{At}}(b))\in I_D
\end{aligned}
\]
With identity maps and composition of maps, configurations form a category.
Indeed, the three implications all hold for the identity map and are closed under
composition.

An operation reading consists of a set $\mathrm{Op}(A,B)$ for each pair of
objects and maps
\[
d_{A,B}:\mathrm{Op}(A,B)\longrightarrow
\mathrm{Hom}(C_A,C_B)
\]
An element $o\in\mathrm{Op}(A,B)$ is called an operation from $A$ to $B$.
An operation keeps its own name and additional data, and $d_{A,B}$ gives its
action on configurations. Even when two operations have the same action, the
operations themselves are kept distinct.
If identity operations and composition are specified so that the unit and
associativity laws hold and $d$ preserves them, the objects and operations form a
category.

\end{definition}

\begin{example}[Named operations with the same action]\label{ex:1.8}

Let $e_1,e_2$ be two operations on the state set $\{0,1\}$, both acting as the
identity map. Taking the operation set to be $\{e_1,e_2\}$, one can distinguish a
condition specified for $e_1$ from a condition specified for $e_2$ even though
their execution results are equal.
In \S\ref{sec:1.10} we actually construct morphisms of configurations whose Atoms
include the operation names.

\end{example}

\begin{definition}[Invariants and signatures]\label{def:1.9}

Values and properties to be preserved by operations are specified as invariants,
and the tuple of values used for comparison is collected into a signature.

A functional invariant is a map $I:\mathrm{ArchObj}(\mathrm{At})\to V_I$, and a
predicate invariant is a predicate $P(A)$ on objects.
Preservation under an operation $o:A\to B$ means
$I(A)=I(B)$ and $P(A)\Rightarrow P(B)$, respectively.
For ordered quantities, the condition $I(B)\leq I(A)$ can also be specified
separately. An invariant reading chooses an index set and assigns to each index a
functional or predicate invariant.

A signature reading consists of a set of axes $\Lambda$, a value range
$V_\lambda$ for each axis, a subset $\Lambda_{\mathrm{sel}}$ of selected axes,
and values $q_\lambda(A)\in V_\lambda$ for each object.
The signature of an object is
$(q_\lambda(A))_{\lambda\in\Lambda_{\mathrm{sel}}}$.
Selecting finitely many axes yields a finite multi-axis representation.

\end{definition}

\section{Local contexts and Laws}\label{sec:1.3}

Instead of handling all the information of an object at once, we consider
extracting part of its structure and observed values. A context specifies at
which positions, along which axes, and what can be read. Expressing an equation
on a context as a residual, so that the equation holds when the value is zero, allows Laws to be
evaluated locally.

\begin{definition}[Contexts and restrictions]\label{def:1.10}

A context $W$ of an object $A$ carries three sets
$\mathrm{Supp}(W)$, $\mathrm{Ax}(W)$, and $\mathrm{Obs}(W)$, together with
\[
\mathrm{reads}_W\subseteq\mathrm{Supp}(W)\times\mathrm{At},
\quad
\mathrm{readAx}_W\subseteq\mathrm{Ax}(W),
\quad
\mathrm{readObs}_W\subseteq\mathrm{Obs}(W)
\]
An element of $\mathrm{Obs}(W)$ is called an observable.
If $\mathrm{reads}_W(s,a)$ holds, we require $a\in F_{C_A}$.
An element of the support is a position at which Atoms are read; its role differs
from that of the set of subjects in Definition~\ref{def:1.2}.
Any additional data needed are also included in the context.

A context reading chooses a small preorder of contexts.
The relation $W'\leq W$ expresses the direction in which data obtained on $W$ can
be restricted to $W'$.
For each $W'\leq W$ we specify maps
\[
\mathrm{Supp}(W')\longrightarrow\mathrm{Supp}(W),\qquad
\mathrm{Ax}(W')\longrightarrow\mathrm{Ax}(W),\qquad
\mathrm{Obs}(W)\longrightarrow\mathrm{Obs}(W')
\]
The support map preserves the reading of the same Atom, the axis map preserves
readable axes, and the restriction of observables sends readable observables to
readable ones.
The three maps are compatible with identities and composition in their respective directions.
We regard this preorder as a category and write $W'\to W$ for the morphism
corresponding to $W'\leq W$.
The context category is denoted $\mathcal{C}_A$.

\end{definition}

\begin{example}[Contexts given by subsets]\label{ex:1.11}

For a finite family $F$, take the subsets $U\subseteq F$ as contexts and the
inclusions as morphisms. Setting $\mathrm{Supp}(U)=U$, with each element read as
its own Atom, taking the axes to be a one-element set, and taking
$\mathrm{Obs}(U)$ to be the set $B^U$ of functions into a set $B$, the
restriction of observables is the restriction of functions.
The intersection of two contexts is given by $U\cap V$.

To handle equations, we choose a ring in which local values can be added and
subtracted. Taking the difference of the two sides of an equation expresses its
validity as the vanishing of a residual. Below we keep both the symbolic
coordinates that present the equations and the residuals evaluated on objects.

\end{example}

\begin{definition}[Architectural equation system]\label{def:1.12}

Fix a context category $\mathcal{C}$ and a set of architecture objects.
An equation system $E$ consists of the following data.

\begin{itemize}
\item A set $K_E$ of equation indices and a role for each index,
\[
\mathrm{role}_E:K_E\to\{\mathrm{required},\mathrm{optional},\mathrm{derived}\}.
\]
\item A presheaf of commutative rings
$O_E:\mathcal{C}^{\mathrm{op}}\to{\mathbf{CommRing}}$.
The restriction along a morphism $j:W'\to W$ is written
$\mathrm{res}_j:O_E(W)\to O_E(W')$.
\item A symbolic coordinate $\nu_{W,i,a}\in O_E(W)$ for each context, index, and
Atom.
\item A residual $\varepsilon_{W,A,i,a}\in O_E(W)$ depending in addition on an
object $A$.
\end{itemize}

By a presheaf we mean here that the rings and restriction homomorphisms satisfy
$\mathrm{res}_{\mathrm{id}}=\mathrm{id}$ and
$\mathrm{res}_{jk}=\mathrm{res}_k\mathrm{res}_j$.
On the two families of coordinates we impose
\begin{equation*}
\mathrm{res}_j(\nu_{W,i,a})=\nu_{W',i,a},
\qquad
\mathrm{res}_j(\varepsilon_{W,A,i,a})=\varepsilon_{W',A,i,a}
\tag{1.1}\label{eq:1.1}
\end{equation*}
The $\nu$ are coordinates for presenting the equations algebraically, and the
$\varepsilon$ are residuals evaluating whether the equations hold on an object.
The Law $L_i$ is the name for the equation with index $i$ together with its
meaning.

\end{definition}

\begin{definition}[Lawfulness]\label{def:1.13}

The validity of an equation and the condition of satisfying all required
equations are defined by
\[
\begin{aligned}
E_i(A)
&\Longleftrightarrow
\forall W\in\mathrm{Ob}(\mathcal{C})\ \forall a\in\mathrm{At},\
\varepsilon_{W,A,i,a}=0,\\
\mathrm{Lawful}_E(A)
&\Longleftrightarrow
\forall i\in K_E^{\mathrm{req}},\ E_i(A),\\
K_E^{\mathrm{req}}
&=\{i\in K_E\mid\mathrm{role}_E(i)=\mathrm{required}\}
\end{aligned}
\]
That is, lawfulness means that for each required equation the residual vanishes
on every context and Atom. Validity at all indices is written
$\mathrm{FullyLawful}_E(A)$.
The roles required, optional, and derived designate an equation as mandatory, as
optional, and as one to be derived, respectively.
To conclude the validity of a derived equation from the other equations, one
proves the implication.

\end{definition}

\section{Finite detection and generation of the core}\label{sec:1.4}

The failure of an equation can sometimes be detected by inspecting a small number
of Atoms and relations. To describe this finite detection, we define queries
about structure and their combinations. We then collect objects, operations,
equations, and detection methods, and construct as the core the totality of
objects reachable by operations.

\begin{definition}[Signed finite queries]\label{def:1.14}

A query on a configuration is one of the following: membership of an Atom $a$,
validity of a relation $R(a,b)$, or validity of an identification $I(a,b)$.
Queries for relations and identifications also require that the Atoms at both
ends belong to the family.
We write $\mathrm{Holds}_A(q)$ when a query $q$ holds on an object $A$.

A signed finite query sequence is $Q=((q_1,b_1),\ldots,(q_n,b_n))$ with
$b_j\in\{0,1\}$, and we define
\[
\mathrm{Matches}(Q,A)
\Longleftrightarrow
\forall j,\quad \mathrm{Holds}_A(q_j)\Longleftrightarrow b_j=1
\]
The sign $0$ allows the absence of an Atom or a relation to be expressed as a
finite condition as well.
For example, attaching the signs $1,1,0$ to the three queries ``$a$ belongs'',
``$b$ belongs'', and ``$R(a,b)$ holds'' expresses the pattern in which the two
Atoms exist but the specified relation between them does not.

\end{definition}

\begin{definition}[Finite detectors and circuits]\label{def:1.15}

A detector code is obtained by finitely many applications of the constructors
$\mathrm{reject}$, which expresses the empty selection, $\mathrm{exact}(Q)$,
which specifies a single sequence, and $\mathrm{any}(c,d)$, the disjunction of
two codes.
The sequences it accepts are, respectively, the empty set, $\{Q\}$, and the union
of the accepted sets.
To each $i\in K_E$ we assign a code $c_i$ and write $T_i$ for its finite
accepted set.

The set of circuits on an object $A$ is
\[
\mathrm{Circ}_E(A,i)=\{Q\in T_i\mid\mathrm{Matches}(Q,A)\}
\]
That is, a circuit is a finite pattern, among those specified by the code, that actually holds
on the object.
To use circuits as evidence for the failure of an equation, we require the
soundness below.
The condition that the failure of every required equation can be detected is
called completeness.
Each condition is required for all objects $A$, indices $i$, and query
sequences $Q$.
\[
\begin{aligned}
Q\in\mathrm{Circ}_E(A,i)&\Longrightarrow \neg E_i(A),\\
i\in K_E^{\mathrm{req}}\ \land\ \neg E_i(A)
&\Longrightarrow\mathrm{Circ}_E(A,i)\ne\varnothing.
\end{aligned}
\]
Soundness states that the failure of the equation follows from an accepted
structural pattern.
Completeness states that a failure of a required equation always has such a
finite pattern.
Both are conditions to be proved for the specified equation system and detector.

\end{definition}

\begin{proposition}[Deciding lawfulness by circuits]\label{prop:1.16}

If the detector is sound, all required circuits of a lawful object are empty.
If it is moreover complete for the required equations, then
\[
\mathrm{Lawful}_E(A)
\Longleftrightarrow
\forall i\in K_E^{\mathrm{req}},\
\mathrm{Circ}_E(A,i)=\varnothing.
\]

\end{proposition}

\begin{proof}

If a lawful object had a required circuit, soundness would make the
corresponding required equation fail, a contradiction.
Conversely, if some required equation fails, completeness provides a
corresponding circuit.

\end{proof}

The same applies when the presence or absence of failures is expressed by quantities
$\omega_i(A)$. Given $E_i(A)\Longleftrightarrow\omega_i(A)=0$ and an aggregation
with $\mathrm{Agg}((v_i)_i)=0\Longleftrightarrow\forall i,\ v_i=0$, the vanishing
of the aggregation over the required indices is equivalent to lawfulness.
This equivalence follows by composing the two equivalences.

\begin{definition}[Core reading]\label{def:1.17}

A core reading $r$ includes the following choices.

\begin{itemize}
\item \textbf{Extraction}: choose a doctrine and a source $(D,s)$, and require
the extracted family $F_r=\mathrm{Atomize}_D(s)$ to be finite.
\item \textbf{Object formation}: choose a composition reading and an object
reading.
\item \textbf{Operations and values for comparison}: choose an operation
reading, an invariant reading, and a signature reading.
\end{itemize}

From these we define the initial configuration and object
\[
C_r=\mathrm{Comp}_r(F_r),\qquad A_r=\mathrm{Form}_r(C_r)
\]
We further choose the context category $\mathcal{C}_r$ of the base object $A_r$,
an equation system $E_r$ on it, and a sound detector.
The residuals are required to be evaluable on every object of
$\mathrm{ArchObj}(\mathrm{At})$, and the context category chosen from the base
object is used in common for evaluating the generated family of objects.

A subset $Z\subseteq\mathrm{ArchObj}(\mathrm{At})$ of objects is closed under
operations if $A\in Z$ and $o\in\mathrm{Op}_r(A,B)$ imply $B\in Z$.
Let $\mathrm{Obj}_r$ be the intersection of all such subsets containing the base
object.
The core $\mathrm{Core}(r)$ is the combined data of this family of objects, the
base object, the context category, the equation system, the family of circuits,
the family of operations, and the invariants and signature.

\end{definition}

\begin{theorem}[Generation of the relative core]\label{thm:1.18}

From a core reading $r$ the core $\mathrm{Core}(r)$ is constructed.
The family of its initial configuration equals $F_r$, and the configuration of
the base object equals $C_r$.
Moreover, $\mathrm{Obj}_r$ is the smallest family of objects containing the base
object and closed under operations, and it is also characterized by the
following finite reachability.
\begin{equation*}
B\in\mathrm{Obj}_r
\Longleftrightarrow
\exists n\in\mathbb{N},\
A_r=A_0\xrightarrow{o_1}A_1\xrightarrow{o_2}
\cdots\xrightarrow{o_n}A_n=B.
\tag{1.2}\label{eq:1.2}
\end{equation*}
Sequences of length $0$ are allowed.

\end{theorem}

\begin{proof}

The extracted family is obtained by Proposition~\ref{prop:1.4}, and the
finiteness condition of the reading allows it to be fed into the composition.
The conditions on the composition and object readings yield the two equalities
of configuration components.
Since the set of all objects contains the base object and is closed under
operations, the family over which the intersection of
Definition~\ref{def:1.17} is taken is nonempty.

The intersection contains the base object. If $A$ belongs to the intersection
and there is an operation $o:A\to B$, then $B$ belongs to every subset defining
the intersection, so the intersection is also closed under operations.
Minimality follows from the definition of the intersection.

Let $Z_{\mathrm{fin}}$ be the set of objects reachable from the base object by
finite sequences.
By the sequences of length $0$ and the appending of an operation to the end of a
sequence, $Z_{\mathrm{fin}}$ contains the base object and is closed under
operations.
Hence $\mathrm{Obj}_r\subseteq Z_{\mathrm{fin}}$.
The reverse inclusion is obtained by induction on the length of the sequence.
Finally, restricting the specified equations, detector, invariants, and
signature to this family of objects and taking the operations for each pair of
objects gives all the components of the core.

\end{proof}

The finiteness in \eqref{eq:1.2} is a condition on the length of each witness of
reachability.
The size of the whole family of objects closed under operations is determined by
the specified operation reading.

\section{From covers to sites}\label{sec:1.5}

To read structure part by part, one needs a family of contexts covering the
whole and a place where the information read on two parts can be compared.
For the subsets of Example~\ref{ex:1.11}, the latter is the intersection.
For general contexts we choose this overlap as a pullback, and take as the
starting point of covers the families that can jointly read the required
information among Atoms, equations, and signatures.

\begin{definition}[Overlaps and coverage requirements]\label{def:1.19}

For each diagram $U\to W\leftarrow V$ of two morphisms in the context category
$\mathcal{C}$, we choose an object $U\times_W V$ and projections $\pi_U,\pi_V$.
The composites of the two projections with $U\to W$ and $V\to W$, respectively,
are required to be equal.
Moreover, any $T\to U$ and $T\to V$ satisfying this commutativity are obtained
by composing a unique $T\to U\times_W V$ with the projections.
This pullback is called the overlap, and its choice is written $\mathrm{Ov}$.
When the preorder of contexts is a partial order, the overlap is the greatest
lower bound of the two contexts below $W$.

Coverage requirements $\mathcal{R}$, with an equation system $E$ and a signature
fixed, choose the following.

\begin{itemize}
\item A subset $A_{\mathrm{req}}\subseteq\mathrm{At}$ of required Atoms.
\item Required equation coordinates
$C_{\mathrm{req}}\subseteq K_E^{\mathrm{req}}\times\mathrm{At}$.
\item A subset $C_{\mathrm{wit}}\subseteq K_E\times\mathrm{At}$ of indices of
the symbolic coordinates $\nu_{W,i,a}$ to be read as witnesses (evidence of
violation).
\item Required signature axes $\Lambda_{\mathrm{req}}\subseteq\Lambda$.
\item Four visibility predicates expressing that these can be read on each
context, and a predicate $B_{\mathcal{R}}(U,W)$ expressing that the required
interaction can be read on the overlap.
\end{itemize}

The selected coordinates and witness indices are kept along the specified
restrictions.
For example, the restriction \eqref{eq:1.1} restricts the value while keeping
the index $(i,a)$.
Visibility is a separate condition for a context to actually read the data of
that index.

A family of morphisms $\mathcal{U}=(u_\alpha:W_\alpha\to W)_{\alpha\in I}$ is
$(E,\mathcal{R},\mathrm{Ov})$-admissible if it satisfies the following
conditions.

\begin{itemize}
\item \textbf{Information read on the parts}: each required Atom and signature
axis is readable on at least one $W_\alpha$.
\item \textbf{Information read on parts or overlaps}: each required equation
coordinate and selected witness is readable on at least one $W_\alpha$ or
$W_{\alpha\beta}=W_\alpha\times_W W_\beta$.
\item \textbf{Interaction on overlaps}: $B_{\mathcal{R}}(W_{\alpha\beta},W)$
holds for all $\alpha,\beta$.
\item \textbf{Correspondence with the object}: the Atoms read by the support map
of each leg belong to the family of the underlying object.
\end{itemize}

A system of covers is required to be closed under restriction to another context
and under further covering of each part.
We define this system using the following sieves.

\end{definition}

\begin{definition}[Sieves and Grothendieck topologies]\label{def:1.20}

A sieve $S$ on $W$ is a collection of morphisms with codomain $W$ such that if
$f:V\to W$ belongs to it, then so does $fg$ for every $g:V'\to V$.
The pullback along a morphism $u:W'\to W$ is
$u^*S=\{g:V\to W'\mid ug\in S\}$.
The sieve $\langle \mathcal{U}\rangle$ generated by a family $\mathcal{U}$
consists of all morphisms that factor through at least one $u_\alpha$.

A Grothendieck topology $J$ assigns to each $W$ a set $J(W)$ of covering sieves
such that the following hold.

\begin{enumerate}
\item The maximal sieve consisting of all morphisms with codomain $W$ belongs
to $J(W)$.
\item If $S\in J(W)$, then $u^*S\in J(W')$.
\item If $S\in J(W)$ and another sieve $T$ satisfies $u^*T\in J(W')$ for every
$u:W'\to W$ with $u\in S$, then $T\in J(W)$.
\end{enumerate}

We use the definition of
\cite[\href{https://stacks.math.columbia.edu/tag/00YW}{\S7.47, Tag 00YW}]{Stacks}.

\end{definition}

\begin{proposition}[Generation of the topology from coverage requirements]\label{prop:1.21}

There exists a smallest Grothendieck topology containing the sieves generated by
all admissible families.
We write it $J_{E,\mathcal{R},\mathrm{Ov}}$.

\end{proposition}

\begin{proof}

The topology in which every sieve on every object is a cover contains all the
specified sieves.
Take, objectwise, the intersection of all topologies containing them.
The maximal-sieve condition and stability under pullback hold in each topology
and hence in the intersection.
As for transitivity, if its hypothesis holds in the intersection, it holds in
each topology, and the concluding sieve belongs to all of them.
The intersection is therefore a topology, and it is smallest by definition.

\end{proof}

A site is a pair of a small category and a Grothendieck topology.
The pair $(\mathcal{C}_r,J_{E_r,\mathcal{R},\mathrm{Ov}})$ obtained from the
base object $A_r$ is called the AAT site.
The choices of the equation system, the signature, the coverage requirements,
and the overlaps are also kept as its construction data.
A covering family of the generated topology that also satisfies the specified
visibility and interaction requirements is called
$(E,\mathcal{R},\mathrm{Ov})$-adequate.
The topology gives the rules for local gluing, and adequacy expresses that the
information used in the theorems is available on the cover.
It is used in Chapter~\ref{chap:2} to connect the local validity of equations
with the consistency of gluing.

\section{Presheaves, sheaves, and geometries with coefficients}\label{sec:1.6}

A presheaf assigns data to each context and has rules for restricting them to
smaller contexts.
A sheaf satisfies in addition the condition that from local data agreeing on the
overlaps, the data on the whole are obtained uniquely.

\begin{definition}[Presheaves and sheaves]\label{def:1.22}

A set-valued presheaf is a contravariant functor
$F:\mathcal{C}^{\mathrm{op}}\to{\mathbf{Set}}$.
An element of $F(W)$ is called a section over $W$.
A matching family on a family $\mathcal{U}=(u_\alpha:W_\alpha\to W)_\alpha$ is a
family of elements $s_\alpha\in F(W_\alpha)$ satisfying
\begin{equation*}
F(\pi_\alpha)(s_\alpha)=F(\pi_\beta)(s_\beta)
\quad\text{in }F(W_{\alpha\beta})
\tag{1.3}\label{eq:1.3}
\end{equation*}
for all $\alpha,\beta$.
The set of matching families is written $\mathrm{Match}(\mathcal{U},F)$.
A presheaf $F$ is a sheaf for $J$ if, for every family with
$\langle \mathcal{U}\rangle\in J(W)$, the map
\begin{equation*}
F(W)\longrightarrow\mathrm{Match}(\mathcal{U},F),
\qquad s\longmapsto(F(u_\alpha)(s))_\alpha
\tag{1.4}\label{eq:1.4}
\end{equation*}
is a bijection. Injectivity is called separatedness, and surjectivity the
existence of gluings.
A morphism of sheaves is a natural transformation of presheaves, and the
category of sheaves is written $\mathrm{Sh}(\mathcal{C},J)$.

A matching family on a sieve $S$ assigns to each $u:V\to W$ with $u\in S$ an
element $s_u\in F(V)$ satisfying $s_{uv}=F(v)(s_u)$.
A matching family on a covering family becomes, by restriction to the morphisms
factoring through the $u_\alpha$, a matching family on the generated sieve.
The values obtained from two factorizations are equal by the universal property
of the pullback and \eqref{eq:1.3}.
Conversely, reading a family on the sieve at each $u_\alpha$ yields
\eqref{eq:1.3}.
The definition therefore also agrees with the definition of sheaves by covering
sieves.

As an example in which the sheaf condition is needed, take the subsets of the
two-point set $\{p,q\}$ as contexts and the families covering by unions as
covers. Consider the presheaf $F$ that takes a one-element set on the empty
context and $\{0,1\}$ on the nonempty contexts, with the identity map as the
restriction between nonempty contexts.
The values $s_p=0$ and $s_q=1$ on the one-point contexts agree on the overlap,
but they glue to neither element over the two points.
Sheafification is the operation that fills such gaps by adding compatible local
data as new sections.

\end{definition}

\begin{proposition}[Sheafification]\label{prop:1.23}

For every presheaf $F$ there exist a sheaf $a_JF$ and a natural transformation
$\eta_F:F\to a_JF$. For every sheaf $G$ and natural transformation
$t:F\to G$ there is a unique $\bar t:a_JF\to G$ with $t=\bar t\eta_F$.

\end{proposition}

\begin{proof}[Construction and proof]

We treat a matching family on a cover as a single section and identify
presentations that agree locally.
Performing this operation twice yields a sheaf equipped with all the required
gluings.

An element of $F^+(W)$ is presented by a matching family on a covering sieve
$S\in J(W)$.
Two presentations are equivalent when they agree on a covering sieve contained
in both.
A finite intersection of covering sieves is again a cover; this follows by
applying the transitivity of Definition~\ref{def:1.20} to the pullbacks along
the morphisms of one of the sieves.
The equivalence relation is therefore transitive.
Pullback defines the restrictions of $F^+$, and the families of restrictions of
global sections give $F\to F^+$.

If two elements of $F^+$ are equal on a cover, we can collect the local covering
sieves expressing this equality. On the cover obtained by the transitivity of
the topology the presentations agree, so the two elements are equal.
Hence $F^+$ is separated.
When $F$ itself is separated, each element of a matching family of $F^+$ is
presented by local sections of $F$. Equality on the overlaps holds after further
localization, and by the separatedness of $F$ it holds on the overlaps themselves.
The family on a covering sieve collecting these presentations gives the gluing.
Thus $F^+$ is a sheaf whenever $F$ is separated.

We therefore set $a_JF=(F^+)^+$.
A transformation $t:F\to G$ sends each presentation to a matching family of $G$
and, by the unique gluing in $G$, determines $F^+\to G$.
Performing the same operation twice gives $\bar t$.
Each element is determined by a local presentation, and since $G$ is separated,
the extension is unique.

\end{proof}

This is the standard sheafification
(\cite[\href{https://stacks.math.columbia.edu/tag/00ZG}{\S7.49, Tag 00ZG}]{Stacks}).
In particular, if the original $F$ is a sheaf, then $\eta_F$ is an isomorphism
by the universal property.

\begin{example}[Sheafification adding local sections]\label{ex:1.24}

The sheafification of the presheaf $F$ on the two-point set above is
$U\mapsto\{0,1\}^U$.
A family taking a different value at each point is also retained, as a single
function.
Indeed, a function is determined uniquely by its restrictions to the single
points, and arbitrary values at the single points can be glued.
From the original presheaf, each value is sent to the constant function.
A natural transformation to any sheaf is determined by its values on the
one-point contexts, and gluing these extends it uniquely from
$U\mapsto\{0,1\}^U$.
The universal property of Proposition~\ref{prop:1.23} is therefore satisfied as
well.

To treat equations algebraically, we give the local data the structure of a
ring.
Taking the coordinates of each context as variables and imposing the structural
relations as polynomial equations produces the ring used on that context.
Specifying restrictions between contexts then yields a ring-valued presheaf.

\end{example}

\begin{definition}[Rings for each context and restrictions]\label{def:1.25}

Choose a coefficient ring $k$. To each context $W$ we assign a set of
coordinates $Z_W$, the kind of each coordinate and the type of its local data,
an index set $H_W$ of structural relations, and polynomials
$r_{W,h}\in k[x_z\mid z\in Z_W]$.
Each element of the polynomial ring is a finite sum in finitely many variables.
The ideal generated by the structural relations and the quotient are
\[
I_W^{\mathrm{str}}=(r_{W,h}\mid h\in H_W),
\qquad
B_{\mathrm{raw}}(W)=k[x_z\mid z\in Z_W]/I_W^{\mathrm{str}}
\]
To each morphism $j:W'\to W$ we assign a $k$-algebra homomorphism
$\rho_j:k[x_z\mid z\in Z_W]\to k[x_z\mid z\in Z_{W'}]$ sending variables to
polynomials, and require
\begin{equation*}
\rho_j(I_W^{\mathrm{str}})\subseteq I_{W'}^{\mathrm{str}},
\qquad
\rho_{\mathrm{id}}=\mathrm{id},
\qquad
\rho_{jk}=\rho_k\rho_j
\tag{1.5}\label{eq:1.5}
\end{equation*}
The coordinates, the relations, and these restrictions of polynomials together
are called a raw restriction system $\mathcal{B}$.
By \eqref{eq:1.5} the restrictions descend to the quotients, and
$B_{\mathrm{raw}}$ becomes a presheaf of commutative $k$-algebras.
This claim follows because the difference of two representatives of the same
residue class stays in the structural ideal after restriction, and because
identities and composition hold at the level of polynomials.

An architecture geometry with coefficients is
\[
G=(r,\mathcal{R},\mathrm{Ov},k,\mathcal{B})
\]
The base object, the contexts, the equations, and the signature come from $r$,
and the topology from Proposition~\ref{prop:1.21}.
The structural relations determine, on each context, the ring representing the
local data.
Chapter~\ref{chap:2} gives a realization connecting this presheaf of rings with
the equation system, and makes explicit the condition under which the
evaluation of the symbolic coordinates corresponds to the residuals of each
object.

\end{definition}

\begin{lemma}[Change of coefficients for raw systems]\label{lem:1.26}

A ring homomorphism $\varphi:k\to k'$ yields a raw system
$\varphi_!\mathcal{B}$ by applying $\varphi$ to the coefficients of all
polynomials while keeping the indices of coordinates and relations.
This construction respects identity homomorphisms and composition of
homomorphisms.

\end{lemma}

\begin{proof}

Send each variable $x_z$ to the same variable and each coefficient $c$ to
$\varphi(c)$.
The new structural ideal is the ideal generated by the images of the structural
relations.
That the original restrictions preserve the structural ideal is expressed by
equations presenting the image of each generating relation as a finite ideal
combination, and these equations remain valid after applying the coefficient
map.
The same applies to the identity and composition equations of \eqref{eq:1.5}.
Successive application of coefficient maps equals application of their
composite, which gives the last claim.

\end{proof}

\section{Morphisms of readings and the three-level projections}\label{sec:1.7}

Changing the reading changes the extracted Atoms, the way objects are formed,
and the choice of equations and locality.
To compare the situations before and after a change, one needs maps connecting
the respective data and the conditions they satisfy.
We define three categories according to the range of data compared.

\begin{xltabular}{\linewidth}{@{}LLL@{}}
\toprule
Category & Data carried by an object & What a morphism compares \\
\midrule
\endhead
$B$ & Extraction doctrine and source & Sources and Atoms \\
\addlinespace[3pt]
$E_{\mathrm{core}}$ & Core reading & In addition to extraction: object formation, operations, equations, detection, invariants, and signatures \\
\addlinespace[3pt]
$E_{\mathrm{geom}}$ & Geometry with coefficients & In addition to the core: covers, overlaps, coefficients, presentations of rings, and the data of contexts \\
\bottomrule
\end{xltabular}

We require the validity of extraction and the validity of equations to be
equivalent before and after the change.
A change equipped with this preservation and reflection is called an exact
change.

\begin{definition}[The category of extraction bases]\label{def:1.27}

Fix the vocabulary of Atoms. An object of the category $B$ is a pair $(D,s)$ of
an extraction doctrine and a selected source. A morphism $(D,s)\to(D',s')$ is a
map $f:\mathrm{Src}_D\to\mathrm{Src}_{D'}$ together with a bijection
$e:\mathrm{At}\xrightarrow{\sim}\mathrm{At}$ satisfying
\begin{equation*}
\begin{aligned}
f(s)&=s',&
N_{D'}f&=fN_D,\\
\mathrm{Extracts}_D(t,a)
&\Longleftrightarrow
\mathrm{Extracts}_{D'}(f(t),e(a))
&&\text{for all }t,a
\end{aligned}
\tag{1.6}\label{eq:1.6}
\end{equation*}
The identity morphism consists of the two identity maps, and composition is
$(f',e')(f,e)=(f'f,e'e)$.
Composing the equalities and equivalences of \eqref{eq:1.6} shows that the same
conditions hold after composition.

The direct image of a family is written $e_*F=\{e(a)\mid a\in F\}$.
From \eqref{eq:1.6} and the bijectivity of $e$ we obtain
\begin{equation*}
\mathrm{Atomize}_{D'}(f(t))=e_*\mathrm{Atomize}_D(t)
\tag{1.7}\label{eq:1.7}
\end{equation*}
Indeed, an element of the right-hand side belongs to the left-hand side by
preservation of extraction.
For an element $b$ of the left-hand side, take $a=e^{-1}(b)$ and use reflection
of extraction.
The source map $f$ is a general map; for example, it may send distinct states
to a single one.

The direct image of a configuration is defined by letting $e$ act on the family
and on both relations.
On finite query sequences and detector codes, $e$ acts on all the Atoms that
occur. These are also written $e_*$.

In the following conditions, the maps are chosen so that, even when the
presentation of objects changes, the validity or failure of the Laws and the
finite patterns witnessing failures correspond.

\end{definition}

\begin{definition}[Morphisms of cores]\label{def:1.28}

The objects of $E_{\mathrm{core}}$ are the core readings with a fixed vocabulary
of Atoms.
A morphism from $r$ to $r'$ consists of a base morphism $(f,e)$ and the
following data and conditions.

\textbf{Object formation and operations.}
We require that the route mapping the Atoms first and then forming the object
agree with the route forming the object first and then mapping it.
We give a map $H:\mathrm{ArchObj}(\mathrm{At})\to\mathrm{ArchObj}(\mathrm{At})$
and require, for every finite family $F$, configuration $C$, and object $A$,
\begin{equation*}
\begin{aligned}
\mathrm{Comp}_{r'}(e_*F)&=e_*\mathrm{Comp}_r(F),\\
H(\mathrm{Form}_r(C))&=\mathrm{Form}_{r'}(e_*C),\\
C_{H(A)}&=e_*C_A
\end{aligned}
\tag{1.8}\label{eq:1.8}
\end{equation*}
On the operation maps
$\Phi_{A,B}:\mathrm{Op}_r(A,B)\to\mathrm{Op}_{r'}(H(A),H(B))$ we impose, for
the action $d(o)$ on configurations,
\begin{equation*}
d(\Phi(o))_{\mathrm{At}}\,e=e\,d(o)_{\mathrm{At}}
\tag{1.9}\label{eq:1.9}
\end{equation*}
Since $e$ itself gives a morphism of configurations $C_A\to C_{H(A)}$,
\eqref{eq:1.9} is also a commutative square of configurations.

\textbf{Equations and detection.}
We match up the contexts, the equation indices, and the rings in which values
are taken, and preserve the coordinates, residuals, and detector codes under
this correspondence.
We give an equivalence of categories
$\theta:\mathcal{C}_r\simeq\mathcal{C}_{r'}$ with a specified quasi-inverse, a
bijection of indices $\lambda:K_{E_r}\xrightarrow{\sim}K_{E_{r'}}$, and ring
isomorphisms $\alpha_W:O_{E_r}(W)\xrightarrow{\sim}O_{E_{r'}}(\theta W)$ for
each context. These are required to satisfy
\begin{equation*}
\begin{aligned}
\mathrm{role}_{E_{r'}}(\lambda i)&=\mathrm{role}_{E_r}(i),\\
\alpha_{W'}\mathrm{res}_j&=\mathrm{res}'_{\theta j}\alpha_W,\\
\alpha_W(\nu_{W,i,a})&=\nu'_{\theta W,\lambda i,e(a)},\\
\alpha_W(\varepsilon_{W,A,i,a})
&=\varepsilon'_{\theta W,H(A),\lambda i,e(a)},\\
c'_{\lambda i}&=e_*c_i
\end{aligned}
\tag{1.10}\label{eq:1.10}
\end{equation*}
The last equality is an equality of detector codes as syntax.

\textbf{Invariants and signatures.}
We give an index map $\mu$ preserving the kind of invariant.
For corresponding functional invariants, we require a bijection $t_i$ of the
value ranges with $t_i(I_i(A))=I'_{\mu i}(H(A))$ for every object $A$.
For predicate invariants we require
$P_i(A)\Longleftrightarrow P'_{\mu i}(H(A))$.
For the signatures we give a map of axes $\sigma:\Lambda_r\to\Lambda_{r'}$ and
bijections $\beta_\ell:V_\ell\xrightarrow{\sim}V'_{\sigma\ell}$ of the value
ranges, and impose
\begin{equation*}
\ell\in\Lambda_{\mathrm{sel}}
\Longleftrightarrow \sigma\ell\in\Lambda'_{\mathrm{sel}},
\qquad
\beta_\ell(q_\ell(A))=q'_{\sigma\ell}(H(A))
\tag{1.11}\label{eq:1.11}
\end{equation*}
A morphism retains all of these constituent maps.
Even morphisms with the same action on configurations are distinct if their
maps of objects, operations, axes, and so on differ.

\end{definition}

\begin{proposition}[Action of exact core changes]\label{prop:1.29}

A morphism as in Definition~\ref{def:1.28} sends $A_r$ to $A_{r'}$ and
satisfies $H(\mathrm{Obj}_r)\subseteq\mathrm{Obj}_{r'}$.
For every object $A$ and equation index $i$,
\begin{equation*}
E_{r,i}(A)\Longleftrightarrow E_{r',\lambda i}(H(A)),
\qquad
\mathrm{Lawful}_{E_r}(A)\Longleftrightarrow
\mathrm{Lawful}_{E_{r'}}(H(A)).
\tag{1.12}\label{eq:1.12}
\end{equation*}
Moreover, $Q\mapsto e_*Q$ gives a bijection between
$\mathrm{Circ}_{E_r}(A,i)$ and $\mathrm{Circ}_{E_{r'}}(H(A),\lambda i)$.

\end{proposition}

\begin{proof}

Using \eqref{eq:1.7} and \eqref{eq:1.8} in turn gives the equality of base objects.
Letting $H,\Phi$ act on a finite sequence of operations from the base
object and using \eqref{eq:1.2} yields the inclusion of generated object
families.

By the residual comparison in \eqref{eq:1.10} and the preservation and
reflection of zero by the ring isomorphisms, the vanishing of a residual on $W$
is equivalent to its vanishing on $\theta W$.
Since $e$ is a bijection, the quantification over all Atoms corresponds as
well.
Every context $V$ after the change is isomorphic to some $\theta W$, and the
restriction along this isomorphism is a ring isomorphism.
Applying \eqref{eq:1.1} to this isomorphism gives the same equivalence for the
residuals on $V$.
Using the bijection of indices and the preservation of roles yields the
equivalence of lawfulness.

By \eqref{eq:1.8}, membership of Atoms and validity of relations and
identifications are preserved and reflected along $e$.
Hence $\mathrm{Matches}(Q,A)\Longleftrightarrow\mathrm{Matches}(e_*Q,H(A))$ for
both truth signs.
From the equality of codes, the accepted sets satisfy $T'_{\lambda i}=e_*T_i$.
Together with the inverse map of query sequences given by $e^{-1}$, we obtain
the bijection of circuits.

\end{proof}

\begin{definition}[Morphisms of geometries]\label{def:1.30}

A morphism from a geometry $G=(r,\mathcal{R},\mathrm{Ov},k,\mathcal{B})$ to
$G'=(r',\mathcal{R}',\mathrm{Ov}',k',\mathcal{B}')$ consists of a morphism of
cores $h:r\to r'$ and the following data and conditions.
The specified quasi-inverse of $\theta$ is written $\bar\theta$.

\textbf{1. Preservation of coverage requirements.} Each requirement is
preserved forward, using $e$ for the required Atoms,
$(i,a)\mapsto(\lambda i,e(a))$ for the equation coordinates and witnesses, and
$\sigma$ for the signature axes.
The four visibility predicates are also preserved along $W\mapsto\theta W$ and
these maps, and
$B_{\mathcal{R}}(U,W)\Rightarrow B_{\mathcal{R}'}(\theta U,\theta W)$ holds.

\textbf{2. Comparison of overlaps.} For each diagram $U\to W\leftarrow V$ in
the context category after the change, we specify an isomorphism between the
carried overlap chosen before the change and the overlap chosen after the
change; that is,
$\theta(\bar\theta U\times_{\bar\theta W}\bar\theta V)\cong U\times_W V$.
The comparison with the projections uses the counit of the equivalence.

\textbf{3. Change of coefficients.} We specify a ring homomorphism
$\varphi:k\to k'$.

\textbf{4. Agreement of ring presentations.} The raw system after the change
equals the raw system before the change with its coefficients changed by
$\varphi$ and reindexed along $\bar\theta$:
\begin{equation*}
\mathcal{B}'=\bar\theta^{\,*}(\varphi_!\mathcal{B}).
\tag{1.13}\label{eq:1.13}
\end{equation*}
The right-hand side places on a context $V$ after the change the coordinates,
kinds, local data types, and relation indices of $\bar\theta V$, with the
coefficients of the polynomials changed.
The restrictions are built from the restrictions along the morphisms sent by
$\bar\theta$.
The equality covers all of these presentation data.

\textbf{5. Comparison of the data contained in contexts.} To each $W$ we
assign comparison maps for supports, axes, and observables
\[
s_W:\mathrm{Supp}(W)\to\mathrm{Supp}'(\theta W),\quad
a_W:\mathrm{Ax}(W)\to\mathrm{Ax}'(\theta W),\quad
o_W:\mathrm{Obs}(W)\to\mathrm{Obs}'(\theta W)
\]
The readings of the support are preserved with the Atoms carried by $e$, and
the readable elements of axes and observables are preserved as well.
Writing $S_j,A_j$ for the covariant structure maps of supports and axes and
$O_j$ for the contravariant restriction of observables along $j:W'\to W$, we
require
\begin{equation*}
S'_{\theta j}s_{W'}=s_WS_j,\qquad
A'_{\theta j}a_{W'}=a_WA_j,\qquad
O'_{\theta j}o_W=o_{W'}O_j
\tag{1.14}\label{eq:1.14}
\end{equation*}
Since the context category is a preorder category, any two parallel morphisms
that exist are equal.
Hence the commutativity with the projections in item~2 and the unit and
composition coherence of the overlap comparisons are determined uniquely.
The comparison maps for supports and the like are retained as actual maps and
satisfy the conditions of item~5 separately.

\end{definition}

\begin{theorem}[The three-level categories and projections]\label{thm:1.31}

The objects and morphisms of Definitions~\ref{def:1.27}, 1.28, and 1.30 form
categories $B$, $E_{\mathrm{core}}$, and $E_{\mathrm{geom}}$, respectively,
with functors
\begin{equation*}
E_{\mathrm{geom}}
\xrightarrow{\,p\,}
E_{\mathrm{core}}
\xrightarrow{\,q\,}
B
\tag{1.15}\label{eq:1.15}
\end{equation*}
The functor $p$ forgets the choice of geometry and its comparisons and extracts
the core, and $q$ extracts the extraction doctrine, the selected source, and
their maps.

\end{theorem}

\begin{proof}

For $B$ this was checked in Definition~\ref{def:1.27}.
Morphisms of cores are composed by composing the maps of sources, Atoms,
objects, indices, and operations in turn.
For the equivalences of context categories we use the composite equivalence,
and for the ring isomorphisms of each context the ring isomorphisms applied in
turn.
Each of \eqref{eq:1.8}--\eqref{eq:1.11} holds after composition by substituting
the two corresponding equalities.
The bijections of value ranges of functional invariants compose, and so do the
equivalences for predicate invariants.
For identity morphisms we use the respective identity maps.
The unit and associativity laws follow from the composition laws of maps and
the uniqueness of comparison morphisms in a preorder category.

The coefficient maps of morphisms of geometries are composed as ring
homomorphisms.
The comparisons of supports and the like are composed, for example, as
$s''_W=s'_{\theta W}s_W$.
The visibility implications and \eqref{eq:1.14} are closed under this
composition.
For raw systems, reindexing commutes with the change of coefficients, and by
Lemma~\ref{lem:1.26} two successive changes of coefficients equal the change by
the composite homomorphism.
Hence the two instances of \eqref{eq:1.13} yield \eqref{eq:1.13} for the
composite morphism.
The identity and associativity laws hold as well in each component: the
coordinates, the relation polynomials, and the restriction maps.

The two projections extract the corresponding components of these composites.
They are therefore defined on objects and morphisms and preserve identities
and composition.

\end{proof}

The diagram \eqref{eq:1.15} is a construction for treating, in a single
diagram, different object formations with the same extraction and different
local geometries with the same core.
Chapter~\ref{chap:4} studies transport, which builds the objects and morphisms
of the upper levels from a change of the base, through its existence and
universal property.

\section{Model synchronization: the semantics of reads and updates}\label{sec:1.8}

In \S\ref{sec:1.7} we compared choices of reading, such as extraction and
equations.
From here on we compare states and operations under semantics defined
independently of AAT.
We take up lenses, which treat the consistency of reads and updates, and
protocols, which treat executions chaining named operations together with their
observations.

In \S\ref{sec:1.10} we construct from both of them objects, operations, and
Laws organized by role, and show that the constructed Laws hold if and only if
the original laws hold.
We further check that typed morphisms between the constructed objects are in
one-to-one correspondence with the original semantics-preserving morphisms.
Through this correspondence, the results obtained in later chapters from the
common theorems, such as the classification of changes and local
reconstruction, are brought back to the original semantics.

For lenses, we first determine a decomposition of states and the maps
preserving reads and updates.
In model synchronization, part of the internal state is displayed, and changes
to the displayed value are reflected back into the internal state.
The displayed part is called the view.
When the consistency of reads and updates is specified by the three lens laws,
a state decomposes into a displayed value and a component preserved by updates.

\begin{definition}[Total lenses and their morphisms]\label{def:1.32}

Fix a set $V$ of views and a reference value $v_0\in V$.
A total lens is a state set $C$, a read $g:C\to V$, and an update
$p:C\times V\to C$ satisfying
\begin{equation*}
p(c,g(c))=c,\qquad
g(p(c,v))=v,\qquad
p(p(c,v),w)=p(c,w)
\tag{1.16}\label{eq:1.16}
\end{equation*}
for all $c,v,w$.
The three equations express, in order, the following.

\begin{enumerate}
\item Writing back the current displayed value unchanged does not change the
state.
\item Reading after an update yields the specified displayed value.
\item Updating twice in succession gives the same result as updating once with
the last displayed value.
\end{enumerate}

In standard lens terminology, this is a total, very well-behaved lens
(\cite[\href{https://www.cis.upenn.edu/~bcpierce/papers/newlenses-full.pdf}{\S3.1}]{FGMPS04}).
For the lenses treated below, the reference fiber
$K_L=\{c\in C\mid g(c)=v_0\}$ is required to be finite.
This is the set of states remaining when the display is fixed at the reference
value.
The set $V$ of displayed values itself may be infinite.
The finiteness of the reference fiber is used later to describe
semantics-preserving morphisms by finite tables.

A morphism $h:L\to L'$ over the fixed view is a map of states $h:C\to C'$
satisfying
\begin{equation*}
g'h=g,\qquad h(p(c,v))=p'(h(c),v)
\tag{1.17}\label{eq:1.17}
\end{equation*}
Identities and composites are the identities and composites of state maps, and
by substitution in \eqref{eq:1.17} they satisfy \eqref{eq:1.17} again.
This category is written $\mathrm{Lens}(V,v_0)$.

\end{definition}

\begin{proposition}[Product decomposition of lenses]\label{prop:1.33}

For a total lens $L=(C,g,p)$, the maps
\begin{equation*}
\begin{aligned}
\eta_L:C&\longrightarrow V\times K_L,
&c&\longmapsto(g(c),p(c,v_0)),\\
\zeta_L:V\times K_L&\longrightarrow C,
&(v,k)&\longmapsto p(k,v)
\end{aligned}
\tag{1.18}\label{eq:1.18}
\end{equation*}
are mutually inverse bijections. In this decomposition the read is the first
projection, and the update is $((v,k),w)\mapsto(w,k)$.

\end{proposition}

\begin{proof}

Since $g(p(c,v_0))=v_0$, the map $\eta_L$ is well defined.
The three laws give
\[
p(p(c,v_0),g(c))=p(c,g(c))=c
\]
Hence $\zeta_L\eta_L=\mathrm{id}$. On the other hand, for $k\in K_L$ we have
$g(p(k,v))=v$ and
$p(p(k,v),v_0)=p(k,v_0)=p(k,g(k))=k$, so
$\eta_L\zeta_L=\mathrm{id}$ holds as well.
The formula for the read follows from the first component. After an update the
second component $p(p(c,w),v_0)=p(c,v_0)$ is unchanged, and the first component
becomes $w$.

\end{proof}

The second component of the product decomposition represents the information
retained when the displayed value changes.
A map preserving reads and updates is likewise determined solely by how it acts
on this component.

\begin{proposition}[Recovery of morphisms from the reference fiber]\label{prop:1.34}

For any two lenses, the restriction
\[
\mathrm{res}:\mathrm{Hom}(L,L')
\longrightarrow\mathrm{Map}(K_L,K_{L'}),\qquad
h\longmapsto h|_{K_L}
\]
is a bijection. Here $\mathrm{Map}(K,K')$ denotes the set of all maps between
the sets.
The inverse is given, for any $t:K_L\to K_{L'}$, by
\begin{equation*}
\mathrm{ext}(t)(c)=p'\bigl(t(p(c,v_0)),g(c)\bigr)
\tag{1.19}\label{eq:1.19}
\end{equation*}
This is the operation of updating once to the reference view, applying $t$ to
that state, and then updating back to the original displayed value.

\end{proposition}

\begin{proof}

A morphism $h$ preserves reads and therefore sends the reference fiber into the
reference fiber.
In the product decomposition of Proposition~\ref{prop:1.33}, $\mathrm{ext}(t)$
is $(v,k)\mapsto(v,t(k))$, which preserves reads and updates.
Hence \eqref{eq:1.19} indeed gives a morphism of lenses.

For $k\in K_L$ we have $p(k,v_0)=k$ and
$p'(t(k),v_0)=t(k)$, so $\mathrm{res}(\mathrm{ext}(t))=t$.
In the other direction, \eqref{eq:1.17} and the three laws give
\[
\begin{aligned}
\mathrm{ext}(\mathrm{res}(h))(c)
&=p'(h(p(c,v_0)),g(c))\\
&=p'(p'(h(c),v_0),g(c))\\
&=p'(h(c),g(c))
=p'(h(c),g'(h(c)))=h(c)
\end{aligned}
\]
That both restriction and extension preserve identities and composition is seen
from the product decomposition.

\end{proof}

When a visible change $u:V\xrightarrow{\sim}V$ is allowed as well, require
the same $h:C\xrightarrow{\sim}C'$ and $u$ to satisfy
\begin{equation*}
g'h=ug,\qquad h(p(c,v))=p'(h(c),u(v))
\tag{1.20}\label{eq:1.20}
\end{equation*}

In the product decompositions the first component is $u(v)$.
Writing the second component as $\phi_v(k)$, the update equation gives
$\phi_w(k)=\phi_v(k)$. Thus the changes satisfying \eqref{eq:1.20} are exactly
\[
h(v,k)=(u(v),\phi(k)),\qquad \phi:K_L\xrightarrow{\sim}K_{L'}
\]
Bijectivity follows from the bijectivity of $u$ and the product decompositions
over the nonempty set $V$.
This description does not require $u(v_0)=v_0$; each lens is expressed using
its own reference fiber.

\begin{example}[Changes preserving only reads]\label{ex:1.35}

Consider the product lens with $V=K=\{0,1\}$, where $\oplus$ denotes addition
modulo $2$.
The bijection $h(v,k)=(v,k\oplus v)$ preserves reads, but
\[
h(p((0,0),1))=(1,1),\qquad
p(h(0,0),1)=(1,0)
\]
The bijections over the fixed view preserving reads number $2\cdot2=4$,
choosing the identity or the swap independently on the two fibers.
Those preserving updates as well number $2$, making the same choice on both
fibers.
By tying the different fibers together, the update constrains the permitted
changes.

\end{example}

\section{Protocols: named operations and all executions}\label{sec:1.9}

A protocol has individual operations and execution sequences chaining them.
When cycles can be repeated, finitely many operations give rise to execution
sequences of arbitrary length.
The meaning of all executions is fixed by specifying the action of each
operation and the pairs of paths to be treated as equal executions.
For this description we use the standard constructions of presenting a category
by a graph with path equations and of realization by set-valued functors
(\cite[\href{https://arxiv.org/pdf/1009.1166v3}{\S\S3.2, 3.4--3.5}]{Spivak12}).

\begin{definition}[Realizations of protocols]\label{def:1.36}

A finite directed multigraph $Q$ is given by a vertex set $Q_0$, a set $Q_1$ of
named edges, and start and end maps.
A path is a composable finite sequence of edges, and the empty sequence at each
vertex is the identity path.
We specify finitely many pairs of paths
$\Pi=\{(\ell_j,r_j)\}_{j\in J_\Pi}$ with the same start and end.

On all paths we impose the smallest equivalence relation that identifies the
specified pairs and is preserved by composition on either side.
The category with the vertices as objects and the equivalence classes of paths
as morphisms is called $\mathcal{C}_Q$.
Composition is induced by concatenation of paths and is well defined on
equivalence classes by congruence.
Associativity and the unit laws follow from concatenation of paths.

Fix a functor $O:\mathcal{C}_Q\to{\mathbf{Set}}$ of observation targets.
A realization of the protocol is a pair of a functor
$X:\mathcal{C}_Q\to{\mathbf{Set}}$ whose state set $X(v)$ at each vertex is
finite and a natural transformation $o_X:X\Rightarrow O$.
For an edge $e:v\to w$, the map $X(e)$ represents the transition of states, and
$o_X(v)$ gives the observed value readable from a state at the vertex $v$.
Naturality of the observation is the condition that transitioning the state
and then observing agrees with carrying the observed value by $O(e)$.

A morphism $a:(X,o_X)\to(Y,o_Y)$ is a natural transformation $a:X\Rightarrow Y$
with $o_Ya=o_X$. Its components $a_v$ are general maps of states.
Naturality preserves executions, and the latter equation preserves
observations.
With componentwise identities and composition we obtain the category
$\mathrm{Prot}(Q,\Pi,O)$.

\end{definition}

\begin{proposition}[Realization from finite operation tables]\label{prop:1.37}

Assign to each vertex a finite set $S_v$, to each edge $e:v\to w$ a map
$T_e:S_v\to S_w$, and to each vertex an observation $b_v:S_v\to O(v)$.
Define the action of paths by composing the edge maps, and assume the
following.
\begin{equation*}
T_{\ell_j}=T_{r_j}\quad(j\in J_\Pi),\qquad
b_wT_e=O(e)b_v\quad(e:v\to w).
\tag{1.21}\label{eq:1.21}
\end{equation*}
Then there is exactly one realization with these states, edges, and
observations.

\end{proposition}

\begin{proof}

Assign the identity map to the empty path and $T_{e_n}\cdots T_{e_1}$ to
$e_n\cdots e_1$.
This is a functor from the free path category to the category of sets.
The pairs of paths inducing the same map form an equivalence relation preserved
by composition on either side and containing the declared relations of
\eqref{eq:1.21}.
The action is therefore well defined on the equivalence classes of paths of
$\mathcal{C}_Q$.

Commutativity of the observations also follows by induction on the length of
the path.
Indeed, for $p:v\to w$ and $e:w\to z$, if $b_wT_p=O(p)b_v$, then
$b_zT_eT_p=O(e)b_wT_p=O(e)O(p)b_v$.
Hence $b$ is a natural transformation.
Since every path is a composite of edges, uniqueness of the extension follows
as well.

\end{proof}

\begin{proposition}[Recovery of semantics-preserving morphisms from vertex maps]\label{prop:1.38}

For realizations $X,Y$, if vertexwise maps $a_v:X(v)\to Y(v)$ satisfy
\begin{equation*}
a_wX(e)=Y(e)a_v,\qquad o_Y(v)a_v=o_X(v)
\tag{1.22}\label{eq:1.22}
\end{equation*}
for all named edges $e:v\to w$ and all vertices, they extend to a unique
morphism of protocols.

\end{proposition}

\begin{proof}

Commutativity for the empty path is a property of identity maps, and for a
path $p:v\to w$ and an edge $e:w\to z$ we have
\[
a_zX(e)X(p)=Y(e)a_wX(p)=Y(e)Y(p)a_v
\]
By induction we obtain naturality with respect to all paths.
Since the actions of both realizations depend on the equivalence class of
a path, naturality holds for all morphisms of the quotient category.
Preservation of observations is exactly \eqref{eq:1.22}.
A natural transformation is determined by its components at the objects, so it
is unique.

\end{proof}

Since the vertices, the named edges, the declared relations, and the state sets
are all finite, the inputs of Propositions~\ref{prop:1.37} and 1.38 are given
by finitely many tables.
On the other hand, paths repeating cycles can have arbitrary length.
Conclusions about all executions are obtained not by enumerating them but from
path induction and the congruence relation.

We regard a semantics-preserving morphism connecting two realizations as an
adapter.
When comparing adapters $q:X\to Y$ and $q':X'\to Y'$, the condition for
changes $a:X\to X'$ and $b:Y\to Y'$ to preserve the adapters is
\begin{equation*}
bq=q'a.
\tag{1.23}\label{eq:1.23}
\end{equation*}
That is, applying the adapter and then the change agrees with applying the
change and then the adapter after the change.
As an equation of natural transformations, this is equivalent to
$b_vq_v=q'_va_v$ at each vertex.
By \eqref{eq:1.22}, both sides commute with all executions.
When studying invertible changes we restrict $a,b$ to isomorphisms.
The adapters $q,q'$ themselves continue to be treated as general
semantics-preserving morphisms.

\section{Constructing Atoms, operations, and Laws from semantics}\label{sec:1.10}

To express lenses and protocols as objects of AAT, one must decide where to
keep the operation names, the state values, the state sets, and the meaning of
the operations.
Here we place the type roles and the operation names in Atoms, and the
individual state values in the source.
The state sets are kept in architecture objects, and the functions such as
reads, updates, and transitions in operations.
This arrangement keeps the vocabulary finite while still handling general maps
sending states to other states.

\begin{construction}[Type roles and named operations]\label{cons:1.39}

We record the difference of object roles in the target of the relations, and
the difference of operation names in the image of a special Atom.
This assigns a morphism of configurations to each operation while still
distinguishing operations that act as the same function on states.

The Atom set of a lens $L=(C,g,p)$ consists of the seven distinct symbols
\[
\mathrm{At}_L=
\{*,\mathrm{state},\mathrm{view},\mathrm{read},
\mathrm{write},\mathrm{get},\mathrm{put}\}
\]
Taking this set as the value range of all five coordinates and the coordinate
maps to be the identity satisfies Definition~\ref{def:1.1}.
The vocabulary of a protocol is built in the same way from the distinct
symbols
\[
\mathrm{At}_Q=
\{*\}\sqcup
\{\mathrm{state}_v,\mathrm{observation}_v,\mathrm{observe}_v
\mid v\in Q_0\}
\sqcup\{\mathrm{edge}_e\mid e\in Q_1\}
\]
Both are nonempty finite sets.

For a selected role $t\in\mathrm{At}$, define the configuration
\[
C_t=(\mathrm{At},R_t,\varnothing),\qquad
R_t(a,b)\ \Longleftrightarrow\ b=t
\]
All Atoms are contained in the family, and the relations point to the selected
role.
Which role a configuration represents can therefore be seen from the target
$t$ of its relations.
The set corresponding to the role is placed in the structure data $S_{A_t}$,
with $Q_{A_t}=v_0$ for lenses and $Q_{A_t}=*$ for protocols.
The sets for the roles are as follows.

\begin{xltabular}{\linewidth}{@{}LLL@{}}
\toprule
Semantics & Role & Set kept \\
\midrule
\endhead
lens & $\mathrm{state}$ & The states $C$ \\
\addlinespace[3pt]
lens & $\mathrm{view}$ & The displayed values $V$ \\
\addlinespace[3pt]
lens & $\mathrm{read}$ & The inputs $C$ of the read \\
\addlinespace[3pt]
lens & $\mathrm{write}$ & The inputs $C\times V$ of the update \\
\addlinespace[3pt]
protocol & $\mathrm{state}_v$ & The states $X(v)$ at the vertex $v$ \\
\addlinespace[3pt]
protocol & $\mathrm{observation}_v$ & The observed values $O(v)$ at the vertex $v$ \\
\bottomrule
\end{xltabular}

At state and read of the lens we place the same set $C$; it serves as the role
of update results and as the role of read inputs, respectively.
At write we place the pairs of a current state and a specified displayed value.

Between these objects we provide the following named operations.

\begin{xltabular}{\linewidth}{@{}LLL@{}}
\toprule
Operation name & Roles of domain and codomain & Function giving the meaning \\
\midrule
\endhead
$\mathrm{get}$ & $\mathrm{read}\to\mathrm{view}$ & $g$ \\
\addlinespace[3pt]
$\mathrm{put}$ & $\mathrm{write}\to\mathrm{state}$ & $(c,v)\mapsto p(c,v)$ \\
\addlinespace[3pt]
$\mathrm{edge}_e$ ($e:v\to w$) & $\mathrm{state}_v\to\mathrm{state}_w$ & $X(e)$ \\
\addlinespace[3pt]
$\mathrm{observe}_v$ & $\mathrm{state}_v\to\mathrm{observation}_v$ & $o_X(v)$ \\
\bottomrule
\end{xltabular}

To an operation $s\to t$ with name $m$ we assign the map on Atoms
\begin{equation*}
d_m(a)=
\begin{cases}
m & a=*,\\
t & a\ne *
\end{cases}
\tag{1.24}\label{eq:1.24}
\end{equation*}
Every domain role $s$ in the table differs from $*$.
Hence $R_s(a,b)$ implies $b=s$, so $d_m(b)=t$, that is,
$R_t(d_m(a),d_m(b))$.
Preservation of the family follows because it contains all Atoms, and
preservation of the identifications because they are empty.
Thus $d_m:C_s\to C_t$ is indeed a morphism of configurations.

We take an operation to be the tuple of its name, the objects at its two ends,
$d_m$, and the function in the table giving its meaning.
This provides the operation reading of Definition~\ref{def:1.7}.
Since $d_m(*)=m$, distinct names are distinguished even as morphisms of
configurations.

\end{construction}

\begin{construction}[Extraction sources carrying state values]\label{cons:1.40}

Take the sources of a lens and of a protocol to be
\begin{equation*}
\begin{aligned}
\mathrm{Src}_L
&=1\sqcup C\sqcup V\sqcup(C\times V),\\
\mathrm{Src}_X
&=1\sqcup\bigsqcup_{v\in Q_0}X(v)
 \sqcup\bigsqcup_{e\in Q_1}X(\mathrm{src}(e))
\end{aligned}
\tag{1.25}\label{eq:1.25}
\end{equation*}
Each summand carries a tag: point, state, view, update input, or edge input.
Normalization is the identity, and the vocabulary, meaning, and resolution
parameters are one-element sets.
The predicates $V_D,M_D,R_D$ of Definition~\ref{def:1.3} are always true, and
$E_D$ is defined by the following table.

\begin{xltabular}{\linewidth}{@{}LL@{}}
\toprule
Tag of the source & Atoms extracted \\
\midrule
\endhead
Point (both semantics) & All Atoms of the vocabulary \\
\addlinespace[3pt]
State $c$ of the lens & $\mathrm{state},\mathrm{read},\mathrm{get}$ \\
\addlinespace[3pt]
View $v$ of the lens & $\mathrm{view}$ \\
\addlinespace[3pt]
Update input $(c,v)$ of the lens & $\mathrm{write},\mathrm{put}$ \\
\addlinespace[3pt]
State $c$ at a vertex $v$ of the protocol & $\mathrm{state}_v,\mathrm{observation}_v,\mathrm{observe}_v$ \\
\addlinespace[3pt]
Input state $c$ of an edge $e$ of the protocol & $\mathrm{edge}_e$ \\
\bottomrule
\end{xltabular}

Taking the point as the selected source, the extracted family is the whole
vocabulary, which is finite.
The state values are kept in the sources \eqref{eq:1.25}, and the extraction
predicate reads their tags.

A morphism of lenses $h:L\to L'$ determines a map of sources using the
identity, $h$, the identity, and $h\times\mathrm{id}_V$ on the respective
summands.
A morphism of protocols $a:X\to Y$ uses $a_v$ on the state summands and
$a_{\mathrm{src}(e)}$ on the edge input summands, and sends the point to the
point.
Together with the identity map on Atoms, these give morphisms as in
Definition~\ref{def:1.27}.
Commutation with normalization is a property of identity maps, and the
equivalence of extraction follows because the tags are preserved.

Identities and composites of source maps agree with identities and composites
on each summand.
Moreover, reading the state summands off the source map recovers $h$ or all
the $a_v$.
The correspondence is therefore faithful on morphisms, and it also retains
non-injective semantics-preserving maps.

\end{construction}

\begin{proposition}[Laws expressing the laws of the semantics]\label{prop:1.41}

Fix the sets $C,V$ of a lens or the family of state sets $(S_v)$ of a protocol.
Consider operation data $d$ on which no laws have yet been imposed:
$d=(g,p)$ for a lens and $d=((T_e),(b_v))$ for a protocol.
From these data one can construct an architecture object $A_d$ and an equation
system such that being Lawful is equivalent to the laws of the respective
semantics.

\end{proposition}

\begin{proof}[Proof and construction]

We split each law into equations indexed by input values and build residuals
that are zero when the equation holds and one when it does not.

Let the configuration of $A_d$ be $C_*$ and its structure data the operation
data $d$.
As the quantity we keep the reference view for a lens and the element of a
one-element set for a protocol.
The indices of the index set $K$ of equations are the pairs of one of the
following equations and an input value.
\begin{equation*}
\begin{array}{ll}
\text{lens:}&
p(c,g(c))=c,\quad
g(p(c,v))=v,\quad
p(p(c,v),w)=p(c,w),\\[2pt]
\text{protocol:}&
T_{\ell_j}(c)=T_{r_j}(c),\quad
b_w(T_e(c))=O(e)(b_v(c)).
\end{array}
\tag{1.26}\label{eq:1.26}
\end{equation*}
For the lens, $c$, $(c,v)$, and $(c,v,w)$ range over the inputs, respectively.
For the protocol, the declared relations together with a state at their start,
and the edges $e:v\to w$ together with $c\in S_v$, range likewise.
All indices are required.

There is a single context $W$, whose support reads all Atoms from a
one-element set.
The axes and observables are also one-element sets.
This is a context of $A_d$, whose family contains all Atoms.
The constant ring presheaf and the equation coordinates are given by
\[
O_E(W)=\mathbb{Z}[x_{(i,a)}\mid (i,a)\in K\times\mathrm{At}],
\qquad \nu_{W,i,a}=x_{(i,a)}
\]
When an object $A$ under evaluation carries operation data of the same type,
let $P_i(A)$ be the equation of index $i$ for those data.
For objects without structure data of the corresponding type, $P_i(A)$ is
defined to be false.
The residuals are the constant polynomials
\begin{equation*}
\varepsilon_{W,A,i,a}=
\begin{cases}
0 & P_i(A),\\
1 & \text{otherwise}
\end{cases}
\tag{1.27}\label{eq:1.27}
\end{equation*}
Since the restrictions are identities, \eqref{eq:1.1} holds.

If all the equations of \eqref{eq:1.26} hold, all residuals of $A_d$ are zero.
Conversely, if lawfulness holds, we evaluate each $i$ at the context $W$ and
the Atom $*$.
Since $0\ne1$ in the polynomial ring over $\mathbb{Z}$, the vanishing of
\eqref{eq:1.27} implies $P_i(A_d)$.
This yields all the equations of \eqref{eq:1.26}.
In the protocol case, by Proposition~\ref{prop:1.37} these agree with the
conditions defining a functor and a natural observation.

\end{proof}

For example, if $C=V=\{0,1\}$, $g(c)=c$, and $p(c,v)=c$, then
$g(p(0,1))=0\ne1$. The residual corresponding to this input is indeed $1$, and
the constructed object is not Lawful.

\begin{definition}[Morphisms between the constructed typed objects]\label{def:1.42}

We write $\mathcal{T}(L)$ or $\mathcal{T}(X)$ for the totality of the roles,
objects, and named operations with their meanings from
Construction~\ref{cons:1.39}.
A typed morphism between them is a family of maps on the sets of the roles
satisfying the following conditions.

\begin{itemize}
\item \textbf{Lens}: the maps at state and read are the same $h:C\to C'$, the
map at view is the identity, and the map at write is $h\times\mathrm{id}_V$.
We require the two commutativities for get and put.
\item \textbf{Protocol}: the maps at state are $a_v:X(v)\to Y(v)$ and the maps
at observation are the identity, and we require the commutativity for each
edge and observe.
\end{itemize}

With componentwise identities and composition, these objects and typed
morphisms form a category.

Construction~\ref{cons:1.40} connects the state maps given by these typed
morphisms with the source maps of the extraction doctrines.

\end{definition}

\begin{proposition}[Recovery of semantics-preserving morphisms]\label{prop:1.43}

Reading the sets of the roles and the meanings of the operations off the
constructed objects recovers the original state sets, reads, updates, edge
actions, and observations. Moreover, the following natural bijections hold.
\begin{equation*}
\begin{aligned}
\mathrm{Hom}_{\mathrm{Lens}(V,v_0)}(L,L')
&\simeq\mathrm{Hom}_{\mathrm{typ}}(\mathcal{T}(L),\mathcal{T}(L')),\\
\mathrm{Hom}_{\mathrm{Prot}(Q,\Pi,O)}(X,Y)
&\simeq\mathrm{Hom}_{\mathrm{typ}}(\mathcal{T}(X),\mathcal{T}(Y)).
\end{aligned}
\tag{1.28}\label{eq:1.28}
\end{equation*}

\end{proposition}

\begin{proof}

For the objects, it suffices to project the sets and functions encoded in
Construction~\ref{cons:1.39}.
The operation names are also recovered from the value of \eqref{eq:1.24} at
$*$.

A semantics-preserving morphism $h$ of lenses determines the maps on the four
roles of Definition~\ref{def:1.42}.
The commutativities for get and put are \eqref{eq:1.17}, so we obtain a typed
morphism.
Conversely, taking the map at state from a typed morphism gives a morphism of
lenses by the two commutativities.
The maps on the other roles are determined by the map at state, so the two
operations are mutually inverse.

For protocols as well, taking the vertex components of a semantics-preserving
morphism as the maps at state gives a typed morphism.
Conversely, the commutativities for edge and observe of a typed morphism are
\eqref{eq:1.22}, and by Proposition~\ref{prop:1.38} a natural transformation
with respect to all paths is recovered uniquely.
These two constructions are also mutually inverse at the components of each
vertex.

Both send identities to identities and composites to componentwise composites.
Hence \eqref{eq:1.28} is natural with respect to pre- and postcomposition at
both ends.

\end{proof}

A typed morphism of lenses can be recovered, by Proposition~\ref{prop:1.34},
from the map between reference fibers alone.
A typed morphism of protocols extends to all executions from finitely many
vertex maps and the commutativities for the generating edges.
For a lens with a visible change $u$, taking the map at view to be $u$ and the
map at write to be $h\times u$ turns the commutativities for the two
operations into \eqref{eq:1.20}.
Commutativity with protocol adapters likewise agrees with \eqref{eq:1.23} upon
reading back the vertex components.

\section*{Summary of the Chapter}

In this chapter we made explicit the choices of extraction, object formation,
operations, Laws, and locality, and constructed relative architecture and its
comparisons. The main results are the following four.

\begin{itemize}
\item \textbf{Generation of objects.} From an extracted finite family of Atoms
we form the base object. The smallest family of objects containing it and
closed under the permitted operations equals the totality of objects reachable
by finite sequences of operations (Theorem~\ref{thm:1.18}).
Under soundness of the detector and completeness for the required equations,
the validity of the Laws is equivalent to the absence of required circuits
(Proposition~\ref{prop:1.16}).
\item \textbf{Construction of local structure.} From contexts and covers we
constructed a site, and from restriction maps compatible with the coordinates
and structural relations a presheaf of rings. We also constructed the
sheafification, which provides the gluing of local data
(\S\S\ref{sec:1.5}--\ref{sec:1.6}).
\item \textbf{Comparison of readings.} We defined objects and morphisms at
each of the stages of extraction, core, and geometry, and constructed the
projection functors connecting them (Theorem~\ref{thm:1.31}).
\item \textbf{Recovery of the semantics of CS.} We described the lenses and
protocols treated in this chapter in the common framework of objects,
operations, and Laws. The constructed Laws hold if and only if the original
laws hold (Proposition~\ref{prop:1.41}), and typed morphisms correspond
one-to-one to the original semantics-preserving morphisms
(Proposition~\ref{prop:1.43}).
\end{itemize}

In the next chapter we construct an ideal from the symbolic equation
coordinates.
Under a realization condition connecting the evaluation of the coordinates
with the residuals of each object, we describe the part satisfying the Laws
inside the local geometry.

\chapter{Geometry of Laws and Local Consistency}\label{chap:2}

\section*{Overview of the Chapter}

The question of this chapter is \textbf{when states that satisfy the Laws on each part can
be glued into a state of the whole}. Even if the Laws hold on each part, the parts need not
represent the same state once they are connected. Suppose, for example, that three services
record the same value against different references. Even if a correction between references
can be specified for each connection, no reference common to all three can be chosen unless
the corrections add up to zero around the cycle through the three. This discrepancy is
invisible from the equations inside the individual services alone.

This chapter treats the problem in two stages. First, we collect the equations of the Laws
into an ideal and construct the space of states that satisfy the equations. Second, we
compare the differences of the states chosen on the parts over the overlaps, and remove the
differences that can be resolved by local corrections. The equivalence class that remains
is the obstruction class that prevents gluing.

\begin{xltabular}{\linewidth}{@{}LLL@{}}
\toprule
Question & Construction & Outcome \\
\midrule
\endhead
Where do the states that satisfy the Laws live? & Construct coordinate rings, ideals of equations, and evaluations & Equivalence of the Laws holding with factorization through the lawful locus (\S\S\ref{sec:2.1}--\ref{sec:2.2}) \\
\addlinespace[3pt]
What is still missing after local success? & Construct differences on overlaps and the \v{C}ech complex & The obstruction class of specified local data, and gluing conditions after correction (\S\S\ref{sec:2.3}--\ref{sec:2.5}) \\
\addlinespace[3pt]
How do covers and coefficients affect obstructions? & Examine cycles, dimensions, differences on boundaries, and the structural and semantic phases & Conditions for vanishing, detection, and comparison (\S\S\ref{sec:2.6}--\ref{sec:2.7}) \\
\addlinespace[3pt]
How are concrete obstructions passed to the diagnosis of the next chapter? & Build integer-coefficient local data from Atoms, generator relations, and an actual cover & Obstruction classes computed from states and transitions (\S\ref{sec:2.8}) \\
\addlinespace[3pt]
Do the semantics of repair and the computation with equations agree? & Construct and compare two coefficients and their local states & Correspondence of the SAGA obstruction classes, and conditions for the existence of a global repair (\S\S\ref{sec:2.9}--\ref{sec:2.10}) \\
\bottomrule
\end{xltabular}

The central conclusions are that, under the necessary conditions on evaluation, the
validity of the Laws can be read as a geometric zero-locus condition, and that, when the
actual repair states form a sheaf, a global repair is obtained from the vanishing of the
obstruction class. The former requires a map connecting the symbolic equations with their
evaluation, and the latter requires the action of local corrections and gluing. We define
each of these maps and conditions before the theorems that use them.

\subsection*{Why we connect to the semantics of repair}

In Chapter~\ref{chap:1} we encoded the independently defined semantics of lenses and
protocols into objects of AAT, and recovered the operations and the semantics-preserving
morphisms from those objects and typed morphisms. We also showed that the validity of the
constructed Laws is equivalent to the validity of the laws of the original semantics. For
local consistency as well, we need to know how the computed obstructions correspond to the
repairs that are actually permitted. To this end, we specify which local changes count as
the same repair and which residuals of the equations are identified, and prove the
correspondence between the two.

This comparison is given by SAGA, the preceding paper
(\cite[\href{https://arxiv.org/abs/2608.21458}{\S\S4--5}]{SAGA}).
In this chapter we construct its coefficients, local states, and comparison maps, and prove
the correspondence of obstruction classes and the existence of repairs. This lets us carry
the zero and nonzero verdicts obtained on the equation side back to repairability in the
chosen semantics.

\section{Local algebras and ideals of equations}\label{sec:2.1}

To solve equations, we first fix the variables and the structural relations that they
satisfy from the outset. Using the ring attached to each context in Chapter~\ref{chap:1},
we incorporate, in order, the relations expressing the structure and the equations of the
Laws to be checked.

Throughout, we fix a reading of the architecture, a site $(\mathcal{C},J)$, and an equation
system $E$. We write $W$ for a context, $i\in K_E$ for a Law index, and
$K_E^{\mathrm{req}}$ for the set of required indices. The two families of coordinates
$\nu_{W,i,a}$ and $\varepsilon_{W,A,i,a}$ of Definition~\ref{def:1.12} represent the
symbolic equations and the residuals at an object $A$, respectively.

\begin{definition}[Presentation of local algebras]\label{def:2.1}

For the coordinate set $Z_W$ and the ideal of structural relations $I_W^{\mathrm{str}}$ of
Definition~\ref{def:1.25}, we choose a $k$-algebra isomorphism
\begin{equation*}
B_{\mathrm{raw}}(W)=k[x_z\mid z\in Z_W]/I_W^{\mathrm{str}},
\qquad
\theta_W:B_{\mathrm{raw}}(W)\xrightarrow{\sim}O_E(W)
\tag{2.1}\label{eq:2.1}
\end{equation*}
We require that $\theta_W$ commute with the restrictions of contexts and send the equation
coordinates and the residual coordinates in the presentation to the corresponding $\nu$ and
$\varepsilon$. We call this a presentation of the local algebras of $E$. The sheaf of rings
$\mathcal{O}=a_JO_E$ is obtained by transferring addition and multiplication to the
sheafification of Proposition~\ref{prop:1.23}. The operations can be computed on the local
presentation, and the ring axioms are preserved as local identities.

\end{definition}

\begin{proposition}[Algebraic presentation of local configurations]\label{prop:2.2}

For a commutative $k$-algebra $R$, let $\mathrm{Conf}_W(R)$ be the set of families of
coordinate values $(c_z)_{z\in Z_W}$ that satisfy all the structural relations. Then there
is a bijection
\[
\mathrm{Conf}_W(R)
\cong\mathrm{Hom}_{k\text{-}\mathrm{Alg}}(B_{\mathrm{raw}}(W),R)
\]
natural in $R$.

\end{proposition}

\begin{proof}

A family of coordinate values determines a unique homomorphism from the polynomial ring to
$R$. That all structural relations map to zero is equivalent to the kernel containing
$I_W^{\mathrm{str}}$, so the homomorphism factors through the quotient. Conversely,
evaluating a homomorphism from the quotient at each $x_z$ recovers the original family.
Composition with a ring homomorphism $R\to R'$ is exactly the transfer of coordinate
values, so naturality follows.

\end{proof}

The relations imposed here are the structural relations that describe one and the same
local configuration. The Laws that such a configuration must satisfy are expressed by the
following ideals.

\begin{definition}[Witness ideals and obstruction ideals]\label{def:2.3}

We define the witness ideal of a Law $i$ and the obstruction ideal of the required Laws by
\begin{equation*}
\begin{aligned}
I_i^E(W)&=(\nu_{W,i,a}\mid a\in\mathrm{At})\subseteq O_E(W),\\
I_{\mathrm{Ob}}^E(W)&=\sum_{i\in K_E^{\mathrm{req}}}I_i^E(W)
\end{aligned}
\tag{2.2}\label{eq:2.2}
\end{equation*}
An element of a generated ideal is a finite sum of the generators with ring coefficients.
Hence, even when the index set is infinite, each individual element is expressed by a
finite sum.

\end{definition}

\begin{proposition}[Restriction and sheafification of ideals]\label{prop:2.4}

A restriction $j:W'\to W$ satisfies $\mathrm{res}_j(I_i^E(W))\subseteq I_i^E(W')$. The same
holds for $I_{\mathrm{Ob}}^E$. The image of the sheafified morphism
$a_JI_i^E\to\mathcal{O}$ is a sheaf of ideals of $\mathcal{O}$. We write it
$\mathcal{I}_i^E$, and write $\mathcal{I}_{\mathrm{Ob}}^E$ for its sum over the required
indices.

\end{proposition}

\begin{proof}

The restriction of a generator is $\nu_{W',i,a}$, and the restriction of any element of the
generated ideal is again a finite sum of generators. Closure under addition and under
multiplication by ring elements can be checked locally, so it is preserved in the image
after sheafification. For the sum, the same argument applies once each section is expressed
locally as a finite sum.

\end{proof}

Enlarging the set of required indices enlarges the ideal and strengthens its zero-locus
condition. When optional Laws are imposed at the same time, their ideals are added to the
sum.

\section{Spaces satisfying the Laws and evaluation}\label{sec:2.2}

An ideal collects the equations, but by itself it does not express that the equations hold
at an object. In this section we read the rings as geometric spaces and provide the
mechanism by which the symbolic equations are evaluated into the residuals of objects.

\begin{definition}[Affine schemes and closed subschemes]\label{def:2.5}

For a commutative ring $B$, $\mathrm{Spec}\,B$ is the affine scheme whose underlying set is
the set of prime ideals, and whose basic open sets and their rings are given by
\begin{equation*}
D(f)=\{\mathfrak{p}\mid f\notin\mathfrak{p}\},
\qquad
\mathcal{O}_{\mathrm{Spec}\,B}(D(f))=B_f
\tag{2.3}\label{eq:2.3}
\end{equation*}
Here $B_f$ is the localization that inverts $f$. Sections over a general open set are
defined by gluing compatible sections over basic open sets. The local ring at each point is
$B_{\mathfrak{p}}$. A space equipped with a sheaf of rings that locally has this form is
called a scheme. These are the usual constructions of the prime spectrum and the structure
sheaf
(\cite[\href{https://stacks.math.columbia.edu/tag/01HR}{\S26.5, Tag~01HR}]{Stacks}).

We write $V_B(I)=\mathrm{Spec}(B/I)$ for the closed subscheme defined by an ideal
$I\subseteq B$. A ring homomorphism $e:B\to R$ factors through $B/I$ if and only if
$e(I)=0$. Hence $V_B(I)$ represents the points that satisfy the equations $I$ in an
arbitrary coefficient ring.

\end{definition}

\subsection*{Evaluating symbolic equations into residuals}

We impose conditions in two stages on a space representing local evaluation data. First, we
make the evaluation of the symbolic equations agree with the evaluation of the residuals of
objects. Second, we set the required residuals to zero. Below, $Y_W$, $X_W$, and
$X_E^{\mathrm{law}}\cap X_W$ denote, in order, the space before the conditions are imposed
and the spaces after each condition is imposed.

\begin{construction}[Geometric realization of the equations]\label{cons:2.6}

In each context, consider local evaluation data $p$ consisting of a reading $A_p$ of an
object and a homomorphism $e_p:O_E(W)\to R$ into a commutative $k$-algebra $R$. We assume
that these data can be transferred along changes of the coefficient ring. We treat the case
in which the following algebraic presentation is given.

\begin{enumerate}
\item The functor of local evaluation data is represented by an affine scheme
$Y_W=\mathrm{Spec}\,D_W$, and a universal ring homomorphism $u_W:O_E(W)\to D_W$ gives the
evaluation.
\item The evaluation of the residuals $p\mapsto e_p(\varepsilon_{W,A_p,i,a})$ is a regular
function, and an element $r_{W,i,a}\in D_W$ representing it has been constructed.
\item The comparisons of the presentations between contexts give isomorphisms on open
parts and preserve structure, coordinates, and residuals on the overlaps. On triple
overlaps the open parts being compared correspond, and the composites of the isomorphisms
agree there.
\item The assignment of charts after gluing, $W\mapsto X_W$ (defined in \eqref{eq:2.4}
below), sends morphisms of contexts to inclusions of open parts and preserves identities,
composites, and overlaps. Covers in $J$ are sent to open covers of charts.
\end{enumerate}

The first condition says that the evaluation data can be described by coordinates, and the
second that the residuals are algebraic functions of those coordinates. The third is the
condition for gluing the local presentations without contradiction. The fourth says that
restrictions, overlaps, and covers of contexts can be handled in the same way on the open
parts of the resulting space. We then impose the relations that equate the evaluation of
the symbolic equations with that of the residuals.
\begin{equation*}
\begin{aligned}
K_W&=(u_W(\nu_{W,i,a})-r_{W,i,a}\mid i,a),\\
B_W&=D_W/K_W,\qquad X_W=\mathrm{Spec}\,B_W,\\
\eta_W&:O_E(W)\longrightarrow B_W.
\end{aligned}
\tag{2.4}\label{eq:2.4}
\end{equation*}
Since the comparison isomorphisms preserve $K_W$, the open parts of the $X_W$ can also be
compared by isomorphisms. Let $X_E$ be the scheme obtained by gluing them. Concretely, we
identify corresponding points in the disjoint union of the $X_W$, and define the functions
on an open set as the families of functions that agree on each chart. The composition
condition on the overlaps makes the identification transitive, and each $X_W$ is embedded
as an open part. The local rings are also those of the charts, so the resulting space is a
scheme
(\cite[\href{https://stacks.math.columbia.edu/tag/01JA}{\S26.14, Tag~01JA}]{Stacks}).

For $s:T\to X_E$, put $T_W=T\times_{X_E}X_W$. We define its evaluation
$\mathrm{ev}_{s,W}:O_E(W)\to\Gamma(T_W,\mathcal{O}_T)$ as the composite of $\eta_W$ with
the pullback of regular functions. We write $A_s$ for the object read off from the
representing functor, and $\varepsilon_{W,i,a}(s)=\mathrm{ev}_{s,W}(\varepsilon_{W,A_s,i,a})$
for its residuals. For a general $T$, we read this formula on affine open sets and glue the
agreeing values.

\end{construction}

\begin{lemma}[Evaluation of generators and localization]\label{lem:2.7}

The evaluation obtained in Construction~\ref{cons:2.6} is compatible with restriction of
contexts and with base change of $T$, and satisfies
\begin{equation*}
\mathrm{ev}_{s,W}(\nu_{W,i,a})=\varepsilon_{W,i,a}(s)
\tag{2.5}\label{eq:2.5}
\end{equation*}
Moreover, putting $J_i(W)=(\eta_W(\nu_{W,i,a})\mid a)$, on a basic open set
$D(f)\subseteq X_W$ this ideal becomes
\begin{equation*}
J_i(W)B_{W,f}
=(\eta_W(\nu_{W,i,a})/1\mid a)
\tag{2.6}\label{eq:2.6}
\end{equation*}
These local ideals glue into a quasi-coherent sheaf of ideals $\mathcal{J}_i$ on $X_E$.

\end{lemma}

\begin{proof}

Equation~\eqref{eq:2.5} is the identity obtained by pulling back along $s$ the fact that
the generators of $K_W$ become zero in $B_W$. The left-hand side is tied to the residual
because, by condition~1, $u_W$ represents the evaluation, and by condition~2, $r_{W,i,a}$
represents the evaluation of the residual. The composition rule for evaluations follows
from that for ring homomorphisms. An element of the localized ideal is a finite sum of
generators written with a denominator, which gives \eqref{eq:2.6}. On each chart we take
the sheaf $\widetilde{J_i(W)}$ associated with the module $J_i(W)$ and glue along the
isomorphisms on the overlaps, which preserve the same generators. Quasi-coherence means
precisely that on each affine open set the sheaf arises in this form from a module.

\end{proof}

By the condition preserving covers and overlaps, $W\mapsto\Gamma(X_W,\mathcal{O}_{X_E})$
is a sheaf with respect to $J$. Hence $\eta$ factors through the sheafification, and we can
compare $\mathcal{I}_i^E$ on the site with $\mathcal{J}_i$ on the scheme. Mapping the local
generators of the former to the charts and taking sums with ring coefficients yields the
$J_i(W)$ of the latter. This comparison is determined by the original generators and
localization.

\begin{definition}[Lawful locus]\label{def:2.8}

Put $\mathcal{J}_{\mathrm{Ob}}=\sum_{i\in K_E^{\mathrm{req}}}\mathcal{J}_i$, and write
$X_E^{\mathrm{law}}=V(\mathcal{J}_{\mathrm{Ob}})$ for the closed subscheme that it defines.
On each chart it is $\mathrm{Spec}(B_W/\sum_iJ_i(W))$. Further,
$s^{*}_{\mathrm{ideal}}\mathcal{J}$ denotes the sheaf of ideals generated inside
$\mathcal{O}_T$ by the image of $\mathcal{J}$; it is the image of the morphism from the
module pullback $s^*\mathcal{J}$ to $\mathcal{O}_T$.

On the ring side, the two-stage conditions correspond to the two quotients
\[
D_W\longrightarrow B_W=D_W/K_W
\longrightarrow B_W/\sum_{i\in K_E^{\mathrm{req}}}J_i(W)
\]
The first quotient equates the evaluation of the equations with that of the residuals, and
the second imposes the validity of the required Laws. This order is what makes the
zero-locus condition of the next theorem test the residuals of objects.

\end{definition}

\begin{theorem}[Correspondence of equations, ideals, and zero loci]\label{thm:2.9}

Under the conditions of Construction~\ref{cons:2.6}, for every $s:T\to X_E$ the following
are equivalent.
\begin{equation*}
\begin{aligned}
&\forall W\ \forall i\in K_E^{\mathrm{req}}\ \forall a,
\quad\varepsilon_{W,i,a}(s)=0;\\
&s^{*}_{\mathrm{ideal}}\mathcal{J}_{\mathrm{Ob}}=0;\\
&s\text{ factors through }X_E^{\mathrm{law}}\text{.}
\end{aligned}
\tag{2.7}\label{eq:2.7}
\end{equation*}

\end{theorem}

\begin{proof}

That the pulled-back ideal is zero is equivalent to all of its generators becoming zero on
$T_W$. By \eqref{eq:2.5}, the vanishing of these generators coincides with the vanishing of
the residuals. Pulling back a sum of ideals and taking the generated ideal is the same as taking the sum of
the pulled-back generators, so the first two conditions are equivalent over all required
indices.

The last condition is obtained, on each affine open set, from the universal property of
homomorphisms factoring through the quotient ring. The local factorizations are unique, so
they agree on overlaps and glue. Conversely, if a factorization exists, the pullbacks of
all generators that become zero in the quotient are zero.

\end{proof}

This factorization is also the standard universal property of closed subschemes
(\cite[\href{https://stacks.math.columbia.edu/tag/01HP}{Lemma~26.4.6, Tag~01HP}]{Stacks}).
We write $\mathrm{EquationLawful}_E(s)$ for the first condition of the theorem.

\begin{example}[Synchronization of two values]\label{ex:2.10}

Let the coordinate ring be $k[x,y]$, and express the Law equating the two values by
$f=y-x$. The residual under the evaluation $x\mapsto u$, $y\mapsto v$ is $v-u$. The ring of
the lawful locus is $k[x,y]/(y-x)\cong k[x]$, and a point of it amounts to assigning one
value to both sides. In this case \eqref{eq:2.5} is verified by the substitution
$f(u,v)=v-u$.

\end{example}

\begin{corollary}[Comparison with lawfulness of objects]\label{cor:2.11}

Suppose that there is a realization $s_A:T_A\to X_E$ of an object $A$, together with, in
each context, a ring isomorphism $\Gamma((T_A)_W,\mathcal{O}_{T_A})\cong O_E(W)$ carrying
the evaluation of the residuals back to the original $\varepsilon_{W,A,i,a}$. Then
$\mathrm{Lawful}_E(A)$ of Definition~\ref{def:1.13} is equivalent to $s_A$ factoring
through $X_E^{\mathrm{law}}$.

\end{corollary}

\begin{proof}

The isomorphisms preserve and reflect zero, so the vanishing of each residual of the
object coincides with the vanishing after evaluation. Apply Theorem~\ref{thm:2.9}.

\end{proof}

\section{Coefficients representing obstructions}\label{sec:2.3}

Adding and subtracting the discrepancies of local states requires an abelian group in
which their values live. We define a way to build quotient groups from the equations and a
way to map finite failure patterns into the coefficients. The choice of coefficients
determines which differences are treated as the same.

\begin{construction}[Quotient coefficients generated from the equations]\label{cons:2.12}

In each context, put
\begin{equation*}
Q_E(W)=O_E(W)/I_{\mathrm{Ob}}^E(W),
\qquad
\overline\varepsilon_{W,A,i,a}=[\varepsilon_{W,A,i,a}]
\tag{2.8}\label{eq:2.8}
\end{equation*}
By Proposition~\ref{prop:2.4} the restrictions descend to the quotients, and $Q_E$ becomes
a presheaf with values in additive groups. The restriction of a residual class equals the
residual of the same object, Law, and Atom, restricted and then passed to the quotient.
Moreover
\[
\overline\varepsilon_{W,A,i,a}=0
\quad\Longleftrightarrow\quad
\varepsilon_{W,A,i,a}\in I_{\mathrm{Ob}}^E(W).
\]
This is the definition of the zero element of a quotient group. If the Law holds and the
residual itself is zero, then the residual class is also zero. Detection in the converse
direction requires that a failing residual not lie in the ideal.

\end{construction}

\begin{example}[Residuals distinguished in the quotient]\label{ex:2.13}

If $O_E(W)=\mathbb{Z}[x]$ and $I_{\mathrm{Ob}}^E(W)=(x)$, then $Q_E(W)\cong\mathbb{Z}$.
The class of the symbolic generator $x$ is zero, while the class of $1$, chosen as the
residual of some object, is nonzero. If the residual is $x$, it is itself nonzero but
becomes zero in the quotient. A verdict in the quotient coefficients therefore reads the
information of the specified residual modulo the ideal.

\end{example}

\begin{definition}[Local circuits and realization in coefficients]\label{def:2.14}

For an object $A$ and a context $W$, we define the Law $i$ to hold on $W$ by
\[
E_{i,W}(A)\quad\Longleftrightarrow\quad
\forall a\in\mathrm{At},\quad\varepsilon_{W,A,i,a}=0
\]
Its failure means that at least one residual is nonzero. The condition $E_i(A)$ of
Chapter~\ref{chap:1} imposes this local validity condition on all contexts.

We choose a configuration read in each context and a detector code of the form of
Definition~\ref{def:1.15}. Let $T_{i,W}$ be its finite accepted set, and write
$\mathrm{Matches}_W(Q,A)$ for the matching condition obtained by evaluating each query of
Definition~\ref{def:1.14} on this local configuration. Among the finite patterns that are
accepted and match, we select those to be used as
\[
\mathrm{Circ}^{\mathrm{loc}}_E(A,i;W)
\subseteq\{Q\in T_{i,W}\mid\mathrm{Matches}_W(Q,A)\}
\]
and call its elements local circuits. We equip this family with maps that preserve
matching and acceptance along the restrictions of the adopted contexts, and impose the
identity and composition laws. Choosing a family that remains usable after restriction in
this way is also part of the data of local detection. Local soundness, and local
completeness for the required Laws, are the conditions
\[
\begin{aligned}
c\in\mathrm{Circ}^{\mathrm{loc}}_E(A,i;W)
&\Longrightarrow\neg E_{i,W}(A),\\
i\in K_E^{\mathrm{req}}\ \land\ \neg E_{i,W}(A)
&\Longrightarrow\mathrm{Circ}^{\mathrm{loc}}_E(A,i;W)\ne\varnothing
\end{aligned}
\]
respectively. We verify these conditions for the objects, contexts, and Laws to which they
are applied.

We further choose a sheaf of abelian groups $F$. To each local circuit $c$ we associate a
witness index $a(c)\in\mathrm{At}$ and assign the generator
$\nu(c)=\nu_{W,i,a(c)}\in I_i^E(W)$. When, in addition, an additive map commuting with
restriction
\begin{equation*}
\rho_{i,W}:I_i^E(W)\longrightarrow F(W),
\qquad \kappa_{i,W}(c)=\rho_{i,W}(\nu(c))
\tag{2.9}\label{eq:2.9}
\end{equation*}
has been constructed, we call it a realization of circuits in the coefficients. The
assignment of generators is also required to commute with the restrictions of local
circuits and of the rings. For example, the sheafified quotient map
$I_i^E\to a_J(I_i^E/(I_i^E)^2)$ is a candidate for such a map. Whether its image detects
the circuits is checked on the chosen generators.

\end{definition}

\begin{proposition}[Detection of finite failures by coefficients]\label{prop:2.15}

Fix an object $A$ and a context $W$. Suppose that local soundness and local completeness
of Definition~\ref{def:2.14} hold for the required Laws, and that the image under
\eqref{eq:2.9} of each local circuit is nonzero. Then at least one required Law fails on
$W$ if and only if there exists a nonzero circuit image for some required Law.

\end{proposition}

\begin{proof}

From a failure, completeness yields a circuit, whose image is nonzero by assumption.
Conversely, if a circuit exists, soundness implies that the corresponding Law fails.

\end{proof}

When several images are aggregated into one, we also need the condition that nonzero
images do not cancel in the sum. For example, if each image is stored in a direct sum with
independent coordinates, the vanishing of the whole coincides with the vanishing of each
coordinate. The coefficients chosen at this stage are also used in the local comparisons
that follow.

\section{Comparison on overlaps and \v{C}ech obstruction classes}\label{sec:2.4}

When comparing differences of local states, we place the differences on double overlaps
and the coherence conditions among the differences on triple overlaps. The \v{C}ech
complex assembles this computation from restriction maps and addition and subtraction.

\begin{definition}[Covers and the \v{C}ech complex]\label{def:2.16}

Fix a finite cover $\mathcal{U}=(U_\alpha\to W)_{\alpha\in A}$. We take each morphism to
be injective and choose a total order on $A$. The overlaps $U_{\alpha\beta}$ and
$U_{\alpha\beta\gamma}$ are pullbacks over $W$. The coefficients $F$ form a presheaf with
values in abelian groups. On empty overlaps we set $F(\varnothing)=0$ and omit those
components from the following products.
\begin{equation*}
\begin{aligned}
C^0(\mathcal{U},F)&=\prod_\alpha F(U_\alpha),\\
C^1(\mathcal{U},F)&=\prod_{\alpha<\beta}F(U_{\alpha\beta}),\\
C^2(\mathcal{U},F)&=\prod_{\alpha<\beta<\gamma}F(U_{\alpha\beta\gamma}).
\end{aligned}
\tag{2.10}\label{eq:2.10}
\end{equation*}
We call these elements cochains. Correction amounts for local states live in degree 0,
and differences on overlaps in degree 1. We define the differentials by
\begin{equation*}
\begin{aligned}
(d^0b)_{\alpha\beta}&=b_\beta|_{U_{\alpha\beta}}-b_\alpha|_{U_{\alpha\beta}},\\
(d^1g)_{\alpha\beta\gamma}&=g_{\beta\gamma}|_{U_{\alpha\beta\gamma}}
-g_{\alpha\gamma}|_{U_{\alpha\beta\gamma}}
+g_{\alpha\beta}|_{U_{\alpha\beta\gamma}}
\end{aligned}
\tag{2.11}\label{eq:2.11}
\end{equation*}
From now on, restriction to the same overlap is abbreviated by $|$. In higher degrees,
likewise, degree $n$ consists of the product of sections on $n+1$-fold intersections,
and the differential is the alternating sum of restrictions.

\end{definition}

\begin{lemma}[Composition of the differentials]\label{lem:2.17}

We have $d^1d^0=0$.

\end{lemma}

\begin{proof}

The component restricted to a triple overlap is
$(b_\gamma-b_\beta)-(b_\gamma-b_\alpha)+(b_\beta-b_\alpha)=0$. That the two ways of
restricting each $b$ agree uses the composition rule of the presheaf.

\end{proof}

\begin{definition}[Obstruction groups on a fixed cover]\label{def:2.18}

A $g$ satisfying $d^1g=0$ is called a 1-cocycle, and an element of the form $d^0b$ a
1-coboundary. The first \v{C}ech cohomology with respect to the cover $\mathcal{U}$ is
\begin{equation*}
\check H^1(\mathcal{U},F)=\ker d^1/\mathrm{im}\,d^0
\tag{2.12}\label{eq:2.12}
\end{equation*}
Here $[g]$ denotes the equivalence class of $g$. Two cocycles have the same class when
they differ by some $d^0b$. In particular, $[g]=0$ means that there exists a local
correction $b$ with $g=d^0b$. Throughout this chapter, $\check H^1$ denotes this group
with respect to the cover and coefficients made explicit.

\end{definition}

\begin{example}[Three local references]\label{ex:2.19}

Suppose that the double overlaps of three charts are all nonempty and the triple overlap
is empty. Take the same abelian group $M$ as the coefficients on each chart and overlap,
with identity maps as restrictions. Then
\begin{equation*}
\begin{aligned}
d^0(b_0,b_1,b_2)&=(b_1-b_0,\ b_2-b_0,\ b_2-b_1),\\
\mathrm{per}(g)&=g_{01}+g_{12}-g_{02}
\end{aligned}
\tag{2.13}\label{eq:2.13}
\end{equation*}
We call the sum around the cycle, $\mathrm{per}(g)$, the period. Since there is no triple
overlap, every 1-cochain is a cocycle. The period of $d^0b$ is zero; conversely, if the
period is zero, then putting $b_0=0$, $b_1=g_{01}$, $b_2=g_{02}$ gives $d^0b=g$. Hence
$\check H^1(\mathcal{U},M)\cong M$.

If $M=\mathbb{Z}$, $g_{01}=1$, and $g_{02}=g_{12}=0$, the period is 1. No matter how
the pairwise differences of references are corrected, this difference around the cycle
remains. An actual cover of this shape is constructed from a finite topological space in
\S\ref{sec:2.8}.

\end{example}

\section{From vanishing of the obstruction class to gluing}\label{sec:2.5}

Equation~\eqref{eq:2.12} is an algebraic condition for finding correction amounts. When
the corrections act on the local states and the corrected states glue, a global state is
obtained. We describe these two conditions in terms of torsors and sheaves.

\begin{definition}[Local states and torsors]\label{def:2.20}

Take a presheaf $P$ of local states and a coefficient presheaf $F$. In this section, the
``intersections of the cover'' include each chart itself. Suppose that $F(V)$ acts on
$P(V)$ on each intersection of the cover, and that these actions commute with the
restrictions between intersections. When, on each nonempty intersection $V$ of the cover,
for any $p,q\in P(V)$ there exists a unique $m\in F(V)$ with $q=m+p$, we call the nonempty
$P(V)$ an $F(V)$-torsor.

A torsor has no designated zero state serving as a reference. Instead, the difference
$q-p=m$ of two states is uniquely determined. Every element of the coefficient group acts
as an actual change of state, so a correction amount found as a cochain can be applied to
the states. Over an empty intersection, we take $P$ to have at most one element. A family of local
states $p_\alpha\in P(U_\alpha)$ is called a local atlas.

Each $p_\alpha$ is given as an element of the state space $P(U_\alpha)$, and no
identification with the coefficient group $F(U_\alpha)$ has been chosen yet. To treat this
family as a 0-cochain of coefficients, we need a reference expressing each state in the
coefficients. Choosing a reference state locally yields such an expression, but these
references may themselves disagree on the overlaps. Whether a common reference compatible
with the restrictions can be chosen is the gluing problem studied here. In \S\ref{sec:2.8}
we make the coordinates in the local references explicit as $p$ and the shifts between
references as $\xi$, and compute the difference of states as $\xi+d^0p$.

\end{definition}

\begin{proposition}[The obstruction class defined by differences of states]\label{prop:2.21}

Defining $g_{\alpha\beta}=p_\beta| -p_\alpha|$ from a local atlas, $g$ is a cocycle.
Changing the atlas to $p'_\alpha=b_\alpha+p_\alpha$ gives
\begin{equation*}
g'=g+d^0b,
\qquad [g']=[g]
\tag{2.14}\label{eq:2.14}
\end{equation*}
We call this $[g]$ the obstruction class for gluing the local states.

\end{proposition}

\begin{proof}

On a triple overlap, the difference that yields $p_\gamma$ directly from $p_\alpha$ equals
the difference obtained via $p_\beta$. By freeness of the action,
$g_{\alpha\gamma}=g_{\beta\gamma}+g_{\alpha\beta}$, and we get $d^1g=0$. Computing the
difference after the change gives $g'_{\alpha\beta}=g_{\alpha\beta}+b_\beta|-b_\alpha|$,
so \eqref{eq:2.14} follows.

\end{proof}

\begin{theorem}[Local corrections and global states]\label{thm:2.22}

Suppose that $P$ of Definition~\ref{def:2.20} is a sheaf on the site $(\mathcal{C},J)$.
Then, for the obstruction class of a local atlas,
\begin{equation*}
P(W)\ne\varnothing
\quad\Longleftrightarrow\quad
[g]=0\text{ in }\check H^1(\mathcal{U},F)
\tag{2.15}\label{eq:2.15}
\end{equation*}
holds. If $g=d^0b$, the corrected family $(-b_\alpha)+p_\alpha$ glues uniquely into a
global state.

\end{theorem}

\begin{proof}

Suppose $g=d^0b$. By Proposition~\ref{prop:2.21} the differences after correction are
zero. On overlaps with the order reversed, the difference merely changes sign and is again
zero. On self-intersections the two projections agree by injectivity of the covering
morphisms, and on empty intersections they agree since there is at most one element. Hence
the corrected family is a matching family agreeing on all overlaps. By the sheaf condition
it glues into a unique global state.

Conversely, given $p\in P(W)$, on each chart we can take a difference with
$p_\alpha=b_\alpha+p|_{U_\alpha}$. Computing on the overlaps gives $g=d^0b$.

\end{proof}

What is unique is the gluing of the fixed corrected family. Different choices of
correction may yield different global states. For torsors under a sheaf of abelian groups,
this argument amounts to the standard criterion for trivialization
(\cite[\href{https://stacks.math.columbia.edu/tag/03AG}{\S21.4, Tag~03AG}]{Stacks}).

\subsection*{Application to local states satisfying the Laws}

\begin{corollary}[Gluing of lawful states]\label{cor:2.23}

Suppose there are a sheaf of states $\mathcal{T}$ and a natural family of required
residuals with values in the sheaf of rings $\mathcal{O}$, and let
$P(V)\subseteq\mathcal{T}(V)$ be the set of states on which all required residuals are
zero. Suppose that the action of $F$ preserves $P$ and satisfies the conditions of
Definition~\ref{def:2.20} on the chosen cover. Then the obstruction class of the local
lawful states is zero if and only if $P(W)\ne\varnothing$.

\end{corollary}

\begin{proof}

$P$ is a sheaf. Indeed, a matching family glues uniquely in $\mathcal{T}$, and each of its
residuals is zero on the cover. By separatedness of $\mathcal{O}$ it is zero globally as
well. Apply Theorem~\ref{thm:2.22}.

\end{proof}

Using the evaluation of Construction~\ref{cons:2.6}, this residual condition coincides
with factorization through the lawful locus of Theorem~\ref{thm:2.9}. To return to
lawfulness of objects, we also use the comparison of evaluations of
Corollary~\ref{cor:2.11}.

When local lawfulness is verified from finite checks, the connection between the checks
and the residuals must be specified. Concretely, we verify the following conditions.

\begin{itemize}
\item \textbf{The information can be read.} The cover is
$(E,\mathcal{R},\mathrm{Ov})$-adequate in the sense of Chapter~\ref{chap:1}, and the
necessary coordinates and interactions can be read.
\item \textbf{Failures can be detected.} Local soundness and local completeness of
Definition~\ref{def:2.14} hold, and all necessary witnesses are checked within the cover.
\item \textbf{Zero verdicts correspond.} The zero verdict on the adopted axes is
equivalent to the vanishing of the corresponding residuals.
\item \textbf{Aggregation loses no failure.} When circuit images are aggregated, each
image is nonzero and they do not cancel in the sum.
\end{itemize}

These conditions tie the local checks to membership in $P(U_\alpha)$, and the torsor
action and the sheaf condition give the global conclusion.

\section{Shape of covers, capacity of obstructions, and detection by cycles}\label{sec:2.6}

The size of the obstruction group expresses how many independent discrepancies it can
record. Whether the local states at hand carry an obstruction, on the other hand, is
determined by examining the class built from those states. In this section we study the
former by dimensions and the latter by sums along cycles.

\begin{proposition}[Dimension of the obstruction space and alternating sums]\label{prop:2.24}

Let $k$ be a field and suppose that $C^0,C^1,C^2$ are finite-dimensional $k$-vector
spaces. Then
\begin{equation*}
\dim_k\check H^1
\geq\dim_k C^1-\dim_k C^0-\dim_k C^2.
\tag{2.16}\label{eq:2.16}
\end{equation*}
Moreover, for a finite-dimensional complex of finite length,
\begin{equation*}
\chi(C^\bullet)
=\sum_n(-1)^n\dim_k C^n
=\sum_n(-1)^n\dim_k H^n(C^\bullet).
\tag{2.17}\label{eq:2.17}
\end{equation*}

\end{proposition}

\begin{proof}

Apply the rank-nullity theorem to
$\dim\check H^1=\dim\ker d^1-\dim\mathrm{im}\,d^0$. From
$\mathrm{rank}\,d^1\leq\dim C^2$ and $\mathrm{rank}\,d^0\leq\dim C^0$ we get
\eqref{eq:2.16}. Equation~\eqref{eq:2.17} follows by taking the alternating sum of
$\dim C^n=\dim\mathrm{im}\,d^{n-1}+\dim H^n+\dim\mathrm{im}\,d^n$, in which the dimensions
of the images cancel between adjacent degrees.

\end{proof}

We call \eqref{eq:2.16} the lower bound on obstruction capacity. A change preserving the
cochain dimensions in each degree preserves $\chi$, but the dimensions of the individual
cohomology groups and a specified class $[g]$ also involve the differentials and $g$
itself.

\begin{proposition}[Constant coefficients and the nerve of a cover]\label{prop:2.25}

Take a finite open cover of a topological space and suppose that all of its nonempty
finite intersections are connected. Take as coefficients the sheaf of locally constant
$k$-valued functions. Let $N(\mathcal{U})$ be the complex with a simplex for each
nonempty intersection $U_{\alpha_0\cdots\alpha_n}$. Then the \v{C}ech complex is
isomorphic to the usual cochain complex of $N(\mathcal{U})$ with coefficients in $k$, and
$\dim_k\check H^1(\mathcal{U},k)=\beta_1(N(\mathcal{U});k)$.

\end{proposition}

\begin{proof}

A locally constant function on a connected intersection is determined by a single value.
Under this identification the restrictions become identity maps, and \eqref{eq:2.11}
coincides with the alternating sum along the faces of a simplex. Taking this
identification in each degree gives the isomorphism of complexes. Moreover,
$\beta_1(N(\mathcal{U});k)$ is the dimension of the first cohomology of this complex.

\end{proof}

When an intersection has several connected components, we place one coordinate per
component. When the components split into finitely many open connected parts, the same
construction can be carried out on the complex whose vertices, edges, and faces are the
components.

\begin{proposition}[Vanishing on forests]\label{prop:2.26}

Suppose that there are no triple overlaps and that the finite graph whose vertices are the
charts and whose edges are the components of the nonempty overlaps is a forest. If the
coefficients decompose along the components and the restriction
$F(U_\alpha)\to F(U_e)$ from each endpoint to each edge component is surjective, then
$\check H^1(\mathcal{U},F)=0$.

\end{proposition}

\begin{proof}

Choose a root in each tree and set the correction at the root to zero. Once the correction
of a parent is determined, choose the correction of each child so that the difference
along the parent-child edge equals the specified $g_e$. Surjectivity of the restriction to
the edges makes this choice possible. Since each child has only one parent, we can proceed
without altering the conditions on edges already satisfied. After processing all vertices
we have $d^0b=g$.

\end{proof}

\begin{example}[Paths and cycles]\label{ex:2.27}

On a three-vertex path with identity restrictions, the two differences can be absorbed in
order from the root. Adding an edge joining the two ends leaves the period of
Example~\ref{ex:2.19}. If moreover there is a triple overlap with the same coefficients
and restrictions, the cocycle condition of \eqref{eq:2.11} makes this period zero. Thus
not only the edges of the overlaps but also the faces filling the space between them enter
the computation of the obstruction group.

\end{example}

\subsection*{Sums along cycles and differences on the boundary}

\begin{proposition}[Periods and the Stokes formula]\label{prop:2.28}

Pair the chains and cochains of a finite complex $N$ with constant coefficients in $k$ by
multiplying the coefficients of simplices with matching orientation and summing. For the
boundary map $\partial$ of chains and the differential $d$ of cochains,
\begin{equation*}
\langle d\omega,z\rangle=\langle\omega,\partial z\rangle
\tag{2.18}\label{eq:2.18}
\end{equation*}
holds. Hence $\langle g,z\rangle$ over a 1-cycle $z$ is unchanged when a coboundary is
added to $g$. If this value is nonzero, then $[g]\ne0$.

\end{proposition}

\begin{proof}

For a single simplex, both sides are the sum of the values on its faces with the same
alternating signs. Linearity extends this to general chains. If $\partial z=0$, then
$\langle d^0b,z\rangle=0$, and the remaining claims follow.

\end{proof}

\begin{proposition}[Obstructions arising from differences on the boundary]\label{prop:2.29}

Split a finite complex into two subcomplexes, $N=N_A\cup N_B$, and put $L=N_A\cap N_B$.
For 0-cochains with constant coefficients, the difference of restrictions
$(u,v)\mapsto u|_L-v|_L$ is surjective. Take $b\in H^0(L,k)$, that is, a 0-cochain with
$db=0$, and choose 0-cochains $u,v$ with $u|_L-v|_L=b$. Then $du,dv$ glue into a
single 1-cocycle $h$, and $\delta(b)=[h]\in H^1(N,k)$ does not depend on the choices.
Moreover,
\begin{equation*}
\begin{aligned}
\delta(b)=0\quad\Longleftrightarrow\quad
&\exists a\in H^0(N_A,k),\ \exists c\in H^0(N_B,k),\\
&b=a|_L-c|_L.
\end{aligned}
\tag{2.19}\label{eq:2.19}
\end{equation*}

\end{proposition}

\begin{proof}

Surjectivity is obtained by extending the value on each simplex of $L$ to $N_A$. Since
$d(u|_L-v|_L)=db=0$, the cochains $du,dv$ agree and glue into $h$. That $dh=0$
follows from $d^2=0$ on both parts. The difference from another lift is the restriction of
a 0-cochain $w$ on $N$, and $h$ changes by $dw$. Moreover, if $h=dw$, then
$a=u-w|_{N_A}$ and $c=v-w|_{N_B}$ are 0-cocycles whose difference is $b$. For the
converse, it suffices to choose $a,c$ as the lifts.

\end{proof}

We can read this $\delta(b)$ as the obstruction created by the difference $b$ on the
boundary. Splitting a cycle $z=z_A+z_B$ into chains on the two parts,
$\partial z_A=-\partial z_B$ is supported on $L$. By Proposition~\ref{prop:2.28},
$\langle h,z\rangle=\langle b,\partial z_A\rangle$. The difference around the whole cycle
coincides with the signed sum of the differences compared on the boundary of the two
parts.

\section{Comparison of the structural and semantic phases}\label{sec:2.7}

Whether a fact belongs to structure or to semantics is determined by what changes under
the permitted changes of semantics. From this classification we split the coefficients and
study conditions under which no obstruction is lost by looking only at the semantic side.

\begin{definition}[Two phases by invariance of extraction]\label{def:2.30}

For the extraction doctrine $D$ of Chapter~\ref{chap:1}, choose a family
$(D_\lambda)_{\lambda\in\Lambda}$ that preserves the vocabulary, the normalization, the
resolution, and the structure of sources, and changes only the semantic reading and its
admissible predicates. We require this family to contain $D$ itself. A pair $(s,a)$ of a
source and an Atom is structural if
\begin{equation*}
\forall\lambda\in\Lambda,\quad
\bigl(a\in\mathrm{Atomize}_{D_\lambda}(s)
\Longleftrightarrow a\in\mathrm{Atomize}_D(s)\bigr)
\tag{2.20}\label{eq:2.20}
\end{equation*}
holds. A pair that is not structural is called semantic. For example, if the existence of
a component is extracted under every reading while only a fact about the interpretation
of values ceases to be extracted under some reading, then the former is structural and
the latter semantic.

\end{definition}

\begin{definition}[The two-phase coefficient complex]\label{def:2.31}

Assign to each coordinate of a finite-dimensional complex $C^\bullet_{\mathrm{all}}$ a
pair of a source and an Atom. Let $C^n_{\mathrm{str}}$ be the subspace spanned by the
coordinates belonging to structural pairs. We write Condition~S for the requirement that the
restrictions and the differentials preserve this part. Under this condition the
differential descends to $C^n_{\mathrm{sem}}=C^n_{\mathrm{all}}/C^n_{\mathrm{str}}$, and
\begin{equation*}
0\longrightarrow C^\bullet_{\mathrm{str}}
\longrightarrow C^\bullet_{\mathrm{all}}
\longrightarrow C^\bullet_{\mathrm{sem}}\longrightarrow0
\tag{2.21}\label{eq:2.21}
\end{equation*}
is an exact sequence in each degree. This $C^\bullet_{\mathrm{sem}}$ is the quotient
complex built from the two phases of extraction. In \S\ref{sec:2.9} we will separately
build coefficients $M_{\mathrm{sem}}$ from the relations of repair operations.

\end{definition}

\begin{proposition}[Retention of obstructions on the semantic side]\label{prop:2.32}

If Condition~S holds and $H^1(C^\bullet_{\mathrm{str}})=0$, then the induced map
\begin{equation*}
H^1(C^\bullet_{\mathrm{all}})
\longrightarrow H^1(C^\bullet_{\mathrm{sem}})
\tag{2.22}\label{eq:2.22}
\end{equation*}
is injective.

\end{proposition}

\begin{proof}

Suppose that a cocycle $g$ becomes a coboundary on the semantic side. Lift the
corresponding semantic 0-cochain to some $b\in C^0_{\mathrm{all}}$ by surjectivity in
degree 0. Then $g-d^0b$ is a cocycle belonging to the structural part. Since $H^1=0$ on
the structural side, there is a structural $c$ with $g-d^0b=d^0c$. Hence $g=d^0(b+c)$.

\end{proof}

If the structural complex satisfies the forest condition of Proposition~\ref{prop:2.26},
the required $H^1=0$ is obtained. The roles of the two conditions are visible in the
following small examples.

\begin{example}[A case where Condition S fails]\label{ex:2.33}

Let $s$ be a pair that is always extracted before and after the semantic changes, and $t$
a pair whose extraction truth value changes. On the complex with two vertices and one
edge, assign $s$ to the left vertex and $t$ to the right vertex and the edge. Then
$C^0_{\mathrm{all}}=k^2$, $C^1_{\mathrm{all}}=k$, and $d^0(a,b)=b-a$; the structural part
is $k(1,0)$ in degree 0 and zero in degree 1. Since $d^0(1,0)=-1$, the structural
part is not preserved, and this differential does not descend to the semantic quotient.

\end{example}

\begin{example}[A case where Condition S alone does not give injectivity]\label{ex:2.34}

Take two copies of the complex joining two vertices by two parallel edges, and assign $s$
to all coordinates of one copy and $t$ to all coordinates of the other. The differential
of each copy is $(a,b)\mapsto(b-a,b-a)$, and degree 2 is zero. The structural part is
exactly the first copy, so Condition~S holds. However, the 1-cochain $(1,0)$ of the
first copy is not a coboundary, because the values on the two edges differ. This nonzero
class maps to zero on the semantic side. It is thus the condition that obstructions
vanish on the structural side that supports the injectivity of
Proposition~\ref{prop:2.32}.

\end{example}

\section{Concrete integer-coefficient obstructions built from Atoms}\label{sec:2.8}

To compare with the diagnosis in the next chapter, we construct coefficients and covers
concretely. The generators of the coefficients are pairs of a Law and a source, and the
points of the geometry represent the ranges read locally. We keep both in a single Atom
family and compute obstruction classes from local states and transitions.

\subsection*{Building coefficients from generator relations}

\begin{definition}[Generators and primitive relations]\label{def:2.35}

Take a finite set of sources $S$ and a finite family of Laws $(\ell)_{\ell\in L}$. Each
Law has a set of values $\mathrm{Val}_\ell$ and an evaluation
$v_\ell:S\to\mathrm{Val}_\ell$. Let the set of generators be $G=L\times S$, and let the
label of $g=(\ell,s)$ be $(\ell,v_\ell(s))$. A primitive relation $R\subseteq G\times G$
is declared as a relation preserving these labels. Let $\mathfrak{B}$ be the quotient set
by the reflexive, symmetric, and transitive closure of $R$, and write
$[g]_R\in\mathfrak{B}$ for the component of a generator $g$. We define the
integer-coefficient group by
\begin{equation*}
M_R=\mathbb{Z}^{(G)}/\langle e_g-e_h\mid gRh\rangle
\tag{2.23}\label{eq:2.23}
\end{equation*}
Here $\mathbb{Z}^{(G)}$ is the free abelian group with basis $e_g$.

\end{definition}

\begin{lemma}[Standard form of the coefficients]\label{lem:2.36}

The map $e_g\mapsto e_{[g]_R}$ induces an isomorphism
$M_R\cong\mathbb{Z}^{(\mathfrak{B})}$.

\end{lemma}

\begin{proof}

Basis elements joined by a primitive relation map to the same component, so the map
descends to the quotient. In the other direction, choose a representative generator of
each component and assign its class in $M_R$. Two generators in the same component are
joined by a finite sequence of primitive relations, and their difference is the sum of
the differences of these relations, so the assignment does not depend on the choice. The
composites of the two maps are the identity on the bases, so they are mutually inverse.

\end{proof}

As a concrete example, let $S=\{0,1\}^2$, $L=\{\ell\}$, and $v_\ell(v,h)=v$. The
primitive relation consists of the two pairs joining $(v,0)$ and $(v,1)$ with the same
$v$. Writing the generators as $g_{00},g_{01},g_{10},g_{11}$, there are two components
$B_0=\{g_{00},g_{01}\}$ and $B_1=\{g_{10},g_{11}\}$, and $M_R\cong\mathbb{Z}^2$. Even
though the second component of the original sources is identified, the two components
with different Law values remain distinguished.

\subsection*{The eight-point space and two covers}

\begin{construction}[Atom input carrying points and generators]\label{cons:2.37}

Take a set $X$ with four vertices $a_0,a_1,b,c$ and four edge points $e_{01},ab,bc,ac$.
The endpoints of the edge points are $e_{01}:(a_0,a_1)$, $ab:(a_1,b)$, $bc:(b,c)$, and
$ac:(a_0,c)$. The order is defined only by the relations placing each vertex below the
edge points incident to it, together with reflexivity. The open sets are the upward
closed subsets.

We attach tags distinguishing point Atoms from generator Atoms, and let the Atom family
be $X\sqcup G$. A point Atom has the point itself as its subject, with fixed values for
the predicate, the payload, and the axis. A generator Atom $(\ell,s)$ has the source $s$,
the Law index $\ell$, and the Law value $v_\ell(s)$ as its subject, predicate, and
payload, respectively, with a single fixed axis. Each Atom is uniquely determined by its
kind and these coordinates. The relation of the configuration is the $R$ declared between
the generator Atoms.

The context $c_V$ corresponding to an open set $V\subseteq X$ reads the point Atoms of
$V$ and all generator Atoms. Inclusions $V'\subseteq V$ are the restriction morphisms,
and the overlaps are given by $c_{V'\cap V}$. For covers we adopt families that cover the
open set and also cover the generators that are read. We use the Grothendieck topology on
the context site generated by these families. In a nonempty covering family each context
reads all generators, so the covering condition on generators also holds. The two covers
below are covers of this context site.

\end{construction}

\begin{lemma}[The three-chart and four-chart covers]\label{lem:2.38}

The minimal open neighborhoods of the vertices are
\begin{equation*}
\begin{aligned}
U_{a_0}&=\{a_0,e_{01},ac\},& U_{a_1}&=\{a_1,e_{01},ab\},\\
U_b&=\{b,ab,bc\},& U_c&=\{c,bc,ac\}.
\end{aligned}
\tag{2.24}\label{eq:2.24}
\end{equation*}
These cover $X$. Setting $U_a=U_{a_0}\cup U_{a_1}$, the family $(U_a,U_b,U_c)$ is also a
cover. In both covers, the charts and the nonempty double overlaps are connected, and the
intersection of three distinct charts is empty.

\end{lemma}

\begin{proof}

Each vertex lies in its own neighborhood and each edge point lies in the neighborhoods of
its two endpoints, so these families are covers. Each minimal neighborhood has a least
element. Any attempt to separate it into two relatively open sets fails, because the side
containing the least element contains all points above it. Since $U_{a_0}$ and $U_{a_1}$
meet in $e_{01}$, their union is also connected. The nonempty double intersections of the
four-chart cover are $\{e_{01}\},\{ab\},\{bc\},\{ac\}$, and those of the three-chart
cover are $\{ab\},\{bc\},\{ac\}$. Each edge point has only two endpoints, so even after
merging the charts as above, no point is shared by three of them.

\end{proof}

\begin{figure}[htbp]
\centering
\begin{tikzpicture}[
    x=1mm, y=1mm,
    vertex/.style={circle, fill, inner sep=0pt, minimum size=4.6pt},
    edgepoint/.style={rectangle, draw, semithick, fill=white, inner sep=0pt, minimum size=5pt},
    order/.style={-{Stealth[length=4.5pt]}, semithick},
    nerve/.style={semithick},
    every label/.style={font=\normalsize},
    panel/.style={font=\small},
    note/.style={font=\small}
  ]
  \begin{scope}
    \node[panel, align=center] at (16,44) {(a) Order on the\\ eight-point space};
    \node[vertex, label=above:{$a_0$}] (a0) at (0,32) {};
    \node[vertex, label=above:{$a_1$}] (a1) at (32,32) {};
    \node[vertex, label=below:{$b$}]   (b)  at (32,0)  {};
    \node[vertex, label=below:{$c$}]   (c)  at (0,0)   {};
    \node[edgepoint, label=above:{$e_{01}$}] (e01) at (16,32) {};
    \node[edgepoint, label=right:{$ab$}]     (ab)  at (32,16) {};
    \node[edgepoint, label=below:{$bc$}]     (bc)  at (16,0)  {};
    \node[edgepoint, label=left:{$ac$}]      (ac)  at (0,16)  {};
    \draw[order] (a0) -- (e01);
    \draw[order] (a1) -- (e01);
    \draw[order] (a1) -- (ab);
    \draw[order] (b)  -- (ab);
    \draw[order] (b)  -- (bc);
    \draw[order] (c)  -- (bc);
    \draw[order] (c)  -- (ac);
    \draw[order] (a0) -- (ac);
    \node[note] at (16,-11) {arrows: vertex $<$ edge point};
  \end{scope}
  \begin{scope}[xshift=50mm]
    \node[panel, align=center] at (16,44) {(b) Nerve of the\\ four-chart cover};
    \node[vertex, label=above:{$U_{a_0}$}] (Ua0) at (0,32) {};
    \node[vertex, label=above:{$U_{a_1}$}] (Ua1) at (32,32) {};
    \node[vertex, label=below:{$U_b$}]     (Ub)  at (32,0)  {};
    \node[vertex, label=below:{$U_c$}]     (Uc)  at (0,0)   {};
    \draw[nerve] (Ua0) -- (Ua1) node[midway, above, note] {$\{e_{01}\}$};
    \draw[nerve] (Ua1) -- (Ub)  node[midway, right, note] {$\{ab\}$};
    \draw[nerve] (Ub)  -- (Uc)  node[midway, below, note] {$\{bc\}$};
    \draw[nerve] (Uc)  -- (Ua0) node[midway, left, note]  {$\{ac\}$};
    \node[note] at (16,-11) {edge labels: double intersections};
  \end{scope}
  \begin{scope}[xshift=100mm]
    \node[panel, align=center] at (16,44) {(c) Nerve of the\\ three-chart cover};
    \node[vertex, label=above:{$U_a$}] (Ua) at (16,32) {};
    \node[vertex, label=below:{$U_b$}] (Vb) at (32,0)  {};
    \node[vertex, label=below:{$U_c$}] (Vc) at (0,0)   {};
    \draw[nerve] (Ua) -- (Vb) node[midway, right=2pt, note] {$\{ab\}$};
    \draw[nerve] (Vb) -- (Vc) node[midway, below, note]     {$\{bc\}$};
    \draw[nerve] (Vc) -- (Ua) node[midway, left=2pt, note]  {$\{ac\}$};
    \node[note] at (16,-11) {$U_a = U_{a_0} \cup U_{a_1}$};
  \end{scope}
\end{tikzpicture}
\caption{\textbf{The eight-point space and the two covers.} The left panel shows the
eight points of the space $X$: circles are vertices, squares are edge points, and arrows
indicate the order from vertices to edge points. The middle and right panels show the
nerves of the covers, with nodes for charts and edges for nonempty double intersections.
When $U_{a_0}$ and $U_{a_1}$ in the middle are merged into $U_a$, their intersection
point $e_{01}$ moves into the interior of $U_a$. The triangle on the right has no face,
and the triple intersection is empty.}\label{fig:2.1}
\end{figure}
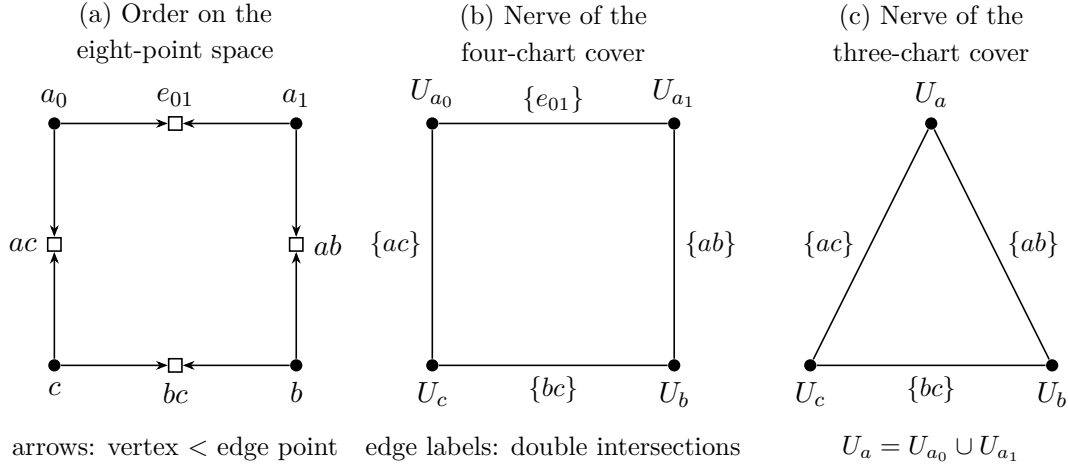

In this computation, the vanishing of degree 2 comes from the actual intersections
being empty.

\begin{construction}[The sheaf of locally constant obstruction coefficients]\label{cons:2.39}

Let $F_R(c_V)$ be the group of locally constant $M_R$-valued functions on $V$, with
restriction of functions as restriction. Functions agreeing on overlaps glue uniquely,
and local constancy is checked locally, so $F_R$ is a sheaf. On the empty open set it is
the zero group. The map $c_V\mapsto V$ from contexts to open sets preserves intersections
and covers, so these coefficients form an actual sheaf on the chosen context site.

By the connectedness of Lemma~\ref{lem:2.38}, a section over a chart or a nonempty
intersection is determined by a single value in $M_R$. In this presentation the
restrictions are identity homomorphisms. Hence the \v{C}ech complexes of both covers are
complexes with $M_R$-values on vertices and edges and zero in degree 2.

\end{construction}

\subsection*{Computing obstructions from local states and affine transitions}

\begin{proposition}[Lawful chart states]\label{prop:2.40}

Through the isomorphism of Lemma~\ref{lem:2.36}, read each chart state $p_\alpha\in M_R$
as componentwise integers $(p_{\alpha,B})_{B\in\mathfrak{B}}$. Define the polynomial ring
and the relation ideal by
\begin{equation*}
\begin{aligned}
P_R&=\mathbb{Z}[X_g\mid g\in G],\\
J_R&=(X_g-X_h\mid gRh),\\
e_\alpha:P_R&\longrightarrow\mathbb{Z},
\qquad X_g\longmapsto p_{\alpha,[g]_R}
\end{aligned}
\tag{2.25}\label{eq:2.25}
\end{equation*}
This evaluation sends $J_R$ to zero and gives an integer point of
$\mathrm{Spec}(P_R/J_R)$. The states on the overlaps obtained by actual restriction are
also lawful. Moreover, the states after translation by an arbitrary $\xi_e\in M_R$ are
lawful.

\end{proposition}

\begin{proof}

If $gRh$, then $[g]_R=[h]_R$, so the evaluation of each generator $X_g-X_h$ is zero.
Restriction to an overlap preserves the integer value of each component. After a
translation, the same $\xi_{e,B}$ is added to the same component, so the identities are
preserved.

\end{proof}

The polynomial equations here require the coordinates to agree along the declared
primitive relations. The Law values of Definition~\ref{def:2.35} retain the origin of
each generator, and in the next chapter we use them to map these coefficients into the
diagnosis.

\begin{proposition}[Obstruction of specified affine local data]\label{prop:2.41}

Write $\mathcal{U}_q$ for either of the covers and choose local data $x=(\xi,p)$. Here
$p\in C^0(\mathcal{U}_q,F_R)$ consists of the chart states, and
$\xi\in C^1(\mathcal{U}_q,F_R)$ is the translation carrying the state on the right into
the coordinates on the left across each overlap. The orientation of the edges is chosen
by the order of the charts. When an orientation is reversed, the signs of the transition
and of the comparison value are reversed. The comparison value on an edge
$e:\alpha\to\beta$ is
\begin{equation*}
\begin{aligned}
o_q(x)_e&=\xi_e+\mathrm{res}_{\beta,e}(p_\beta)
-\mathrm{res}_{\alpha,e}(p_\alpha),\\
o_q(x)&=\xi+d^0p.
\end{aligned}
\tag{2.26}\label{eq:2.26}
\end{equation*}
This is the difference obtained by actually restricting the lawful states on the two
sides, carrying one of them by the transition, and then comparing. Since degree 2 is
zero, it is a cocycle and defines $[o_q(x)]\in\check H^1(\mathcal{U}_q,F_R)$. Changing
the local states to $p+h$ gives
\begin{equation*}
o_q(\xi,p+h)=o_q(\xi,p)+d^0h,
\qquad [o_q(\xi,p+h)]=[o_q(\xi,p)]
\tag{2.27}\label{eq:2.27}
\end{equation*}
\end{proposition}

\begin{proof}

Equation~\eqref{eq:2.26} computes the difference between the transported right state and
the left state, and by Proposition~\ref{prop:2.40} both states being compared are lawful.
Additivity of the differential gives \eqref{eq:2.27}.

This class is also obtained as the gluing obstruction of Proposition~\ref{prop:2.21}.
Place a locally constant $M_R$-valued state on each chart and compare coordinates from
right to left on each overlap by $m\mapsto\xi_e+m$. The reverse direction uses $-\xi_e$,
and self-comparison is zero. Since the intersection of three distinct charts is empty, no
additional condition arises on triple overlaps. Defining the states over an open set as
the families of local states satisfying these comparisons yields a sheaf. On each chart
and each nonempty intersection, the action of the coefficients $F_R$ is free and
transitive, and the difference of the $p$ chosen in chart coordinates equals
\eqref{eq:2.26}. Hence the same cochain yields the same obstruction class.

\end{proof}

\begin{example}[Changing local states and changing transitions]\label{ex:2.42}

Order the three-chart cover by $a<b<c$ and let $u=e_{B_0}\in M_R\cong\mathbb{Z}^2$.
Choose $p=0$, $\xi_{ab}=u$, and $\xi_{ac}=\xi_{bc}=0$. The period of
Example~\ref{ex:2.19} is $u$, so $[o_q(x)]\ne0$. The chart states each satisfy the Laws of
Proposition~\ref{prop:2.40}, but under the fixed transitions they do not glue into a
global state.

Changing only $p$ does not change the class, by \eqref{eq:2.27}. On the other hand,
changing $\xi_{ab}$ itself to zero leaves $p=0$ compatible as it is, and the obstruction
becomes zero. This computation distinguishes the differences removable by changing local
coordinates from those removable only by changing the transitions.

The data passed to Chapter~\ref{chap:3} are the coefficients $F_R$, the cover
$\mathcal{U}_q$, the local data $x$, and the concrete class $[o_q(x)]$ generated from
them.

\end{example}

\section{Semantic local repair and obstruction classes of equations}\label{sec:2.9}

On the repair side, we build coefficients from the available changes and the relations
identifying changes with the same effect. On the equation side, we use the quotient
coefficients of \S\ref{sec:2.3}. After building an obstruction class from the local
states on each side, we compare the two in the next section.

Throughout, we fix a finite cover $\mathcal{U}$ by injective morphisms as in
Definition~\ref{def:2.16}. We call the diagram consisting of the charts, the nonempty
double and triple intersections, and their restriction morphisms the intersection
diagram of the cover. The algebraic conditions needed for the comparison are imposed on
this diagram.

\begin{definition}[Presentation of semantic repair]\label{def:2.43}

To each context $V$ we assign a set $\Lambda(V)$ of Atoms distinguished on the semantic
side, a map $\pi_V:\Lambda(V)\to\mathrm{At}$ giving the underlying Atom, and a subset
$S(V)\subseteq\Lambda(V)$ usable for repair. These carry functorial restrictions; the
restrictions preserve $S$ and commute with $\pi$. Here the underlying Atoms are read as
the same under restriction.

A repair word is a finite integer combination of elements of $S(V)$. We declare, as a subgroup
$R_{\mathrm{rep}}(V)$, the differences of words regarded as locally having the same
effect, and require this subgroup to be preserved by restriction as well.
We define the semantic coefficients by
\begin{equation*}
\begin{aligned}
F_{\mathrm{rep}}(V)&=\mathbb{Z}^{(S(V))},\\
M_{\mathrm{sem}}(V)&=F_{\mathrm{rep}}(V)/R_{\mathrm{rep}}(V)
\end{aligned}
\tag{2.28}\label{eq:2.28}
\end{equation*}
The restrictions extend linearly to the free abelian groups and preserve the relation
subgroups, so they descend to the quotients. The identity and composition rules hold on
generators, so $M_{\mathrm{sem}}$ is a presheaf with values in abelian groups.

\end{definition}

\begin{proposition}[Torsors obtained from the action of repair words]\label{prop:2.44}

Suppose that $F_{\mathrm{rep}}$ acts on a presheaf $P_{\mathrm{sem}}$ of semantic local
repair states and that the action commutes with restriction. On the intersection diagram
we impose the following three conditions.

\begin{itemize}
\item If $w\in R_{\mathrm{rep}}(V)$, then $w+p=p$ for every state.
\item If $w+p=p$, then $w\in R_{\mathrm{rep}}(V)$.
\item For any two states $p,q$, some repair word $w$ satisfies $q=w+p$.
\end{itemize}

Then each nonempty $P_{\mathrm{sem}}(V)$ is an $M_{\mathrm{sem}}(V)$-torsor.

\end{proposition}

\begin{proof}

By the first condition, the action descends to the quotient by the relations. By the
second condition, a repair that is nonzero in the quotient fixes no state, so the action
is free. The third condition gives transitivity.

\end{proof}

The first two conditions state that the repair words declared to have zero effect
coincide with the words that actually leave every state unchanged. The third condition
states that the states being compared can reach one another by the chosen repairs.

\begin{definition}[Two local states and their obstruction classes]\label{def:2.45}

On the semantic side, choose a local atlas $p_\alpha\in P_{\mathrm{sem}}(U_\alpha)$. On
the equation side, choose $Q_E$ of Construction~\ref{cons:2.12} and a presheaf $P_E$ of
local states on which it acts. We require $P_E(V)$ on each nonempty intersection to be a
$Q_E(V)$-torsor, with restrictions commuting with the action. Let its local atlas be
$e_\alpha\in P_E(U_\alpha)$. An element of $P_E$ represents a state that concretely
realizes the chosen local equation data.

With the two coefficients we form the \v{C}ech complexes
$C^\bullet_{\mathrm{rep}}=C^\bullet(\mathcal{U},M_{\mathrm{sem}})$ and
$C^\bullet_E=C^\bullet(\mathcal{U},Q_E)$, and define the residuals by
\begin{equation*}
\begin{aligned}
r_{\mathrm{sem},\alpha\beta}&=p_\beta|-p_\alpha|,\\
r_{E,\alpha\beta}&=e_\beta|-e_\alpha|
\end{aligned}
\tag{2.29}\label{eq:2.29}
\end{equation*}
By Proposition~\ref{prop:2.21}, both are cocycles and their classes do not depend on the
choice of atlas. We abbreviate the semantic group
$\check H^1(\mathcal{U},M_{\mathrm{sem}})$ to $H^1_{\mathrm{sem}}(\mathcal{U})$.

On empty intersections, we take the two coefficients to be the zero groups and the
two state sets to have at most one element. Since the local atlases restrict, the state set on each
nonempty intersection is nonempty.

\end{definition}

\begin{construction}[Correspondence between Atoms and equations]\label{cons:2.46}

To each $\lambda\in S(V)$ we assign a required equation index $i_\lambda$ and a reading
$A_\lambda$ of an object, preserved under restriction. We define the map sending a
repair generator to the residual class of the corresponding equation by
\begin{equation*}
\chi_V(\lambda)
=[\varepsilon_{V,A_\lambda,i_\lambda,\pi_V(\lambda)}]
\in Q_E(V)
\tag{2.30}\label{eq:2.30}
\end{equation*}
By the restriction rule for residuals, $\chi$ commutes with restriction. By the
universal property of free abelian groups, it extends uniquely to a homomorphism
$\widetilde\chi_V:F_{\mathrm{rep}}(V)\to Q_E(V)$. The map $\chi$ specifies which repair
generator represents the residual of which object, equation, and Atom.

\end{construction}

\section{The SAGA comparison and global repair}\label{sec:2.10}

We now connect the two constructions above by a correspondence of local states. To
obtain an isomorphism of coefficients, we need the relations among repair words to hold
on the equation side as well, every repair word lost on the equation side to be
contained in the relations, and the coefficients on the equation side to be expressible
by the repair generators. The first property can be derived from the correspondence of
states.

\begin{definition}[Correspondence of local states and completeness]\label{def:2.47}

On the intersection diagram, we construct maps
$\beta_V:P_{\mathrm{sem}}(V)\to P_E(V)$ commuting with restriction, and require, for
each generator,
\begin{equation*}
\beta_V(\lambda+p)=\chi_V(\lambda)+\beta_V(p)
\tag{2.31}\label{eq:2.31}
\end{equation*}
The $\lambda$ on the left-hand side denotes the action of a basis element of the free
abelian group. Furthermore, on each nonempty intersection $V$ we require
\begin{equation*}
\ker\widetilde\chi_V\subseteq R_{\mathrm{rep}}(V),
\qquad
\mathrm{im}\,\widetilde\chi_V=Q_E(V)
\tag{2.32}\label{eq:2.32}
\end{equation*}
We call the first condition completeness of relations and the second completeness of
generators. The former says that the repair relations account for all words whose effect
becomes zero on the equation side, and the latter that each coefficient on the equation
side can be expressed by a repair word.

\end{definition}

\begin{lemma}[The correspondence of states gives soundness of the relations]\label{lem:2.48}

Under the conditions of Proposition~\ref{prop:2.44} and
Definitions~\ref{def:2.45} and~\ref{def:2.47},
\begin{equation*}
R_{\mathrm{rep}}(V)\subseteq\ker\widetilde\chi_V
\tag{2.33}\label{eq:2.33}
\end{equation*}
holds.

\end{lemma}

\begin{proof}

Extending \eqref{eq:2.31} to finite sums and additive inverses, we get
$\beta_V(w+p)=\widetilde\chi_V(w)+\beta_V(p)$ for every repair word $w$. Let
$w\in R_{\mathrm{rep}}(V)$ and take a state $p$ obtained from the local atlas; then
\[
\beta_V(p)=\beta_V(w+p)
=\widetilde\chi_V(w)+\beta_V(p).
\]
Since the action on the equation side is free, $\widetilde\chi_V(w)=0$.

\end{proof}

\begin{theorem}[The SAGA comparison of coefficients and obstruction classes]\label{thm:2.49}

Under the data and conditions from Definition~\ref{def:2.43} through
Definition~\ref{def:2.47}, we obtain a natural isomorphism on the intersection diagram
\begin{equation*}
\Phi:M_{\mathrm{sem}}\xrightarrow{\sim}Q_E,
\qquad\Phi_V([w])=\widetilde\chi_V(w)
\tag{2.34}\label{eq:2.34}
\end{equation*}
It induces isomorphisms of the complexes in degrees 0 through 2 and of the first
cohomology, and matches the obstruction classes built from the two local atlases.
\begin{equation*}
\begin{aligned}
\kappa^n &: C^n_{\mathrm{rep}}\xrightarrow{\sim}C^n_E,
\qquad (\kappa^nc)_V=\Phi_V(c_V)\quad(n=0,1,2),\\
\kappa^{n+1}d^n_{\mathrm{rep}}&=d^n_E\kappa^n
\quad(n=0,1),
\end{aligned}
\tag{2.35}\label{eq:2.35}
\end{equation*}
\begin{equation*}
\begin{aligned}
\kappa_* &: H^1_{\mathrm{sem}}(\mathcal{U})\xrightarrow{\sim}\check H^1(\mathcal{U},Q_E),\\
\kappa_*[r_{\mathrm{sem}}]&=[r_E].
\end{aligned}
\tag{2.36}\label{eq:2.36}
\end{equation*}

\end{theorem}

\begin{proof}

We first compare the coefficients. By Lemma~\ref{lem:2.48}, $\widetilde\chi_V$ descends
to the quotient and gives $\Phi_V$. If $\Phi_V([w])=0$, then completeness of relations
gives $w\in R_{\mathrm{rep}}(V)$, so $[w]=0$. By completeness of generators, $\Phi_V$ is
also surjective. Commutation with restriction holds for $\chi$ and extends from the
generators to the whole. This gives \eqref{eq:2.34}.

Next we apply $\Phi_V$ to the components on each intersection. By additivity and
commutation with restriction, in degree 0 for example
\[
\begin{aligned}
(\kappa^1d^0_{\mathrm{rep}}b)_{\alpha\beta}
&=\Phi_{U_{\alpha\beta}}(b_\beta|-b_\alpha|)\\
&=\Phi_{U_\beta}(b_\beta)|-\Phi_{U_\alpha}(b_\alpha)|\\
&=(d^0_E\kappa^0b)_{\alpha\beta}.
\end{aligned}
\]
The same computation applies to the alternating sum of three terms in degree 1. Using
the inverse of each $\Phi_V$ componentwise gives the inverse map of complexes, so
cocycles and coboundaries are preserved in both directions. Hence
$[c]\mapsto[\kappa^1c]$ gives the isomorphism of \eqref{eq:2.36}.

Finally we compare the specified residuals. On each chart, take
$h_\alpha\in Q_E(U_\alpha)$ with $e_\alpha=h_\alpha+\beta(p_\alpha)$. Descending
\eqref{eq:2.31} to the action of the quotient and computing the difference on the
overlaps gives
\begin{equation*}
r_E=\kappa^1(r_{\mathrm{sem}})+d^0_Eh
\tag{2.37}\label{eq:2.37}
\end{equation*}
Hence the two residuals correspond to the same cohomology class.

\end{proof}

When the two atlases are aligned by $e_\alpha=\beta(p_\alpha)$, the $h$ in
\eqref{eq:2.37} is zero. Even when they are chosen separately, the difference is an
explicit coboundary, so the correspondence of obstruction classes is preserved. This is
the comparison of
\cite[\href{https://arxiv.org/html/2608.21458v1\#S5}{Theorem~5.1(i)--(ii),
\S\S5.3--5.5}]{SAGA}.

\begin{corollary}[SAGA and global repair]\label{cor:2.50}

In addition to the conditions of Theorem~\ref{thm:2.49}, suppose that the actual repair
states $P_{\mathrm{sem}}$ form a sheaf on the chosen site. Then
\begin{equation*}
P_{\mathrm{sem}}(W)\ne\varnothing
\quad\Longleftrightarrow\quad
[r_{\mathrm{sem}}]=0
\quad\Longleftrightarrow\quad
[r_E]=0.
\tag{2.38}\label{eq:2.38}
\end{equation*}

\end{corollary}

\begin{proof}

The first equivalence is obtained by applying Theorem~\ref{thm:2.22} to the torsor
action derived from the repair words. The second follows from the isomorphism of
Theorem~\ref{thm:2.49} and the correspondence of residual classes. Concretely, finding a
correction with $r_{\mathrm{sem}}=d^0b$ and gluing $(-b_\alpha)+p_\alpha$ by the sheaf
condition yields a global repair.

\end{proof}

This is the global repair statement of
\cite[\href{https://arxiv.org/html/2608.21458v1\#S5.SS6}{Theorems~5.1(iii) and~5.2,
\S5.6}]{SAGA}. The zero class expressing the existence of corrections and the gluing of
the repair states themselves are tied together by the sheaf condition.

\begin{example}[Two presentations by parity]\label{ex:2.51}

We use the four-chart cover of Lemma~\ref{lem:2.38}. On the semantic side, we declare on
each nonempty intersection a repair generator $\sigma$ and the relation $2\sigma=0$. On
the equation side, we take the ring $\mathbb{Z}$, the symbolic generator 2, and one
required residual equal to 1. The coefficients on the two sides are obtained as
\begin{equation*}
M_{\mathrm{sem}}=\mathbb{Z}\sigma/(2\sigma),
\qquad Q_E=\mathbb{Z}/(2),
\qquad\chi(\sigma)=[1]
\tag{2.39}\label{eq:2.39}
\end{equation*}
The restrictions between intersections are identity maps. The coefficients are constant
presheaves on nonempty contexts and zero on the empty context. The action of the
semantic repair words is determined by parity alone, and its stabilizer subgroup is
$2\mathbb{Z}\sigma$. The kernel of the map $n\sigma\mapsto[n]$ is $2\mathbb{Z}\sigma$
and its image is all of $\mathbb{Z}/(2)$, so the two completeness conditions can be
checked directly.

The local states can also be constructed. On the semantic side we place the two states
$\{0,1\}$ with changes by parity, and on the equation side the two residue classes
$\{[0],[1]\}$ with the additive action. Orient the four edges as
$a_0\to a_1\to b\to c\to a_0$, and on both sides choose the comparison that shifts the
reference by one on the first edge only and leaves it unshifted on the other three.
Since the coefficients have characteristic 2, returning the last edge to the
orientation of the total order does not change the values. As in
Proposition~\ref{prop:2.41}, these define a sheaf of states obtained by gluing chart
states along affine transitions. The map sending a semantic state $n$ to $[n]$ on the
equation side is $\beta$, and it commutes with restriction and with the action of
generators.

Choosing the zero state on every chart, the residuals on the two sides become
$(1,0,0,0)$ in their respective presentations. The sum over the four edges is 1, while
the sum of any coboundary is zero, so both obstruction classes are nonzero.
Theorem~\ref{thm:2.49} matches these two nonzero classes, and Corollary~\ref{cor:2.50}
shows that no global repair exists under the fixed transitions. If the first transition
is also changed to zero, the zero states glue as they are, and the obstruction classes
on both sides become zero.

\end{example}

\section*{Summary of the Chapter}

In this chapter we connected equations, comparison of local states, and semantic repair
as follows.

\begin{itemize}
\item \textbf{Geometry of Laws.} We built local algebras from the structural relations
and ideals from the symbolic equations. Under the algebraic presentation of the
evaluation data and the localization condition, simultaneous vanishing of the residuals
coincides with factorization through the lawful locus (Theorem~\ref{thm:2.9}).
\item \textbf{Obstructions to local consistency.} The differences on overlaps form a
cocycle, and changing the local states adds a coboundary. When the corrections are
realized as a torsor action and the states form a sheaf, the zero class is equivalent to
the existence of a global state (Theorem~\ref{thm:2.22}).
\item \textbf{Roles of covers and coefficients.} On a forest, obstructions vanish by
surjectivity of the restrictions, while a nonzero period along a cycle detects a
concrete obstruction. In the comparison that removes the structural phase, the
differential preserving the structural part and the vanishing of first-degree
obstructions on the structural side give injectivity
(Propositions~\ref{prop:2.26}, \ref{prop:2.28}, and~\ref{prop:2.32}).
\item \textbf{Concrete local data.} From an Atom family with points and generators we
built covers and an integer-coefficient sheaf, and obtained $o_q(x)=\xi+d^0p$ as the
comparison of lawful chart states along affine transitions
(Proposition~\ref{prop:2.41}).
\item \textbf{Correspondence with the semantics of repair.} The two coefficients, one
built from a presentation by repair words and the other from the quotient by the
equations, become isomorphic under the correspondence of local states and the
completeness conditions. This isomorphism matches the specified obstruction classes and,
combined with the sheaf condition on the actual repair states, characterizes the
existence of a global repair (Theorem~\ref{thm:2.49}, Corollary~\ref{cor:2.50}).
\end{itemize}

For the three services at the opening of the chapter, the correction summed around the
cycle expresses the discrepancy that remains however the reference of each part is
rechosen. If the obstruction class is zero, the local corrections can be applied so that
the states agree on the overlaps. That the corrected states form a single global state
is guaranteed by the sheaf condition on the actual states.

In the next chapter we study what is preserved in the diagnosis when the resolution or
the cover reading the same Laws is changed. We map the concrete obstruction classes
obtained in \S\ref{sec:2.8} into the diagnosis generated from Law evaluations, and
clarify the conditions under which the zero and nonzero information is preserved.

\part[Diagnosis and Transport]{Diagnosis and Transport\\[0.8em]{\large Chapters 3--4}}
\chapter{Canonical Resolution and Diagnostic Invariance}\label{chap:3}

\section*{Overview of the Chapter}

The question of this chapter is \textbf{under what conditions the same obstruction can be
diagnosed when we merge information or repartition the regions that are read locally}.
In Chapter~\ref{chap:2} we built obstruction classes to gluing from local states and
transitions. To share that verdict across different resolutions, we must specify the
information to retain and the maps that compare local data.

For example, suppose that a state is a pair of a displayed value and an internal value,
and that the chosen Law reads only the displayed value. Merging states with the same
displayed value loses no Law value. If, however, we merge even the regions on which
local states are placed, the cycles of overlaps may change. Preserving Law values and
preserving obstructions on cycles each require conditions to be checked.

In the first half we construct the coarsest resolution that preserves the evaluation of
the Laws, and generate a diagnostic complex from its evaluation values. We then build an
actual cochain map from a comparison of covers, and prove a sufficient condition for it
to be an isomorphism on first cohomology. In the second half we reduce invariance over
all Law families for which both resolutions are adequate to a finite computation, and
connect the integer-coefficient obstructions of Chapter~\ref{chap:2} with the
rational-valued diagnoses of this chapter.

\begin{xltabular}{\linewidth}{@{}LLL@{}}
\toprule
Question & Construction & Outcome \\
\midrule
\endhead
How far can information be merged while the Laws remain readable? & Identify sources on which all Law values agree & Universality of the canonical resolution, and conditions for presentation by specified extraction methods (\S\S\ref{sec:3.1}--\ref{sec:3.2}) \\
\addlinespace[3pt]
How to compare changes of resolution and of local structure? & Build maps from coefficients per Law value and from comparisons of edges and faces & Maps carrying diagnostic classes, and isomorphism under Condition C (\S\S\ref{sec:3.3}--\ref{sec:3.5}) \\
\addlinespace[3pt]
Is the diagnosis preserved whichever Laws are chosen? & Compute kernels and cokernels for each subset of values & A necessary and sufficient condition for uniform invariance, finite decision, and a counterexample for local observation (\S\S\ref{sec:3.6}--\ref{sec:3.7}) \\
\addlinespace[3pt]
Does fixing the structure also fix the obstruction? & Examine the structural support and the chosen coefficients separately & Invariance of the structural nerve, and vanishing for per-source coefficients (\S\ref{sec:3.8}) \\
\addlinespace[3pt]
Does a vanishing diagnosis mean a vanishing actual obstruction? & Send integer generators to Law-value coordinates, and turn rational corrections into integer ones & Correspondence of specified obstruction classes and equivalence of vanishing (\S\S\ref{sec:3.9}--\ref{sec:3.10}) \\
\addlinespace[3pt]
Does refining the cover give the same verdict? & Use an actual refinement from three charts to four & The comparison square, invariance of the obstruction verdict, and concrete zero and nonzero examples (\S\ref{sec:3.11}) \\
\bottomrule
\end{xltabular}

\subsection*{Why compare diagnoses}

The same state of software may be read down to its internal values, or only through its
published values. Local consistency, too, can be examined component by component, or
over regions that group several components together. Comparing results while varying
such choices requires maps that preserve the meaning of diagnostic values.

The diagnosis of this chapter is a first cohomology built from Law values and overlap
information, together with specified classes in it. For the lenses and protocols of
Chapter~\ref{chap:1}, choosing the required readouts and results of operations as the
evaluations determines what the resolution retains. For the gluing problem of
Chapter~\ref{chap:2}, what matters is that the vanishing of a diagnostic class can be
brought back to the existence of local corrections.
\S\S\ref{sec:3.9}--\ref{sec:3.11} provide this way back, with the same inputs and
concrete maps.

\section{The canonical resolution preserving Law evaluations}\label{sec:3.1}

A resolution specifies which sources are distinguished. Here we represent by a
surjection the part of the choices carried by a reading of Chapter~\ref{chap:1} that
concerns this distinction. Local covers and coefficients are added from
\S\ref{sec:3.3} on.

\begin{definition}[Resolutions and Law evaluations]\label{def:3.1}

Let $S$ be the set of sources, and give a resolution as a surjection $q:S\to Q$.
Fix a finite Law index set $L$ and evaluations
$v_\ell:S\to\mathrm{Val}_\ell$.
These are the same kind of input as the Law evaluations used in Definition~\ref{def:2.35}.
When studying the equations of Chapter~\ref{chap:1}, chosen residuals and the like can
serve as evaluations.

We say that $q$ is $L$-adequate if for each $\ell\in L$ there exists
$\bar v_\ell:Q\to\mathrm{Val}_\ell$ with
\begin{equation*}
v_\ell=\bar v_\ell q
\tag{3.1}\label{eq:3.1}
\end{equation*}
This is the condition that ``the value of each Law can be read back from $q(s)$''.
Since the data involved differ from those of the adequacy of Chapter~\ref{chap:1} (that a cover
can read information and interactions), we call this notion $L$-adequacy to distinguish
the two.

For two resolutions we write
\begin{equation*}
q_c\preceq q_f
\quad\Longleftrightarrow\quad
\forall s,t\in S,\
q_f(s)=q_f(t)\Rightarrow q_c(s)=q_c(t)
\tag{3.2}\label{eq:3.2}
\end{equation*}
and say that $q_c$ is coarser than $q_f$.
Two sources identified on the finer side are identified on the coarser side as well.

\end{definition}

\begin{lemma}[Descent of evaluations and comparison maps]\label{lem:3.2}

The following hold.

\begin{enumerate}
\item A map $v:S\to V$ descends through $q$ if and only if
$q(s)=q(t)\Rightarrow v(s)=v(t)$. The descended map is unique.
\item If $q_c\preceq q_f$, there is a unique surjection $\pi:Q_f\to Q_c$ with
$q_c=\pi q_f$.
\item If both sides are $L$-adequate, then $\bar v_{\ell,c}\pi=\bar v_{\ell,f}$.
\end{enumerate}

\end{lemma}

\begin{proof}

The forward direction of 1 follows by composing maps. For the converse, take a
representative $s$ of $u\in Q$ and set $\bar v(u)=v(s)$. By the hypothesis this does
not depend on the representative, and by the surjectivity of $q$ it is unique.
For 2, apply 1 with $v=q_c$; the surjectivity of $q_c$ makes $\pi$ surjective as well.
For 3, both sides equal $v_\ell(s)$ at $q_f(s)$, so they agree by surjectivity.

\end{proof}

\begin{definition}[The canonical resolution]\label{def:3.3}

Define an equivalence relation on sources by
\begin{equation*}
s\sim_L t
\quad\Longleftrightarrow\quad
\forall\ell\in L,\ v_\ell(s)=v_\ell(t)
\tag{3.3}\label{eq:3.3}
\end{equation*}
We call the quotient map $q_L:S\to S/{\sim_L}$ the canonical resolution of this
Law family.

\end{definition}

\begin{theorem}[Universality of the canonical resolution]\label{thm:3.4}

The map $q_L$ is $L$-adequate, and for every $L$-adequate $q:S\to Q$ there is a
unique map $u_q:Q\to S/{\sim_L}$ satisfying
\begin{equation*}
u_q q=q_L
\tag{3.4}\label{eq:3.4}
\end{equation*}
Hence $q_L\preceq q$.
A resolution with this property is unique up to a unique bijection commuting with the
maps from the sources.

\end{theorem}

\begin{proof}

The assignment $[s]\mapsto v_\ell(s)$ does not depend on the representative by
\eqref{eq:3.3}, so $q_L$ is adequate. If $q$ is adequate, then $q(s)=q(t)$ implies
that all Law values agree, hence $s\sim_L t$. By Lemma~\ref{lem:3.2} the map $u_q$
exists and is unique. If two resolutions have the same universal property, we obtain
maps through each other. Both composites commute with the maps from the sources, and
by uniqueness they are the identities.

\end{proof}

The canonical resolution is the coarsest among the resolutions that can read the
chosen Laws. The distinctions to be retained are thus determined before any choice of
implementation or of local covers.

\begin{example}[Displayed and internal values]\label{ex:3.5}

Take $S=\{0,1\}^2$ and a single Law $v_\ell(v,h)=v$.
The canonical resolution has the same equivalence relation as $q_c(v,h)=v$, which
reads the first component. The identity $q_f(v,h)=(v,h)$ is a finer adequate
resolution, and the comparison map is $\pi(v,h)=v$.
The states $(v,0)$ and $(v,1)$ can be merged, while $v=0$ and $v=1$, which the Law
distinguishes, remain apart. Adding a Law that also reads the internal value makes the
canonical resolution agree with the identity.

\end{example}

\section{Finite computation and presentation by extraction methods}\label{sec:3.2}

That the quotient of the canonical resolution can be formed, and that an available
extraction method realizes this quotient, are verified separately. For finite sources,
both can be computed as comparisons of equivalence relations.

\begin{construction}[Computing the finite partition]\label{cons:3.6}

Suppose that $S$ is finite and that equality of evaluation values is decidable.
For each source, compute the finite tuple $(v_\ell(s))_{\ell\in L}$.
Let $\mathcal{P}_L$ be the family of nonempty subsets obtained by grouping sources
according to agreement of these tuples.
The map $s\mapsto\{t\in S\mid t\sim_L s\}$ is a surjection onto $\mathcal{P}_L$.
Equality of two outputs is equivalent to \eqref{eq:3.3}, so this computation presents
the canonical resolution. If the Law index set is empty and the set of sources is not,
everything merges into a single component.

\end{construction}

\begin{definition}[Presentation by permitted extraction methods]\label{def:3.7}

Let $\mathcal{D}$ be a specified family of extraction doctrines of
Chapter~\ref{chap:1} over a common source set $S$.
The resolution induced by $D\in\mathcal{D}$ is the quotient by
\begin{equation*}
s\sim_D t
\quad\Longleftrightarrow\quad
\mathrm{Atomize}_D(s)=\mathrm{Atomize}_D(t)
\tag{3.5}\label{eq:3.5}
\end{equation*}
When $\sim_D=\sim_L$ for some $D\in\mathcal{D}$, we say that the canonical resolution
is presentable by this family.
Given a finite $\mathcal{D}$ and a method for deciding equality of extracted families,
presentability can be decided by comparing \eqref{eq:3.3} and \eqref{eq:3.5} over all
pairs of sources.

\end{definition}

\begin{example}[Adequate extraction methods alone do not yield presentability]\label{ex:3.8}

Take six sources $a_0,a_1,a_2,b_0,b_1,b_2$, with Law value 0 on the $a_i$ and 1 on
the $b_i$. Each of the following rows is a partition of the sources on which the
extraction results agree.

\begin{xltabular}{\linewidth}{@{}LL@{}}
\toprule
Extraction method & Components of sources with the same result \\
\midrule
\endhead
$D_0$ & $\{a_0,a_1,a_2\},\ \{b_0,b_1,b_2\}$ \\
\addlinespace[3pt]
$D_a$ & $\{a_0\},\ \{a_1,a_2\},\ \{b_0,b_1,b_2\}$ \\
\addlinespace[3pt]
$D_b$ & $\{a_0,a_1,a_2\},\ \{b_0\},\ \{b_1,b_2\}$ \\
\bottomrule
\end{xltabular}

Assigning a distinct Atom to each component and taking its singleton as the
extraction result realizes these partitions by actual doctrines. In
Definition~\ref{def:1.3} it suffices to take the normalization to be the identity, to
define the semantics of a source as ``it is the assigned Atom'', and to take the
remaining permission predicates to be true.

All three preserve the Law values. In $\{D_0,D_a\}$, the doctrine $D_0$ presents the
canonical resolution. By contrast, each member of $\{D_a,D_b\}$ retains a distinction
that the Law does not need. The former separates $a_0,a_1$ and the latter
separates $b_0,b_1$, so a family permitting only these two cannot present the
canonical resolution.

\end{example}

\section{The diagnostic complex from Law values}\label{sec:3.3}

From now on, sources are finite and $q:S\to Q$ is $L$-adequate.
For each local region we take as coordinates the distinct Law values readable there.
The point of the construction is that coordinates are not duplicated as more sources
share the same value.

\begin{definition}[Finite nerves with supports]\label{def:3.9}

Let finite sets $N_0,N_1,N_2$ be the sets of charts, edges, and faces, respectively.
An edge $e$ has a start $s(e)$ and an end $t(e)$.
A face $f$ has, for some charts $\alpha,\beta,\gamma$, three edges
$e_0:\alpha\to\beta$, $e_1:\alpha\to\gamma$, and
$e_2:\beta\to\gamma$. Parallel edges, self-loops, and repeated vertices are allowed.

To each chart we assign a nonempty support $S_\alpha\subseteq Q$, and we form the
supports of edges and faces by
\begin{equation*}
S_e=S_{s(e)}\cap S_{t(e)},\qquad
S_f=S_{e_0}\cap S_{e_1}\cap S_{e_2}
\tag{3.6}\label{eq:3.6}
\end{equation*}
The support of an edge or a face may be empty.
This support represents the values of the resolution readable on each cell.
The existence of edges and faces is part of the chosen local comparison data.
When this is applied to a cover of a space $X$, an open set $U_\alpha\subseteq X$
represents the region on which local states are placed, and the support
$S_\alpha\subseteq Q$ represents the values of the resolution readable there.
Cells are taken from intersections of the open sets, and which Law values are read on
a cell is determined by its support.

\end{definition}

\begin{construction}[The Law-value diagnostic complex]\label{cons:3.10}

Let $\Lambda=\{(\ell,v_\ell(s))\mid\ell\in L,\ s\in S\}$ be the set of labels
occurring overall.
Define the labels occurring on a cell $\sigma$ by
\begin{equation*}
\Lambda_\sigma
=\{(\ell,\bar v_\ell(u))\mid \ell\in L,\ u\in S_\sigma\}
\subseteq\Lambda
\tag{3.7}\label{eq:3.7}
\end{equation*}
A label is a pair of a Law index and a value; the element $u$ realizing the value is
not part of it. This pair is the name of a coordinate, distinguished from the rational
number placed in that coordinate.
Define the diagnostic cochain groups by
\begin{equation*}
D_L^n(q,N)=\prod_{\sigma\in N_n}\mathbb{Q}^{\Lambda_\sigma}
\quad(n=0,1,2)
\tag{3.8}\label{eq:3.8}
\end{equation*}
and the differentials, between coordinates with the same label, by
\begin{equation*}
\begin{aligned}
(d^0b)_{e,\lambda}
&=b_{t(e),\lambda}-b_{s(e),\lambda},\\
(d^1z)_{f,\lambda}
&=z_{e_0,\lambda}-z_{e_1,\lambda}+z_{e_2,\lambda}
\end{aligned}
\tag{3.9}\label{eq:3.9}
\end{equation*}
A label occurring on an edge occurs on both endpoints, and a label occurring on a face
occurs on its three edges, so every term on the right-hand side is defined.

\end{construction}

\begin{lemma}[Differentials and the diagnostic group]\label{lem:3.11}

We have $d^1d^0=0$. Hence
\begin{equation*}
H^1_{\mathrm{diag}}(q,N;L)
=\ker d^1/\mathrm{im}\,d^0
\tag{3.10}\label{eq:3.10}
\end{equation*}
is defined.

\end{lemma}

\begin{proof}

Computing \eqref{eq:3.9} on one face and one label gives
$(b_\beta-b_\alpha)-(b_\gamma-b_\alpha)+(b_\gamma-b_\beta)=0$.

\end{proof}

Diagnostic classes arising from concrete local data are constructed as elements of
this group by means of the coefficient comparison of \S\ref{sec:3.9}.

\begin{proposition}[Decomposition by Law value]\label{prop:3.12}

For $\lambda\in\Lambda$, let $N_\lambda$ be the subnerve retaining exactly the cells
with $\lambda\in\Lambda_\sigma$. With its complex of constant rational coefficients
we have
\begin{equation*}
\begin{aligned}
D_L^\bullet(q,N)
&\cong\bigoplus_{\lambda\in\Lambda}C^\bullet(N_\lambda;\mathbb{Q}),\\
H^1_{\mathrm{diag}}(q,N;L)
&\cong\bigoplus_{\lambda\in\Lambda}H^1(N_\lambda;\mathbb{Q})
\end{aligned}
\tag{3.11}\label{eq:3.11}
\end{equation*}
\end{proposition}

\begin{proof}

Reorder the coordinates $(\sigma,\lambda)$ of \eqref{eq:3.8} with $\lambda$ first.
The sets are finite, so products and direct sums agree.
The differentials do not change labels, so the reordering commutes with them.
The cocycle condition and membership in the coboundaries also hold label by label,
giving an isomorphism on the quotients as well.

\end{proof}

By this decomposition, the comparison of diagnoses splits into comparisons of ``the
regions where a single Law value is visible''. For example, taking every chart support
in Example~\ref{ex:3.5} to be the whole set, one copy of the complex of the same nerve
appears for each of the two Law values 0 and 1.

\section{From comparisons of resolutions to cochain maps}\label{sec:3.4}

The map carrying the diagnosis of the coarser side to the finer side can be built
from the correspondence between evaluation values and local regions.
Note that the direction in which resolution values are sent is opposite to the
direction in which cochains are pulled back.
Below we also call the coarser side coarse and the finer side fine.

\begin{definition}[Local comparison data]\label{def:3.13}

Take $q_c\preceq q_f$ and the comparison map $\pi:Q_f\to Q_c$ of Lemma~\ref{lem:3.2}.
For the nerves with supports on the two sides, specify the following.

\begin{itemize}
\item A chart map $r:N_{f,0}\to N_{c,0}$.
\item A map $r_1$ sending each fine edge either to a coarse edge or to a symbol
$\bot$ denoting contraction.
\item A map $r_2$ sending each fine face either to a coarse face or to $\bot$.
\end{itemize}

A mapped edge preserves the start and the end, and a mapped face preserves the
numbering of its three edges.
The two endpoints of a contracted edge map to the same coarse chart.
For a contracted face, its three edges are all contracted as well.
We also assume
\begin{equation*}
\pi(S^f_\alpha)\subseteq S^c_{r(\alpha)}
\tag{3.12}\label{eq:3.12}
\end{equation*}
The corresponding inclusions of supports for edges and faces follow from
\eqref{eq:3.6} and the incidence relations.

\end{definition}

\begin{proposition}[The generated diagnostic comparison]\label{prop:3.14}

From the data of Definition~\ref{def:3.13} and $L$-adequacy on both sides, linear
maps $T^n:D_L^n(q_c,N_c)\to D_L^n(q_f,N_f)$ are given by
\begin{equation*}
\begin{aligned}
(T^0b)_{\alpha,\lambda}&=b_{r(\alpha),\lambda},\\
(T^1z)_{e,\lambda}
&=\begin{cases}
z_{r_1(e),\lambda}&r_1(e)\ne\bot,\\
0&r_1(e)=\bot,
\end{cases}\\
(T^2w)_{f,\lambda}
&=\begin{cases}
w_{r_2(f),\lambda}&r_2(f)\ne\bot,\\
0&r_2(f)=\bot
\end{cases}
\end{aligned}
\tag{3.13}\label{eq:3.13}
\end{equation*}
and satisfy $T^1d_c^0=d_f^0T^0$ and $T^2d_c^1=d_f^1T^1$.
The same maps therefore induce a linear map
\begin{equation*}
T_{\mathrm{diag}}:
H^1_{\mathrm{diag}}(q_c,N_c;L)
\longrightarrow H^1_{\mathrm{diag}}(q_f,N_f;L),
\qquad [z]\longmapsto[T^1z]
\tag{3.14}\label{eq:3.14}
\end{equation*}
Under the decomposition \eqref{eq:3.11}, this is the direct sum of the comparisons of
the individual $N_\lambda$.

\end{proposition}

\begin{proof}

Take $u$ realizing $\lambda=(\ell,v)$ on a fine cell.
On a mapped cell, \eqref{eq:3.12} and Lemma~\ref{lem:3.2} show that $\pi(u)$ realizes
the same label on the coarse side.
The coordinates in \eqref{eq:3.13} therefore exist.

On a mapped edge, $T^1d_c^0=d_f^0T^0$ follows from the condition that the endpoints
are preserved. On a contracted edge, the two chart values agree and their difference
is zero. On a mapped face, the numbering of the three edges is preserved, so
$T^2d_c^1=d_f^1T^1$. On a contracted face, all three edges are contracted and all
three terms are zero. The maps preserve cocycles and coboundaries, hence descend to
the quotients, and since they do not change labels they also commute with the
decomposition.

\end{proof}

The same argument applies to cochain maps with coefficients in abelian groups.
A comparison commuting with the differentials carries classes and sends zero classes
to zero classes. If, moreover, the comparison in each degree is an isomorphism, its
inverse also commutes with the differentials, and the comparison of classes is an
isomorphism. The standard \v{C}ech comparison by refinement of covers and restriction
of sections is given in
\cite[\href{https://stacks.math.columbia.edu/tag/09UY}{Tag 09UY}]{Stacks}.
In \S\ref{sec:3.11} we compute the restriction maps for the actual covers of
Chapter~\ref{chap:2}.

\section{Diagnostic invariance under Condition C}\label{sec:3.5}

Once the comparison map is built, the question is whether it loses no class and
captures every class on the fine side. We prove these two properties as injectivity
and surjectivity, respectively. To this end we examine the structure on the
fine side that gathers inside a single coarse chart.

\begin{definition}[Coordinate fibers and Condition C]\label{def:3.15}

Fix a label $\lambda$ and work in the subnerve of \eqref{eq:3.11}.
The fiber of a coarse chart $v$ consists of the fine charts with $r(\alpha)=v$,
together with all fine edges whose two endpoints lie in this fiber.
A face whose three edges all lie within the fiber is called an internal face.
The edges within a fiber include, besides contracted edges, edges mapping to coarse
self-loops.

Condition C consists of the following seven clauses.

\begin{xltabular}{\linewidth}{@{}LL@{}}
\toprule
Clause & Content \\
\midrule
\endhead
C0: agreement of supports & For each coarse chart $v$, $S^c_v=\bigcup_{r(\alpha)=v}\pi(S^f_\alpha)$ \\
\addlinespace[3pt]
C1: connectedness of fibers & For each chart of the coarse subnerve of each $\lambda$, the fiber is nonempty and connected as a graph with edge orientations forgotten \\
\addlinespace[3pt]
C2: lifting of edges & Each coarse edge and each of its labels admits a fine edge with the same label mapping to that edge \\
\addlinespace[3pt]
C3: cycles within fibers & Every rational 1-cycle within a fiber is a rational linear combination of boundaries of internal faces \\
\addlinespace[3pt]
C4: lifting of faces & Each coarse face and each of its labels admits a fine face with the same label mapping to that face \\
\addlinespace[3pt]
C5: uniqueness of edge lifts & At most one fine edge maps to any given coarse edge \\
\addlinespace[3pt]
C6: reflection of self-loops & A fine edge mapping to a coarse self-loop is itself a self-loop \\
\bottomrule
\end{xltabular}

C0, C5, and C6 are imposed on the whole nerve, and C1--C4 on every
$\lambda\in\Lambda$. The boundaries of chains in C3 are $\partial e=t(e)-s(e)$ and
$\partial f=e_0-e_1+e_2$. A 1-cycle is a finite sum of edges with rational
coefficients in which, at each vertex, the inflowing and the outflowing coefficients
agree.

\end{definition}

\begin{lemma}[Correction within fibers]\label{lem:3.16}

Assume C1 and C3 in the subnerve of a single label. For every 1-cocycle $z$ on the
fine side there is a 0-cochain $p$ such that $z-d_f^0p$ vanishes on all edges within
fibers.

\end{lemma}

\begin{proof}

Choose a root vertex within one fiber.
Let $p(\alpha)$ be the signed sum of $z$ along a path from the root to $\alpha$.
A path exists by C1. The difference of two paths is a 1-cycle, and by C3 a sum of
boundaries of internal faces.
A cocycle vanishes on the boundary of each face, so $p(\alpha)$ does not depend on
the choice of path.
For an edge $e:\alpha\to\beta$, appending $e$ to a path from the root to $\alpha$
gives $p(\beta)-p(\alpha)=z(e)$.
Carrying out this construction in each fiber defines $p$ on all fine vertices.

\end{proof}

\begin{theorem}[Diagnostic invariance]\label{thm:3.17}

If both sides are $L$-adequate and the comparison of Definition~\ref{def:3.13}
satisfies Condition C, then the actual comparison map $T_{\mathrm{diag}}$ of
\eqref{eq:3.14} is an isomorphism.
In particular, for every coarse diagnostic class $\alpha$,
\begin{equation*}
\alpha=0\quad\Longleftrightarrow\quad T_{\mathrm{diag}}(\alpha)=0
\tag{3.15}\label{eq:3.15}
\end{equation*}
\end{theorem}

\begin{proof}

By Propositions~\ref{prop:3.12} and~\ref{prop:3.14} it suffices to fix one label and
prove injectivity and surjectivity.
C0 ensures that the support of each coarse chart is covered by the images of the
supports of the fine charts in its fiber, and is used throughout as the premise for
comparing the two subnerves of the fixed label over the same support.

\textbf{Injectivity.}
Suppose that the image of a coarse cocycle $z$ satisfies $T^1z=d_f^0b$.
If an edge within a fiber is contracted, then $d_f^0b=0$ on it.
If it is not contracted, it maps to a coarse self-loop, and by C6 it is a self-loop
on the fine side as well, so the values of $b$ at its two endpoints again agree.
By C1, $b$ is constant on each fiber and defines values $\bar b$ at the coarse
vertices.
Taking a lift of each coarse edge by C2 gives
$z(e)=\bar b(t(e))-\bar b(s(e))$.
Hence $z=d_c^0\bar b$, and the original class is zero.

\textbf{Surjectivity.}
Take a fine cocycle $w$. By Lemma~\ref{lem:3.16} the cochain
$w'=w-d_f^0p$, representing the same class, can be made zero within fibers.
By C2 and C5, each coarse edge has exactly one lift with the same label.
Let $z(e)$ be the value of $w'$ there.
On mapped fine edges $T^1z=w'$, and contracted edges lie within fibers, so both sides
are zero there.
Lifting each coarse face by C4, the uniqueness of the three edges shows that
$d_c^1z$ equals $d_f^1w'=0$ on the lifted face.
Hence $z$ is a cocycle and $[w]=T_{\mathrm{diag}}[z]$.

\end{proof}

Condition C is the condition for correcting locally removable differences within
fibers and reading the remaining differences back to the coarse side uniquely.
In the next section we also give a decision that computes the comparison itself,
without using Condition C.

\begin{example}[Comparison from three charts to four]\label{ex:3.18}

Use the two resolutions of Example~\ref{ex:3.5}, with every chart support equal to
the whole set of resolution values.
The nerves are taken from the three-chart and four-chart covers of
Lemma~\ref{lem:2.38}; the two covers and the merging of charts were shown in
Figure~\ref{fig:2.1}.
Map $a_0,a_1$ to $a$, and $b,c$ to the charts of the same name.
Contract the edge $e_{01}$, and map $ab,bc,ac$ to the edges of the same name.

For each label, the fiber over $a$ is a tree joining two vertices by a single edge,
and the fibers over $b,c$ are single points.
The supports agree, the lifts of the three coarse edges are unique, and faces and
coarse self-loops are absent on both sides.
Condition C therefore holds.
Each of the two nerves has one independent cycle and there are two labels 0 and 1,
so both diagnostic groups are $\mathbb{Q}^2$.
The comparison sends the values $(z_{ab},z_{bc},z_{ac})$ on the three edges to
$(0,z_{ab},z_{bc},z_{ac})$.
The period of a fine 1-cochain $w$ is
$w_{e_{01}}+w_{ab}+w_{bc}-w_{ac}$.
On the image of the comparison, the value $w_{e_{01}}$ on the internal edge is zero
and the period agrees with the coarse period $z_{ab}+z_{bc}-z_{ac}$, so nonzero
classes are preserved as well.

\end{example}

\begin{example}[Preserving Laws versus preserving diagnoses]\label{ex:3.19}

Let the sources be $\{0,1,2\}$, the coarse resolution
$q_c(0)=q_c(1)=0,\ q_c(2)=1$, and the fine resolution the identity.
The Law $v_0=q_c$ is readable on both sides.
Take every chart support to be the whole set and consider the following two comparisons.

\begin{xltabular}{\linewidth}{@{}LLL@{}}
\toprule
Coarse nerve & Fine nerve and comparison & Comparison of $H^1$ for one Law value \\
\midrule
\endhead
One vertex, one self-loop, no face & Send a single edge joining two vertices to the self-loop & $\mathbb{Q}\to0$; the nonzero coarse class dies \\
\addlinespace[3pt]
Two vertices, one edge, no face & Send two parallel edges between the two vertices to the single coarse edge & $0\to\mathbb{Q}$; the nonzero fine class is not captured \\
\bottomrule
\end{xltabular}

In the first row, any value on the fine edge is a difference of vertex values, and C6
fails. In the second row, a coboundary takes the same value on the two edges, so
their difference detects the nonzero class, and C5 fails.
In both rows, adequacy with respect to $L=\{0\}$ does hold.

Now add a Law reading $v_1(1)=1,\ v_1(0)=v_1(2)=0$; then $q_c$ can no longer read
this Law.
For an arbitrary resolution, one can still build the diagnosis from the subfamily
$L(q)=\{\ell\in L\mid v_\ell\text{ descends through }q\text{}\}$
of Laws that descend. In this example $L(q_c)=\{0\}$ and
$L(q_f)=\{0,1\}$.
The fine diagnosis of the first row is zero even with the added Law, while in the
second row the nonzero class from the original Law remains.
With a subfamily the computation is defined, but one must check both the retained
Laws and the comparison conditions on the covers in order to judge the correspondence
with the original diagnosis.

\end{example}

\section{Deciding invariance for all Laws}\label{sec:3.6}

Having verified invariance for one Law family, we may ask whether the same comparison
geometry works for every Law family for which both resolutions are adequate.
Instead of enumerating all Law families, it suffices to examine subsets of the values
of the coarse resolution.

\begin{definition}[Nerves selected by subsets of values]\label{def:3.20}

Fix the pair of resolutions and the comparison of Definition~\ref{def:3.13}.
For $A\subseteq Q_c$, retain the coarse cells whose supports meet $A$ and the fine
cells whose supports meet $\pi^{-1}(A)$.
Write these as $N_{c,A}$ and $N_{f,\pi^{-1}(A)}$.
If a face remains, so do its three edges; if an edge remains, so do its two
endpoints.
The image of a mapped fine cell also remains, so the same rule as \eqref{eq:3.13}
yields
\begin{equation*}
T_A:H^1(N_{c,A};\mathbb{Q})
\longrightarrow H^1(N_{f,\pi^{-1}(A)};\mathbb{Q})
\tag{3.16}\label{eq:3.16}
\end{equation*}
Define the pair of dimensions of its kernel and cokernel as
\begin{equation*}
J_A=\left(\dim_{\mathbb{Q}}\ker T_A,\
\dim_{\mathbb{Q}}\bigl(H^1(N_{f,\pi^{-1}(A)};\mathbb{Q})/\mathrm{im}\,T_A\bigr)\right)
\tag{3.17}\label{eq:3.17}
\end{equation*}
The first component is the dimension of the coarse classes killed by the comparison,
and the second is the dimension of the fine classes not captured by the image from
the coarse side.

\end{definition}

\begin{definition}[Uniform invariance]\label{def:3.21}

We say that the comparison is uniformly invariant if, for every finite Law family on
the same sources, the map \eqref{eq:3.14} is an isomorphism whenever both resolutions
are adequate.
The quantification here ranges over Law families including the value sets and the
evaluation maps of the individual Laws.
The resolutions, the nerves, the supports, and the cell comparison maps are fixed.

\end{definition}

\begin{theorem}[Reduction of uniform invariance to finite subsets]\label{thm:3.22}

The following are equivalent.

\begin{enumerate}
\item The comparison is uniformly invariant.
\item For every nonempty $A\subseteq Q_c$, the map $T_A$ is an isomorphism.
\item For every nonempty $A\subseteq Q_c$, we have $J_A=(0,0)$.
\end{enumerate}

\end{theorem}

\begin{proof}

We first show that 2 implies 1. Take a Law family for which both resolutions are
adequate and one label $\lambda=(\ell,v)$, and set
\begin{equation*}
A_\lambda=\{u\in Q_c\mid\bar v_{\ell,c}(u)=v\}
\tag{3.18}\label{eq:3.18}
\end{equation*}
This set is nonempty because the label is realized on the sources.
A coarse cell carries $\lambda$ exactly when its support meets $A_\lambda$.
On the fine side, by Lemma~\ref{lem:3.2}, the corresponding condition is that the
support meets $\pi^{-1}(A_\lambda)$.
The complex and the comparison map for each label therefore coincide with those of
\eqref{eq:3.16}.
By 2 and Proposition~\ref{prop:3.12}, the total comparison is an isomorphism as well.

Conversely, assume 1. For a nonempty $A\subseteq Q_c$, define the evaluation of a
single Law by setting $v_A(s)$ equal to 1 when $q_c(s)\in A$ and to 0 otherwise.
This Law is readable on both resolutions, and the comparison of the subnerves for
its value 1 is exactly $T_A$.
The total comparison is the direct sum over Law values, so its bijectivity implies
the bijectivity of each summand.

The equivalence of 2 and 3 is finite-dimensional linear algebra.
Zero kernel dimension means injectivity and zero cokernel dimension means
surjectivity, and conversely.

\end{proof}

\begin{proposition}[Decision from a finite presentation]\label{prop:3.23}

If finite enumerations of the sources, of the values of both resolutions, and of the
cells, the value tables of both readings, membership in the supports, and the
incidence relations and the cell comparison are given in computable form, then
uniform invariance is decidable.

\end{proposition}

\begin{proof}

For each $u\in Q_f$, find a source representative by finite search and compute
$\pi(u)=q_c(s)$.
Lemma~\ref{lem:3.2} guarantees independence of the representative.
For each nonempty subset $A\subseteq Q_c$, enumerate the remaining cells and build
the rational matrices $d_c^0,d_c^1,d_f^0,d_f^1,T^1$ from \eqref{eq:3.9} and \eqref{eq:3.13}.

By rational elimination, compute bases of $Z_c=\ker d_c^1$, $Z_f=\ker d_f^1$,
$B_c=\mathrm{im}\,d_c^0$, and $B_f=\mathrm{im}\,d_f^0$.
Extending the basis of each $B$ to a basis of the corresponding $Z$ yields a basis
of the quotient.
Expressing $T^1$ in these bases and writing $r_A$ for the rank of the induced map,
we have
\begin{equation*}
J_A=(\dim Z_c-\dim B_c-r_A,\
\dim Z_f-\dim B_f-r_A)
\tag{3.19}\label{eq:3.19}
\end{equation*}
Check whether both components are zero for the finitely many $A$.
The generated matrices are exactly the differentials and the comparison defined above, so
Theorem~\ref{thm:3.22} gives the correctness of the decision.

\end{proof}

This decision does not merely compare the dimensions of the groups on the two sides.
A map between groups of the same dimension can still be the zero map, so we compute
the rank $r_A$ of the actual comparison.

\begin{corollary}[Condition C for all subsets]\label{cor:3.24}

Suppose that C0, C5, and C6 hold globally, and that for every nonempty
$A\subseteq Q_c$ the clauses C1--C4 hold on the two selected subnerves.
Then the comparison is uniformly invariant.
These geometric conditions themselves are decidable from the same finite
presentation as in Proposition~\ref{prop:3.23}.

\end{corollary}

\begin{proof}

For an arbitrary Law family for which both resolutions are adequate, take the sets
$A_\lambda$ of \eqref{eq:3.18}.
The identification used in the proof of Theorem~\ref{thm:3.22} preserves vertices,
edges, faces, their incidences, and the partial maps, so the paths of C1, the lifts
of C2 and C4, and the chains and internal faces of C3 transfer directly to the
subnerves of the labels.
C0, C5, and C6 already hold globally. Uniform invariance follows from
Theorem~\ref{thm:3.17}.

For the finite decision, use finite matching for C0, C2, C4, C5, and C6, and graph
reachability for C1.
For C3, compute bases of $\ker\partial_1$ and $\mathrm{im}\,\partial_2$ for each
fiber and check whether the former is contained in the latter.
Since $\partial_1\partial_2=0$, this is also a test of equality of the two spaces.

\end{proof}

\begin{example}[Condition C is sufficient, and its clauses are not necessary]\label{ex:3.25}

The following finite examples show that uniform invariance can hold while individual
clauses fail.
The resolutions are the $q_c,q_f$ of Example~\ref{ex:3.5}, and every chart support is
the whole set.
To both sides of each row we add, as an independent component, a component with one
vertex, one self-loop $h$, and no face, on which the comparison is the identity.
This common component has nonzero $H^1\cong\mathbb{Q}$.
Adding it makes each comparison an example that also carries a nonzero class.
The table records only the parts added beyond it.

\begin{xltabular}{\linewidth}{@{}LL@{}}
\toprule
Failing clause & Added parts on the coarse and fine sides, and the comparison \\
\midrule
\endhead
C0 & Add one isolated vertex on the coarse side, with no corresponding vertex on the fine side \\
\addlinespace[3pt]
C1 & Map two isolated fine vertices to one isolated coarse vertex \\
\addlinespace[3pt]
C2 & Take a triangle with a face on the coarse side and the path formed by two of its edges on the fine side, compared by the inclusion \\
\addlinespace[3pt]
C3 & Add to both sides a component with one vertex, one self-loop, and no face, compared by the identity \\
\addlinespace[3pt]
C4 & On the coarse triangle place two faces with the same boundary; on the fine side place the same three edges and one face \\
\addlinespace[3pt]
C5 & Take a triangle with a face on the coarse side; on the fine side split one edge into two parallel edges and place one face using each. Send the parallel edges to the same edge \\
\addlinespace[3pt]
C6 & On the coarse side place one self-loop $e$ and a face with boundary $e-e+e=e$. The fine side is a single edge joining two vertices, sent to $e$ \\
\bottomrule
\end{xltabular}

The comparisons in the table send the remaining vertices and edges to those of the
same name.
In the C4 row the fine face is sent to one of the two coarse faces, and in the C5
row the two fine faces are sent to the coarse face.
All satisfy Definition~\ref{def:3.13}.

On the added part of the C3 row the comparison is the identity.
In every other row, $H^1$ of the added part is zero on both sides.
The C0 and C1 rows have no edges, and in the path and the faced triangle of the C2
row every cocycle is a difference of vertex values.
The two faces of the C4 row impose the same equation, so the kernel is the same as
with a single face.
In the C5 row, subtracting the equations of the two faces makes the values on the
parallel edges equal, and the remaining triangle equation then writes the cocycle as
a difference of vertex values.
In the C6 row the coarse face equation imposes $z_e=0$, and the fine side is a path,
hence zero as well.

Since all supports are the whole set, the cells of the two nerves are unchanged for any
choice of nonempty $A$.
Each $T_A$ is the identity on the common nonzero component and the isomorphism above
on the rest.
By Theorem~\ref{thm:3.22} each row is uniformly invariant, which shows that the
sufficient condition of Corollary~\ref{cor:3.24} is strictly stronger than uniform
invariance.
The common self-loop without a face cannot be expressed as a boundary of internal
faces, so C3 also fails in every row.
In this way, clauses other than the listed one may fail simultaneously in each row.
Each row is an example showing that the clause listed in the table is not necessary.

\end{example}

\section{An example undecidable from local incidence information alone}\label{sec:3.7}

The finite decision computed the relations among faces on the complex as a whole.
Reading only the kinds and numbers of incidences locally need not allow the same
decision.
We fix which information is read, and compare two examples on which it agrees.

\begin{definition}[Local observation of incidence information]\label{def:3.26}

Let $\mathrm{Obs}_{\mathrm{loc}}$ be the observation given by the following rules.

\begin{enumerate}
\item Record the sources, the values of both resolutions, and both readings.
For each $A\subseteq Q_c$ with $A\ne\varnothing$, form the two subnerves of
Definition~\ref{def:3.20}.
\item Record as the color of each cell its degree, whether it is coarse or fine, and
its support within the chosen subset.
For a fine cell, also record whether the comparison maps it to a cell of the same
degree or contracts it.
\item For each cell, tally the colors of the directly incident cells by kind of
incidence: the start and the end of an edge, and the 0th, 1st, and 2nd edges of a
face.
The count for each item is collapsed to the three values 0, 1, and 2 or more.
The color of the cell itself together with this tally is the local type of that
cell.
\item The numbers of occurrences of the local types on each side are likewise
collapsed to 0, 1, and 2 or more.
The tables obtained for all $A$ form the observation value.
\end{enumerate}

Incidence is read between charts and edges and between edges and faces, and the
proper names of cells are not recorded.
This is an explicit observation rule retaining the kinds of neighbors and their
truncated counts.

\end{definition}

\begin{construction}[Sequences of faces of length 3 and length 6]\label{cons:3.27}

Take the same $S=\{0,1,2\}$ and resolutions as in Example~\ref{ex:3.19}.
Place charts $w_0,w_1$ on both sides, with supports
\begin{equation*}
\begin{array}{c|cc}
&w_0&w_1\\ \hline
\text{coarse}&\{0\}&\{0,1\}\\
\text{fine}&\{0,1\}&\{2\}
\end{array}
\tag{3.20}\label{eq:3.20}
\end{equation*}
Place one self-loop $h$ at $w_0$ and self-loops $e_0,\ldots,e_{n-1}$ at $w_1$.
Let $f_0,\ldots,f_{n-1}$ be the faces over $w_1$, the three edges of $f_j$ being
$(e_j,e_{j+1},e_{j+2})$, with indices read modulo $n$.
The comparison of charts, edges, and faces is the identity onto the cells of the
same name.
Since \eqref{eq:3.20} satisfies \eqref{eq:3.12}, the comparison is defined.
We call the two inputs with $n=3,6$ respectively $P_3,P_6$.

\end{construction}

\begin{proposition}[Same local observation, different uniform invariance]\label{prop:3.28}

We have $\mathrm{Obs}_{\mathrm{loc}}(P_3)=\mathrm{Obs}_{\mathrm{loc}}(P_6)$.
On the other hand, $P_3$ is uniformly invariant and $P_6$ is not.
Hence no predicate of this observation value alone can decide uniform invariance.

\end{proposition}

\begin{proof}

We first compare the observations. The component of $h$ is identical in the two
inputs.
When the component of $w_1$ remains, each $e_j$ has charts of the same color at both
endpoints and is incident to exactly one face in each of the three edge positions of
a face.
Each face is likewise incident to edges of the same color in its three positions.
The number of edges incident to $w_1$, and the numbers of occurrences of the local
types of edges and faces, are all 2 or more for both $n=3,6$.
For each subset, which sides retain this component is determined by \eqref{eq:3.20}
and does not depend on $n$.
All observation tables therefore agree.

We next compute the cohomology. All edges are self-loops, so $d^0=0$.
The cocycle condition over $w_1$ is
\begin{equation*}
z_j-z_{j+1}+z_{j+2}=0
\qquad(j\ \bmod n)
\tag{3.21}\label{eq:3.21}
\end{equation*}
Starting from $z_0=a,z_1=b$, the sequence is
\begin{equation*}
a,\ b,\ b-a,\ -a,\ -b,\ a-b,\ a,\ b,\ldots
\tag{3.22}\label{eq:3.22}
\end{equation*}
Imposing period 3 gives $a=-a,\ b=-b$ and hence $a=b=0$, while with period 6
arbitrary $a,b\in\mathbb{Q}$ are allowed.
The dimension of $H^1$ of this component is therefore 0 for $n=3$ and 2 for $n=6$.

For $A=\{0\}$, both components remain on the coarse side and only the component of
$h$ remains on the fine side.
The map $T_A$ preserves the value on $h$ and discards the component of $w_1$.
For $A=\{1\}$, only the component of $w_1$ remains on both sides, and the comparison
is the identity.
For $A=\{0,1\}$, all components remain on both sides, and the comparison is again
the identity.
Hence
\begin{equation*}
\begin{array}{c|ccc}
&A=\{0\}&A=\{1\}&A=\{0,1\}\\ \hline
P_3:\ J_A&(0,0)&(0,0)&(0,0)\\
P_6:\ J_A&(2,0)&(0,0)&(0,0)
\end{array}
\tag{3.23}\label{eq:3.23}
\end{equation*}
and Theorem~\ref{thm:3.22} gives the conclusion.
If uniform invariance were determined by a predicate of the observation value alone,
these two inputs, whose observations agree, would receive the same truth value.
This contradicts the computation above.

\end{proof}

What produces the difference is the period obtained by following the face relations
around a full turn.
In this example, a relation that cannot be read off from the agreement of local
types is retained by the totality of \eqref{eq:3.21}.

\section{Structural supports and the role of coefficients}\label{sec:3.8}

Chapter~\ref{chap:2} compared the obstructions of the structural and semantic
phases.
Here we directly build the structural support that survives changes of semantic
choices, and study how the choice of coefficients placed on it affects the
diagnosis.
To this end we generate a nerve from all tuples of Atoms holding at a common source,
and also separate the coefficients by source.
In this construction the correction from a reference Atom can be made explicit, and
we can trace the mechanism by which all diagnostic classes become zero.

\begin{definition}[Supports invariant under semantic changes]\label{def:3.29}

For an extraction doctrine of Chapter~\ref{chap:1}, fix the vocabulary, the
resolution, the semantics of sources, and the normalization, and vary only the
semantic reading and its permission predicates. Specify these options by a set
$\Theta$.
Designate a reference choice $\theta_0\in\Theta$, and let $E_\theta(s,a)$ mean that
``Atom $a$ is extracted from source $s$''.
Define the structural support by
\begin{equation*}
E_{\mathrm{str}}(s,a)
\quad\Longleftrightarrow\quad
E_{\theta_0}(s,a)\ \land\
\forall\theta\in\Theta,\
\bigl(E_\theta(s,a)\Longleftrightarrow E_{\theta_0}(s,a)\bigr)
\tag{3.24}\label{eq:3.24}
\end{equation*}
The nerve built from a support $E(s,a)$ has as cells the ordered tuples of Atoms
holding at a common source.
In degree 0 take $(a)$, in degree 1 take $(a,b)$, and in degree 2 take $(a,b,c)$,
the condition being the existence of a source satisfying all entries simultaneously.
Repeated Atoms are allowed, and the incidences of edges and faces are given by the
projections of the tuples.

\end{definition}

\begin{proposition}[Invariance of the structural nerve]\label{prop:3.30}

The nerve obtained from $E_{\mathrm{str}}$ has the same cells and incidence
relations when the reference semantic reading is replaced by another member of the
same family.

\end{proposition}

\begin{proof}

\eqref{eq:3.24} is equivalent to $\forall\theta\in\Theta,\ E_\theta(s,a)$, and hence
does not depend on the reference member. Since the support is the same, the
vertices, edges, and faces defined by the existence of a common source, and the
projections of the tuples, all coincide.

\end{proof}

\begin{example}[The all-Atom nerve can change]\label{ex:3.31}

Consider the case of one source and two Atoms $a,b$.
Suppose that one semantic reading extracts only $a$ and the other extracts $a,b$.
Only $a$ remains in the structural support, while the all-Atom nerve acquires the
vertex $b$ under the latter.
Invariance of the structural part is distinguished from the conclusion that the
whole nerves are equal.

\end{example}

\begin{proposition}[Vanishing with coefficients separated by source]\label{prop:3.32}

Suppose that $S$ and the set of Atoms are finite.
On each cell of the ordered nerve of Definition~\ref{def:3.29}, place one rational
coordinate for each source realizing that cell simultaneously.
The differentials leave the source unchanged and are given by
\begin{equation*}
\begin{aligned}
(d^0c)(a,b;s)&=c(b;s)-c(a;s),\\
(d^1z)(a,b,c;s)&=z(b,c;s)-z(a,c;s)+z(a,b;s)
\end{aligned}
\tag{3.25}\label{eq:3.25}
\end{equation*}
The $H^1$ of this complex is zero.

\end{proposition}

\begin{proof}

For each source $s$ with nonempty support, choose one Atom satisfying $E(s,a_s)$.
For a cocycle $z$, set $c(a;s)=z(a_s,a;s)$.
If the edge $(a,b;s)$ exists, then so does the face $(a_s,a,b;s)$, and its cocycle
condition gives $z(a,b;s)=z(a_s,b;s)-z(a_s,a;s)$.
That is, $z=d^0c$. A source with empty support has no coordinates.

\end{proof}

In this construction, the coordinates sharing a source can each be corrected from
the reference Atom.
To diagnose a nonzero gluing obstruction, therefore, one must examine the relation
between these coefficients separated by source and the actual obstruction
coefficients.
In the next section we give this comparison, using the integer coefficients built
from the generator relations of Chapter~\ref{chap:2}.

\section{Mapping integer obstruction coefficients to the Law-value diagnosis}\label{sec:3.9}

The obstructions of Chapter~\ref{chap:2} arose from differences obtained by actually
restricting local states and comparing them along transitions.
In this section we read the Law values of the generators composing those
differences and transfer them to the diagnostic coordinates.
We construct in turn the map of coefficients, the map of cochains, and the map of
obstruction classes.

\begin{construction}[Coefficient comparison]\label{cons:3.33}

Take the generators $G=L\times S$ of Definition~\ref{def:2.35}, the label-preserving
primitive relation $R$, the set $\mathfrak{B}$ of relation components, and the
integer coefficients $M_R$.
Write $\lambda_B\in\Lambda$ for the common label of a component $B\in\mathfrak{B}$.
Send the class of a generator to the rational delta function of its label:
\begin{equation*}
\varepsilon_R:M_R\longrightarrow \mathbb{Q}^\Lambda,\qquad
\varepsilon_R([e_g])(\lambda)
=\begin{cases}
1&\lambda=\lambda_g,\\
0&\lambda\ne\lambda_g.
\end{cases}
\tag{3.26}\label{eq:3.26}
\end{equation*}
If $gRh$ then the labels are equal and $e_g-e_h$ maps to zero, so this is well
defined as an additive homomorphism on the quotient.
Under the presentation $M_R\cong\mathbb{Z}^{(\mathfrak{B})}$ of
Lemma~\ref{lem:2.36},
\begin{equation*}
\varepsilon_R(m)(\lambda)
=\sum_{\lambda_B=\lambda}m_B
\quad\text{(integers read as rationals via the inclusion)}
\tag{3.27}\label{eq:3.27}
\end{equation*}
The coefficients of the relation components carrying the same Law value are added
into a single diagnostic coordinate.

\end{construction}

\begin{lemma}[Injectivity of the coefficient comparison]\label{lem:3.34}

Suppose that any two generators with the same label are joined by a finite chain of
primitive relations.
Then $B\mapsto\lambda_B$ gives $\mathfrak{B}\cong\Lambda$, and
$\varepsilon_R$ is injective.

\end{lemma}

\begin{proof}

Every label admits a generator, so the map is surjective.
If the labels of two components are equal, the hypothesis joins their representative
generators by relations, and the components are equal as well.
Hence \eqref{eq:3.27} becomes $\varepsilon_R(m)(\lambda_B)=m_B$.
Since the inclusion of the integers into the rationals is injective, all coordinates
can be read back.

\end{proof}

\begin{example}[Different components sent to the same Law value]\label{ex:3.35}

Take two sources, one constant Law, and the empty primitive relation.
Then $M_R=\mathbb{Z}^2$, the set $\Lambda$ is a singleton, and
$\varepsilon_R(m,n)=m+n$.
The nonzero element $(1,-1)$ maps to zero.
Making ``reading the same Law value'' coincide with ``lying in the same component of
the primitive relation'' is the role of Lemma~\ref{lem:3.34}.

\end{example}

\begin{construction}[Comparison with the actual \v{C}ech complex]\label{cons:3.36}

Take a finite cover $\mathcal{U}_q$ consisting of nonempty connected charts, in which the
nonempty double intersections are connected and the intersections of three distinct
charts are empty.
The nerve has as edges the nonempty intersections of distinct pairs of charts,
oriented by the order of the charts.
On this cover place the locally constant sheaf $F_R$ of $M_R$-values from
Construction~\ref{cons:2.39}.

By connectedness, each section is determined by a single $M_R$-value, and the
restrictions preserve that value.
The normalized \v{C}ech complex used in Chapter~\ref{chap:2} can therefore be
presented as
\begin{equation*}
C_{\mathrm{ob}}^0(q)=M_R^{N_{q,0}},\qquad
C_{\mathrm{ob}}^1(q)=M_R^{N_{q,1}},\qquad
C_{\mathrm{ob}}^2(q)=0
\tag{3.28}\label{eq:3.28}
\end{equation*}
Here $d_{\mathrm{ob}}^0p$ is the difference, after the actual restrictions, between
the value on the end side and the value on the start side of an edge.
On the same nerve, specify the chart supports of an $L$-adequate $q$ and build the
diagnostic complex of Construction~\ref{cons:3.10}.
Define the comparisons in degrees 0 and 1 by
\begin{equation*}
(\phi_q^n c)_{\sigma,\lambda}
=\varepsilon_R(c_\sigma)(\lambda)
\quad(n=0,1),\qquad
\phi_q^2:0\longrightarrow0
\tag{3.29}\label{eq:3.29}
\end{equation*}
On the diagnostic side, only the labels occurring on the cell are read.
The additivity of $\varepsilon_R$ and the presentation of the restrictions give
\begin{equation*}
\phi_q^1d_{\mathrm{ob}}^0=d_{\mathrm{diag}}^0\phi_q^0,\qquad
\phi_q^2d_{\mathrm{ob}}^1=d_{\mathrm{diag}}^1\phi_q^1
\tag{3.30}\label{eq:3.30}
\end{equation*}
the latter because degree 2 is zero on both sides.
We therefore obtain an additive homomorphism
\begin{equation*}
\Phi_q:\check H^1(\mathcal{U}_q,F_R)\longrightarrow
H^1_{\mathrm{diag}}(q,N_q;L),\qquad
[z]\longmapsto[\phi_q^1z]
\tag{3.31}\label{eq:3.31}
\end{equation*}
The domain is an abelian group with integer coefficients, and the structure of this
comparison is that of an additive homomorphism.

\end{construction}

\begin{theorem}[Obstruction and diagnostic classes from the same local data]\label{thm:3.37}

For the input of Construction~\ref{cons:3.36}, take any local states and affine
transitions $x=(\xi,p)$.
Define the obstruction cochain of Proposition~\ref{prop:2.41} and the diagnostic
cochain generated from Law values by
\begin{equation*}
\begin{aligned}
o_q(x)&=\xi+d_{\mathrm{ob}}^0p,\\
a_q(x)&=\phi_q^1\xi+d_{\mathrm{diag}}^0\phi_q^0p
\end{aligned}
\tag{3.32}\label{eq:3.32}
\end{equation*}
Then
\begin{equation*}
\phi_q^1o_q(x)=a_q(x),\qquad
\Phi_q([o_q(x)])=[a_q(x)]
\tag{3.33}\label{eq:3.33}
\end{equation*}
In particular, $[o_q(x)]=0$ implies $[a_q(x)]=0$.

\end{theorem}

\begin{proof}

The value of the diagnosis at an edge coordinate $(e,\lambda)$ is the sum of the
Law-value coordinate $\varepsilon_R(\xi_e)(\lambda)$ of the transition and the
difference of the Law-value coordinates of the states on the two sides.
By \eqref{eq:3.30} and additivity, this equals $\varepsilon_R(o_q(x)_e)(\lambda)$.
Degree 2 is zero on both sides, so both cochains are cocycles, and passing to the
quotients gives \eqref{eq:3.33}.

\end{proof}

In this comparison, we first prepare the integer transitions and states, and then
read their Law values to build the diagnosis.
Thus \eqref{eq:3.33} does more than compare two groups abstractly: it determines the
destination of the concrete obstruction classes received from
Chapter~\ref{chap:2}.

\section{From vanishing diagnoses to integer corrections}\label{sec:3.10}

In the reverse direction, a correction seen in the diagnosis must be brought back to
a correction in the original integer coefficients.
For this we use the facts that the relation components can be distinguished by Law
values, and that the required Law values can be read on every chart and every edge.

\begin{definition}[Two conditions reflecting vanishing]\label{def:3.38}

Impose the following conditions on the input of Construction~\ref{cons:3.36}.

\begin{itemize}
\item \textbf{$R_q$: identification of relation components.} Any two generators with
the same label are joined by a finite chain of primitive relations.
\item \textbf{Existence of common representatives.} For each
$\lambda=(\ell,v)\in\Lambda$ there is a single $u_\lambda\in Q$ that belongs to
every chart support and satisfies $\bar v_\ell(u_\lambda)=v$.
\end{itemize}

By the second condition, that representative belongs to the support of every edge as
well.
Each label therefore has one diagnostic coordinate on every chart and every edge.
Even if the representatives are chosen differently, the coordinates of
Construction~\ref{cons:3.10} are treated as the same label.

\end{definition}

\begin{lemma}[Integer corrections from rational ones]\label{lem:3.39}

Let an integer-valued 1-cochain $z$ and a rational-valued 0-cochain $b$ on a
directed graph satisfy $b_{t(e)}-b_{s(e)}=z_e$.
Setting $n_\alpha=\lfloor b_\alpha\rfloor$ gives
$n_{t(e)}-n_{s(e)}=z_e$.

\end{lemma}

\begin{proof}

From $b_{t(e)}=b_{s(e)}+z_e$ and $z_e\in\mathbb{Z}$ it follows that
$\lfloor b_{t(e)}\rfloor=\lfloor b_{s(e)}\rfloor+z_e$.

\end{proof}

For example, the difference of $1/2,3/2$ is 1, and the integer parts $0,1$ have the
same difference.
The integer parts are used to construct a correction preserving these specific edge
differences.

\begin{theorem}[Reflection of vanishing for specified obstruction classes]\label{thm:3.40}

Under the two conditions of Definition~\ref{def:3.38}, for arbitrary local data
$x=(\xi,p)$ we have
\begin{equation*}
[a_q(x)]=0
\quad\Longleftrightarrow\quad
[o_q(x)]=0
\tag{3.34}\label{eq:3.34}
\end{equation*}
\end{theorem}

\begin{proof}

Right to left is Theorem~\ref{thm:3.37}.
Assuming the left side, there is a diagnostic 0-cochain $b$ with
$a_q(x)=d_{\mathrm{diag}}^0b$.
Write $z_{e,B}$ for the integer coefficient of the relation component $B$ in
$o_q(x)_e$.
By the common representatives the coordinate $\lambda_B$ exists on the edges, and by
$R_q$ and Lemma~\ref{lem:3.34} we get
\begin{equation*}
\begin{aligned}
b_{t(e),\lambda_B}-b_{s(e),\lambda_B}
&=a_q(x)_{e,\lambda_B}\\
&=\varepsilon_R(o_q(x)_e)(\lambda_B)
=z_{e,B}
\end{aligned}
\tag{3.35}\label{eq:3.35}
\end{equation*}
Applying Lemma~\ref{lem:3.39} to each component and setting
\begin{equation*}
n_\alpha=\sum_{B\in\mathfrak{B}}
\lfloor b_{\alpha,\lambda_B}\rfloor e_B\in M_R
\tag{3.36}\label{eq:3.36}
\end{equation*}
we obtain $d_{\mathrm{ob}}^0n=o_q(x)$.
Since $\mathfrak{B}$ is finite, this sum defines an element of $M_R$.
On the connected charts of Construction~\ref{cons:3.36}, $n_\alpha$ gives an actual
locally constant section.
The original obstruction class is therefore zero as well.

\end{proof}

This proof produces the correction itself.
Changing the original states to $p-n$ gives
$\xi+d_{\mathrm{ob}}^0(p-n)=0$, and the local states become consistent with respect
to the transitions.
In the sheaf of states of Proposition~\ref{prop:2.41}, the corrected family glues to
a global state.

\begin{example}[Without the reflection conditions the diagnosis loses differences]\label{ex:3.41}

On the three-chart cover of Lemma~\ref{lem:2.38}, place the locally constant
coefficients $M_R=\mathbb{Z}^2$ of Example~\ref{ex:3.35} and chart supports reading
all values of the resolution.
Taking $p=0$, $\xi_{ab}=(1,-1)$, and $\xi_{bc}=\xi_{ac}=0$, the period of the
obstruction is nonzero while the diagnostic cochain is zero. This is an example
lacking $R_q$.

The role of the common representatives can be checked in the same fashion.
This time use two Law values 0 and 1, with an $M_R\cong\mathbb{Z}^2$ satisfying
$R_q$.
Taking each chart support to be a set on which only the value 0 occurs, the
coordinate of value 1 does not appear in the diagnosis.
Placing the basis element $u$ of value 1 in $\xi_{ab}$ and zero elsewhere, the
period of the obstruction is $u\ne0$ while the diagnosis is zero.
Even when the coefficient comparison is injective, the reflection of vanishing can
break if the coordinates reading a component are locally missing.

\end{example}

\section{Refinement of covers and invariance of the obstruction verdict}\label{sec:3.11}

Finally, we tie the preceding comparisons together over the same Atom input.
We use the eight-point space and generator relations of Chapter~\ref{chap:2} and the
two resolutions of Example~\ref{ex:3.5}.
The integer obstruction, the rational diagnosis, and the refinement from three
charts to four then appear as one commutative comparison.

\begin{construction}[Coarse--fine comparison from the same input]\label{cons:3.42}

Take the sources $S=\{0,1\}^2$, the Law $v_\ell(v,h)=v$, and the primitive relation
consisting of the two pairs joining $(v,0)$ and $(v,1)$.
The coefficients are $M_R\cong\mathbb{Z}e_{B_0}\oplus\mathbb{Z}e_{B_1}$.
Use the point Atoms and generator Atoms of Construction~\ref{cons:2.37}; each
context reads the points of the corresponding open set and all generators.

On the coarse side use $q_c(v,h)=v$ and the three-chart cover
$\mathcal{U}_c=(U_a,U_b,U_c)$; on the fine side use $q_f=\mathrm{id}_S$ and the
four-chart cover $\mathcal{U}_f=(U_{a_0},U_{a_1},U_b,U_c)$.
Every chart reads all generators, so the diagnostic chart supports are all of
$Q_c,Q_f$, respectively.
Below we abbreviate the obstruction group as
$H^1_{\mathrm{ob}}(q)=\check H^1(\mathcal{U}_q,F_R)$ and the diagnostic group as
$H^1_{\mathrm{diag}}(q)=H^1_{\mathrm{diag}}(q,N_q;L)$.

The inclusions $U_{a_0},U_{a_1}\subseteq U_a$, together with the identity inclusions
of the remaining charts, give an actual refinement of covers.
In the coordinates of \eqref{eq:3.28}, the maps by restriction of sections are
\begin{equation*}
\begin{aligned}
T_{\mathrm{ob}}^0(p_a,p_b,p_c)&=(p_a,p_a,p_b,p_c),\\
T_{\mathrm{ob}}^1(z_{ab},z_{bc},z_{ac})&=(0,z_{ab},z_{bc},z_{ac}),\\
T_{\mathrm{ob}}^2&:0\longrightarrow0
\end{aligned}
\tag{3.37}\label{eq:3.37}
\end{equation*}
with the fine edges ordered $(e_{01},ab,bc,ac)$.
The value on the internal edge $e_{01}$ is zero, being a self-comparison of the same
coarse chart.
On the remaining edges, restriction to the actual overlaps preserves the same
$M_R$-value.

The fine differential is
\begin{equation*}
d_f^0p=
(p_{a_1}-p_{a_0},\
p_b-p_{a_1},\
p_c-p_b,\
p_c-p_{a_0})
\tag{3.38}\label{eq:3.38}
\end{equation*}
so $T_{\mathrm{ob}}^1d_c^0=d_f^0T_{\mathrm{ob}}^0$ can be checked directly.
This yields $T_{\mathrm{ob}}:H^1_{\mathrm{ob}}(q_c)\to H^1_{\mathrm{ob}}(q_f)$.
On the diagnostic side we use the map $T_{\mathrm{diag}}$ generated from the same
cell comparison as in Example~\ref{ex:3.18}.

\end{construction}

\begin{theorem}[Commutativity of obstruction, diagnosis, and refinement]\label{thm:3.43}

In Construction~\ref{cons:3.42} the following square commutes.
\begin{equation*}
\begin{array}{ccc}
H^1_{\mathrm{ob}}(q_c)&\xrightarrow{\ \Phi_{q_c}\ }&
H^1_{\mathrm{diag}}(q_c)\\
\big\downarrow T_{\mathrm{ob}}&&\big\downarrow T_{\mathrm{diag}}\\
H^1_{\mathrm{ob}}(q_f)&\xrightarrow{\ \Phi_{q_f}\ }&
H^1_{\mathrm{diag}}(q_f)
\end{array}
\tag{3.39}\label{eq:3.39}
\end{equation*}
For any coarse local data $x=(\xi,p)$, setting
$x_f=(T_{\mathrm{ob}}^1\xi,T_{\mathrm{ob}}^0p)$ gives
\begin{equation*}
\begin{aligned}
T_{\mathrm{ob}}[o_{q_c}(x)]&=[o_{q_f}(x_f)],\\
T_{\mathrm{diag}}[a_{q_c}(x)]&=[a_{q_f}(x_f)]
\end{aligned}
\tag{3.40}\label{eq:3.40}
\end{equation*}
\end{theorem}

\begin{proof}

The coefficient comparisons on the two sides send the same generator $[e_g]$ to the
delta function of the same Law value.
On mapped edges and charts, applying \eqref{eq:3.37} first and then reading Law
values yields the same coordinates as reading Law values first and then pulling back
by \eqref{eq:3.13}.
On the internal edge both are zero, and degree 2 is zero on both sides as well.
The comparisons therefore commute in each degree, and passing to classes gives
\eqref{eq:3.39}.

For the local data, commutation with the differentials gives
\begin{equation*}
o_{q_f}(x_f)
=T_{\mathrm{ob}}^1\xi+d_f^0T_{\mathrm{ob}}^0p
=T_{\mathrm{ob}}^1(\xi+d_c^0p)
\tag{3.41}\label{eq:3.41}
\end{equation*}
This gives the first equation of \eqref{eq:3.40}.
The second follows from the first, Theorem~\ref{thm:3.37}, and \eqref{eq:3.39}.

\end{proof}

\begin{theorem}[Invariance of the obstruction verdict along a change of reading]\label{thm:3.44}

For any local data of Construction~\ref{cons:3.42}, the following four statements
are equivalent.
\begin{equation*}
\begin{aligned}
[o_{q_c}(x)]=0
&\quad\Longleftrightarrow\quad[a_{q_c}(x)]=0\\
&\quad\Longleftrightarrow\quad[a_{q_f}(x_f)]=0\\
&\quad\Longleftrightarrow\quad[o_{q_f}(x_f)]=0.
\end{aligned}
\tag{3.42}\label{eq:3.42}
\end{equation*}

\end{theorem}

\begin{proof}

For the coarse and the fine resolution alike, the Law value can be read from the
first component, so both are adequate.
Two generators with the same Law value are joined by one of the declared relations,
so $R_q$ holds.
As a common representative of a value $v$ we may take $v$ on the coarse side and
$(v,0)$ on the fine side; these belong to the supports of all charts.
Theorem~\ref{thm:3.40} therefore applies on both ends.

The diagnostic comparison satisfies Condition C, as verified in
Example~\ref{ex:3.18}, and is an isomorphism by Theorem~\ref{thm:3.17}.
Theorem~\ref{thm:3.43} matches the specified diagnostic classes, so their vanishing
is equivalent as well.
Composing these three equivalences gives \eqref{eq:3.42}.

\end{proof}

In this proof, the merging map $\pi(v,h)=v$ is not injective.
Nevertheless, because the Law values, the relation components, and the comparison of
covers satisfy the conditions above, the obstruction verdict for the same local data
can be shared by the two resolutions.

\begin{example}[A nonzero difference that can be removed, and one that cannot]\label{ex:3.45}

Let $u=e_{B_0}\in M_R$ and fix all chart states to zero.
With the same coefficients, Law, and covers, vary only the transitions as follows.

\begin{xltabular}{\linewidth}{@{}LLLL@{}}
\toprule
Local data & Coarse transitions $(ab,bc,ac)$ & Transitions carried to fine $(e_{01},ab,bc,ac)$ & Obstruction and diagnostic classes \\
\midrule
\endhead
A correctable difference & $(u,-u,0)$ & $(0,u,-u,0)$ & Zero on both sides \\
\addlinespace[3pt]
A difference surviving correction & $(u,0,0)$ & $(0,u,0,0)$ & Nonzero on both sides \\
\bottomrule
\end{xltabular}

The first row is $d^0(0,u,0)$ on the coarse side and
$d^0(0,0,u,0)$ on the fine side.
The difference itself is thus nonzero, but adding the negative of the corresponding
0-cochain to the local states restores consistency.

In the second row the coarse period is $u$, and the fine period is $u$ as well.
The fine period is
$z_{e_{01}}+z_{ab}+z_{bc}-z_{ac}$, and substituting \eqref{eq:3.38} shows that it is
zero on every coboundary.
On the other hand, the coordinate of $\varepsilon_R(u)$ at the value 0 is 1, so the
diagnostic period is nonzero as well.
This gives the nonzero verdicts in the table.

In the presentation by periods, the groups and the comparison of this example are
$H^1_{\mathrm{ob}}\cong\mathbb{Z}^2$, $H^1_{\mathrm{diag}}\cong\mathbb{Q}^2$, and
$\Phi:\mathbb{Z}^2\hookrightarrow\mathbb{Q}^2$.
While keeping the difference between integer and rational coefficients,
\eqref{eq:3.42} decides the same vanishing.

\end{example}

\subsection*{Transporting coefficients and local contexts by isomorphisms}

Another basic comparison is the case in which the chosen local contexts and
coefficients are carried by isomorphisms.
If the covers and their intersection diagrams correspond, and the isomorphisms of
coefficients over each intersection commute with the restrictions, then the
comparison of sections gives isomorphisms of the \v{C}ech complexes in each degree.
The differentials are alternating sums of restrictions, hence commute with the
comparison, and the inverse coefficient isomorphisms yield an inverse cochain map.
The vanishing of the corresponding obstruction classes is therefore equivalent.

When the coefficients are built from a ring and an ideal, a ring isomorphism
matching the ideals that define the coefficients induces an isomorphism of the
quotients as well. Matching also the Laws, the witnesses, and the readouts of the
axes can place the classes built from specified residuals into the same comparison.
The comparison by refinement in this chapter goes beyond the case of direct
degreewise isomorphisms: it gives a way to preserve the obstruction verdict through
Condition C and the integer-correction construction.

\section*{Summary of the Chapter}

In this chapter we connected the retained Laws, the local structure, the
coefficients, and the specified local data, and constructed diagnoses along changes
of resolution.

\begin{itemize}
\item \textbf{Canonical resolution.} The quotient merging sources on which all Law
values agree can be read uniquely through every $L$-adequate resolution. Whether it
can be presented by specified extraction methods is determined by comparison with
the equivalence relations of the extraction results.
\item \textbf{Comparison of diagnoses.} We built coefficients from the distinct Law
values occurring on cells, and generated maps commuting with the differentials from
local comparisons. Under Condition C, the map is an isomorphism on $H^1$.
\item \textbf{Uniform invariance.} That the comparison is an isomorphism for every
Law family for which both resolutions are adequate is equivalent to the kernel and
cokernel dimensions being zero for all nonempty subsets of values, and is decidable
from a finite presentation. Condition C is a sufficient condition, and there are
examples on which a fixed local observation agrees while the verdicts differ.
\item \textbf{The role of coefficients.} Even when the structural support is
invariant, which coefficients read the differences is specified independently. For
the coefficients generating all tuples per source, we showed that $H^1$ is zero, by
an explicit formula for the correction.
\item \textbf{Correspondence of obstruction and diagnosis.} We built a map from the
integer generator relations to the Law-value coefficients, and matched the
obstruction and diagnostic classes of the same local data. Under the conditions of
identification of relation components and of common representatives, rational
corrections can be made integral and vanishing is reflected.
\item \textbf{Concrete invariance.} We constructed the comparison square for the
refinement from three charts to four and, on the same input with the non-injective
comparison map $\pi$, carried zero and nonzero obstruction verdicts to both sides.
\end{itemize}

In the opening example, merging internal values that the Law does not read is a
change of resolution preserving the evaluation.
When the regions on which local states are placed are repartitioned as well, a map
built from the comparison of overlaps also becomes necessary.
The results of this chapter give the conditions under which that map preserves
obstruction classes and vanishing, and delimit the range over which the same verdict
can be shared.

\subsection*{Potential applications to code review}

The results of this chapter provide a starting point for thinking about what to
retain in a summary and at what granularity to check during code review.
Consider, for example, a change to order processing in which the inventory update
and the order persistence are reviewed separately.
Besides the description of each process, we also want to check that values and
assumptions are handed over so that the two handle the same order and the same
processing result.
Can the per-function implementations and a description summarizing the handoffs
between modules capture the same inconsistency?

To treat this question with the constructions of this chapter, fix a version and a
semantics of the target code, and define the sources and the Law evaluations from
states and behavior.
The canonical resolution gives the distinctions that a summary must retain in order
to preserve the chosen evaluations.
Taking the reviewed regions as charts, if their overlaps and the correspondences
between regions can be expressed by the nerves and comparison maps of this chapter,
then Condition C of Theorem~\ref{thm:3.17} lets us confirm the preservation of
diagnostic classes along splits and mergers.

Moreover, when obstruction and diagnosis are built from the same local data and the
constructions and hypotheses of \S\S\ref{sec:3.9}--\ref{sec:3.10} are satisfied, a
vanishing diagnosis on the summary can be brought back to the existence of a
correction that makes the original local states consistent.
What vanishing asserts here is correctability under the chosen Laws and
coefficients.
How to realize that correction as an actual code change is a task to be settled at
application time.

The first thing to verify is the agreement of diagnoses when the same code is read
at different granularities.
Checking concretely the correspondence from code to model and the hypotheses of the
comparison leads to a method for evaluating summaries and partitions for review.
To advance to applications comparing the programs before and after a change, one
adds a construction matching their inputs and the specified obstruction classes.

For the gluing obstructions of Chapter~\ref{chap:2}, this chapter has clarified what
is read, and under which conditions the way of reading can be changed while the
same verdict is obtained. The next chapter treats the construction that transports
objects and morphisms along a change of reading, and studies its universality and
its coherence with units and composition.

\chapter{Transport and Coherence of Composition}\label{chap:4}

\section*{Overview of the Chapter}

The question of this chapter is \textbf{under what conditions structure can be carried along
a change of reading and, when such changes are performed in succession, the same result can
be reached coherently}.
Chapter~\ref{chap:1} defined objects and morphisms at the three levels of extraction, core,
and geometry. In this chapter we construct, from a change of extraction and an object before
the change, the core and the geometry after the change.

For example, consider a refactoring that tidies up the field names and data formats of an
order-processing API. We want to read and write the same orders after the change, and to
keep conditions such as ``the total amount equals the sum of the line items.'' Transport in
the sense of this chapter is a construction that carries over to the target not only the
correspondence of data but also the operations and conditions.

Moreover, if there are two routes, one that moves from the old version to the new one through an
intermediate version and one that moves directly, we want the same order to give the
same result along either route. When the conversions are implemented by different teams,
however, discrepancies can arise, such as the correspondence between shipping and billing
addresses being swapped along one route only. Even if each conversion can be undone, the
results of the two routes need not agree.

In the first half we characterize transport by the universal property that every further
change factors through it uniquely. This uniqueness makes the canonical comparisons for
units and composition coherent. In the second half we express the difference between a
separately specified comparison and the canonical comparison as an automorphism, and prove
that this difference can be removed by reselecting edges if and only if all comparison
diagrams can be made commutative at once.

\begin{xltabular}{\linewidth}{@{}LLL@{}}
\toprule
Question & Construction & Outcome \\
\midrule
\endhead
What can be carried from a change of extraction? & Reindexing by a bijection of Atoms & Transport of the core and unique factorization \\
\addlinespace[3pt]
Can covers and coefficients also be carried? & Images of coverage requirements, comparison of overlaps and local data & Transport of the geometry and conditions for its existence \\
\addlinespace[3pt]
What happens under successive transports? & Composition comparisons built from universality & Coherence for units, triple composites, and between levels \\
\addlinespace[3pt]
Can discrepancies of specified comparisons be removed? & Edge reselection and noncommutative defects & Equivalence with the existence of a coherent reselection \\
\addlinespace[3pt]
Can coherence at the levels be combined? & Projection to the core and its kernel & Decision using lifts and the same selection of edges \\
\addlinespace[3pt]
How to preserve the meaning of operations? & Commutative diagrams for lenses and protocols & Simultaneous preservation of reads, updates, and named executions \\
\bottomrule
\end{xltabular}

The diagnosis of Chapter~\ref{chap:3} compared complexes built from Law values and their
classes. The transport of this chapter compares object formation, operations, equations, and
local geometry all at once. What the two have in common is not the statement that equal
values or isomorphic objects exist, but the study of actual comparison maps built from the
input.

\section{Characterizing transport by unique factorization}\label{sec:4.1}

We write the projections of Chapter~\ref{chap:1} as
\begin{equation*}
E_{\mathrm{geom}}\xrightarrow{p}E_{\mathrm{core}}
\xrightarrow{q}B
\tag{4.1}\label{eq:4.1}
\end{equation*}
An object of the base is a pair of an extraction doctrine and a selected source, and a
morphism of the base is an exact change in the sense of Definition~\ref{def:1.27}. The
vocabulary of Atoms is fixed.

\begin{definition}[Strongly opcartesian morphism]\label{def:4.1}

Take a functor $r:E\to B$ and a morphism $\eta:X\to X'$, and let
$r(\eta)=\sigma:b\to b'$.
If, for every object $Y$, every base morphism $\tau:b'\to r(Y)$, and every morphism
$h:X\to Y$ with $r(h)=\tau\sigma$, we have
\begin{equation*}
\exists!\,k:X'\longrightarrow Y,\qquad
r(k)=\tau,\quad k\eta=h
\tag{4.2}\label{eq:4.2}
\end{equation*}
then $\eta$ is called a strongly opcartesian morphism with respect to $r$.
A morphism whose base image is an identity, as in $r(k)=\mathrm{id}_{b'}$, is called a
vertical morphism.

Equation~\eqref{eq:4.2} expresses that, once the base change $\sigma$ has been carried out
first, any further change $h$ splits uniquely, leaving the remaining morphism $k$. Here $\tau$
is arbitrary, which includes the case where the base changes again afterwards. This is the
dual of a strongly cartesian morphism
(\cite[\href{https://stacks.math.columbia.edu/tag/02XJ}{Tag 02XJ, Definition 4.33.1}]{Stacks}).

\end{definition}

\begin{lemma}[Uniqueness and composition of transport]\label{lem:4.2}

If strongly opcartesian morphisms $\eta:X\to X'$ and $\eta':X\to X''$ lift the same base
morphism $\sigma$ from the same object $X$, then there exists a unique isomorphism
\begin{equation*}
c:X'\xrightarrow{\sim}X'',\qquad
r(c)=\mathrm{id},\quad c\eta=\eta'
\tag{4.3}\label{eq:4.3}
\end{equation*}
Moreover, identity morphisms are strongly opcartesian, and composites of such morphisms are
again strongly opcartesian.

\end{lemma}

\begin{proof}

Applying \eqref{eq:4.2} to the identity morphism of the base, we obtain $c$ and a morphism
$d$ in the opposite direction. Applying uniqueness to $dc\eta=\eta$ and $cd\eta'=\eta'$
gives $dc=\mathrm{id}$ and $cd=\mathrm{id}$.
For composites, a morphism out of $X\xrightarrow{\eta}X'\xrightarrow{\zeta}X''$ is factored
first through $\eta$ and then through $\zeta$. The two applications of uniqueness give
uniqueness for the composite. For an identity morphism, $k=h$.

\end{proof}

\section{Constructing the core along an exact change}\label{sec:4.2}

To obtain the transported core, we reread the names of the Atoms as their names after the
change. An input is first sent back to the original names, the original construction is
used, and the output is sent to the new names. The same procedure applies to operations and
to equations.

\begin{construction}[Reindexing of Atoms]\label{cons:4.3}

Let $\sigma=(f,e):(D,s)\to(D',s')$ be an exact morphism of the base. Here $e$ is a
bijection of Atoms and $s'=f(s)$. For families and configurations we use the direct image
$e_*$ of Chapter~\ref{chap:1}. On objects we define
\begin{equation*}
T_e(A)=(e_*C_A,S_A,Q_A)
\tag{4.4}\label{eq:4.4}
\end{equation*}
The structure data and the observed values are kept, and the configuration is carried. The
map $T_{e^{-1}}$ is inverse to $T_e$.
The Atom map $d$ of a morphism of configurations is carried to $ede^{-1}$.

For a core reading $P$, the reading after the change is defined as follows.
\begin{equation*}
\begin{aligned}
\mathrm{Comp}^{\sigma}(F)
 &=e_*\mathrm{Comp}_P(e_*^{-1}F),\\
\mathrm{Form}^{\sigma}(C)
 &=T_e\mathrm{Form}_P(e_*^{-1}C),\\
\mathrm{Op}^{\sigma}(A,B)
 &=\mathrm{Op}_P(T_{e^{-1}}A,T_{e^{-1}}B).
\end{aligned}
\tag{4.5}\label{eq:4.5}
\end{equation*}
The names of the operations are kept as they are, and their actions on configurations are
conjugated. The evaluations of invariants and signatures are defined by composition with
$T_{e^{-1}}$, and the indices, value ranges, and selected axes are kept.

The sets of supports, axes, and observables of a context $W$ are also kept. The condition
that a support reads an Atom $a'$ is defined as the condition that the original support
reads $e^{-1}(a')$.
This yields an equivalence of context categories $\theta_e:\mathcal{C}_P\simeq \mathcal{C}^\sigma$.
We write $\bar\theta_e$ for the reindexing in the opposite direction.
Keeping the equation indices and roles, we define
\begin{equation*}
\begin{aligned}
O^\sigma(W')&=O_P(\bar\theta_eW'),\\
\nu^\sigma_{W',i,a'}&=\nu_{P,\bar\theta_eW',i,e^{-1}(a')},\\
\varepsilon^\sigma_{W',A',i,a'}
 &=\varepsilon_{P,\bar\theta_eW',T_{e^{-1}}A',i,e^{-1}(a')}
\end{aligned}
\tag{4.6}\label{eq:4.6}
\end{equation*}
The restriction maps are also reindexed along $\bar\theta_e$.
In the detector code, we apply $e$ to each Atom appearing in a query.
The positive and negative signs, the logical operations, and the operation names are
unchanged.

\end{construction}

\begin{proposition}[The transported core and the canonical morphism]\label{prop:4.4}

Construction~\ref{cons:4.3} yields a core $\sigma_!P$ over the base $(D',s')$ and a
morphism of cores
$\eta_{\sigma,P}:P\to\sigma_!P$ over the base morphism $\sigma$.
Its extracted family, base object, and generated object family satisfy
\begin{equation*}
F_{\sigma_!P}=e_*F_P,\qquad
A_{\sigma_!P}=T_eA_P,\qquad
\mathrm{Obj}_{\sigma_!P}=T_e(\mathrm{Obj}_P)
\tag{4.7}\label{eq:4.7}
\end{equation*}
\end{proposition}

\begin{proof}

The first equality follows from preservation and reflection of extraction and from the
bijectivity of $e$. Hence the extracted family is finite, and \eqref{eq:4.5} can be
applied. Applying the object-formation formula next gives the equality for the base
objects. Finite sequences of operations are carried into each other by $T_e$ and
$T_{e^{-1}}$, so the equality for the generated object families follows from
Theorem~\ref{thm:1.18}.

The object map of the canonical morphism is $T_e$, and its operation map is the
correspondence of \eqref{eq:4.5}.
The commutative square for configurations is $(ede^{-1})e=ed$.
Equation~\eqref{eq:4.6} preserves coordinates and residuals, and commutation with the
restrictions follows from the original commutation.
Agreement of queries is preserved and reflected for both the positive and the negative
sign, so acceptance and soundness of the detectors are carried as well.
The conditions on invariants and signatures are obtained by composing with the inverse
maps. This verifies all the conditions of Definition~\ref{def:1.28}.

\end{proof}

\begin{theorem}[Universality of core transport]\label{thm:4.5}

The morphism $\eta_{\sigma,P}$ is strongly opcartesian with respect to
$q:E_{\mathrm{core}}\to B$.

\end{theorem}

\begin{proof}

Take $h:P\to R$ with $q(h)=\tau\sigma$.
We build the factor $k$ by sending each datum of $\sigma_!P$ back to its original names
and then applying $h$. For example, its object map is
\begin{equation*}
H_k=H_hT_{e^{-1}}
\tag{4.8}\label{eq:4.8}
\end{equation*}
and the operation map likewise returns to the original operations by the correspondence of
\eqref{eq:4.5} and then applies $\Phi_h$.
The comparison of equations is reindexed along $\bar\theta_e$, and $e^{-1}$ is substituted
into the Atoms of the coordinates, the residuals, and the detectors.
For the maps of invariant indices and of signature axes and value ranges we use those of
$h$. For the base morphism we use $\tau$.

Substituting these into \eqref{eq:4.5}--\eqref{eq:4.6}, the preservation equations for $h$
become exactly the preservation equations for $k$. Hence $k$ is a morphism of cores.
From $T_{e^{-1}}T_e=\mathrm{id}$ and the inverse correspondences of the indices we obtain
$k\eta_{\sigma,P}=h$.

Conversely, the object map of any morphism satisfying this composition equation is
determined to be \eqref{eq:4.8} by the surjectivity of $T_e$.
The components for operations, contexts, coordinates of equations, and invariants and
signatures are likewise determined uniquely through the invertible correspondences of the
canonical morphism.
Since the base is fixed to $\tau$, the whole morphism is unique.

\end{proof}

Bijectivity of the source map $f$ is not needed for this construction.
What is needed is that extraction is preserved and reflected by the specified
correspondence of Atoms.

\begin{example}[Forward preservation of extraction is not enough]\label{ex:4.6}

Let the Atoms be $\{a,b,c\}$, with a single source and identity normalization.
Let the extracted family before the change be $\{a,b\}$ and the one after the change be
$\{a,b,c\}$, and let the correspondence of Atoms be the bijection that swaps $a,b$ and
fixes $c$.
Extraction is preserved in the forward direction, but reflection fails for $c$.
The family after the change is not the direct image of the original family, so this is not
an exact transport satisfying \eqref{eq:4.7}.

Object formation that uses the added Atom requires data about the added part.
The next construction makes this role explicit.

\end{example}

\begin{construction}[Additional data for a change that enlarges extraction]\label{cons:4.7}

Take a change $(f,e)$ that preserves extraction in the forward direction, with $e$ a
bijection.
Finiteness of the extracted family $F'$ after the change is given as part of the data.
Let $A'$ be the base object built from $F'$ by the object formation of \eqref{eq:4.5}, and
add the following data.

\begin{enumerate}
\item An actual operation $A'\to T_eA_P$ in the transported operation reading.
\item An equation system on the contexts of $A'$, together with a bijection to the original
equation indices.
\item Soundness, with respect to that equation system, of the code obtained by carrying the
original detector code along $e$.
\end{enumerate}

From these data the core after the change can be constructed.
Moreover, the images of the original generated objects are contained in the generated
objects after the change.
Indeed, it suffices to follow the added operation by the image of an original finite
sequence of operations.
On the images of the original objects, agreement of queries and acceptance by the
detectors are preserved.
This follows from the commutation of the evaluation of queries and codes with the
reindexing of Atoms.
Operations, invariants, and signatures are also preserved by the corresponding maps.
What this construction provides is a comparison equipped with the forward preservation
just described.

\end{construction}

\section{Lifting the transport of the core to the geometry}\label{sec:4.3}

After the core has been carried, we carry the choice of which local data are viewed
together. For the coverage requirements we use their images, and for the coefficients and
the presentations of rings we use reindexing. The supports, axes, and observables inside a
context are compared together with their readable elements and restriction maps.

\begin{construction}[Transport of the components of a geometry]\label{cons:4.8}

Take a geometry $G$ and a morphism of cores $h:pG\to P'$.
Let $\theta$ be its equivalence of contexts and $\bar\theta$ a specified quasi-inverse.
The coverage requirements after the change are defined as the images under existential
quantification along the maps of Definition~\ref{def:1.30}.
For example, the required supports and their visibility are
\begin{equation*}
\begin{aligned}
\mathrm{Req}'(a')
 &\Longleftrightarrow \exists a,\ \mathrm{Req}(a)\land e(a)=a',\\
\mathrm{Vis}'(W',a')
 &\Longleftrightarrow
 \exists W,a,\ \mathrm{Vis}(W,a)\land\theta W=W'\land e(a)=a'
\end{aligned}
\tag{4.9}\label{eq:4.9}
\end{equation*}
The equation coordinates, the witnesses, the axes, and the visibility between two contexts
are carried by the same rule.
The overlaps are
\begin{equation*}
\mathrm{Ov}'(U'\to W'\leftarrow V')
=\theta\bigl(\mathrm{Ov}(\bar\theta U'\to
\bar\theta W'\leftarrow\bar\theta V')\bigr)
\tag{4.10}\label{eq:4.10}
\end{equation*}
and the projections are sent to $U',V'$ using the counit.
The universal property of the overlaps follows from the equivalence $\theta,\bar\theta$.
The coefficient ring $k$ is kept, and the raw system is
$\mathcal{B}'=\bar\theta^*\mathcal{B}$.
The coordinate types, the local data types, and the relation polynomials and restrictions
are all reindexed.
We write $h_*G$ for this geometry.

\end{construction}

\begin{definition}[Transport condition for local realizations]\label{def:4.9}

Define $H_{\mathrm{geom}}(G,h)$ to be a family, over the contexts, of three comparison maps
\begin{equation*}
\begin{aligned}
s_W&:\mathrm{Supp}(W)\to\mathrm{Supp}'(\theta W),\\
a_W&:\mathrm{Ax}(W)\to\mathrm{Ax}'(\theta W),\\
o_W&:\mathrm{Obs}(W)\to\mathrm{Obs}'(\theta W)
\end{aligned}
\tag{4.11}\label{eq:4.11}
\end{equation*}
that preserves readable elements in the sense of Definition~\ref{def:1.30} and satisfies
the naturality of \eqref{eq:1.14}.
The three sets on the target side and their readouts are those carried by the contexts of
the target core.

\end{definition}

\begin{proposition}[Existence condition for morphisms of geometries]\label{prop:4.10}

The data of $H_{\mathrm{geom}}(G,h)$ exist if and only if a morphism of geometries
$G\to h_*G$ exists over $h$.
Moreover, any morphism of geometries over $h$ yields $H_{\mathrm{geom}}(G,h)$.

\end{proposition}

\begin{proof}

Equation~\eqref{eq:4.9} gives forward preservation of the coverage requirements.
Equation~\eqref{eq:4.10} gives the comparison of overlaps; for the coefficients we use the
identity map, and for the raw system the reindexing equality.
The remaining comparisons are \eqref{eq:4.11} together with their preservation and
naturality, and these are exactly the data of $H_{\mathrm{geom}}(G,h)$.
In the opposite direction, one extracts this family of three comparisons from a morphism
of geometries.

\end{proof}

\begin{theorem}[Canonical geometry transport and universality]\label{thm:4.11}

For the canonical morphism of cores $h=\eta_{\sigma,pG}$, the data
$H_{\mathrm{geom}}(G,h)$ can be constructed.
We write
$\sigma_!^{\mathrm g}G$ and $\widetilde\eta_{\sigma,G}:G\to\sigma_!^{\mathrm g}G$
for the resulting geometry and morphism. This morphism is strongly opcartesian with
respect to $p:E_{\mathrm{geom}}\to E_{\mathrm{core}}$, and it satisfies
\begin{equation*}
p(\sigma_!^{\mathrm g}G)=\sigma_!(pG),\qquad
p(\widetilde\eta_{\sigma,G})=\eta_{\sigma,pG}
\tag{4.12}\label{eq:4.12}
\end{equation*}
\end{theorem}

\begin{proof}

In the context transport of Construction~\ref{cons:4.3}, the sets of supports, axes, and
observables were kept. We use these identity correspondences in \eqref{eq:4.11}. Since the
readout of supports was defined by $e^{-1}$, the condition of mapping Atoms by $e$ is
satisfied, and the three naturalities follow from the original structure maps.
Hence Proposition~\ref{prop:4.10} gives a morphism of geometries.

To prove universality, take an arbitrary morphism of cores
$t:\sigma_!(pG)\to pK$ and a morphism of geometries $F:G\to K$ with
$p(F)=t\eta_{\sigma,pG}$.
We send each context back by $\bar\theta_e$, invert the three invertible correspondences
of the canonical morphism, and then apply the comparison maps of $F$.
This defines the comparisons of supports, axes, and observables of the factor.
The coefficient map is set equal to the coefficient map of $F$.

Each coverage requirement on the target has, by \eqref{eq:4.9}, a witness coming from an
original requirement.
Applying the preservation laws of $F$ to that witness yields the preservation laws of the
factor.
The comparison of overlaps is built by composing the two equivalences of categories, and
the equality of raw systems is obtained from the commutation of reindexing with the change
of coefficients.
We thus obtain a morphism of geometries whose base is $t$, and its composite with the
canonical morphism is $F$.

The three comparisons of the canonical morphism are bijections and its coefficient map is
the identity, so each map of the factor is unique.
The comparisons of overlaps inside the context category are determined by the uniqueness
of parallel morphisms in a preorder category.
This establishes the existence and uniqueness of \eqref{eq:4.2}.
Equation~\eqref{eq:4.12} follows from the construction.

\end{proof}

For a general morphism of cores, Proposition~\ref{prop:4.10} gives the existence
condition; for the canonical core transport, both that condition and the universal
property have now been constructed.

\section{Units, composition, and projections of transport}\label{sec:4.4}

The universal property of transport determines not only how objects are carried but also
how comparison morphisms are carried. Carrying morphisms between objects over the same
base yields functors between the fibers.

\begin{lemma}[Universality in the tower]\label{lem:4.12}

If a morphism $v:G\to G'$ is strongly opcartesian with respect to $p$, and $p(v)$ is
strongly opcartesian with respect to $q$, then
$v$ is strongly opcartesian with respect to $qp$.

\end{lemma}

\begin{proof}

A factorization problem over $qp$ is first split by the universal property of $q$ into a
factor at the level of cores, and the universal property of $p$ is then applied with that
core factor as the base morphism.
Existence and uniqueness at the two levels give the required existence and uniqueness.

\end{proof}

\begin{construction}[Transport functors between fibers]\label{cons:4.13}

For a functor $r:E\to B$, the fiber $E_b$ is the category consisting of the
objects satisfying $r(X)=b$ and the vertical morphisms between them.
For each $\sigma:b\to b'$ and $X\in E_b$, choose a strongly opcartesian lift
$\eta_{\sigma,X}:X\to T_\sigma X$.
The image of a vertical morphism $v:X\to Y$ is defined to be the unique vertical morphism
\begin{equation*}
T_\sigma(v):T_\sigma X\longrightarrow T_\sigma Y,\qquad
T_\sigma(v)\eta_{\sigma,X}=\eta_{\sigma,Y}v
\tag{4.13}\label{eq:4.13}
\end{equation*}
The right-hand side lies over the base $\sigma$, so this morphism exists.
Identity morphisms and composites of two morphisms satisfy the same equation, so
uniqueness shows that $T_\sigma:E_b\to E_{b'}$ is a functor.
By Theorems~\ref{thm:4.5} and~\ref{thm:4.11} and Lemma~\ref{lem:4.12}, this
construction applies to cores and to geometries with base $B$.

\end{construction}

\begin{construction}[Composition comparisons and unit comparisons]\label{cons:4.14}

For $b\xrightarrow{\sigma}b'\xrightarrow{\tau}b''$, applying Lemma~\ref{lem:4.2} to the
twofold lift and the single lift yields an isomorphism
\begin{equation*}
\begin{aligned}
c_{\tau,\sigma,X}&:T_\tau T_\sigma X\xrightarrow{\sim}T_{\tau\sigma}X,\\
c_{\tau,\sigma,X}\eta_{\tau,T_\sigma X}\eta_{\sigma,X}
 &=\eta_{\tau\sigma,X}
\end{aligned}
\tag{4.14}\label{eq:4.14}
\end{equation*}
For identity morphisms we similarly obtain
\begin{equation*}
\epsilon_X:T_{\mathrm{id}}X\xrightarrow{\sim}X,\qquad
\epsilon_X\eta_{\mathrm{id},X}=\mathrm{id}_X
\tag{4.15}\label{eq:4.15}
\end{equation*}
We call these the compositor and the unitor, respectively.

\end{construction}

\begin{theorem}[Coherence of composition comparisons]\label{thm:4.15}

The families $c_{\tau,\sigma}$ and $\epsilon$ are natural isomorphisms.
Moreover, for composable morphisms $\sigma,\tau,\rho$,
\begin{equation*}
\begin{aligned}
c_{\rho,\tau\sigma,X}\,T_\rho(c_{\tau,\sigma,X})
 &=c_{\rho\tau,\sigma,X}\,c_{\rho,\tau,T_\sigma X},\\
c_{\mathrm{id},\sigma,X}&=\epsilon_{T_\sigma X},\\
c_{\sigma,\mathrm{id},X}&=T_\sigma(\epsilon_X)
\end{aligned}
\tag{4.16}\label{eq:4.16}
\end{equation*}
Associativity and units of composed base morphisms are identified through the equalities
of the category $B$.
The structure consisting of the fibers, the transport functors, and these natural
isomorphisms is called a pseudofunctor.

\end{theorem}

\begin{proof}

We compose the two routes of the naturality square for $c$ with the twofold lift from the
starting point.
By \eqref{eq:4.13}--\eqref{eq:4.14}, both become $\eta_{\tau\sigma,Y}v$.
The twofold lift is strongly opcartesian, so the two routes are equal.
Naturality of $\epsilon$ follows by the same argument.

Both sides of the first equation of \eqref{eq:4.16} are vertical morphisms from
$T_\rho T_\tau T_\sigma X$ to $T_{\rho\tau\sigma}X$.
Composed with the threefold lift, both become $\eta_{\rho\tau\sigma,X}$.
Again they are equal by uniqueness. The two unit equations are checked in the same way by
composing with the corresponding twofold lifts.

\end{proof}

This canonical construction is the dual of the cartesian case treated in
\cite[\href{https://stacks.math.columbia.edu/tag/02XJ}{Tag 02XJ, Lemmas 4.33.2 and 4.33.7}]{Stacks}.
The reason for the coherence lies in the uniqueness of the factorizations that determined
the comparisons.

\begin{theorem}[Compatibility with the projection to the core]\label{thm:4.16}

Let $p_b$ be the projection from the fiber of geometries to the fiber of cores.
Write $T^{\mathrm g}_\sigma$ and $T^{\mathrm c}_\sigma$ for the transport of geometries
and the transport of cores, respectively.
There is a natural isomorphism
\begin{equation*}
\chi_\sigma:p_{b'}T^{\mathrm g}_\sigma
\xRightarrow{\sim}T^{\mathrm c}_\sigma p_b
\tag{4.17}\label{eq:4.17}
\end{equation*}
and, for composition and units, it satisfies
\begin{equation*}
\begin{aligned}
\chi_{\tau\sigma,G}\,p(c^{\mathrm g}_{\tau,\sigma,G})
 &=
c^{\mathrm c}_{\tau,\sigma,pG}\,
T^{\mathrm c}_\tau(\chi_{\sigma,G})\,
\chi_{\tau,T^{\mathrm g}_\sigma G},\\
p(\epsilon^{\mathrm g}_G)
 &=\epsilon^{\mathrm c}_{pG}\,\chi_{\mathrm{id},G}
\end{aligned}
\tag{4.18}\label{eq:4.18}
\end{equation*}
\end{theorem}

\begin{proof}

The morphisms $p(\widetilde\eta_{\sigma,G})$ and $\eta_{\sigma,pG}$ are strongly
opcartesian morphisms lifting the same base morphism from the same object.
The unique comparison connecting them is $\chi_{\sigma,G}$.
In the canonical construction the two lifts correspond by \eqref{eq:4.12}.
Equation~\eqref{eq:4.13} still holds after projecting a vertical morphism of geometries,
so the comparison is natural.
The two routes of \eqref{eq:4.18} also become the same morphism when composed with the
lift from the original core.
Uniqueness gives the two equations.

\end{proof}

Hence the route that carries the geometry and then reads off the core and the route that
reads off the core and then carries it are compatible, including objects, morphisms, and
composition comparisons.

\section{The difference between specified and canonical comparisons}\label{sec:4.5}

The canonical comparisons were coherent by uniqueness.
In practice, however, a correspondence of state names or of local data is sometimes
specified separately.
To measure how far such a correspondence differs from the canonical comparison, we read
both as automorphisms of the same object.

Henceforth, let $r:E\to B$ be a functor and choose strongly opcartesian
morphisms on the required edges.
For cores we may use $r=q$, and for full geometries, $r=qp$.

\begin{definition}[Finite comparison data]\label{def:4.17}

On each vertex $i$ of a finite directed graph we place an object $X_i$, and on each edge
$e:i\to j$ a base morphism $\sigma_e:r(X_i)\to r(X_j)$ and a strongly opcartesian
morphism $L_e:X_i\to X_j$ over it.
We write $L_\pi$ for the composite along a path $\pi$; the composite along the empty path
is the identity.

We further specify finitely many faces $f$.
A face is a pair of paths $\ell_f,\rho_f:i\to j$ with the same start and end whose
composites in the base are equal.
To each face we assign a fiber automorphism of the end,
$u_f\in\mathrm{Aut}_r(X_j)$, where
\begin{equation*}
\mathrm{Aut}_r(X)
=\{a:X\xrightarrow{\sim}X\mid r(a)=\mathrm{id}_{r(X)}\}
\tag{4.19}\label{eq:4.19}
\end{equation*}
The automorphism $u_f$ represents the comparison we wish to adopt on that face.
At this stage we do not assume $u_fL_{\ell_f}=L_{\rho_f}$.

\end{definition}

\begin{construction}[Canonical comparisons and raw defects]\label{cons:4.18}

By Lemma~\ref{lem:4.2}, each face carries a unique fiber automorphism
\begin{equation*}
\phi_fL_{\ell_f}=L_{\rho_f},\qquad
\phi_f\in\mathrm{Aut}_r(X_j)
\tag{4.20}\label{eq:4.20}
\end{equation*}
The difference from the specified comparison is defined as
\begin{equation*}
\delta_f=u_f\phi_f^{-1}
\tag{4.21}\label{eq:4.21}
\end{equation*}
and called the raw defect.
Thus $\delta_f=1$ means that the specified comparison agrees with the canonical one.
Products are composites of morphisms, and the right-hand factor acts first.

If the automorphism group is noncommutative, the result depends on the order of the
comparisons.
For this reason a discrepancy cannot in general be treated as a numerical difference or as
an element of a commutative group.
Equation~\eqref{eq:4.21} is a difference that also records the order of the comparisons.

\end{construction}

\section{Removing discrepancies by reselecting edges}\label{sec:4.6}

We now keep the comparisons fixed and change the lift adopted on each edge.
By Lemma~\ref{lem:4.2}, lifts of the same base morphism with the same start and end are
carried into one another by composing with a fiber automorphism of the end.

\begin{definition}[Edge reselection]\label{def:4.19}

For each edge $e:i\to j$ choose
$a_e\in\mathrm{Aut}_r(X_j)$ and set
\begin{equation*}
L_e^a=a_eL_e,\qquad
\mathcal{A}=\prod_{e:i\to j}\mathrm{Aut}_r(X_j)
\tag{4.22}\label{eq:4.22}
\end{equation*}
We call $a=(a_e)_e$ an edge reselection.
A vertical isomorphism $a_e$ is strongly opcartesian, since the factor in \eqref{eq:4.2}
is uniquely determined as $k=ha_e^{-1}$. By Lemma~\ref{lem:4.2}, the composite $a_eL_e$ is
then strongly opcartesian as well.
For the composites $L_\pi^a$ along paths we define \eqref{eq:4.20}--\eqref{eq:4.21} in the
same way and write $\phi_f(a)$ and $\delta_f(a)$.
The specified comparisons $u_f$ stay fixed.

\end{definition}

\begin{example}[A comparison of two independent edges can be absorbed into one of them]\label{ex:4.20}

Regard a group as a one-object category, with the terminal category as base.
All morphisms are invertible and strongly opcartesian.
For the group product we use composition acting from the right.
Take two parallel edges $x,y$ and one face $x\Rightarrow y$.
Let both original lifts be the identity, and let the specified comparison be an arbitrary
group element $u$.
For edge selections $a_x,a_y$ we have
\begin{equation*}
\phi(a)=a_ya_x^{-1},\qquad
\delta(a)=u\,a_xa_y^{-1}
\tag{4.23}\label{eq:4.23}
\end{equation*}
Setting $a_x=1$ and $a_y=u$ makes the defect vanish.
The two edges can be reselected independently, so the specified comparison can be absorbed
into one of the edges.

In a general comparison diagram, the selection at each
edge acts on several paths and faces.
We therefore examine the change arising at the end of a path, and determine how it alters
the comparison on each face.

\end{example}

\begin{lemma}[The change at the end of a path]\label{lem:4.21}

If a reselection $a$ is followed by $b$, each path $\pi:i\to j$ carries a unique
automorphism
\begin{equation*}
A_\pi(a,b)L_\pi^a=L_\pi^{ba}
\tag{4.24}\label{eq:4.24}
\end{equation*}
where $(ba)_e=b_ea_e$.
This automorphism satisfies
\begin{equation*}
\begin{aligned}
A_\pi(a,1)&=1,\\
A_\pi(ba,c)A_\pi(a,b)&=A_\pi(a,cb)
\end{aligned}
\tag{4.25}\label{eq:4.25}
\end{equation*}
and, for each face,
\begin{equation*}
\begin{aligned}
\phi_f(ba)
 &=A_{\rho_f}(a,b)\phi_f(a)A_{\ell_f}(a,b)^{-1},\\
\delta_f(ba)
 &=\bigl(u_fA_{\ell_f}(a,b)u_f^{-1}\bigr)
   \delta_f(a)A_{\rho_f}(a,b)^{-1}
\end{aligned}
\tag{4.26}\label{eq:4.26}
\end{equation*}
\end{lemma}

\begin{proof}

Existence and uniqueness in \eqref{eq:4.24} follow from the composite along a path being
strongly opcartesian.
Composing both sides of \eqref{eq:4.25} with $L_\pi^a$ gives the same morphism, so
uniqueness applies again.
The first equation of \eqref{eq:4.26} is obtained by applying uniqueness to
$\phi_f(ba)A_{\ell_f}(a,b)L_{\ell_f}^a=A_{\rho_f}(a,b)\phi_f(a)L_{\ell_f}^a$.
Substituting its inverse into \eqref{eq:4.21} yields the second.

\end{proof}

\begin{proposition}[The action of reselections]\label{prop:4.22}

Let $\mathcal{Z}=\prod_f\mathrm{Aut}_r(X_{\mathrm{end}(f)})$ be the family of
automorphisms at the ends of the faces.
On $\mathcal{A}\times \mathcal{Z}$, a left action is defined by
\begin{equation*}
\Gamma_b(a,z)=
\left(ba,\
\left(
u_fA_{\ell_f}(a,b)u_f^{-1}\,
z_f\,A_{\rho_f}(a,b)^{-1}
\right)_f\right)
\tag{4.27}\label{eq:4.27}
\end{equation*}
In particular, the orbit of $(1,\delta(1))$ is
$\{(a,\delta(a))\mid a\in \mathcal{A}\}$.

\end{proposition}

\begin{proof}

Substituting \eqref{eq:4.25} gives $\Gamma_1=\mathrm{id}$ and
$\Gamma_c\Gamma_b=\Gamma_{cb}$.
The description of the orbit is by \eqref{eq:4.26}.

\end{proof}

This action also records the current edge selection $a$.
The reason is that, for a path through several edges, the change $A_\pi(a,b)$ at the end
depends on the current selection.

For example, take the functor from a one-object group category to the terminal
category, and let both initial edge lifts be identities.
Writing $12$ for the path that traverses edge 1 and then edge 2, we have
\[
\begin{aligned}
L_{12}^{a}&=a_2a_1,\\
L_{12}^{ba}&=b_2a_2b_1a_1,\\
A_{12}(a,b)&=L_{12}^{ba}(L_{12}^{a})^{-1}
=b_2a_2b_1a_2^{-1}.
\end{aligned}
\]
The current selection $a_2$ remains in the last expression.
In a noncommutative group, even reselection by the same $b_1,b_2$ can produce
different changes at the end depending on $a_2$.

\begin{theorem}[Vanishing of the obstruction and coherent selections]\label{thm:4.23}

We call the orbit of Proposition~\ref{prop:4.22} the obstruction class of the finite
comparison data.
When the orbit meets $\mathcal{A}\times\{1\}$, we say that the obstruction vanishes.
This vanishing is equivalent to the existence of an edge reselection $a$ such that, for
all faces simultaneously,
\begin{equation*}
u_fL_{\ell_f}^a=L_{\rho_f}^a
\tag{4.28}\label{eq:4.28}
\end{equation*}
\end{theorem}

\begin{proof}

Vanishing of the obstruction amounts to $\delta_f(a)=1$ for all faces, which by
\eqref{eq:4.21} is equivalent to $u_f=\phi_f(a)$.
This equality gives \eqref{eq:4.28}.
Conversely, if \eqref{eq:4.28} holds, then $u_f$ and $\phi_f(a)$ both send
$L_{\ell_f}^a$ to $L_{\rho_f}^a$.
They are equal by the uniqueness for strongly opcartesian morphisms.

\end{proof}

\section{Additional conditions for pasting faces}\label{sec:4.7}

Performing the comparison of one face and then that of another produces a route that
rewrites paths.
For example, when three paths are compared, there are the direct comparison and the
comparison through the third path.
Agreement of these two routes must also be examined as a separate condition.

\begin{lemma}[Transport of automorphisms along a subsequent path]\label{lem:4.24}

Fix an edge selection $a$.
For a path $\beta:j\to k$ and $v\in\mathrm{Aut}_r(X_j)$, a unique automorphism
$\beta_*v$ satisfies
\begin{equation*}
(\beta_*v)L_\beta^a=L_\beta^av
\tag{4.29}\label{eq:4.29}
\end{equation*}
The map $\beta_*$ is a group homomorphism.
\end{lemma}

\begin{proof}

We use the universal property of the strongly opcartesian morphism $L_\beta^a$.
The morphism obtained from $v^{-1}$ is the inverse.
For identities and products, apply \eqref{eq:4.29} and uniqueness as well.

\end{proof}

\begin{construction}[Pasting of comparisons]\label{cons:4.25}

Consider a single step that rewrites the path $\alpha\ell_f\beta$ into
$\alpha \rho_f\beta$.
Paths proceed from left to right. Since composition of morphisms acts from the right,
\[
L_{\alpha\ell_f\beta}^{a}
=L_\beta^{a}L_{\ell_f}^{a}L_\alpha^{a}
\]
The specified comparison of this step is $\beta_*u_f$, and its canonical comparison is
$\beta_*\phi_f(a)$.
For a rewriting in the opposite direction we use the respective inverses.

Along a sequence of rewritings $P$, we write $U_P(a)$ and $\Phi_P(a)$ for the composites
of these in order.
If the initial path is $\pi$ and the final path is $\rho$, then
\begin{equation*}
\Phi_P(a)L_\pi^a=L_\rho^a,\qquad
D_P(a)=U_P(a)\Phi_P(a)^{-1}
\tag{4.30}\label{eq:4.30}
\end{equation*}
We call $D_P(a)$ the pasted raw defect.
The canonical comparison $\Phi_P(a)$ is uniquely determined by the initial and final
paths.

\end{construction}

For two successive rewrites, write $u_1,u_2$ for the specified comparisons,
$\phi_1,\phi_2$ for the canonical comparisons, and $\delta_1,\delta_2$ for the defects.
Then
\begin{equation*}
\begin{aligned}
U_P\Phi_P^{-1}
 &=u_2u_1(\phi_2\phi_1)^{-1}\\
 &=(u_2\delta_1u_2^{-1})\delta_2.
\end{aligned}
\tag{4.31}\label{eq:4.31}
\end{equation*}

The later comparison conjugates the earlier defect.
Thus the defect of a pasting cannot be replaced by the simple product of the defects
of its faces.

\begin{proposition}[Coherence condition among relations]\label{prop:4.26}

A pair of rewriting sequences $P,Q$ from the same path to the same path is called a
syzygy.
Specify finitely many syzygies, and suppose that for the edge selection $a$
\begin{equation*}
U_P(a)=U_Q(a)
\tag{4.32}\label{eq:4.32}
\end{equation*}
holds for each of them. Then
$D_P(a)=D_Q(a)$.

\end{proposition}

\begin{proof}

The comparisons $\Phi_P(a)$ and $\Phi_Q(a)$ send the same strongly opcartesian morphism
to the same morphism, hence are equal. Multiply \eqref{eq:4.32} by their common inverse
on the right.

\end{proof}

A raw defect is defined by \eqref{eq:4.21} alone.
To obtain the cocycle condition $D_P(a)=D_Q(a)$ along the specified syzygies,
\eqref{eq:4.32} is required.
Condition~\eqref{eq:4.32} is a relation among the specified comparisons; it does not
assume the vanishing of the defect on each face.

\begin{proposition}[Obstruction for closed comparisons]\label{prop:4.27}

For any syzygy, writing $\Phi$ for the common canonical comparison,
\begin{equation*}
D_Q^{-1}D_P
=\Phi\,(U_Q^{-1}U_P)\,\Phi^{-1}
\tag{4.33}\label{eq:4.33}
\end{equation*}
In particular, the left-hand side being the identity is equivalent to \eqref{eq:4.32}.

\end{proposition}

\begin{proof}

Substitute $D_P=U_P\Phi^{-1}$ and $D_Q=U_Q\Phi^{-1}$.

\end{proof}

Equation~\eqref{eq:4.33} shows that, once the edge selection is fixed, a mismatch of the
specified pastings persists through conjugation.
If every face satisfies \eqref{eq:4.28}, then every specified pasting agrees with the
canonical comparison.
Failure of \eqref{eq:4.32} therefore obstructs overall coherence for that edge selection.

\section{Removable and irremovable discrepancies}\label{sec:4.8}

In Example~\ref{ex:4.20}, the comparison of a single face could be made coherent by
reselecting one of two independent edges.
When several faces share edges, each face may be made coherent on its own, and yet not all of them at
once by the same selection.
Again, regard a group as a one-object category, with the terminal category as base.

\begin{example}[Different requirements on the same two edges]\label{ex:4.28}

Let the group be the permutation group $S_3$ on three points, and place two faces on the
same two edges.
Let the specified comparisons be $s=(12)$ and $t=(23)$.
The canonical comparison of both faces is the same $a_ya_x^{-1}$, so coherence requires
\begin{equation*}
a_ya_x^{-1}=s,\qquad a_ya_x^{-1}=t
\tag{4.34}\label{eq:4.34}
\end{equation*}
Since $s\ne t$, no selection satisfies both at once.
Each face on its own can be made coherent by Example~\ref{ex:4.20}.
Making the two faces coherent by the same edge selection is the new condition.

\end{example}

\begin{example}[Three comparisons and order]\label{ex:4.29}

Take three parallel edges $x_0,x_1,x_2$, with all original lifts the identity.
Specify the comparison of $x_0\Rightarrow x_1$ to be $s$,
that of $x_1\Rightarrow x_2$ to be $t$, and
that of $x_0\Rightarrow x_2$ to be $st$.
The indirect comparison is $ts$ and the direct comparison is $st$.
For $s=(12)$ and $t=(23)$, the two differ.

Indeed, if the first two faces are coherent, then
$a_1=sa_0$ and $a_2=ta_1=tsa_0$.
The third face requires $a_2=sta_0$, so the three faces are not coherent simultaneously.
The corresponding syzygy has no subsequent path, and the two specified routes do not
depend on the edge selection.
The inner discrepancy in \eqref{eq:4.33} is
\begin{equation*}
(st)^{-1}ts=t^{-1}s^{-1}ts\ne1
\tag{4.35}\label{eq:4.35}
\end{equation*}
and it persists for every selection.
In this example, besides each correspondence being a bijection, the orders of composition
must be brought into line.

\end{example}

\section{Achieving coherence of the core and the geometry with the same selection}\label{sec:4.9}

Projecting a comparison of geometries to the core yields a comparison of cores.
After correcting the discrepancies visible at the core, we want to correct the
discrepancies that remain only in the geometry.
The key that connects these two stages is to lift the changes selected at the core to the
geometry, and to examine the remaining conditions using that same selection.

\begin{definition}[The projection and its kernel]\label{def:4.30}

For an object $G$ and $P=pG$, set
\begin{equation*}
\begin{aligned}
C_G&=\mathrm{Aut}_{qp}(G),&
B_P&=\mathrm{Aut}_q(P),\\
\pi_G:C_G&\longrightarrow B_P,&
\pi_G(a)&=p(a),\\
H_G&=\ker\pi_G=\mathrm{Aut}_p(G)
\end{aligned}
\tag{4.36}\label{eq:4.36}
\end{equation*}
Here $C_G$ represents the changes of the geometry that fix the extraction, $B_P$ the
changes of the core that fix the extraction, and $H_G$ the changes of the geometry that
fix the entire core.
The map $\pi_G$ is a group homomorphism, and $H_G$ is a normal subgroup.

Place $G_i$ on each vertex of Definition~\ref{def:4.17}, and assume that the
edges $L_e$ of geometries and their projections $p(L_e)$ are strongly opcartesian
with respect to $qp$ and $q$, respectively.
The transport constructed in \S\ref{sec:4.4} satisfies this condition.
The specified comparisons are $u_f\in C_{G_j}$, and for the specified comparisons of
cores we use $p(u_f)$.
We write $\phi_f$ for the canonical comparison of geometries and $\bar\phi_f$ for the
canonical comparison of cores.

\end{definition}

\begin{proposition}[Projection of defects]\label{prop:4.31}

For every edge reselection $c$ of the geometries,
\begin{equation*}
p(\phi_f(c))=\bar\phi_f(p(c)),\qquad
p(\delta_f(c))=\bar\delta_f(p(c))
\tag{4.37}\label{eq:4.37}
\end{equation*}
\end{proposition}

\begin{proof}

Project $\phi_f(c)L_{\ell_f}^c=L_{\rho_f}^c$.
The paths of cores are also strongly opcartesian, so the first equation follows from the
uniqueness of comparisons.
The second follows because $p$ preserves composites and inverses.

\end{proof}

\begin{definition}[Edgewise lifts and alignment]\label{def:4.32}

For an edge reselection $a_e\in B_{P_j}$ of the cores, choose
$\widetilde a_e\in C_{G_j}$ with $p(\widetilde a_e)=a_e$.
These finitely many edgewise lifts are called an edge section.
The section is said to be aligned when, for each face, the composites of the core paths
satisfy
\begin{equation*}
p(u_f)\,\bar L_{\ell_f}^{\,a}
=\bar L_{\rho_f}^{\,a}
\tag{4.38}\label{eq:4.38}
\end{equation*}
Here $\bar L^{\,a}$ is the path whose projected edges have been reselected by $a$.

Alignment is the condition that all comparisons visible at the core are coherent for that
edge selection.
As in Definition~\ref{def:4.32}, an edge section is the datum that provides, together
with the reselection at the core, a lift realizing it on the side of the geometry.

\end{definition}

\begin{theorem}[Order-preserving decomposition of the defect]\label{thm:4.33}

Take an aligned edge section, and write $\phi_f^0=\phi_f(1)$ for the comparison before
reselection and $m_f=\phi_f(\widetilde a)$ for the comparison at the chosen section. Then
\begin{equation*}
p(m_f)=p(u_f)
\tag{4.39}\label{eq:4.39}
\end{equation*}
and the original defect decomposes as
\begin{equation*}
\underbrace{u_f(\phi_f^0)^{-1}}_{\delta_f}
=
\underbrace{u_fm_f^{-1}}_{\delta_f^{\mathrm{in}}}
\underbrace{m_f(\phi_f^0)^{-1}}_{\widetilde\delta_f^{\mathrm{out}}}
\tag{4.40}\label{eq:4.40}
\end{equation*}
where
$\delta_f^{\mathrm{in}}\in H_{G_j}$ and
$p(\widetilde\delta_f^{\mathrm{out}})=\bar\delta_f(1)$.

\end{theorem}

\begin{proof}

By Proposition~\ref{prop:4.31} we have $p(m_f)=\bar\phi_f(a)$.
From \eqref{eq:4.38} and the uniqueness of canonical comparisons we obtain
$\bar\phi_f(a)=p(u_f)$, which is \eqref{eq:4.39}.
Hence $p(u_fm_f^{-1})=1$.
Equation~\eqref{eq:4.40} is checked by cancelling the middle $m_f^{-1}m_f$.
The last claim follows from
$p(m_f(\phi_f^0)^{-1})=p(u_f)\bar\phi_f(1)^{-1}$.

\end{proof}

In this decomposition, the factor that lifts the defect visible at the core acts first,
and the factor invisible to the core acts next.
In general the order of the two factors cannot be interchanged.

\begin{theorem}[Condition for simultaneous vanishing]\label{thm:4.34}

Fix an aligned edge section $\widetilde a$.
Then there exists, over its core selection $a$, an edge selection of the geometries that
is coherent on all faces if and only if there exists a family of kernel elements
$h_e\in H_{G_j}$ with
\begin{equation*}
u_f=\phi_f(h\widetilde a)\qquad\text{(for every face }f\text{)}
\tag{4.41}\label{eq:4.41}
\end{equation*}
Consequently, the obstruction for the geometries vanishes if and only if an aligned edge
section can be chosen for which \eqref{eq:4.41} can be satisfied.

\end{theorem}

\begin{proof}

Since $h$ lies in the kernel, $p(h\widetilde a)=a$.
By \eqref{eq:4.41} and Theorem~\ref{thm:4.23}, $h\widetilde a$ is the
required coherent selection.
Conversely, given a coherent selection $c$ with $p(c)=a$, set
$h_e=c_e\widetilde a_e^{-1}$ on each edge.
Its projection is $a_ea_e^{-1}=1$, so $h_e\in H_{G_j}$, and
$c=h\widetilde a$ gives \eqref{eq:4.41}.
For the forward direction of the final equivalence, the coherent selection of the
geometries itself can be used as the edge section.

\end{proof}

Here the same edge selection is used for the core and the geometry, and all faces are
examined at once.
Even if a coherent selection is found at each level on its own, the question remains whether
that selection can be lifted to the next level.

\begin{example}[Four axes and a solution that does not lift]\label{ex:4.35}

We illustrate the distinction by a finite tower in which groups are regarded
as one-object categories.
Regard the four indices $1,2,3,4$ as axes, and take their permutation group $S_4$
as the group of core changes. At the level of geometries, select $\{1,2\}$
among the four axes. From the permutations preserving this selection and a
two-valued extra component, form
\begin{equation*}
\begin{aligned}
K&=\{\gamma\in S_4\mid\gamma\{1,2\}=\{1,2\}\}\times C_2,\\
\pi:K&\longrightarrow S_4,\qquad(\gamma,z)\longmapsto\gamma
\end{aligned}
\tag{4.42}\label{eq:4.42}
\end{equation*}
The group $C_2=\mathbb{Z}/2\mathbb{Z}$ is written additively.
The kernel is $\{1\}\times C_2$, a degree of freedom that changes only the geometry.
Take the terminal category as the base category.

Take one vertex and two loops $e,g$, with the original lifts the identity.
The faces and the specified comparisons are
\begin{equation*}
e^2\Rightarrow\varnothing:\quad u=(b,0),\quad b=(12)(34),
\qquad
g\Rightarrow\varnothing:\quad v=(1,1)
\tag{4.43}\label{eq:4.43}
\end{equation*}
Here $\varnothing$ is the empty path.
The coherence conditions at the core are $a_e^2=b$ and $a_g=1$, and
$a_e=(1324)$, $a_g=1$ is a solution.

On the other hand, the permutations allowed at the geometry preserve $\{1,2\}$ and $\{3,4\}$
separately.
The permutation part is therefore $S_2\times S_2$, in which every element squares to the
identity.
No selection $c_e$ of the geometry can satisfy $c_e^2=(b,0)$.
The square root $(1324)$ at the core does not preserve the two selected axes.

The second face, belonging only to the geometry, is coherent with the kernel selection $c_g=(1,1)$.
Thus all core conditions and the conditions belonging only to the geometry each
have solutions.
Still, no overall solution exists.
The obstruction is the absence of an edge section lifting the core solution.

\end{example}

\section{Carrying the reads and updates of a lens by the same map}\label{sec:4.10}

To connect the comparisons of the preceding sections with software operations, we must
determine which changes preserve the meaning of the operations.
In the lenses of Chapter~\ref{chap:1}, one state appeared both in reads and in updates.
Through this sharing, the maps of the two commutative diagrams are tied together.

\begin{definition}[Morphisms of lenses including a change of view]\label{def:4.36}

A relative morphism between lenses $L=(C,V,g,p)$ and $L'=(C',V',g',p')$
satisfying the three laws is a pair of maps $h:C\to C'$ and $u:V\to V'$ such that
\begin{equation*}
g'h=ug,\qquad
h\,p=p'(h\times u)
\tag{4.44}\label{eq:4.44}
\end{equation*}

Identities and composition are componentwise and preserve \eqref{eq:4.44}.
Taking $u=\mathrm{id}_V$ recovers the fixed-view morphisms of Chapter~\ref{chap:1}.

\end{definition}

Under the lens laws of this chapter, preservation of updates implies preservation of reads.
Indeed, for any state $c$, applying the first law before the change
(\eqref{eq:1.16}), preservation of updates, and the second law after the change gives
\[
\begin{aligned}
g'(h(c))
&=g'\bigl(h(p(c,g(c)))\bigr)\\
&=g'\bigl(p'(h(c),u(g(c)))\bigr)\\
&=u(g(c)).
\end{aligned}
\]
This derivation does not require $h,u$ to be invertible.
The following construction expresses preservation of reads and updates by the same
maps as one commutative diagram.

\begin{proposition}[A square combining the two operations]\label{prop:4.37}

Combine reads and updates into a single map of sum types,
\begin{equation*}
c_L:C\sqcup(C\times V)\longrightarrow V\sqcup C,\qquad
c_L|_C=g,\quad c_L|_{C\times V}=p
\tag{4.45}\label{eq:4.45}
\end{equation*}

where the right-hand side uses the inclusion into the corresponding summand.
The following equation is equivalent to \eqref{eq:4.44}:
\begin{equation*}
(u\sqcup h)c_L
=c_{L'}\bigl(h\sqcup(h\times u)\bigr).
\tag{4.46}\label{eq:4.46}
\end{equation*}

\end{proposition}

\begin{proof}

Restriction to the left input summand $C$ gives the read equation, and restriction
to the right input summand $C\times V$ gives the update equation.
Conversely, equality on both summands gives equality of maps on the sum.

\end{proof}

In \eqref{eq:4.46}, the same $h$ is used for the input of reads, the state
component of the input of updates, and the output of updates.
This is the preservation condition for the shared state.

\begin{proposition}[Invertible changes preserving both operations]\label{prop:4.38}

The group of invertible changes of a lens is the following subgroup of
$\mathrm{Aut}(C)\times\mathrm{Aut}(V)$:
\begin{equation*}
\begin{aligned}
G_L
 &=G_{\mathrm{get}}\cap G_{\mathrm{put}}
 =G_{\mathrm{put}}\subseteq G_{\mathrm{get}},\\
G_{\mathrm{get}}&=\{(h,u)\mid gh=ug\},\\
G_{\mathrm{put}}&=\{(h,u)\mid hp=p(h\times u)\}
\end{aligned}
\tag{4.47}\label{eq:4.47}
\end{equation*}

Equation~\eqref{eq:4.46} says that the input and output changes constructed from
this same pair $(h,u)$ commute with $c_L$.

\end{proposition}

\begin{proof}

Both conditions are closed under identities, composition, and inverses.
Their intersection consists exactly of the invertible morphisms satisfying
\eqref{eq:4.44}. Since preservation of updates implies preservation of reads,
this intersection equals $G_{\mathrm{put}}$.
The last assertion follows from Proposition~\ref{prop:4.37}.

\end{proof}

In Example~\ref{ex:1.35} of Chapter~\ref{chap:1}, of the four bijections
preserving reads, only two also preserved updates.
Thus the inclusion in \eqref{eq:4.47} is strict in general.
Preserving updates further constrains the allowed correspondence of states.

\begin{example}[A semantics-preserving morphism that merges information]\label{ex:4.39}

For product lenses $C=V\times K$ and $C'=V\times K'$ over a nonempty view set $V$,
any map $t:K\to K'$ satisfies \eqref{eq:4.44} through
\begin{equation*}
h(v,k)=(v,t(k)),\qquad u=\mathrm{id}_V
\tag{4.48}\label{eq:4.48}
\end{equation*}

Indeed, both reads give $v$, and an update to the displayed value $w$ gives
$(w,t(k))$ in either order.
In particular, if $V=\{*\}$, $K=\{0,1\}$, and $K'=\{*\}$, then $h$ is not injective.
General semantics-preserving morphisms include such aggregation of information;
taking the invertible ones yields the group in \eqref{eq:4.47}.

\end{example}

\section{Carrying protocol operations and adapters}\label{sec:4.11}

In a protocol, what is shared are the state sets at the vertices.
The same state map is used for the operations that enter a vertex and for those that
leave it.
Furthermore, if an adapter connects two protocols, that connection is preserved at the
same time.

\begin{proposition}[Preservation from named operations to all executions]\label{prop:4.40}

Take realizations $X,X'$ of $\mathrm{Prot}(Q,\Pi,O)$ of Chapter~\ref{chap:1} and vertex
maps $a_v:X(v)\to X'(v)$.
If, for each named edge $e:v\to w$ and each observation,
\begin{equation*}
a_wX(e)=X'(e)a_v,\qquad
o_{X'}(v)a_v=o_X(v)
\tag{4.49}\label{eq:4.49}
\end{equation*}
then $a$ determines a unique semantics-preserving morphism of protocols.
Moreover, for adapters $q:X\to Y$ and $q':X'\to Y'$ and semantics-preserving morphisms
$a:X\to X'$ and $b:Y\to Y'$,
\begin{equation*}
bq=q'a
\quad\Longleftrightarrow\quad
b_vq_v=q'_va_v\quad\text{(for every vertex }v\text{)}
\tag{4.50}\label{eq:4.50}
\end{equation*}
\end{proposition}

\begin{proof}

Condition~\eqref{eq:4.49} is the condition of Proposition~\ref{prop:1.38} and gives a
unique semantics-preserving morphism.
Reading the preservation condition for adapters in \eqref{eq:1.23} componentwise as a
natural transformation gives \eqref{eq:4.50}.

\end{proof}

In this check, operations with different names are treated as separate generating edges
even when their actions coincide.
When adapters are changed in succession, pasting the two commutative squares shows that
\eqref{eq:4.50} also holds for the composed change.

\section{Comparison of semantics and typed constructions}\label{sec:4.12}

Invertible changes that preserve a semantics-preserving morphism of lenses or a
protocol adapter can be recovered after passing to the typed constructions.
Common to both is the following correspondence through a fully faithful functor.

\begin{theorem}[Preserving comparisons between semantics and typed constructions]\label{thm:4.41}

For a fully faithful functor $F:\mathcal{C}\to \mathcal{D}$ and a morphism $c:X\to Y$,
define the group preserving the comparison by
\begin{equation*}
\mathrm{Cmp}(c)=
\{(a,b)\in\mathrm{Aut}_{\mathcal{C}}(X)\times\mathrm{Aut}_{\mathcal{C}}(Y)
\mid bc=ca\}
\tag{4.51}\label{eq:4.51}
\end{equation*}
Then
\begin{equation*}
\mathrm{Cmp}(c)\xrightarrow{\sim}\mathrm{Cmp}(F(c)),
\qquad(a,b)\longmapsto(F(a),F(b))
\tag{4.52}\label{eq:4.52}
\end{equation*}
is a group isomorphism.
If $c$ is an isomorphism, the inverse of the first projection,
$a\mapsto(a,cac^{-1})$, also commutes with this correspondence.

\end{theorem}

\begin{proof}

A functor preserves the equation $bc=ca$, so the map is defined and is a group
homomorphism.
By fullness, the automorphisms of $F(X),F(Y)$ and their inverses lift to the original
category.
By faithfulness, the composites of the lifted morphisms are identities, and
$bc=ca$ also follows from $F(b)F(c)=F(c)F(a)$.
Hence the map is surjective. Injectivity also follows from faithfulness.
When $c$ is an isomorphism, $b=cac^{-1}$, and the last claim follows because a functor
preserves composites and inverses.

\end{proof}

Proposition~\ref{prop:1.43} of Chapter~\ref{chap:1} gives fully faithful functors from
the semantics of fixed-view lenses and of fixed protocols to the categories of typed
objects constructed from each.
Therefore the invertible changes that preserve semantics-preserving morphisms of
lenses or protocol adapters can be recovered exactly from the typed constructions.
For lenses whose views also change, take the view component in
Definition~\ref{def:1.42} to be $u$ and the write component to be $h\times u$.
By \eqref{eq:4.44} and Proposition~\ref{prop:4.37}, extracting and assembling
the role maps are mutually inverse, giving the same conclusion.

Once a category with these semantics-preservation conditions and a functor to the
base have been defined, choose strongly opcartesian edge lifts and specified comparisons.
The defects of \S\S\ref{sec:4.5}--\ref{sec:4.7} then express discrepancies
between changes that preserve the actual operations.
To connect further with transport of readings equipped with geometry, supply
the core morphism and the comparisons of local realizations of Definition~\ref{def:4.9}.
Providing these data in this order connects the meaning of operations, changes of
reading, and lifts of geometries.

\section*{Summary of the Chapter}

In this chapter we constructed transport along changes of reading, and gave conditions
under which several transports and specified comparisons become coherent for the same
selection of edges.

\begin{itemize}
\item \textbf{Existence and universality of transport.} We constructed the core from an
exact change of extraction and showed that its canonical morphism is strongly
opcartesian.
We also constructed the canonical transport of geometries and, for general morphisms of
cores, extracted the three comparisons of local realizations as the existence condition
(\S\S\ref{sec:4.2}--\ref{sec:4.3}).
\item \textbf{Coherence of units and composition.} From the uniqueness of factorizations
we obtained the functors between fibers and the natural isomorphisms for units and
composition. Triple composites and the projection to the core
are coherent as well (\S\ref{sec:4.4}).
\item \textbf{Obstructions for specified comparisons.} We expressed the difference from
the canonical comparison as a noncommutative automorphism and showed that vanishing of the obstruction is equivalent to an edge
reselection making all faces coherent.
Pasting faces requires additional coherence conditions among the specified comparisons
(\S\S\ref{sec:4.5}--\ref{sec:4.8}).
\item \textbf{Simultaneous vanishing across levels.} After the selection at the core has
been lifted to the geometry, being able to remove the remainder by kernel changes for the
same selection corresponds to overall coherence.
Even if each level has a solution independently, without a lift there is no overall
solution (\S\ref{sec:4.9}).
\item \textbf{Comparisons preserving the meaning of operations.} For lenses,
reads and updates use the same state map, and preservation of updates also implies
preservation of reads. For protocols, all named operations and adapters are preserved
simultaneously. The invertible changes preserving these comparisons can be recovered
across the constructed typed categories (\S\S\ref{sec:4.10}--\ref{sec:4.12}).
\end{itemize}

For the order-processing API of the opening, the discrepancy between the route that moves
directly from the old version to the new one and the route through the intermediate
version is extracted as a defect.
Even a discrepancy that could be absorbed into one of two independent edges must be made
coherent simultaneously once that edge is shared with another face.
Theorem~\ref{thm:4.23} characterizes the existence of a family of transformations
realizing the specified comparisons on every face under consideration as the vanishing
of the obstruction.
When both levels, core and geometry, are treated, the lifts and the kernel changes are
again performed for the same selection.

In the next chapter we compare the route that changes the base of the input before
constructing and the route that constructs first and then changes.

\part[Comparison and Normalization]{Comparison and Normalization\\[0.8em]{\large Chapters 5--6}}
\chapter{Base Change and Generated Comparisons}\label{chap:5}

\section*{Overview of the Chapter}

The question of this chapter is \textbf{whether the operation that carries structure and the
operation that changes the input under consideration can be interchanged}.
Chapter~\ref{chap:4} treated transport along a single change and the coherence of successive
transports. In this chapter we construct the two routes formed from these two kinds of
operations and study the comparison between them.

For example, suppose that a monolith with order processing and inventory management is split
into an order service and an inventory service. In a design review we first want to examine
the part related to order cancellation. That part contains the operations of cancellation
request, release of an inventory reservation, and cancellation completion, together with the
condition that if a cancellation is complete then the corresponding inventory reservation has
been released. Under the same splitting policy, we compare two routes that yield the design of
this part.

\begin{xltabular}{\linewidth}{@{}LLL@{}}
\toprule
Procedure & First step & Next step \\
\midrule
\endhead
Split the whole, then select the scope & Split the whole monolith into the order and inventory
services & Extract the part related to order cancellation from the design after the split \\
\addlinespace[3pt]
Select the scope, then split & Extract the part related to order cancellation from the
original design & Apply the same splitting policy to that part \\
\bottomrule
\end{xltabular}

Does the splitting proposal that the person in charge drew up by focusing on the order-cancellation
part remain consistent with the design obtained after splitting the whole? For example,
consider a design in which, after the split, the order service asks the inventory service to
release the reservation and proceeds to cancellation completion only after confirming the
release. Between the designs obtained by the two routes, we want to verify that this
interaction and the condition above correspond. Extracting only the processing placed in the
order service leaves the release of inventory reservations outside the scope, and we lose the
dependencies needed to check the condition.

We therefore look for the inputs on the original that correspond to the scope selected on the
target. In this example, this means tracing the operations and dependencies needed for the
cancellation condition back to the design before the split. In this chapter we construct that
correspondence as pairs of inputs and pull the structure of objects, operations, and Laws back
onto it. This is base change. Moreover, from the universal properties of transport and
pullback, we build a map that compares the results of the two routes. We call this map the
canonical comparison.

\begin{xltabular}{\linewidth}{@{}LLL@{}}
\toprule
Question & Construction & Outcome \\
\midrule
\endhead
Can two inputs be matched under the same condition? & Pairs of sources with a common image &
Fiber products of extraction doctrines \\
\addlinespace[3pt]
Can the structure on the target be read on the original? & Backward reindexing of Atoms &
Cartesian pullback and its equivalence with transport \\
\addlinespace[3pt]
Can the order of transport and pullback be exchanged? & A mate built from the two universal properties & A canonical invertible comparison \\
\addlinespace[3pt]
Can several comparisons and diagnoses also be carried? & Commutative base diagrams and the
correspondence of end groups & Preservation and reflection of coherence, vanishing of defects,
and reselection orbits \\
\addlinespace[3pt]
Can structure still be carried back even along a change that enlarges the extracted content? & Reflection of extraction at
inputs that actually carry structure & A necessary and sufficient condition for backward
transport along refinements \\
\addlinespace[3pt]
What is compared when geometry and normalization are included? & Two routes of full geometries
built from the same input & The generated
comparison analyzed in Chapter~\ref{chap:6} \\
\bottomrule
\end{xltabular}

For the exact changes of the first half, the Atoms extracted from the selected sources agree,
including the change of names. For the refinements of the second half, only forward
preservation of the extracted content is required. This difference determines which structure
can be carried backwards. When treating the split of a monolith as well, we specify whether we
read the business operations and conditions or also communication, failures, and retries, and
we check the preservation conditions for extraction and structure in that reading.

\section{Constructing corresponding inputs by fiber products}\label{sec:5.1}

An extraction doctrine is a reading that normalizes a source and then extracts Atoms. The
category $B$ of Chapter~\ref{chap:1} has objects that in addition select one source. We first
carry out the construction in the category of doctrines with the selected source forgotten,
and then restore the selection.

\begin{construction}[Fiber product of exact doctrines]\label{cons:5.1}

Over a common Atom set $\mathrm{At}$, take two exact morphisms.
\begin{equation*}
D_1\xrightarrow{\sigma_1}D_0\xleftarrow{\sigma_2}D_2,
\qquad \sigma_i=(f_i,e_i).
\tag{5.1}\label{eq:5.1}
\end{equation*}
Here $f_i$ is the source map and $e_i$ is the bijection of Atoms. We define the sources and
the normalization of the new doctrine $D_{12}$ by
\begin{equation*}
\begin{aligned}
\mathrm{Src}_{12}
 &=\{(s_1,s_2)\mid f_1(s_1)=f_2(s_2)\},\\
N_{12}(s_1,s_2)&=(N_1s_1,N_2s_2)
\end{aligned}
\tag{5.2}\label{eq:5.2}
\end{equation*}
The normalization preserves this set because $f_iN_i=N_0f_i$. The choices of vocabulary,
semantics, and resolution are taken from $D_1$, and the predicates that read a source receive
the first component. Hence
\begin{equation*}
\mathrm{Extracts}_{12}((s_1,s_2),a)
\Longleftrightarrow \mathrm{Extracts}_1(s_1,a).
\tag{5.3}\label{eq:5.3}
\end{equation*}
The two projections are
\begin{equation*}
\pi_1=(\mathrm{pr}_1,\mathrm{id}_{\mathrm{At}}),
\qquad
\pi_2=(\mathrm{pr}_2,e_2^{-1}e_1)
\tag{5.4}\label{eq:5.4}
\end{equation*}
The Atom map of the second projection carries the first names to the names of the common base
and returns from there to the second names. Exactness follows from
\begin{equation*}
\begin{aligned}
\mathrm{Extracts}_1(s_1,a)
&\Longleftrightarrow\mathrm{Extracts}_0(f_1(s_1),e_1(a))\\
&\Longleftrightarrow\mathrm{Extracts}_2(s_2,e_2^{-1}e_1(a))
\end{aligned}
\tag{5.5}\label{eq:5.5}
\end{equation*}
In both the source and the Atom components we have $\sigma_1\pi_1=\sigma_2\pi_2$.

\end{construction}

\begin{proposition}[Universal property with respect to all cones]\label{prop:5.2}

If exact morphisms $h_i:T\to D_i$ satisfy $\sigma_1h_1=\sigma_2h_2$, then there is a unique
exact morphism $h:T\to D_{12}$ with $\pi_i h=h_i$. Hence $D_{12}$ is the fiber product
of~\eqref{eq:5.1}.

\end{proposition}

\begin{proof}

Write $h_i=(g_i,d_i)$, and let the source map of $h$ be $t\mapsto(g_1(t),g_2(t))$ and its
Atom map be $d_1$. The pair of sources belongs to~\eqref{eq:5.2} by the commutativity of the
cone. Commutation with the normalizations holds componentwise, and the equivalence of
extraction is the exactness of $h_1$. For the second projection, from $e_1d_1=e_2d_2$ we
obtain $e_2^{-1}e_1d_1=d_2$. The equations with the two projections determine the source map
uniquely, and the equation with the first projection determines the Atom map uniquely.

\end{proof}

When the selected sources $s_i$ satisfy $f_1(s_1)=f_2(s_2)$, we select $(s_1,s_2)$ in
$D_{12}$. If the two morphisms of a cone with specified selected sources (a pointed cone)
preserve the selected sources, then so does the universal morphism above. In this way the
same fiber product is obtained in $B$.

\begin{example}[Inputs corresponding to a scope on the target]\label{ex:5.3}

We check Construction~\ref{cons:5.1} in a case where the sources form small finite sets. Let
the sources of the original be $\{a,b,c\}$ and the sources of the target be $\{0,1\}$, with
the correspondence $f(a)=f(b)=0$ and $f(c)=1$. The fiber product along the inclusion
$j:\{0\}\to\{0,1\}$, which selects the object $\{0\}$ after the change, is
\begin{equation*}
\{a,b,c\}\times_{\{0,1\}}\{0\}
=\{(a,0),(b,0)\}.
\tag{5.6}\label{eq:5.6}
\end{equation*}
If every source extracts the same finite Atom set and the normalizations and the Atom maps
are identities, this becomes a concrete instance of Construction~\ref{cons:5.1}. Even though
$a,b$ are sent to the same source of the target, both remain as corresponding inputs of the
original. When applying this to the split of a monolith, we take design documents and code as
sources, specify the reading used for the design change and the selection of the scope, and
check~\eqref{eq:5.5}.

\end{example}

\section{Pulling back cores from the target}\label{sec:5.2}

The transport $\sigma_!$ of Chapter~\ref{chap:4} started from a core on the original. The
pullback $\sigma^*$ starts from a core on the target. Both correspond to the same base
morphism $\sigma:b\to b'$, but the direction in which the object is constructed is reversed.

\begin{definition}[Strongly cartesian morphism]\label{def:5.4}

Take a functor $r:E\to B$ and a morphism $\kappa:Y\to Q$ with $r(\kappa)=\sigma:b\to b'$. We
call $\kappa$ a strongly cartesian morphism when, for every object $R$, base morphism
$\tau:r(R)\to b$, and morphism $h:R\to Q$ with $r(h)=\sigma\tau$,
\begin{equation*}
\exists!\,k:R\longrightarrow Y,\qquad
r(k)=\tau,\quad \kappa k=h
\tag{5.7}\label{eq:5.7}
\end{equation*}
This is the condition that a morphism into $Q$ splits uniquely into a final step
corresponding to $\sigma$ on the base and a morphism up to that point
(\cite[\href{https://stacks.math.columbia.edu/tag/02XJ}{Tag 02XJ, Definition 4.33.1}]{Stacks}).

\end{definition}

\begin{construction}[Backward reindexing]\label{cons:5.5}

Let $\sigma=(f,e):(D,s)\to(D',s')$ be an exact morphism and let $Q$ be a core on $(D',s')$.
Writing $F_D(s)$ for the selected extracted family, we have
\begin{equation*}
F_{D'}(s')=e_*F_D(s),\qquad
F_D(s)=(e^{-1})_*F_{D'}(s').
\tag{5.8}\label{eq:5.8}
\end{equation*}
Since the right-hand side is finite, we obtain a finite family on the original as well that
can serve as input to a core. We place the base at $(D,s)$ and define composition, object
formation, and operations as follows.
\begin{equation*}
\begin{aligned}
\mathrm{Comp}_{\sigma^*Q}(F)
 &=(e^{-1})_*\mathrm{Comp}_Q(e_*F),\\
\mathrm{Form}_{\sigma^*Q}(C)
 &=T_{e^{-1}}\mathrm{Form}_Q(e_*C),\\
\mathrm{Op}_{\sigma^*Q}(A,B)
 &=\mathrm{Op}_Q(T_eA,T_eB).
\end{aligned}
\tag{5.9}\label{eq:5.9}
\end{equation*}
Here $T_e$ is the reindexing of objects from Construction~\ref{cons:4.3}. We conjugate the
configuration maps of the operations and compose the invariants and signatures with $T_e$.
The contexts, equations, residuals, and detector code are reindexed by the same $e^{-1}$.
By~\eqref{eq:5.8} and the formulas for composition and object formation, the generated base
object coincides with the base object of $T_{e^{-1}}Q$. This coincidence aligns the base
objects of the equation reading.

The resulting morphism $\kappa_{\sigma,Q}:\sigma^*Q\to Q$ has $\sigma$ as its base. The
upper-level reading map $u$ has a two-sided inverse $u^{-1}$ given by the inverse reindexing.
The information about sources used by this construction is $f(s)=s'$ and the equivalence of
extraction.

\end{construction}

\begin{theorem}[Cartesian lifts along arbitrary exact morphisms]\label{thm:5.6}

For every exact base morphism $\sigma:b\to b'$ and every core $Q$ on $b'$, the morphism
$\kappa_{\sigma,Q}$ of Construction~\ref{cons:5.5} is strongly cartesian with respect to
$q:E_{\mathrm{core}}\to B$.

\end{theorem}

\begin{proof}

For $h:R\to Q$ as in~\eqref{eq:5.7}, let the base of $k$ be $\tau$ and its upper level be
$u^{-1}h_{\mathrm{up}}$. The Atom component agrees with the Atom component of $\tau$, by the fact that
the base of $h$ is $\sigma\tau$ and by $u^{-1}$. The other components are composites of
morphisms of readings, so they preserve object formation, operations, equations, invariants,
and signatures. From $uu^{-1}=1$ we get $\kappa_{\sigma,Q}k=h$. Any other factor also has
base $\tau$, and composing its upper-level equation with $u^{-1}$ gives the same $k$.

\end{proof}

\begin{lemma}[Pullback functors and composition comparisons]\label{lem:5.7}

Choosing a cartesian lift for each $\sigma,Q$ gives a functor $\sigma^*:E_{b'}\to E_b$ between fiber
categories. The image of a vertical morphism $v:Q\to Q'$ is determined uniquely by
\begin{equation*}
\kappa_{\sigma,Q'}\,\sigma^*(v)
=v\,\kappa_{\sigma,Q}
\tag{5.10}\label{eq:5.10}
\end{equation*}
For $\sigma:b_0\to b_1$ and $\tau:b_1\to b_2$ there are natural isomorphisms of unit and
composition
\begin{equation*}
1_{E_b}\xrightarrow{\sim}(1_b)^*,
\qquad
\sigma^*\tau^*\xrightarrow{\sim}(\tau\sigma)^*
\tag{5.11}\label{eq:5.11}
\end{equation*}
satisfying the coherence for units and triple composites.

\end{lemma}

\begin{proof}

Applying~\eqref{eq:5.10} to identities and to composites of morphisms and using uniqueness
yields the functor laws. A composite of cartesian morphisms is cartesian, by
applying~\eqref{eq:5.7} twice. Hence between the two-step lift and the direct lift there is a
unique isomorphism making the triangle into their common target commute. The two comparisons
arising from a three-step lift also satisfy the same triangle and are therefore equal. For
the unit we use the same uniqueness.

\end{proof}

\begin{proposition}[Adjoint equivalence between transport and pullback]\label{prop:5.8}

For an exact morphism $\sigma:b\to b'$, the transport of Chapter~\ref{chap:4} and the
pullback of this section form an adjunction
\begin{equation*}
\sigma_!:E_b\rightleftarrows E_{b'}:\sigma^*,
\qquad \sigma_!\dashv\sigma^*
\tag{5.12}\label{eq:5.12}
\end{equation*}
whose unit $\eta^\sigma:1\to\sigma^*\sigma_!$ and counit
$\epsilon^\sigma:\sigma_!\sigma^*\to1$ are invertible.

\end{proposition}

\begin{proof}

Write $\iota_{\sigma,P}:P\to\sigma_!P$ for the opcartesian morphism in the same direction as
in Chapter~\ref{chap:4}. The two universal properties give the following natural bijections.
\begin{equation*}
\begin{aligned}
\mathrm{Hom}_{E_{b'}}(\sigma_!P,Q)
&\cong\{h:P\to Q\mid q(h)=\sigma\}\\
&\cong\mathrm{Hom}_{E_b}(P,\sigma^*Q).
\end{aligned}
\tag{5.13}\label{eq:5.13}
\end{equation*}
The first correspondence is $v\mapsto v\iota_{\sigma,P}$, and the inverse of the second is
$w\mapsto\kappa_{\sigma,Q}w$. The unit and the counit are produced by transposing identity
morphisms. The triangle identities follow from the fact that both composites become identities
after composition with the respective lifts, together with uniqueness.

Moreover, the upper levels of the two lifts are built from mutually inverse Atom reindexings.
Using the proof of Theorem~\ref{thm:5.6} for factorizations in the opposite direction, each
lift is both cartesian and opcartesian. The unit is the comparison of two cartesian lifts
with the same base morphism and the same end, and the counit is the comparison of two
opcartesian lifts with the same base morphism and the same start. Both are invertible by
uniqueness.

\end{proof}

\begin{example}[Fibers become equivalent even when the base is not invertible]\label{ex:5.9}

Take the constant map from the source set $\{0,1\}$ to $\{*\}$ and select $0$ in the former.
Every source extracts the same finite family $F$, and the normalizations and Atom maps are
identities. This is an exact morphism, but it is not an isomorphism of bases, since the
source map has no two-sided inverse. Proposition~\ref{prop:5.8} shows that the fiber
categories over the selected sources are nevertheless equivalent. What travels back and
forth is the object formation, the operations, and the Laws read at that source. Recovering
the distinctions among all sources is a different question.

\end{example}

\section{The canonical comparison exchanging transport and pullback}\label{sec:5.3}

We express the two splitting plans from the opening as routes with the same start and end.
Let $V$ be the route that splits the whole and then selects the scope, and $D$ the route that
selects the scope and then splits; from the universal properties of transport and pullback
we build a comparison between them.

\begin{construction}[Two routes with matched types]\label{cons:5.10}

Write the pointed fiber product of Construction~\ref{cons:5.1} as
\begin{equation*}
\begin{array}{ccc}
b_{12}&\xrightarrow{\ \pi_2\ }&b_2\\
{\scriptstyle\pi_1}\downarrow&&\downarrow{\scriptstyle\sigma_2}\\
b_1&\xrightarrow{\ \sigma_1\ }&b_0
\end{array}
\qquad \sigma_1\pi_1=\sigma_2\pi_2
\tag{5.14}\label{eq:5.14}
\end{equation*}
Each vertex is a pair of a doctrine and a selected source, and in the opening example the
vertices play the following roles.

\begin{xltabular}{\linewidth}{@{}LL@{}}
\toprule
Vertex & How the input is read \\
\midrule
\endhead
$b_1$ & Reads the input on the monolith side before the split \\
\addlinespace[3pt]
$b_0$ & Reads the whole input after the split \\
\addlinespace[3pt]
$b_2$ & Reads the scope of input selected for order cancellation after the split \\
\addlinespace[3pt]
$b_{12}$ & Reads the pairs of sources, on the monolith side and in the selected scope,
that point to the same target \\
\bottomrule
\end{xltabular}

The morphism $\sigma_1$ is the correspondence given by the splitting policy, and $\sigma_2$
is the correspondence from the selected scope into the whole. The projections $\pi_1,\pi_2$
extract the components from a corresponding pair of sources. The selection of the scope here
is performed on sources, and is distinguished from the operation of removing Atoms inside a
single core. To express these as exact base morphisms, we check the equivalence of
extraction of~\eqref{eq:5.5} for each correspondence.

The two routes from cores on $b_1$ to cores on $b_2$ are
\begin{equation*}
D=(\pi_2)_!\pi_1^*,\qquad
V=\sigma_2^*(\sigma_1)_!,\qquad
D,V:E_{b_1}\longrightarrow E_{b_2}.
\tag{5.15}\label{eq:5.15}
\end{equation*}
The route $D$ pulls back to the corresponding inputs and then carries them over; the route
$V$ carries them to the common base and then pulls back. From the composition comparisons of
transport in Chapter~\ref{chap:4} and~\eqref{eq:5.14} we obtain a natural isomorphism
\begin{equation*}
\Sigma:(\sigma_2)_!(\pi_2)_!
\xrightarrow{\sim}(\sigma_1)_!(\pi_1)_!
\tag{5.16}\label{eq:5.16}
\end{equation*}
It is built by comparing both with the direct transport along the same composite base
morphism.

\end{construction}

\begin{theorem}[Canonical Beck--Chevalley mate]\label{thm:5.11}

The following three-step composite defines a natural isomorphism
$\alpha:D\xrightarrow{\sim}V$.
\begin{equation*}
\begin{aligned}
DP&\xrightarrow{\ \eta^{\sigma_2}_{DP}\ }
 \sigma_2^*(\sigma_2)_!DP\\
&\xrightarrow{\ \sigma_2^*(\Sigma_{\pi_1^*P})\ }
 \sigma_2^*(\sigma_1)_!(\pi_1)_!\pi_1^*P\\
&\xrightarrow{\ \sigma_2^*(\sigma_1)_!(\epsilon^{\pi_1}_P)\ }
 VP.
\end{aligned}
\tag{5.17}\label{eq:5.17}
\end{equation*}
We call this comparison the canonical Beck--Chevalley mate. Moreover, if the choice of
pullbacks is changed, the mates whose two ends are aligned by the unique lift comparisons
coincide.

\end{theorem}

\begin{proof}

The three arrows are, respectively, the images under functors of the unit, the natural
isomorphism of transports, and the counit. Hence the composite is a natural transformation.
By Proposition~\ref{prop:5.8} the unit and the counit are invertible, and functors preserve
isomorphisms, so all three arrows are invertible.

Independence of the choices follows because the next equation characterizes the comparison
itself.
\begin{equation*}
\kappa_{\sigma_2,(\sigma_1)_!P}\,
\alpha_P\,\iota_{\pi_2,\pi_1^*P}
=\iota_{\sigma_1,P}\,\kappa_{\pi_1,P}.
\tag{5.18}\label{eq:5.18}
\end{equation*}
Factoring the right-hand side first through $\iota_{\pi_2,\pi_1^*P}$ and then through
$\kappa_{\sigma_2,(\sigma_1)_!P}$ yields $\alpha_P$ uniquely. Substituting the
factorizations of the unit and counit of the adjunction shows that this morphism
is~\eqref{eq:5.17}. Aligning a different choice of lifts by the canonical isomorphisms
preserves~\eqref{eq:5.18}, so the comparisons coincide.

\end{proof}

The reason for invertibility lies not only in the shape of the fiber product but in the fact
that the unit and counit of the actual transport of cores are invertible. The proof also
shows which components of the comparison have to be checked.

\begin{proposition}[Testing agreement with a specified comparison]\label{prop:5.12}

A separately specified morphism $a_P:DP\to VP$ satisfies \eqref{eq:5.18}
if and only if $a_P=\alpha_P$.
In particular, if $a_P$ is invertible, then
\begin{equation*}
\Delta_P=a_P\alpha_P^{-1}\in\mathrm{Aut}_{E_{b_2}}(VP),
\qquad
a_P=\alpha_P\Longleftrightarrow\Delta_P=1.
\tag{5.19}\label{eq:5.19}
\end{equation*}

\end{proposition}

\begin{proof}

The first assertion follows from the two applications of uniqueness in \eqref{eq:5.18}.
For the second, compose on the right with $\alpha_P$.

\end{proof}

\begin{example}[A mix-up in implementing the comparison]\label{ex:5.13}

Suppose that, after separating orders and inventory, in comparing the designs obtained by
the two routes the correspondence between two inventory reservations is mixed up. If this
swap is an automorphism $s$ in the chosen reading, set $a_P=s\alpha_P$. Writing the mismatch
of this example as $\Delta_P=a_P\alpha_P^{-1}$, we have $\Delta_P=s$. Here both $a_P$ and
$\alpha_P$ are invertible. Invertibility, and correctly matching the two specified routes,
must each be checked by its own equation.

\end{example}

\section{Existence of the constructions and the scope of finite presentation}\label{sec:5.4}

The pullbacks and canonical comparisons of \S\S\ref{sec:5.2}--\ref{sec:5.3}
exist even when all sources cannot be described by a finite table.
Which base morphisms a finite input format can represent needs a separate classification.
Here we specify one concrete format also used in the subsequent reconstruction.
The discussion of comparisons and diagnoses continues in~\S\ref{sec:5.5}.

\begin{definition}[Codes by finite exception tables]\label{def:5.14}

The sources are enumerated as a finite set, and the normalization map is given by a table.
The Atom predicate used at each normalized source is represented by a default truth value
and a finite exception set that flips it. A default value of false represents a finite set,
and a default of true a cofinite set, that is, a set with finite complement. A pointed code
adds the selected source. A morphism between codes uses a finite table of sources and an
Atom permutation table with finite support, and we require commutation with the
normalizations, preservation of the selected sources, and the following equality of codes.
The extraction code obtained by normalizing an original source and transporting it by the
Atom permutation is equal to the extraction code itself obtained by normalizing the
corresponding source of the target. This transport preserves the default value and sends the
finite exception set by the Atom permutation. The equality is therefore required both for
the default values and for the finite exception sets.

An exact morphism $\sigma:b\to b'$ is presentable in this format if there are a morphism
$\widehat\sigma:\widehat b\to\widehat b'$ decoded from a code and isomorphisms of bases
$i:\widehat b\xrightarrow{\sim}b$ and $j:\widehat b'\xrightarrow{\sim}b'$ such that
\begin{equation*}
\sigma i=j\widehat\sigma
\tag{5.20}\label{eq:5.20}
\end{equation*}
The presentation of the whole morphism, the isomorphisms $i,j$ included, is what is being
compared.

\end{definition}

\begin{theorem}[Presentability by this code]\label{thm:5.15}

Assume that Atom equality is decidable. An exact morphism $\sigma=(f,e):b\to b'$ is
presentable in the sense of Definition~\ref{def:5.14} if and only if the following two
conditions hold.

\begin{enumerate}
\item The source sets of both endpoints are finite.
\item For every source $t$ of the target, the extraction set $F_{D'}(t)$ is finite or
cofinite.
\end{enumerate}

\end{theorem}

\begin{proof}

If a presentation exists, the enumeration of the sources and the endpoint isomorphisms give
finiteness of both endpoints. The set represented by a finite exception table is finite or
cofinite, and this property is preserved by bijections of Atoms. The endpoint isomorphisms
match up all sources, so condition~2 holds for every extraction set of the target.

Conversely, assume the two conditions. By exactness, each extraction set of the original is
$e^{-1}_*F_{D'}(f(s))$ and hence also finite or cofinite. We enumerate the sources of both
endpoints and express the normalization and the source map in that enumeration. At sources
in the image of the normalization map, the extraction table encodes the conjunction of the
original predicates. That conjunction equals the extraction predicate of the inputs that
normalize there. The table outside the image may be set to false, because decoding always
normalizes before reading the table. This procedure does not require idempotence of the
normalization.

Choose the Atom components of the endpoint isomorphisms to be the identity on the original
and $e$ on the target. Rereading the tables of the target along $e$ as well, the identity
can be used for the Atom permutation inside the code. For equal extraction sets we choose
the same default value and exception set, making the normalized tables of the original agree
with the corresponding tables of the target. In this way the equality of codes holds in
addition to the equivalence of predicates. The pair with the table of sources is a morphism
in the sense of Definition~\ref{def:5.14} and satisfies~\eqref{eq:5.20}.

\end{proof}

In this conclusion the presentations of the endpoints are also rechosen. Fullness of the Hom
sets---whether, after fixing two codes in advance, every morphism between the decoded
objects can be presented by a morphism of codes---is a further question about those fixed
presentations.

\begin{example}[Finitely many sources are not enough]\label{ex:5.16}

Let the Atom set be $\mathbb{N}$, take a single source, let the normalization be the
identity, and let the extraction set be the even numbers. The identity morphism of this
doctrine is exact, but since both the even and the odd numbers are infinite, it cannot be
presented by a finite exception table. On the other hand, with the same source and
extraction set $\mathbb{N}\setminus\{1,3\}$, it can be presented with default value true and
exceptions $\{1,3\}$. The selected family of the latter is infinite, and no core exists at
that point, since cores of Chapter~\ref{chap:1} require a finite family. The presentation of
the base by finite codes and the existence of a core at the point each have their own
conditions.

\end{example}

\section{Assembling commutative squares into whole diagrams}\label{sec:5.5}

In an actual change we treat not just one square but several APIs, conversions, and
inspection routes at the same time. This requires building correspondences of paths from the
correspondences of the individual edges, and preserving the relations between pairs of
paths.

\begin{definition}[Base diagrams and their changes]\label{def:5.17}

Fix a shape with finitely many vertices, generating edges, and specified pairs of parallel
paths. A base diagram $X$ assigns $X_v\in B$ to each vertex $v$ and $X_a:X_v\to X_w$ to each
edge $a:v\to w$. For the two paths $\ell_f,\rho_f$ of each face $f$ we require
$X_{\ell_f}=X_{\rho_f}$.

A change between diagrams $X,Y$ of the same shape is a family of exact morphisms
$i_v:X_v\to Y_v$, one for each vertex, satisfying for each generating edge
\begin{equation*}
Y_a i_v=i_w X_a
\tag{5.21}\label{eq:5.21}
\end{equation*}
Then $Y_\gamma i_v=i_wX_\gamma$ holds for every path $\gamma:v\to w$ as well. This is proved
by using the identity law for the empty path and substituting~\eqref{eq:5.21} each time the
path is extended by one edge.

What a diagram specifies here is the objects and morphisms of the base and their relations.
Cores, lifts of edges, and automorphisms for comparisons are placed on top of them next.

\end{definition}

\begin{construction}[Building upper-level edges from the same squares]\label{cons:5.18}

Suppose given a core $P_v$ at each vertex and a strongly opcartesian lift $L_a:P_v\to P_w$
for each edge. At the vertices set $P'_v=(i_v)_!P_v$ and write $I_v:P_v\to P'_v$ for the
canonical lift. The edge $L'_a$ on the target is constructed as the unique morphism
satisfying
\begin{equation*}
L'_a I_v=I_wL_a,\qquad q(L'_a)=Y_a
\tag{5.22}\label{eq:5.22}
\end{equation*}
Its existence is given by opcartesianness of $I_v$ and by~\eqref{eq:5.21}. Moreover, since
$I_wL_a$ is opcartesian, $L'_a$ is opcartesian as well. Indeed, compose a morphism out of
the target of $L'_a$ with $I_v$ and factor it through $I_wL_a$. Eliminating $I_v$ from the
resulting equation by the universal property yields the factorization through $L'_a$ and
its uniqueness.

Composing~\eqref{eq:5.22} along a path gives $L'_\gamma I_v=I_wL_\gamma$. For identities,
composites of changes of diagrams, and horizontal and vertical pasting of squares, the two
constructions satisfy the same factorization equations and therefore agree under the
canonical comparisons.

\end{construction}

\begin{proposition}[Conditions for obtaining relations from a family of squares]\label{prop:5.19}

Suppose given only the diagram $X$ on the original, the vertices and edges of the target,
the vertex morphisms $i_v$, and~\eqref{eq:5.21}. The equation obtained for the two paths of
a face $f:v\Rightarrow w$ is
\begin{equation*}
Y_{\ell_f}i_v=i_wX_{\ell_f}
=i_wX_{\rho_f}=Y_{\rho_f}i_v.
\tag{5.23}\label{eq:5.23}
\end{equation*}
If $i_v$ is epi, then $Y_{\ell_f}=Y_{\rho_f}$ follows. Hence, if $i_v$ is epi at every
vertex that appears as the start of a specified face, the family of squares yields a diagram
on the target and a change of diagrams.

Moreover, for a single $i:X\to Y$, the property
\begin{equation*}
ui=vi\Longrightarrow u=v
\tag{5.24}\label{eq:5.24}
\end{equation*}
for all objects $Z$ and all parallel morphisms $u,v:Y\to Z$ is equivalent to $i$ being epi.

\end{proposition}

\begin{proof}

From~\eqref{eq:5.21} for paths and the relations on the original we
obtain~\eqref{eq:5.23}. Applying the cancellation property of epis establishes the relation
for each face. The final equivalence is precisely the definition of an epi.

\end{proof}

The quantification in~\eqref{eq:5.24}, which ranges over all parallel morphisms, differs
from the agreement of only the finitely many paths selected here. For a particular diagram
to be coherent, it is not necessary that each $i_v$ be epi.

\begin{example}[Discrepancy at inputs that are not read]\label{ex:5.20}

Use doctrines that extract the same finite family at every source, with identity
normalizations and Atom maps. Consider the source maps
\begin{equation*}
i:\{*\}\to\{0,1\},\quad i(*)=0,\qquad
u=\mathrm{id}_{\{0,1\}},\quad v(0)=v(1)=0
\tag{5.25}\label{eq:5.25}
\end{equation*}
Taking every selected source to be $*$ or $0$, these become base morphisms. If the two
parallel edges on the original are both identities, the edges $u,v$ on the target
individually form commutative squares. But even though $ui=vi$, we have $u(1)\ne v(1)$.
Agreement at the input $0$ alone does not yield agreement at the input $1$.

If both edges on the target are taken to be $u$, the relation holds even with the same
non-epi $i$. These two examples show the difference between a uniform cancellation
condition and the coherence of an individual diagram.

\end{example}

\section{Carrying diagnoses and edge reselections back and forth}\label{sec:5.6}

Chapter~\ref{chap:4} expressed the difference between a specified comparison and the
canonical comparison as a defect. When the base diagram is changed, we want to carry this
difference as well from the same input. By the fiber equivalence of
Proposition~\ref{prop:5.8}, not only the values of the diagnosis but also the edge
reselections that change those values correspond.

\begin{construction}[End groups, cochains, and the correspondence of reselections]\label{cons:5.21}

Fix the change of diagrams of Definition~\ref{def:5.17} and the cores and edges of
Construction~\ref{cons:5.18}. Let $G_v=\mathrm{Aut}_{E_{X_v}}(P_v)$ be the group of vertical
automorphisms at a vertex $v$, and $G'_v$ its counterpart on the target. The transport
functor induces a group isomorphism
\begin{equation*}
\Phi_v:G_v\xrightarrow{\sim}G'_v,
\qquad \Phi_v(g)=(i_v)_!(g)
\tag{5.26}\label{eq:5.26}
\end{equation*}
Surjectivity comes from fullness and injectivity from faithfulness. A cochain
$\omega=(\omega_f)_f$ on faces and an edge reselection $a=(a_e)_e$ are each sent by applying
this isomorphism at the ends.
\begin{equation*}
(\Phi_C\omega)_f=\Phi_{t(f)}(\omega_f),
\qquad
(\Phi_R a)_e=\Phi_{t(e)}(a_e).
\tag{5.27}\label{eq:5.27}
\end{equation*}
Here $t(f)$ and $t(e)$ are the ends of the face and of the edge. Using each $\Phi_v^{-1}$
componentwise, both assignments have two-sided inverses. The value of a specified comparison
$u_f$ on the target is likewise taken to be $u'_f=\Phi_{t(f)}(u_f)$.

\end{construction}

\begin{theorem}[Preservation and reflection of coherence and reselection orbits]\label{thm:5.22}

Suppose that the base diagram on the target also satisfies the specified path relations. For
the construction above, the canonical comparisons $\phi_f$ of Chapter~\ref{chap:4} and the
raw defects satisfy
\begin{equation*}
\begin{aligned}
\phi'_f(\Phi_Ra)&=\Phi_{t(f)}(\phi_f(a)),\\
\delta'_f(\Phi_Ra)&=\Phi_{t(f)}(\delta_f(a))
\end{aligned}
\tag{5.28}\label{eq:5.28}
\end{equation*}
Consequently, the following equivalences hold.

\begin{enumerate}
\item All faces are coherent at $a$ if and only if all faces are coherent at $\Phi_Ra$.
\item A reselection making all faces coherent exists before the change if and only if one
exists after the change.
\item $\omega$ belongs to the reselection orbit of the raw defect if and only if
$\Phi_C\omega$ belongs to the orbit after the change.
\end{enumerate}

For an arbitrary cochain after the change as well, the same verdicts hold for the value
brought back by $\Phi_C^{-1}$.

\end{theorem}

\begin{proof}

A reselected edge is $L_e^a=a_eL_e$. By the action of transport on morphisms
and~\eqref{eq:5.22}, its image is $\Phi_{t(e)}(a_e)L'_e$. Extending to paths, the morphism
obtained by carrying the canonical comparison of the original satisfies the equation comparing the
two paths after the change. By uniqueness of strongly opcartesian lifts, this is the
canonical comparison after the change. Since group homomorphisms preserve products and
inverses, the second equation of~\eqref{eq:5.28} follows from $\delta_f=u_f\phi_f^{-1}$.

Each $\Phi_v$ preserves and reflects the identity element, so the defect being the identity
in all components is equivalent on the two sides. Moreover $\Phi_R$ is a bijection, so every
reselection after the change can be brought back as well. Carrying the witness
$\omega=\delta(a)$ of orbit membership by~\eqref{eq:5.28}, and using $\Phi_R^{-1}$ and
$\Phi_C^{-1}$ in the reverse direction, the three conclusions follow.

\end{proof}

This diagnosis concerns the fixed diagram and the comparisons generated from it. When new
faces or new Laws are added after the change, a diagram including them is specified anew.
For geometries involving a coefficient ring, this section keeps the same coefficient ring
fixed.

\begin{example}[Correspondence of reselections that cancel the defect]\label{ex:5.23}

In the one-object group category example of Chapter~\ref{chap:4}, take the automorphism
group to be the permutation group $S_3$ on three letters. On two parallel identity edges,
place the specified comparison $u=(12)$. Reselecting the left edge to $a=1$ and the right
edge to $b=(12)$, the canonical comparison becomes $ba^{-1}=(12)$ and the defect vanishes.

Under the group isomorphism $\Phi(g)=cgc^{-1}$ given by the renaming $c=(123)$, the
specified comparison and the reselection of the right edge are both carried to $(23)$. Hence
the same face is coherent after the change. Not only is the truth value ``vanished''
preserved; the reselection that achieved it is carried along.

\end{example}

\section{Conditions for backward transport along refinements}\label{sec:5.7}

When the extraction rules are made more detailed, Atoms that were not extracted before may
appear. Such a change cannot always be carried backwards in the way an exact morphism can.
What is needed is that, at the sources where cores are actually placed, the extraction on
the target can be brought back to the original.

\begin{definition}[Refinements and the realized locus]\label{def:5.24}

A refinement of bases $r:(D,s)\to(D',s')$ has a source map $f$ and a bijection $e$ of
Atoms, and satisfies $f(s)=s'$, $fN_D=N_{D'}f$, and the forward implication
\begin{equation*}
\mathrm{Extracts}_D(t,a)
\Longrightarrow\mathrm{Extracts}_{D'}(f(t),e(a))
\tag{5.29}\label{eq:5.29}
\end{equation*}
Exactness is the condition that adds the backward implication to this.

The collection of objects of the base that carry at least one core is called the realized
locus. We say that $r$ reflects extraction on the realized locus if, whenever $E_{b'}$ has
an object, for every Atom $a$ we have
\begin{equation*}
\mathrm{Extracts}_{D'}(s',e(a))
\Longrightarrow\mathrm{Extracts}_D(s,a)
\tag{5.30}\label{eq:5.30}
\end{equation*}
We write $R(r)$ for this conditional property, namely ``if $E_{b'}$ has an object,
then~\eqref{eq:5.30} holds''.

For core morphisms over refinements, only the base is relaxed to~\eqref{eq:5.29}, while the
upper level is subject to the same preservation conditions for readings as in
Definition~\ref{def:1.28}. In particular, the comparison of selected families is still
required as an equality. These morphisms and objects are closed under composition and
define a category $E_{\mathrm{core}}^{\mathrm{ref}}$ with a projection $q_{\mathrm{ref}}$
to the base.

\end{definition}

\begin{theorem}[Classification of cartesian lifts by the realized locus]\label{thm:5.25}

Fix one pointed refinement $r:b\to b'$. That a strongly cartesian lift with respect to
$q_{\mathrm{ref}}$ can be chosen for every core on $b'$ is equivalent to $R(r)$.

\end{theorem}

\begin{proof}

Assume $R(r)$ and take a core $Q$ on $b'$. Combining~\eqref{eq:5.29} and \eqref{eq:5.30} gives
$F_{D'}(s')=e_*F_D(s)$. From this equality we bring finiteness back and build a core $P$ by
the same inverse reindexing as in Construction~\ref{cons:5.5}. The upper level of the
morphism $P\to Q$ with base $r$ has a two-sided inverse. The same factorization as in
Theorem~\ref{thm:5.6} works after an arbitrary refinement base morphism as well.

Conversely, suppose that lifts can be chosen, and take one core $Q$ on $b'$. The upper
level of its lift $P\to Q$ satisfies $F_Q=e_*F_P$ for the selected families. If $e(a)$ is
extracted on the target, then $a\in F_P$ by injectivity of $e$, so it is extracted on the
original as well. This is~\eqref{eq:5.30}. When there is no core on $b'$, there is no
object for which to choose a lift, and both conditions hold vacuously.

\end{proof}

\begin{construction}[Forward base change of refinements]\label{cons:5.26}

Take an exact cospan $D_1\to D_0\leftarrow D_2$ and a refinement $r:D'_1\to D_1$. Define
the pairs of corresponding sources by
\begin{equation*}
\mathrm{Src}'_{12}
=\{(s'_1,s_2)\mid f_1(f_r(s'_1))=f_2(s_2)\}
\tag{5.31}\label{eq:5.31}
\end{equation*}
and define extraction from the first component in $D'_1$. The first projection
$\pi'_1:D'_{12}\to D'_1$ is exact, and $(s'_1,s_2)\mapsto(f_r(s'_1),s_2)$ together with the
Atom map $e_r$ defines a refinement $r_{12}:D'_{12}\to D_{12}$. Preservation of extraction
follows from~\eqref{eq:5.29}, and the following square commutes.
\begin{equation*}
\begin{array}{ccc}
D'_{12}&\xrightarrow{\ r_{12}\ }&D_{12}\\
{\scriptstyle\pi'_1}\downarrow&&\downarrow{\scriptstyle\pi_1}\\
D'_1&\xrightarrow{\ r\ }&D_1.
\end{array}
\tag{5.32}\label{eq:5.32}
\end{equation*}
This construction does not require $R(r)$. Choosing one pair in~\eqref{eq:5.31} fixes the
sources of the four vertices and a pointed square.

\end{construction}

\begin{proposition}[Transfer of the realized locus and the two backward routes]\label{prop:5.27}

Choose a compatible pair in~\eqref{eq:5.31} and suppose that $R(r)$ holds at that point.
Then $R(r_{12})$ holds as well, and we obtain two functors from cores on $D_1$ to cores on
$D'_{12}$ and a natural comparison
\begin{equation*}
(\pi'_1)^*r^*
\longrightarrow r_{12}^*\pi_1^*
\tag{5.33}\label{eq:5.33}
\end{equation*}
That both refinement lifts can be chosen at every compatible pair is equivalent to $R(r)$
holding at all such pairs. Moreover, pointed refinements that reflect extraction on the
realized locus are closed under identities and composition.

\end{proposition}

\begin{proof}

If there is a core at the selected point of $D_{12}$, it can be transported forwards along
the exact $\pi_1$. This yields a core at the corresponding point of $D_1$ as well, and
$R(r)$ can be used. The extraction of both fiber products is that of the first component,
so~\eqref{eq:5.30} carries over to $r_{12}$ as it stands. Theorem~\ref{thm:5.25} gives the
two refinement lifts, and Theorem~\ref{thm:5.6} the two exact lifts. Factoring the two-step
lifts along the commutative square gives~\eqref{eq:5.33}. Its naturality follows because
the two factors obtained by composing with a vertical morphism satisfy the same triangle.
For the equivalence over all pairs we apply Theorem~\ref{thm:5.25} pointwise.

For a composite $b_0\xrightarrow{r}b_1\xrightarrow{t}b_2$, a core on $b_1$ is built by
pulling a core on $b_2$ back along $t$. This also satisfies the realization condition of
$R(r)$. Reflecting extraction along $t$ and then along $r$ yields $R(tr)$. In the case of
an identity, \eqref{eq:5.30} becomes the identity implication.

\end{proof}

\begin{example}[A forward-only change and the scope that can be carried back]\label{ex:5.28}

Let the Atoms be $\{A,B,C\}$ and the sources $\{\mathrm{partial},\mathrm{all}\}$. Take the
identity normalization and define the extractions of the original and the target as
follows.

\begin{xltabular}{\linewidth}{@{}LLL@{}}
\toprule
Source & Original & Target \\
\midrule
\endhead
partial & A, B & A, B, C \\
\addlinespace[3pt]
all & A, B, C & A, B, C \\
\bottomrule
\end{xltabular}

The source map is the identity, and the Atom map is the bijection that swaps $A,B$ and
fixes $C$. This is a refinement. All extracted families are finite, and a core can be
placed at each point. For instance, set $\mathrm{Comp}(F)=(F,\varnothing,\varnothing)$, and
let object formation preserve the configuration and place the element of a one-element set
in the structure data and in the quantity. If the operations are only the identities
between equal objects, and the equation indices, invariants, and signature axes are empty,
we can construct a core satisfying the conditions of the reading.

If partial is selected, the Atom $C$ on the target cannot be brought back to the
original. By Theorem~\ref{thm:5.25}, no lift to a core on the target exists. If both cospan
legs are the identity of the target doctrine, the forward square of
Construction~\ref{cons:5.26} can still be built.

On the other hand, let the second cospan leg be the exact morphism into all on the target
from the doctrine that has a single source and extracts all Atoms. In a compatible pair
the first source is then restricted to all, where~\eqref{eq:5.30} holds. Hence that square
has both backward transports. The original refinement still enlarges the extraction at
partial. The reason backward transport became possible is that the inputs matched this
time are narrowed to the scope of all.

\end{example}

\begin{example}[Contrast with the case where no object exists]\label{ex:5.29}

Consider the identity refinement with one source, Atoms $\mathbb{N}$, and the whole of
$\mathbb{N}$ as the extracted family. Since the selected family is infinite, no core
exists. Both conditions of Theorem~\ref{thm:5.25} hold, but this is not an example of cores
actually traveling back and forth. Example~\ref{ex:5.28} compares the two behaviors after
confirming that objects exist.

\end{example}

\section{Comparisons including covers and local data}\label{sec:5.8}

A comparison of cores includes the objects, the operations, and the Laws. To compare local
inspections under the same conditions, we must also carry which contexts read what and
where they overlap. The full geometries of Chapter~\ref{chap:1} are the objects that add
these covers, overlaps, coefficients, and raw systems to a core.

\begin{definition}[Morphisms of geometries over refinements]\label{def:5.30}

For a morphism of geometries $G\to H$, take its lower level to be a morphism of
$E_{\mathrm{core}}^{\mathrm{ref}}$, and take the correspondences of contexts, equations,
Atoms, and axes from its upper-level reading map. On the components of the geometries we
impose the same five conditions as in Definition~\ref{def:1.30}: forward preservation of
coverage requirements, comparison of overlaps, a coefficient homomorphism, the transport
equations for raw systems, and the readouts of support, axes, and observables together with
their naturality. These define a category $E_{\mathrm{geom}}^{\mathrm{ref}}$ and
projections
\begin{equation*}
E_{\mathrm{geom}}^{\mathrm{ref}}
\xrightarrow{\ p_{\mathrm{ref}}\ }
E_{\mathrm{core}}^{\mathrm{ref}}
\xrightarrow{\ q_{\mathrm{ref}}\ }B^{\mathrm{ref}}
\tag{5.34}\label{eq:5.34}
\end{equation*}
The exact three-level categories enter here by restricting the base to exact morphisms.

\end{definition}

\begin{construction}[Pullback of geometries along the generated backward transport]\label{cons:5.31}

Take a lift of cores $K:P\to Q$ built by Theorem~\ref{thm:5.6} or 5.25, and a geometry $H$
on $Q$. Let $u,u^{-1}$ be the upper level of $K$ and its generated inverse, and
$\theta,\bar\theta$ the correspondence of contexts. We build a geometry $K^*H$ on $P$ as
follows.

\begin{xltabular}{\linewidth}{@{}LL@{}}
\toprule
Data & How it is pulled back \\
\midrule
\endhead
Coverage requirements and visibility & Send Atoms, equations, and axes forward, then read
the predicates of $H$ \\
\addlinespace[3pt]
Overlap of two contexts & Send them to $H$ by $\theta$, take the overlap in $H$, and return
by the specified quasi-inverse \\
\addlinespace[3pt]
Coefficient ring & Use the same ring $k$ as $H$ \\
\addlinespace[3pt]
Raw system & In a context $W$, read the coordinates, relations, and local data of $H$ at
$\theta W$ \\
\addlinespace[3pt]
Support, axes, observables & Use the same correspondence of carriers that accompanies the
reindexing \\
\bottomrule
\end{xltabular}

For the restriction maps of the raw system we use the morphisms sent by $\theta$. The
identity and composition laws and the preservation of relation polynomials follow from the
original raw system. The universal property of overlaps is brought back by the equivalence
of context categories. In this way we obtain a morphism with identity coefficient
component,
\begin{equation*}
\widehat K:K^*H\longrightarrow H
\tag{5.35}\label{eq:5.35}
\end{equation*}
This morphism is strongly cartesian with respect to $p_{\mathrm{ref}}$. Indeed, for a
morphism passing through $K$ on the lower level, the cover conditions return to the
preimage conditions just defined, and the raw-system equations return by the reindexing of
$u^{-1}$. The comparison maps of local data are also brought back by the generated
two-sided inverses. These constitute the factor, and uniqueness follows by composing with
the same inverses.

\end{construction}

\begin{theorem}[Two backward routes built from the same geometry]\label{thm:5.32}

Fix the point and the realized-locus condition of Proposition~\ref{prop:5.27}, and place a
geometry $H$ on a core $Q$ on $D_1$. Repeating Construction~\ref{cons:5.31} along the two
routes yields geometries $H_B,H_P$ and morphisms $K_B:H_B\to H$ and $K_P:H_P\to H$. Between
them there is a natural comparison $m_H:H_B\to H_P$ satisfying
\begin{equation*}
K_Pm_H=K_B,
\qquad p_{\mathrm{ref}}(m_H)=m_Q
\tag{5.36}\label{eq:5.36}
\end{equation*}
Here $m_Q$ is the comparison of cores obtained by aligning the endpoints of the two routes
of~\eqref{eq:5.33} by the canonical isomorphisms. For the generated routes $m_H$ is
invertible, and its coefficient component is $1_k$.

\end{theorem}

\begin{proof}

The core morphisms of the two routes are each composites of two strongly cartesian
morphisms. Since the square of bases commutes, the comparison $m_Q$ aligning their starts
and ends, and its inverse, are obtained uniquely. The two routes of geometries are likewise
composites of cartesian morphisms from Construction~\ref{cons:5.31}. We first build $m_Q$
on the lower level and lift its factorization to the upper level to obtain $m_H$.
Performing the same procedure in the opposite direction and applying uniqueness gives a
two-sided inverse. The coefficient map at each step is the identity, so the coefficient map
of the comparison is the identity as well.

\end{proof}

To compare a whole diagram, we take a finite rooted diagram with a path to each vertex, and
fix a diagram of cores inside the fiber category at the same point of the base, geometries
on it, and a coefficient ring $k$. We further give the two-level transport data of
Chapter~\ref{chap:4}: each edge of cores and of geometries is a strongly opcartesian
morphism with respect to its projection, and the coefficient components of the edges and of
the specified comparisons are identities. Applying Theorem~\ref{thm:5.32} at each vertex,
we pull the edges and the specified comparisons of the same source back to both routes. As
a result, the generated comparison matches up not only the edges but also the specified
comparisons of the faces.
\begin{equation*}
\begin{aligned}
m_w L^B_e&=L^P_e m_v,\\
m_{t(f)}u^B_f&=u^P_fm_{t(f)}.
\end{aligned}
\tag{5.37}\label{eq:5.37}
\end{equation*}
Composed with $K_P$, both sides of the second equation become the same morphism, obtained by carrying
a single source comparison. The equation is obtained by reflecting it first through
cartesianness of cores and then through cartesianness of geometries. The first equation is
similar, and naturality for paths follows by induction from the generating edges. When the
specified comparisons are chosen separately on the two routes, the second equation becomes
an additional condition to be checked.

\begin{lemma}[Transfer of solutions by endpoint isomorphisms]\label{lem:5.33}

Suppose that between the diagrams of the two routes and other presentations of them there
are natural isomorphisms
$b_v:\widetilde H_{B,v}\xrightarrow{\sim}H_{B,v}$ and
$c_v:\widetilde H_{P,v}\xrightarrow{\sim}H_{P,v}$.
Assume the isomorphisms fix the coefficients and match up the morphisms of the routes and
the specified comparisons. The families of comparisons $s_v$ satisfying the fixed
lower-level comparison, the triangles, and~\eqref{eq:5.37} then correspond in both
directions by
\begin{equation*}
\widetilde s_v=c_v^{-1}s_vb_v,
\qquad
s_v=c_v\widetilde s_vb_v^{-1}
\tag{5.38}\label{eq:5.38}
\end{equation*}
The endpoints of the lower-level comparisons and of the triangles are aligned by the same
isomorphisms.

\end{lemma}

\begin{proof}

Substitute the naturality of $b,c$ into the naturality for edges and cancel adjacent
inverses. The same computation applies to the equations for the specified comparisons and
to the triangles. Using the identity law for the empty path and associativity for
concatenation of paths and pasting of faces, all conditions are preserved. The two formulas
of~\eqref{eq:5.38} are mutually inverse.

\end{proof}

Using the same conjugation at the ends for reselections and defects, vanishing and
reselection orbits correspond in both directions, for the same reason as
in~\eqref{eq:5.28}. What this correspondence uses is routes generated from the same source
and endpoint isomorphisms that preserve the components. For an arbitrarily given morphism
of geometries, the preservation of coverage requirements and local data may be forward
only, and transferring solutions backwards requires the inverse morphisms that
form~\eqref{eq:5.38}.

\section{Reading the exchange of order in lenses and protocols}\label{sec:5.9}

When the view of a lens or the operations of a protocol are restricted, we ask whether
semantics-preserving changes can also be restricted to that scope. For lenses we show that
after the restriction the updates stay within the scope; for protocols, that the
comparisons of executions and adapters are preserved.

\begin{proposition}[Restriction of the view of a lens]\label{prop:5.34}

For a total lens $g:C\to V$, $p:C\times V\to C$ satisfying the three laws and a subset
$W\subseteq V$, set
\begin{equation*}
C_W=\{c\in C\mid g(c)\in W\}
\tag{5.39}\label{eq:5.39}
\end{equation*}
The restrictions of read and update define a total lens $C_W\to W$.
Moreover, for a lens isomorphism $(h,u):(C,V)\to(C',V')$ and $W'\subseteq V'$,
restriction of $h$ gives a lens isomorphism
\begin{equation*}
(C_{u^{-1}(W')},u^{-1}(W'))
\xrightarrow{\sim}(C'_{W'},W')
\tag{5.40}\label{eq:5.40}
\end{equation*}

\end{proposition}

\begin{proof}

The read after an update is $g(p(c,w))=w$, so if $w\in W$ the state after the update also
belongs to $C_W$. Reading the equations of the three laws on this subset gives the three
laws after the restriction. Next, $g'h=ug$ gives
\begin{equation*}
h^{-1}(C'_{W'})=C_{u^{-1}(W')}.
\tag{5.41}\label{eq:5.41}
\end{equation*}

Hence $h$ and $h^{-1}$ match the state sets.
The update equation $h(p(c,v))=p'(h(c),u(v))$ restricts as well.

\end{proof}

\begin{example}[Restriction and update with four states]\label{ex:5.35}

We check Proposition~\ref{prop:5.34} on a product lens with four states. Let the displayed values be $V=\{0,1\}$ and the component preserved by
updates be $R=\{0,1\}$, and use the product lens $C=V\times R$. Read returns the first
component, and update rewrites only the first component. Let the displayed values of the
target be $V'=\{A,B\}$, with the bijection $u$ sending 0 to A and 1 to B. The
correspondence of states $h$ converts the displayed value by $u$ and swaps 0 and 1 in the
second component. Let $W'$ be the set of displayed values to select, keeping only A on the
target. That is, we set
\begin{equation*}
h(v,r)=(u(v),1-r),\qquad W'=\{A\}
\tag{5.42}\label{eq:5.42}
\end{equation*}
Selecting the states with displayed value 0 on the original and then applying the
correspondence, or selecting the states with displayed value A on the target, reaches the
same two states. The restricted update is closed in the selected view, and $h$ preserves
updates as well. In this example, comparing the two routes specifies at the same time
which states to keep and which updates to allow.

\end{example}

This restriction takes place within the semantics of lenses.
To use it as a base change of cores, construct the rolewise Atoms, operations, and Laws
of Chapter~\ref{chap:1} and check exactness at the base and the operation-preservation
diagrams. Preservation of semantics in \eqref{eq:5.40} alone does not uniquely
determine the choices of covers or local data.

\begin{proposition}[Restriction of protocols and adapters]\label{prop:5.36}

Let $\mathcal{P}$ be the presented protocol category of Chapter~\ref{chap:1}, and let
$j:\mathcal{Q}\to\mathcal{P}$ be the inclusion of the protocol consisting of the vertices
and operations to be kept and the relations between them. Two realizations
$F,F':\mathcal{P}\to{\mathbf{Set}}$ and a semantics-preserving adapter $a:F\Rightarrow F'$
determine restrictions $Fj,F'j$ and an adapter $aj:Fj\Rightarrow F'j$. Comparison triangles
and commutative adapter squares are also preserved by the restriction.

\end{proposition}

\begin{proof}

The restrictions $Fj,F'j$ are given by composition of functors. The naturality of $a$ for
an arbitrary morphism $e:x\to y$ of $\mathcal{Q}$, namely
$a_{j(y)}F(j(e))=F'(j(e))a_{j(x)}$, gives the naturality of $aj$. Preservation of
observations and the equations of the triangles and squares are likewise restricted at
each vertex.

\end{proof}

In the example of splitting the monolith, from the protocol of the whole we keep the
cancellation request, the release of inventory reservations, and cancellation completion,
together with the relations between them. Giving realizations before and after the split
and checking the naturality of the adapter for each operation, we obtain a correspondence
for the execution of a cancellation that chains them. In selecting this part, we include
the operations carried by the inventory service. Only then does the cancellation process
that crosses services become an object of comparison. The Laws on the states of
cancellation completion and reservation release are checked together on the corresponding
states.

When communication failures and retries are read as new operations and states, we specify
a protocol and a correspondence that include them. Coherence of the comparison is a
property of the selected operations and relations, and for diagrams that also contain
operations outside the selection, it must be checked separately, as in
Example~\ref{ex:5.20}.

\section{The generated comparison including the two routes and normalization}\label{sec:5.10}

The canonical comparison $\alpha$ gives the correspondence of structure along the two
routes. The generated comparison $\beta$ additionally incorporates a normalization
selected according to the diagnosis. Even when a normalization preserves the operations and
the Laws, the distinctions among the original representations may not be recoverable. We
construct this comparison from the same input, and Chapter~\ref{chap:6} analyzes its
invertibility and the information that remains.

\begin{construction}[Two routes of full geometries]\label{cons:5.37}

Fix the exact pointed pullback square of \eqref{eq:5.14}, a finite comparison
diagram, a face $z$ and a cochain $\omega$, the corresponding core $P_z$ over $b_1$,
a commutative coefficient ring $k$, and a geometry $G_z$ over $P_z$.
The geometry is specified together with its coverage requirements, overlaps, and raw system.
Using the canonical geometry transport of Chapter~\ref{chap:4} and
Construction~\ref{cons:5.31}, form
\begin{equation*}
\overline D(G_z)=(\pi_2)_!\pi_1^*G_z,
\qquad
\overline V(G_z)=\sigma_2^*(\sigma_1)_!G_z
\tag{5.43}\label{eq:5.43}
\end{equation*}
The lower and upper indices denote the operations on entire geometries. Both routes keep
the coefficient ring $k$.

Construct the canonical comparison of full geometries
$\overline\alpha_z:\overline D(G_z)\xrightarrow{\sim}\overline V(G_z)$ as the morphism
satisfying the triangle corresponding to \eqref{eq:5.18}.
The lifts of Construction~\ref{cons:5.31} and canonical geometry transport have
two-sided inverses in their upper-level and local-data comparisons, so the units
and counits for geometries are invertible as well.
Thus the three-step construction of \eqref{eq:5.17} performed for geometries is invertible.

This comparison also arises from the comparison of the two backward routes of
\S\ref{sec:5.8}. Setting $H_0=(\sigma_1)_!G_z$, the comparison that
traces the same square backwards is
\[
m_{H_0}:\pi_1^*\sigma_1^*H_0
\xrightarrow{\sim}\pi_2^*\sigma_2^*H_0
\]
Aligning the choices at the endpoints, the following three steps can be composed using the
unit and the counit.
\[
\begin{aligned}
\overline D(G_z)
&\xrightarrow{\ (\pi_2)_!\pi_1^*(\eta^{\sigma_1}_{G_z})\ }
(\pi_2)_!\pi_1^*\sigma_1^*H_0\\
&\xrightarrow{\ (\pi_2)_!(m_{H_0})\ }
(\pi_2)_!\pi_2^*\sigma_2^*H_0\\
&\xrightarrow{\ \epsilon^{\pi_2}_{\sigma_2^*H_0}\ }
\overline V(G_z).
\end{aligned}
\]
The first morphism is the unit pulled back and then transported, and the last is the
counit. When a different choice of lifts is used, the endpoint isomorphisms built from
Construction~\ref{cons:5.31} are inserted before and after the middle comparison. The order
is: unit, isomorphism at the start, comparison of the backward routes, isomorphism at the
end, counit. Substituting~\eqref{eq:5.36} and the unit-counit triangles yields the
geometric version of~\eqref{eq:5.18}, so uniqueness gives the same $\overline\alpha_z$.

\end{construction}

\begin{proposition}[Compatibility with the projection to cores]\label{prop:5.38}

Write $p$ for the functor that forgets the geometry, and
$j_D:p\overline D(G_z)\xrightarrow{\sim}DP_z$ and
$j_V:p\overline V(G_z)\xrightarrow{\sim}VP_z$ for the generated endpoint isomorphisms.
Then
\begin{equation*}
j_V\,p(\overline\alpha_z)=\alpha_{P_z}\,j_D.
\tag{5.44}\label{eq:5.44}
\end{equation*}

\end{proposition}

\begin{proof}

Projecting the lifts of geometries yields the lifts of cores used in the generation. Even
when the choices differ, $j_D,j_V$ are their unique comparisons. Projecting the geometric
version of~\eqref{eq:5.18} and aligning it by these comparisons gives~\eqref{eq:5.18} for
cores. That morphism is unique, so it coincides with the mate of Theorem~\ref{thm:5.11}.

\end{proof}

So far, the structures obtained by the two routes have been matched by an invertible
canonical comparison. Next we choose which differences of representation to keep in the
comparison. For example, we may want to align several design representations with the same
configuration after the split to a common standard representation. To add this processing,
we must preserve not only the presentation of objects but also the operations and the
Laws. We first extract the conditions under which such a normalization can be defined.

\begin{definition}[Normalization preserving configurations]\label{def:5.39}

A core $P$ selects an object $\mathrm{Form}_P(C)$ for each configuration $C$. Let the map
sending an arbitrary object $A$ to the selected object on the same configuration be
\begin{equation*}
n_P(A)=\mathrm{Form}_P(C_A)
\tag{5.45}\label{eq:5.45}
\end{equation*}
By the conditions on object formation, $C_{n_P(A)}=C_A$ and $n_P^2(A)=n_P(A)$. To make
this map a morphism of cores, we let $\mathrm{Ad}(P)$ be the following conditions.

\begin{itemize}
\item In every context, object, equation index, and Atom, the residual is unchanged under
$A\mapsto n_P(A)$.
\item $\mathrm{Op}_P(A,B)$ and $\mathrm{Op}_P(n_P(A),n_P(B))$ have the same type, and this
identification preserves the configuration maps.
\item The indices and kinds of the invariants are preserved, and on every object the truth
values of predicate-type invariants and the values of function-type invariants are
unchanged by the normalization.
\item On every signature axis, the coordinate values are unchanged under
$A\mapsto n_P(A)$.
\end{itemize}

The function-type condition is equivalent to the correspondence by bijections of the value
ranges in Definition~\ref{def:1.28}. Indeed, abbreviate $n=n_P$ and suppose there are
bijections $t_i$ with $t_i(I_i(A))=I_i(n(A))$ for all objects. By idempotence $n^2=n$ of
the object map,
\[
t_i(I_i(A))=I_i(n(A))=I_i(n(n(A)))=t_i(I_i(n(A))).
\]
Injectivity of $t_i$ gives $I_i(A)=I_i(n(A))$. Conversely, if the values are unchanged, we
may choose $t_i=\mathrm{id}$.

We then define an endomorphism $N_P$ of the core by using $n_P$ on objects, the type
identification above on operations, and identities on Atoms, contexts, equation indices,
and axes. Since configurations are preserved, the detector code is also kept identical.
For a geometry $G$ we further define an endomorphism $N_G$ using the identity maps on
coefficients, support, axes, and observables. The equations for covers, overlaps, and raw
systems hold with identity comparisons.

This normalization is incorporated into the comparison of the two routes according to a
specified diagnosis. In the next construction, the normalization is selected when a
component of the cochain is not the identity and the preservation conditions above hold.
In this way, which diagnosis was responded to, and on which route the representations were
aligned, can be expressed as a single comparison morphism.

\end{definition}

\begin{construction}[Generated comparisons selected by the same diagnosis]\label{cons:5.40}

For the fixed $z,\omega,P_z,G_z$, select the endomorphism used on the original core by
\begin{equation*}
S_{z,\omega}=
\begin{cases}
N_{P_z},&\omega_z\ne1\ \text{and}\ \mathrm{Ad}(P_z),\\
1_{P_z},&\text{otherwise}
\end{cases}
\tag{5.46}\label{eq:5.46}
\end{equation*}
This selection is a mathematical case distinction and does not assume a finite decision
procedure for $\mathrm{Ad}(P_z)$. For the full geometry we likewise select $N_{G_z}$ or
the identity under the same condition and write $\overline S_{z,\omega}$. Next, this
endomorphism is carried along the actual second route.
\begin{equation*}
\begin{aligned}
E_{z,\omega}&=V(S_{z,\omega}):VP_z\longrightarrow VP_z,\\
\overline d_{z,\omega}
 &=\overline V(\overline S_{z,\omega}):
 \overline V(G_z)\longrightarrow\overline V(G_z),\\
\beta_{z,\omega}&=E_{z,\omega}\alpha_{P_z},\\
\overline\beta_{z,\omega}
 &=\overline d_{z,\omega}\overline\alpha_z.
\end{aligned}
\tag{5.47}\label{eq:5.47}
\end{equation*}
This is one generation rule that incorporates the permitted normalization when the
component of the specified cochain is not the identity. The current raw defect may also be
chosen as the cochain. The square, the face, the cochain, the coefficients, and the
original geometry used in the construction are common to the two routes. Under the same
projections and endpoint isomorphisms,
\begin{equation*}
j_V\,p(\overline d_{z,\omega})=E_{z,\omega}j_V,
\qquad
j_V\,p(\overline\beta_{z,\omega})=\beta_{z,\omega}j_D
\tag{5.48}\label{eq:5.48}
\end{equation*}
The first equation follows from having selected the same normalization on the original
object and from the projection compatibility of transport. The second is the composite of
the first with~\eqref{eq:5.44}.

Three kinds of comparisons are now in place.

\begin{xltabular}{\linewidth}{@{}LLL@{}}
\toprule
Comparison & What determines it & What to check \\
\midrule
\endhead
The canonical $\alpha$ & The universal properties of transport and pullback & It satisfies
the triangle of the two routes and is invertible \\
\addlinespace[3pt]
A separately specified $a$ & A comparison chosen by an implementation or a way of reading &
Whether it satisfies the triangle, and what its difference from $\alpha$ is \\
\addlinespace[3pt]
The generated comparison $\beta$ & The $\alpha$ of the same input, and the normalization
selected by the diagnosis & Invertibility and image after the normalization is
incorporated \\
\bottomrule
\end{xltabular}

Chapter~\ref{chap:6} proves the idempotence of $N_P$ and $N_G$ as morphisms and studies
when this generated comparison is invertible and, when it is not, what it leaves as its
image.

\end{construction}

\section*{Summary of the Chapter}

In this chapter we built two routes from transport and changes of input, and extended
their comparison to diagnoses, geometries, and normalization.

\begin{itemize}
\item \textbf{Base change and the exchange of order.} We built the inputs corresponding to
a common condition as fiber products and pulled structure back. Along an exact base
morphism a cartesian lift to any core on the target is obtained, and transport and
pullback form an adjoint equivalence. The Beck--Chevalley mate built from its unit and
counit is invertible (\S\S\ref{sec:5.1}--\ref{sec:5.3}).
\item \textbf{Correspondence of diagrams and diagnoses.} In diagrams whose base relations
are aligned, the end groups, cochains, and edge reselections correspond, preserving and
reflecting the vanishing of defects and membership in reselection orbits. In aligning the
relations of a whole diagram, we distinguish conditions cancellable by epis from relations
imposed individually (\S\S\ref{sec:5.5}--\ref{sec:5.6}).
\item \textbf{The scope of backward transport.} For refinements, reflection of extraction
on the realized locus is necessary and sufficient for backward transport. Even when the
doctrine as a whole is not exact, backward transport is possible if the extractions agree
at the sources actually matched (\S\ref{sec:5.7}).
\item \textbf{Extension to geometries and operations.} By carrying the covers and local
data as well from the same original geometry, we raised the comparison of the two routes
to full geometries. For lenses we showed the condition that updates stay within the
selected scope, and for protocols the correspondence of all executions of the kept
operations (\S\S\ref{sec:5.8}--\ref{sec:5.9}).
\item \textbf{Generated comparisons including normalization.} We selected a normalization
according to the diagnosis and composed it with the canonical comparison to form
$\beta=E\alpha$. Since the same normalization is selected for the geometry and the core,
this comparison is also compatible with the projection (\S\ref{sec:5.10}).
\end{itemize}

For the split of the monolith from the opening, we compare the splitting proposal built
from the order-cancellation part with the cancellation part extracted after splitting the
whole. Including the cancellation request, the release of inventory reservations, and
cancellation completion in the scope makes the operations and Laws of the two routes
correspond. If, on top of this, several design representations are aligned to a common
representative, the generated comparison provides the correspondence including that
normalization.

The next chapter examines the normalization factor of the generated comparison and
clarifies its invertibility and the information that remains.

\chapter{Idempotent Normalization and Realization}\label{chap:6}

\section*{Overview of the Chapter}

The question of this chapter is \textbf{what a comparison that includes a normalization
retains, and when one can return to the original}.
Chapter~\ref{chap:5} built two routes from the transport of structure and the selection of
the input, and obtained the invertible canonical comparison $\alpha$ between them.
It further incorporated a normalization $E$ chosen according to the diagnosis, and
constructed the generated comparison $\beta=E\alpha$.
In this chapter we examine how this last step changes the meaning of the comparison.

For example, consider a situation in which a design split into an order service and an
inventory service is brought together into a common design model.
We want to keep the operations of cancellation request, release of an inventory
reservation, and cancellation completion together with their dependencies, and to align the
class structure inside the services and the auxiliary data representations with a chosen
standard representation.
If we can check that the operations, Laws, and evaluation values that are kept are
unchanged by this processing, then the two design proposals can be compared on the standard
representation.
Whether the original class structure can be recovered from that comparison, however, must
be checked separately.

A normalization is idempotent when applying the alignment to the standard representation a
second time does not change the result.
Idempotence is the property that ``standardization is stable under repetition,'' and it
differs from invertibility, the property that ``the original representation can be
recovered.''
In this chapter we explain this difference at three stages: the map of objects, the
structure-preserving morphism, and the image after normalization.

\begin{xltabular}{\linewidth}{@{}LLL@{}}
\toprule
Question & Construction & Outcome \\
\midrule
\endhead
Which objects remain after normalization? & The object selected for each configuration & Correspondence between fixed points and configurations, unique factorization of readouts \\
\addlinespace[3pt]
Can normalization be repeated including operations and Laws? & Endomorphisms of cores and full geometries & Idempotence of the entire morphism \\
\addlinespace[3pt]
When is a comparison with a normalization added invertible? & Factorization into the canonical comparison and an idempotent factor & Equivalence with the normalization factor being the identity \\
\addlinespace[3pt]
How is a noninvertible comparison to be read between the pieces of information that remain? & The image in the idempotent completion & Isomorphism between the images given by the same generated comparison \\
\addlinespace[3pt]
Can the order of a normalization and an arbitrary design change be interchanged? & The normalization functor and the condition of naturality & A one-sided equation that always holds, and a counterexample from additional information on operations \\
\addlinespace[3pt]
Can a comparison itself be made an object of change? & The arrow category and idempotent completion & A category for classifying comparison-preserving changes in Chapter~\ref{chap:7} \\
\bottomrule
\end{xltabular}

In \S\S\ref{sec:6.1}--\ref{sec:6.4} we analyze a single generated comparison and proceed as
far as the classification of invertibility for full geometries.
In \S\S\ref{sec:6.5}--\ref{sec:6.7} we construct the changes between objects and the
category of comparisons, and in \S\ref{sec:6.8} we return to the semantics of lenses and
protocols.

\section{Selecting one representation for each configuration}\label{sec:6.1}

The architecture object of Chapter~\ref{chap:1} was an object obtained by adding structure
data and a quantity to a configuration.
What remains when the objects with the same configuration are aligned to a single one?
We first answer this question at the level of the map of objects, before operations and
Laws are added.

\begin{construction}[Projection to configurations and selection]\label{cons:6.1}

We fix the vocabulary of Atoms and write $\mathcal{A}$ for the totality of architecture
objects and $\mathcal{C}$ for the totality of configurations. For a core $P$ we set
\begin{equation*}
\pi:\mathcal{A}\longrightarrow\mathcal{C},\quad \pi(A)=C_A,
\qquad
s_P:\mathcal{C}\longrightarrow\mathcal{A},\quad s_P(C)=\mathrm{Form}_P(C)
\tag{6.1}\label{eq:6.1}
\end{equation*}
Here $\mathcal{A},\mathcal{C}$ are collections of objects, and \eqref{eq:6.1} defines maps
between them.
By the conditions on object formation, $\pi s_P=1_{\mathcal{C}}$.
Hence the normalization of Definition~\ref{def:5.39} satisfies
\begin{equation*}
n_P=s_P\pi,\qquad \pi n_P=\pi,\qquad n_P^2=n_P
\tag{6.2}\label{eq:6.2}
\end{equation*}
The last equation is the computation $s_P\pi s_P\pi=s_P\pi$.

\end{construction}

\begin{proposition}[Fixed points and configurations]\label{prop:6.2}

Between the collection of fixed points $\mathrm{Fix}(n_P)=\{A\mid n_P(A)=A\}$ and
$\mathcal{C}$ there are mutually inverse correspondences
\begin{equation*}
\begin{aligned}
\pi:\mathrm{Fix}(n_P)&\longrightarrow\mathcal{C},& A&\longmapsto C_A,\\
s_P:\mathcal{C}&\longrightarrow\mathrm{Fix}(n_P),& C&\longmapsto s_P(C)
\end{aligned}
\tag{6.3}\label{eq:6.3}
\end{equation*}
Moreover, if we define $A\sim B$ by $C_A=C_B$, then
$\mathcal{A}/{\sim}\cong\mathcal{C}$.

\end{proposition}

\begin{proof}

Since $n_Ps_P=s_P$, the object $s_P(C)$ is a fixed point.
The equation $\pi s_P=1$ and the equation $s_P\pi(A)=A$ at fixed points give a two-sided
inverse.
For the quotient, $[A]\mapsto C_A$ is well defined, and the inverse is $C\mapsto[s_P(C)]$.
The configurations of an object and of its selected object are equal, so this is also a
two-sided inverse.

\end{proof}

\begin{theorem}[Readouts unchanged by normalization]\label{thm:6.3}

For any set of values $Y$ and any map $f:\mathcal{A}\to Y$,
\begin{equation*}
fn_P=f
\quad\Longleftrightarrow\quad
\exists!\,\widetilde f:\mathcal{C}\to Y,
\qquad f=\widetilde f\pi.
\tag{6.4}\label{eq:6.4}
\end{equation*}

\end{theorem}

\begin{proof}

When the left-hand side holds, setting $\widetilde f=fs_P$ gives
$\widetilde f\pi=fn_P=f$.
If another factor $g$ satisfies $g\pi=f$, composing with $s_P$ on the right gives
$g=fs_P$. Conversely, if $f=\widetilde f\pi$, then
$fn_P=f$ by $\pi n_P=\pi$.

\end{proof}

This theorem characterizes the information that a normalization retains, from the point of
view of readouts.
A value that is unchanged before and after normalization can be read uniquely from the
configuration alone.
This theorem can also be applied to the invariance of residuals, invariants, and signatures
required in Definition~\ref{def:5.39}.
In that case we read with the arguments other than the object, such as the context, the
equation index, and the axis, held fixed.

\begin{example}[A finite model with two representations]\label{ex:6.4}

We take the set of configurations to be $\{c_0,c_1\}$, the tags of the representations
to be $\{0,1\}$, the objects to be $(c_i,t)$, and the selected objects to be $(c_i,0)$.
\begin{equation*}
n(c_i,t)=(c_i,0).
\tag{6.5}\label{eq:6.5}
\end{equation*}
Two of the four objects are fixed points.
The readout $f(c_i,t)=i$ is unchanged by the normalization and can be read from the
configuration.
The readout $g(c_i,t)=t$, on the other hand, distinguishes the original representations, so
the same value cannot be recovered after the normalization.
In the software example, one may take the selected operations and dependencies to
correspond to the former, and which auxiliary data represent that structure to correspond
to the latter.

\end{example}

\section{Lifting object normalization to a structure-preserving idempotent}\label{sec:6.2}

Even if a map of objects is idempotent, that alone does not make it a morphism preserving
operations and Laws.
The condition $\mathrm{Ad}(P)$ of Definition~\ref{def:5.39} supplies the preservation laws
needed for this.
In this section we show that the entire morphism built from that condition is idempotent.

\begin{theorem}[Idempotence of the normalization of a core]\label{thm:6.5}

For a core $P$ satisfying $\mathrm{Ad}(P)$, the endomorphism
$N_P:P\to P$ of Definition~\ref{def:5.39} satisfies
\begin{equation*}
N_P^2=N_P,\qquad q(N_P)=1_{q(P)}
\tag{6.6}\label{eq:6.6}
\end{equation*}
Here $q:E_{\mathrm{core}}\to B$ is the projection to the base.

\end{theorem}

\begin{proof}

For the object map we use \eqref{eq:6.2}.
The maps of Atoms, the base, contexts, equation indices, invariant indices, and signature
axes are identities, so composing them twice gives the same maps.
The correspondence of the rings of the equations and the correspondence of the value ranges
of the signatures are identities as well, and preservation of the residuals is by
$\mathrm{Ad}(P)$.

For the operations we use the identification along the equality of the types
$\mathrm{Op}_P(A,B)$ and $\mathrm{Op}_P(n_P(A),n_P(B))$.
Composing the identification twice is equal to the identification taken once, after the
endpoints have been aligned by $n_P^2=n_P$.
Its configuration map is the same as well, by the operation-preservation condition of
$\mathrm{Ad}(P)$.
This shows that all the data of a morphism in Definition~\ref{def:1.28} agree.
The base morphism is the identity by construction.

\end{proof}

\begin{lemma}[Naturality for the object map]\label{lem:6.6}

Let $H$ be the object map of an arbitrary morphism of cores $h:P\to Q$. Then
\begin{equation*}
Hn_P=n_QH.
\tag{6.7}\label{eq:6.7}
\end{equation*}
Neither $\mathrm{Ad}(P)$ nor $\mathrm{Ad}(Q)$ is needed for this conclusion.

\end{lemma}

\begin{proof}

Let $e$ be the bijection of Atoms of $h$. By the preservation of object formation and of
configurations in Definition~\ref{def:1.28}, we obtain
\[
H(n_P(A))
=H(\mathrm{Form}_P(C_A))
=\mathrm{Form}_Q(e_*C_A)
=\mathrm{Form}_Q(C_{H(A)})
=n_Q(H(A))
\]

\end{proof}

Equation~\eqref{eq:6.7} is an equality about the correspondence of objects.
The condition $hN_P=N_Qh$, which includes the correspondence of the operations as well, is examined in \S\ref{sec:6.6}.

\begin{proposition}[Distinct objects over the same configuration]\label{prop:6.7}

Over any configuration $C$ there exist at least two distinct architecture objects.
Hence, for every core $P$, the map $n_P$ is not injective, and if $\mathrm{Ad}(P)$ holds,
then $N_P$ is not an isomorphism.

\end{proposition}

\begin{proof}

Take the types of the structure data to be the one-element set $\{*\}$ and the two-element
set $\{0,1\}$, respectively, and the values of the structure data to be $*$ and $0$,
respectively.
For the quantities, in both objects we take the type to be a one-element set and the value
to be its unique element.
These are objects with the same $C$, but they are not equal, since the cardinalities of the
types of their structure data differ.
Both are normalized to $\mathrm{Form}_P(C)$, so $n_P$ is not injective.
If a morphism of cores had an inverse, its object map would have a two-sided inverse as
well, contradicting the noninjectivity of $n_P$. Hence $N_P$ is not an isomorphism.

\end{proof}

What we treat here is not only the selected representations but the totality of
architecture objects of Definition~\ref{def:1.6}.
This collection contains objects that represent the same configuration with different
structure data.
Collecting that distinction into a single selected object is what produces the
noninvertibility of $N_P$.

\begin{theorem}[Normalization does not split within the category of cores]\label{thm:6.8}

Suppose $\mathrm{Ad}(P)$ holds.
There exist no core $Q$ and morphisms of cores $r:P\to Q$ and $i:Q\to P$ satisfying
\begin{equation*}
ri=1_Q,\qquad ir=N_P
\tag{6.8}\label{eq:6.8}
\end{equation*}

\end{theorem}

\begin{proof}

Suppose they exist, and let the object maps of $r,i$ be $R,I$, respectively.
Since $RI=1$, the map $I$ is injective.
By Lemma~\ref{lem:6.6} and $IR=n_P$, for every object $A$ we obtain
\[
I(n_Q(A))=n_P(I(A))=IR(I(A))=I(A)
\]
Injectivity gives $n_Q(A)=A$, contradicting Proposition~\ref{prop:6.7}.

\end{proof}

For the collections of objects, \eqref{eq:6.2} gave a splitting through the configurations.
A core, however, is not the set of configurations itself: it carries the readings of object
formation, operations, and Laws for all architecture objects. Merely extracting the set of
configurations does not yield the intermediate core of \eqref{eq:6.8}.
To treat the image of a normalization as an object of a category, in the next section we
extend the objects and the morphisms.

\section{Making the remaining information into objects by idempotent completion}\label{sec:6.3}

For an idempotent map on a set, the set of its fixed points can be extracted as its image.
The idempotent completion is what makes it possible to treat objects playing the same role
in a general category.
By this construction, the information that the original comparison loses and the
correspondence preserved between the images can be expressed by the same equations.

\begin{definition}[Karoubi envelope]\label{def:6.9}

An object of the idempotent completion $\mathrm{Kar}(E)$ of a category $E$ is a pair
$(X,e)$ of an object $X\in E$ and an idempotent $e:X\to X$.
A morphism $a:(X,e)\to(Y,d)$ is a morphism $a:X\to Y$ of $E$ satisfying
\begin{equation*}
dae=a
\quad\Longleftrightarrow\quad
da=a=ae
\tag{6.9}\label{eq:6.9}
\end{equation*}
For the composition we use the composition of $E$, and we take the identity morphism of
$(X,e)$ to be $e$.
These form a category, by the category laws of $E$ and \eqref{eq:6.9}.

\end{definition}

\begin{construction}[Embedding of the original category and splitting of idempotents]\label{cons:6.10}

Define $J:E\to\mathrm{Kar}(E)$ by $J(X)=(X,1_X)$ and $J(f)=f$.
It is fully faithful. For any $(X,e)$ there are two morphisms, each of which is $e$ as a
morphism of the original category,
\begin{equation*}
J(X)\xrightarrow{\ r_X\ }(X,e)
\xrightarrow{\ i_X\ }J(X)
\tag{6.10}\label{eq:6.10}
\end{equation*}
and
\begin{equation*}
r_Xi_X=1_{(X,e)},\qquad i_Xr_X=J(e).
\tag{6.11}\label{eq:6.11}
\end{equation*}

\end{construction}

\begin{proof}

Between objects $(X,1_X)$, condition \eqref{eq:6.9} holds for every morphism, so $J$ is
fully faithful.
Checking \eqref{eq:6.9} for $r_X,i_X$, we find that both amount to $e^2=e$.
The underlying morphism of both composites is $e$, which is the identity on $(X,e)$ and is
$J(e)$ on $J(X)$.

\end{proof}

\begin{proposition}[Realization of the image of an idempotent]\label{prop:6.11}

In $\mathrm{Kar}(E)$ every idempotent splits.
Moreover, when there is a splitting in $E$ satisfying $ri=1_Z$ and $ir=e$, we have
$(X,e)\cong J(Z)$.

\end{proposition}

\begin{proof}

An idempotent morphism $a:(X,e)\to(X,e)$ satisfies $a^2=a$ and $ea=a=ae$.
Forming the object $(X,a)$ and taking the underlying morphism in both directions to be $a$
gives its splitting.
For the second claim, we regard the given $r,i$ as morphisms between $(X,e)$ and $J(Z)$.
The equations $re=r$ and $ei=i$ and the two-sided inverse equations follow from the
equations of the splitting.

\end{proof}

Hence $(P,N_P)$ becomes a new object that handles the information retained by the
normalization of a core.
Theorem~\ref{thm:6.8} shows that this is not an object obtained by splitting the same
normalization inside $E_{\mathrm{core}}$.
In the idempotent completion, taking the identity morphism to be $N_P$ specifies the
morphisms appropriate to that image.

\begin{theorem}[Normalization factorization of a comparison and isomorphism of images]\label{thm:6.12}

In an arbitrary category $E$, take an isomorphism $\alpha:X\xrightarrow{\sim}Y$ and an
idempotent $d:Y\to Y$, and set
\begin{equation*}
\beta=d\alpha,\qquad e=\alpha^{-1}d\alpha
\tag{6.12}\label{eq:6.12}
\end{equation*}
Then $e^2=e$, and the following hold.
\begin{equation*}
\begin{gathered}
\beta\text{ is invertible in }E\text{}
\quad\Longleftrightarrow\quad d=1_Y,\\
\beta:(X,e)\xrightarrow{\sim}(Y,d)
\quad\text{in }\mathrm{Kar}(E).
\end{gathered}
\tag{6.13}\label{eq:6.13}
\end{equation*}
The underlying morphism of the inverse between the images is $\gamma=\alpha^{-1}d$.

\end{theorem}

\begin{proof}

The computation with conjugation gives $e^2=\alpha^{-1}d^2\alpha=e$.
Since $\alpha$ is invertible, invertibility of $\beta$ is equivalent to invertibility of
$d$.
For an invertible idempotent, cancelling one $d$ in $d^2=d$ gives $d=1_Y$.
The converse is clear.

Next we check $d\beta e=\beta$ and $e\gamma d=\gamma$ using $d^2=d$.
The two composites are
\begin{equation*}
\gamma\beta=e=1_{(X,e)},\qquad
\beta\gamma=d=1_{(Y,d)}
\tag{6.14}\label{eq:6.14}
\end{equation*}
so $\gamma$ is the inverse morphism in $\mathrm{Kar}(E)$.

\end{proof}

What becomes isomorphic in this result are the images that remain after the normalization.
The morphism used for the isomorphism of the images is the same $\beta$ that was not
invertible in the original category, and what has changed are the objects at the two ends
and their identity morphisms.

\begin{example}[From four points to an image with two points]\label{ex:6.13}

Take the sets $X=\{a,b,c,d\}$ and $Y=\{0,1,2,3\}$, and give the isomorphism $\alpha$ and
the normalization $d_Y$ by the following table.

\begin{xltabular}{\linewidth}{@{}LLLL@{}}
\toprule
$x$ & $\alpha(x)$ & $\beta(x)=d_Y(\alpha(x))$ & $e_X(x)$ \\
\midrule
\endhead
$a$ & $1$ & $0$ & $c$ \\
\addlinespace[3pt]
$b$ & $3$ & $2$ & $d$ \\
\addlinespace[3pt]
$c$ & $0$ & $0$ & $c$ \\
\addlinespace[3pt]
$d$ & $2$ & $2$ & $d$ \\
\bottomrule
\end{xltabular}

Here $d_Y(0)=d_Y(1)=0$, $d_Y(2)=d_Y(3)=2$, and
$e_X=\alpha^{-1}d_Y\alpha$.
Since $\beta$ identifies $a,c$ with each other and $b,d$ with each other, it is not
invertible between the four points.
Between the sets of fixed points $\{c,d\}$ and $\{0,2\}$, on the other hand, it is a
bijection.
Equation~\eqref{eq:6.14} expresses this correspondence of images in a general category.

\end{example}

\section{Classifying the generated comparisons of full geometries}\label{sec:6.4}

In the generated comparison of Chapter~\ref{chap:5}, whether to normalize was determined by
the diagnosis and the preservation conditions, and that endomorphism was carried along the
actual route.
To apply Theorem~\ref{thm:6.12} to this construction, we verify idempotence for morphisms
that include the covers, the overlaps, the coefficients, and the local data, in addition to
the operations and Laws of a core.

\begin{theorem}[Idempotent normalization of a full geometry]\label{thm:6.14}

For a full geometry $G$ whose core $P=p(G)$ satisfies $\mathrm{Ad}(P)$, the endomorphism
$N_G:G\to G$ of Definition~\ref{def:5.39} satisfies
\begin{equation*}
N_G^2=N_G,\qquad p(N_G)=N_P,\qquad
qp(N_G)=1_{qp(G)}
\tag{6.15}\label{eq:6.15}
\end{equation*}
The map to the coefficient ring is the identity.

\end{theorem}

\begin{proof}

Idempotence of the core component is by Theorem~\ref{thm:6.5}.
In the remaining components of a morphism of geometries, the maps of contexts, Atoms, and
axes are identities, the coverage requirements are preserved as they are, and the
comparison of overlaps is the identity isomorphism.
The coefficient homomorphism and the comparison maps of supports, axes, and observables are
identities as well, so the twofold composite equals the single one.
The raw system is preserved by the identity reindexing and the identity change of
coefficients.
Hence all the components of Definition~\ref{def:1.30} agree. The equations for the
projections follow from the construction.

\end{proof}

\begin{lemma}[Commutation of generated transport and normalization]\label{lem:6.15}

The canonical transport and pullback of Chapters~\ref{chap:4} and~\ref{chap:5}
along an exact base morphism $\sigma:b\to b'$ preserve $\mathrm{Ad}$.
Moreover, for admissible cores and full geometries,
\begin{equation*}
\sigma_!(N_P)=N_{\sigma_!P},\qquad
\sigma^*(N_Q)=N_{\sigma^*Q},
\tag{6.16}\label{eq:6.16}
\end{equation*}

The same equations hold for full geometries with $G,H$ in place of $P,Q$.
The actual generated lifts $\iota_{\sigma,P}:P\to\sigma_!P$ and
$\kappa_{\sigma,Q}:\sigma^*Q\to Q$ satisfy
\begin{equation*}
N_{\sigma_!P}\iota_{\sigma,P}=\iota_{\sigma,P}N_P,
\qquad
N_Q\kappa_{\sigma,Q}=\kappa_{\sigma,Q}N_{\sigma^*Q}
\tag{6.17}\label{eq:6.17}
\end{equation*}

and the same equations hold for the lifts of geometries.

\end{lemma}

\begin{proof}

Write $e$ for the Atom map of the base morphism and $T_e$ for the reindexing of objects.
From the definition of transported object formation,
\begin{equation*}
n_{\sigma_!P}(T_eA)=T_e(n_P(A))
\tag{6.18}\label{eq:6.18}
\end{equation*}

Since $T_e$ has inverse $T_{e^{-1}}$, the preservation laws for every object
after transport can be brought back to those for objects before transport.
Carrying the invariance of residuals, invariants, and signatures through this
correspondence gives the same conditions after transport.
The equality of operation types and the commutativity of configuration maps are
carried by the same reindexing. This is preservation of $\mathrm{Ad}$.
For pullback, use $e^{-1}$.

The operation map of a generated lift is built from reindexing and identification
by equality of the endpoint types. Moving the normalization identification before
or after it yields the same original operation once the endpoints are aligned by
\eqref{eq:6.18}. The other core components are identity maps or the same reindexing,
giving \eqref{eq:6.17}. For geometries, the identity local comparisons of
Theorem~\ref{thm:6.14} are composed before and after, so the same equations hold
in all components. Finally, by the universal property of the lifts, the vertical
morphism satisfying \eqref{eq:6.17} is unique, giving \eqref{eq:6.16}.

\end{proof}

This lemma concerns transport and pullback generated from the same reindexing.
For general morphisms having their own operation maps beyond objects,
\S\ref{sec:6.6} examines the interchange condition.

\begin{theorem}[Normalization factorization of the generated comparison and classification of invertibility]\label{thm:6.16}

Fix the same exact pointed pullback square, face $z$ of the comparison diagram,
cochain $\omega$, source core $P_z$, commutative coefficient ring $k$, and
full geometry $G_z$ as in Constructions~\ref{cons:5.37} and~\ref{cons:5.40}.
We abbreviate the ends of the two routes and the canonical comparison as
\begin{equation*}
X=\overline D(G_z),\qquad Y=\overline V(G_z),\qquad
\overline\alpha=\overline\alpha_z:X\xrightarrow{\sim}Y
\tag{6.19}\label{eq:6.19}
\end{equation*}
The condition for selecting the normalization and the generated morphisms are taken to be
\begin{equation*}
\begin{aligned}
\chi&\Longleftrightarrow
   (\omega_z\ne1)\ \text{and}\ \mathrm{Ad}(P_z),\\
\overline d&=\overline V(\overline S_{z,\omega}),\qquad
\overline\beta=\overline d\,\overline\alpha,\qquad
\overline e=\overline\alpha^{-1}\overline d\,\overline\alpha
\end{aligned}
\tag{6.20}\label{eq:6.20}
\end{equation*}
Then $\overline d^2=\overline d$ and $\overline e^2=\overline e$, and
\begin{equation*}
\overline\beta\text{ is invertible in }E_{\mathrm{geom}}\text{}
\quad\Longleftrightarrow\quad
\overline d=1_Y
\quad\Longleftrightarrow\quad
\neg\chi.
\tag{6.21}\label{eq:6.21}
\end{equation*}
Moreover, the same morphism $\overline\beta$ gives an isomorphism
\begin{equation*}
(X,\overline e)\xrightarrow{\ \overline\beta\ }(Y,\overline d)
\quad\text{in }\mathrm{Kar}(E_{\mathrm{geom}})
\tag{6.22}\label{eq:6.22}
\end{equation*}
and the underlying morphism of its inverse is $\overline\alpha^{-1}\overline d$.
When $\chi$ holds, the cores at both ends satisfy $\mathrm{Ad}$ as well, and
\begin{equation*}
\overline d=N_Y,\qquad \overline e=N_X.
\tag{6.23}\label{eq:6.23}
\end{equation*}
Hence, when the normalization is selected, the same generated comparison matches the
normalization images $(X,N_X)$ and $(Y,N_Y)$, determined at the respective ends of the
two routes, by an isomorphism.
The idempotent at the start was defined by conjugation with the comparison, and we see
further that it agrees with the normalization $N_X$ of the start itself.
When $\chi$ does not hold, both idempotents are identities.

\end{theorem}

\begin{proof}

The endomorphism on the source in Construction~\ref{cons:5.40} is $N_{G_z}$ if $\chi$ holds
and the identity otherwise.
By Theorem~\ref{thm:6.14} and functoriality, its image $\overline d$ is idempotent.
Since $\overline\alpha$ of Construction~\ref{cons:5.37} is an isomorphism,
Theorem~\ref{thm:6.12} gives the first equivalence of \eqref{eq:6.21}, the idempotence of
$\overline e$, and \eqref{eq:6.22}.

Suppose $\chi$ holds. Applying Lemma~\ref{lem:6.15} to the transports and pullbacks along
the four edges gives $\mathrm{Ad}$ at both ends and $\overline d=N_Y$.
Furthermore, into the geometric version of \eqref{eq:5.18}, which characterizes the
canonical comparison,
\begin{equation*}
\kappa_{\sigma_2,(\sigma_1)_!G_z}\,
\overline\alpha\,\iota_{\pi_2,\pi_1^*G_z}
=\iota_{\sigma_1,G_z}\,\kappa_{\pi_1,G_z}
\tag{6.24}\label{eq:6.24}
\end{equation*}
we substitute \eqref{eq:6.17} for each lift.
After composing with the lifts on both outer sides, $N_Y\overline\alpha$ and
$\overline\alpha N_X$ become the same morphism.
Using in turn the uniqueness of each lift for cores and for geometries gives
$N_Y\overline\alpha=\overline\alpha N_X$, which is the second equation of \eqref{eq:6.23}.

By Proposition~\ref{prop:6.7}, $N_{p(Y)}$ is not the identity.
Since $p(N_Y)=N_{p(Y)}$, we have $\overline d=N_Y\ne1_Y$.
If, on the other hand, $\neg\chi$, then both the endomorphism on the source and its image
are identities, and the conjugated $\overline e$ is the identity as well. This gives the
remaining equivalence of \eqref{eq:6.21}.

\end{proof}

Equation~\eqref{eq:6.21} is a classification for the generation rule chosen in
Construction~\ref{cons:5.40}.
Even when the diagnosis is not the identity, this rule selects the identity morphism unless
the preservation conditions for the normalization are satisfied.
Moreover, selecting a normalization and removing the value of the diagnosis are
separate statements.
What this theorem clarifies is the invertibility and the image of the comparison generated
from that selection.

\begin{proposition}[Projection to the comparison of cores]\label{prop:6.17}

For the input of Theorem~\ref{thm:6.16} we write $\alpha=\alpha_{P_z}$,
$\beta=\beta_{z,\omega}$, and $e=\alpha^{-1}E_{z,\omega}\alpha$.
Under the endpoint isomorphisms $j_D:pX\xrightarrow{\sim}DP_z$ and
$j_V:pY\xrightarrow{\sim}VP_z$ of Proposition~\ref{prop:5.38},
\begin{equation*}
\begin{aligned}
j_Vp(\overline\alpha)&=\alpha j_D,&
j_Vp(\overline d)&=E_{z,\omega}j_V,\\
j_Dp(\overline e)&=e j_D,&
j_Vp(\overline\beta)&=\beta j_D.
\end{aligned}
\tag{6.25}\label{eq:6.25}
\end{equation*}
Hence, projecting the isomorphism of images of geometries to the core gives the isomorphism
of images constructed at the core.

\end{proposition}

\begin{proof}

The first two equations and the last one are \eqref{eq:5.44} and \eqref{eq:5.48}.
The remaining equation is obtained by projecting
$\overline e=\overline\alpha^{-1}\overline d\,\overline\alpha$, using that $p$ preserves
composites and inverses.
In the idempotent completion, the underlying morphisms of the isomorphisms that align the
endpoints are $e j_D$ and $E_{z,\omega} j_V$, respectively.
By \eqref{eq:6.25} and idempotence, these define isomorphisms between the images and make
the square of the comparisons commute.

\end{proof}

\begin{example}[A case in which normalization is selected]\label{ex:6.18}

We prepare an exchange of axes to produce a nonidentity cochain and a normalization
that merges object representations over the same configuration.
The exchange of axes is used to satisfy the selection condition of Construction~\ref{cons:5.40}.
The selected normalization merges object representations and leaves the axes unchanged.
First, define the core data as follows.

\begin{itemize}
\item \textbf{Atoms and sources.} We take one Atom and one source, let the extracted family
be that one-element set of Atoms, and take the source normalization to be the identity.
\item \textbf{Composition and object formation.} We set
$\mathrm{Comp}(F)=(F,\varnothing,\varnothing)$.
For every configuration we take the types of the structure data and of the quantity to be
one-element sets and their values to be the unique elements.
\item \textbf{Operations.} We set $\mathrm{Op}_P(A,B)=\mathrm{Hom}(C_A,C_B)$.
\item \textbf{Signatures.} We take two axes and let both values be the integer 0. Both axes
are taken to be selected axes.
\item \textbf{Contexts and local readouts.} A context is a pair $(U,T)$ of a subset $U$ of
the extracted family and a subset $T$ of the signature axes, and the morphisms are the
componentwise inclusions. The support reads each Atom of $U$ as it is, and all of $T$ can
be read as axes.
The observables form a one-element set that can be read, and for the restrictions we use
the inclusions on supports and axes and the identity on observables.
\item \textbf{Equations and invariants.} We take both index sets to be empty, the ring of
the equations to be the constant $\mathbb{Z}$, and the restrictions to be identities.
Since the indices are empty, the conditions on the residuals and the specification of the
detectors are empty.
\end{itemize}

This core $P$ satisfies $\mathrm{Ad}(P)$.
The morphism $\tau:P\to P$ that exchanges the two axes and fixes the other
components is an automorphism with $\tau^2=1$ and $\tau\ne1$.
Take an identity square of bases and a comparison diagram with one specified face,
and choose $\omega_z=\tau$.
Identifying each endpoint with $P$ by the unit comparisons of identity reindexing,
the two routes and the canonical comparison become identities.
The generation rule selects $N_P$, so $\beta=N_P$ is not invertible.

For the full geometry we add the following data.

\begin{itemize}
\item \textbf{Overlaps and covers.} On the contexts above we take the overlaps to be the
componentwise intersections and the coverage requirements to be empty.
\item \textbf{Coefficients and the raw system.} We take the coefficients to be $\mathbb{Z}$,
and use a single raw coordinate $x$ for each context together with the relation $x^2-x$.
We take the kind of the coordinates to be semantic, the type of the local data to be a
one-element set, and all the restrictions to be identities.
\end{itemize}

The restrictions preserve the relations and respect identities and composition, so a full
geometry $G$ is obtained.
Aligning the geometry endpoints by the same unit comparisons gives
$\overline\beta=N_G$. This is not invertible in $E_{\mathrm{geom}}$, but is
the identity morphism of $(G,N_G)$.

\end{example}

\section{Collecting normalized objects and changes into a single category}\label{sec:6.5}

Having constructed the image for a single comparison, we next treat several cores and the
changes between them at the same time.
The aim is to build the morphisms between the images consistently from the original
morphisms, rather than to choose them anew at each design change.

\begin{lemma}[One-sided absorption law]\label{lem:6.19}

Let $\mathcal{C}_{\mathrm{ad}}$ be the full subcategory of cores satisfying
$\mathrm{Ad}$. For every morphism $f:P\to Q$ of this subcategory,
\begin{equation*}
N_QfN_P=fN_P.
\tag{6.26}\label{eq:6.26}
\end{equation*}

\end{lemma}

\begin{proof}

Let $F$ be the object map. By Lemma~\ref{lem:6.6},
$n_QFn_P=Fn_P^2=Fn_P$.
An operation $o\in\mathrm{Op}_P(A,B)$ is first identified with one at the normalized
endpoints and is then sent by the operation map of $f$.
Its target objects $F(n_P(A)),F(n_P(B))$ are already fixed points by the equality of object
maps just obtained.
Hence the operation map of the final $N_Q$ is the identity identification on the same type.
For the components of the base, Atoms, contexts, equations, invariant indices, axes, and
value ranges, only an identity is composed at the end.
All the components of the morphisms are equal, so we obtain \eqref{eq:6.26}.

\end{proof}

\begin{construction}[The category of normalized cores]\label{cons:6.20}

For the objects of $\mathcal{C}_{\mathrm{nor}}$ we use the same cores as in
$\mathcal{C}_{\mathrm{ad}}$. Its morphisms and identity morphisms are defined by
\begin{equation*}
\begin{aligned}
\mathrm{Hom}_{\mathcal{C}_{\mathrm{nor}}}(P,Q)
 &=\{a:P\to Q\text{ in }\mathcal{C}_{\mathrm{ad}}\mid N_QaN_P=a\},\\
1_P^{\mathrm{nor}}&=N_P
\end{aligned}
\tag{6.27}\label{eq:6.27}
\end{equation*}
The composition is the composition of the original morphisms of cores.
By \eqref{eq:6.9}, this is the category whose morphisms are all the Karoubi morphisms between the objects
$(P,N_P)$.
Hence
\begin{equation*}
K:\mathcal{C}_{\mathrm{nor}}\longrightarrow\mathrm{Kar}(\mathcal{C}_{\mathrm{ad}}),
\qquad K(P)=(P,N_P),\quad K(a)=a
\tag{6.28}\label{eq:6.28}
\end{equation*}
is a fully faithful functor. Even though the objects are denoted by the same symbol $P$,
the identity morphisms and the sets of morphisms have changed.

\end{construction}

\begin{theorem}[Fullness of the normalization functor]\label{thm:6.21}

The following assignment defines a full functor.
\begin{equation*}
\begin{gathered}
\mathsf{N}_{\mathrm{core}}:\mathcal{C}_{\mathrm{ad}}\longrightarrow\mathcal{C}_{\mathrm{nor}},\\
\mathsf{N}_{\mathrm{core}}(P)=P,\qquad
\mathsf{N}_{\mathrm{core}}(f)=fN_P.
\end{gathered}
\tag{6.29}\label{eq:6.29}
\end{equation*}

\end{theorem}

\begin{proof}

By \eqref{eq:6.26} and $N_P^2=N_P$, the morphism $fN_P$ is a morphism of \eqref{eq:6.27}.
The identity morphism is sent to $1_PN_P=N_P=1_P^{\mathrm{nor}}$.
For $f:P\to Q$ and $g:Q\to R$,
\begin{equation*}
(gN_Q)(fN_P)=g(N_QfN_P)=gfN_P
\tag{6.30}\label{eq:6.30}
\end{equation*}
so the composition is preserved.
Every morphism $a:P\to Q$ in $\mathcal{C}_{\mathrm{nor}}$ satisfies $aN_P=a$ by
\eqref{eq:6.9}.
Choosing the underlying morphism of cores $a$ itself as a preimage, we have
$\mathsf{N}_{\mathrm{core}}(a)=a$.

\end{proof}

Fullness is the assertion that every morphism permitted after the normalization has an
original morphism of cores.
Distinct morphisms of cores can have the same image, and \S\ref{sec:6.6} gives an example.

\begin{proposition}[Extension to full geometries and compatibility with the projection]\label{prop:6.22}

Let $\mathcal{G}_{\mathrm{ad}}$ be the full subcategory of full geometries whose cores
satisfy $\mathrm{Ad}$. For every morphism $f:G\to H$,
\begin{equation*}
N_HfN_G=fN_G.
\tag{6.31}\label{eq:6.31}
\end{equation*}
Hence we obtain the category $\mathcal{G}_{\mathrm{nor}}$ whose morphisms are defined by
$N_HaN_G=a$, and the full functor $\mathsf{N}_{\mathrm{geom}}(f)=fN_G$.
Writing $p_{\mathrm{ad}},p_{\mathrm{nor}}$ for the functors induced by the projection to
cores,
\begin{equation*}
p_{\mathrm{nor}}\mathsf{N}_{\mathrm{geom}}
=\mathsf{N}_{\mathrm{core}}p_{\mathrm{ad}}.
\tag{6.32}\label{eq:6.32}
\end{equation*}
Moreover, the base and the coefficient homomorphism of a morphism after the normalization
are equal to those of the original morphism.

\end{proposition}

\begin{proof}

For the core component we use Lemma~\ref{lem:6.19}.
In the components of the geometry, the final $N_H$ composes identity maps onto the
coefficients, supports, axes, and observables, so the result equals the original $fN_G$.
This gives \eqref{eq:6.31}.
Functoriality and fullness are the same computation as in Theorem~\ref{thm:6.21}.
By $p(N_G)=N_{p(G)}$, equation \eqref{eq:6.32} holds both on objects and on morphisms.
For the base and the coefficients, the components of $N_G$ are identities, so composing
with them does not change the original morphism.

\end{proof}

\section{Conditions for interchanging the order of a normalization and a change}\label{sec:6.6}

That the normalization is a functor does not mean that the order of an arbitrary change and
the normalization can be interchanged.
What the former required was \eqref{eq:6.26}, and the latter is a stronger equation.
We examine in which component that difference appears.

\begin{proposition}[Naturality of the inclusion and the condition for the projection]\label{prop:6.23}

We use $K$ of Construction~\ref{cons:6.20} and the embedding $J$ obtained by applying
Construction~\ref{cons:6.10} to $\mathcal{C}_{\mathrm{ad}}$.
The inclusion $i_P:(P,N_P)\to J(P)$ whose underlying morphism is $N_P$ defines a natural
transformation
\begin{equation*}
i:K\mathsf{N}_{\mathrm{core}}\Longrightarrow J
\tag{6.33}\label{eq:6.33}
\end{equation*}
The morphism $r_P:J(P)\to(P,N_P)$ in the opposite direction satisfies
$r_Pi_P=1_{(P,N_P)}$ objectwise, and naturality of that family with respect to a morphism
$f:P\to Q$ is equivalent to
\begin{equation*}
fN_P=N_Qf
\tag{6.34}\label{eq:6.34}
\end{equation*}

\end{proposition}

\begin{proof}

The underlying morphisms of the two sides of the naturality of the inclusion are $fN_P$ and
$N_QfN_P$, which are equal by Lemma~\ref{lem:6.19}.
The objectwise splitting is by \eqref{eq:6.11}.
Naturality of the family in the opposite direction is the equation $(fN_P)N_P=N_Qf$, which
becomes \eqref{eq:6.34} on using $N_P^2=N_P$.

\end{proof}

\begin{proposition}[Deciding naturality by the correspondence of operations]\label{prop:6.24}

Suppose $\mathrm{Ad}(P)$ and $\mathrm{Ad}(Q)$ hold, and let $F$ be the object map of
$f:P\to Q$.
We write the operation map of the normalization as
\begin{equation*}
c^P_{A,B}:\mathrm{Op}_P(A,B)
\longrightarrow\mathrm{Op}_P(n_P(A),n_P(B))
\tag{6.35}\label{eq:6.35}
\end{equation*}
Then \eqref{eq:6.34} is equivalent to the statement that, for all $A,B,o$,
\begin{equation*}
f^{\mathrm{op}}_{n_P(A),n_P(B)}(c^P_{A,B}(o))
=c^Q_{F(A),F(B)}(f^{\mathrm{op}}_{A,B}(o))
\tag{6.36}\label{eq:6.36}
\end{equation*}
holds. The endpoints of the two sides are identified by Lemma~\ref{lem:6.6}.

\end{proposition}

\begin{proof}

Equation~\eqref{eq:6.36} is the equality of the operation components of two composite
morphisms.
The object components are equal by Lemma~\ref{lem:6.6}, and the components of the base,
Atoms, contexts, equations, invariant indices, axes, and value ranges are equal because the
identity maps of the normalization are composed either before or after.
Hence agreement of the operation components is equivalent to agreement of the entire
morphisms.

\end{proof}

In \eqref{eq:6.36} we compare not only the action of an operation on configurations but the
operation itself.
As in Example~\ref{ex:1.8} of Chapter~\ref{chap:1}, operations with the same action can
differ in their names or in additional information.

\begin{example}[The order changes according to the tag of an operation]\label{ex:6.25}

We replace the operations of the core of Example~\ref{ex:6.18} by
\begin{equation*}
\mathrm{Op}_{P^{\mathrm{tag}}}(A,B)
=\mathrm{Hom}(C_A,C_B)\times\{0,1\}
\tag{6.37}\label{eq:6.37}
\end{equation*}
For the action on configurations we use only the first component, and the other readings
are kept as they are.
The second component is a tag that distinguishes two operations with the same action.
This core also satisfies $\mathrm{Ad}$, and the normalization keeps the configuration map
and the tag as they are.

Over a configuration $C$, let $A_\bullet$ be an object whose type of structure data is a
two-element set.
The type of the structure data of its selected object $n(A_\bullet)$ is a one-element set,
so $n(A_\bullet)\ne A_\bullet$.
Keeping the objects, the base, and the other readings identical, we apply to the operations
alone
\begin{equation*}
f^{\mathrm{op}}_{A,B}(u,t)=
\begin{cases}
(u,1-t),&A=A_\bullet,\\
(u,t),&A\ne A_\bullet
\end{cases}
\tag{6.38}\label{eq:6.38}
\end{equation*}
Since the first component $u$ is unchanged, the commutative square for configurations in
Definition~\ref{def:1.28} is satisfied.
The other components are identities as well, and $f$ is an actual endomorphism of cores.
Applying it twice returns to the original, so $f^2=1$ as well.

Taking the start and the end to be $A_\bullet$ and the operation to be $(1_C,0)$, the
following difference arises.

\begin{xltabular}{\linewidth}{@{}LLL@{}}
\toprule
Order of application & Start and end after normalization & Final tag \\
\midrule
\endhead
Normalize and then apply $f$ & Both $n(A_\bullet)$ & 0 \\
\addlinespace[3pt]
Apply $f$ and then normalize & Both $n(A_\bullet)$ & 1 \\
\bottomrule
\end{xltabular}

Hence $fN\ne Nf$, and the family in the opposite direction in Proposition~\ref{prop:6.23}
is not natural.
The object map and the action on configurations agree along both routes.
What detects the difference of order is the tag kept in the operations.

Furthermore, all the targets of the normalization have one-element structure data, so none
of them is $A_\bullet$.
From this, $fN=N=1N$.
Two distinct morphisms $f\ne1$ are sent to the same morphism after the normalization, so
the full functor of Theorem~\ref{thm:6.21} is not faithful in general.

Read in terms of design changes, even when the correspondence of operations after the
alignment to the standard representation is the same, the specification of operations that
depends on the original representation can differ.
For example, for the same cancellation operation one may consider processing that changes
the label for auditing according to the original implementation variant.
If the distinction of implementation variants is collected first, the operations whose
labels are to be changed can no longer be distinguished.
When one wants to interchange the order before and after the normalization freely, a review
checks \eqref{eq:6.36} in addition to the correspondence of objects.

\end{example}

\section{The category of comparisons themselves and the three-level projections}\label{sec:6.7}

Chapter~\ref{chap:7} studies the changes that preserve a single comparison.
For this it is convenient to take the comparison morphisms themselves as objects and to
treat the changes at the two ends as morphisms.
The idempotent completion is compatible with this category of comparisons as well.

\begin{definition}[The arrow category and the category of comparisons]\label{def:6.26}

The arrow category $\mathrm{Arr}(E)$ of a category $E$ has the morphisms $c:X\to Y$ as its
objects.
A morphism from $c$ to $c':X'\to Y'$ is a pair of $a:X\to X'$ and $b:Y\to Y'$ satisfying
\begin{equation*}
bc=c'a
\tag{6.39}\label{eq:6.39}
\end{equation*}
Identities and composition are defined componentwise at the two ends.
We take the category that handles the comparisons after normalization to be
\begin{equation*}
\mathcal{M}(E)=\mathrm{Arr}(\mathrm{Kar}(E))
\tag{6.40}\label{eq:6.40}
\end{equation*}

\end{definition}

\begin{theorem}[Interchange of comparison and idempotent completion]\label{thm:6.27}

For every category $E$ there is a natural equivalence of categories
\begin{equation*}
\Phi_E:\mathrm{Kar}(\mathrm{Arr}(E))
\simeq\mathrm{Arr}(\mathrm{Kar}(E))
\tag{6.41}\label{eq:6.41}
\end{equation*}
Representing an object of the left-hand side by a morphism $c:X\to Y$ and an idempotent
commutative square $(e,d)$, that is, by $e^2=e$, $d^2=d$, and $dc=ce$, the functor $\Phi_E$
sends it to
\begin{equation*}
(X,e)\xrightarrow{\ dce\ }(Y,d)
\tag{6.42}\label{eq:6.42}
\end{equation*}
This correspondence is shown in Figure~\ref{fig:6.1}.

\end{theorem}

\begin{figure}[htbp]
\centering
\begin{tikzpicture}[
    x=1mm, y=1mm,
    arr/.style={-{Stealth[length=4.5pt]}, semithick},
    mapsto/.style={{Bar[width=5pt]}-{Stealth[length=4.5pt]}, semithick},
    panel/.style={font=\small},
    note/.style={font=\small}
  ]
  \node[panel] at (15,31) {Idempotent commutative square};
  \node (XT) at (0,22)  {$X$};
  \node (YT) at (30,22) {$Y$};
  \node (XB) at (0,0)   {$X$};
  \node (YB) at (30,0)  {$Y$};
  \draw[arr] (XT) -- (YT) node[midway, above] {$c$};
  \draw[arr] (XB) -- (YB) node[midway, below] {$c$};
  \draw[arr] (XT) -- (XB) node[midway, left]  {$e$};
  \draw[arr] (YT) -- (YB) node[midway, right] {$d$};
  \node[note] at (15,-10) {$e^2 = e$, \ $d^2 = d$};
  \node[note] at (15,-17) {$dc = ce$};
  \draw[mapsto] (44,11) -- (60,11) node[midway, above] {$\Phi_E$};
  \node[panel] at (95,31) {Comparison between images};
  \node (L) at (75,11)  {$(X,e)$};
  \node (R) at (115,11) {$(Y,d)$};
  \draw[arr] (L) -- (R) node[midway, above] {$dce$};
  \node[note] at (95,-10) {$d(dce)e = dce$};
\end{tikzpicture}
\caption{\textbf{From an idempotent change of a comparison to a comparison between the images.}
On the left we require $e^2=e$, $d^2=d$, and the commutativity $dc=ce$.
On the right the two ends are changed to the idempotent images, and the comparison between
them is $dce$.
In general $dce=c$ need not hold, so the round trip in the other direction returns to the
original by the isomorphism in the proof.}\label{fig:6.1}
\end{figure}

\begin{proof}

From $dc=ce$ and idempotence, $dce$ satisfies \eqref{eq:6.9}.
Let $a,b$ be the two ends of a morphism of the left-hand side. In addition to the original
commutative square, $e'ae=a$ and $d'bd=b$ hold.
Hence, regarding the same $a,b$ as Karoubi morphisms,
$b(dce)=(d'c'e')a$ holds.
Identities and composition are the same computation at the two ends, so this gives a
functor.

In the opposite direction, from a Karoubi morphism $u:(X,e)\to(Y,d)$ we take the underlying
morphism $u:X\to Y$ and the idempotent square $(e,d)$. The square commutes by $du=u=ue$.
For morphisms as well we take the underlying morphisms at the two ends.
Applying $\Phi_E$ after this returns to the original, since $due=u$.

In the round trip in the other direction, the comparison of $(c,e,d)$ changes to $dce$.
Placing the squares whose two ends are $(e,d)$ in both directions, we obtain an
isomorphism inside the idempotent completion.
The reason is that the two ends of both composites are $(e,d)$, which are the identity
morphisms there.
Naturality of this isomorphism also follows from the equations $e'ae=a$ and $d'bd=b$ at the
two ends of a morphism.

\end{proof}

\begin{proposition}[Compatibility with functors and projections]\label{prop:6.28}

A functor $F:E\to E'$ induces $\mathrm{Kar}(F)$ by $(X,e)\mapsto(FX,F(e))$ and its action
on morphisms.
This action and \eqref{eq:6.41} commute under a natural isomorphism.
In particular, the three-level projections induce functors
\begin{equation*}
\mathcal{M}(E_{\mathrm{geom}})
\longrightarrow\mathcal{M}(E_{\mathrm{core}})
\longrightarrow\mathcal{M}(B)
\tag{6.43}\label{eq:6.43}
\end{equation*}
and the projection obtained by following two levels agrees with the composite projection.

\end{proposition}

\begin{proof}

A functor preserves idempotence and \eqref{eq:6.9}.
The comparison obtained by applying the functor to \eqref{eq:6.42} is $F(dce)$, and the
comparison obtained by sending each component first is $F(d)F(c)F(e)$. The two are equal
because a functor preserves composites.
The endpoints and the components at the two ends of a morphism are the same as well, and we
obtain a natural isomorphism whose components are the identity morphisms at the endpoints.
Carrying out the same computation of components for the composite of two functors yields
the last claim.

\end{proof}

\begin{construction}[Invertible changes of comparisons]\label{cons:6.29}

Keeping all the objects of $\mathcal{M}(E)$ and taking as morphisms only the isomorphisms
between them yields the maximal groupoid.
We do not ask for invertibility of the comparison $c$ itself, which serves as an object.
What is required to be invertible are the changes at the two ends that match the
comparisons.
Indeed, an isomorphism of the arrow category is equivalent to a commutative square whose
two ends are isomorphisms.
One direction follows by evaluating at the two ends, and the other follows because the
inverse morphisms at the two ends form the inverse commutative square.

There are two ways of reading the generated comparison $\beta$ as an object.
They are $J(\beta):J(X)\to J(Y)$, the embedding of the original morphism, and
$\beta:(X,e)\to(Y,d)$, the morphism between the images of Theorem~\ref{thm:6.12}.
Even when the former is not invertible, the latter is an isomorphism.
By making these endpoints explicit, Chapter~\ref{chap:7} can specify ``which change
preserves which comparison.''

\end{construction}

\section{What remains in the image for lenses and protocols}\label{sec:6.8}

We examine whether reads, updates, and executions can be continued with the remaining
states after the states have been collected by an idempotent.
For fixed-view lenses and protocols with observations, the image of a semantics-preserving
idempotent can be realized inside each semantics.

\begin{proposition}[Splitting of the idempotents of fixed-view lenses]\label{prop:6.30}

The category $\mathrm{Lens}(V,v_0)$ of Definition~\ref{def:1.32} is idempotent complete.
That is, every idempotent semantics-preserving morphism $h:L\to L$ splits within the same
category.

\end{proposition}

\begin{proof}

By Propositions~\ref{prop:1.33} and~\ref{prop:1.34}, under the product decomposition
$C\cong V\times K_L$ of a lens $L$, an endomorphism can be written as
\[
h(v,k)=(v,t(k)),\qquad t:K_L\to K_L
\]
If $h^2=h$, then $t^2=t$, so we take
\[
K_t=\{k\in K_L\mid t(k)=k\},\qquad C_t=V\times K_t
\]
and take the read to be the first projection and the update to be
$((v,k),w)\mapsto(w,k)$.
Under the product decomposition, setting $r:C\to C_t$ to be $r(v,k)=(v,t(k))$ and
$i:C_t\to C$ to be the inclusion $i(v,k)=(v,k)$, both morphisms preserve reads and updates.
By idempotence $t(k)\in K_t$, and at fixed points $t(k)=k$, so we obtain
$ri=1_{C_t}$ and $ir=h$.
The reference fiber is in one-to-one correspondence with $K_t$, a subset of the finite set
$K_L$, so this is a splitting within the same category of lenses.

\end{proof}

We check the splitting of Proposition~\ref{prop:6.30} on a lens with four states.
We take the product lens of the two-element sets $V=R=\{0,1\}$, with states $C=V\times R$.
We take the reference view to be 0, the read to be $g(v,r)=v$, and the update to be
$p((v,r),w)=(w,r)$.
The map that aligns the second component to 0,
\begin{equation*}
h(v,r)=(v,0)
\tag{6.44}\label{eq:6.44}
\end{equation*}
defines an endomorphism of $\mathrm{Lens}(V,0)$. Indeed,
\begin{equation*}
gh=g,\qquad
h(p((v,r),w))=(w,0)=p(h(v,r),w),\qquad h^2=h.
\tag{6.45}\label{eq:6.45}
\end{equation*}
It forgets the distinction of the second component while preserving reads and updates.
The state set of the fixed points $C_0=V\times\{0\}$ is closed under updates and becomes a
lens by the same equations.
Taking $C\to C_0$ to be $h$ and the map in the opposite direction to be the inclusion, and
using the identity on views in both cases, we see that this idempotent splits within the
category of
lenses.

Hence the image obtained by applying Theorem~\ref{thm:6.12} to the category of lenses can,
in this example, be expressed as the actual lens $C_0\to V$.
Processing that uses only reads and updates can be continued on it, while processing that
asks for the original second component needs the original states.

In this way, for lenses over a fixed view, the image of a normalization can be realized
again as a lens.
For the normalization of cores treated in Theorem~\ref{thm:6.8}, what was required of an intermediate object was a core carrying the readings for all
the objects, together with its
morphisms.
Even in the same discussion of idempotents, the conclusion changes according to the
category in which the image is realized.

\begin{proposition}[Images of protocols preserving executions]\label{prop:6.31}

Fix the category of paths $\mathcal{C}_Q$ of Definition~\ref{def:1.36} and the observation
targets $O$, and take a realization $F:\mathcal{C}_Q\to{\mathbf{Set}}$ and an observation
$o_F:F\Rightarrow O$.
The state set at each vertex is taken to be finite.
Suppose a semantics-preserving adapter $n:F\Rightarrow F$ satisfies $n^2=n$.
In particular, we are assuming naturality with respect to all executions and preservation
of observations, $o_Fn=o_F$.
At each vertex we set
\begin{equation*}
F_n(x)=\{s\in F(x)\mid n_x(s)=s\}
\tag{6.46}\label{eq:6.46}
\end{equation*}
Restricting the executions and the observations then yields a realization $(F_n,o_{F_n})$
of the same protocol.
The maps $r_x(s)=n_x(s)$ and the inclusions $i_x$ are adapters preserving executions and
observations, and
\begin{equation*}
ri=1_{F_n},\qquad ir=n.
\tag{6.47}\label{eq:6.47}
\end{equation*}

\end{proposition}

\begin{proof}

For an execution $a:x\to y$ and a fixed point $s\in F_n(x)$, naturality gives
\begin{equation*}
n_y(F(a)(s))=F(a)(n_x(s))=F(a)(s)
\tag{6.48}\label{eq:6.48}
\end{equation*}
so the result after the execution again belongs to the fixed points.
For identities and composition it suffices to restrict the functor laws of $F$.
Naturality of $r$ is naturality of $n$, and naturality of $i$ is by the definition of the
restriction.
Setting the observation to be $o_{F_n}=o_Fi$, we have $o_{F_n}r=o_Fn=o_F$, so $r$ preserves
observations as well.
Each set of fixed points is a subset of the original finite set, so finiteness is preserved
as well.
Equation~\eqref{eq:6.47} follows from the definition of fixed points and idempotence.

\end{proof}

Having an idempotent map of states at each vertex alone does not give \eqref{eq:6.48}.
For example, realize a protocol with one vertex, one operation, and no declared relations
by two states and a constant observation.
Take the operation to be the exchange of 0 and 1, and the normalization to be the constant
map to 0.
Then the execution from the fixed point 0 goes out to 1.
In this case the normalization of states and the operation do not commute.

Equations \eqref{eq:6.45} for lenses and \eqref{eq:6.48} for protocols give the condition
under which operations can be continued with only the information retained after the
normalization.
Given a semantics-preserving isomorphism $\alpha$ and such an idempotent adapter $d$, the
morphism $d\alpha$ of Theorem~\ref{thm:6.12} matches the images at the two ends by an
isomorphism.
The reason why Chapter~\ref{chap:1} kept the names and the actions of operations separately
also appears here.
By checking the correspondence of the selected operations in addition to the agreement of
the actions on states and configurations, one can make clear which meaning the
normalization preserves.

\section*{Summary of the Chapter}

In this chapter we analyzed the generated comparison through its normalization factor and
clarified its invertibility and the structure that remains in the image.

\begin{itemize}
\item \textbf{Normalization and the information retained.} The map to the selected object
for each configuration is idempotent, and the fixed points are in one-to-one correspondence
with the configurations.
A readout unchanged by normalization factors uniquely through the configurations.
Under the conditions preserving operations, residuals, invariants, and signatures, this map
becomes an idempotent of full geometries (\S\S\ref{sec:6.1}--\ref{sec:6.2},
\S\ref{sec:6.4}).
\item \textbf{Invertibility and image of the generated comparison.} The generated
comparison, obtained by composing an invertible canonical comparison with a normalization,
is invertible if and only if the normalization factor is the identity.
The same comparison morphism gives an isomorphism between the images in the idempotent
completion even when the normalization is not the identity, and is compatible with the
projection to cores (\S\S\ref{sec:6.3}--\ref{sec:6.4}).
\item \textbf{Normalized changes.} The objects and morphisms after normalization form a category, and the normalization functor obtained from the original changes is full.
Whether the order of an original change and the normalization can be interchanged is
determined by the correspondence of the operations themselves, and it can fail to hold even
when only the actions on objects and configurations agree
(\S\S\ref{sec:6.5}--\ref{sec:6.6}).
\item \textbf{The category of comparisons themselves.} Combining the arrow category with
idempotent completion, comparisons and their changes can be handled.
The interchange of the two constructions is compatible with the three-level projections of
extraction, core, and geometry (\S\ref{sec:6.7}).
\item \textbf{Images on which operations can be continued.} The idempotents of fixed-view
lenses split within the same category.
For protocols as well, executions and observations are restricted to the image of an
idempotent adapter that preserves executions and observations (\S\ref{sec:6.8}).
\end{itemize}

For the design review of the opening, we select a standard representation that preserves
the operations, the dependencies, and the Laws of order cancellation.
Even when the two proposals after normalization can be compared by an isomorphism, the
original class structure and the auxiliary representations need not be recoverable.
The classification of this chapter answers separately what remains in the image and whether
one can return to the original representation.
To move a further design change across the normalization, one checks that the
correspondence of the operations commutes with the normalization.

In the next chapter we use the category of comparisons constructed here to classify the
changes that preserve a single comparison.
The next question is how far the changes that look the same after normalization can be
distinguished on the original objects.

\part[Classification and Reconstruction]{Classification and Reconstruction\\[0.8em]{\large Chapters 7--8}}
\chapter{Comparison-Preserving Changes and Information}\label{chap:7}

\section*{Overview of the Chapter}

The question of this chapter is \textbf{by what information the changes preserving a
comparison can be distinguished, and how they can be constructed}.
Chapter~\ref{chap:6} used normalization to gather together the differences between
representations, and obtained a way of comparing the remaining images.
Now we consider the situation of a refactoring, which changes the internal design while
keeping the correspondence of processing and data.

For example, we extract the update logic from the processing that edits an order into a
shared component, and split the shipping address and the payment information contained in a
single order record into separate data types.
That the shipping address can still be displayed correctly after the change does not tell us
that the payment information is not mixed up when the shipping address is updated.
What is needed is that moving the old data into the new format and then updating, and updating
in the old format and then moving, give the same state of the order.
When only the display on the screen and the call relations are compared, this carrying over
of data can drop out of the check.

In this chapter we single out the changes under which chosen states and structures can be put
into one-to-one correspondence, and study the conditions under which this correspondence
preserves reads, updates, and the carrying over by the processing.

There are two tasks here. One is to \textbf{determine} whether a given change preserves the
comparison.
The other is to \textbf{construct}, from a change of the summarized design, a change of the
detailed design that preserves the comparison.
In this chapter we exhibit, for the normalization of AAT, cases where compatible changes can
be constructed and yet determining a given change requires the information lost in the
normalization.
On that basis, we classify all the options for compatible changes by a group action.

\begin{xltabular}{\linewidth}{@{}LLL@{}}
\toprule
Question & Construction & Outcome \\
\midrule
\endhead
What are the changes that preserve a comparison? & The comparison-preserving group and the
projections to the endpoints & Existence conditions and freedom for a change following a
change on one side (\S\S\ref{sec:7.1}--\ref{sec:7.2}) \\
\addlinespace[3pt]
Can compatibility be determined by observations alone? & The kernel of the observation
homomorphism & Conditions on the information necessary and sufficient for the determination,
and classification of the changes with the same observation (\S\ref{sec:7.3}) \\
\addlinespace[3pt]
Can a change after normalization be lifted to the original? & Restriction to the idempotent
image and a section of it & General conditions for the determination, and construction of
lifts for the generated comparisons of AAT (\S\S\ref{sec:7.4}--\ref{sec:7.6}) \\
\addlinespace[3pt]
How do the operations of software constrain changes? & Connected components of the graph of
operations & A split short exact sequence common to lenses and protocols, and the number of changes
(\S\S\ref{sec:7.7}--\ref{sec:7.8}) \\
\bottomrule
\end{xltabular}

\section{Changes preserving a single comparison}\label{sec:7.1}

We first confirm that, even when the call relations of the processing are the same, a
refactoring can break the data that ought to be carried over.
We then gather the condition of preserving the correspondence with operations and adapters
into a group.

\begin{example}[Extraction into a shared component changes what is carried over]\label{ex:7.1}

We extract the update of the shipping address of an order into a shared component.
Suppose there is a condition that this update does not change the payment category.
Focusing on the internal code $\{0,1\}$ of the payment category, we gather the processing
into a small state machine.
We take a single control point, and let the operation that returns to the same control point
before and after the update be a named self-loop $a$.
We define the correct implementation $X$ and the implementation $Y$ whose conversion in the
extracted component inverts this code by
\begin{equation*}
T_a^X(k)=k,\qquad T_a^Y(k)=1-k
\tag{7.1}\label{eq:7.1}
\end{equation*}
If we read only the vertices and the edges, both have the same update operation.
Updating an order with code 0 once, however, leaves it 0 in $X$ and turns it into 1 in $Y$.
Reading the Law ``an update of the shipping address keeps the payment category'' lets us
distinguish this difference.
In addition to the agreement of the call relations, we need to read the conditions on the
data that the processing carries over.

Likewise for changes, the part remaining in the observation can agree while the
compatibility with operations and adapters differs.
To describe this difference, we fix a single comparison.

\end{example}

\begin{definition}[Comparison-preserving group]\label{def:7.2}

Take a morphism $c:X\to Y$ of a category $E$ and subgroups
$G_X\le\mathrm{Aut}_E(X)$, $G_Y\le\mathrm{Aut}_E(Y)$ of the permitted endpoint changes.
We take the group of changes preserving $c$ to be
\begin{equation*}
\Gamma_c=\{(u,v)\in G_X\times G_Y\mid vc=cu\}
\tag{7.2}\label{eq:7.2}
\end{equation*}
A change satisfying this equation is said to be \textbf{compatible} with the comparison.
The product of the group is composition of morphisms, and $u_2u_1$ applies $u_2$ after
$u_1$.

In a refactoring as well, once a single compatible invertible reference correspondence
linking the old and the new states is fixed, the difference from another invertible
correspondence of states can be expressed as an automorphism of the original set of states.
In \S\ref{sec:7.8} we use this reduction to move the conditions for preserving reads and
operations to conditions on automorphisms.

In AAT we take $E=E_{\mathrm{geom}}$ and write the composite of the two-level projections of
Chapter~\ref{chap:1} as $\pi:E_{\mathrm{geom}}\to B$.
Besides the case where all automorphisms are permitted, we use the changes fixing the base
\begin{equation*}
G_X^0=\ker\bigl(\mathrm{Aut}(X)\xrightarrow{\pi}
\mathrm{Aut}(\pi X)\bigr),\qquad
\Gamma_c^0=\Gamma_c\cap(G_X^0\times G_Y^0)
\tag{7.3}\label{eq:7.3}
\end{equation*}
In this application to AAT, the unadorned $\Gamma_c$ denotes the case where Definition~\ref{def:7.2} is used
with the full endpoint automorphism groups.
What is fixed is the image in the base of the changes $u,v$, and $\pi(c)$ may remain a
general comparison.
Moreover, the condition of fixing the base and the condition of fixing the coefficients can
be specified separately.

\end{definition}

\begin{proposition}[The group obtained from a comparison object]\label{prop:7.3}

Equation~\eqref{eq:7.2} defines a subgroup of $G_X\times G_Y$.
With the full endpoint groups, it is the automorphism group of $c$ in the arrow
category $\mathrm{Arr}(E)$ of Chapter~\ref{chap:6}.
Via the embedding $J:E\to\mathrm{Kar}(E)$ by identity idempotents, it is also
isomorphic to the automorphism group of $J(c)$ in the category of comparisons
$\mathcal{M}(E)=\mathrm{Arr}(\mathrm{Kar}(E))$.
These identifications commute with the endpoint projections and, in AAT, restrict
to the subgroups fixing the base.

\end{proposition}

\begin{proof}

The identity pair satisfies \eqref{eq:7.2}.
For two compatible pairs, $v_2v_1c=v_2cu_1=cu_2u_1$, so their product is compatible.
Composing inverses on both sides of $vc=cu$ gives $v^{-1}c=cu^{-1}$.
An automorphism in the arrow category is precisely a commutative square with
isomorphisms at both ends. Moreover, morphisms between $J(X)=(X,1_X)$ carry no
additional constraint from idempotents, and $J$ is fully faithful.
Both directions of the correspondence retain the endpoint morphisms, so they also
preserve the projections and the conditions on the base.

\end{proof}

When only one side is changed, the other need not be able to follow.
When $(u,v)$ preserves the comparison, we say that $v$ follows $u$.
When following is possible, all the candidates can be described by the changes that leave
the comparison as it is.

\begin{theorem}[Projections to the endpoints and following changes]\label{thm:7.4}

Take the projections to the endpoints $p_X:\Gamma_c\to G_X$ and
$p_Y:\Gamma_c\to G_Y$.
Their kernels are, respectively,
\begin{equation*}
\ker p_X\cong T_c:=\{v\in G_Y\mid vc=c\},\qquad
\ker p_Y\cong S_c:=\{u\in G_X\mid cu=c\}
\tag{7.4}\label{eq:7.4}
\end{equation*}
A necessary and sufficient condition for the existence of a $v$ following $u\in G_X$ is
$u\in\mathrm{im}\,p_X$.
If one following change $v_0$ exists, all the candidates are $T_cv_0$.
Likewise, if a $u_0$ following $v\in G_Y$ exists, all the candidates are $S_cu_0$.

\end{theorem}

\begin{proof}

The first kernel is obtained by substituting $u=1_X$ into \eqref{eq:7.2}. The second is
similar.
From $v_0c=cu$ we get $v_0^{-1}c=cu^{-1}$. Using this together with $vc=cu$, we obtain
\begin{equation*}
(vv_0^{-1})c=vc\,u^{-1}=c
\tag{7.5}\label{eq:7.5}
\end{equation*}
Hence we may write it uniquely as $v=tv_0$, $t=vv_0^{-1}\in T_c$.
Conversely, if $t\in T_c$ then $tv_0c=tcu=cu$.
For the other side the computation is the same, using $cu_0^{-1}=v^{-1}c$.

\end{proof}

When the action of a group on a nonempty set is free and transitive, the set is called a
\textbf{torsor} under the group.
That is, between two candidates there is exactly one element of the group carrying one to
the other.
The candidate sets of Theorem~\ref{thm:7.4} are torsors under $T_c$ and $S_c$ respectively,
by composition on the left.
Choosing one candidate as a reference identifies the set with the group, but the candidate
set itself needs no specified reference.

\begin{corollary}[Changes of an isomorphism comparison]\label{cor:7.5}

If $c$ is an isomorphism and conjugation $u\mapsto cuc^{-1}$ matches $G_X$ with $G_Y$, then
\begin{equation*}
\Gamma_c=\{(u,cuc^{-1})\mid u\in G_X\},\qquad
\Gamma_c\cong G_X\cong G_Y.
\tag{7.6}\label{eq:7.6}
\end{equation*}

\end{corollary}

\begin{proof}

Solve \eqref{eq:7.2} as $v=cuc^{-1}$.
The conjugation condition holds automatically for the full endpoint groups.
It also holds for the groups fixing the base, since
$\pi(cuc^{-1})=\pi(c)1\pi(c)^{-1}=1$.

\end{proof}

\section{Carrying changes from the same generator}\label{sec:7.2}

In Chapter~\ref{chap:5} we constructed two routes and a comparison from the same generator.
Sharing this generator also answers which changes are to be combined at the two endpoints.

\begin{theorem}[Compatibility of generated changes]\label{thm:7.6}

Take the common original object $S$ of Chapter~\ref{chap:5}, \S\ref{sec:5.8}, the ends
$X,Y$ of the two routes, and the generated comparison $c:X\to Y$.
Let $G_S^0$ be the group of automorphisms of the original object $S$ fixing the base.
Writing the homomorphisms induced by the transports and the pullbacks along the two routes
as $F_*:G_S^0\to G_X^0$ and $G_*:G_S^0\to G_Y^0$, we have
\begin{equation*}
G_*(a)c=cF_*(a)\qquad(a\in G_S^0).
\tag{7.7}\label{eq:7.7}
\end{equation*}
Setting further $J=G_*^{-1}(T_c)$, for independently chosen $a,b\in G_S^0$,
\begin{equation*}
\bigl(F_*(a),G_*(b)\bigr)\in\Gamma_c^0
\quad\Longleftrightarrow\quad ba^{-1}\in J.
\tag{7.8}\label{eq:7.8}
\end{equation*}
Hence, once the change $a$ on one side is fixed, the changes of the source usable on the
other side are $Ja$.

\end{theorem}

\begin{proof}

Write the cartesian legs of the two routes as $q_X:X\to S$ and $q_Y:Y\to S$.
Here, as in Chapter~\ref{chap:5}, we compose the legs after moving to the refinement side
when necessary.
The factorizations at the time of generation and the triangle of the comparison are
$q_XF_*(a)=aq_X$, $q_YG_*(a)=aq_Y$, and $q_Yc=q_X$.
Composing both sides of \eqref{eq:7.7} with $q_Y$ therefore gives $aq_X$ in either case.
Both sides lie over the same morphism of the base. Using the uniqueness coming from
cartesianness of the core and then from cartesianness of the geometry, we obtain
\eqref{eq:7.7}.

By \eqref{eq:7.7} and Theorem~\ref{thm:7.4}, compatibility of $G_*(b)$ with $F_*(a)$ is
equivalent to $G_*(b)G_*(a)^{-1}\in T_c$.
Since these are homomorphisms, this becomes \eqref{eq:7.8}.

\end{proof}

Carrying the same change along both routes gives compatibility, and when different changes
are chosen the remaining difference must lie in $J$.
This description is preserved when the presentation of the generator is changed.

\begin{proposition}[Change of presentation of the input and regeneration]\label{prop:7.7}

Write the diagram consisting of the common original objects of Chapter~\ref{chap:5} as
$(S_v,L_e,\kappa_f)$.
Here $L_e$ is the morphism of an edge and $\kappa_f$ is the comparison specified on a face.
Give at each vertex an isomorphism $\theta_v:S'_v\xrightarrow{\sim}S_v$ fixing the base and
the coefficients, and form the new input by
\begin{equation*}
L'_e=\theta_{t(e)}^{-1}L_e\theta_{s(e)},\qquad
\kappa'_f=\theta_{t(f)}^{-1}\kappa_f\theta_{t(f)}
\tag{7.9}\label{eq:7.9}
\end{equation*}
The choices of transport of Chapter~\ref{chap:4} are also carried through these
isomorphisms, and the two routes are constructed from the new input.
Then, with the endpoint isomorphisms
$b_v:X'_v\xrightarrow{\sim}X_v$ and $d_v:Y'_v\xrightarrow{\sim}Y_v$ given by universality,
\begin{equation*}
\begin{aligned}
c'_v&=d_v^{-1}c_vb_v,\\
F'_*(\theta_v^{-1}a\theta_v)&=b_v^{-1}F_*(a)b_v,\\
G'_*(\theta_v^{-1}a\theta_v)&=d_v^{-1}G_*(a)d_v
\end{aligned}
\tag{7.10}\label{eq:7.10}
\end{equation*}
The comparison-preserving group, the residual subgroup of \eqref{eq:7.8}, and the set of
following changes are carried along this correspondence.

\end{proposition}

\begin{proof}

The edges carried through the isomorphisms are again strongly opcartesian morphisms.
Indeed, composing $\theta$ and its inverse before and after the original factorization
carries over existence and uniqueness.
The same holds for the cartesian morphisms of the pullback.
Hence the universality of each route, old and new, yields $b_v,d_v$ uniquely.
Abbreviating the two legs as $q_X,q_Y$, these satisfy
$q'_X=\theta^{-1}q_Xb$ and $q'_Y=\theta^{-1}q_Yd$.
The candidate $d^{-1}cb$ satisfies the triangle of the new comparison, hence coincides with
$c'$ by uniqueness.
The formulas for the two changes are obtained likewise, by substituting into the new
factorizations.
The conditions on edges and faces are preserved because the intermediate inverses in
\eqref{eq:7.9} cancel.
Finally, substituting \eqref{eq:7.10} into \eqref{eq:7.2} and \eqref{eq:7.8} yields the
correspondence of the groups and of the candidates.

\end{proof}

\section{Conditions for distinguishing by observation}\label{sec:7.3}

To distinguish the changes compatible with a single comparison, it is not always necessary
to read all the components of a change.
Which components may be forgotten can be stated precisely by studying the changes that look
like the identity under the observation.

\begin{definition}[Observation of a change]\label{def:7.8}

Take a group $Q$ of changes, a compatible subgroup $\Gamma\le Q$, and a group homomorphism
$O:Q\to R$.
We call $O(q)$ the observation of the change $q$ and set
\begin{equation*}
K=\ker O,\qquad L=K\cap\Gamma
\tag{7.11}\label{eq:7.11}
\end{equation*}
Here $K$ consists of all the changes that do not appear in the observation, and $L$ of
those among them that are also compatible with the comparison.
The observation here is defined as a map preserving the composition of changes.

\end{definition}

\begin{theorem}[Determination by the observation alone]\label{thm:7.9}

The following conditions are equivalent.

\begin{enumerate}
\item There is a predicate $D:R\to\{\text{true},\text{false}\}$ such that
$q\in\Gamma\Longleftrightarrow D(O(q))$ for every $q\in Q$.
\item $K\subseteq\Gamma$.
\end{enumerate}

Moreover,
\begin{equation*}
O^{-1}(O(\Gamma))=\Gamma K
\tag{7.12}\label{eq:7.12}
\end{equation*}
always holds. Surjectivity of $O$ is not assumed.

\end{theorem}

\begin{proof}

Assume 1. Then $k\in K$ has the same observation as the identity element.
The identity element belongs to $\Gamma$, so $k\in\Gamma$, which gives 2.
Conversely, assume 2 and define $D(r)$ to be $r\in O(\Gamma)$.
If $O(q)=O(\gamma)$ with $\gamma\in\Gamma$, then
$k=\gamma^{-1}q\in K$ and $q=\gamma k\in\Gamma$.
This decomposition shows both inclusions of \eqref{eq:7.12} without assuming 2.

\end{proof}

What we have obtained here is the condition for compatibility to depend only on the observed
value.
To actually build a program that determines it, a way of computing $D$ is needed as well.

\begin{proposition}[Differences remaining within the same observation]\label{prop:7.10}

For $\gamma\in\Gamma$, the fiber of the observation and its compatible part are
\begin{equation*}
O^{-1}(O(\gamma))=\gamma K,\qquad
O^{-1}(O(\gamma))\cap\Gamma=\gamma L
\tag{7.13}\label{eq:7.13}
\end{equation*}
Specifying the base point $L$ in the set of left cosets $K/L=\{kL\mid k\in K\}$, we have
\begin{equation*}
\gamma k\in\Gamma\quad\Longleftrightarrow\quad kL=L.
\tag{7.14}\label{eq:7.14}
\end{equation*}
Hence $K/L$ being a single point is equivalent to the condition for the determination in
Theorem~\ref{thm:7.9}.

\end{proposition}

\begin{proof}

Agreement of the observations is equivalent to $\gamma^{-1}q\in K$.
Moreover, since $\gamma\in\Gamma$, the condition $\gamma k\in\Gamma$ is equivalent to
$k\in\Gamma$.
This gives \eqref{eq:7.13} and \eqref{eq:7.14}.
The set of cosets being a single point is equivalent to $K=L$, that is, to
$K\subseteq\Gamma$.

\end{proof}

$L$ is a normal subgroup of $\Gamma$, but need not be a normal subgroup of $K$.
What is needed here is therefore the set of cosets with a base point.

\begin{corollary}[The case of reading the coefficients only]\label{cor:7.11}

Fix an isomorphism comparison $c:X\xrightarrow{\sim}Y$ of full geometries.
From the groups fixing the base at the two endpoints, we take the homomorphisms reading the
coefficient component to be
$o_X:G_X^0\to\mathrm{Aut}(k_X)$ and $o_Y:G_Y^0\to\mathrm{Aut}(k_Y)$.
Write the coefficient isomorphism of the comparison as $\kappa:k_X\xrightarrow{\sim}k_Y$.
The conjugation $T(u)=cuc^{-1}$ satisfies
$o_Y(T(u))=\kappa o_X(u)\kappa^{-1}$.
Setting $K_X=\ker o_X$, $K_Y=\ker o_Y$, and $O=o_X\times o_Y$, we have
\begin{equation*}
\begin{gathered}
K=K_X\times K_Y,\qquad L=\{(a,T(a))\mid a\in K_X\},\\
K/L\xrightarrow{\sim}K_Y,\qquad [(a,b)]\longmapsto bT(a)^{-1}.
\end{gathered}
\tag{7.15}\label{eq:7.15}
\end{equation*}
The last correspondence is a bijection of sets sending the base point to the identity
element.
A necessary and sufficient condition for compatibility with the comparison to be
determinable from the observation of the coefficients alone is $K_Y=\{1\}$.

\end{corollary}

\begin{proof}

By the composition rule for the coefficient maps and Corollary~\ref{cor:7.5}, $T$ gives
$K_X\cong K_Y$, and $L$ is its graph.
Even if the representative of a coset is changed to $(a,b)(x,T(x))$, we have
$bT(x)T(ax)^{-1}=bT(a)^{-1}$, so \eqref{eq:7.15} is well defined.
The inverse is $y\mapsto[(1,y)]$.
Indeed, $(a,b)(a^{-1},T(a)^{-1})=(1,bT(a)^{-1})$, and both composites are the identity.
The condition for the determination follows from Proposition~\ref{prop:7.10}.

\end{proof}

Cases in which this set of cosets cannot be treated as a quotient group also occur in concrete cases.
If $K_X=K_Y=S_3$ (the group of permutations of three points) and $T=\mathrm{id}$, then $L$
is the diagonal subgroup.
For $s=(12)$ and $t=(23)$, the element
$(s,1)(t,t)(s,1)^{-1}=(sts^{-1},t)$ does not belong to the diagonal.
The bijection of \eqref{eq:7.15} is still valid in this case.

When the presentation of the input is carried as in Proposition~\ref{prop:7.7}, the
conjugations of the endpoint groups and of the coefficient groups commute with the
observation homomorphism.
Hence $K,L$ and the pointed cosets correspond as well.
Changing the presentation and regenerating does not change the conclusion as to whether
compatibility can be determined from the observation.

\section{Restriction to the normalization, reflection, and lifting}\label{sec:7.4}

When normalization is used as an observation, we study three properties separately.
That a change compatible in the original is compatible in the image as well is called
\textbf{preservation}, and that compatibility in the image implies compatibility in the
original is called \textbf{reflection}. Constructing an original of a compatible change
given in the image is \textbf{lifting}.

\begin{proposition}[Preservation of a comparison by a functor]\label{prop:7.12}

A functor $F:E\to D$ and a comparison $c:X\to Y$ induce the homomorphism
\begin{equation*}
\begin{gathered}
r_F:\mathrm{Aut}(X)\times\mathrm{Aut}(Y)
\longrightarrow\mathrm{Aut}(FX)\times\mathrm{Aut}(FY),\\
(u,v)\longmapsto(Fu,Fv)
\end{gathered}
\tag{7.16}\label{eq:7.16}
\end{equation*}
and we have $r_F(\Gamma_c)\subseteq\Gamma_{F(c)}$.
Applying this to the core normalization functor $\mathcal N$ of Chapter~\ref{chap:6} gives
preservation of an arbitrary comparison.
Moreover, the equality $\pi_N\mathcal N=\pi$ of the projections to the base in
Theorem~\ref{thm:6.21} yields a homomorphism between the comparison-preserving groups that
fix the base as well.
By Proposition~\ref{prop:6.22}, the same holds for the normalization functor of full
geometries.

\end{proposition}

\begin{proof}

Functors preserve isomorphisms, identities, and composition, so \eqref{eq:7.16} is a
homomorphism.
Applying $F$ to $vc=cu$ gives preservation.
Applying the equality on the base to $u,v$ shows that the condition of fixing the base is
preserved as well.

\end{proof}

Separately, a restriction can also be formed when idempotents are specified in an arbitrary
category.
In this case we choose the changes commuting with the normalization as the domain.

\begin{construction}[Restriction to the idempotent image]\label{cons:7.13}

Assume $dc=ce$ for a morphism $c:X\to Y$ and idempotents $e:X\to X$ and $d:Y\to Y$.
We take the centralizer groups and the comparison of the images to be
\begin{equation*}
\begin{aligned}
C_e&=\{u\in\mathrm{Aut}(X)\mid ue=eu\},\quad
C_d=\{v\in\mathrm{Aut}(Y)\mid vd=dv\},\\
H&=C_e\times C_d,\qquad
a=dce:(X,e)\longrightarrow(Y,d)
\end{aligned}
\tag{7.17}\label{eq:7.17}
\end{equation*}
Here $a$ is a morphism of $\mathrm{Kar}(E)$. The restriction is given by
\begin{equation*}
\begin{gathered}
r:H\longrightarrow\mathrm{Aut}_{\mathrm{Kar}(E)}(X,e)
\times\mathrm{Aut}_{\mathrm{Kar}(E)}(Y,d),\\
r(u,v)=(eue,dvd)
\end{gathered}
\tag{7.18}\label{eq:7.18}
\end{equation*}
The inverse of $eue$ is $eu^{-1}e$, and both composites are the identity morphism $e$ of
$(X,e)$.
By the commutation condition, the restriction preserves products as well.
Setting $\Gamma_0=H\cap\Gamma_c$ and $\Delta=\Gamma_a$, we have $r(\Gamma_0)\subseteq\Delta$.
Indeed, it suffices to sandwich $vc=cu$ between the idempotents and use $ue=eu$ and $vd=dv$.

Proposition~\ref{prop:7.12} treats all the automorphisms on which the normalization functor
is defined, and Construction~\ref{cons:7.13} treats the centralizer groups.
For the normalization functor of Chapter~\ref{chap:6}, the two homomorphisms agree on the centralizing changes.

In Theorem~\ref{thm:7.9} membership in the set of observed compatible changes could be used
for the determination.
Here the compatible subgroup on the side of the image is determined independently.
Besides the condition on the kernel, it is therefore also necessary that the compatible
changes on the side of the image that are actually reached be obtained from changes
compatible in the original. The second condition of the next theorem expresses this.

\end{construction}

\begin{theorem}[General determination of reflection and lifting]\label{thm:7.14}

Assume $r(\Gamma_0)\subseteq\Delta$ for a group homomorphism $r:H\to R$ and subgroups
$\Gamma_0\le H$ and $\Delta\le R$. Then
\begin{equation*}
r^{-1}(\Delta)=\Gamma_0
\quad\Longleftrightarrow\quad
\begin{cases}
\ker r\subseteq\Gamma_0,\\
r(\Gamma_0)=\Delta\cap\mathrm{im}\,r.
\end{cases}
\tag{7.19}\label{eq:7.19}
\end{equation*}
We take the restricted homomorphism to be $\bar r:\Gamma_0\to\Delta$ and its kernel to be
$L=\ker\bar r$.
A compatible lift of $\delta\in\Delta$ exists if and only if $\delta\in r(\Gamma_0)$.
Taking one of them to be $\gamma_0$, all the compatible lifts are
\begin{equation*}
\bar r^{-1}(\delta)=\gamma_0L
\tag{7.20}\label{eq:7.20}
\end{equation*}
and form a torsor under $L$ by multiplication on the right.
If $\bar r$ is surjective, the inclusion $L\hookrightarrow\Gamma_0$ gives the short exact
sequence
\begin{equation*}
1\longrightarrow L\longrightarrow\Gamma_0
\xrightarrow{\bar r}\Delta\longrightarrow1
\tag{7.21}\label{eq:7.21}
\end{equation*}
Here short exact means that the first map is injective, its image is the kernel of the next
map, and the last map is surjective.
When there is a group homomorphism $\sigma:\Delta\to\Gamma_0$ satisfying
$\bar r\sigma=\mathrm{id}_\Delta$, the sequence is said to split and $\sigma$ is called a
section.

\end{theorem}

\begin{proof}

If reflection holds, the preimage of the identity element lies in $\Gamma_0$, and every
image of $r$ belonging to $\Delta$ comes from $\Gamma_0$.
Conversely, assume the right-hand side and let $r(q)\in\Delta$.
By the condition on the image there is $\gamma\in\Gamma_0$ with $r(\gamma)=r(q)$, and
$\gamma^{-1}q\in\ker r\subseteq\Gamma_0$ gives $q\in\Gamma_0$.

If the lifts $\gamma_0,\gamma$ have the same observation, then $\gamma_0^{-1}\gamma\in L$.
This gives \eqref{eq:7.20}. The right action satisfies
$(\gamma\ell_1)\ell_2=\gamma(\ell_1\ell_2)$, and the element between two lifts is uniquely
determined as $\gamma_0^{-1}\gamma$.
Short exactness is just the definition of the kernel and surjectivity.

\end{proof}

It is $L=\ker\bar r$ that expresses the freedom of lifting.
What obstructs the reflection of compatibility, on the other hand, are those elements of the
kernel $\ker r$ of all endpoint changes that do not lie in $\Gamma_0$.
The two can be distinguished by the following examples.

\begin{example}[Lifts exist even when reflection fails]\label{ex:7.15}

In the category of finite sets we set $X=Y=\{0,1,2\}$ and $c=1_X$, and take $e=d$ to be the
constant map $x\mapsto0$.
The permutation $\tau=(12)$ commutes with $e$, but the endpoint pair $(\tau,1)$ does not
preserve the identity comparison.
Since the image is a one-element set, after the restriction both endpoints become the
identity, which preserves the comparison of the images.
The comparison-preserving group of the image is the trivial group, so its unique element
lifts to $(1,1)$.
Moreover, $(\tau,\tau)$ is a compatible lift of the same element as well.

\end{example}

\begin{example}[A case where a change of the image does not lift]\label{ex:7.16}

Using the same three-element set and $c=1_X$, we now take
\begin{equation*}
e(0)=0,\qquad e(1)=1,\qquad e(2)=1,
\qquad d=e
\tag{7.22}\label{eq:7.22}
\end{equation*}
Exchanging 0 and 1 at both endpoints of the image $\{0,1\}$ preserves the identity
comparison of the images.
A permutation $u$ commuting with $e$, however, puts $e^{-1}(0)$ and $e^{-1}(u(0))$ in
bijection.
The former has one element and $e^{-1}(1)$ has two, so $u(0)=1$ is impossible.
Since $e^{-1}(2)$ is empty, $u(0)=2$ is impossible as well.
Hence this exchange in the image has no lift inside the centralizer group.

These examples show that preservation, reflection, and the existence of lifts each require
their own proof.
In the next section we construct lifts using the concrete form of object formation and
normalization in AAT.

\end{example}

\section{Constructing lifts for full geometries}\label{sec:7.5}

A general idempotent can have fibers of different sizes, as in Example~\ref{ex:7.16}.
In the canonical normalization of AAT, an object splits into a configuration and the
structure data accompanying it.
Using this common form, a change given on the normalized image can be extended consistently
to all the original objects.

In this section we assume that the core $P$ of a full geometry $G$ satisfies
$\mathrm{Ad}(P)$ of Definition~\ref{def:5.39}.
That is, the normalization leaves the residuals of the equations and the coordinates of the
invariants and of the signature unchanged, and the identification of the types of operations
preserves the configuration maps.
We write $\mathcal A$ for the full subcategory of full geometries satisfying this assumption.
We use the normalization functor of Chapter~\ref{chap:6} as
$\mathcal N:\mathcal A\to\mathrm{Kar}(\mathcal A)$,
$\mathcal N(G)=(G,N_G)$, $\mathcal N(f)=fN_G$.

\subsection*{Carrying the data accompanying an object}

\begin{construction}[Extension preserving the selected objects]\label{cons:7.17}

We split an architecture object $A$ of Chapter~\ref{chap:1} into a configuration $C_A$ and
accompanying data $(S_A,Q_A,s_A,q_A)$. We write the set of all accompanying data, in a
universe of the required size, as
\begin{equation*}
D=\coprod_{S,Q\in\mathbb{U}}(S\times Q)
\tag{7.23}\label{eq:7.23}
\end{equation*}
Since the types $S,Q$ themselves are included in the data, all the objects are in one-to-one
correspondence with $\mathcal C\times D$.
Here $\mathcal C$ is the set of configurations.
We write the object selected by the core as $s_P(C)=(C,d_C)$ and fix a reference element
$d_*$ of $D$.
Such a $d_*$ is obtained using two one-element sets and their elements.

Let $\theta_C$ be the permutation exchanging $d_C$ and $d_*$.
When the two are equal we use the identity. Then
\begin{equation*}
\theta_C^2=1,\qquad \theta_C(d_C)=d_*,\qquad
\theta_C(d_*)=d_C.
\tag{7.24}\label{eq:7.24}
\end{equation*}
Let $\sigma$ be the Atom component of an automorphism $a$ of the normalized image, and
abbreviate the transport of a configuration as $\sigma C$.
If we simply take $(C,d)\mapsto(\sigma C,d)$, the selected object $(C,d_C)$ goes to
$(\sigma C,d_C)$, which need not agree with the selected object $(\sigma C,d_{\sigma C})$ at
the target.
We therefore combine the two exchanges, at the start and at the target, and define the
extension to all the objects by
\begin{equation*}
\widehat a(C,d)=
\bigl(\sigma C,\theta_{\sigma C}\theta_C(d)\bigr)
\tag{7.25}\label{eq:7.25}
\end{equation*}
This first aligns the selected value at the start with the common reference, and returns to
the selected value at the target.
Hence $\widehat a(s_P(C))=s_P(\sigma C)$, so object formation is preserved.
The same formula can be defined for a self-map of Atoms as well.

With this construction, the intermediate permutations cancel when changes are performed in
succession.
For the two extensions corresponding to the Atom components $\sigma,\rho$, we have
\begin{equation*}
\theta_{\rho\sigma C}\theta_{\sigma C}
\theta_{\sigma C}\theta_C
=\theta_{\rho\sigma C}\theta_C.
\tag{7.26}\label{eq:7.26}
\end{equation*}
Hence the map on objects preserves identities and composition.
That this equality holds not only for the selected objects but for all the accompanying data
is the reason we obtain a lift as a group homomorphism.

\end{construction}

\subsection*{Extending the components of the operations and of the geometry}

\begin{lemma}[A section for the automorphisms of a full geometry]\label{lem:7.18}

For every $G\in\mathcal A$ there exists a group homomorphism
\begin{equation*}
\sigma_G:\mathrm{Aut}(\mathcal N G)\longrightarrow\mathrm{Aut}(G),
\qquad \mathcal N(\sigma_G(a))=a
\tag{7.27}\label{eq:7.27}
\end{equation*}
This section keeps the base and the coefficient components, and its values commute with
$N_G$.

\end{lemma}

\begin{proof}

We represent a morphism of the normalized image by a morphism $f:G\to G$ of the original
category with $N_GfN_G=f$.
For objects we use Construction~\ref{cons:7.17}. We abbreviate the normalization of objects
of Chapter~\ref{chap:6} as $n=n_P$ and write the identification of the types of operations as
$\nu_{A,B}:\mathrm{Op}(A,B)\xrightarrow{\sim}\mathrm{Op}(nA,nB)$.
These identifications, which follow the equalities of the types, are the identity on the
selected objects and are compatible with the composition of successive identifications.
Object formation gives $f(nA)=n\widehat f(A)$, so we define the action on operations by
\begin{equation*}
S_G(f)_{A,B}^{\mathrm{op}}
=\nu_{\widehat f(A),\widehat f(B)}^{-1}
\,f_{nA,nB}^{\mathrm{op}}\,\nu_{A,B}
\tag{7.28}\label{eq:7.28}
\end{equation*}
For the identification of the intermediate objects we use the equality of object formation
above.
This is the map that normalizes at the start, applies the given morphism, and returns to the
type at the target of the extension.
Since $\nu$ and $f$ each preserve the action on configurations, so does this composite.

For the maps of the Atoms, the contexts, the equation indices, the invariant indices, and
the axes and their value ranges we use the components of $f$.
Preservation of the residuals of the equations follows by replacing $A$ with $nA$ at the
start, applying the preservation law of $f$, and returning $n\widehat f(A)$ to
$\widehat f(A)$ at the target.
The same three-step computation is carried out for the coordinates of the invariants and of
the signature.
The base and the coefficients, the covers and the overlaps, and the local maps of the
supports, the axes, and the observables are taken over from those of $f$.
Since the maps of Atoms and contexts to which these refer are unchanged, the equalities of
the raw system and the commutativity of the local restrictions are preserved as well.
Hence $S_G(f)$ is a morphism of full geometries.

By \eqref{eq:7.26} for objects and the cancellation of the intermediate $\nu\nu^{-1}$ in
\eqref{eq:7.28} for operations, we obtain, for endomorphisms $f,g$ of the normalized image,
\begin{equation*}
S_G(N_G)=1_G,\qquad
S_G(gf)=S_G(g)S_G(f),\qquad
S_G(f)N_G=N_GS_G(f)=f
\tag{7.29}\label{eq:7.29}
\end{equation*}
In the last equality we also use that $f=N_GfN_G$ gives $fN_G=N_Gf=f$.
The other components are taken over from $f,g$, so they satisfy the same equalities.
In particular, applying $S_G$ to the morphisms in both directions of an automorphism of
the normalized image, we find that both composites are $1_G$.
Taking this to be $\sigma_G$, we obtain \eqref{eq:7.27} and the homomorphism property.
Keeping the base and the coefficients follows from the construction, and commutation with
$N_G$ follows from \eqref{eq:7.29}.

\end{proof}

\subsection*{Combining the two endpoints so as to preserve the comparison}

Sections built separately at the two endpoints of a comparison need not be compatible with
that comparison.
We therefore lift only one side and determine the other by conjugation with the original
comparison.

\begin{theorem}[Compatible lifts for an isomorphism comparison]\label{thm:7.19}

Take $X,Y\in\mathcal A$ and an isomorphism $c:X\xrightarrow{\sim}Y$.
The restriction
$\bar r_N:\Gamma_c\to\Gamma_{\mathcal N(c)}$ of Proposition~\ref{prop:7.12} has the section
\begin{equation*}
\sigma_c(a,b)=\bigl(\sigma_X(a),\ c\sigma_X(a)c^{-1}\bigr)
\tag{7.30}\label{eq:7.30}
\end{equation*}
The base and coefficient components at the two endpoints agree with those of the given
$a,b$.
Hence the restriction to the subgroups fixing the base has a section as well.

\end{theorem}

\begin{proof}

Since the second component is built by conjugation, \eqref{eq:7.30} preserves the original
comparison $c$.
The normalization of the first component is $a$.
By $(a,b)\in\Gamma_{\mathcal N(c)}$ and functoriality, the normalization of the second
component is
\begin{equation*}
\mathcal N(c)\,a\,\mathcal N(c)^{-1}=b
\tag{7.31}\label{eq:7.31}
\end{equation*}
This gives $\bar r_N\sigma_c=\mathrm{id}$.
Both the section on the first component and the conjugation are group homomorphisms, so
$\sigma_c$ is a group homomorphism.
Since the normalization does not change the base and coefficient components, keeping them
for the second component follows from \eqref{eq:7.31}.

\end{proof}

\section{Information remaining in a generated comparison, and all the lifts}\label{sec:7.6}

Given a section, a compatible change in the original can be built from a compatible change
after normalization.
Not every change with the same value after normalization is compatible, however.
This difference can be confirmed by building an actual automorphism on all the objects of
AAT.

\begin{construction}[A nonidentity change that does not appear in the normalization]\label{cons:7.20}

Let $G\in\mathcal A$ and use the presentation $(C,d)$ of Construction~\ref{cons:7.17}.
The set $D$ has at least three distinct elements.
For instance, it suffices to fix $S=\{0,1,2\}$ and $Q=\{*\}$ and use the structure values
$s=0,1,2$.
For each $C$, choose two elements different from the selected value $d_C$ and let $\rho_C$
be the permutation exchanging them.
The map on objects
\begin{equation*}
\tau_G(C,d)=(C,\rho_C(d))
\tag{7.32}\label{eq:7.32}
\end{equation*}
fixes the configurations and all the selected objects, and applying it twice gives the
identity.
For instance, a configuration specified by the core has two objects that are exchanged, so
this map is not the identity.

This map can be extended to an automorphism of the full geometry.
For the operations we use $\nu_{\tau_G A,\tau_G B}^{-1}\nu_{A,B}$, and for the maps of the
Atoms, the base, the contexts, the equation indices, the invariant indices, the axes, the
coefficients, and the local geometry we use the identity.
The normalizations of the two objects agree, so the preservation conditions follow by the
same computation through the normalization as in Lemma~\ref{lem:7.18}.
The exchange of the accompanying data and the identification of the types of operations
return after two applications. Hence
\begin{equation*}
\begin{gathered}
\tau_G\ne1_G,\qquad \tau_G^2=1_G,\\
N_G\tau_G=\tau_GN_G=N_G,\qquad
\mathcal N(\tau_G)=1_{\mathcal N G}.
\end{gathered}
\tag{7.33}\label{eq:7.33}
\end{equation*}
This change fixes the base and the coefficients.
While it moves the distinction between objects, it does not appear in the components read
through the normalization under Ad.

\end{construction}

\begin{theorem}[Distinguishing and lifting for a generated comparison]\label{thm:7.21}

From the same exact square of bases, face, original core, coefficients, and
original geometry of Construction~\ref{cons:5.37}, generate the canonical comparison
$\alpha:X\xrightarrow{\sim}Y$ of full geometries.
We assume Ad for the source core and use the canonical normalizations at the two endpoints.
By Lemma~\ref{lem:6.15} and Theorem~\ref{thm:6.16} of Chapter~\ref{chap:6}, the two
endpoints satisfy Ad as well, and $N_Y\alpha=\alpha N_X$.

We set $\Gamma=\Gamma_\alpha$ and $\Delta=\Gamma_{\mathcal N(\alpha)}$, take
$r_N$ to be the homomorphism on the full endpoint groups, and set $L=\ker(r_N|_\Gamma)$. Then:

\begin{enumerate}
\item $r_N|_\Gamma$ has a section $\sigma_\alpha$, and we obtain the split short exact
sequence \eqref{eq:7.34} below.
\item For every $\delta\in\Delta$, all the compatible lifts are
$\sigma_\alpha(\delta)L$, a right $L$-torsor.
\item The full endpoint groups also contain incompatible changes with the same value
$\delta$ after normalization.
Hence compatibility with $\alpha$ cannot be determined from the value after normalization
alone.
\end{enumerate}
\begin{equation*}
1\longrightarrow L\longrightarrow\Gamma
\xrightarrow{\bar r_N}\Delta\longrightarrow1
\tag{7.34}\label{eq:7.34}
\end{equation*}
These also hold for the groups fixing the base at the two endpoints.

\end{theorem}

\begin{proof}

1 and 2 follow from Theorems~\ref{thm:7.19} and 7.14.
For 3, set $q_-=(\tau_X,1_Y)$. By Construction~\ref{cons:7.20}, $r_N(q_-)=1$.
If $q_-\in\Gamma$, then $\alpha=\alpha\tau_X$, and invertibility of $\alpha$ gives the
contradiction $\tau_X=1_X$.
Hence
\begin{equation*}
q_+=\sigma_\alpha(\delta),\qquad
q_{\mathrm{bad}}=\sigma_\alpha(\delta)q_-
\tag{7.35}\label{eq:7.35}
\end{equation*}
have the same value $\delta$, and only the former is compatible.
If the latter were compatible as well, their difference $q_-$ would lie in the subgroup
$\Gamma$.
The section keeps the base component and $q_-$ fixes the base, so the same proof applies to
the base-fixing version.

\end{proof}

In the case of reading the coefficients only, Construction~\ref{cons:7.20} also gives an
actual incompatible pair.
The pair $(\tau_X,\alpha\tau_X\alpha^{-1})$ is compatible and $(\tau_X,1_Y)$ is not, yet the
observations of the base and the coefficients of both pairs are the same identity pair.
Hence the loss of information in Corollary~\ref{cor:7.11} occurs for comparisons of full
geometries generated from the same input as well.

\begin{corollary}[Selection of the normalization by the diagnosis]\label{cor:7.22}

Take the same input of Construction~\ref{cons:5.40} and write its selection condition as $\chi$.
In the notation of Chapter~\ref{chap:6} we set $\beta=d\alpha=\alpha e$ and let $H$ be the
group of endpoint changes commuting with $e,d$.
The restriction of Construction~\ref{cons:7.13}, from $\Gamma_\alpha\cap H$ to
the comparison-preserving group $\Gamma_{\beta:(X,e)\to(Y,d)}$ of the image, has a
section.
Preservation and reflection of compatibility with $\alpha$ under this restriction are as
follows.

\begin{xltabular}{\linewidth}{@{}LLLL@{}}
\toprule
Normalization & Preservation of the comparison & Reflection of compatibility with $\alpha$ &
Lifting of the compatible changes of the image \\
\midrule
\endhead
$\chi$ is false, $e=1_X,d=1_Y$ & Holds & Holds & A section exists \\
\addlinespace[3pt]
$\chi$ is true, $e=N_X,d=N_Y$ & Holds & Fails & A section exists \\
\addlinespace[3pt]
The canonical normalization is chosen directly under Ad & Holds & Fails & A section exists \\
\bottomrule
\end{xltabular}

Each row holds for the subgroups fixing the base as well.

\end{corollary}

\begin{proof}

If the selection condition is false, the restriction is to the identity idempotents, so the
group is the same as the original comparison group.
In the true case, Theorem~\ref{thm:6.16} gives $e=N_X,d=N_Y$.
The element $\sigma_X(a)$ of Lemma~\ref{lem:7.18} commutes with $N_X$.
From $\alpha N_X=N_Y\alpha$, the conjugate $\alpha\sigma_X(a)\alpha^{-1}$ commutes with
$N_Y$ as well.
Hence \eqref{eq:7.30} gives a section inside $H$.
The failure of reflection follows from $(\tau_X,1_Y)$, which belongs to $H$ as well.
The proof is the same when the canonical normalization is chosen directly. The condition on
the base is kept in every construction.

\end{proof}

What is not reflected here is compatibility with $\alpha$ before normalization.
On the centralizer group $H$, preserving $\beta$ in the image is equivalent to preserving
$\beta$ as a morphism of the original category, and
\begin{equation*}
r^{-1}\bigl(\Gamma_{\beta:(X,e)\to(Y,d)}\bigr)
=H\cap\Gamma_{\beta:X\to Y}
\tag{7.36}\label{eq:7.36}
\end{equation*}
Indeed, by $d\beta=\beta=\beta e$ and the centralizing condition, the equality
$(dvd)\beta=\beta(eue)$ in the image reduces to $v\beta=\beta u$.
The normalization moves to a comparison that has forgotten part of the condition for
preserving the original comparison.

\section{Classification of changes by the connections of operations}\label{sec:7.7}

We bring the classification so far back to the operations of software.
An operation that carries part of the state unchanged ties together the changes that are
invisible to the readout.
The update of a lens, and a protocol that proceeds while keeping an internal state, have
this same form.

\begin{definition}[Systems of operations keeping hidden states]\label{def:7.23}

Take a directed multigraph $Q=(V,E,s,t)$, a set $K$, and a group
$H\le\mathrm{Aut}(Q)$ of the permitted visible changes.
An automorphism $u$ of the graph consists of a permutation $u_V$ of the vertices and a
permutation $u_E$ of the named edges, and we take it to preserve starts and ends.
We define the states, the observation, and the operation of each edge by
\begin{equation*}
S=V\times K,\qquad o(v,k)=v,\qquad
T_e(s(e),k)=(t(e),k)
\tag{7.37}\label{eq:7.37}
\end{equation*}
Here $K$ is the internal state that the operations do not change.
We take a state change following a visible change $u\in H$ to be a bijection
$h:S\xrightarrow{\sim}S$ satisfying $oh=u_Vo$.
The condition of preserving the named operations is, moreover, $hT_e=T_{u_E(e)}h$ on the
fiber over the start of each edge.
The directions and the names of the edges are kept as the input of this equality.

We write $B_Q$ for the group of all pairs $(u,h)$ following the readout, and $A_Q$ for the
subgroup of those that also preserve the named operations.
The observation of a change is the visible projection $(u,h)\mapsto u$.
These correspond to all the changes, the compatible changes, and the observation of
Theorem~\ref{thm:7.9}.
In what follows we study between which vertices the condition of preserving the operations
propagates, and determine this subgroup concretely.

\end{definition}

\begin{theorem}[Preservation of operations and connected components]\label{thm:7.24}

A state change following a fixed $u$ can be written uniquely, using permutations for each
vertex ($\mathrm{Sym}(K)$ is the group of all permutations of $K$), as
\begin{equation*}
h(v,k)=(u_V(v),\phi_v(k)),\qquad \phi_v\in\mathrm{Sym}(K)
\tag{7.38}\label{eq:7.38}
\end{equation*}
A necessary and sufficient condition for preserving the operations is
\begin{equation*}
\phi_{s(e)}=\phi_{t(e)}\qquad(e\in E)
\tag{7.39}\label{eq:7.39}
\end{equation*}
Hence the changes preserving the operations are in one-to-one correspondence with the
choices of one permutation for each element of the set $\pi_0(Q)$ of connected components
with the directions forgotten.

\end{theorem}

\begin{proof}

By the condition on the observation, $h$ sends $\{v\}\times K$ to $\{u_V(v)\}\times K$.
The inverse map also returns to the corresponding fiber, so the second component is a
permutation $\phi_v$.
Evaluating both sides of the preservation of operations at $(s(e),k)$, the first components
are $u_V(t(e))$ in either case, and the second components are $\phi_{t(e)}(k)$ and
$\phi_{s(e)}(k)$ respectively.
This gives \eqref{eq:7.39}.

The quotient of $V$ by the equivalence relation generated by the relation $s(e)\sim t(e)$
between vertices is $\pi_0(Q)$.
Since \eqref{eq:7.39} is preserved under reflexivity, symmetry, and transitivity, the family
of permutations descends uniquely to this quotient.
Conversely, a family of permutations indexed by the connected components satisfies
\eqref{eq:7.39} once pulled back to the vertices.

\end{proof}

Since the operation of an edge carries the internal state as it is, the same permutation is
required at its two ends.
The connected components express the range over which this equality propagates regardless of
the directions of the edges.

\begin{theorem}[Split short exact sequence and the freedom of changes]\label{thm:7.25}

We take the projection from the group $A_Q$ of changes preserving the operations to the
visible changes to be $p:A_Q\to H$.
The composition is $(u,h)(u',h')=(uu',hh')$, and for the families of \eqref{eq:7.38} it is
\begin{equation*}
(\phi\star\phi')_v=\phi_{u'_V(v)}\phi'_v
\tag{7.40}\label{eq:7.40}
\end{equation*}
Setting $M_Q=\prod_{j\in\pi_0(Q)}\mathrm{Sym}(K)$, we obtain the split short exact sequence
\begin{equation*}
1\longrightarrow M_Q\longrightarrow A_Q
\xrightarrow{p}H\longrightarrow1
\tag{7.41}\label{eq:7.41}
\end{equation*}
The section is $\sigma(u)=(u,h_u)$ with $h_u(v,k)=(u_V(v),k)$.
The set of changes over each $u\in H$ is a right torsor under $M_Q$.
For finite $V,K$, the number of such changes is
\begin{equation*}
|p^{-1}(u)|=(|K|!)^{|\pi_0(Q)|}
\tag{7.42}\label{eq:7.42}
\end{equation*}
Moreover, writing the action of $H$ on the connected components as $u\cdot j$, we have
\begin{equation*}
A_Q\cong M_Q\rtimes H,\qquad
(u\cdot\psi)_j=\psi_{u^{-1}\cdot j}
\tag{7.43}\label{eq:7.43}
\end{equation*}

The product on the right is $(\psi,u)(\eta,u')=(\psi(u\cdot\eta),uu')$.

\end{theorem}

\begin{proof}

The equalities for the observation and the operations are preserved under composition and
inverses, so $A_Q$ is a group.
By Theorem~\ref{thm:7.24}, the fiber over $u=1$ is isomorphic as a group to $M_Q$.
Since $\sigma(u)$ preserves the operations and satisfies $\sigma(uu')=\sigma(u)\sigma(u')$
and $p\sigma=\mathrm{id}$, the short exact sequence splits.
The right torsor structure of each fiber is obtained by applying Theorem~\ref{thm:7.14} to
$p$.
Counting the families of permutations of Theorem~\ref{thm:7.24} gives \eqref{eq:7.42}.

For the semidirect product, index the family by the component at the destination
rather than at the source. Sending $(\psi,u)$ to
$h(v,k)=(u_V(v),\psi_{[u_V(v)]}(k))$ turns \eqref{eq:7.40} into the product
of \eqref{eq:7.43}. Every change has a unique such expression, giving a group isomorphism.

\end{proof}

\begin{corollary}[The version preserving the selected states]\label{cor:7.26}

Fix $k_0\in K$ and take $s_0(v)=(v,k_0)$ as the selected state.
If $hs_0=s_0u_V$ is also required, replace $\mathrm{Sym}(K)$ in
Theorems~\ref{thm:7.24} and~\ref{thm:7.25} by
$\mathrm{Sym}(K)_{k_0}=\{\phi\mid\phi(k_0)=k_0\}$.
In the finite case, the number of elements of each fiber is $((|K|-1)!)^{|\pi_0(Q)|}$.

\end{corollary}

\begin{proof}

Evaluating \eqref{eq:7.38} at $(v,k_0)$, the additional condition is $\phi_v(k_0)=k_0$.
The identity action on the hidden state satisfies this condition, so the same section can be
used.

\end{proof}

Using this classification, we find the condition under which preservation of the operations
can be determined from the visible change alone.
The kernel of the visible projection of the group $B_Q$ of changes following the readout is
$\prod_{v\in V}\mathrm{Sym}(K)$, and the part of it preserving the operations is $M_Q$.
Hence, by Theorem~\ref{thm:7.9}, preservation of the operations can be determined from the
visible change alone exactly when these two agree.
If there is an edge joining distinct vertices and $K$ has two or more elements, choosing
different permutations at its two ends gives a counterexample.
Reading the operations brings out the constraint forgotten by the visible projection.

\section{Consequences for lenses and protocols}\label{sec:7.8}

In Chapter~\ref{chap:1} we defined the states, the operations, and the
semantics-preserving morphisms of lenses and protocols independently, and put them in
correspondence with the construction of AAT having the roles of objects and typed operations.
Here we use that construction for the one-to-one correspondence of states linking the
situations before and after a refactoring.

Write the states before and after the change as $C,C'$, the reads as $g,g'$, and the
operations as $T_e,T'_e$.
Once the view and the names of the operations are fixed, what we require of a state
correspondence $h:C\xrightarrow{\sim}C'$ is $g'h=g$ and $hT_e=T'_eh$.
The former expresses the agreement of the results of the reads, and the latter the agreement
of ``convert and then operate'' with ``operate and then convert.''

Once one compatible correspondence $h_0$ has been obtained, every other correspondence can
be written uniquely as $h=h_0a$ using $a=h_0^{-1}h\in\mathrm{Sym}(C)$.
The condition on the reads then becomes $ga=g$ and the condition on the operations becomes
$aT_e=T_ea$.
Indeed, it suffices to compose $h_0^{-1}$ on the left of $h_0aT_e=T'_eh_0a=h_0T_ea$.
Once the states before and after are expressed in the same form through a reference
correspondence in this way, the automorphism groups of the previous section classify the
options for the state correspondence that remain in the refactoring.

\subsection*{Lenses: the total update ties all the views together}

\begin{proposition}[A common classification of lens changes]\label{prop:7.27}

Take a total lens $L=(C,g,p)$ of Definition~\ref{def:1.32} and its finite
reference fiber $K_L$. Using the product decomposition $C\cong V\times K_L$
of Proposition~\ref{prop:1.33}, form the graph whose vertices are views and whose
edge for each ordered pair $(v,w)$ is the update to $w$.
Any admissible visible group $H\le\mathrm{Sym}(V)$ moves edges by
$(v,w)\mapsto(u(v),u(w))$. This input satisfies Definition~\ref{def:7.23},
and its $A_Q$ is isomorphic as a group to the invertible lens changes satisfying
\begin{equation*}
gh=ug,\qquad h(p(c,w))=p(h(c),u(w))
\tag{7.44}\label{eq:7.44}
\end{equation*}

The set $V$ is nonempty because it has a reference view, and the graph is connected.
Thus we obtain
\begin{equation*}
1\longrightarrow\mathrm{Sym}(K_L)\longrightarrow A_L
\longrightarrow H\longrightarrow1,
\qquad
h(v,k)=(u(v),\phi(k))
\tag{7.45}\label{eq:7.45}
\end{equation*}

Here the action on connected components is trivial, so the semidirect product
is a direct product. There are $|K_L|!$ changes following each visible change.

\end{proposition}

\begin{proof}

In the product decomposition, $g(v,k)=v$ and $p((v,k),w)=(w,k)$.
The first part of \eqref{eq:7.44} is exactly the observation condition of
Definition~\ref{def:7.23}, and the second is preservation of operations on every edge.
The correspondence through the product decomposition works in both directions
and preserves identities and composition.
By Theorem~\ref{thm:7.24}, the necessary and sufficient condition is to use the
same permutation $\phi$ at every vertex.
The rest follows from Theorem~\ref{thm:7.25}.

\end{proof}

Requiring a selected section $s:V\to C$ to satisfy $gs=\mathrm{id}$ and
$p(s(v),w)=s(w)$ makes it $s(v)=(v,k_0)$ in the product decomposition.
Changes preserving this section as well correspond to Corollary~\ref{cor:7.26}.

\begin{example}[Carrying over the payment data even when the editing of the shipping
address is separated]\label{ex:7.28}

In the refactoring of the opening, we separate the processing that edits the shipping
address from the keeping of the payment data.
The \texttt{Order(address, payment)} before the change is moved to the pair
\texttt{OrderDetails(address)} and \texttt{PaymentDetails(payment)} after the change. This
repackaging is the reference state correspondence.
We take the shipping addresses to be $V=\{0,1\}$ and the internal code of the payment data,
which is not displayed, to be $K=\{0,1,2\}$.
Views 0 and 1 are the two shipping addresses, home and work, and $k_0=0$ represents the
unset value of the payment data.
The correspondence between the old and the new codes is also part of the proposed change,
and the correspondence of the unset value 0 is fixed.
This is the product lens whose read returns the shipping address and whose update rewrites
only the shipping address and carries over the payment data.

We fix the visible change to be the identity and align the states before and after into the
same product decomposition.
The correspondences preserving the read of the shipping address and the unset value are the
four given by choosing at each view whether to exchange 1 and 2.
The correspondences that also preserve the update are the two that use the same
correspondence of codes at both shipping addresses.

For instance, suppose the conversion branches on the shipping address, keeping the code at
view 0 while exchanging 1 and 2 only at view 1. For this correspondence $h$,
\begin{equation*}
h(p((0,1),1))=(1,2),\qquad
p(h((0,1)),1)=(1,1)
\tag{7.46}\label{eq:7.46}
\end{equation*}
Changing the shipping address first and then moving to the new format gives code 2, while
moving to the new format and then changing the shipping address gives code 1, so the payment
data carried over does not match.
A test of the display of the shipping address and a test on an order whose payment data is
unset both miss this difference.
What is needed is to check the update and the conversion of states in both orders for an
order that already has payment data.

\end{example}

\subsection*{Protocols: independent changes remain on independent components}

\begin{proposition}[Protocols keeping an internal state]\label{prop:7.29}

Take the finite protocol graph $Q$ of Definition~\ref{def:1.36} and a finite set $K$, and
take the state at each vertex to be $F(v)=K$ and the action of each generating edge to be
$F(e)=\mathrm{id}_K$.
We take the target of the observation to be the constant one-element set. All parallel paths
then give the same identity map, so we obtain a realization satisfying the specified path
equations $\Pi$.

Take the group $H\le\mathrm{Aut}(Q)$ of visible changes preserving the path equivalence
generated by $\Pi$.
The invertible changes of the realization following a fixed $u\in H$ are the vertexwise
bijections $\phi_v:K\to K$ satisfying
\begin{equation*}
\phi_{t(e)}F(e)=F(u_E(e))\phi_{s(e)}
\tag{7.47}\label{eq:7.47}
\end{equation*}
and are the same as the operation-preserving changes of Definition~\ref{def:7.23}.
The classification of Theorem~\ref{thm:7.25} by the split short exact sequence, the section,
and the torsor under the kernel therefore applies to the group of changes.

\end{proposition}

\begin{proof}

We gather all the states of the realization, keeping the control points distinct, into
$\coprod_{v\in V}F(v)=V\times K$.
The readout of the control point is the observation $o(v,k)=v$ of the previous section.
Equation~\eqref{eq:7.47} is equivalent to $\phi_{t(e)}=\phi_{s(e)}$.
By the same induction on paths as in Proposition~\ref{prop:1.38}, this equality extends to
all the executions.
The visible changes preserve the path equivalence as well, so the correspondence is
determined on the quotient category too.
Conversely, an invertible change of the realization returns to this family of permutations
once the vertex components are read.
Both directions preserve composition, so the groups and the fibers over each visible change
correspond.

\end{proof}

The changes with $u=1$ over a fixed graph are the automorphisms of this realization in
$\mathrm{Prot}(Q,\Pi,1)$, where the observation is constant to a one-element set.
When $u$ is moved as well, we record at the same time the isomorphisms following the
reindexing of the control points and of the names of the operations.
For protocols with general state transitions, the actual edge maps remain inside
\eqref{eq:7.47}.
The classification above by the connected components alone is a consequence obtained from
the input that each edge carries the internal state by the identity.

\begin{example}[Gathering the execution contexts of workers]\label{ex:7.30}

Suppose two workers A and B with the same role each receive and execute jobs independently.
Consider a refactoring that gathers the arguments scattered inside each worker into a data
type \texttt{JobContext} and tidies up the handoff from the queue to the execution.
Even when a job proceeds to Running, the context in which it runs must not change.

We represent by the internal code $K=\{0,1\}$ one component of the context whose handoff we examine.
A permutation is the choice of how to match the old and the new representations of the same
context.
Among the four control points $V=\{0,1,2,3\}$, we take 0 and 1 to be Queued and Running for
A, and 2 and 3 to be Queued and Running for B.
The edges $a:0\to1$ and $b:2\to3$ representing the start of execution carry this code as it
is.
In this model there is no edge handing a context between the workers, so it has two
connected components.

\end{example}

\begin{figure}[htbp]
\centering
\begin{tikzpicture}[
    x=1mm, y=1mm,
    state/.style={circle, draw, semithick, minimum size=8mm, inner sep=0pt, fill=white},
    box/.style={rounded corners=2mm, draw=black!45, semithick, fill=black!4},
    arr/.style={-{Stealth[length=5pt]}, semithick},
    lbl/.style={font=\small},
    note/.style={font=\small}
  ]
  \draw[box] (0,0) rectangle (62,36);
  \node[font=\small\bfseries] at (31,32) {Worker A};
  \node[state] (a0) at (12,20) {$0$};
  \node[state] (a1) at (50,20) {$1$};
  \draw[arr] (a0) -- (a1) node[midway, above, lbl] {$a$: start job}
                          node[midway, below, lbl] {preserves $k$};
  \node[lbl] at (12,11) {Queued};
  \node[lbl] at (50,11) {Running};
  \node[lbl] at (12,6)  {$\varphi_0$};
  \node[lbl] at (50,6)  {$\varphi_1$};
  \node[lbl] at (31,2.5) {Required: $\varphi_0 = \varphi_1$};
  \begin{scope}[xshift=68mm]
    \draw[box] (0,0) rectangle (62,36);
    \node[font=\small\bfseries] at (31,32) {Worker B};
    \node[state] (b2) at (12,20) {$2$};
    \node[state] (b3) at (50,20) {$3$};
    \draw[arr] (b2) -- (b3) node[midway, above, lbl] {$b$: start job}
                            node[midway, below, lbl] {preserves $k$};
    \node[lbl] at (12,11) {Queued};
    \node[lbl] at (50,11) {Running};
    \node[lbl] at (12,6)  {$\varphi_2$};
    \node[lbl] at (50,6)  {$\varphi_3$};
    \node[lbl] at (31,2.5) {Required: $\varphi_2 = \varphi_3$};
  \end{scope}
  \node[note] at (65,-7)
    {$K = \{0,1\}$: identity or swap per worker $\to$ $2 \times 2 = 4$ choices};
\end{tikzpicture}
\caption{\textbf{Aligning the correspondence at both ends of a handoff.} If the visible
change is the identity, $\phi_0=\phi_1$ is required for A and $\phi_2=\phi_3$ for B.
If, for instance, the codes are exchanged on the queue side only while the execution side
uses the previous correspondence, then it can still be confirmed that a job proceeds from
Queued to Running, but the correspondence of the contexts breaks.
Since there is no handoff between the two workers, the internal representation of A and that
of B can be chosen independently.}\label{fig:7.1}
\end{figure}

We compare the case where the placement of the workers is fixed with the case where A and B,
which have the same role, are exchanged.

\begin{xltabular}{\linewidth}{@{}LLL@{}}
\toprule
Condition checked & Placement of the workers fixed & Workers A and B exchanged \\
\midrule
\endhead
Following the readout of the control points & $2^4=16$ & $2^4=16$ \\
\addlinespace[3pt]
Named operations preserved as well & $2^2=4$ & $2^2=4$ \\
\bottomrule
\end{xltabular}

The exchange moves the vertices $0\leftrightarrow2$ and $1\leftrightarrow3$ and the edges
$a\leftrightarrow b$ at the same time.
The exchange leaving the internal state unchanged gives a section, and choosing
the permutations of the two components after it gives all candidates.
For the visible group consisting of the identity and the worker exchange, the kernel
is $S_2\times S_2$. The exchange interchanges these factors, so the full change
group is $(S_2\times S_2)\rtimes S_2$, of order 8.

\subsection*{Connection with the comparison-preserving group}

For lenses, restricting the correspondence of typed morphisms of
Proposition~\ref{prop:1.43} to invertible morphisms identifies the group of
semantics-preserving automorphisms for a fixed view with the group of typed automorphisms on
the side of AAT.
For protocols as well, the same proposition makes the invertible morphisms preserving the
vertex maps, the named operations, and the observation correspond.
When visible changes are allowed, specify the same reindexing of views or control
points and operation names on both sides.
Equations~\eqref{eq:7.44} and~\eqref{eq:7.47} are the commutativity conditions
for these typed operations.

Fix also an adapter $c:X\to Y$ and apply Theorem~\ref{thm:4.41} to the fully
faithful functor of Proposition~\ref{prop:1.43}.
Then the group of endpoint pairs satisfying the compatibility condition $vc=cu$
is identified between the original semantics and the typed construction.
Since the identification matches the endpoint components, it also matches the
projections, their kernels, and the torsors of nonempty fibers.
In this application take $E$ of Definition~\ref{def:7.2} to be the category of
typed objects, with the typed changes preserving the specified operations and
observations as admissible groups.
The classifications of Theorems~\ref{thm:7.4} and~\ref{thm:7.14} then become
classifications of the original semantics.

The equalities for the operations determine which subgroup of the comparison-preserving
group is chosen, and restrict the freedom of lifting.
For lenses the total update ties all the views together, and in the example of protocols a
change remains for each independent worker.
Their difference is thus explained by different inputs to the same theorem on connected components.

\section*{Summary of the Chapter}

In this chapter we distinguished the determination, the construction, and the classification
of all the candidates for changes preserving a comparison, and found the information needed
for each.

\begin{itemize}
\item \textbf{The group of changes preserving a comparison.} For a single comparison $c$, we
constructed the group of endpoint pairs satisfying $vc=cu$.
Its projections and kernels give the condition for following a change on one side
and the freedom among all following candidates.
Changes generated from a common original object preserve the comparison and are compatible
with the re-presentation of the input by isomorphisms
(\S\S\ref{sec:7.1}--\ref{sec:7.2}).
\item \textbf{Determination by observation.} A necessary and sufficient condition for
compatibility to be determinable from the observation $O$ of changes alone is
$\ker O\subseteq\Gamma$.
When a change that does not appear in the observation breaks the comparison, a compatible
change and an incompatible change have the same observation (\S\ref{sec:7.3}).
\item \textbf{Construction of lifts and reflection.} For the canonical normalization of AAT,
a section can be constructed from the common presentation of the accompanying data.
For the same generated comparison, even when compatible lifts exist, the compatibility of a
given change cannot in some cases be determined from the value after normalization alone
(\S\S\ref{sec:7.4}--\ref{sec:7.5}).
\item \textbf{The freedom among all the lifts.} A section gives one compatible lift, and the
kernel of the restricted homomorphism describes the other candidates.
Each fiber is a torsor under this kernel, and the loss of information and the options for
lifting can be read from the same restriction map (\S\ref{sec:7.6}).
\item \textbf{Classification by operations.} Using the connected components of the
operation graph, we described lens and protocol changes by a common split short exact
sequence. Correspondences agree along operations within each component, while independent
changes remain on independent components (\S\S\ref{sec:7.7}--\ref{sec:7.8}).
\end{itemize}

In the refactoring of the opening, we require that the payment data be carried over by the
same correspondence after the shipping address has been updated.
In the example of a lens, the four candidates preserving the display of the shipping address
and the unset value are narrowed down to the two that also preserve updates.
In the example of workers as well, the sixteen candidates preserving the correspondence of
the control points are narrowed down to the four that preserve the handoff of contexts.
These numbers show how much freedom of change remains once each operation ties the
correspondences between states together.

In Chapter~\ref{chap:8} we move on to the problem of assembling the original structure and
the structure-preserving morphisms from the local data read off from objects and changes.
The classification of changes in this chapter gives concretely the morphisms that this
reconstruction must keep and the freedom remaining in the observation.

\chapter{Presentation and Local Reconstruction}\label{chap:8}

\section*{Overview of the Chapter}

The question of this chapter is \textbf{when the overall geometry and the
structure-preserving morphisms can be recovered from local descriptions}.
For the changes preserving structure treated in Chapter~\ref{chap:7}, we treat the problem of
constructing a single morphism from the descriptions of its individual components.

As an example, consider the case where the modernization of an intrabank transfer system is
divided among the account, payment, and accounting teams.
The descriptions of the teams must agree on the shared transactions, amounts, and
completion conditions.
For the correspondence before and after the modernization as well, the correspondence
of transactions, the actions of the operations, and the evaluation of the Laws are
aligned across the teams.
In this chapter we construct the overall model from such divided descriptions and explain the
conditions under which the changes on the parts are assembled into a single
structure-preserving morphism.

We impose three conditions on reconstruction:
that the local data distinguish different overall morphisms,
that an overall morphism can be constructed from compatible correspondence tables,
and that an object can be constructed from a local description of the structure.
In this chapter we formulate these as separation, assembly of morphisms, and assembly of
objects, respectively, and prove them.
For full geometries we give these constructions from the primitive data of types, operations,
Laws, local geometry, and realization.
The main theorem shows that the category of full geometries and all their
structure-preserving morphisms is equivalent to the category of local models defined from
primitive data and compatibility conditions.

In the gluing of Chapter~\ref{chap:2} we fixed a geometry and a cover of contexts and glued
the states chosen on each context after correcting them.
There ``local'' refers to the range of the context in which a state is read.
In this chapter ``local'' refers to the readout of the finitely many types, values, and
equations that describe a geometry and a morphism.
The covers and the contexts themselves are included in the data that are read, and we assemble
the geometries equipped with them together with the structure-preserving morphisms.

What is used in the general reconstruction is a compatible family over all finite fragments.
Reconstruction requires specifying readout items sufficient to determine the objects and
morphisms, together with their compatibility conditions.
After the main theorem we treat the application to design changes and the cases where, by the laws of the operations, a finite
description suffices.
For finite models as well, we distinguish the choice of the necessary information from the
computational cost of checking and of reconstruction.

\begin{xltabular}{\linewidth}{@{}LLL@{}}
\toprule
Question & Construction & Outcome \\
\midrule
\endhead
When can objects and morphisms be finitely presented? & Finite exception codes and morphisms between fixed endpoints & Conditions for finite presentation and its limits \\
\addlinespace[3pt]
When can finite fragments be glued into a single table or function? & Restrictions and graphs of typed functions & Gluing of tables, and the conditions for being a function \\
\addlinespace[3pt]
What are the conditions for recovering the whole from local data? & Separation, assembly of morphisms, assembly of objects & Equivalence of categories given by the readout functor \\
\addlinespace[3pt]
From what are the geometries and morphisms of AAT assembled? & Primitive data of operations, Laws, geometry, and realization & Reconstruction of objects and of all structure-preserving morphisms \\
\addlinespace[3pt]
Can a design change be constructed from divided descriptions? & Local descriptions and correspondences for an intrabank transfer & Construction of the model and of the change from divided descriptions \\
\addlinespace[3pt]
Can an overall morphism be recovered from a finite table? & Unique extension by the laws of the operations & Conditions for recovering an overall morphism from a finite description \\
\addlinespace[3pt]
How can one check in a finite model? & Choice of readout items, enumeration and decision of equality & Distinction between the information sufficient for reconstruction and the computational procedure \\
\bottomrule
\end{xltabular}

The diagnostic invariance of Chapter~\ref{chap:3} guarantees that the verdict of a chosen
diagnosis does not change when the reading is changed.
The reflection of Chapter~\ref{chap:7} is the property that compatibility of the original
change follows from compatibility in the normalized image.
The reconstruction of this chapter recovers the objects and morphisms themselves.
Each of these properties holds for specified objects, morphisms, and observed quantities.

\section{Recovery of objects and morphisms by finite presentation}\label{sec:8.1}

In Chapter~\ref{chap:5} we showed that, if the presentations of the endpoints may be chosen,
a base morphism can be presented from a finite source and a finite or cofinite extraction.
When the presentations of the endpoints are fixed, not every semantic morphism can be
presented as a morphism between them.
In this section we use finite exception codes to make this difference explicit.

\subsection*{Finite exception tables and the one-point compactification}

\begin{definition}[Finite exception code]\label{def:8.1}

For a set $D$, we take a code to be a pair of a default value $b\in\{0,1\}$ and a finite set
$E\subseteq D$. The evaluation is
\begin{equation*}
\mathrm{ev}(b,E)(x)=
\begin{cases}
1-b,&x\in E,\\
b,&x\notin E
\end{cases}
\tag{8.1}\label{eq:8.1}
\end{equation*}
The default value 0 represents the characteristic function of a finite set, and the default
value 1 that of a cofinite set.

The default value specifies the value of the evaluation outside the finite exception set.
To represent it by a single point, we consider $D^+=D\sqcup\{\infty\}$.
We take each point of $D$ to be isolated, and we take a set containing $\infty$ to be open
if its complement is a finite subset of $D$.
This is the one-point compactification of the discrete space $D$.
When $D$ is finite, $\infty$ is isolated as well.

\end{definition}

\begin{proposition}[Topological meaning of codes]\label{prop:8.2}

Giving $\{0,1\}$ the discrete topology, we have the following bijection.
\begin{equation*}
\begin{aligned}
\mathrm{Code}(D)&\ \cong\ C(D^+,\{0,1\}),\\
(b,E)&\longmapsto
\left[
x\longmapsto
\begin{cases}
\mathrm{ev}(b,E)(x),&x\in D,\\
b,&x=\infty.
\end{cases}
\right]
\end{aligned}
\tag{8.2}\label{eq:8.2}
\end{equation*}
If $D$ is infinite, the code is determined uniquely by its evaluation on $D$ alone.
If $D$ is finite, then for every function $a:D\to\{0,1\}$ there are two codes, one with
default value 0 and one with default value 1.

\end{proposition}

\begin{proof}

The function obtained from a code is constant on the neighborhood $D^+\setminus E$ of
$\infty$ and is therefore continuous.
Conversely, for a continuous function $a$, put $b=a(\infty)$.
By continuity $a^{-1}(\{b\})$ is a neighborhood of $\infty$, so
$E=\{x\in D\mid a(x)\ne b\}$ is finite. The two constructions are mutually inverse.

For infinite $D$, a point can be taken outside the union of two finite exception sets.
If the evaluations agree, the default values agree there, and the exception sets agree as
well.
For finite $D$, the set $E_b=\{x\mid a(x)\ne b\}$ is finite for each $b$, and these give the
two codes.

\end{proof}

Theorem~\ref{thm:5.15} of Chapter~\ref{chap:5} therefore gives the following topological
characterization. Finiteness of the sources at both ends, together with the extension of each
extraction predicate of the target to a continuous function on $D^+$, characterizes the
finite presentation of a base morphism including the endpoint isomorphisms.
Finiteness of a discrete source is also equivalent, topologically, to compactness.
Indeed, that the open cover consisting of the individual points has a finite subcover means
that the total number of points is finite.
This characterization permits meaning-preserving changes of the enumeration of the
endpoints and of the correspondence of Atoms.

\subsection*{Morphisms between fixed presentations}

Next we investigate which semantic morphisms can be presented between fixed presentations.
We take as objects the codes of Chapter~\ref{chap:5} with a finite source, a normalization
table, and a finite exception table.
A morphism of presentations records a finite table of sources and a permutation moving only
finitely many Atoms.
We require that the normalization and the designated point be preserved and that the
transported extraction code be equal to the code of the target.
We identify the descriptions of permutations that give the same semantic morphism, and we
write $D_0$ for the decoding functor of the category obtained by this quotient.

We fix two presentations $P,Q$ and write $D_0P,D_0Q$ for their meanings.
We take $b_P(s)$ to be the default value of the extraction code used after normalizing the
source $s$.

\begin{proposition}[Existence condition for morphisms between fixed presentations]\label{prop:8.3}

Let $s_f$ be the source map and $\sigma_f$ the permutation of Atoms of a semantic base
morphism $f:D_0P\to D_0Q$. Then
\begin{equation*}
\exists h:P\to Q,\ D_0(h)=f
\quad\Longleftrightarrow\quad
\left\{
\begin{aligned}
&\{a\in D\mid \sigma_f(a)\ne a\}\text{ is finite},\\
&b_Q(s_f(s))=b_P(s)\quad\text{for all }s.
\end{aligned}
\right.
\tag{8.3}\label{eq:8.3}
\end{equation*}

\end{proposition}

\begin{proof}

A presented permutation moves only finitely many Atoms.
Moreover, the transport of a code does not change the default value, so the left-hand side
implies the right-hand side.

Conversely, assume the right-hand side.
Enumerating the finite set moved by $\sigma_f$ and writing the permutation on it as a table,
this permutation extends to $\sigma_f$ itself.
For the source map we use $s_f$.
Preservation of the normalization and of the designated point is a condition on $f$.
By exactness of the extraction, the evaluations of the transported code and of the target
code are equal.
Since the default values are equal as well, \eqref{eq:8.1} shows that the exception sets are
equal, and we obtain the equality of codes.

\end{proof}

The semantic morphisms not captured by this presentation are therefore only those that move
infinitely many Atoms and those whose default values after normalization do not agree at
corresponding sources.
The next example shows that the latter occurs even for a finite set of Atoms.

\begin{example}[Non-fullness between fixed presentations]\label{ex:8.4}

We take $D=\{a\}$, a single source, and the identity normalization.
If the extraction codes are taken to be $(0,\varnothing)$ and $(1,\{a\})$, then both
extractions are always false.
The semantic identity morphism exists, but since the default values differ, that identity
morphism cannot be presented between these two codes.

This example shows that recovery of the Hom-sets between fixed presentations does not follow
from agreement of the meanings of the objects.
The information about the default values that remains in a presentation constrains the
presentability of semantic morphisms.

\end{example}

\begin{corollary}[Normalization of presentations over finitely many Atoms]\label{cor:8.5}

If $D$ is finite, each extraction code can be replaced by
\begin{equation*}
R_{\mathrm{fin}}(b,E)
=
\bigl(0,\{a\in D\mid\mathrm{ev}(b,E)(a)=1\}\bigr)
\tag{8.4}\label{eq:8.4}
\end{equation*}
On the full subcategory of the objects for which all the default values of the extraction
codes used after normalization are 0, the functor $D_0$ is fully faithful.
Moreover $D_0R_{\mathrm{fin}}\cong D_0$, and the semantic objects and morphisms are
preserved.

\end{corollary}

\begin{proof}

Equation~\eqref{eq:8.4} does not change the evaluation.
A permutation of a finite set has finite support and the default values at both ends are 0,
so Proposition~\ref{prop:8.3} applies to every semantic morphism.
Faithfulness comes from having identified the descriptions that give the same semantic
morphism.
The construction $R_{\mathrm{fin}}$ also acts on morphisms, retaining the source map and the
Atom permutation.
After evaluation each component is an identity isomorphism, and naturality with respect to
morphisms holds as well.

\end{proof}

\section{Compatibility of finite fragments and gluing}\label{sec:8.2}

In the previous section we saw that, even when the meanings of objects agree, the morphisms
between fixed presentations need not be recoverable.
In what follows we read the individual components of morphisms in addition to the objects,
and investigate the conditions under which the overall objects and morphisms are constructed
from compatible local descriptions.
In the gluing of finite fragments we distinguish the agreement of values on the common part
from the conditions under which the glued table defines a function or a structure.
We define the former as compatibility of finite fragments.
The latter includes the existence and uniqueness of an output for each input, as well as
preservation of the operations.

\subsection*{Indices and restrictions of the readout}

\begin{definition}[Compatible family of finite fragments]\label{def:8.6}

We take $\Lambda$ to be the set of the items to be read and $V_q$ to be the type of the value
of an item $q$.
For a finite subset $S\subseteq\Lambda$, we take a finite fragment to be an element of
\begin{equation*}
T_S=\prod_{q\in S}V_q
\tag{8.5}\label{eq:8.5}
\end{equation*}
For $S\subseteq T$ we use the restriction map $r_{S,T}:T_T\to T_S$ that forgets components.
A family $(a_S)_S$ over all the finite subsets is said to be \textbf{compatible} if
\begin{equation*}
r_{S,T}(a_T)=a_S
\quad\text{whenever }S\subseteq T
\tag{8.6}\label{eq:8.6}
\end{equation*}
holds.

What is finite here is the number of items contained in a single fragment.
The types of the values may include Booleans, elements of a ring, finite polynomials,
references to declared sets, and so on.
A finite fragment therefore does not necessarily represent a bit string of finite length.

\end{definition}

\begin{lemma}[Gluing from single points]\label{lem:8.7}

Between the overall table and the compatible families of finite fragments there is the
following bijection.
\begin{equation*}
\prod_{q\in\Lambda}V_q
\ \cong\
\left\{(a_S)_S\ \middle|\ r_{S,T}(a_T)=a_S\right\}.
\tag{8.7}\label{eq:8.7}
\end{equation*}

\end{lemma}

\begin{proof}

From the overall table we take the restrictions to the finite subsets.
In the other direction, we define the value at $q$ to be the single-point fragment
$a_{\{q\}}(q)$.
By \eqref{eq:8.6}, each component of any finite fragment agrees with this value.
Restriction and gluing compose to the identity in both directions.

\end{proof}

\subsection*{Representation by graphs of typed functions}

\begin{definition}[Graph of a typed function]\label{def:8.8}

We describe a function between declared sets $A,B$ by a Boolean table $R(x,y)$ on the pairs
of an input and an output.
When the candidate pairs of types are also included in the indices, we set the value to 0 on
the pairs other than the selected $(A,B)$.
Over the valid types we require the following.
\begin{equation*}
\forall x\in A,\quad\exists!y\in B,\quad R(x,y)=1.
\tag{8.8}\label{eq:8.8}
\end{equation*}
That is, each row has exactly one output.
We take the readout of a function $f:A\to B$ to be
$R_f(x,y)=1\Longleftrightarrow f(x)=y$.

\end{definition}

\begin{lemma}[Readout and assembly of graphs of functions]\label{lem:8.9}

The functions $A\to B$ and the tables satisfying Definition~\ref{def:8.8} are in one-to-one
correspondence.
This correspondence preserves identities and composition.

\end{lemma}

\begin{proof}

We take $f_R(x)$ to be the unique output in \eqref{eq:8.8}.
By definition $R_{f_R}=R$ and $f_{R_f}=f$.
The table of the identity is the diagonal $x=y$.
Composition is given by
\begin{equation*}
(R_g\star R_f)(x,z)=1
\quad\Longleftrightarrow\quad
\exists y,\ R_f(x,y)=1\ \land\ R_g(y,z)=1
\tag{8.9}\label{eq:8.9}
\end{equation*}
which is equivalent to $g(f(x))=z$.
By the uniqueness in each row, \eqref{eq:8.8} holds after composition as well.

\end{proof}

\begin{example}[Compatibility and the conditions for a graph of a function]\label{ex:8.10}

When $A$ is nonempty, the finite fragments of the Boolean table all of whose values are 0 are
compatible.
No row, however, has an output, so \eqref{eq:8.8} is not satisfied.
Moreover, placing two different outputs on one input satisfies existence but loses
uniqueness.
Compatibility of finite fragments alone therefore does not guarantee the conditions for a
graph of a function.

\end{example}

\begin{definition}[Local conditions by primitive evaluation]\label{def:8.11}

A formula of primitive evaluation is an equation or an implication assembled from
finitely many components of types, values, and graphs of functions.
We take the support of a formula to be the finite set of the indices referred to in it.
We take a local condition to be such a formula quantified over the specified inputs.

For example, for the actions of operations $d:A\to B$ and $d':A'\to B'$, the condition that
correspondences $h_A:A\to A'$ and $h_B:B\to B'$ preserve the actions,
\begin{equation*}
h_B(d(a))=d'(h_A(a))
\tag{8.10}\label{eq:8.10}
\end{equation*}
becomes, once $a$ and the intermediate output are fixed, a finite formula evaluating four
applications of functions.
Requiring this condition for all $a$ yields preservation of the actions of the operations.

Likewise, \eqref{eq:8.8} takes an output $y$ for each $x$ and requires its value to be 1 and
the value at every other candidate $z$ to be 0.
The formula of the evaluation for each candidate is finite, but the condition expressing the
totality of a row requires quantification.
When a finite carrier is required, the existence of a finite list covering all the elements
is also included in the conditions.
That each formula of evaluation has finite support is distinguished from the whole condition being verifiable by finitely many checks.

\end{definition}

\section{Reconstruction principle by separation and assembly}\label{sec:8.3}

Before turning to individual structures, we state a sufficient condition for local
reconstruction.
Let $N:\mathcal{R}\to\mathcal{M}$ be a functor that reads the overall structures and
morphisms into local models.

\begin{theorem}[Local reconstruction principle]\label{thm:8.12}

If $N$ satisfies the following three conditions, we obtain an equivalence of categories
$\mathcal{R}\simeq\mathcal{M}$ whose forward functor is $N$.

\begin{enumerate}
\item \textbf{Separation.} For $f,g$ with the same endpoints, $N(f)=N(g)$ implies $f=g$.
\item \textbf{Assembly of morphisms.} From any $u:NX\to NY$ we can construct
$\mathrm{asm}(u):X\to Y$ with $N(\mathrm{asm}(u))=u$.
\item \textbf{Assembly of objects.} From any $m\in\mathcal{M}$ we can construct an object
$A(m)\in\mathcal{R}$ and an isomorphism
$\varepsilon_m:N(A(m))\xrightarrow{\sim}m$.
\end{enumerate}

Then for each Hom-set we have
\begin{equation*}
\mathrm{Hom}_{\mathcal{R}}(X,Y)
\ \underset{\mathrm{asm}}{\overset{N}{\rightleftarrows}}\
\mathrm{Hom}_{\mathcal{M}}(NX,NY),\qquad
N\mathrm{asm}=1,\quad
\mathrm{asm}N=1
\tag{8.11}\label{eq:8.11}
\end{equation*}
and to each local morphism there corresponds exactly one overall morphism.

\end{theorem}

\begin{proof}

For a morphism $f:X\to Y$, assembly of morphisms gives $N(\mathrm{asm}(Nf))=Nf$.
Using separation we obtain $\mathrm{asm}(Nf)=f$, and \eqref{eq:8.11} holds.
Hence $N$ is fully faithful.
Assembly of objects gives essential surjectivity, so we obtain an equivalence of categories.
This is the standard characterization of an equivalence of categories
(\cite[\href{https://stacks.math.columbia.edu/tag/02C3}{Lemma 4.2.19, Tag 02C3}]{Stacks}).

We construct the inverse functor as follows.
To an object we assign $A(m)$, and for $u:m\to n$ we put
\begin{equation*}
A(u)=
\mathrm{asm}
\bigl(\varepsilon_n^{-1}\,u\,\varepsilon_m\bigr)
\tag{8.12}\label{eq:8.12}
\end{equation*}
Preservation of identities and composition holds in the image of $N$, and by separation it
holds in $\mathcal{R}$ as well.
By \eqref{eq:8.12}, $\varepsilon$ is a natural isomorphism.
The other natural isomorphism is given by the unique assembly of
$\varepsilon_{NX}^{-1}:NX\to N(A(NX))$.
Naturality of this natural isomorphism and the triangle identity on the $\mathcal{R}$ side
are checked by applying $N$ to both sides.
Using \eqref{eq:8.12} and the inverse laws of assembly and reducing the composites of
$\varepsilon$ with its inverse, the morphisms being compared agree.
By separation we obtain the original equality. The triangle identity on the $\mathcal{M}$
side follows from the fact that the composite of $\varepsilon_{NX}$ with its inverse is the
identity morphism.

\end{proof}

In this principle, separation guarantees the uniqueness of morphisms.
The existence of the morphisms and the objects requires proofs that construct the respective
structures from the local conditions.
In what follows we define local objects and local morphisms from types and equations and
prove these three conditions.

\section{Local models of full geometries}\label{sec:8.4}

\subsection*{The category of realizations and the variants of morphisms}

\begin{construction}[Input for full geometries and the category of realizations]\label{cons:8.13}

We fix a set of Atoms and a variant of morphisms defined below, and we take $\Theta$ to be
this pair.
We write $\mathcal{R}_\Theta$ for the category whose objects are the full geometries of
Chapter~\ref{chap:1} and whose morphisms are all the structure-preserving morphisms of the
chosen variant.
The local model is defined from the primitive data that constitute these objects and
morphisms.

Changes of full geometries arise in two ways: comparing the presentations after
transport by an equality, and retaining the concrete correspondence maps between
the two presentations. We use the following two variants, one for each.

\end{construction}

\begin{definition}[The two variants of morphisms of full geometries]\label{def:8.14}

In the \textbf{representative variant} we use the morphisms of Definition~\ref{def:1.30}.
They have maps of the base and of the core, transport of covers and overlaps, and a
homomorphism of coefficient rings, and we require the raw system of the codomain to be equal
to the raw system of the domain with its coefficients and contexts transported.
The comparison maps for supports, axes, and observables are natural with respect to the
restrictions chosen there.

In the \textbf{explicit-map variant} the components of the base, the core, the covers, the
overlaps, and the coefficients are the same, and the equality of raw systems is replaced by
the following concrete correspondence.
On each context after transport, the raw coordinates, the local data, and the relation
indices are put in bijection.
This correspondence preserves the evaluation of polynomials using the map of coefficients,
the relations, and the restrictions.
The image of each context morphism is also specified, and restriction morphisms are required
to be sent to restriction morphisms.
Bijections are specified for the supports, axes, and observables of each context as well, and
the predicates of the readout and naturality along these context morphisms are required.
For the raw coordinates, for example, we specify which coordinate a given coordinate is sent
to, and apply that correspondence to the relations and to the polynomials of the
restrictions.

In either variant, the maps of the base and of architecture objects and the map of
coefficients need not be invertible.
The identity is the identity in each component, and composition is the composition of the
corresponding components.

\end{definition}

\subsection*{Primitive data and local compatibility conditions}

\begin{definition}[Local objects and local morphisms of full geometries]\label{def:8.15}

We fix $\Theta$.
The following table collects the data that are read in order to reconstruct the objects and
morphisms of full geometries.
Each datum is read as a value or as one point of the graph of a function.
The sets that serve as the domain and the codomain of a function are specified by the type
reference of the corresponding role.

\begin{xltabular}{\linewidth}{@{}LLL@{}}
\toprule
Layer read & Primitive data read from an object & Primitive data read from a morphism \\
\midrule
\endhead
Base and Atoms & Types of the sources, the designated point, the graph of the normalization, the extraction predicates & Graphs of the source map and of the bijection of Atoms \\
\addlinespace[3pt]
architecture & Composition relations of finite Atom families, object formation, types and endpoints of the operations and their actions on Atoms & Graphs of the object map and of the operation map \\
\addlinespace[3pt]
Laws and invariants & Indices, both sides, and evaluations of the Laws, detectors, types and values of the invariants, signatures & Correspondence of indices, preservation of evaluations and of actions, correspondence of signatures \\
\addlinespace[3pt]
Contexts and observables & Types and relations of the contexts, restrictions, ring operations and readouts of the observables & Correspondence of contexts, ring isomorphisms of the observables and preservation of restrictions \\
\addlinespace[3pt]
Selected geometry & Coverage requirements, overlaps, coefficient ring, raw coordinates, relations, polynomials of the restrictions & Preservation of covers and overlaps, map of coefficients, and, according to the variant, the equality of raw systems or the graph of the correspondence \\
\addlinespace[3pt]
Realization & Types of supports, axes, and observables, readout predicates, actions of the context morphisms & Comparisons of each realization component and naturality \\
\bottomrule
\end{xltabular}

For example, one item of the action of an operation $u:A\to B$ represents the truth value of
the proposition that its morphism of configurations sends an Atom $a$ to an Atom $b$.
One item of the multiplication of the coefficient ring represents the truth value of the
equation $a\cdot b=c$.
The restrictions of the raw system are written as finite polynomials, and the types of their
variables and coefficients are matched up.
For a local value we specify these primitive components, and the geometry or the morphism as
a whole is constructed by assembling them.

In the readout of a morphism we distinguish the items of the domain object, of the codomain
object, and of the morphism itself.
Their common index set is
\begin{equation*}
\Lambda_\Theta
=
\Lambda_\Theta^{\mathrm{src}}
\sqcup
\Lambda_\Theta^{\mathrm{tgt}}
\sqcup
\Lambda_\Theta^{\mathrm{hom}}.
\tag{8.13}\label{eq:8.13}
\end{equation*}
We assign a type of values to each item and define the finite fragments and restrictions of
Definition~\ref{def:8.6}.
In the case of an object, only the items of the object are used.
For the domain and codomain parts of a morphism we require agreement with the specified local
objects at the two ends.

We take a \textbf{local object} to be a compatible family of finite fragments of an object
satisfying the following conditions.

\begin{itemize}
\item The types declared for each role agree, and each graph of a function satisfies
Definition~\ref{def:8.8}.
\item The laws of each structure constituting a full geometry of Chapter~\ref{chap:1} hold as
formulas of primitive evaluation.
\item The family of Atoms selected in the core is finite.
\end{itemize}

We take a \textbf{local morphism} to be a compatible family of finite fragments of a morphism
satisfying the agreement of the types and of the two ends, the totality and uniqueness of the
graphs of functions, and the preservation conditions listed in the table.
For the components that must be bijections we specify the graph in the opposite direction and
the two inverse laws.
On a ring homomorphism we impose preservation of 0, 1, addition, and multiplication, and on a
Law we impose the correspondence of the indices and preservation of the evaluations of both
sides.
Naturality is a componentwise equation for each specified restriction or context morphism.
The conditions on the polynomials of the raw system are evaluated from the correspondence of
finitely many coefficients and variables.

For example, writing $e$ for the correspondence of Atoms, $\Phi$ for the correspondence of
operations, and $d(u)$ for the morphism of configurations of an operation $u$, the condition that operations and their actions on configurations be preserved reads
\begin{equation*}
e\bigl(d(u)_{\mathrm{At}}(a)\bigr)
=
d(\Phi(u))_{\mathrm{At}}\bigl(e(a)\bigr)
\tag{8.14}\label{eq:8.14}
\end{equation*}
Naturality for the restrictions and the observables requires that applying the restriction and
the component maps in either order give equal values.
Expanding each formula at a fixed input gives finite support.
Requiring the formulas at all inputs yields the preservation conditions for the structure as
a whole.

In the invariant condition of Definition~\ref{def:1.28}, the bijection of value ranges is only
required to exist and is not included among the components retained by a morphism.
To express such an existence condition, we attach to the local presentation auxiliary graphs
representing the maps in both directions.
On these too we impose the types, the inverse laws, preservation of evaluations, and
compatibility with the restrictions of finite fragments.
After constructing a compatible presentation, we identify presentations that differ only in
the choice of the auxiliary graphs.
This quotient leaves the components of a morphism that are to be retained, such as the source,
the operations, the contexts, the coefficients, and the raw system.

\end{definition}

\begin{lemma}[The local category and the readout functor]\label{lem:8.16}

The local objects and local morphisms of Definition~\ref{def:8.15} form a category
$\mathcal{M}_\Theta$.
The readout of the primitive data defines a functor
\begin{equation*}
N_\Theta:\mathcal{R}_\Theta\longrightarrow\mathcal{M}_\Theta
\tag{8.15}\label{eq:8.15}
\end{equation*}

\end{lemma}

\begin{proof}

The identity of a local morphism is defined by the diagonal graph, and composition
componentwise by \eqref{eq:8.9}.
Each graph is again total and unique.
The preservation equation for a composite follows from two preservation equations, and the
graphs in the opposite direction preserve the inverse laws by being composed in the reverse
order.
The implications required for the covers and the restrictions are preserved as well, by
chaining two implications.
In the formulas of the raw system we compose the map of coefficients and the substitution of
variables in turn.

The correspondences in both directions can also be composed for the auxiliary presentations
of the invariants.
Presentations with equal retained components give composites with equal retained components,
so composition is still defined after the identification.
The unit and associativity laws follow from the composition of graphs and from componentwise
composition.

If each component of a realization is read, the types and laws hold by the definition of that
realization.
Since the readout of graphs of functions preserves identities and composition, these readouts
form a functor.

\end{proof}

\section{The local reconstruction theorem for full geometries}\label{sec:8.5}

From the primitive data and the conditions of Definition~\ref{def:8.15} we construct full
geometries and their structure-preserving morphisms.
With this construction we prove the conditions of separation and assembly of
Theorem~\ref{thm:8.12} and obtain the equivalence between the category of full geometries and
the category of local models.

\subsection*{Assembly of full geometries}

\begin{lemma}[Assembly and separation for full geometries]\label{lem:8.17}

For both variants of full geometries, a full geometry can be assembled from a local object.
Moreover, between any full geometries $G,H$, the readout and the assembly relating the local
morphisms $N_\Theta G\to N_\Theta H$ to the morphisms of full geometries $G\to H$ are
mutually inverse.

\end{lemma}

\begin{proof}

We first construct the objects in the order of dependence of the types and then assemble the
morphisms between those objects.
Finally we check that these constructions and the readout of the primitive data are mutually
inverse.

We glue the finite fragments by Lemma~\ref{lem:8.7}.
We obtain the types of the sources, the Atoms, the architecture objects, the configurations,
and the operations, construct from each graph, by Lemma~\ref{lem:8.9}, the normalization and
the actions of the operations on Atoms, and extract the values of the extraction, of the
composition relations of finite Atom families, and of the object formation.
The designated point belongs to the sources by the conditions on the types.
The conditions on the normalization and the extraction, the finiteness of the selected family
of Atoms, the conditions on the object formation and the composition relations, and the
conditions on the endpoints of the operations and on the morphisms of configurations give the
structures of the base and of the architecture of Chapter~\ref{chap:1}.

Next we extract the invariants and the signatures and the indices of the Laws.
We then construct the contexts and their relations and the ring operations and restrictions
of the observables of each context, and check their respective axioms. Using these contexts
and rings, we construct both sides of the Laws and their evaluations.
Rereading the primitive conditions as equations about the values of the assembled functions,
the conditions for the Laws and for detection hold.
This determines the core.

Over this core we take the coverage requirements and the overlaps and construct the selected
site.
The type and the operations of the coefficient ring satisfy the ring axioms.
We assemble the raw coordinates, the local data, the relation indices, and the polynomials of
the restrictions of each context.
The primitive conditions include preservation of the relations and the equations for the
identities and the composition of the restrictions, so these form a raw system.
Finally we combine the supports, axes, and observables with their readouts and with the
actions of the context morphisms, and obtain a full geometry.
By this order, the types referred to by the later data are already determined by the data
constructed earlier.

For morphisms as well, we first assemble the graph of each retained component into a function.
For the components that are bijections, the two inverse laws give the inverse map.
The local conditions for the sources, the normalizations, the extractions, the operations, the
Laws, and the observables give the preservation equations for the assembled maps as they
stand.
The same holds for preservation of the covers and the overlaps and for the homomorphism of
coefficients.
In the representative variant we obtain the equality of transported raw systems.
In the explicit-map variant we obtain the correspondence of raw systems and naturality along
the actual context morphisms.
The auxiliary graphs of the invariants give the existence of the required bijections of
values.
Changing the choice of the auxiliary graphs does not change the retained components of the
assembled morphism.

The graph of each function in this construction agrees with the original table by
Lemma~\ref{lem:8.9}.
Agreement of the finite fragments follows from Lemma~\ref{lem:8.7} as well.
Conversely, assembling the readout of a morphism of full geometries, the value of each
function agrees with the value of the original morphism.
Equality of morphisms is determined by the equality of these retained components, so the
result of the assembly is equal to the original morphism.
For objects as well, each structure is recovered, and we obtain a canonical isomorphism
through the identification of the types.

\end{proof}

\subsection*{The main theorem}

\begin{theorem}[Local reconstruction of full geometries]\label{thm:8.18}

We fix the set of Atoms and the variant of morphisms of Construction~\ref{cons:8.13}.
The primitive readout functor $N_\Theta$ into the local model of Definition~\ref{def:8.15}
gives an equivalence of categories.
\begin{equation*}
\mathcal{R}_\Theta\simeq\mathcal{M}_\Theta.
\tag{8.16}\label{eq:8.16}
\end{equation*}
Its forward functor is $N_\Theta$ of \eqref{eq:8.15} itself.
For any full geometries $X,Y$, the maps
\begin{equation*}
\mathrm{Hom}_{\mathcal{R}_\Theta}(X,Y)
\ \underset{\mathrm{asm}_\Theta}
{\overset{N_\Theta}{\rightleftarrows}}\
\mathrm{Hom}_{\mathcal{M}_\Theta}(N_\Theta X,N_\Theta Y)
\tag{8.17}\label{eq:8.17}
\end{equation*}
are bijections, and the readout and the assembly compose to the identity in both directions.

\end{theorem}

\begin{proof}

By Lemma~\ref{lem:8.17}, for either variant of morphisms the primitive readout separates
morphisms and all local morphisms and local objects can be assembled.
The three conditions of Theorem~\ref{thm:8.12} are therefore satisfied, and we obtain an
equivalence of categories whose forward functor is $N_\Theta$.

\end{proof}

The Hom-sets of the main theorem consist of all the structure-preserving morphisms,
including the
noninvertible ones permitted by the chosen variant.
Their existence is derived from the primitive types and the preservation conditions by
Lemma~\ref{lem:8.17}.
Objects are recovered up to isomorphism, and the morphisms between fixed endpoints are
recovered uniquely.

\begin{corollary}[Recovery of evaluations, composition, and isomorphisms]\label{cor:8.19}

For $u:N_\Theta X\to N_\Theta Y$ and
$v:N_\Theta Y\to N_\Theta Z$ we have
\begin{equation*}
\begin{aligned}
\mathrm{asm}_\Theta(1)&=1,\\
\mathrm{asm}_\Theta(vu)
&=\mathrm{asm}_\Theta(v)\mathrm{asm}_\Theta(u).
\end{aligned}
\tag{8.18}\label{eq:8.18}
\end{equation*}
The evaluation of each primitive component agrees with the value of the original local table.
Moreover, $N_\Theta(f)$ is an isomorphism if and only if $f$ is an isomorphism.

\end{corollary}

\begin{proof}

Reading both sides of \eqref{eq:8.18}, they become equal by functoriality of the readout and
by \eqref{eq:8.17}.
By separation the original morphisms are equal as well.
The agreement of the evaluations follows from the definition of the output chosen in the
assembly of graphs.
If a local morphism has an inverse, assembling that inverse and using \eqref{eq:8.18} gives
the inverse of the overall morphism.
The converse direction is functoriality.

\end{proof}

\section{Application to design changes}\label{sec:8.6}

\subsection*{Modernization of an intrabank transfer system}

We make the intrabank transfer of the opening concrete as an example explaining the
reconstruction of full geometries.
The division of business responsibilities that distinguishes account management, payment
execution, and accounting is also found in the
\href{https://bian.org/wp-content/uploads/2024/12/BIAN-Service-Landscape-V9_0-Value-Chain-View.pdf}{BIAN Service Domain Landscape 9.0}.
In this example we treat a transfer in a single currency without fees, and we set the
following states and Laws.

Suppose that the old design kept the information about the transfer, the update of the
accounts, and the posting in a single transaction record.
In the new design these are separated into a transfer record, an account statement, and a
journal entry, linked by a shared transaction reference.
Each team describes the following structures and the correspondence from the old design.

\begin{xltabular}{\linewidth}{@{}LLL@{}}
\toprule
Team & Structures described & Matters made compatible with the other descriptions \\
\midrule
\endhead
Account & Accounts, balances, debit and credit operations & Correspondence between transactions and balance changes, sending and receiving accounts of a transfer, amount \\
\addlinespace[3pt]
Payment & Transfer requests, amounts, processing states, acceptance and completion operations & The transaction referred to by each record, the completion condition of a transfer \\
\addlinespace[3pt]
Accounting & Journal entries, transaction references, posting and correction operations & Correspondence between transactions and journal entries, agreement of amounts, balance of debits and credits \\
\bottomrule
\end{xltabular}

The division among the teams is a division of descriptions, and a finite fragment of this
chapter consists of the finitely many types, values, and equations that each team describes.
After declaring the shared types of transactions and amounts, we specify the endpoints and
actions of the operations, the evaluations of the Laws, the contexts and their restrictions,
and the data of the selected geometry and of the realization, following
Definition~\ref{def:8.15}.
When the same readout item appears in several descriptions, its values must agree.

We exhibit part of the compatibility conditions through a Law about the completion of a
transfer.
We fix a transfer identifier $q$ and a positive amount $m$ in a single currency.
We take $c,d,i,j\in\{0,1\}$ to be the values representing, respectively, the completion
verdict, the record of the debit from the sending account, the record of the credit to the receiving account, and the existence of the corresponding pair of debit and credit journal
entries.
For the last three, the determination also requires that the transaction reference and
the amount agree with $q,m$, respectively.
The completion condition of this example is $c=1\Longrightarrow d=i=j=1$.
Evaluating the values as integers, this can be expressed as the condition that the three
residuals $c(1-d)$, $c(1-i)$, and $c(1-j)$ are all 0.
Since each residual refers to finitely many values, it can be described as a primitive
evaluation in the sense of Definition~\ref{def:8.11}.

Suppose, for example, that for a completed transaction of the old design the payment team and
the accounting team specify the completed and posted records of the new design, while the
account team specifies the records before the debit and the credit.

\begin{xltabular}{\linewidth}{@{}LLLL@{}}
\toprule
Readout item & Transaction in the old design & Incompatible candidate correspondence & Candidate correspondence satisfying this Law \\
\midrule
\endhead
Completion verdict $c$ & 1 & 1 & 1 \\
\addlinespace[3pt]
Debit record $d$ & 1 & 0 & 1 \\
\addlinespace[3pt]
Credit record $i$ & 1 & 0 & 1 \\
\addlinespace[3pt]
Pair of journal entries $j$ & 1 & 1 & 1 \\
\bottomrule
\end{xltabular}

For the incompatible candidate, the first two residuals are 1.
Even though the types of the references to each record are correct, this candidate does
not preserve the evaluation of the Law required for a completed transfer.
The rightmost candidate satisfies this Law. To construct an overall morphism we further
require that the actions of the operations and all the preservation conditions listed in
Definition~\ref{def:8.15} be satisfied.

In the assembly of objects we glue these primitive data by Lemma~\ref{lem:8.7} and construct
the full geometry of the new design by Lemma~\ref{lem:8.17}.
Once the necessary data are collected and a local object is defined, its realization is
obtained together with an isomorphism to the readout.
In the assembly of morphisms we fix the set of Atoms and the variant of morphisms as a common
input $\Theta$.
We write $G_{\mathrm{old}},G_{\mathrm{new}}$ for the full geometries of the old and the new
design and specify the correspondences of the teams as a local morphism
$u:N_\Theta G_{\mathrm{old}}\to N_\Theta G_{\mathrm{new}}$.
Theorem~\ref{thm:8.18} gives a unique structure-preserving morphism
$h:G_{\mathrm{old}}\to G_{\mathrm{new}}$ whose readout is this morphism.
The objects and morphisms constructed here are those of an AAT model in which the data of
types, operations, Laws, local geometry, and realization are made explicit.

\begin{figure}[htbp]
\centering
\begin{tikzpicture}[
    x=1mm, y=1mm,
    team/.style={rounded corners=1.5mm, draw=black!45, semithick, fill=black!3,
                 align=center, font=\footnotesize, inner sep=2mm, text width=44mm,
                 minimum height=26mm,
                 execute at begin node={\hyphenpenalty=10000 \exhyphenpenalty=10000 }},
    wide/.style={rounded corners=1.5mm, draw=black!60, semithick, fill=white,
                 align=center, font=\footnotesize, inner sep=2mm, text width=136mm},
    result/.style={rounded corners=1.5mm, draw=black!45, semithick, fill=black!3,
                align=center, font=\footnotesize, inner sep=2mm, text width=62mm,
                minimum height=17mm},
    arr/.style={-{Stealth[length=5pt]}, semithick},
    head/.style={font=\small\bfseries},
    note/.style={font=\small}
  ]
  \node[head] at (75,97) {Local descriptions for an internal transfer};
  \node[note] at (75,90) {Declared structures and old-to-new correspondences};
  \node[team] (acct) at (25,72)
    {\textbf{Account management}\\[1pt]
     Accounts and balance changes\\ Debit / credit operations\\ Account correspondences};
  \node[team] (pay) at (75,72)
    {\textbf{Payment management}\\[1pt]
     Requests, amounts and states\\ Acceptance / completion\\ Transaction correspondences};
  \node[team] (jrnl) at (125,72)
    {\textbf{Accounting}\\[1pt]
     Journal entries and references\\ Posting / correction operations\\ Journal-entry correspondences};
  \node[wide] (cond) at (75,44)
    {\textbf{All required primitive data and compatibility conditions}\\[1pt]
     Shared readings agree on types, transaction references, amounts and completion\\
     Actions, laws, geometry and realizations satisfy the declared conditions};
  \node[result] (obj) at (39,17)
    {\textbf{Object assembly}\\[1pt] An AAT model, up to isomorphism};
  \node[result] (mor) at (111,17)
    {\textbf{Morphism assembly}\\[1pt]
     A unique structure-preserving map\\ between fixed old and new AAT models};
  \draw[arr] (acct.south) -- (acct.south |- cond.north);
  \draw[arr] (pay.south)  -- (pay.south  |- cond.north);
  \draw[arr] (jrnl.south) -- (jrnl.south |- cond.north);
  \draw[arr] (obj.north |- cond.south) -- (obj.north);
  \draw[arr] (mor.north |- cond.south) -- (mor.north);
  \node[note] at (75,3) {AAT local reconstruction: Theorem 8.18};
\end{tikzpicture}
\caption{\textbf{Descriptions of the primitive data and the correspondences by the account,
payment, and accounting teams.}
From all the required data and the compatibility conditions, objects are reconstructed up to
isomorphism and the morphisms between fixed endpoints are reconstructed uniquely.
In addition to the agreement of the types of the individual records, we require preservation of the actions of the operations, of the evaluations of the Laws,
and of the selected geometry and the realization.}\label{fig:8.1}
\end{figure}
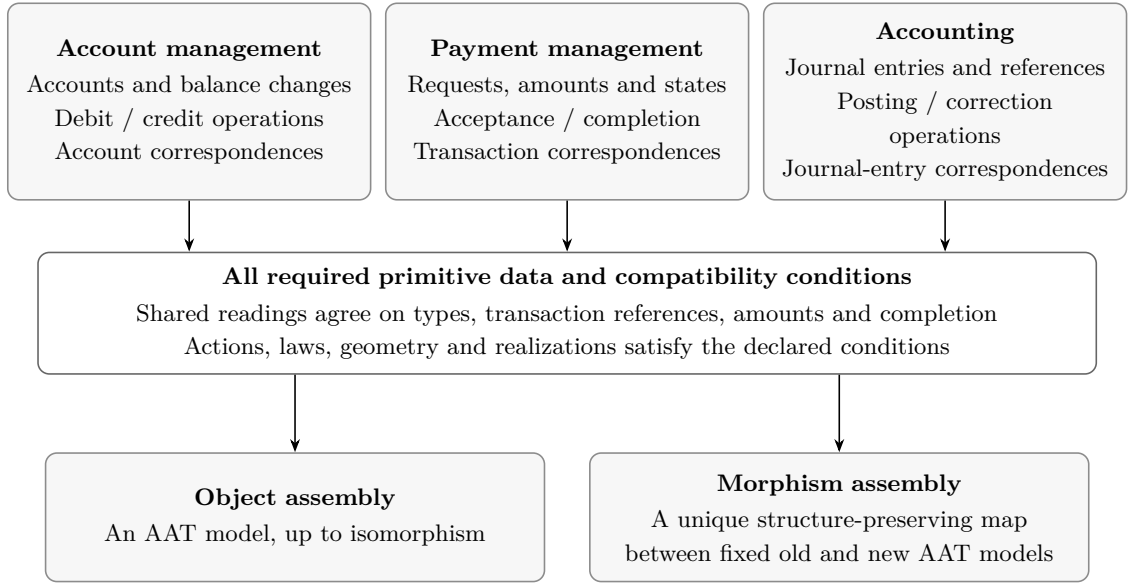

By this construction, the compatibility of a design change divided among teams can be
described as the existence of an overall structure-preserving morphism.
For another modernization proposal as well, we can fix the same old design and the structures
to be retained and compare the proposals by checking whether the local correspondences to that
proposal satisfy the preservation conditions.

\subsection*{Application to normalization and generated comparisons}

\begin{example}[Tagged operations]\label{ex:8.20}

The operation $(u,t)$ of Example~\ref{ex:6.25} carries an action $u$ on configurations
together with a tag $t\in\{0,1\}$ that does not affect that action.
We take $\Omega$ to be the set of all architecture objects and put, for
$\chi:\Omega\to\{0,1\}$,
\begin{equation*}
T_\chi(u,t)=(u,t+\chi(A))
\quad\text{for }u:A\to B
\tag{8.19}\label{eq:8.19}
\end{equation*}
Addition is modulo 2.
Since the Laws, the invariants, and the geometry of this example do not depend on the value of
the tag, $T_\chi$ is an automorphism of the full geometry in the explicit-map variant.
The uniform tag flip is $\chi=1$.
We call $\chi$ a selection of flips at each starting object.

Theorem~\ref{thm:8.18} recovers all these $T_\chi$, together with the canonical normalization
of the same example, as morphisms of the original full geometry.
By reading the tag component, two morphisms with equal actions on configurations can also be
distinguished.
The equations for the composition with the normalization hold in the local category as they
stand as well, by Corollary~\ref{cor:8.19}.

\end{example}

\begin{example}[Objects, morphisms, and changes of endpoints for generated comparisons]\label{ex:8.21}

Using the input of Constructions~\ref{cons:5.37}--\ref{cons:5.40},
we take $G_{\mathrm d},G_{\mathrm v}$ to be the full geometries of the two routes and
$\overline\alpha:G_{\mathrm d}\xrightarrow{\sim}G_{\mathrm v}$ to be the canonical
comparison.
Writing $e,d$ for the normalizations of Chapter~\ref{chap:6} at the two ends, the generated
comparison satisfies
\begin{equation*}
\overline\beta=d\overline\alpha,\qquad
e^2=e,\quad d^2=d,\qquad
d\overline\alpha=\overline\alpha e
\tag{8.20}\label{eq:8.20}
\end{equation*}
Hence $d\overline\beta=\overline\beta=\overline\beta e$.

These are objects and morphisms of the representative variant, and Theorem~\ref{thm:8.18}
recovers all of the original geometry used in the generation, the geometries of the two
routes, and $\overline\alpha,\overline\beta,e,d$.
It also recovers all the automorphisms at the two ends and the commutative squares between
them.
By Corollary~\ref{cor:8.19}, \eqref{eq:8.20} is preserved in the local model as well.

As an example, take a comparison diagram over a core satisfying $\mathrm{Ad}(P_z)$,
with two faces having the same endpoints and with both routes acting as identities.
The input of Construction~\ref{cons:5.37} includes the choice of face and cochain
used for the comparison. Take the first face to be the identity, and for the second
use an input exchanging two of three signature axes.
Take coefficients in $\mathbb{Z}$ and apply Constructions~\ref{cons:5.37}--\ref{cons:5.40}
to the second face.

For a choice whose initial defect exchanges those two axes, the defect is not the
identity, so the normalization selection condition holds.
Theorem~\ref{thm:6.16} gives a nontrivial idempotent normalization and a noninvertible
$\overline\beta$. On the other hand, choosing the identity cochain on the same
comparison diagram makes the normalization the identity and gives
$\overline\beta=\overline\alpha$.
In both cases, reconstruction applies to the selected comparison and all admissible
morphisms at its two endpoints.

\end{example}

\section{Application of the reconstruction principle to the semantics of CS}\label{sec:8.7}

We apply the local reconstruction principle of Theorem~\ref{thm:8.12} to the lenses and the
protocols with observations introduced in Chapter~\ref{chap:1}.
We first recover the objects and the general morphisms from the primitive data and the
preservation conditions.
Next, using the respective laws, we exhibit the reconstruction from a finite description, the
reference fiber or the table of generating edges.

\subsection*{Local models and assembly}

\begin{construction}[Local models for the semantics of CS]\label{cons:8.22}

We choose the input $\Theta$ from the following table and write $\mathcal{R}_\Theta$ for the
category of realizations.

\begin{xltabular}{\linewidth}{@{}LLL@{}}
\toprule
Input & What is fixed & Category of realizations $\mathcal{R}_\Theta$ \\
\midrule
\endhead
lens & The set $V$ of views and a reference value $v_0\in V$ & The lenses that satisfy the three laws and have a finite reference fiber, and all the maps preserving get and put \\
\addlinespace[3pt]
protocol & A finite presentation $Q$ and an observation functor $O$ on the quotient path category & The realizations whose state set at each vertex is finite, and all the natural transformations preserving the observations \\
\bottomrule
\end{xltabular}

The objects and morphisms are defined from the semantics of Definitions~\ref{def:1.32}
and~\ref{def:1.36}, respectively.
Their operations and Laws are connected to descriptions by Atoms and operations through the
typed construction of \S\ref{sec:1.10}.
An arbitrary map between reference fibers, and a noninjective state map, are also morphisms if
they satisfy the respective preservation conditions.

For the local description we use the following primitive data.

\begin{xltabular}{\linewidth}{@{}LLL@{}}
\toprule
Input & Data read from an object & Data read from a morphism \\
\midrule
\endhead
lens & The type of the states, the graphs of get and of each put & The graph of the state map \\
\addlinespace[3pt]
protocol & The type of the states at each vertex, the graphs of the generating edges and of the observations & The graphs of the maps at each vertex \\
\bottomrule
\end{xltabular}

Types are specified by the reference for their role, and we take each value of a graph of a
function to be a readout item.
Among the items of a morphism we distinguish the source, the target, and the morphism itself,
and we use the finite fragments and restrictions of \S\ref{sec:8.2}.
We take a local object to be a compatible family of finite fragments satisfying the agreement
of the types and the totality and uniqueness of each graph.
On lenses we impose the three laws and the finiteness of the reference fiber, and on protocols
the relations of paths, the naturality of the observations, and the finiteness of the state
set at each vertex, as conditions of primitive evaluation.
Finiteness is expressed by the existence of a finite list covering all the elements of the set
in question.

On a local morphism we require agreement with the specified objects at the two ends and the
totality and uniqueness of the graphs.
For lenses we impose preservation of get and put, and for protocols preservation of the
generating edges and of the observations, by componentwise equations.
The preservation equations for the operations and the observations have finite support once
the input is fixed.
The diagonal graph and the composition of \eqref{eq:8.9} preserve these conditions, so we
obtain a local category $\mathcal{M}_\Theta$.
We write $N_\Theta:\mathcal{R}_\Theta\to\mathcal{M}_\Theta$ for the functor that reads each
component of a realization.

\end{construction}

\begin{lemma}[Assembly and separation for lenses]\label{lem:8.23}

From the local objects and local morphisms for lenses, the lenses and their general morphisms
can be assembled.
The readout and the assembly are mutually inverse on each Hom-set, and the objects are
recovered up to isomorphism as well.

\end{lemma}

\begin{proof}

From a local object we take the state set $C$ and assemble the graph of get and the graph of
put at each view $v$.
We write $g:C\to V$ and $p:C\times V\to C$ for the resulting maps.
The three pointwise conditions of the local object are
\begin{equation*}
p(c,g(c))=c,\qquad
g(p(c,v))=v,\qquad
p(p(c,v),w)=p(c,w)
\tag{8.21}\label{eq:8.21}
\end{equation*}
Hence $(C,g,p)$ is a total lens.
The local object also carries the condition that a finite list covering $g^{-1}(v_0)$ exists,
so its reference fiber is finite.

We assemble the graph of a local morphism into $h:C\to C'$.
The two local conditions of a morphism give
\begin{equation*}
g'h=g,\qquad
h(p(c,v))=p'(h(c),v)
\tag{8.22}\label{eq:8.22}
\end{equation*}
so $h$ is a morphism of lenses.
By Lemmas~\ref{lem:8.7} and~\ref{lem:8.9}, the two directions are mutually inverse for
all the graphs and finite fragments.

\end{proof}

\begin{lemma}[Assembly and separation for protocols]\label{lem:8.24}

From the local objects and local morphisms for protocols with observations, the realizations
and their general morphisms can be assembled.
The readout and the assembly are mutually inverse on each Hom-set, and the objects are
recovered up to isomorphism as well.

\end{lemma}

\begin{proof}

For each vertex $v$ of the finite presentation $Q$ we obtain the state set $X_v$.
By the condition that a finite list covers all the states, $X_v$ is finite.
We assemble the graph of a generating edge $e:v\to w$ into $X_e:X_v\to X_w$ and the graph of
the observation into $o_v:X_v\to O(v)$.

A local object satisfies the equality of actions for each specified relation $p=q$.
Moreover, the local condition at each generating edge gives
\begin{equation*}
o_wX_e=O(e)o_v
\tag{8.23}\label{eq:8.23}
\end{equation*}
By Proposition~\ref{prop:1.37}, there is therefore a unique realization $(X,o)$ with these
finite states, edge maps, and observations.

From a local morphism we obtain the vertex maps $h_v:X_v\to Y_v$.
The local conditions are
\begin{equation*}
h_wX_e=Y_eh_v,\qquad
o_v^Yh_v=o_v^X
\tag{8.24}\label{eq:8.24}
\end{equation*}
where the first equation holds at all the generating edges and the second at all the vertices.
By Proposition~\ref{prop:1.38}, these vertex maps determine a unique semantics-preserving
morphism $h:X\Rightarrow Y$.
All the components are obtained uniquely from the graphs, and the extension is unique as
well.
Since the maps of Lemmas~\ref{lem:8.7} and~\ref{lem:8.9} are mutually inverse, so are
the readout and the assembly.

\end{proof}

\begin{corollary}[Local reconstruction for the semantics of CS]\label{cor:8.25}

For each input of Construction~\ref{cons:8.22}, the primitive readout functor $N_\Theta$ gives
an equivalence of categories $\mathcal{R}_\Theta\simeq\mathcal{M}_\Theta$.
The objects are recovered up to isomorphism, and between any fixed realizations the readout
and the assembly are mutually inverse, including for noninvertible morphisms.
This reconstruction preserves identities, composition, and the primitive evaluations, and
reflects isomorphisms.

\end{corollary}

\begin{proof}

Lemmas~\ref{lem:8.23} and~\ref{lem:8.24} each give the three conditions of
Theorem~\ref{thm:8.12}.
We therefore obtain an equivalence of categories whose forward functor is the primitive
readout.
Preservation of identities and composition and reflection of isomorphisms follow from
functoriality of the readout and from the mutual inverses on each Hom-set.
Agreement of the primitive evaluations follows from the mutual inverses of the readout and the
assembly in Lemmas~\ref{lem:8.7} and~\ref{lem:8.9}.

\end{proof}

\subsection*{Reconstruction of lenses from the reference fiber}

We restate Proposition~\ref{prop:1.34} of Chapter~\ref{chap:1} as the table used for the
reconstruction of morphisms and its extension.

\begin{proposition}[Tables on the reference fiber and general morphisms of lenses]\label{prop:8.26}

Let $L=(C,g,p)$ and $L'=(C',g',p')$ be lenses with the same views $V$ and the same reference
value $v_0$.
Put $K=g^{-1}(v_0)$ and $K'=(g')^{-1}(v_0)$.
Every map $t:K\to K'$ extends uniquely to a morphism of lenses by
\begin{equation*}
\mathrm{ext}(t)(c)
=
p'\bigl(t(p(c,v_0)),\,g(c)\bigr)
\tag{8.25}\label{eq:8.25}
\end{equation*}
Here $p(c,v_0)$ is read as an element of $K$.
This extension and the restriction to the reference fiber are mutually inverse.

\end{proposition}

\begin{proof}

Equation~\eqref{eq:8.25} is the extension \eqref{eq:1.19} of Proposition~\ref{prop:1.34}.
By the same proposition, this extension preserves get and put and is inverse to the
restriction to the reference fiber.

\end{proof}

The table needed for the reconstruction of a morphism is a map between $K,K'$, and finiteness
of the set $V$ of views is not assumed.
The correspondence at each view is determined as the unique extension by the law for put.

\begin{example}[Merging complementary information]\label{ex:8.27}

We take the state of a product lens to be $(a,k)\in V\times K$, take get to read the first
component, and take put to replace the first component by the specified view.
From a map $t:K\to K'$ of reference fibers we obtain
\begin{equation*}
h(a,k)=(a,t(k))
\tag{8.26}\label{eq:8.26}
\end{equation*}
For example, taking $K=\{0,1,2\}$ and $K'=\{0,1\}$ with $t(0)=t(1)=0$ and $t(2)=1$ gives a
noninjective morphism merging two values of the complementary information.
Equation~\eqref{eq:8.26} preserves get and put and, by Proposition~\ref{prop:8.26}, is
determined uniquely by this table on the reference fiber.

\end{example}

\subsection*{Reconstruction of protocols from the generating edges}

\begin{proposition}[Unique extension of an adapter from the generating edges]\label{prop:8.28}

For two realizations $(X,o^X),(Y,o^Y)$ of a protocol with observations, finite tables
$h_v:X_v\to Y_v$ at each vertex extend to a general morphism if and only if they satisfy the
first equation of \eqref{eq:8.24} at all the generating edges and the second at all the
vertices.
The extension is unique, and restriction and extension are mutually inverse.

\end{proposition}

\begin{proof}

Necessity follows from the naturality and the preservation of observations of
Definition~\ref{def:1.36}.
Sufficiency and uniqueness of the extension are obtained by applying
Proposition~\ref{prop:1.38} to the vertex tables $h_v$.
The vertex components of the resulting morphism are the original tables themselves, so
restriction and extension are mutually inverse.

\end{proof}

Preservation of observations is required for isolated vertices with no generating edges as
well.

For lenses, a map on the reference fiber determines, by the law for put, a morphism on all the
states.
For protocols, the maps at each vertex determine, by naturality at the generating edges, a
natural transformation on all the paths.
In both cases the laws of the operations are used to extend a local map to an overall
morphism.

\begin{example}[Preservation of observations and naturality]\label{ex:8.29}

Consider a protocol with two vertices and one generating edge $e:v\to w$.
We take the state set at each vertex of both realizations to be $\{0,1\}\times\{0,1\}$, and
the observation functor assigns $\{0,1\}$ to both vertices and the identity map to the edge.
We take the observation of each state to be $(a,k)\mapsto a$.
We define the actions of the edge on states by $X_e(a,k)=(a,k)$ and $Y_e(a,k)=(a,0)$.
Taking the identity map at both vertices as the candidate correspondence, the observations are
preserved.
At $k=1$, however, $h_wX_e(a,1)=(a,1)$ while $Y_eh_v(a,1)=(a,0)$, so naturality fails.
The preservation condition at the generating edges thus detects a difference of actions that
the observations alone cannot distinguish.

\end{example}

\subsection*{Compatibility of finite presentations with the primitive readout}

\begin{corollary}[Compatibility with the finite decoding functor]\label{cor:8.30}

For lenses and for protocols there are, respectively, the following category
$\mathcal{P}_\Theta$ of finite tables and decoding functor
$D_\Theta:\mathcal{P}_\Theta\to\mathcal{R}_\Theta$.

We take the presentation objects for lenses to be the natural numbers $n$ and the morphisms to
be arbitrary tables $\mathrm{Fin}(n)\to\mathrm{Fin}(m)$.
The decoding functor sends $n$ to the product lens $V\times\mathrm{Fin}(n)$ and a table $t$ to
$(v,k)\mapsto(v,t(k))$.

We take the presentation objects for protocols to be a natural number $n_v$ at each vertex, a
table for each generating edge, and a table of the observations, satisfying the relations and
the naturality of the observations.
A morphism is a vertex table preserving the generating edges and the observations.
The decoding functor extends the tables of the generating edges to the composites of paths.

Both decoding functors are equivalences of categories, and the composite with the primitive
readout of Corollary~\ref{cor:8.25},
\begin{equation*}
\mathcal{P}_\Theta
\xrightarrow{D_\Theta}
\mathcal{R}_\Theta
\xrightarrow{N_\Theta}
\mathcal{M}_\Theta
\tag{8.27}\label{eq:8.27}
\end{equation*}
is an equivalence of categories as well.
Moreover, the route that assembles the local model from the primitive data and then reads the
reference fiber or the table of generating edges agrees, up to natural isomorphism, with the
route that reads them directly from the realization.

\end{corollary}

\begin{proof}

For lenses, by Proposition~\ref{prop:8.26} the decoding functor is a bijection on each
Hom-set.
For objects, combining the isomorphism of Proposition~\ref{prop:1.33},
\begin{equation*}
C\xrightarrow{\sim}V\times K,\qquad
c\longmapsto\bigl(g(c),p(c,v_0)\bigr)
\tag{8.28}\label{eq:8.28}
\end{equation*}
with an enumeration of the finite set $K$, every lens becomes isomorphic to an image of the
decoding functor.
For protocols we enumerate the finite state set at each vertex and transport the generating
edges and the observations to that enumeration.
Proposition~\ref{prop:8.28} gives the bijection of the Hom-sets, so we obtain an equivalence
of categories in the same way.
Equation~\eqref{eq:8.27} is the composite of these with Corollary~\ref{cor:8.25}.

To exhibit the last natural isomorphism, we take $F_\Theta$ to be the functor that reads the
reference fiber or the table of generating edges and $A_\Theta$ to be the inverse functor of
the equivalence of Corollary~\ref{cor:8.25}.
Applying $F_\Theta$ to the unit isomorphism $1\cong A_\Theta N_\Theta$ of
Theorem~\ref{thm:8.12}, we obtain
\begin{equation*}
F_\Theta
\ \cong\
F_\Theta A_\Theta N_\Theta
\tag{8.29}\label{eq:8.29}
\end{equation*}
Hence the two routes agree not only on the objects but also on the general morphisms.

\end{proof}

The procedure for obtaining a morphism from a finite presentation differs between the two
semantics.
For lenses we take as input a table between the finite reference fibers and construct by
\eqref{eq:8.25} a map on all the states.
Preservation of get and put is guaranteed on all the states by Proposition~\ref{prop:8.26}.
When the set of views is infinite, the preservation conditions for an arbitrarily given map on all
the states need not be checkable on finitely many states alone.

For protocols we enumerate the finite states at each vertex and check, for a candidate vertex
table, the naturality of \eqref{eq:8.24} at all the generating edges and the preservation of
observations at all the vertices.
When the presentation object itself is also checked, in addition to the tables of the edges
and the naturality of the observations, we compare the actions of paths of finite length for
the specified finitely many relations.
If the evaluation of the necessary tables and the decision of equality are computable, these
can be carried out as finite checks.

\begin{proposition}[Connection with retracts, the Karoubi envelope, and comparison squares]\label{prop:8.31}

For both inputs, lenses and protocols, the category of realizations is idempotent complete.
Writing $D_\Theta$ for the finite decoding functor, \eqref{eq:8.27} extends to
\begin{equation*}
\mathrm{Kar}(\mathcal{P}_\Theta)
\simeq
\mathcal{R}_\Theta
\simeq
\mathcal{M}_\Theta
\tag{8.30}\label{eq:8.30}
\end{equation*}
and agrees, up to natural isomorphism, with the original decoding functor on the inclusion of
the finite presentations.
Moreover, these equivalences extend to the arrow categories and to the commutative squares of
comparisons, and they are compatible with the evaluation at the endpoints and with the
exchange of the Karoubi envelope and the arrow category of Chapter~\ref{chap:6}.

\end{proposition}

\begin{proof}

Idempotent completeness of lenses over a fixed view is given by Proposition~\ref{prop:6.30}.
For protocols, for an idempotent semantics-preserving adapter $p:X\Rightarrow X$ we take
\begin{equation*}
X_v^p=\{x\in X_v\mid p_v(x)=x\}
\tag{8.31}\label{eq:8.31}
\end{equation*}
By Proposition~\ref{prop:6.31}, this image, with the executions and observations restricted,
is a realization of the same protocol, and $p_v$ together with the inclusion gives its
splitting.

By Corollary~\ref{cor:8.30}, every realization is isomorphic to an image of the finite decoding
functor and is in particular a retract of that image.
Using the fully faithful decoding functor and the splitting above, sending $(P,e)$ to the split
image of $D_\Theta(e)$ yields \eqref{eq:8.30} from the Karoubi envelope of
Chapter~\ref{chap:6}.
When $e=1$, the result is isomorphic to the decoded original object.

The action on the arrow categories is defined by applying the equivalence functors at both ends
of a morphism and of a commutative square.
Compatibility with the evaluation at the endpoints follows from this definition.
The equivalence
$\mathrm{Kar}(\mathrm{Arr}(\mathcal{C}))\simeq\mathrm{Arr}(\mathrm{Kar}(\mathcal{C}))$ of
Chapter~\ref{chap:6} also uses the same idempotents at both ends and the same squares.
Taking the arrow category after the Karoubi envelope and taking the Karoubi envelope after
the arrow category give the same data of endpoints, morphisms, and squares.
The differences coming from the choice of the splittings agree under that canonical
isomorphism.

\end{proof}

\section{Description and reconstruction in finite models}\label{sec:8.8}

\subsection*{Resource constraints of implementations and finiteness of models}

In an implementation whose storage is fixed at $b$ bits, the states that can be stored in that
memory number at most $2^b$.
Integers of fixed bit width, and strings with an upper bound on the character set and on the
stored length, also form finite sets.
Under such resource constraints, the problem of reconstruction becomes the question of which
information a specified structure and change can be recovered from.

To use the finiteness of an implementation for reconstruction from finite tables, the types,
coefficients, and contexts adopted by the model and the range of the readout items must also
be specified.
In what follows we separate the case of tabulating all the required items from the case of
recovering by the laws from a partial readout.

\subsection*{The overall table and partial readouts}

If the index set $\Lambda$ in Definition~\ref{def:8.6} is finite, a compatible family consists
of the table $a_\Lambda$ containing all the items and its restrictions.
The gluing of Lemma~\ref{lem:8.7} is the recovery of this overall table.
When each value can be encoded finitely, the overall table also has a finite description.

Even when the items are finitely many, a chosen part alone need not distinguish the
candidates.
In the intrabank transfer example of \S\ref{sec:8.6}, if only the completion verdict $c$ and
the existence $j$ of journal entries are read, the incompatible candidate and the candidate
satisfying the Law both return $(c,j)=(1,1)$.
Adding the debit record $d$ and the credit record $i$, we distinguish $(d,i)=(0,0)$ for the
former from $(d,i)=(1,1)$ for the latter and can check the completion condition.
This shows that, even for finite candidates, the choice of the items read governs the outcome
of the check.

Theorem~\ref{thm:8.18} recovers the overall structure and morphisms using all the necessary
primitive data and preservation conditions.
To recover a morphism between fixed endpoints from fewer items, the values of the omitted
components must be determined uniquely by the laws.
The reconstructions from the reference fiber and from the table of generating edges in the
previous section are examples of such a unique extension.

\subsection*{Finite checking procedures and computational cost}

\begin{xltabular}{\linewidth}{@{}LLL@{}}
\toprule
Property & Content & Counterpart in this chapter \\
\midrule
\endhead
Reconstruction from all the required data & Recovers objects up to isomorphism and morphisms between fixed endpoints uniquely from descriptions satisfying the compatibility conditions & Theorem~\ref{thm:8.18} \\
\addlinespace[3pt]
Determination by a chosen finite table & The table extends uniquely to an overall morphism & Reconstruction from the reference fiber and from the table of generating edges \\
\addlinespace[3pt]
Finite executable procedure & Computes the check of the conditions and the extension using explicit enumerations and decisions of equality & The finite presentations of \S\ref{sec:8.7} \\
\bottomrule
\end{xltabular}

If the tables in question and the conditions to be checked can be enumerated finitely and each
evaluation and decision of equality can be computed, then whether the specified conditions hold
can be decided by finitely many checks.
The existence of a finite set alone does not amount to giving these enumerations or
computational procedures.

Moreover, a check terminating in finitely many steps is different from its being executable
within given time and memory resources.
A practical check requires examining the size of the input tables and the cost of evaluating
each condition.
The reconstruction theorems of this chapter show the existence and uniqueness of the
reconstruction and do not give upper bounds on the computational cost.

\section{Supplement: presentation and reconstruction when infinite objects are allowed}\label{sec:8.9}

As a mathematical generalization, we treat the case where infinite sets are allowed for the
Atoms and for the indices of the readout.
Specification models that abstract away resource bounds sometimes use such sets.
The two counterexamples below are statements about the morphisms and the readouts that such a
model permits.
They do not mean that presentation or reconstruction is impossible for an implementation under
fixed finite resources.

\subsection*{Limits of presentation by a countable syntax}

In the first counterexample we consider all the automorphisms of an infinite set and
impose no computability restriction.
Using the cardinality of that set, we show that a decoding map from a countable syntax is not
surjective.

\begin{proposition}[Impossibility of a surjective presentation by a countable syntax]\label{prop:8.32}

The base category $B$ of Chapter~\ref{chap:1} has an object whose set of automorphisms does not
fit into the image of the decoding map from any countable set of syntax.

\end{proposition}

\begin{proof}

We take the Atoms to be $\mathbb{N}$, the source to be a single point, the normalization to be
the identity, and the extraction to be always true.
We write $b_\infty$ for this base object.
For any $S\subseteq \mathbb{N}$ we define the permutation $\sigma_S$ that exchanges $2n$ and
$2n+1$ exactly when $n\in S$.
Since the extraction is constant, each $\sigma_S$ is an automorphism of $b_\infty$.
The set of exchanged pairs determines $S$ uniquely, so
\begin{equation*}
\mathcal P(\mathbb{N})\hookrightarrow\mathrm{Aut}_{B}(b_\infty),
\qquad S\longmapsto\sigma_S
\tag{8.32}\label{eq:8.32}
\end{equation*}
is injective. The left-hand side is uncountable, so there is no surjection from a countable set
onto the right-hand side.

\end{proof}

For example, the set of all finite strings over countably many symbols is countable, and this
conclusion applies to it.
This argument by cardinality depends on imposing no computability condition and on allowing all
subsets $S\subseteq \mathbb{N}$.
From this proposition one cannot conclude the impossibility of presentation when the scope is restricted to computable changes described by finite codes.

Moreover, the permutation $\sigma_{\mathbb{N}}$ exchanging all the adjacent pairs has infinite
support, so by Proposition~\ref{prop:8.3} it cannot be presented as a morphism of finite
exception codes.
On the other hand, this permutation can be described by the finite rule of adding 1 to even
numbers and subtracting 1 from odd numbers.
Presentability in a specified tabular format and describability by a finite rule are different.

This example holds for an object of the base category. The selected extraction is infinite, and the family of Atoms selected in the core is
therefore not finite.
In the local descriptions from \S\ref{sec:8.2} on, sets are retained as type references and the
whole index set is allowed to be infinite.
Reconstruction by this readout and surjective presentation by a countable syntax are different
properties.

\subsection*{Limits of finite fragments for infinite indices}

The second counterexample uses the tag change of Example~\ref{ex:8.20}.
For each $A$ in the set $\Omega$ of architecture objects, whether the tags of the operations
starting at $A$ are flipped is determined independently of the choices at the other objects.
We recover the map $\chi:\Omega\to\{0,1\}$ representing this choice from all the finite
fragments.

\begin{proposition}[Recovery of a tag change from all the finite fragments]\label{prop:8.33}

The selections of flips at each starting object $\chi:\Omega\to\{0,1\}$ of Example~\ref{ex:8.20} and the
compatible families of Boolean tables $a_S$ on the finite sets $S\subseteq\Omega$ are in
one-to-one correspondence.
With addition modulo 2, this becomes the group isomorphism
\begin{equation*}
\{0,1\}^{\Omega}
\ \cong\
\varprojlim_{S\subseteq\Omega,\ S\text{ finite}}\{0,1\}^{S}
\tag{8.33}\label{eq:8.33}
\end{equation*}
If $\Omega$ is infinite, the readout on any single finite set $S$ cannot distinguish all these
selections of flips at each starting object.

\end{proposition}

\begin{proof}

The bijection is Lemma~\ref{lem:8.7} applied to $V_q=\{0,1\}$.
Addition and restriction are defined pointwise, so the bijection preserves the group
operation.
Moreover, \eqref{eq:8.19} gives $T_\psi T_\chi=T_{\chi+\psi}$.
Since $\chi(A)$ is determined by the action on the operation obtained by attaching the tag 0
to the identity action of each $A$, this family is also represented faithfully as a subgroup of
the actual automorphisms.

If $\Omega$ is infinite and $S$ is finite, take $A\notin S$ and compare the function that is
identically 0 with the function that is 1 only at $A$.
Their values on $S$ are equal, but their actions on the tags of the operations starting at $A$
differ.

\end{proof}

If $\Omega$ is finite, this family can be determined by reading the whole of it with
$S=\Omega$.
The counterexample in the infinite case uses the fact that outside every finite fragment there
remains a place that can be changed.
Unlike the first counterexample, which uses cardinality, this argument needs only a change that
differs at a single point outside the range of the readout.

When applying these counterexamples, therefore, one must specify the infinite sets in the
model, the range of the changes allowed, and the readout scheme.
The questions of sufficiency of information and of computational cost for finite models are
treated by the conditions stated in \S\ref{sec:8.8}.

\section*{Summary of the Chapter}

In this chapter we gave a method for recovering objects and all the structure-preserving
morphisms between them from local descriptions of geometries and morphisms.

\begin{itemize}
\item \textbf{The scope of finite presentation.} With finite exception codes, even when the
meaning of an object can be recovered, the morphisms between fixed presentations need not
represent the whole semantic Hom-set.
Over finitely many Atoms, normalization of the presentations resolves this non-fullness
(\S\ref{sec:8.1}).
\item \textbf{Separation and assembly.} From typed readout items, compatibility of finite
fragments, and the totality and uniqueness of graphs of functions, we assembled tables and
functions.
Once separation of morphisms, assembly of morphisms, and assembly of objects are all in place,
the readout functor gives an equivalence of categories (\S\S\ref{sec:8.2}--\ref{sec:8.3}).
\item \textbf{Recovery of full geometries.} We proved the three conditions for the primitive
data of types, operations, Laws, local geometry, and realization.
The equivalence of categories of the main theorem recovers the objects and all the
structure-preserving morphisms of the chosen variant, including the noninvertible ones, and
preserves identities and composition (\S\S\ref{sec:8.4}--\ref{sec:8.5}).
\item \textbf{Finite descriptions by the laws of the operations.} The same reconstruction
principle applies to lenses and protocols as well.
From a finite reference fiber or a table of generating edges, a general morphism extends
uniquely by the laws of the operations (\S\ref{sec:8.7}).
\item \textbf{Sufficiency of information and computation.} For finite models we distinguish the
information that determines objects and morphisms, the procedure based on enumeration and
decision of equality, and the computational cost.
When infinite objects are allowed, a difference appears between reconstruction from all the
finite fragments and presentation by a single finite table or a countable syntax
(\S\S\ref{sec:8.8}--\ref{sec:8.9}).
\end{itemize}

In the modernization of an intrabank transfer system, once the descriptions of the account, payment,
and accounting teams define a local object that supplies the necessary primitive data
and satisfies the conditions, the whole design is constructed up to isomorphism.
Between a fixed old design and a fixed new design, when the correspondences of the teams
satisfy all the preservation conditions, they can be assembled into a unique
structure-preserving morphism.
Hence, if the same old design and the structures to be retained are fixed, several
modernization proposals can be compared from the point of view of whether a change that is
compatible overall exists.

\bookmarksetup{startatroot}
\lettersections{R}
\chapter*{Related Work}
\addcontentsline{toc}{chapter}{Related Work}

AAT is a mathematical theory that constructs relative architecture from Atoms and Laws.
It defines objects and structure-preserving morphisms relative to a reading, and develops
their geometry, transport, comparison, and local reconstruction. Below we compare objects,
assumptions, morphisms, and conclusions, focusing on algebraic specification, sheaf
semantics, abstract interpretation, data migration, and bidirectional transformation, and
situate the constructions and consequences proper to AAT.

\section{Semantics of specifications and architecture}\label{sec:R.1}

When the description of a specification changes, its meaning must change accordingly.
The institutions of Goguen and Burstall define this relation independently of the logic.
An institution carries signatures, sentences, categories of models, and a satisfaction
relation, and requires that the translation of sentences and the reduction of models along
a change of signature preserve the satisfaction relation. Morphisms between models are also
part of this framework
(\cite[\href{https://courses.grainger.illinois.edu/cs522/sp2016/InstitutionsAbstractModelTheory.pdf}{\S2.1, Definition 1}]{GB92}).
Moreover, colimits of theories can be constructed from colimits of signatures.
This result makes it possible to state the principle of assembling large specifications from
small ones on a foundation common to different logics
(\cite[\href{https://courses.grainger.illinois.edu/cs522/sp2016/InstitutionsAbstractModelTheory.pdf}{Theorem 11 and its consequences, p. 108}]{GB92}).

Once the connection of parts is taken as the object of study, compatibility of behavior
becomes the issue. Allen and Garlan's Wright describes the ports of components and the roles
and glue of connectors in CSP. If a connector is conservative and deadlock-free, and
the ports attached to each role satisfy the compatibility condition, then the connector
instantiated with those ports is deadlock-free as well
(\cite[\href{https://ix.cs.uoregon.edu/~michal/Classes/f01/fsv/papers/Allen-Garlan.pdf}{\S\S8.1--8.2, Definition 8.1.2, Theorem 8.2.3}]{AG97}).

AAT fixes a reading of Atoms and determines the relative core from the values of the Laws.
Adding covers, coefficients, and evaluation, it constructs a geometry and, through equations
and ideals, expresses the satisfaction of Laws as a geometric zero condition and defines the
lawful locus (Theorem~\ref{thm:2.9}).
This geometry is the object on which obstructions to repair, transport along readings, and
comparison of construction routes are treated.

For protocols, that correspondence can be captured by preservation equations for operations
and observations. Once named operations, the relations among paths, and the observations are
fixed, a morphism between realizations is a vertexwise state map that preserves the
operations and the observations. AAT constructs a typed input from this independent
semantics and puts the morphisms on the two sides in correspondence
(Propositions~\ref{prop:1.38} and~\ref{prop:1.43}).

\section{Local consistency and cohomological obstructions}\label{sec:R.2}

Behaviors given part by part are read as the behavior of a single system.
Goguen's sheaf semantics is a precedent for this local-to-global description.
It treats objects as sheaves and interactions as diagrams, and, under suitable completeness,
obtains the behavior of a system as the limit of a diagram
(\cite[\href{https://ncatlab.org/nlab/files/Goguen-SheafSemantics.pdf}{public manuscript, \S\S2, 3.1--3.3, Proposition 26}]{Goguen92}).
Gibson's preprint, which treats design views, likewise takes the assignment of values to
each parameter as a presheaf and glues views that agree on the shared parameters. What is at
work here is the sheaf condition given by agreement on pairwise intersections
(\cite[\href{https://arxiv.org/html/2605.08609v1}{Definition 3.2, Theorem 4.1, \S\S3--5}]{Gibson26}).

Before gluing there is the question of whether the mismatches can be resolved.
The pattern of asking for the existence of a global extension also appears in the study of
contextuality. Abramsky and Brandenburger treat the distribution of outcomes for each
measurement context as a compatible empirical model, and put the extension to a global
distribution in correspondence with the existence of a factorizable hidden-variable model
(\cite[\href{https://arxiv.org/pdf/1102.0264v7}{\S\S2.2--2.5, Theorem 8.1}]{AB11}).

Abramsky, Mansfield, and Barbosa go further and linearize the support over a free abelian
group, constructing a \v{C}ech obstruction class for the extension of a local section.
A nonzero obstruction shows non-extendability. When the obstruction vanishes, on the other
hand, what one obtains is a compatible family after linearization, and an extension to an
actual assignment of outcomes requires an additional condition.
The Hardy model is an example that exhibits this difference
(\cite[\href{https://arxiv.org/pdf/1111.3620v2}{\S4, Propositions 4.2--4.3, \S5}]{AMB12}).

AAT constructs a global repair from the vanishing of the specified obstruction class, under
the torsor structure of repair states given by abelian-group coefficients, the equivariance
of restriction and action, and the sheaf condition (Theorem~\ref{thm:2.22}).
This is a construction that ties the standard fact---that a section of a torsor
trivializes it---to the evaluation of Laws and to repair states
(\cite[\href{https://stacks.math.columbia.edu/tag/03AG}{Tag 03AG}]{Stacks}).

The choice of coefficients also determines the meaning of a diagnosis.
Young's proposal for program analysis takes the substructures of a program as a site and
uses presheaves whose values are complete lattices of semantic properties
(\cite[\href{https://arxiv.org/html/2603.27015v1}{\S\S3.1--3.4}]{Young26}).

AAT compares abelian-group coefficients generated independently from the relations among
repair words and from the equations of the Laws.
We showed this comparison in the earlier work SAGA.
From an equivariant correspondence of local states we derive the soundness of the relations,
and under two completeness conditions we obtain a coefficient isomorphism and a
correspondence of cochains and residual classes. Adding the sheaf condition on repair states
leads to a judgment of global repair
(\cite[\href{https://arxiv.org/html/2608.21458v1}{\S\S3.4--3.7, 4--5, Theorems 5.1--5.2}]{SAGA}).

On the basis of this comparison from SAGA (Theorem~\ref{thm:2.49},
Corollary~\ref{cor:2.50}), AAT compares obstructions and diagnoses along a change of
reading. Under conditions on the relation between coefficients and on common
representatives, it reflects the vanishing of the original obstruction class from the
vanishing of the diagnosis (Theorem~\ref{thm:3.40}).
It further compares the two along a refinement of covers, and carries diagnostic invariance
back to the verdict on obstructions
(Theorems~\ref{thm:3.43} and~\ref{thm:3.44}).

\section{Abstraction and preservation of information}\label{sec:R.3}

The information that a readout must retain varies with the question one wants to answer.
In the abstract interpretation of Cousot and Cousot, the concrete and the abstract domains are connected by an order structure, and the correctness of a global fixpoint computation is
derived from approximations of local semantic operations. The relation between abstraction
and concretization underlies approximations that contain the actual behavior
(\cite[\href{https://www.di.ens.fr/~cousot/publications.www/CousotCousot-POPL-77-ACM-p238--252-1977.pdf}{\S\S5--7, pp. 240--243}]{CC77}).
Giacobazzi et al. study in more detail the form of completeness in which abstraction
commutes with the semantic operations, and distinguish it from completeness for the fixpoint result
(\cite[\href{https://www.sci.unich.it/~scozzari/paper/JACM00.pdf}{\S3, p. 372}]{GRS00}).
Besides what is approximated, what is preserved exactly for the specified operations becomes
the issue.

AAT constructs, as the canonical resolution, the coarsest reading that retains the values of
the Laws exactly (Theorem~\ref{thm:3.4}).
Preserving diagnoses requires, in addition, comparisons of coefficients and covers.
The isomorphism on first cohomology under Condition~C (Theorem~\ref{thm:3.17}) and the
reflection of the vanishing of the specified obstruction class (Theorem~\ref{thm:3.40})
capture the sufficiency of this information with different strengths.

AAT characterizes the information needed to judge a change by the kernel of the observation.
For a comparison $c$, let $\Gamma_c$ be the subgroup of changes that preserve it, and let
$O$ be the observation homomorphism. Membership in $\Gamma_c$ can be determined by
observations alone
if and only if $\ker O\subseteq\Gamma_c$ (Theorem~\ref{thm:7.9}).
If every change that observations do not distinguish preserves the comparison, then the same
observation yields the same judgment.

The sufficiency here concerns the judgment of a fixed property.
Local reconstruction, which recovers objects and all structure-preserving morphisms,
requires in addition the separation and assembly of local data (Chapter~\ref{chap:8}).

\section{Transport and base change}\label{sec:R.4}

If the schema changes, the data must be moved along with it.
Spivak's functorial data migration treats a schema as a category presented by a generating
graph and relations among paths, an instance as a set-valued functor, and a morphism of
instances as a natural transformation. Restriction along a functor of schemas has both a
left and a right adjoint, which yields three transfer functors
(\cite[\href{https://arxiv.org/pdf/1009.1166v3}{\S\S3.2, 3.4--3.5, 4, Propositions 4.2.1, 4.3.1}]{Spivak12}).
The algebraic databases of Schultz et al. incorporate types, operations, and equations into
this description, and construct instances that link entities with algebras on the type side,
together with restriction along a schema mapping and its left and right adjoints
(\cite[\href{https://arxiv.org/pdf/1602.03501v3}{\S\S6--7, especially \S6.19, Propositions 7.3--7.4}]{SSVW17}).

In this theory, schemas and their morphisms, instances, and queries fit into a common
double-categorical structure called a proarrow equipment. Transfers and the evaluation of
queries are also described using composition and adjunctions in that structure
(\cite[\href{https://arxiv.org/pdf/1602.03501v3}{Definition 8.13, Proposition 8.14, Lemma 8.18, \S\S8.25--8.26, 9}]{SSVW17}).
Treating objects and transfers in the same framework is a point of contact with AAT at
the level of the overall system.

AAT builds a reindexing from an exact change of reading, and constructs the transport and
the pullback of the relative core.
Its universal property rests on the standard theory of fibered categories
(\cite[\href{https://stacks.math.columbia.edu/tag/02XJ}{Tag 02XJ}]{Stacks}).
In contrast with the adjunctions of general data migration, it is this exactness condition
that makes the transport and the pullback of the core an adjoint equivalence
(Proposition~\ref{prop:5.8}).

Once two transfers are combined, a comparison of routes is needed in turn.
The method of forming a mate from the units and counits of adjunctions and of capturing its
invertibility as the Beck--Chevalley condition is made explicit in Shulman's study of framed
bicategories and fibrations
(\cite[\href{https://tac.mta.ca/tac/volumes/20/18/20-18.pdf}{\S13, Definition 13.11}]{Shulman08}).
AAT forms this mate from a square of readings and relates the comparison of the
selected transports to the canonical comparison
(Theorem~\ref{thm:5.11}, Proposition~\ref{prop:5.12}).

From this transport and base change, AAT builds a generated comparison of full
geometries and factors it into a composite of an invertible morphism and an idempotent.
Restricting to the image of the idempotent yields an invertible comparison
(Theorem~\ref{thm:6.16}).

\section{Lenses, updates, and comparison-preserving changes}\label{sec:R.5}

Even when the shipping address of an order is updated, we want the payment information to be
carried over unchanged. In updating a view, what matters is how to treat information that,
as here, does not appear in the readout.
Bancilhon and Spyratos specify that information as a complement, and define the translation
of an update by the condition that a fixed complement be preserved.
When a lift exists, that translation is unique.
For a family of updates that is closed under composition and in which updates can be undone
at each state, if all updates can be lifted under this condition, one obtains a translator
that preserves composition
(\cite[\href{https://www.inf.unibz.it/franconispace/lib/exe/fetch.php?media=organisation\%3Asakt\%3A2013\%3A5.pdf}{\S\S3, 5, 7, Theorems 5.6, 7.1}]{BS81}).

The lenses of Foster et al. treat bidirectional transformations through the laws of the read
operation get and the update operation put. The total lenses that AAT treats correspond to the very
well-behaved and total lenses of the original paper
(\cite[\href{https://www.cis.upenn.edu/~bcpierce/papers/newlenses-full.pdf}{\S3.1, Definitions 3.1.2--3.1.3, PutPut}]{FGMPS04}).
Johnson et al. captured the relation between these laws and complements
category-theoretically. In a category with finite products, if the morphism from the view to
the terminal object splits, then the category of lenses is equivalent to the category of
complements. In the case of sets, once a reference point of the view is chosen, states can
be expressed as the product of the view and its reference fiber
(\cite[\href{https://mta.ca/~rrosebru/articles/Lens2Cambridge.pdf}{\S3, Propositions 3.1--3.2}]{JRW12}).

Using this product decomposition, AAT describes any morphism that preserves get and put as
the restriction and extension of a map of reference fibers
(Propositions~\ref{prop:1.33} and~\ref{prop:1.34}).
We connect this description to typed morphisms and to comparisons.

As for the universal property of lifting updates, Johnson et al. relate categorical lenses
to split opfibrations
(\cite[\href{https://mta.ca/~rrosebru/articles/Lens2Cambridge.pdf}{\S4, Remark 4.1, Corollary 4.1, Proposition 4.3}]{JRW12}).
AAT classifies the invertible changes that preserve a comparison under a fixed semantics.
For models in which the operations carry hidden states unchanged, AAT classifies the changes
of those states by the connected components of the operation graph, and expresses their
relation to visible changes by a split short exact sequence
(Theorems~\ref{thm:7.24} and~\ref{thm:7.25}).
For lenses, updates connect the views, so changes on the fibers must agree.
For protocols, they may be chosen separately for each independent connected component
(Propositions~\ref{prop:7.27} and~\ref{prop:7.29}).
Applying the same classification theorem to two independently defined semantics
shows how the way operations tie states together determines the freedom of change.

\section{Presentation and local reconstruction}\label{sec:R.6}

To treat objects and their changes by local presentations, structure-preserving morphisms
must be recoverable as well.
AAT assembles objects from local data and compatibility equations defined independently for
each input family, and puts local morphisms in correspondence with global
structure-preserving morphisms (Chapter~\ref{chap:8}).

At the basis of obtaining a category of models from operations and equations lies Lawvere's
functorial semantics. It treats an algebraic theory as a category with finite products, an
algebra as a product-preserving functor, and a homomorphism as a natural transformation
(\cite[\href{https://www.sas.rochester.edu/mth/sites/doug-ravenel/otherpapers/lawvere.pdf}{pp. 869--870}]{Lawvere63}).
The sketches of Barr and Wells give a graph together with diagrams to be made commutative
and specified cones. A model is an interpretation that makes the diagrams commute and sends
the cones to limit cones, and a morphism is a natural transformation
(\cite[\href{https://tac.mta.ca/tac/reprints/articles/12/tr12.pdf}{Chapter 4, \S1.2}]{BW85}).

The nerve theorem of Berger et al. constructs a fully faithful nerve from the category of
algebras to a presheaf category, under conditions on a dense generator and on a monad having
it as arities. Its essential image is characterized by a condition on restriction
(\cite[\href{https://math.univ-cotedazur.fr/~cberger/arities.pdf}{Theorem 1.10}]{BMW12}).

Here it also matters how presentations with the same meaning are identified.
The 2025 third version of \emph{Algebraic Databases} corrects the equivalence it had stated
between the category of presentations and the category of meanings, on the ground that
distinct syntactic morphisms can represent the same semantic morphism
(\cite[\href{https://arxiv.org/pdf/1602.03501v3}{Appendix B.1--B.3}]{SSVW17}).
AAT proves separately that a global morphism can be built from local maps and that morphisms
with the same readout coincide, and assembles local objects as well, up to isomorphism.
Deriving these three properties from the independent local conditions on each input family,
we obtain the equivalence between the category of realizations and the category of local
models. It is standard to derive an equivalence of categories from full faithfulness and
essential surjectivity
(\cite[\href{https://stacks.math.columbia.edu/tag/02C3}{Tag 02C3, Lemma 4.2.19}]{Stacks}).

Among the local conditions appear the graphs of get and put for lenses together with the
three laws, the graphs of generating edges and observations for protocols together with the
relations among paths, and the preservation conditions on coefficients and evaluation for
full geometries. For lenses we use finite reference fibers, and for protocols the finite
state set at each vertex and its finite list cover.

For the extension to images of idempotents we use the standard theory of Cauchy completion.
Kelly shows that, in the case enriched over sets, the small projectives of the presheaf
category over a small category are the retracts of representable presheaves, and that the
Cauchy completion corresponds to the completion that adds splittings of idempotents
(\cite[\href{https://www.sas.rochester.edu/mth/sites/doug-ravenel/otherpapers/kelly-book.pdf}{\S5.8, Theorem 5.36}]{Kelly82}).
AAT factors a generated comparison through the idempotent of normalization and builds the
restriction and extension to the image
(Theorems~\ref{thm:6.12}, \ref{thm:6.16}, and~\ref{thm:6.27}). The reconstruction of lenses
and protocols from reference fibers and generating tables also extends to the Karoubi
envelope and to the category of morphisms.

\section{Situating AAT}\label{sec:R.7}

The contribution of AAT lies in the theorems that connect its constructions and in the
correspondence with the semantics of individual problems.
The \textbf{Rising Sea} that AAT takes as its approach is to build foundations of this
kind and thereby to see individual problems as special cases of common constructions.

\chapter*{Conclusions and Further Directions}
\addcontentsline{toc}{chapter}{Conclusions and Further Directions}

This paper has constructed architecture relative to a reading from Atoms and Laws, and has
developed geometry, local consistency, diagnosis, transport, comparison, and reconstruction
as one system. We review the questions that each group of results answered, and state the
mathematical problems that arise next from this system.

\section*{Conclusions}

\begin{itemize}
\item \textbf{Foundational geometry.} We made explicit, as a reading and on the side of the objects, the choice of what to
take as objects and which operations and laws to preserve.
From a finite family of Atoms we generated a core carrying a family of objects closed under
the operations (Theorem~\ref{thm:1.18}), from contexts, covers, and coefficients we
constructed a geometry equipped with a site and a presheaf of rings, and we connected the
three levels of extraction, core, and geometry by projections (Theorem~\ref{thm:1.31}). The
independently defined semantics of lenses and protocols is encoded in this framework, and
the validity of the laws and the semantics-preserving morphisms correspond on both sides
(Propositions~\ref{prop:1.41} and~\ref{prop:1.43}).
\item \textbf{Obstructions to local consistency.} We answered, by an obstruction class, when
states that satisfy the Laws on each part glue into a whole. Satisfaction of the Laws reads
as factorization through the space defined by the ideal of the equations
(Theorem~\ref{thm:2.9}), and when the corrections act as a torsor and the states form a
sheaf, the vanishing of the first obstruction class is equivalent to the existence of a
global state (Theorem~\ref{thm:2.22}). The repair defined on the side of the semantics and
the computation on the side of the equations are connected by a comparison isomorphism of
coefficients, which gives a condition for the existence of a global repair
(Theorem~\ref{thm:2.49}, Corollary~\ref{cor:2.50}).
\item \textbf{Diagnostic invariance.} We gave conditions under which the same obstruction
can be diagnosed even after information is summarized and the ranges read locally are
repartitioned. The canonical resolution that preserves the evaluation of Laws is
characterized by a universal property (Theorem~\ref{thm:3.4}), and the comparison map of
covers, coefficients, and resolutions is an isomorphism on first cohomology under
Condition~C (Theorem~\ref{thm:3.17}). Invariance for all families of Laws for which both
resolutions are adequate can be decided by a finite computation (Theorem~\ref{thm:3.22},
Proposition~\ref{prop:3.23}). Under conditions on the identification of relation components
and on common representatives, the vanishing of the integer-coefficient obstruction and that
of the rational-valued diagnosis agree
(Theorems~\ref{thm:3.37} and~\ref{thm:3.40}). Moreover, for the refinement from three charts
to four charts built from the same Atom input, we constructed the comparison square and
showed that the verdicts, zero or nonzero, agree
(Theorems~\ref{thm:3.43} and~\ref{thm:3.44}).
\item \textbf{Transport and coherence of routes.} We characterized the construction that
carries structure along a change of reading by the universal property that every further
change factors uniquely (Theorems~\ref{thm:4.5} and~\ref{thm:4.11}). The removal of the
discrepancy between a specified comparison and the canonical comparison is equivalent to the
existence of an edge reselection that makes all faces commute at once
(Theorem~\ref{thm:4.23}). For exact pointed pullback squares, the exchange of the order
of transport and base change is realized as an invertible canonical comparison
(Theorem~\ref{thm:5.11}), and the range over which one can carry backward is characterized
as the reflection of extraction on the realized locus (Theorem~\ref{thm:5.25}).
\item \textbf{Normalization of comparisons and information about changes.} We factored a
generated comparison involving normalization into an invertible comparison and an idempotent
normalization factor, and showed that invertibility is equivalent to that factor being the
identity (Theorem~\ref{thm:6.16}). Even when it is not the identity, an isomorphism between
the images in the idempotent completion remains (Theorem~\ref{thm:6.12}). For
comparison-preserving changes we gave a necessary and sufficient condition for determination
by observations alone (Theorem~\ref{thm:7.9}), the construction of compatible lifts and the
classification of their totality as a torsor under the kernel
(\S\S\ref{sec:7.4}--\ref{sec:7.6}, Theorems~\ref{thm:7.19} and~\ref{thm:7.21}), and a split
short exact sequence for the operation-preserving changes in systems of operations that keep
hidden states (Theorems~\ref{thm:7.24} and~\ref{thm:7.25}).
\item \textbf{Presentation and local reconstruction.} We showed that, once separation of
morphisms, assembly of morphisms, and assembly of objects are available, the readout functor
gives an equivalence of categories (Theorem~\ref{thm:8.12}), and proved that the category of
full geometries and all their structure-preserving morphisms is equivalent to the category
of local models given by primitive data and independently defined compatibility conditions
(Theorem~\ref{thm:8.18}). The same principle applies to lenses, protocols, and design
descriptions divided among teams, and gives the unique extension from finite tables
(Propositions~\ref{prop:8.26} and~\ref{prop:8.28}) and the distinction between information,
procedure, and computational cost for finite models (\S\ref{sec:8.8}).
\end{itemize}

The questions of the introduction, the freedom of change and the information needed for a
judgment, received the following answers.
The degrees of freedom of the permitted changes are determined by how far the operations
tie the correspondences between states together, and for systems of operations that keep
hidden states the number of such changes is given by the families of permutations indexed
by connected components together with the torsor under the kernel (\S\S\ref{sec:7.6}--\ref{sec:7.8}). The information needed for a
judgment is characterized by the inclusion of the kernel of the observation
homomorphism in the comparison-preserving group (Theorem~\ref{thm:7.9}).
Even when compatible lifts can be constructed from the normalized information, judging a
given change may require information that normalization has lost
(\S\S\ref{sec:7.4}--\ref{sec:7.5}).

These results were proved starting from constructions on the same input and using the output
of an earlier result as the input of the next. The answers to the questions of local consistency, diagnosis,
transport, comparison, and reconstruction are connected to one another as theorems about
common objects.

\section*{Further directions}

From the objects and constructions of this paper, the following problems can be formulated.

\begin{enumerate}
\item \textbf{Coherence of reconstruction and comparison groups.}
Chapters~\ref{chap:6}--\ref{chap:7} constructed the normalization of generated comparisons,
the group of comparison-preserving changes, and a section of the restriction homomorphism.
Chapter~\ref{chap:8} gave the recovery of objects and all structure-preserving morphisms.
The next question is how to give, for an arbitrary generated comparison, the
correspondence
between its comparison-preserving group and the comparison-preserving group after
normalization in a form naturally compatible with the projections to the base, the
observations, and the coefficients, and to derive it all at once from the recovery given by
the main equivalence (Theorem~\ref{thm:8.18}), including compatibility with the subgroup
that fixes the base, with idempotent completion, and with the category of morphisms. The
sections, split short exact sequences, and descriptions of kernels and fibers constructed
individually in Chapter~\ref{chap:7} become checkpoints for this general correspondence.
\item \textbf{A common definition of distinction, extension, and effectiveness.} In
\S\ref{sec:8.8} we distinguished the information that determines objects and morphisms, the
checking procedure based on enumeration and equality tests, and the computational cost. The
next task is to give a common definition for these and to establish it as a necessary and
sufficient condition in terms of the connected components of the system of operations
(\S\ref{sec:7.7}) and as a computation of determination and extension over explicit finite
inputs.
\item \textbf{Unification of finite determinacy.} Proposition~\ref{prop:8.32} showed the
limits of a countable syntax with respect to automorphisms of the extraction base.
Proposition~\ref{prop:8.33} showed that, even when a family of tags is recovered from all
finite fragments, it is not determined by a single finite fragment over an infinite index.
For lenses and protocols we obtained the unique extension from finite tables
(Propositions~\ref{prop:8.26} and~\ref{prop:8.28}) and the connection with retracts,
idempotent completion, and comparison squares (Proposition~\ref{prop:8.31}). The question is
to unify these as recovery by the main equivalence under the same readout, and to state, as
a single theory of finite determinacy, the judgment of invertible changes by connected
components together with the recovery of comparison groups, sections, kernels, and fibers.
\item \textbf{Descent for families and global representation.} The gluing of
Chapter~\ref{chap:2} and the obstruction to composition of Chapter~\ref{chap:4} treated
obstruction classes for each fixed input. Once sites, coefficients, and transport are
indexed by families, one can ask for the relation between the vanishing of the obstruction
for each family and globalization, the correspondence between the first cohomology that
classifies lifts of local changes and the existing obstruction classes, and Morita-type
equivalences between categories of comparisons. Beyond that we place the problem of
representability: whether the functor of realizations, natural in the coefficient algebra,
is represented by a geometric object such as a scheme or a stack.
If representability holds, the totality of realizations for a fixed input becomes a single
geometric object, individual realizations can be treated as its points, and families of
realizations as morphisms into that object. The range of repairable realizations, and the places
where the value of the diagnosis jumps, can then be stated globally as computations of
subsets of that object.
These objects and assumptions are formulated separately from the problems treated by the
foundational constructions of this paper.
\item \textbf{Structure in the time direction.} The reading used in this paper concerns the
structure of a single version. Introducing traces subordinate to a sequence of measurements,
a product site, and coefficients in the time direction, one can ask under which assumptions
transport and gluing along time can be compared with the constructions of transport and
descent of this paper.
\item \textbf{Connection with measurement and with a theory of evolution.} The input of this
paper consists of observed Atoms, a specification of Laws and coefficients, and a fixed
reading. ArchSig, the author's measurement tool, computes diagnoses from two streams of
input: observation records of source code, and a specification of laws and equations. To
connect this measurement with the present paper, one must formulate as verifiable conditions
that the extracted observations satisfy the conditions on extraction of
Chapter~\ref{chap:1}, and that the computed diagnoses satisfy the assumptions on comparison
of Chapter~\ref{chap:3}. Once this connection is complete, the judgments of each chapter of
this paper become measurements with evidence, returning the same result from the same input
on actual source code. Besides the detection of inconsistencies that are correct locally but
break as a whole (Chapter~\ref{chap:2}) and the sharing of that verdict along a replacement
of the unit of review (Chapter~\ref{chap:3}), the scope of measurement extends to the
computation of the discrepancies between several migration routes and the decision whether a
reselection that removes them exists (Chapter~\ref{chap:4}); the check that the procedure of
splitting the whole and then selecting a scope agrees with the procedure of selecting a
scope and then splitting (Chapter~\ref{chap:5}); the decision of what a normalization to the
standard representation preserves and whether one can return to the original
(Chapter~\ref{chap:6}); the determination by observations whether a proposed change
preserves a comparison, and the number of all compatible candidate changes
(Chapter~\ref{chap:7}); and the finite check whether design descriptions divided among teams
assemble into a single model and a structure-preserving morphism (Chapter~\ref{chap:8}).
For this last check, Chapter~\ref{chap:8} even distinguishes the information
required, the procedure, and the computational cost.
Furthermore, for the connection with a theory that treats how artifacts and practices change
the reachable future of software evolution (Software Field Theory), the question is to
formulate a correspondence that takes the freedom given by the classification of changes of
Chapter~\ref{chap:7} and by the reconstruction of Chapter~\ref{chap:8} as the input of a
description of evolution. Once this correspondence is obtained, the process of development
is described as motion in the space formed by all realizations, and one can ask, as an
object of computation, which changes and practices widen the range of reachable designs and
which narrow it.
\end{enumerate}

Every one of these problems can be stated from the objects and maps constructed in this
paper. The Rising Sea approach, of enveloping individual problems by raising the water level
of the theory, continues on this foundation. The higher the water rises, the more the problems
of development practice acquire coordinates, become theorems, and descend to measurement.

\bibliographystyle{unsrturl}
\bibliography{references}

\appendix
\chaptersections
\chapter{Correspondence with the Lean Formalization}\label{app:A}

This appendix gives the Lean declarations that correspond to the definitions, constructions, and results in the text, together with their objects, hypotheses, and conclusions.
For applications to individual equations or to specific constructions, we record the relevant place in the text and the conditions under which they hold.
Only the last component of each declaration name is displayed, and each link points to the definition and the proof.

For the Lean sources of AAT, each declaration links to its fixed version.
General transport, lenses, base change, and comparison groups use \href{https://github.com/iroha1203/AlgebraicArchitectureTheoryV2/tree/7e68ec6e77ef0249ede6a4c875715cfaf1cb3ee6}{commit \texttt{7e68ec6e77ef}}; the other declarations use \href{https://github.com/iroha1203/AlgebraicArchitectureTheoryV2/tree/719f81f47d410701fd82c2bc88613cc140c59377}{commit \texttt{719f81f47d41}}.
Mathlib is at \href{https://github.com/leanprover-community/mathlib4/tree/8f9d9cff6bd728b17a24e163c9402775d9e6a365}{\texttt{8f9d9cff6bd7}}, and Lean is \texttt{v4.28.0}.

\section{Chapter 1: Construction of Relative Architecture}\label{sec:A.1}

\begin{xltabular}{\linewidth}{@{}LLL@{}}
\toprule
Text & Lean declaration & Objects and conditions \\
\midrule
\endhead
Definition~\ref{def:1.1} & \href{https://github.com/iroha1203/AlgebraicArchitectureTheoryV2/blob/719f81f47d410701fd82c2bc88613cc140c59377/Formal/AG/Atom/Atom.lean\#L28}{\texttt{Atom\allowbreak{}Carrier}} / \href{https://github.com/iroha1203/AlgebraicArchitectureTheoryV2/blob/719f81f47d410701fd82c2bc88613cc140c59377/Formal/AG/Atom/Axioms.lean\#L164}{\texttt{Atom\allowbreak{}Axiom\allowbreak{}System}} & Takes nonemptiness, and extensionality of the five coordinates, as input. \\
\addlinespace[3pt]
Definition~\ref{def:1.2} & \href{https://github.com/iroha1203/AlgebraicArchitectureTheoryV2/blob/719f81f47d410701fd82c2bc88613cc140c59377/Formal/AG/Atom/Family.lean\#L8}{\texttt{Atom\allowbreak{}Family}} & A family is a predicate on Atoms. \\
\addlinespace[3pt]
Definition~\ref{def:1.3} & \href{https://github.com/iroha1203/AlgebraicArchitectureTheoryV2/blob/719f81f47d410701fd82c2bc88613cc140c59377/Formal/AG/Atom/Axioms.lean\#L29}{\texttt{Extraction\allowbreak{}Doctrine}} / \href{https://github.com/iroha1203/AlgebraicArchitectureTheoryV2/blob/719f81f47d410701fd82c2bc88613cc140c59377/Formal/AG/Atom/Axioms.lean\#L59}{\texttt{extracts}} & Evaluates the four admissibility conditions on the source after normalization. \\
\addlinespace[3pt]
Proposition~\ref{prop:1.4} & \href{https://github.com/iroha1203/AlgebraicArchitectureTheoryV2/blob/719f81f47d410701fd82c2bc88613cc140c59377/Formal/AG/Atom/Axioms.lean\#L117}{\texttt{atomize\_\allowbreak{}holds}} / \href{https://github.com/iroha1203/AlgebraicArchitectureTheoryV2/blob/719f81f47d410701fd82c2bc88613cc140c59377/Formal/AG/Atom/Axioms.lean\#L126}{\texttt{atomize\_\allowbreak{}unique}} & Constructs a family from the extraction predicate and proves uniqueness by extensionality of predicates. \\
\addlinespace[3pt]
Definition~\ref{def:1.5} & \href{https://github.com/iroha1203/AlgebraicArchitectureTheoryV2/blob/719f81f47d410701fd82c2bc88613cc140c59377/Formal/AG/Atom/Configuration.lean\#L8}{\texttt{Atom\allowbreak{}Configuration}} & Holds a family, binary relations, and identifications. \\
\addlinespace[3pt]
Definition~\ref{def:1.6} & \href{https://github.com/iroha1203/AlgebraicArchitectureTheoryV2/blob/719f81f47d410701fd82c2bc88613cc140c59377/Formal/AG/Atom/ArchitectureObject.lean\#L14}{\texttt{Architecture\allowbreak{}Object}} & Holds a configuration, structure data, and a selected quantity. \\
\addlinespace[3pt]
Definition~\ref{def:1.7} & \href{https://github.com/iroha1203/AlgebraicArchitectureTheoryV2/blob/719f81f47d410701fd82c2bc88613cc140c59377/Formal/AG/Atom/ObjectAlgebra.lean\#L16}{\texttt{Configuration\allowbreak{}Hom}} / \href{https://github.com/iroha1203/AlgebraicArchitectureTheoryV2/blob/719f81f47d410701fd82c2bc88613cc140c59377/Formal/AG/Atom/ObjectAlgebra.lean\#L71}{\texttt{Operation\allowbreak{}Reading}} & Separates the type of named operations from their configuration maps. \\
\addlinespace[3pt]
Definition~\ref{def:1.9} & \href{https://github.com/iroha1203/AlgebraicArchitectureTheoryV2/blob/719f81f47d410701fd82c2bc88613cc140c59377/Formal/AG/Atom/Invariant.lean\#L22}{\texttt{Invariant\allowbreak{}Family}} / \href{https://github.com/iroha1203/AlgebraicArchitectureTheoryV2/blob/719f81f47d410701fd82c2bc88613cc140c59377/Formal/AG/Atom/ObjectAlgebra.lean\#L82}{\texttt{Architecture\allowbreak{}Signature}} & Function-typed and predicate-typed invariants, and an indexed signature. \\
\addlinespace[3pt]
Definition~\ref{def:1.10} & \href{https://github.com/iroha1203/AlgebraicArchitectureTheoryV2/blob/719f81f47d410701fd82c2bc88613cc140c59377/Formal/AG/Site/Context.lean\#L29}{\texttt{Architecture\allowbreak{}Context}} / \href{https://github.com/iroha1203/AlgebraicArchitectureTheoryV2/blob/719f81f47d410701fd82c2bc88613cc140c59377/Formal/AG/Site/ContextCategory.lean\#L22}{\texttt{Context\allowbreak{}Preorder\allowbreak{}Category}} & Readouts of support, axes, and observables, and a selected preorder. \\
\addlinespace[3pt]
Definition~\ref{def:1.12}, Definition~\ref{def:1.13} & \href{https://github.com/iroha1203/AlgebraicArchitectureTheoryV2/blob/719f81f47d410701fd82c2bc88613cc140c59377/Formal/AG/Equation/Basic.lean\#L42}{\texttt{Architectural\allowbreak{}Equation\allowbreak{}System}} / \href{https://github.com/iroha1203/AlgebraicArchitectureTheoryV2/blob/719f81f47d410701fd82c2bc88613cc140c59377/Formal/AG/Equation/Basic.lean\#L132}{\texttt{Equation\allowbreak{}Lawful}} & Separates symbolic violation coordinates from object-dependent residuals, and uses only the required indices for lawfulness. \\
\addlinespace[3pt]
Definition~\ref{def:1.14}, Definition~\ref{def:1.15} & \href{https://github.com/iroha1203/AlgebraicArchitectureTheoryV2/blob/719f81f47d410701fd82c2bc88613cc140c59377/Formal/AG/Atom/Obstruction.lean\#L32}{\texttt{Circuit\allowbreak{}Query}} / \href{https://github.com/iroha1203/AlgebraicArchitectureTheoryV2/blob/719f81f47d410701fd82c2bc88613cc140c59377/Formal/AG/Atom/Obstruction.lean\#L85}{\texttt{Finite\allowbreak{}Circuit\allowbreak{}Datum}} / \href{https://github.com/iroha1203/AlgebraicArchitectureTheoryV2/blob/719f81f47d410701fd82c2bc88613cc140c59377/Formal/AG/Atom/Obstruction.lean\#L99}{\texttt{Circuit\allowbreak{}Detector\allowbreak{}Code}} / \href{https://github.com/iroha1203/AlgebraicArchitectureTheoryV2/blob/719f81f47d410701fd82c2bc88613cc140c59377/Formal/AG/Atom/Obstruction.lean\#L144}{\texttt{Equation\allowbreak{}Circuit\allowbreak{}Reading}} & Signed query sequences, finite detector syntax, and circuits that match and are accepted. \\
\addlinespace[3pt]
Proposition~\ref{prop:1.16} & \href{https://github.com/iroha1203/AlgebraicArchitectureTheoryV2/blob/719f81f47d410701fd82c2bc88613cc140c59377/Formal/AG/Atom/ThreeReading.lean\#L36}{\texttt{equation\allowbreak{}Lawful\_\allowbreak{}iff\_\allowbreak{}no\allowbreak{}Required\allowbreak{}Equation\allowbreak{}Circuit}} & Sound and RequiredComplete agree with the two hypotheses in the text. \\
\addlinespace[3pt]
Definition~\ref{def:1.17} & \href{https://github.com/iroha1203/AlgebraicArchitectureTheoryV2/blob/719f81f47d410701fd82c2bc88613cc140c59377/Formal/AG/Atom/AATCore.lean\#L40}{\texttt{Core\allowbreak{}Reading}} / \href{https://github.com/iroha1203/AlgebraicArchitectureTheoryV2/blob/719f81f47d410701fd82c2bc88613cc140c59377/Formal/AG/Atom/AATCore.lean\#L62}{\texttt{AATCore\allowbreak{}Package}} & Generated from a finite extracted family, a family-preserving composition, and a configuration-preserving object reading. \\
\addlinespace[3pt]
Theorem~\ref{thm:1.18} & \href{https://github.com/iroha1203/AlgebraicArchitectureTheoryV2/blob/719f81f47d410701fd82c2bc88613cc140c59377/Formal/AG/Atom/AATCore.lean\#L153}{\texttt{generate}} / \href{https://github.com/iroha1203/AlgebraicArchitectureTheoryV2/blob/719f81f47d410701fd82c2bc88613cc140c59377/Formal/AG/Atom/AATCore.lean\#L223}{\texttt{configuration\_\allowbreak{}family\_\allowbreak{}eq}} / \href{https://github.com/iroha1203/AlgebraicArchitectureTheoryV2/blob/719f81f47d410701fd82c2bc88613cc140c59377/Formal/AG/Atom/AATCore.lean\#L258}{\texttt{object\_\allowbreak{}configuration\_\allowbreak{}eq}} / \href{https://github.com/iroha1203/AlgebraicArchitectureTheoryV2/blob/719f81f47d410701fd82c2bc88613cc140c59377/Formal/AG/Atom/AATCore.lean\#L298}{\texttt{algebra\_\allowbreak{}object\_\allowbreak{}nonempty\_\allowbreak{}iff\_\allowbreak{}reachable}} / \href{https://github.com/iroha1203/AlgebraicArchitectureTheoryV2/blob/719f81f47d410701fd82c2bc88613cc140c59377/Formal/AG/Atom/ObjectAlgebra.lean\#L275}{\texttt{Reachable}} & The finite sequences of operations in the text are the inductive presentation given by base/step of Reachable. Minimality is expressed by this induction principle. \\
\addlinespace[3pt]
Definition~\ref{def:1.19} & \href{https://github.com/iroha1203/AlgebraicArchitectureTheoryV2/blob/719f81f47d410701fd82c2bc88613cc140c59377/Formal/AG/Site/ContextCategory.lean\#L529}{\texttt{Context\allowbreak{}Overlap\allowbreak{}Pullback}} / \href{https://github.com/iroha1203/AlgebraicArchitectureTheoryV2/blob/719f81f47d410701fd82c2bc88613cc140c59377/Formal/AG/Site/Coverage.lean\#L54}{\texttt{Coverage\allowbreak{}Requirements}} / \href{https://github.com/iroha1203/AlgebraicArchitectureTheoryV2/blob/719f81f47d410701fd82c2bc88613cc140c59377/Formal/AG/Site/Coverage.lean\#L78}{\texttt{Admissible\allowbreak{}Cover}} & Selected pullbacks, and conditions on visibility of four kinds, on interaction, and on support in the original family. \\
\addlinespace[3pt]
Definition~\ref{def:1.20} & \href{https://github.com/leanprover-community/mathlib4/blob/8f9d9cff6bd728b17a24e163c9402775d9e6a365/Mathlib/CategoryTheory/Sites/Grothendieck.lean\#L76}{\texttt{Grothendieck\allowbreak{}Topology}} & The three axioms of the maximal sieve, pullbacks, and transitivity. \\
\addlinespace[3pt]
Proposition~\ref{prop:1.21} & \href{https://github.com/iroha1203/AlgebraicArchitectureTheoryV2/blob/719f81f47d410701fd82c2bc88613cc140c59377/Formal/AG/Site/Topology.lean\#L83}{\texttt{AATGrothendieck\allowbreak{}Topology}} / \href{https://github.com/iroha1203/AlgebraicArchitectureTheoryV2/blob/719f81f47d410701fd82c2bc88613cc140c59377/Formal/AG/Site/Topology.lean\#L122}{\texttt{generate\_\allowbreak{}mem}} / \href{https://github.com/leanprover-community/mathlib4/blob/8f9d9cff6bd728b17a24e163c9402775d9e6a365/Mathlib/CategoryTheory/Sites/PrecoverageToGrothendieck.lean\#L61}{\texttt{to\allowbreak{}Grothendieck}} & Builds the generated topology from admissible presieves. \\
\addlinespace[3pt]
Definition~\ref{def:1.22} & \href{https://github.com/iroha1203/AlgebraicArchitectureTheoryV2/blob/719f81f47d410701fd82c2bc88613cc140c59377/Formal/AG/Site/Sheaf.lean\#L48}{\texttt{AATSheaf\allowbreak{}Condition}} / \href{https://github.com/iroha1203/AlgebraicArchitectureTheoryV2/blob/719f81f47d410701fd82c2bc88613cc140c59377/Formal/AG/Site/Sheaf.lean\#L67}{\texttt{iff\_\allowbreak{}presieve\_\allowbreak{}is\allowbreak{}Sheaf}} / \href{https://github.com/iroha1203/AlgebraicArchitectureTheoryV2/blob/719f81f47d410701fd82c2bc88613cc140c59377/Formal/AG/Site/SheafCategory.lean\#L17}{\texttt{AATSh}} & Puts the matching families on covers in correspondence with the sheaf condition given by presieves in Mathlib. \\
\addlinespace[3pt]
Proposition~\ref{prop:1.23} & \href{https://github.com/leanprover-community/mathlib4/blob/8f9d9cff6bd728b17a24e163c9402775d9e6a365/Mathlib/CategoryTheory/Sites/Sheafification.lean\#L73}{\texttt{sheafification\allowbreak{}Adjunction}} / \href{https://github.com/leanprover-community/mathlib4/blob/8f9d9cff6bd728b17a24e163c9402775d9e6a365/Mathlib/CategoryTheory/Sites/Sheafification.lean\#L185}{\texttt{sheafify\allowbreak{}Lift\_\allowbreak{}unique}} & The adjunction of the standard sheafification of set-valued presheaves and the uniqueness of the extension. Uses Mathlib's HasSheafify. \\
\addlinespace[3pt]
Definition~\ref{def:1.25} & \href{https://github.com/iroha1203/AlgebraicArchitectureTheoryV2/blob/719f81f47d410701fd82c2bc88613cc140c59377/Formal/AG/LawAlgebra/StructureSheaf.lean\#L46}{\texttt{Raw\allowbreak{}Ambient\allowbreak{}Restriction\allowbreak{}System}} / \href{https://github.com/iroha1203/AlgebraicArchitectureTheoryV2/blob/719f81f47d410701fd82c2bc88613cc140c59377/Formal/AG/LawAlgebra/StructureSheaf.lean\#L253}{\texttt{to\allowbreak{}Presheaf}} & Constructs the presheaf of quotients from the polynomial restrictions that preserve the structural relations. \\
\addlinespace[3pt]
Lemma~\ref{lem:1.26} & \href{https://github.com/iroha1203/AlgebraicArchitectureTheoryV2/blob/719f81f47d410701fd82c2bc88613cc140c59377/Formal/AG/ReadingFunctoriality/Coefficient.lean\#L437}{\texttt{base\allowbreak{}Change}} / \href{https://github.com/iroha1203/AlgebraicArchitectureTheoryV2/blob/719f81f47d410701fd82c2bc88613cc140c59377/Formal/AG/ReadingFunctoriality/Coefficient.lean\#L736}{\texttt{base\allowbreak{}Change\_\allowbreak{}comp}} / \href{https://github.com/iroha1203/AlgebraicArchitectureTheoryV2/blob/719f81f47d410701fd82c2bc88613cc140c59377/research/lean/ResearchLean/AG/GeometryTransport/Basic.lean\#L381}{\texttt{raw\allowbreak{}Transport\_\allowbreak{}id}} & Applies the coefficient map to the generating relations and the restrictions, and is coherent with identities and composition. \\
\addlinespace[3pt]
Definition~\ref{def:1.27} & \href{https://github.com/iroha1203/AlgebraicArchitectureTheoryV2/blob/719f81f47d410701fd82c2bc88613cc140c59377/research/lean/ResearchLean/AG/AtomFoundation/Doctrine.lean\#L39}{\texttt{Exact\allowbreak{}Doctrine\allowbreak{}Hom}} / \href{https://github.com/iroha1203/AlgebraicArchitectureTheoryV2/blob/719f81f47d410701fd82c2bc88613cc140c59377/research/lean/ResearchLean/AG/AtomFoundation/Doctrine.lean\#L100}{\texttt{atomize\_\allowbreak{}naturality}} / \href{https://github.com/iroha1203/AlgebraicArchitectureTheoryV2/blob/719f81f47d410701fd82c2bc88613cc140c59377/research/lean/ResearchLean/AG/AtomFoundation/Categories.lean\#L25}{\texttt{Extraction\allowbreak{}Instance}} & The source map is a general map, and the Atom map is an equivalence. It commutes with normalization and extraction. \\
\addlinespace[3pt]
Definition~\ref{def:1.28} & \href{https://github.com/iroha1203/AlgebraicArchitectureTheoryV2/blob/719f81f47d410701fd82c2bc88613cc140c59377/Formal/AG/ReadingFunctoriality/Core.lean\#L1005}{\texttt{Signed\allowbreak{}Exact\allowbreak{}Core\allowbreak{}Reading\allowbreak{}Hom}} & Compares object formation, operations, ring isomorphisms, residuals, detector syntax, invariants, and signatures. \\
\addlinespace[3pt]
Definition~\ref{def:1.30} & \href{https://github.com/iroha1203/AlgebraicArchitectureTheoryV2/blob/719f81f47d410701fd82c2bc88613cc140c59377/research/lean/ResearchLean/AG/GeometryTransport/Categories.lean\#L380}{\texttt{Geom\allowbreak{}Read\allowbreak{}Hom}} & Holds covers, overlaps, coefficients, the equality of raw presentations, and the support/axis/observable comparisons. \\
\addlinespace[3pt]
Theorem~\ref{thm:1.31} & \href{https://github.com/iroha1203/AlgebraicArchitectureTheoryV2/blob/719f81f47d410701fd82c2bc88613cc140c59377/research/lean/ResearchLean/AG/AtomFoundation/Categories.lean\#L232}{\texttt{package\allowbreak{}Total\allowbreak{}Category}} / \href{https://github.com/iroha1203/AlgebraicArchitectureTheoryV2/blob/719f81f47d410701fd82c2bc88613cc140c59377/research/lean/ResearchLean/AG/AtomFoundation/Categories.lean\#L271}{\texttt{package\allowbreak{}Projection}} / \href{https://github.com/iroha1203/AlgebraicArchitectureTheoryV2/blob/719f81f47d410701fd82c2bc88613cc140c59377/research/lean/ResearchLean/AG/GeometryTransport/Categories.lean\#L598}{\texttt{geometry\allowbreak{}Projection}} & The total category, and the projection functors to core and doctrine. \\
\addlinespace[3pt]
Definition~\ref{def:1.32} & \href{https://github.com/iroha1203/AlgebraicArchitectureTheoryV2/blob/719f81f47d410701fd82c2bc88613cc140c59377/research/lean/ResearchLean/AG/RealizationReconstruction/LensSemantics.lean\#L59}{\texttt{Lens\allowbreak{}Realization}} / \href{https://github.com/iroha1203/AlgebraicArchitectureTheoryV2/blob/719f81f47d410701fd82c2bc88613cc140c59377/research/lean/ResearchLean/AG/RealizationReconstruction/LensSemantics.lean\#L92}{\texttt{Hom}} & The three laws and a finite reference fiber. Finiteness of the set of views itself is not required. \\
\addlinespace[3pt]
Proposition~\ref{prop:1.33} & \href{https://github.com/iroha1203/AlgebraicArchitectureTheoryV2/blob/719f81f47d410701fd82c2bc88613cc140c59377/research/lean/ResearchLean/AG/RealizationReconstruction/LensSemantics.lean\#L314}{\texttt{canonical\allowbreak{}Normal\allowbreak{}Form\allowbreak{}Equiv}} / \href{https://github.com/iroha1203/AlgebraicArchitectureTheoryV2/blob/719f81f47d410701fd82c2bc88613cc140c59377/research/lean/ResearchLean/AG/RealizationReconstruction/LensSemantics.lean\#L358}{\texttt{canonical\allowbreak{}Normal\allowbreak{}Form\allowbreak{}Iso}} & Sends c to (get c, put c v0), and proves from the three laws that the maps are mutually inverse and preserve the operations. \\
\addlinespace[3pt]
Proposition~\ref{prop:1.34} & \href{https://github.com/iroha1203/AlgebraicArchitectureTheoryV2/blob/719f81f47d410701fd82c2bc88613cc140c59377/research/lean/ResearchLean/AG/RealizationReconstruction/LensSemantics.lean\#L208}{\texttt{hom\allowbreak{}Equiv\allowbreak{}Fiber\allowbreak{}Map}} / \href{https://github.com/iroha1203/AlgebraicArchitectureTheoryV2/blob/719f81f47d410701fd82c2bc88613cc140c59377/research/lean/ResearchLean/AG/RealizationReconstruction/LensSemantics.lean\#L218}{\texttt{res\_\allowbreak{}id}} / \href{https://github.com/iroha1203/AlgebraicArchitectureTheoryV2/blob/719f81f47d410701fd82c2bc88613cc140c59377/research/lean/ResearchLean/AG/RealizationReconstruction/LensSemantics.lean\#L226}{\texttt{res\_\allowbreak{}comp}} / \href{https://github.com/iroha1203/AlgebraicArchitectureTheoryV2/blob/719f81f47d410701fd82c2bc88613cc140c59377/research/lean/ResearchLean/AG/RealizationReconstruction/LensSemantics.lean\#L234}{\texttt{ext\_\allowbreak{}comp}} & Restriction to the reference fiber and extension are mutually inverse, and preserve identities and composition. \\
\addlinespace[3pt]
After Proposition~\ref{prop:1.34}, \eqref{eq:1.20} & \href{https://github.com/iroha1203/AlgebraicArchitectureTheoryV2/blob/7e68ec6e77ef0249ede6a4c875715cfaf1cb3ee6/research/lean/ResearchLean/AG/RealizationReconstruction/GeneralRelativeLens.lean\#L343}{\texttt{invertible\allowbreak{}Normal\allowbreak{}Form}} / \href{https://github.com/iroha1203/AlgebraicArchitectureTheoryV2/blob/7e68ec6e77ef0249ede6a4c875715cfaf1cb3ee6/research/lean/ResearchLean/AG/RealizationReconstruction/GeneralRelativeLens.lean\#L427}{\texttt{iso\allowbreak{}Equiv\allowbreak{}View\allowbreak{}Fiber\allowbreak{}Equiv}} & Uses independent reference views of two lenses to obtain the normal form from a visible change and an equivalence of reference fibers. The reference views need not be preserved. \\
\addlinespace[3pt]
Example~\ref{ex:1.35} & \href{https://github.com/iroha1203/AlgebraicArchitectureTheoryV2/blob/719f81f47d410701fd82c2bc88613cc140c59377/research/lean/ResearchLean/AG/RealizationReconstruction/FixedFFiniteExamples.lean\#L220}{\texttt{bool\allowbreak{}Lens\allowbreak{}Twist\_\allowbreak{}does\_\allowbreak{}not\_\allowbreak{}preserve\_\allowbreak{}put}} / \href{https://github.com/iroha1203/AlgebraicArchitectureTheoryV2/blob/719f81f47d410701fd82c2bc88613cc140c59377/research/lean/ResearchLean/AG/RealizationReconstruction/FixedFFiniteExamples.lean\#L227}{\texttt{bool\allowbreak{}Lens\_\allowbreak{}get\_\allowbreak{}count\_\allowbreak{}identity}} / \href{https://github.com/iroha1203/AlgebraicArchitectureTheoryV2/blob/719f81f47d410701fd82c2bc88613cc140c59377/research/lean/ResearchLean/AG/RealizationReconstruction/FixedFFiniteExamples.lean\#L243}{\texttt{bool\allowbreak{}Lens\_\allowbreak{}get\_\allowbreak{}put\_\allowbreak{}count\_\allowbreak{}identity}} & An example given by the xor map on the Bool product lens. The mismatch in the order of updates, and the counts 4 and 2. \\
\addlinespace[3pt]
Definition~\ref{def:1.36} & \href{https://github.com/iroha1203/AlgebraicArchitectureTheoryV2/blob/719f81f47d410701fd82c2bc88613cc140c59377/research/lean/ResearchLean/AG/RealizationReconstruction/ProtocolSchema.lean\#L37}{\texttt{Protocol\allowbreak{}Schema}} / \href{https://github.com/iroha1203/AlgebraicArchitectureTheoryV2/blob/719f81f47d410701fd82c2bc88613cc140c59377/research/lean/ResearchLean/AG/RealizationReconstruction/ProtocolSemantics.lean\#L33}{\texttt{Protocol\allowbreak{}Realization}} & The quotient category of a finite graph by the declared path relations, a finite-state functor, and natural observations. \\
\addlinespace[3pt]
Proposition~\ref{prop:1.38} & \href{https://github.com/iroha1203/AlgebraicArchitectureTheoryV2/blob/719f81f47d410701fd82c2bc88613cc140c59377/research/lean/ResearchLean/AG/RealizationReconstruction/ProtocolReconstruction.lean\#L60}{\texttt{generator\_\allowbreak{}path\_\allowbreak{}naturality}} / \href{https://github.com/iroha1203/AlgebraicArchitectureTheoryV2/blob/719f81f47d410701fd82c2bc88613cc140c59377/research/lean/ResearchLean/AG/RealizationReconstruction/ProtocolReconstruction.lean\#L118}{\texttt{hom\allowbreak{}Equiv\allowbreak{}Generator\allowbreak{}Map}} & Constructs naturality on paths from commutativity on edges, mutually inverse with the restriction to the components. \\
\addlinespace[3pt]
Construction~\ref{cons:1.39} & \href{https://github.com/iroha1203/AlgebraicArchitectureTheoryV2/blob/719f81f47d410701fd82c2bc88613cc140c59377/research/lean/ResearchLean/AG/RealizationReconstruction/CSAATArchitectureObjects.lean\#L66}{\texttt{typed\allowbreak{}Role\allowbreak{}Configuration}} / \href{https://github.com/iroha1203/AlgebraicArchitectureTheoryV2/blob/719f81f47d410701fd82c2bc88613cc140c59377/research/lean/ResearchLean/AG/RealizationReconstruction/CSAATArchitectureObjects.lean\#L119}{\texttt{lens\allowbreak{}Named\allowbreak{}Configuration\allowbreak{}Hom}} / \href{https://github.com/iroha1203/AlgebraicArchitectureTheoryV2/blob/719f81f47d410701fd82c2bc88613cc140c59377/research/lean/ResearchLean/AG/RealizationReconstruction/CSAATArchitectureObjects.lean\#L479}{\texttt{protocol\allowbreak{}Named\allowbreak{}Configuration\allowbreak{}Hom}} & A construction recording the roles as the ends of relations and the operation names as the image of the point. \\
\addlinespace[3pt]
Definition~\ref{def:1.42} & \href{https://github.com/iroha1203/AlgebraicArchitectureTheoryV2/blob/719f81f47d410701fd82c2bc88613cc140c59377/research/lean/ResearchLean/AG/RealizationReconstruction/CSAATForwardMorphisms.lean\#L33}{\texttt{Lens\allowbreak{}AATForward\allowbreak{}Morphism}} / \href{https://github.com/iroha1203/AlgebraicArchitectureTheoryV2/blob/719f81f47d410701fd82c2bc88613cc140c59377/research/lean/ResearchLean/AG/RealizationReconstruction/CSAATForwardMorphisms.lean\#L195}{\texttt{Protocol\allowbreak{}AATForward\allowbreak{}Morphism}} & Uses the role maps and the commutativity of get/put and of edge/observe. \\
\addlinespace[3pt]
Proposition~\ref{prop:1.43} & \href{https://github.com/iroha1203/AlgebraicArchitectureTheoryV2/blob/719f81f47d410701fd82c2bc88613cc140c59377/research/lean/ResearchLean/AG/RealizationReconstruction/CSAATForwardMorphisms.lean\#L81}{\texttt{semantic\allowbreak{}Hom\allowbreak{}Equiv}} (lens) / \href{https://github.com/iroha1203/AlgebraicArchitectureTheoryV2/blob/719f81f47d410701fd82c2bc88613cc140c59377/research/lean/ResearchLean/AG/RealizationReconstruction/CSAATForwardMorphisms.lean\#L258}{\texttt{semantic\allowbreak{}Hom\allowbreak{}Equiv}} (protocol) / \href{https://github.com/iroha1203/AlgebraicArchitectureTheoryV2/blob/719f81f47d410701fd82c2bc88613cc140c59377/research/lean/ResearchLean/AG/RealizationReconstruction/CSAATTypedOperationTranslation.lean\#L88}{\texttt{lens\allowbreak{}Get\_\allowbreak{}square}} / \href{https://github.com/iroha1203/AlgebraicArchitectureTheoryV2/blob/719f81f47d410701fd82c2bc88613cc140c59377/research/lean/ResearchLean/AG/RealizationReconstruction/CSAATTypedOperationTranslation.lean\#L216}{\texttt{protocol\allowbreak{}Edge\_\allowbreak{}square}} & Mutually inverse constructions of semantics-preserving morphisms from typed morphisms. The naming and role components are preserved. \\
\bottomrule
\end{xltabular}

\subsection*{Aggregation, source maps, and concrete lens examples}

\begin{xltabular}{\linewidth}{@{}LLL@{}}
\toprule
Text & Lean declaration & Objects and conditions \\
\midrule
\endhead
\S\ref{sec:1.4} (aggregation) & \href{https://github.com/iroha1203/AlgebraicArchitectureTheoryV2/blob/719f81f47d410701fd82c2bc88613cc140c59377/Formal/AG/Atom/LawfulnessZero.lean\#L68}{\texttt{equation\allowbreak{}Lawful\_\allowbreak{}iff\_\allowbreak{}omega\allowbreak{}E\_\allowbreak{}zero}} & Under a zero-reflecting aggregation, soundness, and completeness, lawfulness agrees with the vanishing of the aggregation. \\
\addlinespace[3pt]
Construction~\ref{cons:1.40} (source maps) & \href{https://github.com/iroha1203/AlgebraicArchitectureTheoryV2/blob/719f81f47d410701fd82c2bc88613cc140c59377/research/lean/ResearchLean/AG/RealizationReconstruction/CSAATArchitectureObjects.lean\#L248}{\texttt{lens\allowbreak{}AATSource\allowbreak{}Map}} / \href{https://github.com/iroha1203/AlgebraicArchitectureTheoryV2/blob/719f81f47d410701fd82c2bc88613cc140c59377/research/lean/ResearchLean/AG/RealizationReconstruction/CSAATArchitectureObjects.lean\#L623}{\texttt{protocol\allowbreak{}AATSource\allowbreak{}Map}} & Corresponds to maps of tagged sources. The placement of the tag table inside the doctrine is not included in the objects. \\
\addlinespace[3pt]
Example just after Proposition~\ref{prop:1.41} & \href{https://github.com/iroha1203/AlgebraicArchitectureTheoryV2/blob/719f81f47d410701fd82c2bc88613cc140c59377/research/lean/ResearchLean/AG/RealizationReconstruction/CSAATLawSystems.lean\#L202}{\texttt{ignored\allowbreak{}Bool\allowbreak{}Lens\allowbreak{}Law\allowbreak{}Structure\_\allowbreak{}not\_\allowbreak{}get\allowbreak{}Put}} & A lens candidate over Bool whose update ignores the requested value does not satisfy the get-put law. \\
\bottomrule
\end{xltabular}

\section{Chapter 2: Geometry of Laws and Local Consistency}\label{sec:A.2}

\begin{xltabular}{\linewidth}{@{}LLL@{}}
\toprule
Text & Lean declaration & Objects and conditions \\
\midrule
\endhead
Definition~\ref{def:2.1} & \href{https://github.com/iroha1203/AlgebraicArchitectureTheoryV2/blob/719f81f47d410701fd82c2bc88613cc140c59377/Formal/AG/LawAlgebra/StructuralRelation.lean\#L25}{\texttt{Structural\allowbreak{}Relation\allowbreak{}Family}} / \href{https://github.com/iroha1203/AlgebraicArchitectureTheoryV2/blob/719f81f47d410701fd82c2bc88613cc140c59377/Formal/AG/LawAlgebra/StructuralRelation.lean\#L41}{\texttt{JStruct}} & Divides the polynomial ring in typed coordinates by the ideal generated by the structural relations. \\
\addlinespace[3pt]
Proposition~\ref{prop:2.2} & \href{https://github.com/iroha1203/AlgebraicArchitectureTheoryV2/blob/719f81f47d410701fd82c2bc88613cc140c59377/Formal/AG/LawAlgebra/StructuralRelation.lean\#L215}{\texttt{configuration\allowbreak{}Representability}} / \href{https://github.com/iroha1203/AlgebraicArchitectureTheoryV2/blob/719f81f47d410701fd82c2bc88613cc140c59377/Formal/AG/LawAlgebra/StructuralRelation.lean\#L251}{\texttt{configuration\allowbreak{}Representability\_\allowbreak{}natural}} & The natural bijection between the coordinate values satisfying the relations and the algebra homomorphisms out of the quotient. \\
\addlinespace[3pt]
Definition~\ref{def:2.3} & \href{https://github.com/iroha1203/AlgebraicArchitectureTheoryV2/blob/719f81f47d410701fd82c2bc88613cc140c59377/Formal/AG/LawAlgebra/LawEquation.lean\#L25}{\texttt{witness\allowbreak{}Ideal}} / \href{https://github.com/iroha1203/AlgebraicArchitectureTheoryV2/blob/719f81f47d410701fd82c2bc88613cc140c59377/Formal/AG/LawAlgebra/LawEquation.lean\#L49}{\texttt{obstruction\allowbreak{}Ideal}} & The witness ideal and the supremum over the required indices. \\
\addlinespace[3pt]
Definition~\ref{def:2.5} & \href{https://github.com/iroha1203/AlgebraicArchitectureTheoryV2/blob/719f81f47d410701fd82c2bc88613cc140c59377/Formal/AG/LawAlgebra/StructuralRelation.lean\#L41}{\texttt{JStruct}} & The definition of the affine presentation that forms Spec from the quotient ring. \\
\addlinespace[3pt]
Definition~\ref{def:2.8} & \href{https://github.com/iroha1203/AlgebraicArchitectureTheoryV2/blob/719f81f47d410701fd82c2bc88613cc140c59377/Formal/AG/LawAlgebra/ClosedEquationalGeometry.lean\#L3554}{\texttt{equation\allowbreak{}Generated\allowbreak{}Lawful\allowbreak{}Closed\allowbreak{}Subscheme}} & The closed subscheme corresponding to the generated sheaf of Law ideals. \\
\addlinespace[3pt]
Construction~\ref{cons:2.12} & \href{https://github.com/iroha1203/AlgebraicArchitectureTheoryV2/blob/719f81f47d410701fd82c2bc88613cc140c59377/Formal/AG/LawAlgebra/LawEquation.lean\#L165}{\texttt{obstruction\allowbreak{}Quotient\allowbreak{}Coefficient}} / \href{https://github.com/iroha1203/AlgebraicArchitectureTheoryV2/blob/719f81f47d410701fd82c2bc88613cc140c59377/Formal/AG/LawAlgebra/LawEquation.lean\#L213}{\texttt{quotient\_\allowbreak{}mk\_\allowbreak{}eq\_\allowbreak{}zero\_\allowbreak{}iff\_\allowbreak{}mem\_\allowbreak{}obstruction\allowbreak{}Ideal}} / \href{https://github.com/iroha1203/AlgebraicArchitectureTheoryV2/blob/719f81f47d410701fd82c2bc88613cc140c59377/Formal/AG/LawAlgebra/LawEquation.lean\#L482}{\texttt{interpret\_\allowbreak{}ne\_\allowbreak{}zero\_\allowbreak{}detects\_\allowbreak{}displayed\_\allowbreak{}required\_\allowbreak{}law\_\allowbreak{}failure}} & Restriction of the quotient coefficients, and the verdict whether the residual is zero. The converse direction requires a condition that distinguishes residuals. \\
\addlinespace[3pt]
Example~\ref{ex:2.13} & \href{https://github.com/leanprover-community/mathlib4/blob/8f9d9cff6bd728b17a24e163c9402775d9e6a365/Mathlib/RingTheory/Polynomial/Quotient.lean\#L32}{\texttt{quotient\allowbreak{}Span\allowbreak{}XSub\allowbreak{}CAlg\allowbreak{}Equiv}} / \href{https://github.com/iroha1203/AlgebraicArchitectureTheoryV2/blob/719f81f47d410701fd82c2bc88613cc140c59377/Formal/AG/LawAlgebra/LawEquation.lean\#L213}{\texttt{quotient\_\allowbreak{}mk\_\allowbreak{}eq\_\allowbreak{}zero\_\allowbreak{}iff\_\allowbreak{}mem\_\allowbreak{}obstruction\allowbreak{}Ideal}} & An example of the quotient, specializing the standard $R[X]/(X-x)\simeq R$ to $R=\mathbb{Z}$ and $x=0$. \\
\addlinespace[3pt]
Definition~\ref{def:2.16} & \href{https://github.com/iroha1203/AlgebraicArchitectureTheoryV2/blob/719f81f47d410701fd82c2bc88613cc140c59377/Formal/AG/SemanticRepair/Saga/CechThreeTerm.lean\#L39}{\texttt{Intersection\allowbreak{}Coefficient\allowbreak{}Data}} / \href{https://github.com/iroha1203/AlgebraicArchitectureTheoryV2/blob/719f81f47d410701fd82c2bc88613cc140c59377/Formal/AG/SemanticRepair/Saga/CechThreeTerm.lean\#L115}{\texttt{delta0}} / \href{https://github.com/iroha1203/AlgebraicArchitectureTheoryV2/blob/719f81f47d410701fd82c2bc88613cc140c59377/Formal/AG/SemanticRepair/Saga/CechThreeTerm.lean\#L128}{\texttt{delta1}} & Builds the three-term \v{C}ech complex from coefficients on intersections and face restrictions. \\
\addlinespace[3pt]
Lemma~\ref{lem:2.17} & \href{https://github.com/iroha1203/AlgebraicArchitectureTheoryV2/blob/719f81f47d410701fd82c2bc88613cc140c59377/Formal/AG/SemanticRepair/Saga/CechThreeTerm.lean\#L142}{\texttt{delta1\_\allowbreak{}delta0}} & The cancellation of terms by the composition rule for restrictions gives $d^1d^0=0$. \\
\addlinespace[3pt]
Definition~\ref{def:2.18} & \href{https://github.com/iroha1203/AlgebraicArchitectureTheoryV2/blob/719f81f47d410701fd82c2bc88613cc140c59377/Formal/AG/SemanticRepair/Saga/CechThreeTerm.lean\#L175}{\texttt{inc\allowbreak{}Complex}} / \href{https://github.com/iroha1203/AlgebraicArchitectureTheoryV2/blob/719f81f47d410701fd82c2bc88613cc140c59377/Formal/AG/SemanticRepair/Saga/CechThreeTerm.lean\#L182}{\texttt{Inc\allowbreak{}H1}} & Bundles the three-term complex and defines the first \v{C}ech cohomology on a fixed cover as a quotient. \\
\addlinespace[3pt]
Definition~\ref{def:2.20} & \href{https://github.com/iroha1203/AlgebraicArchitectureTheoryV2/blob/719f81f47d410701fd82c2bc88613cc140c59377/Formal/AG/SemanticRepair/Saga/EquationLift.lean\#L52}{\texttt{Affine\allowbreak{}Coefficient\allowbreak{}Lift\allowbreak{}System}} / \href{https://github.com/iroha1203/AlgebraicArchitectureTheoryV2/blob/719f81f47d410701fd82c2bc88613cc140c59377/Formal/AG/SemanticRepair/Saga/EquationLift.lean\#L74}{\texttt{Coefficient\allowbreak{}Lift\allowbreak{}Atlas}} & A free and transitive action of coefficients on intersections, and a choice of local lifts. \\
\addlinespace[3pt]
Proposition~\ref{prop:2.21} & \href{https://github.com/iroha1203/AlgebraicArchitectureTheoryV2/blob/719f81f47d410701fd82c2bc88613cc140c59377/Formal/AG/SemanticRepair/Saga/EquationLift.lean\#L172}{\texttt{residual\_\allowbreak{}cocycle}} / \href{https://github.com/iroha1203/AlgebraicArchitectureTheoryV2/blob/719f81f47d410701fd82c2bc88613cc140c59377/Formal/AG/SemanticRepair/Saga/EquationLift.lean\#L251}{\texttt{residual\_\allowbreak{}choice}} / \href{https://github.com/iroha1203/AlgebraicArchitectureTheoryV2/blob/719f81f47d410701fd82c2bc88613cc140c59377/Formal/AG/SemanticRepair/Saga/EquationLift.lean\#L300}{\texttt{residual\allowbreak{}Class\_\allowbreak{}choice\_\allowbreak{}independent}} & Obtains the cocycle from freeness of the action, and the coboundary between choices from transitivity. \\
\addlinespace[3pt]
Definition~\ref{def:2.30} & \href{https://github.com/iroha1203/AlgebraicArchitectureTheoryV2/blob/719f81f47d410701fd82c2bc88613cc140c59377/research/lean/ResearchLean/AG/TwoPhase/DependencyProfile.lean\#L94}{\texttt{Declared\allowbreak{}Semantic\allowbreak{}Family}} / \href{https://github.com/iroha1203/AlgebraicArchitectureTheoryV2/blob/719f81f47d410701fd82c2bc88613cc140c59377/research/lean/ResearchLean/AG/TwoPhase/DependencyProfile.lean\#L112}{\texttt{Structural}} & Invariance of the truth value of the extraction across all of the declared semantic variants. \\
\addlinespace[3pt]
Definition~\ref{def:2.31} & \href{https://github.com/iroha1203/AlgebraicArchitectureTheoryV2/blob/719f81f47d410701fd82c2bc88613cc140c59377/research/lean/ResearchLean/AG/TwoPhase/CoefficientComplex.lean\#L331}{\texttt{Atom\allowbreak{}Indexed\allowbreak{}Coefficient\allowbreak{}Complex}} / \href{https://github.com/iroha1203/AlgebraicArchitectureTheoryV2/blob/719f81f47d410701fd82c2bc88613cc140c59377/research/lean/ResearchLean/AG/TwoPhase/CoefficientComplex.lean\#L496}{\texttt{Condition\allowbreak{}E}} / \href{https://github.com/iroha1203/AlgebraicArchitectureTheoryV2/blob/719f81f47d410701fd82c2bc88613cc140c59377/research/lean/ResearchLean/AG/TwoPhase/CoefficientComplex.lean\#L636}{\texttt{degreewise\_\allowbreak{}short\allowbreak{}Exact}} & Condition S is ConditionE in Lean. Exactness in each degree for the structural part that preserves the differential and for the semantic quotient. \\
\addlinespace[3pt]
Proposition~\ref{prop:2.32} & \href{https://github.com/iroha1203/AlgebraicArchitectureTheoryV2/blob/719f81f47d410701fd82c2bc88613cc140c59377/research/lean/ResearchLean/AG/TwoPhase/CohomologyComparison.lean\#L239}{\texttt{standard\allowbreak{}Semantic\allowbreak{}H1\allowbreak{}Map\_\allowbreak{}injective}} & Derives injectivity on $H^1$ of the canonical quotient map from ConditionE and the vanishing of $H^1$ on the structural side. \\
\addlinespace[3pt]
Definition~\ref{def:2.35} & \href{https://github.com/iroha1203/AlgebraicArchitectureTheoryV2/blob/719f81f47d410701fd82c2bc88613cc140c59377/research/lean/ResearchLean/AG/ObstructionDiagnosticBridge/GeneratorPresentation.lean\#L56}{\texttt{Generator\allowbreak{}Presentation}} / \href{https://github.com/iroha1203/AlgebraicArchitectureTheoryV2/blob/719f81f47d410701fd82c2bc88613cc140c59377/research/lean/ResearchLean/AG/ObstructionDiagnosticBridge/GeneratorPresentation.lean\#L72}{\texttt{Related}} & Law/source generators, and the equivalence closure of the primitive relations. \\
\addlinespace[3pt]
Lemma~\ref{lem:2.36} & \href{https://github.com/iroha1203/AlgebraicArchitectureTheoryV2/blob/719f81f47d410701fd82c2bc88613cc140c59377/research/lean/ResearchLean/AG/ObstructionDiagnosticBridge/PresentationGroup.lean\#L157}{\texttt{presentation\allowbreak{}Group\allowbreak{}Equiv\allowbreak{}Blocks}} & Identifies the relation quotient of a free abelian group with the finitely supported integer functions on the relation components by mutually inverse maps. \\
\addlinespace[3pt]
Construction~\ref{cons:2.37} & \href{https://github.com/iroha1203/AlgebraicArchitectureTheoryV2/blob/719f81f47d410701fd82c2bc88613cc140c59377/research/lean/ResearchLean/AG/ObstructionDiagnosticBridge/PointGeneratorAtomInput.lean\#L64}{\texttt{carrier}} / \href{https://github.com/iroha1203/AlgebraicArchitectureTheoryV2/blob/719f81f47d410701fd82c2bc88613cc140c59377/research/lean/ResearchLean/AG/ObstructionDiagnosticBridge/CombinedAtomContextSupport.lean\#L37}{\texttt{open\allowbreak{}Context}} / \href{https://github.com/iroha1203/AlgebraicArchitectureTheoryV2/blob/719f81f47d410701fd82c2bc88613cc140c59377/research/lean/ResearchLean/AG/ObstructionDiagnosticBridge/CombinedAtomContextSupport.lean\#L58}{\texttt{open\allowbreak{}Context\_\allowbreak{}reads\_\allowbreak{}point}} / \href{https://github.com/iroha1203/AlgebraicArchitectureTheoryV2/blob/719f81f47d410701fd82c2bc88613cc140c59377/research/lean/ResearchLean/AG/ObstructionDiagnosticBridge/CombinedAtomContextSupport.lean\#L68}{\texttt{open\allowbreak{}Context\_\allowbreak{}reads\_\allowbreak{}generator}} & Carries points and generators as Atoms with distinct tags, and constructs the context that reads the points of an open set and all the generators. \\
\addlinespace[3pt]
Lemma~\ref{lem:2.38} & \href{https://github.com/iroha1203/AlgebraicArchitectureTheoryV2/blob/719f81f47d410701fd82c2bc88613cc140c59377/research/lean/ResearchLean/AG/ObstructionDiagnosticBridge/FiniteCoverGeometry.lean\#L193}{\texttt{fine\_\allowbreak{}cover}} / \href{https://github.com/iroha1203/AlgebraicArchitectureTheoryV2/blob/719f81f47d410701fd82c2bc88613cc140c59377/research/lean/ResearchLean/AG/ObstructionDiagnosticBridge/FiniteCoverGeometry.lean\#L206}{\texttt{coarse\_\allowbreak{}cover}} / \href{https://github.com/iroha1203/AlgebraicArchitectureTheoryV2/blob/719f81f47d410701fd82c2bc88613cc140c59377/research/lean/ResearchLean/AG/ObstructionDiagnosticBridge/FiniteCoverGeometry.lean\#L136}{\texttt{patch\_\allowbreak{}is\allowbreak{}Preconnected}} / \href{https://github.com/iroha1203/AlgebraicArchitectureTheoryV2/blob/719f81f47d410701fd82c2bc88613cc140c59377/research/lean/ResearchLean/AG/ObstructionDiagnosticBridge/FiniteCoverGeometry.lean\#L366}{\texttt{fine\_\allowbreak{}distinct\_\allowbreak{}triple\_\allowbreak{}empty}} & An eight-point finite space, the covers by three charts and by four charts, their connectedness, and the absence of triple intersections. \\
\addlinespace[3pt]
Construction~\ref{cons:2.39} & \href{https://github.com/iroha1203/AlgebraicArchitectureTheoryV2/blob/719f81f47d410701fd82c2bc88613cc140c59377/research/lean/ResearchLean/AG/ObstructionDiagnosticBridge/AATLocallyConstantObstruction.lean\#L65}{\texttt{aat\allowbreak{}Locally\allowbreak{}Constant\allowbreak{}Add\allowbreak{}Comm\allowbreak{}Grp\allowbreak{}Presheaf\_\allowbreak{}is\allowbreak{}Sheaf}} / \href{https://github.com/iroha1203/AlgebraicArchitectureTheoryV2/blob/719f81f47d410701fd82c2bc88613cc140c59377/research/lean/ResearchLean/AG/ObstructionDiagnosticBridge/CombinedAtomContextContinuity.lean\#L114}{\texttt{support\allowbreak{}Functor\_\allowbreak{}is\allowbreak{}Continuous}} & By the continuity of the support functor, pulls back the locally constant coefficient sheaf. \\
\addlinespace[3pt]
Proposition~\ref{prop:2.40} & \href{https://github.com/iroha1203/AlgebraicArchitectureTheoryV2/blob/719f81f47d410701fd82c2bc88613cc140c59377/research/lean/ResearchLean/AG/ObstructionDiagnosticBridge/ExistingObstructionBridge.lean\#L75}{\texttt{chart\allowbreak{}Lawful\allowbreak{}Section\_\allowbreak{}lawful}} / \href{https://github.com/iroha1203/AlgebraicArchitectureTheoryV2/blob/719f81f47d410701fd82c2bc88613cc140c59377/research/lean/ResearchLean/AG/ObstructionDiagnosticBridge/ExistingObstructionBridge.lean\#L136}{\texttt{restricted\allowbreak{}Lawful\allowbreak{}Section\allowbreak{}Data\_\allowbreak{}lawful}} / \href{https://github.com/iroha1203/AlgebraicArchitectureTheoryV2/blob/719f81f47d410701fd82c2bc88613cc140c59377/research/lean/ResearchLean/AG/ObstructionDiagnosticBridge/ExistingObstructionBridge.lean\#L247}{\texttt{translated\allowbreak{}Right\allowbreak{}Lawful\allowbreak{}Section\allowbreak{}Data\_\allowbreak{}lawful}} & Builds lawful sections from chart-wise evaluations that satisfy the relations, and preserves them under restriction and translation. \\
\addlinespace[3pt]
Definition~\ref{def:2.43} & \href{https://github.com/iroha1203/AlgebraicArchitectureTheoryV2/blob/719f81f47d410701fd82c2bc88613cc140c59377/Formal/AG/SemanticRepair/Saga/Presentation.lean\#L155}{\texttt{Semantic\allowbreak{}Repair\allowbreak{}Presentation}} / \href{https://github.com/iroha1203/AlgebraicArchitectureTheoryV2/blob/719f81f47d410701fd82c2bc88613cc140c59377/Formal/AG/SemanticRepair/Saga/Presentation.lean\#L221}{\texttt{m\allowbreak{}Sem\allowbreak{}Presheaf}} & Takes the quotient by the subgroup generated by the primitive repair relations, and builds the natural coefficient presheaf. \\
\addlinespace[3pt]
Proposition~\ref{prop:2.44} & \href{https://github.com/iroha1203/AlgebraicArchitectureTheoryV2/blob/719f81f47d410701fd82c2bc88613cc140c59377/Formal/AG/SemanticRepair/Saga/RepairTorsor.lean\#L131}{\texttt{mact\_\allowbreak{}free}} / \href{https://github.com/iroha1203/AlgebraicArchitectureTheoryV2/blob/719f81f47d410701fd82c2bc88613cc140c59377/Formal/AG/SemanticRepair/Saga/RepairTorsor.lean\#L142}{\texttt{mact\_\allowbreak{}transitive}} / \href{https://github.com/iroha1203/AlgebraicArchitectureTheoryV2/blob/719f81f47d410701fd82c2bc88613cc140c59377/Formal/AG/SemanticRepair/Saga/RepairTorsor.lean\#L165}{\texttt{to\allowbreak{}Lift\allowbreak{}System}} & Builds the free transitive action of the quotient group from triviality of the relation action, agreement of the stabilizers, and transitivity. \\
\addlinespace[3pt]
Definition~\ref{def:2.45} & \href{https://github.com/iroha1203/AlgebraicArchitectureTheoryV2/blob/719f81f47d410701fd82c2bc88613cc140c59377/Formal/AG/SemanticRepair/Saga/EquationLift.lean\#L52}{\texttt{Affine\allowbreak{}Coefficient\allowbreak{}Lift\allowbreak{}System}} / \href{https://github.com/iroha1203/AlgebraicArchitectureTheoryV2/blob/719f81f47d410701fd82c2bc88613cc140c59377/Formal/AG/SemanticRepair/Saga/EquationLift.lean\#L74}{\texttt{Coefficient\allowbreak{}Lift\allowbreak{}Atlas}} & Uses the equation lift system and the residual on the actual intersections. \\
\addlinespace[3pt]
Construction~\ref{cons:2.46} & \href{https://github.com/iroha1203/AlgebraicArchitectureTheoryV2/blob/719f81f47d410701fd82c2bc88613cc140c59377/Formal/AG/SemanticRepair/Saga/EquationRealization.lean\#L95}{\texttt{chi\allowbreak{}E}} / \href{https://github.com/iroha1203/AlgebraicArchitectureTheoryV2/blob/719f81f47d410701fd82c2bc88613cc140c59377/Formal/AG/SemanticRepair/Saga/EquationRealization.lean\#L122}{\texttt{chi\allowbreak{}E\_\allowbreak{}natural}} & Sends the actual equation residual of a generator to the quotient coefficients, and makes it commute with the face restrictions. \\
\addlinespace[3pt]
Definition~\ref{def:2.47} & \href{https://github.com/iroha1203/AlgebraicArchitectureTheoryV2/blob/719f81f47d410701fd82c2bc88613cc140c59377/Formal/AG/SemanticRepair/Saga/Exactness.lean\#L105}{\texttt{Primary\allowbreak{}State\allowbreak{}Correspondence}} / \href{https://github.com/iroha1203/AlgebraicArchitectureTheoryV2/blob/719f81f47d410701fd82c2bc88613cc140c59377/Formal/AG/SemanticRepair/Saga/Exactness.lean\#L215}{\texttt{Relation\allowbreak{}Complete}} / \href{https://github.com/iroha1203/AlgebraicArchitectureTheoryV2/blob/719f81f47d410701fd82c2bc88613cc140c59377/Formal/AG/SemanticRepair/Saga/Exactness.lean\#L219}{\texttt{Generator\allowbreak{}Complete}} & Makes explicit the comparison of states on generators, and the relation-completeness and generator-completeness of the coefficient map. \\
\addlinespace[3pt]
Lemma~\ref{lem:2.48} & \href{https://github.com/iroha1203/AlgebraicArchitectureTheoryV2/blob/719f81f47d410701fd82c2bc88613cc140c59377/Formal/AG/SemanticRepair/Saga/EquationRealization.lean\#L169}{\texttt{equation\allowbreak{}Relation\allowbreak{}Sound}} / \href{https://github.com/iroha1203/AlgebraicArchitectureTheoryV2/blob/719f81f47d410701fd82c2bc88613cc140c59377/Formal/AG/SemanticRepair/Saga/Exactness.lean\#L190}{\texttt{relation\allowbreak{}Sound\_\allowbreak{}of\_\allowbreak{}state\allowbreak{}Correspondence}} & Extends the comparison of the actions of generators to words, and derives soundness of the relations from freeness of the state action. \\
\addlinespace[3pt]
Theorem~\ref{thm:2.49} & \href{https://github.com/iroha1203/AlgebraicArchitectureTheoryV2/blob/719f81f47d410701fd82c2bc88613cc140c59377/Formal/AG/SemanticRepair/Saga/KappaComparison.lean\#L92}{\texttt{kappa1\_\allowbreak{}delta0}} / \href{https://github.com/iroha1203/AlgebraicArchitectureTheoryV2/blob/719f81f47d410701fd82c2bc88613cc140c59377/Formal/AG/SemanticRepair/Saga/KappaComparison.lean\#L103}{\texttt{kappa2\_\allowbreak{}delta1}} / \href{https://github.com/iroha1203/AlgebraicArchitectureTheoryV2/blob/719f81f47d410701fd82c2bc88613cc140c59377/Formal/AG/SemanticRepair/Saga/KappaComparison.lean\#L345}{\texttt{kappa\allowbreak{}Star\allowbreak{}Add\allowbreak{}Equiv}} / \href{https://github.com/iroha1203/AlgebraicArchitectureTheoryV2/blob/719f81f47d410701fd82c2bc88613cc140c59377/Formal/AG/SemanticRepair/Saga/KappaComparison.lean\#L477}{\texttt{residual\_\allowbreak{}correspondence\_\allowbreak{}class}} & Constructs isomorphisms of complexes and of $H^1$ from the actual coefficient isomorphism, and also treats the difference between distinct atlas choices as a coboundary. \\
\addlinespace[3pt]
Corollary~\ref{cor:2.50} & \href{https://github.com/iroha1203/AlgebraicArchitectureTheoryV2/blob/719f81f47d410701fd82c2bc88613cc140c59377/Formal/AG/SemanticRepair/Saga/TrueSheafDescent.lean\#L277}{\texttt{saga\allowbreak{}Grounded\allowbreak{}Gluing}} / \href{https://github.com/iroha1203/AlgebraicArchitectureTheoryV2/blob/719f81f47d410701fd82c2bc88613cc140c59377/Formal/AG/SemanticRepair/Saga/TrueSheafDescent.lean\#L257}{\texttt{global\allowbreak{}Repair\_\allowbreak{}nonempty\_\allowbreak{}iff}} & For SAGA inputs that assume the actual state sheaf and the normalization of empty intersections, matches the existence of a global repair with the vanishing of the specified class. \\
\bottomrule
\end{xltabular}

\subsection*{Restriction of ideals, dimension lower bounds, and obstruction classes}

\begin{xltabular}{\linewidth}{@{}LLL@{}}
\toprule
Text & Lean declaration & Objects and conditions \\
\midrule
\endhead
Proposition~\ref{prop:2.4} (restriction of ideals) & \href{https://github.com/iroha1203/AlgebraicArchitectureTheoryV2/blob/719f81f47d410701fd82c2bc88613cc140c59377/Formal/AG/LawAlgebra/LawEquation.lean\#L76}{\texttt{map\_\allowbreak{}witness\allowbreak{}Ideal\_\allowbreak{}le}} / \href{https://github.com/iroha1203/AlgebraicArchitectureTheoryV2/blob/719f81f47d410701fd82c2bc88613cc140c59377/Formal/AG/LawAlgebra/LawEquation.lean\#L99}{\texttt{map\_\allowbreak{}obstruction\allowbreak{}Ideal\_\allowbreak{}le}} & The witness ideal and the obstruction ideal are preserved under restriction. \\
\addlinespace[3pt]
Proposition~\ref{prop:2.24} and \eqref{eq:2.16} & \href{https://github.com/iroha1203/AlgebraicArchitectureTheoryV2/blob/719f81f47d410701fd82c2bc88613cc140c59377/Formal/AG/Cohomology/CoverNerve.lean\#L223}{\texttt{topological\allowbreak{}Debt\allowbreak{}Capacity\_\allowbreak{}from\allowbreak{}Complex}} & A dimension lower bound for the three-term complex. \\
\addlinespace[3pt]
Proposition~\ref{prop:2.41} (comparison of obstruction classes) & \href{https://github.com/iroha1203/AlgebraicArchitectureTheoryV2/blob/719f81f47d410701fd82c2bc88613cc140c59377/research/lean/ResearchLean/AG/ObstructionDiagnosticBridge/SpecifiedAffineObstruction.lean\#L90}{\texttt{actual\_\allowbreak{}class\_\allowbreak{}adjust\_\allowbreak{}local\_\allowbreak{}state}} / \href{https://github.com/iroha1203/AlgebraicArchitectureTheoryV2/blob/719f81f47d410701fd82c2bc88613cc140c59377/research/lean/ResearchLean/AG/ObstructionDiagnosticBridge/ExistingObstructionBridge.lean\#L385}{\texttt{gluing\allowbreak{}Mismatch\allowbreak{}Cochain\_\allowbreak{}eq\_\allowbreak{}actual\allowbreak{}Mismatch}} / \href{https://github.com/iroha1203/AlgebraicArchitectureTheoryV2/blob/719f81f47d410701fd82c2bc88613cc140c59377/research/lean/ResearchLean/AG/ObstructionDiagnosticBridge/ExistingObstructionBridge.lean\#L427}{\texttt{existing\allowbreak{}Descent\allowbreak{}Additive\allowbreak{}Class\_\allowbreak{}eq\_\allowbreak{}actual\allowbreak{}Class}} & Agreement of the class of the mismatch obtained from local states with the obstruction class, and invariance under correction of the local states. \\
\addlinespace[3pt]
Example~\ref{ex:2.42} (nonvanishing) & \href{https://github.com/iroha1203/AlgebraicArchitectureTheoryV2/blob/719f81f47d410701fd82c2bc88613cc140c59377/research/lean/ResearchLean/AG/ObstructionDiagnosticBridge/SelectedFiniteObstructionExamples.lean\#L177}{\texttt{coarse\_\allowbreak{}nonzero\_\allowbreak{}actual}} & Nonvanishing of the specified obstruction class with integer coefficients. \\
\bottomrule
\end{xltabular}

\section{Chapter 3: Canonical Resolution and Diagnostic Invariance}\label{sec:A.3}

\begin{xltabular}{\linewidth}{@{}LLL@{}}
\toprule
Text & Lean declaration & Objects and conditions \\
\midrule
\endhead
Definition~\ref{def:3.1} & \href{https://github.com/iroha1203/AlgebraicArchitectureTheoryV2/blob/719f81f47d410701fd82c2bc88613cc140c59377/research/lean/ResearchLean/AG/CanonicalResolution/Reading.lean\#L22}{\texttt{Reading}} / \href{https://github.com/iroha1203/AlgebraicArchitectureTheoryV2/blob/719f81f47d410701fd82c2bc88613cc140c59377/research/lean/ResearchLean/AG/CanonicalResolution/Reading.lean\#L156}{\texttt{Adequate}} & Surjections from the same source, and the condition that the evaluation of each Law descends. \\
\addlinespace[3pt]
Lemma~\ref{lem:3.2} & \href{https://github.com/iroha1203/AlgebraicArchitectureTheoryV2/blob/719f81f47d410701fd82c2bc88613cc140c59377/research/lean/ResearchLean/AG/CanonicalResolution/Reading.lean\#L65}{\texttt{factors\_\allowbreak{}iff\_\allowbreak{}kernel}} / \href{https://github.com/iroha1203/AlgebraicArchitectureTheoryV2/blob/719f81f47d410701fd82c2bc88613cc140c59377/research/lean/ResearchLean/AG/ResolutionInvariance/ComparisonData.lean\#L43}{\texttt{comparison\allowbreak{}Factor\_\allowbreak{}unique}} / \href{https://github.com/iroha1203/AlgebraicArchitectureTheoryV2/blob/719f81f47d410701fd82c2bc88613cc140c59377/research/lean/ResearchLean/AG/ResolutionInvariance/ComparisonData.lean\#L55}{\texttt{comparison\allowbreak{}Factor\_\allowbreak{}surjective}} & The equivalence of constancy on the kernel with descent, and the uniqueness and surjectivity of the canonical factor. \\
\addlinespace[3pt]
Definition~\ref{def:3.3} & \href{https://github.com/iroha1203/AlgebraicArchitectureTheoryV2/blob/719f81f47d410701fd82c2bc88613cc140c59377/research/lean/ResearchLean/AG/CanonicalResolution/JointKernel.lean\#L34}{\texttt{joint\allowbreak{}Kernel\allowbreak{}Reading}} & Builds the quotient reading by taking the agreement of the values of all Laws as the equivalence relation. \\
\addlinespace[3pt]
Theorem~\ref{thm:3.4} & \href{https://github.com/iroha1203/AlgebraicArchitectureTheoryV2/blob/719f81f47d410701fd82c2bc88613cc140c59377/research/lean/ResearchLean/AG/CanonicalResolution/JointKernel.lean\#L61}{\texttt{joint\allowbreak{}Kernel\_\allowbreak{}adequate}} / \href{https://github.com/iroha1203/AlgebraicArchitectureTheoryV2/blob/719f81f47d410701fd82c2bc88613cc140c59377/research/lean/ResearchLean/AG/CanonicalResolution/JointKernel.lean\#L114}{\texttt{joint\allowbreak{}Kernel\_\allowbreak{}universal}} & Constructs the factor from an arbitrary adequate reading to the canonical quotient, and proves its uniqueness. \\
\addlinespace[3pt]
Construction~\ref{cons:3.6} & \href{https://github.com/iroha1203/AlgebraicArchitectureTheoryV2/blob/719f81f47d410701fd82c2bc88613cc140c59377/research/lean/ResearchLean/AG/CanonicalResolution/Effective.lean\#L69}{\texttt{computed\allowbreak{}Class\_\allowbreak{}eq\_\allowbreak{}iff}} / \href{https://github.com/iroha1203/AlgebraicArchitectureTheoryV2/blob/719f81f47d410701fd82c2bc88613cc140c59377/research/lean/ResearchLean/AG/CanonicalResolution/Effective.lean\#L106}{\texttt{computed\_\allowbreak{}adequate}} / \href{https://github.com/iroha1203/AlgebraicArchitectureTheoryV2/blob/719f81f47d410701fd82c2bc88613cc140c59377/research/lean/ResearchLean/AG/CanonicalResolution/Effective.lean\#L114}{\texttt{computed\_\allowbreak{}kernel\allowbreak{}Equivalent\_\allowbreak{}joint\allowbreak{}Kernel}} & Agreement of the partition computed from a finite table with the joint kernel. \\
\addlinespace[3pt]
Definition~\ref{def:3.7} & \href{https://github.com/iroha1203/AlgebraicArchitectureTheoryV2/blob/719f81f47d410701fd82c2bc88613cc140c59377/research/lean/ResearchLean/AG/CanonicalResolution/Admissible.lean\#L68}{\texttt{Admissible\allowbreak{}Class}} / \href{https://github.com/iroha1203/AlgebraicArchitectureTheoryV2/blob/719f81f47d410701fd82c2bc88613cc140c59377/research/lean/ResearchLean/AG/CanonicalResolution/Admissible.lean\#L89}{\texttt{Representable}} & Presentation of the canonical kernel by the extraction readings of specified finite doctrine candidates. \\
\addlinespace[3pt]
Example~\ref{ex:3.8} & \href{https://github.com/iroha1203/AlgebraicArchitectureTheoryV2/blob/719f81f47d410701fd82c2bc88613cc140c59377/research/lean/ResearchLean/AG/CanonicalResolution/PositiveWitness.lean\#L176}{\texttt{representable}} / \href{https://github.com/iroha1203/AlgebraicArchitectureTheoryV2/blob/719f81f47d410701fd82c2bc88613cc140c59377/research/lean/ResearchLean/AG/CanonicalResolution/NegativeWitness.lean\#L171}{\texttt{not\_\allowbreak{}representable}} & A presentable family of candidates and a non-presentable one, built from two proper refinements of six sources. \\
\addlinespace[3pt]
Definition~\ref{def:3.9}, Construction~\ref{cons:3.10} & \href{https://github.com/iroha1203/AlgebraicArchitectureTheoryV2/blob/719f81f47d410701fd82c2bc88613cc140c59377/research/lean/ResearchLean/AG/ResolutionInvariance/LawGeneratedComplex.lean\#L48}{\texttt{Target\allowbreak{}Supported\allowbreak{}Nerve}} / \href{https://github.com/iroha1203/AlgebraicArchitectureTheoryV2/blob/719f81f47d410701fd82c2bc88613cc140c59377/research/lean/ResearchLean/AG/ResolutionInvariance/LawGeneratedComplex.lean\#L485}{\texttt{law\allowbreak{}Generated\allowbreak{}Complex}} & Generates the coordinates from pairs of a support and a Law value, and does not make witnesses into additional coordinates. \\
\addlinespace[3pt]
Lemma~\ref{lem:3.11} & \href{https://github.com/iroha1203/AlgebraicArchitectureTheoryV2/blob/719f81f47d410701fd82c2bc88613cc140c59377/research/lean/ResearchLean/AG/ResolutionInvariance/LawGeneratedComplex.lean\#L457}{\texttt{law\allowbreak{}Generated\_\allowbreak{}d1\_\allowbreak{}comp\_\allowbreak{}d0}} & The composite of the two generated differentials is zero. \\
\addlinespace[3pt]
Proposition~\ref{prop:3.12} & \href{https://github.com/iroha1203/AlgebraicArchitectureTheoryV2/blob/719f81f47d410701fd82c2bc88613cc140c59377/research/lean/ResearchLean/AG/ResolutionInvariance/LawValueBlockCohomology.lean\#L338}{\texttt{law\allowbreak{}Generated\allowbreak{}H1\allowbreak{}Block\allowbreak{}Equiv}} & Constructs the actual isomorphism on $H^1$ from the cochain isomorphism onto the Law-value blocks and the coherence with the differentials. \\
\addlinespace[3pt]
Definition~\ref{def:3.13} & \href{https://github.com/iroha1203/AlgebraicArchitectureTheoryV2/blob/719f81f47d410701fd82c2bc88613cc140c59377/research/lean/ResearchLean/AG/ResolutionInvariance/SupportedNerveMorphism.lean\#L38}{\texttt{Target\allowbreak{}Supported\allowbreak{}Nerve\allowbreak{}Morphism}} & A chart map, partial maps on edges and faces that allow contraction, and inclusions of supports. \\
\addlinespace[3pt]
Proposition~\ref{prop:3.14} & \href{https://github.com/iroha1203/AlgebraicArchitectureTheoryV2/blob/719f81f47d410701fd82c2bc88613cc140c59377/research/lean/ResearchLean/AG/ResolutionInvariance/GeneratedComparisonMap.lean\#L430}{\texttt{generated\allowbreak{}Pullback\_\allowbreak{}comm0}} / \href{https://github.com/iroha1203/AlgebraicArchitectureTheoryV2/blob/719f81f47d410701fd82c2bc88613cc140c59377/research/lean/ResearchLean/AG/ResolutionInvariance/GeneratedComparisonMap.lean\#L477}{\texttt{generated\allowbreak{}Pullback\_\allowbreak{}comm1}} / \href{https://github.com/iroha1203/AlgebraicArchitectureTheoryV2/blob/719f81f47d410701fd82c2bc88613cc140c59377/research/lean/ResearchLean/AG/ResolutionInvariance/GeneratedComparisonMap.lean\#L606}{\texttt{generated\allowbreak{}Comparison\allowbreak{}H1\allowbreak{}Map}} & The comparison on the actual coordinates commutes with both differentials and descends to $H^1$. \\
\addlinespace[3pt]
Definition~\ref{def:3.15} & \href{https://github.com/iroha1203/AlgebraicArchitectureTheoryV2/blob/719f81f47d410701fd82c2bc88613cc140c59377/research/lean/ResearchLean/AG/ResolutionInvariance/ResolutionInvarianceConditions.lean\#L549}{\texttt{Condition\allowbreak{}C}} & C0, C5, and C6, together with C1--C4 for each Law-value block. \\
\addlinespace[3pt]
Lemma~\ref{lem:3.16} & \href{https://github.com/iroha1203/AlgebraicArchitectureTheoryV2/blob/719f81f47d410701fd82c2bc88613cc140c59377/research/lean/ResearchLean/AG/ResolutionInvariance/LawValueBlockComparisonFiberPrimitive.lean\#L261}{\texttt{law\allowbreak{}Value\allowbreak{}Block\allowbreak{}Cycle\_\allowbreak{}exists\_\allowbreak{}coordinate\allowbreak{}Fiber\allowbreak{}Primitive}} & Constructs, from the connectedness of the fibers and the rational chain condition, a primitive that kills the cocycle on the internal edges. \\
\addlinespace[3pt]
Theorem~\ref{thm:3.17} & \href{https://github.com/iroha1203/AlgebraicArchitectureTheoryV2/blob/719f81f47d410701fd82c2bc88613cc140c59377/research/lean/ResearchLean/AG/ResolutionInvariance/LawValueBlockComparisonBijectivity.lean\#L105}{\texttt{generated\allowbreak{}Comparison\allowbreak{}H1\allowbreak{}Map\_\allowbreak{}bijective}} & Uses C0--C6, and proves bijectivity of the actual comparison through the block decomposition and its naturality. \\
\addlinespace[3pt]
Definition~\ref{def:3.20} & \href{https://github.com/iroha1203/AlgebraicArchitectureTheoryV2/blob/719f81f47d410701fd82c2bc88613cc140c59377/research/lean/ResearchLean/AG/UniformInvariance/DefectSemantics.lean\#L33}{\texttt{block\allowbreak{}Defect}} / \href{https://github.com/iroha1203/AlgebraicArchitectureTheoryV2/blob/719f81f47d410701fd82c2bc88613cc140c59377/research/lean/ResearchLean/AG/UniformInvariance/DefectSemantics.lean\#L115}{\texttt{a\allowbreak{}Subnerve\allowbreak{}Defect}} & The dimensions of the kernel and the cokernel of the actual comparison map of each subnerve. \\
\addlinespace[3pt]
Definition~\ref{def:3.21} & \href{https://github.com/iroha1203/AlgebraicArchitectureTheoryV2/blob/719f81f47d410701fd82c2bc88613cc140c59377/research/lean/ResearchLean/AG/UniformInvariance/UniformityReduction.lean\#L254}{\texttt{Uniform\allowbreak{}Invariance}} & Quantifies over all adequate finite families of Laws for the same comparison geometry. \\
\addlinespace[3pt]
Theorem~\ref{thm:3.22} & \href{https://github.com/iroha1203/AlgebraicArchitectureTheoryV2/blob/719f81f47d410701fd82c2bc88613cc140c59377/research/lean/ResearchLean/AG/UniformInvariance/UniformityReduction.lean\#L272}{\texttt{uniform\allowbreak{}Invariance\_\allowbreak{}iff\_\allowbreak{}all\allowbreak{}Nonempty\allowbreak{}ASubnerve\allowbreak{}H1\allowbreak{}Bijective}} / \href{https://github.com/iroha1203/AlgebraicArchitectureTheoryV2/blob/719f81f47d410701fd82c2bc88613cc140c59377/research/lean/ResearchLean/AG/UniformInvariance/DefectSemantics.lean\#L134}{\texttt{uniform\allowbreak{}Invariance\_\allowbreak{}iff\_\allowbreak{}all\allowbreak{}Nonempty\allowbreak{}ASubnerve\allowbreak{}Defect\_\allowbreak{}eq\_\allowbreak{}zero}} & Passes from the fibers of a Law to subnerves, and detects the converse direction with indicator Laws. \\
\addlinespace[3pt]
Proposition~\ref{prop:3.23} & \href{https://github.com/iroha1203/AlgebraicArchitectureTheoryV2/blob/719f81f47d410701fd82c2bc88613cc140c59377/research/lean/ResearchLean/AG/UniformInvariance/UniformPresentationDecider.lean\#L117}{\texttt{uniform\allowbreak{}Presentation\allowbreak{}Check\_\allowbreak{}eq\_\allowbreak{}true\_\allowbreak{}iff}} & The equivalence of uniformity with the decision of the defect for all nonempty subsets by rational matrices. \\
\addlinespace[3pt]
Corollary~\ref{cor:3.24} & \href{https://github.com/iroha1203/AlgebraicArchitectureTheoryV2/blob/719f81f47d410701fd82c2bc88613cc140c59377/research/lean/ResearchLean/AG/UniformInvariance/AtlasPositioning.lean\#L43}{\texttt{uniform\allowbreak{}Invariance\_\allowbreak{}of\_\allowbreak{}condition\allowbreak{}CAll\allowbreak{}A}} & If Condition C holds on every nonempty subset, then uniform invariance follows. \\
\addlinespace[3pt]
Construction~\ref{cons:3.27} & \href{https://github.com/iroha1203/AlgebraicArchitectureTheoryV2/blob/719f81f47d410701fd82c2bc88613cc140c59377/research/lean/ResearchLean/AG/UniformInvariance/GLocalV1T3T6Witnesses.lean\#L273}{\texttt{t3\allowbreak{}Presentation}} / \href{https://github.com/iroha1203/AlgebraicArchitectureTheoryV2/blob/719f81f47d410701fd82c2bc88613cc140c59377/research/lean/ResearchLean/AG/UniformInvariance/GLocalV1T3T6Witnesses.lean\#L284}{\texttt{t6\allowbreak{}Presentation}} & Finite comparison inputs of period 3 and of period 6. \\
\addlinespace[3pt]
Definition~\ref{def:3.29} & \href{https://github.com/iroha1203/AlgebraicArchitectureTheoryV2/blob/719f81f47d410701fd82c2bc88613cc140c59377/research/lean/ResearchLean/AG/StructuralCover/NerveGeneration.lean\#L180}{\texttt{structural\allowbreak{}Support}} / \href{https://github.com/iroha1203/AlgebraicArchitectureTheoryV2/blob/719f81f47d410701fd82c2bc88613cc140c59377/research/lean/ResearchLean/AG/StructuralCover/NerveGeneration.lean\#L63}{\texttt{nerve\allowbreak{}Of}} & Takes the invariant part of the extraction as support, and takes as cells the ordered tuples of Atoms realized on a common source. \\
\addlinespace[3pt]
Proposition~\ref{prop:3.32} & \href{https://github.com/iroha1203/AlgebraicArchitectureTheoryV2/blob/719f81f47d410701fd82c2bc88613cc140c59377/research/lean/ResearchLean/AG/StructuralCover/GeneratedH1Vanishing.lean\#L155}{\texttt{generated\allowbreak{}D0\_\allowbreak{}cocycle\allowbreak{}Primitive}} / \href{https://github.com/iroha1203/AlgebraicArchitectureTheoryV2/blob/719f81f47d410701fd82c2bc88613cc140c59377/research/lean/ResearchLean/AG/StructuralCover/GeneratedH1Vanishing.lean\#L195}{\texttt{generated\allowbreak{}All\allowbreak{}Complex\_\allowbreak{}h1\allowbreak{}Zero}} & Builds the actual primitive from a reference Atom for each source, and turns every cocycle into a coboundary. \\
\addlinespace[3pt]
Construction~\ref{cons:3.33} & \href{https://github.com/iroha1203/AlgebraicArchitectureTheoryV2/blob/719f81f47d410701fd82c2bc88613cc140c59377/research/lean/ResearchLean/AG/ObstructionDiagnosticBridge/CoefficientComparison.lean\#L130}{\texttt{coefficient\allowbreak{}Comparison}} / \href{https://github.com/iroha1203/AlgebraicArchitectureTheoryV2/blob/719f81f47d410701fd82c2bc88613cc140c59377/research/lean/ResearchLean/AG/ObstructionDiagnosticBridge/CoefficientComparison.lean\#L136}{\texttt{coefficient\allowbreak{}Comparison\_\allowbreak{}generator\allowbreak{}Class\_\allowbreak{}apply}} & Maps from the integer quotient of generators to the finitely supported rational delta functions on Law labels, preserving the primitive relations. \\
\addlinespace[3pt]
Lemma~\ref{lem:3.34} & \href{https://github.com/iroha1203/AlgebraicArchitectureTheoryV2/blob/719f81f47d410701fd82c2bc88613cc140c59377/research/lean/ResearchLean/AG/ObstructionDiagnosticBridge/CoefficientComparison.lean\#L149}{\texttt{coefficient\allowbreak{}Comparison\_\allowbreak{}injective}} & Derives injectivity through the component presentation, from the connectedness by the primitive relations of the same label. \\
\addlinespace[3pt]
Example~\ref{ex:3.35} & \href{https://github.com/iroha1203/AlgebraicArchitectureTheoryV2/blob/719f81f47d410701fd82c2bc88613cc140c59377/research/lean/ResearchLean/AG/ObstructionDiagnosticBridge/CoefficientComparison.lean\#L204}{\texttt{disconnected\_\allowbreak{}coefficient\allowbreak{}Comparison\_\allowbreak{}not\_\allowbreak{}injective}} & For an input with two generators, the same label, and empty relations, the difference is nonzero and the image vanishes. \\
\addlinespace[3pt]
Construction~\ref{cons:3.36} & \href{https://github.com/iroha1203/AlgebraicArchitectureTheoryV2/blob/719f81f47d410701fd82c2bc88613cc140c59377/research/lean/ResearchLean/AG/ObstructionDiagnosticBridge/FaceEmptyCechNormalization.lean\#L165}{\texttt{face\allowbreak{}Empty\allowbreak{}Cech\allowbreak{}Complex}} / \href{https://github.com/iroha1203/AlgebraicArchitectureTheoryV2/blob/719f81f47d410701fd82c2bc88613cc140c59377/research/lean/ResearchLean/AG/ObstructionDiagnosticBridge/FaceEmptyCechNormalization.lean\#L305}{\texttt{face\allowbreak{}Empty\allowbreak{}Cech\_\allowbreak{}d0\_\allowbreak{}normalizes}} / \href{https://github.com/iroha1203/AlgebraicArchitectureTheoryV2/blob/719f81f47d410701fd82c2bc88613cc140c59377/research/lean/ResearchLean/AG/ObstructionDiagnosticBridge/FaceEmptyCechNormalization.lean\#L429}{\texttt{actual\allowbreak{}Cech\allowbreak{}Coefficient\allowbreak{}Cochain\allowbreak{}Map}} & Identifies the actual sections on connected charts and intersections with constant coordinates, and constructs the comparison of face-empty \v{C}ech complexes. \\
\addlinespace[3pt]
Theorem~\ref{thm:3.37} & \href{https://github.com/iroha1203/AlgebraicArchitectureTheoryV2/blob/719f81f47d410701fd82c2bc88613cc140c59377/research/lean/ResearchLean/AG/ObstructionDiagnosticBridge/ActualCechH1Comparison.lean\#L73}{\texttt{actual\allowbreak{}Cech\allowbreak{}Diagnostic\allowbreak{}H1\allowbreak{}Map}} / \href{https://github.com/iroha1203/AlgebraicArchitectureTheoryV2/blob/719f81f47d410701fd82c2bc88613cc140c59377/research/lean/ResearchLean/AG/ObstructionDiagnosticBridge/SpecifiedAffineObstruction.lean\#L120}{\texttt{actual\_\allowbreak{}cech\_\allowbreak{}coefficient\_\allowbreak{}actual\_\allowbreak{}mismatch\_\allowbreak{}eq\_\allowbreak{}diagnostic\_\allowbreak{}mismatch}} / \href{https://github.com/iroha1203/AlgebraicArchitectureTheoryV2/blob/719f81f47d410701fd82c2bc88613cc140c59377/research/lean/ResearchLean/AG/ObstructionDiagnosticBridge/SpecifiedAffineObstruction.lean\#L145}{\texttt{h1\_\allowbreak{}map\_\allowbreak{}actual\_\allowbreak{}class\_\allowbreak{}eq\_\allowbreak{}diagnostic\_\allowbreak{}class}} & Equates, at the cochain stage, the actual mismatch of the same local data with the independently generated diagnostic mismatch, and descends it to $H^1$. \\
\addlinespace[3pt]
Definition~\ref{def:3.38} & \href{https://github.com/iroha1203/AlgebraicArchitectureTheoryV2/blob/719f81f47d410701fd82c2bc88613cc140c59377/research/lean/ResearchLean/AG/ObstructionDiagnosticBridge/SpecifiedClassReflection.lean\#L52}{\texttt{Common\allowbreak{}Label\allowbreak{}Chart\allowbreak{}Support}} / \href{https://github.com/iroha1203/AlgebraicArchitectureTheoryV2/blob/719f81f47d410701fd82c2bc88613cc140c59377/research/lean/ResearchLean/AG/ObstructionDiagnosticBridge/GeneratorPresentation.lean\#L146}{\texttt{Reflection\allowbreak{}Condition}} & A common representative across all charts for each label, and relation-connectedness of the generators with the same label. \\
\addlinespace[3pt]
Lemma~\ref{lem:3.39} & \href{https://github.com/iroha1203/AlgebraicArchitectureTheoryV2/blob/719f81f47d410701fd82c2bc88613cc140c59377/research/lean/ResearchLean/AG/ObstructionDiagnosticBridge/IntegralReflection.lean\#L46}{\texttt{floor\allowbreak{}Correction\_\allowbreak{}edge\allowbreak{}Difference}} / \href{https://github.com/iroha1203/AlgebraicArchitectureTheoryV2/blob/719f81f47d410701fd82c2bc88613cc140c59377/research/lean/ResearchLean/AG/ObstructionDiagnosticBridge/IntegralReflection.lean\#L66}{\texttt{exists\_\allowbreak{}integral\_\allowbreak{}correction}} & If the edge differences of a rational primitive are integers, taking the floor preserves the same edge differences. \\
\addlinespace[3pt]
Theorem~\ref{thm:3.40} & \href{https://github.com/iroha1203/AlgebraicArchitectureTheoryV2/blob/719f81f47d410701fd82c2bc88613cc140c59377/research/lean/ResearchLean/AG/ObstructionDiagnosticBridge/SpecifiedClassReflection.lean\#L383}{\texttt{diagnostic\_\allowbreak{}class\_\allowbreak{}eq\_\allowbreak{}zero\_\allowbreak{}iff\_\allowbreak{}actual\_\allowbreak{}class\_\allowbreak{}eq\_\allowbreak{}zero}} & Constructs an integer primitive for the specified class from adequacy, common representatives, and relation-connectedness, and reflects vanishing. \\
\addlinespace[3pt]
Construction~\ref{cons:3.42} & \href{https://github.com/iroha1203/AlgebraicArchitectureTheoryV2/blob/719f81f47d410701fd82c2bc88613cc140c59377/research/lean/ResearchLean/AG/ObstructionDiagnosticBridge/PointAtomActualNerve.lean\#L36}{\texttt{coarse\allowbreak{}Reading}} / \href{https://github.com/iroha1203/AlgebraicArchitectureTheoryV2/blob/719f81f47d410701fd82c2bc88613cc140c59377/research/lean/ResearchLean/AG/ObstructionDiagnosticBridge/PointAtomActualNerve.lean\#L42}{\texttt{fine\allowbreak{}Reading}} / \href{https://github.com/iroha1203/AlgebraicArchitectureTheoryV2/blob/719f81f47d410701fd82c2bc88613cc140c59377/research/lean/ResearchLean/AG/ObstructionDiagnosticBridge/CombinedAtomReadingNaturality.lean\#L49}{\texttt{actual\allowbreak{}H1\allowbreak{}Map}} / \href{https://github.com/iroha1203/AlgebraicArchitectureTheoryV2/blob/719f81f47d410701fd82c2bc88613cc140c59377/research/lean/ResearchLean/AG/ObstructionDiagnosticBridge/CombinedAtomReadingNaturality.lean\#L57}{\texttt{diagnostic\allowbreak{}H1\allowbreak{}Map}} & The actual restriction comparison on the same point and generator Atoms and on the three-chart and four-chart covers. \\
\addlinespace[3pt]
Theorem~\ref{thm:3.43} & \href{https://github.com/iroha1203/AlgebraicArchitectureTheoryV2/blob/719f81f47d410701fd82c2bc88613cc140c59377/research/lean/ResearchLean/AG/ObstructionDiagnosticBridge/CombinedAtomReadingNaturality.lean\#L132}{\texttt{h1\_\allowbreak{}comparison\_\allowbreak{}square}} / \href{https://github.com/iroha1203/AlgebraicArchitectureTheoryV2/blob/719f81f47d410701fd82c2bc88613cc140c59377/research/lean/ResearchLean/AG/ObstructionDiagnosticBridge/CombinedAtomReadingNaturality.lean\#L172}{\texttt{actual\allowbreak{}H1\allowbreak{}Map\_\allowbreak{}existing\allowbreak{}Obstruction\allowbreak{}Class}} / \href{https://github.com/iroha1203/AlgebraicArchitectureTheoryV2/blob/719f81f47d410701fd82c2bc88613cc140c59377/research/lean/ResearchLean/AG/ObstructionDiagnosticBridge/CombinedAtomReadingNaturality.lean\#L180}{\texttt{diagnostic\allowbreak{}H1\allowbreak{}Map\_\allowbreak{}diagnostic\allowbreak{}Class}} & The commutative square of the actual $H^1$ maps, and the transport of the specified class built from the same local data. \\
\addlinespace[3pt]
Theorem~\ref{thm:3.44} & \href{https://github.com/iroha1203/AlgebraicArchitectureTheoryV2/blob/719f81f47d410701fd82c2bc88613cc140c59377/research/lean/ResearchLean/AG/ObstructionDiagnosticBridge/SelectedReadingConditionC.lean\#L450}{\texttt{diagnostic\allowbreak{}H1\allowbreak{}Map\_\allowbreak{}bijective}} / \href{https://github.com/iroha1203/AlgebraicArchitectureTheoryV2/blob/719f81f47d410701fd82c2bc88613cc140c59377/research/lean/ResearchLean/AG/ObstructionDiagnosticBridge/SelectedReadingConditionC.lean\#L476}{\texttt{existing\_\allowbreak{}obstruction\_\allowbreak{}class\_\allowbreak{}eq\_\allowbreak{}zero\_\allowbreak{}iff\_\allowbreak{}mapped\_\allowbreak{}existing\_\allowbreak{}obstruction\_\allowbreak{}class\_\allowbreak{}eq\_\allowbreak{}zero}} & Connects Condition C for the actual refinement with the reflection of vanishing at both ends. \\
\bottomrule
\end{xltabular}

\subsection*{Refinements, uniform invariance, and concrete examples of obstruction classes}

\begin{xltabular}{\linewidth}{@{}LLL@{}}
\toprule
Text & Lean declaration & Objects and conditions \\
\midrule
\endhead
Example~\ref{ex:3.18} (comparison of refinements) & \href{https://github.com/iroha1203/AlgebraicArchitectureTheoryV2/blob/719f81f47d410701fd82c2bc88613cc140c59377/research/lean/ResearchLean/AG/ObstructionDiagnosticBridge/SelectedReadingConditionC.lean\#L437}{\texttt{condition\allowbreak{}C}} / \href{https://github.com/iroha1203/AlgebraicArchitectureTheoryV2/blob/719f81f47d410701fd82c2bc88613cc140c59377/research/lean/ResearchLean/AG/ObstructionDiagnosticBridge/SelectedReadingConditionC.lean\#L450}{\texttt{diagnostic\allowbreak{}H1\allowbreak{}Map\_\allowbreak{}bijective}} & Condition C for the specified refinement, and bijectivity of the comparison map. \\
\addlinespace[3pt]
Proposition~\ref{prop:3.28} (uniformity) & \href{https://github.com/iroha1203/AlgebraicArchitectureTheoryV2/blob/719f81f47d410701fd82c2bc88613cc140c59377/research/lean/ResearchLean/AG/UniformInvariance/GLocalV1T3T6Uniformity.lean\#L915}{\texttt{t3\_\allowbreak{}uniform\allowbreak{}Presentation}} / \href{https://github.com/iroha1203/AlgebraicArchitectureTheoryV2/blob/719f81f47d410701fd82c2bc88613cc140c59377/research/lean/ResearchLean/AG/UniformInvariance/GLocalV1T3T6Uniformity.lean\#L1002}{\texttt{t6\_\allowbreak{}not\_\allowbreak{}uniform\allowbreak{}Presentation}} & Uniformity of T3 and non-uniformity of T6. Obs\_loc is not included in the objects. \\
\addlinespace[3pt]
Example~\ref{ex:3.45} (four conditions for vanishing) & \href{https://github.com/iroha1203/AlgebraicArchitectureTheoryV2/blob/719f81f47d410701fd82c2bc88613cc140c59377/research/lean/ResearchLean/AG/ObstructionDiagnosticBridge/SelectedFiniteObstructionExamples.lean\#L426}{\texttt{existing\_\allowbreak{}zero\_\allowbreak{}example\_\allowbreak{}outcomes}} / \href{https://github.com/iroha1203/AlgebraicArchitectureTheoryV2/blob/719f81f47d410701fd82c2bc88613cc140c59377/research/lean/ResearchLean/AG/ObstructionDiagnosticBridge/SelectedFiniteObstructionExamples.lean\#L436}{\texttt{existing\_\allowbreak{}nonzero\_\allowbreak{}example\_\allowbreak{}outcomes}} & For the local data of the table in the text, the four vanishing conditions for the obstruction class and the diagnostic class on the coarse side and on the fine side. \\
\bottomrule
\end{xltabular}

\section{Chapter 4: Transport and Coherence of Composition}\label{sec:A.4}

\begin{xltabular}{\linewidth}{@{}LLL@{}}
\toprule
Text & Lean declaration & Objects and conditions \\
\midrule
\endhead
Definition~\ref{def:4.1} & \href{https://github.com/leanprover-community/mathlib4/blob/8f9d9cff6bd728b17a24e163c9402775d9e6a365/Mathlib/CategoryTheory/FiberedCategory/Cocartesian.lean\#L71}{\texttt{Is\allowbreak{}Strongly\allowbreak{}Cocartesian}} & Existence and uniqueness quantified over arbitrary subsequent morphisms in the base. \\
\addlinespace[3pt]
Lemma~\ref{lem:4.2} & \href{https://github.com/iroha1203/AlgebraicArchitectureTheoryV2/blob/719f81f47d410701fd82c2bc88613cc140c59377/research/lean/ResearchLean/AG/CrossStageCoherence/Pseudofunctor.lean\#L59}{\texttt{strong\allowbreak{}Lift\allowbreak{}Comparison\allowbreak{}Iso}} / \href{https://github.com/leanprover-community/mathlib4/blob/8f9d9cff6bd728b17a24e163c9402775d9e6a365/Mathlib/CategoryTheory/FiberedCategory/Cocartesian.lean\#L291}{\texttt{comp}} / \href{https://github.com/leanprover-community/mathlib4/blob/8f9d9cff6bd728b17a24e163c9402775d9e6a365/Mathlib/CategoryTheory/FiberedCategory/Cocartesian.lean\#L333}{\texttt{of\_\allowbreak{}iso}} & Vertical isomorphisms for a general functor, and closure under identity and composition. \\
\addlinespace[3pt]
Construction~\ref{cons:4.3} & \href{https://github.com/iroha1203/AlgebraicArchitectureTheoryV2/blob/719f81f47d410701fd82c2bc88613cc140c59377/research/lean/ResearchLean/AG/AtomFoundation/Transport.lean\#L51}{\texttt{transport\allowbreak{}Architecture\allowbreak{}Object\allowbreak{}Equiv}} / \href{https://github.com/iroha1203/AlgebraicArchitectureTheoryV2/blob/719f81f47d410701fd82c2bc88613cc140c59377/research/lean/ResearchLean/AG/AtomFoundation/Transport.lean\#L230}{\texttt{transport\allowbreak{}Operation\allowbreak{}Reading}} / \href{https://github.com/iroha1203/AlgebraicArchitectureTheoryV2/blob/719f81f47d410701fd82c2bc88613cc140c59377/research/lean/ResearchLean/AG/AtomFoundation/Transport.lean\#L660}{\texttt{transport\allowbreak{}Equation\allowbreak{}System\allowbreak{}Exact}} / \href{https://github.com/iroha1203/AlgebraicArchitectureTheoryV2/blob/719f81f47d410701fd82c2bc88613cc140c59377/research/lean/ResearchLean/AG/AtomFoundation/Transport.lean\#L843}{\texttt{transport\allowbreak{}Along}} & Reindexes objects, operations, contexts, equations, and detectors along a bijection of Atoms. \\
\addlinespace[3pt]
Theorem~\ref{thm:4.5} & \href{https://github.com/iroha1203/AlgebraicArchitectureTheoryV2/blob/719f81f47d410701fd82c2bc88613cc140c59377/research/lean/ResearchLean/AG/AtomFoundation/Opcartesian.lean\#L101}{\texttt{transport\allowbreak{}Along\allowbreak{}Hom\_\allowbreak{}factor\_\allowbreak{}exists\allowbreak{}Unique}} / \href{https://github.com/iroha1203/AlgebraicArchitectureTheoryV2/blob/719f81f47d410701fd82c2bc88613cc140c59377/research/lean/ResearchLean/AG/AtomFoundation/Opcartesian.lean\#L122}{\texttt{transport\allowbreak{}Along\allowbreak{}Hom\_\allowbreak{}is\allowbreak{}Strongly\allowbreak{}Cocartesian}} & Constructs the factor for an arbitrary exact subsequent morphism, and proves the uniqueness of all its components. \\
\addlinespace[3pt]
Construction~\ref{cons:4.7} & \href{https://github.com/iroha1203/AlgebraicArchitectureTheoryV2/blob/719f81f47d410701fd82c2bc88613cc140c59377/research/lean/ResearchLean/AG/AtomFoundation/RefinementSupply.lean\#L79}{\texttt{Refinement\allowbreak{}Extension\allowbreak{}Supply}} / \href{https://github.com/iroha1203/AlgebraicArchitectureTheoryV2/blob/719f81f47d410701fd82c2bc88613cc140c59377/research/lean/ResearchLean/AG/AtomFoundation/RefinementSupply.lean\#L289}{\texttt{refinement\allowbreak{}Lift\allowbreak{}Of\allowbreak{}Supply}} / \href{https://github.com/iroha1203/AlgebraicArchitectureTheoryV2/blob/719f81f47d410701fd82c2bc88613cc140c59377/research/lean/ResearchLean/AG/AtomFoundation/RefinementSupply.lean\#L128}{\texttt{refinement\allowbreak{}Query\allowbreak{}Map\_\allowbreak{}matches}} / \href{https://github.com/iroha1203/AlgebraicArchitectureTheoryV2/blob/719f81f47d410701fd82c2bc88613cc140c59377/research/lean/ResearchLean/AG/AtomFoundation/RefinementSupply.lean\#L138}{\texttt{refinement\allowbreak{}Query\allowbreak{}Map\_\allowbreak{}accepts}} & Constructs the forward comparison from the supply of finiteness, of the actual operations out of the extended base object, and of the equations and detectors. \\
\addlinespace[3pt]
Construction~\ref{cons:4.8} & \href{https://github.com/iroha1203/AlgebraicArchitectureTheoryV2/blob/719f81f47d410701fd82c2bc88613cc140c59377/research/lean/ResearchLean/AG/GeometryTransport/Transport.lean\#L30}{\texttt{push\allowbreak{}Coverage}} / \href{https://github.com/iroha1203/AlgebraicArchitectureTheoryV2/blob/719f81f47d410701fd82c2bc88613cc140c59377/research/lean/ResearchLean/AG/GeometryTransport/Transport.lean\#L75}{\texttt{push\allowbreak{}Overlap}} / \href{https://github.com/iroha1203/AlgebraicArchitectureTheoryV2/blob/719f81f47d410701fd82c2bc88613cc140c59377/research/lean/ResearchLean/AG/GeometryTransport/Transport.lean\#L135}{\texttt{push\allowbreak{}Geometry\allowbreak{}Package}} & The existentially quantified image of the coverage requirements, the overlap given by the equivalence of categories, and the reindexing of the raw system. \\
\addlinespace[3pt]
Definition~\ref{def:4.9} & \href{https://github.com/iroha1203/AlgebraicArchitectureTheoryV2/blob/719f81f47d410701fd82c2bc88613cc140c59377/research/lean/ResearchLean/AG/GeometryTransport/Categories.lean\#L369}{\texttt{HGeom}} & The three comparisons of support, axes, and observables, together with the readouts and naturality. \\
\addlinespace[3pt]
Proposition~\ref{prop:4.10} & \href{https://github.com/iroha1203/AlgebraicArchitectureTheoryV2/blob/719f81f47d410701fd82c2bc88613cc140c59377/research/lean/ResearchLean/AG/GeometryTransport/Supply.lean\#L190}{\texttt{h\allowbreak{}Geom\_\allowbreak{}iff\_\allowbreak{}nonempty\_\allowbreak{}geom\allowbreak{}Read\allowbreak{}Hom}} / \href{https://github.com/iroha1203/AlgebraicArchitectureTheoryV2/blob/719f81f47d410701fd82c2bc88613cc140c59377/research/lean/ResearchLean/AG/GeometryTransport/Supply.lean\#L180}{\texttt{h\allowbreak{}Geom\_\allowbreak{}necessary}} & Constructs in both directions the existence of a morphism to the specified pushed target and the existence of the three comparisons. \\
\addlinespace[3pt]
Theorem~\ref{thm:4.11} & \href{https://github.com/iroha1203/AlgebraicArchitectureTheoryV2/blob/719f81f47d410701fd82c2bc88613cc140c59377/research/lean/ResearchLean/AG/GeometryTransport/Supply.lean\#L203}{\texttt{canonical\allowbreak{}HGeom}} / \href{https://github.com/iroha1203/AlgebraicArchitectureTheoryV2/blob/719f81f47d410701fd82c2bc88613cc140c59377/research/lean/ResearchLean/AG/GeometryTransport/Factorization.lean\#L406}{\texttt{geom\allowbreak{}Transport\allowbreak{}Along\allowbreak{}Hom\_\allowbreak{}factor\_\allowbreak{}exists\allowbreak{}Unique}} / \href{https://github.com/iroha1203/AlgebraicArchitectureTheoryV2/blob/719f81f47d410701fd82c2bc88613cc140c59377/research/lean/ResearchLean/AG/GeometryTransport/Factorization.lean\#L427}{\texttt{geom\allowbreak{}Transport\allowbreak{}Along\allowbreak{}Hom\_\allowbreak{}is\allowbreak{}Strongly\allowbreak{}Cocartesian}} & Actually builds the three comparisons of the canonical core transport, and makes the geometric factor unique for an arbitrary subsequent core morphism. \\
\addlinespace[3pt]
Lemma~\ref{lem:4.12} & \href{https://github.com/iroha1203/AlgebraicArchitectureTheoryV2/blob/719f81f47d410701fd82c2bc88613cc140c59377/research/lean/ResearchLean/AG/CrossStageCoherence/Basic.lean\#L158}{\texttt{strongly\allowbreak{}Cocartesian\_\allowbreak{}comp\_\allowbreak{}projection}} & In a tower of two general functors, derives the universal property of the composite projection from the universal properties of the two levels. \\
\addlinespace[3pt]
Theorem~\ref{thm:4.16} & \href{https://github.com/iroha1203/AlgebraicArchitectureTheoryV2/blob/719f81f47d410701fd82c2bc88613cc140c59377/research/lean/ResearchLean/AG/CrossStageCoherence/TowerCompatibility.lean\#L171}{\texttt{tower\allowbreak{}Transport\allowbreak{}Comparison}} / \href{https://github.com/iroha1203/AlgebraicArchitectureTheoryV2/blob/719f81f47d410701fd82c2bc88613cc140c59377/research/lean/ResearchLean/AG/CrossStageCoherence/TowerCompatibility.lean\#L318}{\texttt{tower\allowbreak{}Transport\allowbreak{}Comparison\_\allowbreak{}compositor}} / \href{https://github.com/iroha1203/AlgebraicArchitectureTheoryV2/blob/719f81f47d410701fd82c2bc88613cc140c59377/research/lean/ResearchLean/AG/CrossStageCoherence/TowerCompatibility.lean\#L392}{\texttt{tower\allowbreak{}Transport\allowbreak{}Comparison\_\allowbreak{}unitor}} & The natural isomorphism between the fiber functors of the canonical core and of the canonical geometry, and the coherence of composition and units. \\
\addlinespace[3pt]
Definition~\ref{def:4.17} & \href{https://github.com/iroha1203/AlgebraicArchitectureTheoryV2/blob/7e68ec6e77ef0249ede6a4c875715cfaf1cb3ee6/research/lean/ResearchLean/AG/TransportCoherence/ArbitraryFinitePresentation.lean\#L112}{\texttt{Lift\allowbreak{}Data}} / \href{https://github.com/iroha1203/AlgebraicArchitectureTheoryV2/blob/7e68ec6e77ef0249ede6a4c875715cfaf1cb3ee6/research/lean/ResearchLean/AG/TransportCoherence/ArbitraryFinitePresentation.lean\#L221}{\texttt{Transport\allowbreak{}Data}} & Over an arbitrary functor, places objects, strong edge lifts, equalities of base paths, and specified endpoint fiber automorphisms on finitely many vertices, edges, and faces. \\
\addlinespace[3pt]
Construction~\ref{cons:4.18} & \href{https://github.com/iroha1203/AlgebraicArchitectureTheoryV2/blob/7e68ec6e77ef0249ede6a4c875715cfaf1cb3ee6/research/lean/ResearchLean/AG/TransportCoherence/ArbitraryFinitePresentation.lean\#L440}{\texttt{canonical\allowbreak{}Face\allowbreak{}Comparator}} / \href{https://github.com/iroha1203/AlgebraicArchitectureTheoryV2/blob/7e68ec6e77ef0249ede6a4c875715cfaf1cb3ee6/research/lean/ResearchLean/AG/TransportCoherence/ArbitraryFinitePresentation.lean\#L463}{\texttt{canonical\allowbreak{}Face\allowbreak{}Comparator\_\allowbreak{}fac}} / \href{https://github.com/iroha1203/AlgebraicArchitectureTheoryV2/blob/7e68ec6e77ef0249ede6a4c875715cfaf1cb3ee6/research/lean/ResearchLean/AG/TransportCoherence/ArbitraryFinitePresentation.lean\#L476}{\texttt{raw\allowbreak{}Face\allowbreak{}Defect}} & Constructs the unique canonical comparison of the same two paths; its inverse composed with the specified comparison gives the raw defect. \\
\addlinespace[3pt]
Definition~\ref{def:4.19} & \href{https://github.com/iroha1203/AlgebraicArchitectureTheoryV2/blob/7e68ec6e77ef0249ede6a4c875715cfaf1cb3ee6/research/lean/ResearchLean/AG/TransportCoherence/ArbitraryFinitePresentation.lean\#L238}{\texttt{Edge\allowbreak{}Reselection}} / \href{https://github.com/iroha1203/AlgebraicArchitectureTheoryV2/blob/7e68ec6e77ef0249ede6a4c875715cfaf1cb3ee6/research/lean/ResearchLean/AG/TransportCoherence/ArbitraryFinitePresentation.lean\#L247}{\texttt{reselected\allowbreak{}Edge\allowbreak{}Lift}} / \href{https://github.com/iroha1203/AlgebraicArchitectureTheoryV2/blob/7e68ec6e77ef0249ede6a4c875715cfaf1cb3ee6/research/lean/ResearchLean/AG/TransportCoherence/ArbitraryFinitePresentation.lean\#L273}{\texttt{reselected\allowbreak{}Edge\allowbreak{}Lift\_\allowbreak{}is\allowbreak{}Strongly\allowbreak{}Cocartesian}} & Postcomposes each edge with an endpoint fiber automorphism, preserving its base morphism and strong opcartesianness. \\
\addlinespace[3pt]
Lemma~\ref{lem:4.24} and Construction~\ref{cons:4.25} & \href{https://github.com/iroha1203/AlgebraicArchitectureTheoryV2/blob/7e68ec6e77ef0249ede6a4c875715cfaf1cb3ee6/research/lean/ResearchLean/AG/TransportCoherence/ArbitraryObstruction.lean\#L894}{\texttt{whisker\allowbreak{}Fiber\allowbreak{}Aut\allowbreak{}Hom}} / \href{https://github.com/iroha1203/AlgebraicArchitectureTheoryV2/blob/7e68ec6e77ef0249ede6a4c875715cfaf1cb3ee6/research/lean/ResearchLean/AG/TransportCoherence/ArbitraryObstruction.lean\#L804}{\texttt{whisker\allowbreak{}Fiber\allowbreak{}Aut\_\allowbreak{}fac}} / \href{https://github.com/iroha1203/AlgebraicArchitectureTheoryV2/blob/7e68ec6e77ef0249ede6a4c875715cfaf1cb3ee6/research/lean/ResearchLean/AG/TransportCoherence/ArbitraryObstruction.lean\#L1153}{\texttt{canonical\allowbreak{}Pasting\allowbreak{}Comparator\_\allowbreak{}fac}} / \href{https://github.com/iroha1203/AlgebraicArchitectureTheoryV2/blob/7e68ec6e77ef0249ede6a4c875715cfaf1cb3ee6/research/lean/ResearchLean/AG/TransportCoherence/ArbitraryObstruction.lean\#L1194}{\texttt{canonical\allowbreak{}Pasting\allowbreak{}Comparator\_\allowbreak{}unique}} / \href{https://github.com/iroha1203/AlgebraicArchitectureTheoryV2/blob/7e68ec6e77ef0249ede6a4c875715cfaf1cb3ee6/research/lean/ResearchLean/AG/TransportCoherence/ArbitraryObstruction.lean\#L1255}{\texttt{pasting\allowbreak{}Raw\allowbreak{}Defect\_\allowbreak{}two\_\allowbreak{}steps}} & Transport along a subsequent path is a group homomorphism. Constructs pastings of oriented faces; the two-step defect includes conjugation by the later specified comparison. \\
\addlinespace[3pt]
Definition~\ref{def:4.30} & \href{https://github.com/iroha1203/AlgebraicArchitectureTheoryV2/blob/719f81f47d410701fd82c2bc88613cc140c59377/research/lean/ResearchLean/AG/CrossStageCoherence/ObstructionGroups.lean\#L150}{\texttt{composite\allowbreak{}Fiber\allowbreak{}Pushforward}} / \href{https://github.com/iroha1203/AlgebraicArchitectureTheoryV2/blob/719f81f47d410701fd82c2bc88613cc140c59377/research/lean/ResearchLean/AG/CrossStageCoherence/ObstructionGroups.lean\#L188}{\texttt{inner\allowbreak{}Fiber\allowbreak{}Aut\allowbreak{}Subgroup\_\allowbreak{}eq\_\allowbreak{}ker}} & A group homomorphism into the core, defined on the geometric automorphisms that fix the extraction; its kernel is the group that fixes the core. \\
\addlinespace[3pt]
Definition~\ref{def:4.36} and Proposition~\ref{prop:4.37} & \href{https://github.com/iroha1203/AlgebraicArchitectureTheoryV2/blob/7e68ec6e77ef0249ede6a4c875715cfaf1cb3ee6/research/lean/ResearchLean/AG/RealizationReconstruction/GeneralRelativeLens.lean\#L18}{\texttt{General\allowbreak{}Lens}} / \href{https://github.com/iroha1203/AlgebraicArchitectureTheoryV2/blob/7e68ec6e77ef0249ede6a4c875715cfaf1cb3ee6/research/lean/ResearchLean/AG/RealizationReconstruction/GeneralRelativeLens.lean\#L31}{\texttt{Hom}} / \href{https://github.com/iroha1203/AlgebraicArchitectureTheoryV2/blob/7e68ec6e77ef0249ede6a4c875715cfaf1cb3ee6/research/lean/ResearchLean/AG/RealizationReconstruction/GeneralRelativeLens.lean\#L38}{\texttt{get\_\allowbreak{}comm\_\allowbreak{}of\_\allowbreak{}put\_\allowbreak{}comm}} / \href{https://github.com/iroha1203/AlgebraicArchitectureTheoryV2/blob/7e68ec6e77ef0249ede6a4c875715cfaf1cb3ee6/research/lean/ResearchLean/AG/RealizationReconstruction/GeneralRelativeLens.lean\#L48}{\texttt{operation}} / \href{https://github.com/iroha1203/AlgebraicArchitectureTheoryV2/blob/7e68ec6e77ef0249ede6a4c875715cfaf1cb3ee6/research/lean/ResearchLean/AG/RealizationReconstruction/GeneralRelativeLens.lean\#L54}{\texttt{operation\_\allowbreak{}square\_\allowbreak{}iff}} & Two lenses with arbitrary state and view types. Derives preservation of reads from preservation of updates and identifies the combined sum-type operation square with the two preservation equations. \\
\addlinespace[3pt]
Proposition~\ref{prop:4.38} & \href{https://github.com/iroha1203/AlgebraicArchitectureTheoryV2/blob/7e68ec6e77ef0249ede6a4c875715cfaf1cb3ee6/research/lean/ResearchLean/AG/RealizationReconstruction/GeneralRelativeLens.lean\#L207}{\texttt{aut\allowbreak{}Mul\allowbreak{}Equiv\allowbreak{}Put\allowbreak{}Group}} / \href{https://github.com/iroha1203/AlgebraicArchitectureTheoryV2/blob/7e68ec6e77ef0249ede6a4c875715cfaf1cb3ee6/research/lean/ResearchLean/AG/RealizationReconstruction/GeneralRelativeLens.lean\#L172}{\texttt{put\allowbreak{}Group\_\allowbreak{}le\_\allowbreak{}get\allowbreak{}Group}} / \href{https://github.com/iroha1203/AlgebraicArchitectureTheoryV2/blob/7e68ec6e77ef0249ede6a4c875715cfaf1cb3ee6/research/lean/ResearchLean/AG/RealizationReconstruction/GeneralRelativeLens.lean\#L177}{\texttt{get\allowbreak{}Group\_\allowbreak{}inf\_\allowbreak{}put\allowbreak{}Group}} / \href{https://github.com/iroha1203/AlgebraicArchitectureTheoryV2/blob/7e68ec6e77ef0249ede6a4c875715cfaf1cb3ee6/research/lean/ResearchLean/AG/RealizationReconstruction/GeneralRelativeLens.lean\#L450}{\texttt{bool\allowbreak{}Fiber\allowbreak{}Twist\_\allowbreak{}get\_\allowbreak{}not\_\allowbreak{}put}} & The group of invertible relative lens morphisms is the put-preserving subgroup, equal to its intersection with the get-preserving group. The Bool example makes the inclusion strict. \\
\addlinespace[3pt]
Example~\ref{ex:4.39} & \href{https://github.com/iroha1203/AlgebraicArchitectureTheoryV2/blob/7e68ec6e77ef0249ede6a4c875715cfaf1cb3ee6/research/lean/ResearchLean/AG/RealizationReconstruction/GeneralRelativeLens.lean\#L464}{\texttt{product\allowbreak{}Map}} / \href{https://github.com/iroha1203/AlgebraicArchitectureTheoryV2/blob/7e68ec6e77ef0249ede6a4c875715cfaf1cb3ee6/research/lean/ResearchLean/AG/RealizationReconstruction/GeneralRelativeLens.lean\#L471}{\texttt{product\allowbreak{}Map\_\allowbreak{}state}} / \href{https://github.com/iroha1203/AlgebraicArchitectureTheoryV2/blob/7e68ec6e77ef0249ede6a4c875715cfaf1cb3ee6/research/lean/ResearchLean/AG/RealizationReconstruction/GeneralRelativeLens.lean\#L477}{\texttt{bool\allowbreak{}To\allowbreak{}Point\_\allowbreak{}not\_\allowbreak{}injective}} & An arbitrary map of hidden components gives a product-lens morphism. For a singleton view, the map from Bool to a singleton gives a noninjective state map. \\
\addlinespace[3pt]
Proposition~\ref{prop:4.40} & \href{https://github.com/iroha1203/AlgebraicArchitectureTheoryV2/blob/719f81f47d410701fd82c2bc88613cc140c59377/research/lean/ResearchLean/AG/RealizationReconstruction/ProtocolReconstruction.lean\#L60}{\texttt{generator\_\allowbreak{}path\_\allowbreak{}naturality}} / \href{https://github.com/iroha1203/AlgebraicArchitectureTheoryV2/blob/719f81f47d410701fd82c2bc88613cc140c59377/research/lean/ResearchLean/AG/RealizationReconstruction/ProtocolReconstruction.lean\#L118}{\texttt{hom\allowbreak{}Equiv\allowbreak{}Generator\allowbreak{}Map}} / \href{https://github.com/iroha1203/AlgebraicArchitectureTheoryV2/blob/719f81f47d410701fd82c2bc88613cc140c59377/research/lean/ResearchLean/AG/RealizationReconstruction/CSAATProtocolAdapterSquares.lean\#L134}{\texttt{of\allowbreak{}Semantic\allowbreak{}Hom\_\allowbreak{}directed\allowbreak{}Adapter\allowbreak{}Square\_\allowbreak{}iff}} & Inducts from the commutativity of the named generating edges to all paths, and reads the adapter square with the same vertex components. \\
\addlinespace[3pt]
Theorem~\ref{thm:4.41} & \href{https://github.com/iroha1203/AlgebraicArchitectureTheoryV2/blob/719f81f47d410701fd82c2bc88613cc140c59377/research/lean/ResearchLean/AG/RealizationReconstruction/CSAATFullyFaithfulComparisonTransport.lean\#L93}{\texttt{generated\allowbreak{}Arrow\allowbreak{}Comparison\allowbreak{}Mul\allowbreak{}Equiv\allowbreak{}Of\allowbreak{}Fully\allowbreak{}Faithful}} / \href{https://github.com/iroha1203/AlgebraicArchitectureTheoryV2/blob/719f81f47d410701fd82c2bc88613cc140c59377/research/lean/ResearchLean/AG/RealizationReconstruction/CSAATFullyFaithfulComparisonTransport.lean\#L116}{\texttt{generated\allowbreak{}Arrow\allowbreak{}Comparison\_\allowbreak{}section\_\allowbreak{}compatibility}} & A group isomorphism for an arbitrary category, fully faithful functor, and comparison morphism. If the comparison morphism is an isomorphism, it also commutes with the canonical section. \\
\addlinespace[3pt]
Theorem~\ref{thm:4.41} (relative lenses) & \href{https://github.com/iroha1203/AlgebraicArchitectureTheoryV2/blob/7e68ec6e77ef0249ede6a4c875715cfaf1cb3ee6/research/lean/ResearchLean/AG/RealizationReconstruction/GeneralRelativeLensAAT.lean\#L383}{\texttt{aat\allowbreak{}Constructed\allowbreak{}Functor}} / \href{https://github.com/iroha1203/AlgebraicArchitectureTheoryV2/blob/7e68ec6e77ef0249ede6a4c875715cfaf1cb3ee6/research/lean/ResearchLean/AG/RealizationReconstruction/GeneralRelativeLensComparison.lean\#L20}{\texttt{aat\allowbreak{}Comparison\allowbreak{}Mul\allowbreak{}Equiv}} / \href{https://github.com/iroha1203/AlgebraicArchitectureTheoryV2/blob/7e68ec6e77ef0249ede6a4c875715cfaf1cb3ee6/research/lean/ResearchLean/AG/RealizationReconstruction/GeneralRelativeLensComparison.lean\#L28}{\texttt{aat\allowbreak{}Comparison\_\allowbreak{}source\_\allowbreak{}compatibility}} & The typed AAT construction for lenses allowing different state and view types gives a fully faithful functor and matches comparison groups and the source projection. \\
\bottomrule
\end{xltabular}

\subsection*{Canonical transport and coherence over the core}

\begin{xltabular}{\linewidth}{@{}LLL@{}}
\toprule
Text & Lean declaration & Objects and conditions \\
\midrule
\endhead
Proposition~\ref{prop:4.4} (canonical core transport) & \href{https://github.com/iroha1203/AlgebraicArchitectureTheoryV2/blob/719f81f47d410701fd82c2bc88613cc140c59377/research/lean/ResearchLean/AG/AtomFoundation/Transport.lean\#L2098}{\texttt{transport\allowbreak{}Along\allowbreak{}Hom}} / \href{https://github.com/iroha1203/AlgebraicArchitectureTheoryV2/blob/719f81f47d410701fd82c2bc88613cc140c59377/research/lean/ResearchLean/AG/AtomFoundation/Transport.lean\#L2128}{\texttt{transport\allowbreak{}Along\_\allowbreak{}family\_\allowbreak{}eq}} / \href{https://github.com/iroha1203/AlgebraicArchitectureTheoryV2/blob/719f81f47d410701fd82c2bc88613cc140c59377/research/lean/ResearchLean/AG/AtomFoundation/Transport.lean\#L2143}{\texttt{transport\allowbreak{}Along\_\allowbreak{}object\_\allowbreak{}eq}} & The construction of the canonical core and of the morphism, and the equalities for the extracted family and the base object. \\
\addlinespace[3pt]
Construction~\ref{cons:4.13}--Theorem~\ref{thm:4.15} & \href{https://github.com/iroha1203/AlgebraicArchitectureTheoryV2/blob/7e68ec6e77ef0249ede6a4c875715cfaf1cb3ee6/research/lean/ResearchLean/AG/CrossStageCoherence/ArbitraryStrongLiftPseudofunctor.lean\#L38}{\texttt{Strong\allowbreak{}Lift\allowbreak{}Selection}} / \href{https://github.com/iroha1203/AlgebraicArchitectureTheoryV2/blob/7e68ec6e77ef0249ede6a4c875715cfaf1cb3ee6/research/lean/ResearchLean/AG/CrossStageCoherence/ArbitraryStrongLiftPseudofunctor.lean\#L120}{\texttt{transport}} / \href{https://github.com/iroha1203/AlgebraicArchitectureTheoryV2/blob/7e68ec6e77ef0249ede6a4c875715cfaf1cb3ee6/research/lean/ResearchLean/AG/CrossStageCoherence/ArbitraryStrongLiftPseudofunctor.lean\#L244}{\texttt{compositor}} / \href{https://github.com/iroha1203/AlgebraicArchitectureTheoryV2/blob/7e68ec6e77ef0249ede6a4c875715cfaf1cb3ee6/research/lean/ResearchLean/AG/CrossStageCoherence/ArbitraryStrongLiftPseudofunctor.lean\#L312}{\texttt{unitor}} / \href{https://github.com/iroha1203/AlgebraicArchitectureTheoryV2/blob/7e68ec6e77ef0249ede6a4c875715cfaf1cb3ee6/research/lean/ResearchLean/AG/CrossStageCoherence/ArbitraryStrongLiftPseudofunctor.lean\#L490}{\texttt{compositor\_\allowbreak{}assoc}} / \href{https://github.com/iroha1203/AlgebraicArchitectureTheoryV2/blob/7e68ec6e77ef0249ede6a4c875715cfaf1cb3ee6/research/lean/ResearchLean/AG/CrossStageCoherence/ArbitraryStrongLiftPseudofunctor.lean\#L633}{\texttt{compositor\_\allowbreak{}left\_\allowbreak{}unit}} / \href{https://github.com/iroha1203/AlgebraicArchitectureTheoryV2/blob/7e68ec6e77ef0249ede6a4c875715cfaf1cb3ee6/research/lean/ResearchLean/AG/CrossStageCoherence/ArbitraryStrongLiftPseudofunctor.lean\#L558}{\texttt{compositor\_\allowbreak{}right\_\allowbreak{}unit}} & For an arbitrary functor and selected strongly opcartesian lifts, constructs fiber transport and natural unit and composition isomorphisms, and derives their associativity and unit laws. \\
\addlinespace[3pt]
Lemma~\ref{lem:4.21} & \href{https://github.com/iroha1203/AlgebraicArchitectureTheoryV2/blob/7e68ec6e77ef0249ede6a4c875715cfaf1cb3ee6/research/lean/ResearchLean/AG/TransportCoherence/ArbitraryObstruction.lean\#L113}{\texttt{path\allowbreak{}Reselection\allowbreak{}Transition\_\allowbreak{}fac}} / \href{https://github.com/iroha1203/AlgebraicArchitectureTheoryV2/blob/7e68ec6e77ef0249ede6a4c875715cfaf1cb3ee6/research/lean/ResearchLean/AG/TransportCoherence/ArbitraryObstruction.lean\#L150}{\texttt{path\allowbreak{}Reselection\allowbreak{}Transition\_\allowbreak{}mul}} / \href{https://github.com/iroha1203/AlgebraicArchitectureTheoryV2/blob/7e68ec6e77ef0249ede6a4c875715cfaf1cb3ee6/research/lean/ResearchLean/AG/TransportCoherence/ArbitraryObstruction.lean\#L265}{\texttt{canonical\allowbreak{}Face\allowbreak{}Comparator\_\allowbreak{}transition}} / \href{https://github.com/iroha1203/AlgebraicArchitectureTheoryV2/blob/7e68ec6e77ef0249ede6a4c875715cfaf1cb3ee6/research/lean/ResearchLean/AG/TransportCoherence/ArbitraryObstruction.lean\#L284}{\texttt{raw\allowbreak{}Face\allowbreak{}Defect\_\allowbreak{}transition}} & Over an arbitrary functor, derives path transitions depending on the current selection, the noncommutative product order, and transformation laws for comparisons and defects. \\
\addlinespace[3pt]
Proposition~\ref{prop:4.22} & \href{https://github.com/iroha1203/AlgebraicArchitectureTheoryV2/blob/7e68ec6e77ef0249ede6a4c875715cfaf1cb3ee6/research/lean/ResearchLean/AG/TransportCoherence/ArbitraryObstruction.lean\#L420}{\texttt{reselection\allowbreak{}Step\_\allowbreak{}one}} / \href{https://github.com/iroha1203/AlgebraicArchitectureTheoryV2/blob/7e68ec6e77ef0249ede6a4c875715cfaf1cb3ee6/research/lean/ResearchLean/AG/TransportCoherence/ArbitraryObstruction.lean\#L431}{\texttt{reselection\allowbreak{}Step\_\allowbreak{}mul}} / \href{https://github.com/iroha1203/AlgebraicArchitectureTheoryV2/blob/7e68ec6e77ef0249ede6a4c875715cfaf1cb3ee6/research/lean/ResearchLean/AG/TransportCoherence/ArbitraryObstruction.lean\#L471}{\texttt{initial\allowbreak{}Reselection\allowbreak{}Orbit}} / \href{https://github.com/iroha1203/AlgebraicArchitectureTheoryV2/blob/7e68ec6e77ef0249ede6a4c875715cfaf1cb3ee6/research/lean/ResearchLean/AG/TransportCoherence/ArbitraryObstruction.lean\#L608}{\texttt{two\allowbreak{}Edge\allowbreak{}Endpoint\allowbreak{}Transition\_\allowbreak{}formula}} / \href{https://github.com/iroha1203/AlgebraicArchitectureTheoryV2/blob/7e68ec6e77ef0249ede6a4c875715cfaf1cb3ee6/research/lean/ResearchLean/AG/TransportCoherence/ArbitraryObstruction.lean\#L615}{\texttt{fin\allowbreak{}Three\_\allowbreak{}two\allowbreak{}Edge\allowbreak{}Transition\_\allowbreak{}depends\_\allowbreak{}on\_\allowbreak{}current}} & Constructs the left action on edge selections and cochains and its initial orbit. The two-edge formula and a three-point permutation example show dependence on the current selection. \\
\addlinespace[3pt]
Theorem~\ref{thm:4.23} & \href{https://github.com/iroha1203/AlgebraicArchitectureTheoryV2/blob/7e68ec6e77ef0249ede6a4c875715cfaf1cb3ee6/research/lean/ResearchLean/AG/TransportCoherence/ArbitraryObstruction.lean\#L715}{\texttt{transport\allowbreak{}Obstruction\allowbreak{}Vanishes\_\allowbreak{}iff\_\allowbreak{}coherentizable}} & For the same comparison diagram over an arbitrary functor, obstruction vanishing is equivalent to an edge reselection making all faces coherent simultaneously. \\
\addlinespace[3pt]
Propositions~\ref{prop:4.26}--\ref{prop:4.27} & \href{https://github.com/iroha1203/AlgebraicArchitectureTheoryV2/blob/7e68ec6e77ef0249ede6a4c875715cfaf1cb3ee6/research/lean/ResearchLean/AG/TransportCoherence/ArbitraryObstruction.lean\#L1453}{\texttt{raw\allowbreak{}Defect\_\allowbreak{}cocycle\_\allowbreak{}of\_\allowbreak{}syzygy}} / \href{https://github.com/iroha1203/AlgebraicArchitectureTheoryV2/blob/7e68ec6e77ef0249ede6a4c875715cfaf1cb3ee6/research/lean/ResearchLean/AG/TransportCoherence/ArbitraryObstruction.lean\#L1492}{\texttt{closed\allowbreak{}Pasting\allowbreak{}Raw\allowbreak{}Obstruction\_\allowbreak{}eq\_\allowbreak{}conjugate}} / \href{https://github.com/iroha1203/AlgebraicArchitectureTheoryV2/blob/7e68ec6e77ef0249ede6a4c875715cfaf1cb3ee6/research/lean/ResearchLean/AG/TransportCoherence/ArbitraryObstruction.lean\#L1508}{\texttt{closed\allowbreak{}Pasting\allowbreak{}Raw\allowbreak{}Obstruction\_\allowbreak{}eq\_\allowbreak{}one\_\allowbreak{}iff}} & Over an arbitrary functor, derives the defect equation along a specified syzygy, the conjugation formula for a closed discrepancy, and the equivalence of its being the identity. \\
\addlinespace[3pt]
Proposition~\ref{prop:4.31} & \href{https://github.com/iroha1203/AlgebraicArchitectureTheoryV2/blob/7e68ec6e77ef0249ede6a4c875715cfaf1cb3ee6/research/lean/ResearchLean/AG/CrossStageCoherence/CompositeQualification.lean\#L36}{\texttt{strongly\allowbreak{}Cocartesian\_\allowbreak{}of\_\allowbreak{}comp\_\allowbreak{}projection}} / \href{https://github.com/iroha1203/AlgebraicArchitectureTheoryV2/blob/7e68ec6e77ef0249ede6a4c875715cfaf1cb3ee6/research/lean/ResearchLean/AG/CrossStageCoherence/CompositeQualification.lean\#L196}{\texttt{Two\allowbreak{}Layer\allowbreak{}Transport\allowbreak{}Data.\allowbreak{}of\allowbreak{}Composite\allowbreak{}Strong}} / \href{https://github.com/iroha1203/AlgebraicArchitectureTheoryV2/blob/7e68ec6e77ef0249ede6a4c875715cfaf1cb3ee6/research/lean/ResearchLean/AG/CrossStageCoherence/UpperObstruction.lean\#L437}{\texttt{pushforward\_\allowbreak{}upper\allowbreak{}Canonical\allowbreak{}Two\allowbreak{}Cell\allowbreak{}Comparator}} / \href{https://github.com/iroha1203/AlgebraicArchitectureTheoryV2/blob/7e68ec6e77ef0249ede6a4c875715cfaf1cb3ee6/research/lean/ResearchLean/AG/CrossStageCoherence/UpperObstruction.lean\#L491}{\texttt{pushforward\_\allowbreak{}upper\allowbreak{}Raw\allowbreak{}Two\allowbreak{}Cell\allowbreak{}Defect}} & Derives strongness at the middle level from strongness for the composite projection and for the projected edge, then projects canonical comparisons and raw defects. \\
\addlinespace[3pt]
Definition~\ref{def:4.32} & \href{https://github.com/iroha1203/AlgebraicArchitectureTheoryV2/blob/719f81f47d410701fd82c2bc88613cc140c59377/research/lean/ResearchLean/AG/CrossStageCoherence/SectionDecomposition.lean\#L38}{\texttt{Edge\allowbreak{}Section\allowbreak{}Family}} / \href{https://github.com/iroha1203/AlgebraicArchitectureTheoryV2/blob/719f81f47d410701fd82c2bc88613cc140c59377/research/lean/ResearchLean/AG/CrossStageCoherence/SectionDecomposition.lean\#L87}{\texttt{Core\allowbreak{}Alignment\allowbreak{}At}} & In the same two-level diagram, pairs a selection of core edges with its lift to the geometry, and requires that the projection agree with the original selection on each edge. Coherence of the core is checked on all faces. \\
\addlinespace[3pt]
Theorem~\ref{thm:4.33} & \href{https://github.com/iroha1203/AlgebraicArchitectureTheoryV2/blob/719f81f47d410701fd82c2bc88613cc140c59377/research/lean/ResearchLean/AG/CrossStageCoherence/SectionDecomposition.lean\#L100}{\texttt{section\allowbreak{}Cell\allowbreak{}Comparator\_\allowbreak{}pushforward\_\allowbreak{}eq\_\allowbreak{}authored}} / \href{https://github.com/iroha1203/AlgebraicArchitectureTheoryV2/blob/719f81f47d410701fd82c2bc88613cc140c59377/research/lean/ResearchLean/AG/CrossStageCoherence/SectionDecomposition.lean\#L198}{\texttt{section\allowbreak{}Inner\allowbreak{}Obstruction}} / \href{https://github.com/iroha1203/AlgebraicArchitectureTheoryV2/blob/719f81f47d410701fd82c2bc88613cc140c59377/research/lean/ResearchLean/AG/CrossStageCoherence/SectionDecomposition.lean\#L221}{\texttt{section\allowbreak{}Lift\allowbreak{}Term\_\allowbreak{}projection}} / \href{https://github.com/iroha1203/AlgebraicArchitectureTheoryV2/blob/719f81f47d410701fd82c2bc88613cc140c59377/research/lean/ResearchLean/AG/CrossStageCoherence/SectionDecomposition.lean\#L234}{\texttt{total\allowbreak{}Obstruction\_\allowbreak{}kernel\_\allowbreak{}decomposition}} & For the same aligned edge section, decomposes the mismatch of the geometry into the product of a factor in the kernel and a lifted outer factor. The inner obstruction takes values in the kernel of the projection, and the projection of the outer factor agrees with the raw defect of the core. \\
\addlinespace[3pt]
Theorem~\ref{thm:4.34} & \href{https://github.com/iroha1203/AlgebraicArchitectureTheoryV2/blob/719f81f47d410701fd82c2bc88613cc140c59377/research/lean/ResearchLean/AG/CrossStageCoherence/RelativeObstruction.lean\#L48}{\texttt{relative\allowbreak{}Upper\allowbreak{}Reselection\_\allowbreak{}projects}} / \href{https://github.com/iroha1203/AlgebraicArchitectureTheoryV2/blob/719f81f47d410701fd82c2bc88613cc140c59377/research/lean/ResearchLean/AG/CrossStageCoherence/RelativeObstruction.lean\#L220}{\texttt{inner\allowbreak{}Vanishes\allowbreak{}At\_\allowbreak{}iff\_\allowbreak{}section\allowbreak{}Relative\allowbreak{}Coherentizable}} / \href{https://github.com/iroha1203/AlgebraicArchitectureTheoryV2/blob/719f81f47d410701fd82c2bc88613cc140c59377/research/lean/ResearchLean/AG/CrossStageCoherence/RelativeObstruction.lean\#L240}{\texttt{section\allowbreak{}Replacement\allowbreak{}Gauge}} / \href{https://github.com/iroha1203/AlgebraicArchitectureTheoryV2/blob/719f81f47d410701fd82c2bc88613cc140c59377/research/lean/ResearchLean/AG/CrossStageCoherence/RelativeObstruction.lean\#L255}{\texttt{section\allowbreak{}Replacement\allowbreak{}Gauge\_\allowbreak{}mul\_\allowbreak{}first}} / \href{https://github.com/iroha1203/AlgebraicArchitectureTheoryV2/blob/719f81f47d410701fd82c2bc88613cc140c59377/research/lean/ResearchLean/AG/CrossStageCoherence/GlobalVanishing.lean\#L266}{\texttt{joint\allowbreak{}Vanishes\_\allowbreak{}iff\_\allowbreak{}aligned\allowbreak{}Section\allowbreak{}Vanishes}} & Under a fixed edge section, the vanishing of the inner obstruction by a reselection in the kernel is equivalent to the coherence of the geometry. Constructs the difference in the kernel that changes the section, and obtains the two-level equivalence of simultaneous vanishing quantified over the existence of a section. \\
\bottomrule
\end{xltabular}

\section{Chapter 5: Base Change and Generated Comparisons}\label{sec:A.5}

\begin{xltabular}{\linewidth}{@{}LLL@{}}
\toprule
Text & Lean declaration & Objects and conditions \\
\midrule
\endhead
Construction~\ref{cons:5.1} & \href{https://github.com/iroha1203/AlgebraicArchitectureTheoryV2/blob/719f81f47d410701fd82c2bc88613cc140c59377/research/lean/ResearchLean/AG/DoctrineFiberProduct/DoctrinePullback.lean\#L40}{\texttt{doctrine\allowbreak{}Pullback}} / \href{https://github.com/iroha1203/AlgebraicArchitectureTheoryV2/blob/719f81f47d410701fd82c2bc88613cc140c59377/research/lean/ResearchLean/AG/DoctrineFiberProduct/DoctrinePullback.lean\#L248}{\texttt{doctrine\allowbreak{}Pullback\_\allowbreak{}is\allowbreak{}Pullback}} & From an arbitrary exact doctrine cospan, constructs the pullback with the compatible sources, the vocabulary of the first leg, componentwise normalization, and the two projections. Neither finiteness nor a presentation by codes is assumed. \\
\addlinespace[3pt]
Proposition~\ref{prop:5.2} & \href{https://github.com/iroha1203/AlgebraicArchitectureTheoryV2/blob/719f81f47d410701fd82c2bc88613cc140c59377/research/lean/ResearchLean/AG/DoctrineFiberProduct/PointedDoctrinePullback.lean\#L35}{\texttt{pointed\allowbreak{}Pullback}} / \href{https://github.com/iroha1203/AlgebraicArchitectureTheoryV2/blob/719f81f47d410701fd82c2bc88613cc140c59377/research/lean/ResearchLean/AG/DoctrineFiberProduct/PointedDoctrinePullback.lean\#L159}{\texttt{pointed\allowbreak{}Pullback\_\allowbreak{}is\allowbreak{}Pullback}} & For a pointed cospan that assumes compatibility of the designated points, proves the existence and uniqueness of the pointed lift. \\
\addlinespace[3pt]
Definition~\ref{def:5.4} & \href{https://github.com/leanprover-community/mathlib4/blob/8f9d9cff6bd728b17a24e163c9402775d9e6a365/Mathlib/CategoryTheory/FiberedCategory/Cartesian.lean\#L76}{\texttt{Is\allowbreak{}Strongly\allowbreak{}Cartesian}} & The definition by lifts of morphisms along p and by the unique g-lift of an arbitrary morphism covering $g\gg f$. \\
\addlinespace[3pt]
Construction~\ref{cons:5.5} & \href{https://github.com/iroha1203/AlgebraicArchitectureTheoryV2/blob/719f81f47d410701fd82c2bc88613cc140c59377/research/lean/ResearchLean/AG/DoctrineFiberProduct/CartesianTarget.lean\#L289}{\texttt{inverse\allowbreak{}Core\allowbreak{}Package}} / \href{https://github.com/iroha1203/AlgebraicArchitectureTheoryV2/blob/719f81f47d410701fd82c2bc88613cc140c59377/research/lean/ResearchLean/AG/DoctrineFiberProduct/CartesianTarget.lean\#L1452}{\texttt{inverse\allowbreak{}Core\allowbreak{}Package\allowbreak{}Hom}} & From an arbitrary semantic exact hom and a target CoreFiber, transports the extracted family backwards and constructs a core with operations, equations, invariants, and a signature, together with the covering morphism. \\
\addlinespace[3pt]
Theorem~\ref{thm:5.6} & \href{https://github.com/iroha1203/AlgebraicArchitectureTheoryV2/blob/719f81f47d410701fd82c2bc88613cc140c59377/research/lean/ResearchLean/AG/DoctrineFiberProduct/CartesianTarget.lean\#L1550}{\texttt{inverse\allowbreak{}Core\allowbreak{}Package\allowbreak{}Hom\_\allowbreak{}is\allowbreak{}Strongly\allowbreak{}Cartesian}} / \href{https://github.com/iroha1203/AlgebraicArchitectureTheoryV2/blob/719f81f47d410701fd82c2bc88613cc140c59377/research/lean/ResearchLean/AG/DoctrineFiberProduct/ExactBottomGlobalLift.lean\#L28}{\texttt{exact\_\allowbreak{}bottom\_\allowbreak{}semantic\_\allowbreak{}global\_\allowbreak{}strong\_\allowbreak{}cartesian\_\allowbreak{}lift}} & A strong cartesian lift for CartSemanticInput U and an arbitrary target package. Uses the exact equivalence of extractions on all sources, and does not assume a coverage by finite codes. \\
\addlinespace[3pt]
Proposition~\ref{prop:5.8} & \href{https://github.com/iroha1203/AlgebraicArchitectureTheoryV2/blob/719f81f47d410701fd82c2bc88613cc140c59377/research/lean/ResearchLean/AG/DiagnosticConservativity/TransportEquivalence.lean\#L24}{\texttt{semantic\allowbreak{}Global\allowbreak{}Transport\allowbreak{}Equivalence}} & For an arbitrary pointed exact semantic hom, the forward core transport and the canonical reindexing give an equivalence of categories. Neither injectivity nor surjectivity of the source map is assumed. \\
\addlinespace[3pt]
Construction~\ref{cons:5.10} & \href{https://github.com/iroha1203/AlgebraicArchitectureTheoryV2/blob/7e68ec6e77ef0249ede6a4c875715cfaf1cb3ee6/research/lean/ResearchLean/AG/DoctrineFiberProduct/SemanticCoreBeckChevalleyMate.lean\#L38}{\texttt{semantic\allowbreak{}Pointed\allowbreak{}Pullback\allowbreak{}Square\_\allowbreak{}is\allowbreak{}Pullback}} / \href{https://github.com/iroha1203/AlgebraicArchitectureTheoryV2/blob/7e68ec6e77ef0249ede6a4c875715cfaf1cb3ee6/research/lean/ResearchLean/AG/DoctrineFiberProduct/SemanticCoreBeckChevalleyMate.lean\#L48}{\texttt{semantic\allowbreak{}Core\allowbreak{}Transport\allowbreak{}Square\allowbreak{}Iso}} / \href{https://github.com/iroha1203/AlgebraicArchitectureTheoryV2/blob/7e68ec6e77ef0249ede6a4c875715cfaf1cb3ee6/research/lean/ResearchLean/AG/DoctrineFiberProduct/SemanticCoreBeckChevalleyMate.lean\#L58}{\texttt{semantic\allowbreak{}Core\allowbreak{}Beck\allowbreak{}Chevalley\allowbreak{}Mate}} & Builds the pointed square, two routes, and transport comparison from an arbitrary semantic exact cospan and compatible points. Requires neither finite codes nor decidable Atom equality. \\
\addlinespace[3pt]
Theorem~\ref{thm:5.11} & \href{https://github.com/iroha1203/AlgebraicArchitectureTheoryV2/blob/7e68ec6e77ef0249ede6a4c875715cfaf1cb3ee6/research/lean/ResearchLean/AG/DoctrineFiberProduct/SemanticCoreBeckChevalleyMate.lean\#L73}{\texttt{semantic\allowbreak{}Core\allowbreak{}Beck\allowbreak{}Chevalley\allowbreak{}Mate\_\allowbreak{}app}} / \href{https://github.com/iroha1203/AlgebraicArchitectureTheoryV2/blob/7e68ec6e77ef0249ede6a4c875715cfaf1cb3ee6/research/lean/ResearchLean/AG/DoctrineFiberProduct/SemanticCoreBeckChevalleyMate.lean\#L94}{\texttt{semantic\allowbreak{}Core\allowbreak{}Beck\allowbreak{}Chevalley\allowbreak{}Mate\_\allowbreak{}is\allowbreak{}Iso}} / \href{https://github.com/iroha1203/AlgebraicArchitectureTheoryV2/blob/7e68ec6e77ef0249ede6a4c875715cfaf1cb3ee6/research/lean/ResearchLean/AG/DoctrineFiberProduct/SemanticCoreBeckChevalleyFactorization.lean\#L36}{\texttt{semantic\allowbreak{}Core\allowbreak{}Beck\allowbreak{}Chevalley\allowbreak{}Mate\_\allowbreak{}lift\_\allowbreak{}fac}} & For the same general semantic square, constructs the mate from the unit, transport comparison, and counit, and proves invertibility and the triangle for the selected lifts. \\
\addlinespace[3pt]
Proposition~\ref{prop:5.12} & \href{https://github.com/iroha1203/AlgebraicArchitectureTheoryV2/blob/7e68ec6e77ef0249ede6a4c875715cfaf1cb3ee6/research/lean/ResearchLean/AG/DoctrineFiberProduct/SemanticCoreBeckChevalleyFactorization.lean\#L112}{\texttt{semantic\allowbreak{}Core\allowbreak{}Beck\allowbreak{}Chevalley\allowbreak{}Mate\_\allowbreak{}lift\_\allowbreak{}fac\_\allowbreak{}iff}} / \href{https://github.com/iroha1203/AlgebraicArchitectureTheoryV2/blob/7e68ec6e77ef0249ede6a4c875715cfaf1cb3ee6/research/lean/ResearchLean/AG/DoctrineFiberProduct/SemanticCoreBeckChevalleyMate.lean\#L197}{\texttt{semantic\allowbreak{}Core\allowbreak{}Beck\allowbreak{}Chevalley\allowbreak{}Discrepancy\allowbreak{}Aut\_\allowbreak{}eq\_\allowbreak{}one\_\allowbreak{}iff}} & A specified comparison satisfies the lift triangle exactly when it equals the canonical mate. For an invertible specified comparison, this is also equivalent to the defect being the identity. \\
\addlinespace[3pt]
Definition~\ref{def:5.14} & \href{https://github.com/iroha1203/AlgebraicArchitectureTheoryV2/blob/719f81f47d410701fd82c2bc88613cc140c59377/research/lean/ResearchLean/AG/DoctrineFiberProduct/Schema.lean\#L480}{\texttt{Cart\allowbreak{}Presentation\allowbreak{}Between}} & A finite presentation that requires equality of the extraction codes themselves after transport along the normalization and the Atom permutation. Not only is the evaluation of the extraction predicates equivalent, but equality of codes is a condition on the input. \\
\addlinespace[3pt]
Theorem~\ref{thm:5.15} & \href{https://github.com/iroha1203/AlgebraicArchitectureTheoryV2/blob/719f81f47d410701fd82c2bc88613cc140c59377/research/lean/ResearchLean/AG/DoctrineFiberProduct/ExactBottomCoverageClassification.lean\#L183}{\texttt{finite\allowbreak{}Cofinite\allowbreak{}Extraction\allowbreak{}Code\_\allowbreak{}eq\_\allowbreak{}of\_\allowbreak{}mem\_\allowbreak{}range}} / \href{https://github.com/iroha1203/AlgebraicArchitectureTheoryV2/blob/719f81f47d410701fd82c2bc88613cc140c59377/research/lean/ResearchLean/AG/DoctrineFiberProduct/ExactBottomCoverageClassification.lean\#L428}{\texttt{endpoint\allowbreak{}Finite\allowbreak{}Target\allowbreak{}Cofinite\allowbreak{}Presentation}} / \href{https://github.com/iroha1203/AlgebraicArchitectureTheoryV2/blob/719f81f47d410701fd82c2bc88613cc140c59377/research/lean/ResearchLean/AG/DoctrineFiberProduct/ExactBottomCoverageClassification.lean\#L679}{\texttt{covered\allowbreak{}Object\allowbreak{}Witness\_\allowbreak{}necessary}} & Under decidable equality of Atoms, the finiteness of the Sources at both ends and the finiteness or cofiniteness of all the extractions of the target are equivalent to presentability up to the endpoint isomorphism of Definition~\ref{def:5.14}. Constructs the equality of extraction codes on the sources in the image of the normalization. \\
\addlinespace[3pt]
Definition~\ref{def:5.17} & \href{https://github.com/iroha1203/AlgebraicArchitectureTheoryV2/blob/719f81f47d410701fd82c2bc88613cc140c59377/research/lean/ResearchLean/AG/DoctrineFiberProduct/IndexedBaseDiagram.lean\#L133}{\texttt{Indexed\allowbreak{}Base\allowbreak{}Diagram}} & Corresponds to the definition of the vertices, the edge bases, and the path relations of the faces over a finite graph presentation. \\
\addlinespace[3pt]
Construction~\ref{cons:5.18} & \href{https://github.com/iroha1203/AlgebraicArchitectureTheoryV2/blob/719f81f47d410701fd82c2bc88613cc140c59377/research/lean/ResearchLean/AG/DoctrineFiberProduct/IndexedDiagnosticAssembly.lean\#L33}{\texttt{Indexed\allowbreak{}Diagnostic\allowbreak{}Interpretation}} / \href{https://github.com/iroha1203/AlgebraicArchitectureTheoryV2/blob/719f81f47d410701fd82c2bc88613cc140c59377/research/lean/ResearchLean/AG/DoctrineFiberProduct/IndexedDiagnosticAssembly.lean\#L141}{\texttt{transported\allowbreak{}Package}} / \href{https://github.com/iroha1203/AlgebraicArchitectureTheoryV2/blob/719f81f47d410701fd82c2bc88613cc140c59377/research/lean/ResearchLean/AG/DoctrineFiberProduct/IndexedDiagnosticAssembly.lean\#L159}{\texttt{transported\allowbreak{}Edge\allowbreak{}Lift}} / \href{https://github.com/iroha1203/AlgebraicArchitectureTheoryV2/blob/719f81f47d410701fd82c2bc88613cc140c59377/research/lean/ResearchLean/AG/DoctrineFiberProduct/IndexedDiagnosticAssembly.lean\#L275}{\texttt{transported\allowbreak{}Comparator}} & Constructs the vertex packages, the edge lifts, and the comparisons from an IndexedBaseDiagramHom and a diagnostic interpretation. The transports of edges and paths are generated from the semantic indexed action, and preserve the strong cocartesian qualification. \\
\addlinespace[3pt]
Proposition~\ref{prop:5.19} & \href{https://github.com/iroha1203/AlgebraicArchitectureTheoryV2/blob/719f81f47d410701fd82c2bc88613cc140c59377/research/lean/ResearchLean/AG/DoctrineFiberProduct/IndexedRawFamilyClassification.lean\#L183}{\texttt{target\allowbreak{}Relation\_\allowbreak{}of\_\allowbreak{}epi}} / \href{https://github.com/iroha1203/AlgebraicArchitectureTheoryV2/blob/719f81f47d410701fd82c2bc88613cc140c59377/research/lean/ResearchLean/AG/DoctrineFiberProduct/IndexedRawFamilyClassification.lean\#L56}{\texttt{uniform\allowbreak{}Target\allowbreak{}Base\allowbreak{}Liftable\allowbreak{}At\_\allowbreak{}iff\_\allowbreak{}epi}} & If the index morphism at the start of each face is an epi, the relations descend by naturality along paths. The uniform cancellation condition for arbitrary parallel morphisms is equivalent to being an epi. \\
\addlinespace[3pt]
Example~\ref{ex:5.20} & \href{https://github.com/iroha1203/AlgebraicArchitectureTheoryV2/blob/719f81f47d410701fd82c2bc88613cc140c59377/research/lean/ResearchLean/AG/DoctrineFiberProduct/IndexedBaseChangeTwoCellNoGo.lean\#L103}{\texttt{finite\allowbreak{}One\allowbreak{}To\allowbreak{}Two\_\allowbreak{}comp\_\allowbreak{}identity\_\allowbreak{}eq\_\allowbreak{}constant}} / \href{https://github.com/iroha1203/AlgebraicArchitectureTheoryV2/blob/719f81f47d410701fd82c2bc88613cc140c59377/research/lean/ResearchLean/AG/DoctrineFiberProduct/IndexedBaseChangeTwoCellNoGo.lean\#L114}{\texttt{finite\allowbreak{}Two\allowbreak{}Source\allowbreak{}Identity\_\allowbreak{}ne\_\allowbreak{}constant}} / \href{https://github.com/iroha1203/AlgebraicArchitectureTheoryV2/blob/719f81f47d410701fd82c2bc88613cc140c59377/research/lean/ResearchLean/AG/DoctrineFiberProduct/IndexedBaseChangeTwoCellNoGo.lean\#L129}{\texttt{finite\allowbreak{}One\allowbreak{}To\allowbreak{}Two\allowbreak{}Index\_\allowbreak{}not\_\allowbreak{}epi}} & Under the finite extraction of all Atoms and the identity normalization, the map $\mathrm{Fin}\,1\to\mathrm{Fin}\,2$ that selects 0 is not an epi. The identity map of Fin 2 and the constant map at 0 are different, but their composites with this inclusion are equal. \\
\addlinespace[3pt]
Construction~\ref{cons:5.21} & \href{https://github.com/iroha1203/AlgebraicArchitectureTheoryV2/blob/719f81f47d410701fd82c2bc88613cc140c59377/research/lean/ResearchLean/AG/DiagnosticConservativity/EndpointExactness.lean\#L47}{\texttt{indexed\allowbreak{}Diagnostic\allowbreak{}Endpoint\allowbreak{}Equivalence}} & The endpoint core transport of an arbitrary indexed semantic hom carries PackageFiberAut by a group isomorphism. \\
\addlinespace[3pt]
Theorem~\ref{thm:5.22} & \href{https://github.com/iroha1203/AlgebraicArchitectureTheoryV2/blob/719f81f47d410701fd82c2bc88613cc140c59377/research/lean/ResearchLean/AG/DiagnosticConservativity/CoherenceExactness.lean\#L23}{\texttt{indexed\allowbreak{}Coherent\allowbreak{}At\_\allowbreak{}transport\_\allowbreak{}iff}} / \href{https://github.com/iroha1203/AlgebraicArchitectureTheoryV2/blob/719f81f47d410701fd82c2bc88613cc140c59377/research/lean/ResearchLean/AG/DiagnosticConservativity/ObstructionExactness.lean\#L26}{\texttt{indexed\allowbreak{}Transport\allowbreak{}Obstruction\allowbreak{}Vanishes\_\allowbreak{}iff}} / \href{https://github.com/iroha1203/AlgebraicArchitectureTheoryV2/blob/719f81f47d410701fd82c2bc88613cc140c59377/research/lean/ResearchLean/AG/DiagnosticConservativity/OrbitExactness.lean\#L27}{\texttt{indexed\allowbreak{}Diagnostic\allowbreak{}In\allowbreak{}Reselection\allowbreak{}Orbit\_\allowbreak{}iff}} & Before and after the transport by the same indexed diagram hom, coherence, the vanishing of the obstruction class, and the orbits of edge reselections correspond respectively. Reselections can also be transported in the reverse direction. \\
\addlinespace[3pt]
Definition~\ref{def:5.24} & \href{https://github.com/iroha1203/AlgebraicArchitectureTheoryV2/blob/719f81f47d410701fd82c2bc88613cc140c59377/research/lean/ResearchLean/AG/DoctrineFiberProduct/RefinementBaseChange/RealizedSupport.lean\#L20}{\texttt{Realized\allowbreak{}Locus\allowbreak{}Extraction\allowbreak{}Reflecting}} & A definition that requires the reverse extraction implication only on the sources where a CoreFiber of the target exists. \\
\addlinespace[3pt]
Theorem~\ref{thm:5.25} & \href{https://github.com/iroha1203/AlgebraicArchitectureTheoryV2/blob/719f81f47d410701fd82c2bc88613cc140c59377/research/lean/ResearchLean/AG/DoctrineFiberProduct/RefinementBaseChange/RealizedSupport.lean\#L159}{\texttt{refinement\allowbreak{}Cartesian\allowbreak{}Cleavage\_\allowbreak{}iff\_\allowbreak{}realized\allowbreak{}Reflection}} & Extraction reflection on the realizable sources is equivalent to the existence of a strong cartesian lift to every target core. \\
\addlinespace[3pt]
Construction~\ref{cons:5.26} & \href{https://github.com/iroha1203/AlgebraicArchitectureTheoryV2/blob/719f81f47d410701fd82c2bc88613cc140c59377/research/lean/ResearchLean/AG/DoctrineFiberProduct/RefinementBaseChange/Configuration.lean\#L18}{\texttt{Refinement\allowbreak{}BCConfiguration}} & Defines the compatible sources and the pulled refinement from a general exact cospan and a refinement on the left. \\
\addlinespace[3pt]
Proposition~\ref{prop:5.27} & \href{https://github.com/iroha1203/AlgebraicArchitectureTheoryV2/blob/719f81f47d410701fd82c2bc88613cc140c59377/research/lean/ResearchLean/AG/DoctrineFiberProduct/RefinementBaseChange/Regime.lean\#L99}{\texttt{refinement\allowbreak{}BCRegime\_\allowbreak{}iff\_\allowbreak{}configuration\allowbreak{}Condition}} / \href{https://github.com/iroha1203/AlgebraicArchitectureTheoryV2/blob/719f81f47d410701fd82c2bc88613cc140c59377/research/lean/ResearchLean/AG/DoctrineFiberProduct/RefinementBaseChange/Qualification.lean\#L20}{\texttt{realized\allowbreak{}Reflection\_\allowbreak{}id}} / \href{https://github.com/iroha1203/AlgebraicArchitectureTheoryV2/blob/719f81f47d410701fd82c2bc88613cc140c59377/research/lean/ResearchLean/AG/DoctrineFiberProduct/RefinementBaseChange/Qualification.lean\#L81}{\texttt{realized\allowbreak{}Reflection\_\allowbreak{}comp}} & The reflection condition, which requires extraction to be reflected at each compatible source whenever a target core exists there, is equivalent to the existence of a regime, and regimes are closed under identity and composition. \\
\addlinespace[3pt]
Definition~\ref{def:5.30} & \href{https://github.com/iroha1203/AlgebraicArchitectureTheoryV2/blob/719f81f47d410701fd82c2bc88613cc140c59377/research/lean/ResearchLean/AG/DoctrineFiberProduct/RefinementGeometry.lean\#L360}{\texttt{Refinement\allowbreak{}Geom\allowbreak{}Read\allowbreak{}Hom}} / \href{https://github.com/iroha1203/AlgebraicArchitectureTheoryV2/blob/719f81f47d410701fd82c2bc88613cc140c59377/research/lean/ResearchLean/AG/DoctrineFiberProduct/RefinementGeometry.lean\#L500}{\texttt{Refinement\allowbreak{}Geometry\allowbreak{}Hom}} & Corresponds to the definition of a geometric morphism with a refinement base, an upper core map, and coverage/\allowbreak{}overlap/\allowbreak{}coefficient/\allowbreak{}raw/\allowbreak{}realization. \\
\addlinespace[3pt]
Construction~\ref{cons:5.31} & \href{https://github.com/iroha1203/AlgebraicArchitectureTheoryV2/blob/719f81f47d410701fd82c2bc88613cc140c59377/research/lean/ResearchLean/AG/DoctrineFiberProduct/UpperGeometryCleavage.lean\#L244}{\texttt{pull\allowbreak{}Geometry\allowbreak{}Package\allowbreak{}Along\allowbreak{}Upper\allowbreak{}Pair}} / \href{https://github.com/iroha1203/AlgebraicArchitectureTheoryV2/blob/719f81f47d410701fd82c2bc88613cc140c59377/research/lean/ResearchLean/AG/DoctrineFiberProduct/UpperGeometryCleavage.lean\#L284}{\texttt{exact\allowbreak{}Source\allowbreak{}Geometry}} / \href{https://github.com/iroha1203/AlgebraicArchitectureTheoryV2/blob/719f81f47d410701fd82c2bc88613cc140c59377/research/lean/ResearchLean/AG/DoctrineFiberProduct/UpperGeometryCleavage.lean\#L345}{\texttt{refinement\allowbreak{}Source\allowbreak{}Geometry}} & Pulls back the full geometry along an upper inverse pair. Constructs the geometries over the two sources, including the raw equalities and the coefficient and realization components. \\
\addlinespace[3pt]
Theorem~\ref{thm:5.32} & \href{https://github.com/iroha1203/AlgebraicArchitectureTheoryV2/blob/719f81f47d410701fd82c2bc88613cc140c59377/research/lean/ResearchLean/AG/DoctrineFiberProduct/UpperGeometryMate.lean\#L317}{\texttt{generated\allowbreak{}Route\allowbreak{}Refinement\allowbreak{}Mate}} / \href{https://github.com/iroha1203/AlgebraicArchitectureTheoryV2/blob/719f81f47d410701fd82c2bc88613cc140c59377/research/lean/ResearchLean/AG/DoctrineFiberProduct/UpperGeometryMate.lean\#L333}{\texttt{generated\allowbreak{}Route\allowbreak{}Refinement\allowbreak{}Mate\_\allowbreak{}fac}} / \href{https://github.com/iroha1203/AlgebraicArchitectureTheoryV2/blob/719f81f47d410701fd82c2bc88613cc140c59377/research/lean/ResearchLean/AG/DoctrineFiberProduct/UpperGeometryCompatibleGlobalMate.lean\#L26}{\texttt{generated\allowbreak{}Compatible\allowbreak{}Upper\allowbreak{}Geometry\allowbreak{}Mate\allowbreak{}At\_\allowbreak{}comparator\_\allowbreak{}intertwining}} & The mate between the two routes for an ActiveRefinementBCContext and a compatible source geometry. The triangular factorization and its uniqueness, and the coherence of the edges and the comparison morphisms. \\
\addlinespace[3pt]
Proposition~\ref{prop:5.36} & \href{https://github.com/leanprover-community/mathlib4/blob/8f9d9cff6bd728b17a24e163c9402775d9e6a365/Mathlib/CategoryTheory/Whiskering.lean\#L46}{\texttt{whisker\allowbreak{}Left}} / \href{https://github.com/leanprover-community/mathlib4/blob/8f9d9cff6bd728b17a24e163c9402775d9e6a365/Mathlib/CategoryTheory/Whiskering.lean\#L77}{\texttt{whiskering\allowbreak{}Left}} / \href{https://github.com/leanprover-community/mathlib4/blob/8f9d9cff6bd728b17a24e163c9402775d9e6a365/Mathlib/CategoryTheory/Whiskering.lean\#L192}{\texttt{whisker\allowbreak{}Left\_\allowbreak{}comp}} & Left whiskering by an arbitrary functor j preserves composition of natural transformations and commutative diagrams. It is not restricted to finite schemas or inclusion functors. \\
\addlinespace[3pt]
Construction~\ref{cons:5.37} & \href{https://github.com/iroha1203/AlgebraicArchitectureTheoryV2/blob/7e68ec6e77ef0249ede6a4c875715cfaf1cb3ee6/research/lean/ResearchLean/AG/FullGeometryNormalization/SemanticDerivedDiagnosticEndpointBridge.lean\#L29}{\texttt{semantic\allowbreak{}Derived\allowbreak{}Diagnostic\allowbreak{}Direct\allowbreak{}Geometry\allowbreak{}At}} / \href{https://github.com/iroha1203/AlgebraicArchitectureTheoryV2/blob/7e68ec6e77ef0249ede6a4c875715cfaf1cb3ee6/research/lean/ResearchLean/AG/FullGeometryNormalization/SemanticDerivedDiagnosticEndpointBridge.lean\#L44}{\texttt{semantic\allowbreak{}Derived\allowbreak{}Diagnostic\allowbreak{}Via\allowbreak{}Base\allowbreak{}Geometry\allowbreak{}At}} / \href{https://github.com/iroha1203/AlgebraicArchitectureTheoryV2/blob/7e68ec6e77ef0249ede6a4c875715cfaf1cb3ee6/research/lean/ResearchLean/AG/FullGeometryNormalization/SemanticDerivedDiagnosticEndpointBridge.lean\#L79}{\texttt{semantic\allowbreak{}Derived\allowbreak{}Diagnostic\allowbreak{}Bar\allowbreak{}Alpha\allowbreak{}Iso\allowbreak{}At\_\allowbreak{}triangle}} / \href{https://github.com/iroha1203/AlgebraicArchitectureTheoryV2/blob/7e68ec6e77ef0249ede6a4c875715cfaf1cb3ee6/research/lean/ResearchLean/AG/FullGeometryNormalization/SemanticDerivedDiagnosticEndpointBridge.lean\#L116}{\texttt{semantic\allowbreak{}Derived\allowbreak{}Diagnostic\allowbreak{}Bar\allowbreak{}Alpha\allowbreak{}Iso\allowbreak{}At\_\allowbreak{}generated\allowbreak{}Five\allowbreak{}Factor\_\allowbreak{}hom}} & Generates both routes from a general semantic square, a diagnostic face, the equality placing its endpoint core over the specified point, a coefficient ring, and the same full geometry. Gives the comparison triangle and the five factors: unit, endpoint isomorphism, mate, endpoint isomorphism, counit. \\
\addlinespace[3pt]
Proposition~\ref{prop:5.38} & \href{https://github.com/iroha1203/AlgebraicArchitectureTheoryV2/blob/7e68ec6e77ef0249ede6a4c875715cfaf1cb3ee6/research/lean/ResearchLean/AG/FullGeometryNormalization/SemanticCoreBeckChevalleyMateIdentification.lean\#L34}{\texttt{semantic\allowbreak{}Core\allowbreak{}Beck\allowbreak{}Chevalley\allowbreak{}Mate\_\allowbreak{}eq\_\allowbreak{}doctrine}} / \href{https://github.com/iroha1203/AlgebraicArchitectureTheoryV2/blob/7e68ec6e77ef0249ede6a4c875715cfaf1cb3ee6/research/lean/ResearchLean/AG/FullGeometryNormalization/SemanticCoreBeckChevalleyMateIdentification.lean\#L44}{\texttt{semantic\allowbreak{}Derived\allowbreak{}Bar\allowbreak{}Alpha\allowbreak{}Iso\allowbreak{}At\_\allowbreak{}projection\_\allowbreak{}doctrine\allowbreak{}Mate}} & Under endpoint isomorphisms generated from the same input, projects the full-geometry comparison to the actual core mate of Theorem~\ref{thm:5.11}. \\
\addlinespace[3pt]
Definition~\ref{def:5.39} (normalization endomorphisms) & \href{https://github.com/iroha1203/AlgebraicArchitectureTheoryV2/blob/719f81f47d410701fd82c2bc88613cc140c59377/research/lean/ResearchLean/AG/DoctrineFiberProduct/BCAuthoredCanonicalObjectNormalization.lean\#L28}{\texttt{canonical\allowbreak{}Object\allowbreak{}Normalization}} / \href{https://github.com/iroha1203/AlgebraicArchitectureTheoryV2/blob/719f81f47d410701fd82c2bc88613cc140c59377/research/lean/ResearchLean/AG/DoctrineFiberProduct/BCAuthoredCanonicalObjectNormalization.lean\#L98}{\texttt{Canonical\allowbreak{}Object\allowbreak{}Normalization\allowbreak{}Admissible}} / \href{https://github.com/iroha1203/AlgebraicArchitectureTheoryV2/blob/719f81f47d410701fd82c2bc88613cc140c59377/research/lean/ResearchLean/AG/DoctrineFiberProduct/BCAuthoredCanonicalObjectNormalization.lean\#L208}{\texttt{canonical\allowbreak{}Object\allowbreak{}Normalization\allowbreak{}Total}} & Constructs the upper and total endomorphisms from object normalization and admissibility. On the Lean side the condition on function-typed invariants is a comparison by a bijection of the codomain. \\
\addlinespace[3pt]
Construction~\ref{cons:5.40} & \href{https://github.com/iroha1203/AlgebraicArchitectureTheoryV2/blob/7e68ec6e77ef0249ede6a4c875715cfaf1cb3ee6/research/lean/ResearchLean/AG/FullGeometryNormalization/SemanticExactCoreBarSelector.lean\#L92}{\texttt{semantic\allowbreak{}Exact\allowbreak{}Bar\allowbreak{}Source\allowbreak{}Selector\_\allowbreak{}projection}} / \href{https://github.com/iroha1203/AlgebraicArchitectureTheoryV2/blob/7e68ec6e77ef0249ede6a4c875715cfaf1cb3ee6/research/lean/ResearchLean/AG/FullGeometryNormalization/SemanticExactBarBetaProjection.lean\#L19}{\texttt{semantic\allowbreak{}Exact\allowbreak{}Bar\allowbreak{}DAt\_\allowbreak{}eq\_\allowbreak{}source\allowbreak{}Selector\_\allowbreak{}map}} / \href{https://github.com/iroha1203/AlgebraicArchitectureTheoryV2/blob/7e68ec6e77ef0249ede6a4c875715cfaf1cb3ee6/research/lean/ResearchLean/AG/FullGeometryNormalization/SemanticExactBarBetaProjection.lean\#L46}{\texttt{semantic\allowbreak{}Exact\allowbreak{}Bar\allowbreak{}DAt\_\allowbreak{}projection}} / \href{https://github.com/iroha1203/AlgebraicArchitectureTheoryV2/blob/7e68ec6e77ef0249ede6a4c875715cfaf1cb3ee6/research/lean/ResearchLean/AG/FullGeometryNormalization/SemanticExactBarBetaProjection.lean\#L143}{\texttt{semantic\allowbreak{}Exact\allowbreak{}Bar\allowbreak{}Beta\allowbreak{}At\_\allowbreak{}projection\_\allowbreak{}doctrine\allowbreak{}Mate}} & Selects normalization or identity from the same diagnosis, core, and geometry over a general semantic square and carries it along the routes. Aligns the core and geometry selectors, projectors, and generated comparisons under projection. \\
\bottomrule
\end{xltabular}

\subsection*{Reindexing and coherence of canonical lifts}

\begin{xltabular}{\linewidth}{@{}LLL@{}}
\toprule
Text & Lean declaration & Objects and conditions \\
\midrule
\endhead
Lemma~\ref{lem:5.7} (a single morphism) & \href{https://github.com/iroha1203/AlgebraicArchitectureTheoryV2/blob/7e68ec6e77ef0249ede6a4c875715cfaf1cb3ee6/research/lean/ResearchLean/AG/DoctrineFiberProduct/CartesianRegimeReindexingCleavage.lean\#L172}{\texttt{reindex\allowbreak{}Functor}} / \href{https://github.com/iroha1203/AlgebraicArchitectureTheoryV2/blob/7e68ec6e77ef0249ede6a4c875715cfaf1cb3ee6/research/lean/ResearchLean/AG/DoctrineFiberProduct/CartesianRegimeReindexingCleavage.lean\#L191}{\texttt{reindex\allowbreak{}Functor\_\allowbreak{}map\_\allowbreak{}fac}} / \href{https://github.com/iroha1203/AlgebraicArchitectureTheoryV2/blob/7e68ec6e77ef0249ede6a4c875715cfaf1cb3ee6/research/lean/ResearchLean/AG/DoctrineFiberProduct/CartesianRegimeReindexingCleavage.lean\#L200}{\texttt{reindex\allowbreak{}Functor\_\allowbreak{}map\_\allowbreak{}unique}} & Any choice of a strongly cartesian lift to each object gives the reindexing functor and the factorization equation and uniqueness for morphisms. \\
\addlinespace[3pt]
Lemma~\ref{lem:5.7} (arbitrary cleavage) & \href{https://github.com/iroha1203/AlgebraicArchitectureTheoryV2/blob/7e68ec6e77ef0249ede6a4c875715cfaf1cb3ee6/research/lean/ResearchLean/AG/DoctrineFiberProduct/ArbitrarySemanticCleavageCoherence.lean\#L264}{\texttt{arbitrary\_\allowbreak{}semantic\_\allowbreak{}compositor}} / \href{https://github.com/iroha1203/AlgebraicArchitectureTheoryV2/blob/7e68ec6e77ef0249ede6a4c875715cfaf1cb3ee6/research/lean/ResearchLean/AG/DoctrineFiberProduct/ArbitrarySemanticCleavageCoherence.lean\#L383}{\texttt{arbitrary\_\allowbreak{}semantic\_\allowbreak{}unitor}} / \href{https://github.com/iroha1203/AlgebraicArchitectureTheoryV2/blob/7e68ec6e77ef0249ede6a4c875715cfaf1cb3ee6/research/lean/ResearchLean/AG/DoctrineFiberProduct/ArbitrarySemanticCleavageCoherence.lean\#L630}{\texttt{arbitrary\_\allowbreak{}semantic\_\allowbreak{}pentagon}} / \href{https://github.com/iroha1203/AlgebraicArchitectureTheoryV2/blob/7e68ec6e77ef0249ede6a4c875715cfaf1cb3ee6/research/lean/ResearchLean/AG/DoctrineFiberProduct/ArbitrarySemanticCleavageCoherence.lean\#L853}{\texttt{arbitrary\_\allowbreak{}semantic\_\allowbreak{}triangle}} & Derives unit and composition comparisons and their pentagon and triangle from arbitrary choices of semantic cartesian lifts. Coherence is not part of the chosen data. \\
\bottomrule
\end{xltabular}

\section{Chapter 6: Idempotent Normalization and Realization}\label{sec:A.6}

\begin{xltabular}{\linewidth}{@{}LLL@{}}
\toprule
Text & Lean declaration & Objects and conditions \\
\midrule
\endhead
Construction~\ref{cons:6.1} & \href{https://github.com/iroha1203/AlgebraicArchitectureTheoryV2/blob/719f81f47d410701fd82c2bc88613cc140c59377/research/lean/ResearchLean/AG/DoctrineFiberProduct/BCAuthoredCanonicalObjectNormalization.lean\#L28}{\texttt{canonical\allowbreak{}Object\allowbreak{}Normalization}} / \href{https://github.com/iroha1203/AlgebraicArchitectureTheoryV2/blob/719f81f47d410701fd82c2bc88613cc140c59377/research/lean/ResearchLean/AG/DoctrineFiberProduct/BCAuthoredCanonicalObjectNormalization.lean\#L34}{\texttt{canonical\allowbreak{}Object\allowbreak{}Normalization\_\allowbreak{}configuration}} & \texttt{N\_P(A)=\allowbreak{}objectReading.\allowbreak{}object(A.\allowbreak{}configuration)} preserves the configuration. Admissibility is not needed for the construction of this map on objects. \\
\addlinespace[3pt]
Proposition~\ref{prop:6.2} & \href{https://github.com/iroha1203/AlgebraicArchitectureTheoryV2/blob/719f81f47d410701fd82c2bc88613cc140c59377/research/lean/ResearchLean/AG/DoctrineFiberProduct/BCAuthoredCanonicalObjectNormalization.lean\#L52}{\texttt{canonical\allowbreak{}Object\allowbreak{}Normalization\_\allowbreak{}idempotent}} / \href{https://github.com/iroha1203/AlgebraicArchitectureTheoryV2/blob/719f81f47d410701fd82c2bc88613cc140c59377/research/lean/ResearchLean/AG/DoctrineFiberProduct/ConfigurationDescent.lean\#L29}{\texttt{canonical\allowbreak{}Normalization\allowbreak{}Fixed\allowbreak{}Equiv}} & For an arbitrary core P the fixed points of N are the selected objects, and an explicit bijection with configurations is constructed. \\
\addlinespace[3pt]
Theorem~\ref{thm:6.3} & \href{https://github.com/iroha1203/AlgebraicArchitectureTheoryV2/blob/719f81f47d410701fd82c2bc88613cc140c59377/research/lean/ResearchLean/AG/DoctrineFiberProduct/ConfigurationDescent.lean\#L45}{\texttt{canonical\allowbreak{}Object\allowbreak{}Normalization\_\allowbreak{}factorization\_\allowbreak{}iff}} / \href{https://github.com/iroha1203/AlgebraicArchitectureTheoryV2/blob/719f81f47d410701fd82c2bc88613cc140c59377/research/lean/ResearchLean/AG/DoctrineFiberProduct/ConfigurationDescent.lean\#L88}{\texttt{architecture\allowbreak{}Object\allowbreak{}Configuration\allowbreak{}Quotient\allowbreak{}Equiv}} & For an arbitrary value type Y and a function f, $f\circ N=f$ is equivalent to a unique factorization through the configuration. The bijection between the quotient by the configuration equivalence relation and the configurations is also proved. \\
\addlinespace[3pt]
Theorem~\ref{thm:6.5} & \href{https://github.com/iroha1203/AlgebraicArchitectureTheoryV2/blob/719f81f47d410701fd82c2bc88613cc140c59377/research/lean/ResearchLean/AG/DoctrineFiberProduct/IdempotentExchangeNormalization.lean\#L124}{\texttt{canonical\allowbreak{}Object\allowbreak{}Normalization\allowbreak{}Total\_\allowbreak{}comp}} & The total normalization morphism built from Ad(P) satisfies $N\gg N=N$ as morphisms. \\
\addlinespace[3pt]
Lemma~\ref{lem:6.6} & \href{https://github.com/iroha1203/AlgebraicArchitectureTheoryV2/blob/719f81f47d410701fd82c2bc88613cc140c59377/research/lean/ResearchLean/AG/DoctrineFiberProduct/CanonicalObjectNormalizationNaturality.lean\#L33}{\texttt{canonical\allowbreak{}Object\allowbreak{}Normalization\_\allowbreak{}natural}} & For an arbitrary PackageTotalHom, the map on objects commutes with the canonical object normalizations at both ends. Naturality in the object component. \\
\addlinespace[3pt]
Proposition~\ref{prop:6.7} & \href{https://github.com/iroha1203/AlgebraicArchitectureTheoryV2/blob/719f81f47d410701fd82c2bc88613cc140c59377/research/lean/ResearchLean/AG/DoctrineFiberProduct/DistinctArchitectureObjects.lean\#L69}{\texttt{exists\_\allowbreak{}distinct\_\allowbreak{}architecture\allowbreak{}Objects\_\allowbreak{}over\_\allowbreak{}configuration}} / \href{https://github.com/iroha1203/AlgebraicArchitectureTheoryV2/blob/719f81f47d410701fd82c2bc88613cc140c59377/research/lean/ResearchLean/AG/DoctrineFiberProduct/BCAuthoredCanonicalObjectNormalization.lean\#L229}{\texttt{canonical\allowbreak{}Object\allowbreak{}Normalization\allowbreak{}Total\_\allowbreak{}not\_\allowbreak{}is\allowbreak{}Iso}} & From the distinct structural decorations by Unit and by Bool over each configuration, the object normalization is not injective. Under Ad(P) the total normalization morphism is not an isomorphism. \\
\addlinespace[3pt]
Theorem~\ref{thm:6.8} & \href{https://github.com/iroha1203/AlgebraicArchitectureTheoryV2/blob/719f81f47d410701fd82c2bc88613cc140c59377/research/lean/ResearchLean/AG/DoctrineFiberProduct/InternalNormalizationSplitNoGo.lean\#L52}{\texttt{canonical\allowbreak{}Object\allowbreak{}Normalization\allowbreak{}Total\_\allowbreak{}not\_\allowbreak{}internal\_\allowbreak{}split}} & For arbitrary P and Q and under Ad(P), there is no pair of morphisms $r:P\to Q$ and $i:Q\to P$ satisfying $r\gg i=N$ and $i\gg r=\mathrm{id}$ at the same time. \\
\addlinespace[3pt]
Definition~\ref{def:6.9} & \href{https://github.com/leanprover-community/mathlib4/blob/8f9d9cff6bd728b17a24e163c9402775d9e6a365/Mathlib/CategoryTheory/Idempotents/Karoubi.lean\#L44}{\texttt{Karoubi}} / \href{https://github.com/leanprover-community/mathlib4/blob/8f9d9cff6bd728b17a24e163c9402775d9e6a365/Mathlib/CategoryTheory/Idempotents/Karoubi.lean\#L72}{\texttt{Hom}} & The objects $(X,e,e^2=e)$ of the Karoubi category, the morphisms $e\gg f\gg d=f$, and identities and composition. \\
\addlinespace[3pt]
Construction~\ref{cons:6.10} & \href{https://github.com/leanprover-community/mathlib4/blob/8f9d9cff6bd728b17a24e163c9402775d9e6a365/Mathlib/CategoryTheory/Idempotents/Karoubi.lean\#L139}{\texttt{to\allowbreak{}Karoubi}} / \href{https://github.com/leanprover-community/mathlib4/blob/8f9d9cff6bd728b17a24e163c9402775d9e6a365/Mathlib/CategoryTheory/Idempotents/Karoubi.lean\#L256}{\texttt{decomp\allowbreak{}Id}} / \href{https://github.com/leanprover-community/mathlib4/blob/8f9d9cff6bd728b17a24e163c9402775d9e6a365/Mathlib/CategoryTheory/Idempotents/Karoubi.lean\#L260}{\texttt{decomp\_\allowbreak{}p}} & The inclusion by the identity idempotent is fully faithful, and each Karoubi object is a retract of an included object. \\
\addlinespace[3pt]
Theorem~\ref{thm:6.14} & \href{https://github.com/iroha1203/AlgebraicArchitectureTheoryV2/blob/719f81f47d410701fd82c2bc88613cc140c59377/research/lean/ResearchLean/AG/FullGeometryNormalization/CanonicalNormalization.lean\#L78}{\texttt{canonical\allowbreak{}Geometry\allowbreak{}Normalization}} / \href{https://github.com/iroha1203/AlgebraicArchitectureTheoryV2/blob/719f81f47d410701fd82c2bc88613cc140c59377/research/lean/ResearchLean/AG/FullGeometryNormalization/CanonicalNormalization.lean\#L136}{\texttt{canonical\allowbreak{}Geometry\allowbreak{}Normalization\_\allowbreak{}idem}} & Constructs the normalization morphism of the full geometry under Ad(G.core), and obtains idempotence. The coefficient, support, axis, and observable components are identities, and the raw component is the equality for the identity base change. \\
\addlinespace[3pt]
Lemma~\ref{lem:6.15} (core and lifts) & \href{https://github.com/iroha1203/AlgebraicArchitectureTheoryV2/blob/7e68ec6e77ef0249ede6a4c875715cfaf1cb3ee6/research/lean/ResearchLean/AG/FullGeometryNormalization/SemanticExactCoreNormalizationNaturality.lean\#L37}{\texttt{canonical\allowbreak{}Core\allowbreak{}Normalization\allowbreak{}Admissible\_\allowbreak{}semantic\allowbreak{}Transport}} / \href{https://github.com/iroha1203/AlgebraicArchitectureTheoryV2/blob/7e68ec6e77ef0249ede6a4c875715cfaf1cb3ee6/research/lean/ResearchLean/AG/FullGeometryNormalization/SemanticExactCoreNormalizationNaturality.lean\#L127}{\texttt{canonical\allowbreak{}Core\allowbreak{}Normalization\allowbreak{}Admissible\_\allowbreak{}semantic\allowbreak{}Pull}} / \href{https://github.com/iroha1203/AlgebraicArchitectureTheoryV2/blob/7e68ec6e77ef0249ede6a4c875715cfaf1cb3ee6/research/lean/ResearchLean/AG/FullGeometryNormalization/SemanticExactCoreNormalizationNaturality.lean\#L49}{\texttt{semantic\allowbreak{}Core\allowbreak{}Fiber\allowbreak{}Lift\_\allowbreak{}normalization\_\allowbreak{}natural}} / \href{https://github.com/iroha1203/AlgebraicArchitectureTheoryV2/blob/7e68ec6e77ef0249ede6a4c875715cfaf1cb3ee6/research/lean/ResearchLean/AG/FullGeometryNormalization/SemanticExactCoreNormalizationNaturality.lean\#L140}{\texttt{semantic\allowbreak{}Core\allowbreak{}Inverse\allowbreak{}Lift\_\allowbreak{}normalization\_\allowbreak{}natural}} & For every exact base morphism, preserves core admissibility and commutes normalization with the generated transport and pullback lifts. \\
\addlinespace[3pt]
Lemma~\ref{lem:6.15} (fiber functors) & \href{https://github.com/iroha1203/AlgebraicArchitectureTheoryV2/blob/7e68ec6e77ef0249ede6a4c875715cfaf1cb3ee6/research/lean/ResearchLean/AG/FullGeometryNormalization/SemanticExactCoreNormalizationNaturality.lean\#L66}{\texttt{semantic\allowbreak{}Core\allowbreak{}Fiber\allowbreak{}Transport\allowbreak{}Functor\_\allowbreak{}map\_\allowbreak{}normalization}} / \href{https://github.com/iroha1203/AlgebraicArchitectureTheoryV2/blob/7e68ec6e77ef0249ede6a4c875715cfaf1cb3ee6/research/lean/ResearchLean/AG/FullGeometryNormalization/SemanticExactCoreNormalizationNaturality.lean\#L160}{\texttt{semantic\allowbreak{}Core\allowbreak{}Inverse\allowbreak{}Reindex\allowbreak{}Functor\_\allowbreak{}map\_\allowbreak{}normalization}} / \href{https://github.com/iroha1203/AlgebraicArchitectureTheoryV2/blob/7e68ec6e77ef0249ede6a4c875715cfaf1cb3ee6/research/lean/ResearchLean/AG/FullGeometryNormalization/SemanticExactNormalizationPush.lean\#L59}{\texttt{semantic\allowbreak{}Geom\allowbreak{}Fiber\allowbreak{}Transport\allowbreak{}Functor\_\allowbreak{}map\_\allowbreak{}normalization}} / \href{https://github.com/iroha1203/AlgebraicArchitectureTheoryV2/blob/7e68ec6e77ef0249ede6a4c875715cfaf1cb3ee6/research/lean/ResearchLean/AG/FullGeometryNormalization/SemanticExactNormalizationNaturality.lean\#L82}{\texttt{semantic\allowbreak{}Geometry\allowbreak{}Pull\allowbreak{}Functor\_\allowbreak{}map\_\allowbreak{}normalization}} & The core and full-geometry transport and pullback functors along the same general exact morphism preserve normalization morphisms. \\
\addlinespace[3pt]
Lemma~\ref{lem:6.15} (geometry lifts) & \href{https://github.com/iroha1203/AlgebraicArchitectureTheoryV2/blob/7e68ec6e77ef0249ede6a4c875715cfaf1cb3ee6/research/lean/ResearchLean/AG/FullGeometryNormalization/SemanticExactNormalizationNaturality.lean\#L37}{\texttt{canonical\allowbreak{}Geometry\allowbreak{}Normalization\allowbreak{}Admissible\_\allowbreak{}semantic\allowbreak{}Exact\allowbreak{}Pull}} / \href{https://github.com/iroha1203/AlgebraicArchitectureTheoryV2/blob/7e68ec6e77ef0249ede6a4c875715cfaf1cb3ee6/research/lean/ResearchLean/AG/FullGeometryNormalization/SemanticExactNormalizationPush.lean\#L40}{\texttt{semantic\allowbreak{}Geom\allowbreak{}Fiber\allowbreak{}Lift\_\allowbreak{}normalization\_\allowbreak{}natural}} / \href{https://github.com/iroha1203/AlgebraicArchitectureTheoryV2/blob/7e68ec6e77ef0249ede6a4c875715cfaf1cb3ee6/research/lean/ResearchLean/AG/FullGeometryNormalization/SemanticExactNormalizationNaturality.lean\#L53}{\texttt{semantic\allowbreak{}Geometry\allowbreak{}Pull\allowbreak{}Lift\_\allowbreak{}normalization\_\allowbreak{}natural}} & Preservation of admissibility and interchange equations for generated lifts of full geometries. No finite-code presentation is required. \\
\addlinespace[3pt]
Theorem~\ref{thm:6.16} & \href{https://github.com/iroha1203/AlgebraicArchitectureTheoryV2/blob/7e68ec6e77ef0249ede6a4c875715cfaf1cb3ee6/research/lean/ResearchLean/AG/FullGeometryNormalization/SemanticExactBarBetaClassification.lean\#L123}{\texttt{semantic\allowbreak{}Exact\allowbreak{}Bar\allowbreak{}DAt\_\allowbreak{}idem}} / \href{https://github.com/iroha1203/AlgebraicArchitectureTheoryV2/blob/7e68ec6e77ef0249ede6a4c875715cfaf1cb3ee6/research/lean/ResearchLean/AG/FullGeometryNormalization/SemanticExactBarBetaClassification.lean\#L414}{\texttt{semantic\allowbreak{}Exact\allowbreak{}Bar\allowbreak{}EAt\_\allowbreak{}idem}} / \href{https://github.com/iroha1203/AlgebraicArchitectureTheoryV2/blob/7e68ec6e77ef0249ede6a4c875715cfaf1cb3ee6/research/lean/ResearchLean/AG/FullGeometryNormalization/SemanticExactBarBetaClassification.lean\#L474}{\texttt{semantic\allowbreak{}Exact\allowbreak{}Bar\allowbreak{}Beta\allowbreak{}Karoubi\allowbreak{}Iso\allowbreak{}At}} / \href{https://github.com/iroha1203/AlgebraicArchitectureTheoryV2/blob/7e68ec6e77ef0249ede6a4c875715cfaf1cb3ee6/research/lean/ResearchLean/AG/FullGeometryNormalization/SemanticExactBarBetaClassification.lean\#L672}{\texttt{semantic\allowbreak{}Exact\allowbreak{}Bar\allowbreak{}Beta\allowbreak{}At\_\allowbreak{}is\allowbreak{}Iso\_\allowbreak{}iff\_\allowbreak{}not\_\allowbreak{}selected}} / \href{https://github.com/iroha1203/AlgebraicArchitectureTheoryV2/blob/7e68ec6e77ef0249ede6a4c875715cfaf1cb3ee6/research/lean/ResearchLean/AG/FullGeometryNormalization/SemanticExactBarBetaClassification.lean\#L605}{\texttt{semantic\allowbreak{}Exact\allowbreak{}Bar\allowbreak{}Projectors\allowbreak{}At\_\allowbreak{}eq\_\allowbreak{}endpoint\_\allowbreak{}normalizations}} & For the same endpoints generated from a general semantic square, proves idempotence, the Karoubi isomorphism, invertibility classified by selection, and agreement with the endpoint normalizations. \\
\addlinespace[3pt]
Proposition~\ref{prop:6.17} & \href{https://github.com/iroha1203/AlgebraicArchitectureTheoryV2/blob/7e68ec6e77ef0249ede6a4c875715cfaf1cb3ee6/research/lean/ResearchLean/AG/FullGeometryNormalization/SemanticExactBarBetaKaroubiProjection.lean\#L84}{\texttt{semantic\allowbreak{}Exact\allowbreak{}Bar\allowbreak{}EAt\_\allowbreak{}projection}} / \href{https://github.com/iroha1203/AlgebraicArchitectureTheoryV2/blob/7e68ec6e77ef0249ede6a4c875715cfaf1cb3ee6/research/lean/ResearchLean/AG/FullGeometryNormalization/SemanticExactBarBetaKaroubiProjection.lean\#L167}{\texttt{semantic\allowbreak{}Exact\allowbreak{}Bar\allowbreak{}Core\allowbreak{}EAt\_\allowbreak{}idem}} / \href{https://github.com/iroha1203/AlgebraicArchitectureTheoryV2/blob/7e68ec6e77ef0249ede6a4c875715cfaf1cb3ee6/research/lean/ResearchLean/AG/FullGeometryNormalization/SemanticExactBarBetaKaroubiProjection.lean\#L404}{\texttt{semantic\allowbreak{}Exact\allowbreak{}Bar\allowbreak{}Beta\allowbreak{}Karoubi\allowbreak{}Projection\allowbreak{}Alignment\allowbreak{}At}} & Projects the canonical comparison, both projectors, and the generated comparison to cores. The projected geometry Karoubi isomorphism agrees with the one constructed for cores under the generated endpoint isomorphisms. \\
\addlinespace[3pt]
Lemma~\ref{lem:6.19} & \href{https://github.com/iroha1203/AlgebraicArchitectureTheoryV2/blob/719f81f47d410701fd82c2bc88613cc140c59377/research/lean/ResearchLean/AG/RealizationComparisonIdempotents/CanonicalNormalizationAbsorption.lean\#L80}{\texttt{canonical\allowbreak{}Package\allowbreak{}Normalization\_\allowbreak{}absorption}} & For an arbitrary PackageTotalHom between ends that satisfy Ad, $N_P\gg f\gg N_Q=N_P\gg f$ as morphisms. \\
\addlinespace[3pt]
Construction~\ref{cons:6.20} & \href{https://github.com/iroha1203/AlgebraicArchitectureTheoryV2/blob/719f81f47d410701fd82c2bc88613cc140c59377/research/lean/ResearchLean/AG/RealizationComparisonIdempotents/NormalizationCategory.lean\#L34}{\texttt{Normalized\allowbreak{}Package\allowbreak{}Object}} / \href{https://github.com/iroha1203/AlgebraicArchitectureTheoryV2/blob/719f81f47d410701fd82c2bc88613cc140c59377/research/lean/ResearchLean/AG/RealizationComparisonIdempotents/NormalizationCategory.lean\#L51}{\texttt{normalized\allowbreak{}Package\allowbreak{}Karoubi\allowbreak{}Object}} / \href{https://github.com/iroha1203/AlgebraicArchitectureTheoryV2/blob/719f81f47d410701fd82c2bc88613cc140c59377/research/lean/ResearchLean/AG/RealizationComparisonIdempotents/NormalizationCategory.lean\#L101}{\texttt{normalized\allowbreak{}Package\allowbreak{}Karoubi\allowbreak{}Functor}} & Constructs the normalization category from admissible packages and the Karoubi morphisms of the specified idempotents. The inclusion into the Karoubi category is fully faithful. \\
\addlinespace[3pt]
Theorem~\ref{thm:6.21} & \href{https://github.com/iroha1203/AlgebraicArchitectureTheoryV2/blob/719f81f47d410701fd82c2bc88613cc140c59377/research/lean/ResearchLean/AG/RealizationComparisonIdempotents/NormalizationCategory.lean\#L142}{\texttt{package\allowbreak{}Normalization\allowbreak{}Functor}} / \href{https://github.com/iroha1203/AlgebraicArchitectureTheoryV2/blob/719f81f47d410701fd82c2bc88613cc140c59377/research/lean/ResearchLean/AG/RealizationComparisonIdempotents/NormalizationCategory.lean\#L192}{\texttt{package\allowbreak{}Normalization\allowbreak{}Functor\_\allowbreak{}full}} & Fullness of the normalization functor that sends a morphism $f$ to $N(f)=N_P\gg f$. \\
\addlinespace[3pt]
Proposition~\ref{prop:6.22} & \href{https://github.com/iroha1203/AlgebraicArchitectureTheoryV2/blob/719f81f47d410701fd82c2bc88613cc140c59377/research/lean/ResearchLean/AG/FullGeometryNormalization/CanonicalNormalization.lean\#L204}{\texttt{canonical\allowbreak{}Admissible\allowbreak{}Geometry\allowbreak{}Normalization\_\allowbreak{}absorption}} / \href{https://github.com/iroha1203/AlgebraicArchitectureTheoryV2/blob/719f81f47d410701fd82c2bc88613cc140c59377/research/lean/ResearchLean/AG/FullGeometryNormalization/CanonicalNormalization.lean\#L325}{\texttt{geometry\allowbreak{}Normalization\allowbreak{}Functor}} / \href{https://github.com/iroha1203/AlgebraicArchitectureTheoryV2/blob/719f81f47d410701fd82c2bc88613cc140c59377/research/lean/ResearchLean/AG/FullGeometryNormalization/CanonicalNormalization.lean\#L385}{\texttt{geometry\allowbreak{}Normalization\allowbreak{}Functor\_\allowbreak{}map\_\allowbreak{}coefficient\allowbreak{}Hom}} & Constructs the normalization functor from the absorption law for morphisms between admissible geometries. It preserves the projections to the coefficients and to the base of the package. \\
\addlinespace[3pt]
Proposition~\ref{prop:6.23} & \href{https://github.com/iroha1203/AlgebraicArchitectureTheoryV2/blob/719f81f47d410701fd82c2bc88613cc140c59377/research/lean/ResearchLean/AG/RealizationComparisonIdempotents/NormalizationNaturalityFailure.lean\#L41}{\texttt{normalized\allowbreak{}Package\allowbreak{}Inclusion}} / \href{https://github.com/iroha1203/AlgebraicArchitectureTheoryV2/blob/719f81f47d410701fd82c2bc88613cc140c59377/research/lean/ResearchLean/AG/RealizationComparisonIdempotents/NormalizationNaturalityFailure.lean\#L62}{\texttt{normalized\allowbreak{}Package\allowbreak{}Retraction\allowbreak{}App}} / \href{https://github.com/iroha1203/AlgebraicArchitectureTheoryV2/blob/719f81f47d410701fd82c2bc88613cc140c59377/research/lean/ResearchLean/AG/RealizationComparisonIdempotents/NormalizationNaturalityFailure.lean\#L80}{\texttt{normalized\allowbreak{}Package\allowbreak{}Inclusion\_\allowbreak{}retraction}} & The natural inclusion, and the splitting identity given by the retraction at each object. \\
\addlinespace[3pt]
Proposition~\ref{prop:6.24} & \href{https://github.com/iroha1203/AlgebraicArchitectureTheoryV2/blob/719f81f47d410701fd82c2bc88613cc140c59377/research/lean/ResearchLean/AG/RealizationComparisonIdempotents/NormalizationNaturalityFailure.lean\#L144}{\texttt{normalized\allowbreak{}Package\allowbreak{}Retraction\allowbreak{}Natural\allowbreak{}At\_\allowbreak{}iff\_\allowbreak{}operation\allowbreak{}Coherent}} / \href{https://github.com/iroha1203/AlgebraicArchitectureTheoryV2/blob/719f81f47d410701fd82c2bc88613cc140c59377/research/lean/ResearchLean/AG/DoctrineFiberProduct/LaxDiagnosticProjectorModificationBlocker.lean\#L30}{\texttt{Canonical\allowbreak{}Normalization\allowbreak{}Operation\allowbreak{}Coherent}} & Naturality of the retraction at a specified morphism f is equivalent to coherence of the normalization of operations. \\
\addlinespace[3pt]
Definition~\ref{def:6.26} & \href{https://github.com/leanprover-community/mathlib4/blob/8f9d9cff6bd728b17a24e163c9402775d9e6a365/Mathlib/CategoryTheory/Comma/Arrow.lean\#L38}{\texttt{Arrow}} / \href{https://github.com/iroha1203/AlgebraicArchitectureTheoryV2/blob/719f81f47d410701fd82c2bc88613cc140c59377/research/lean/ResearchLean/AG/RealizationComparisonIdempotents/MaximalSubgroupoid.lean\#L30}{\texttt{Realization\allowbreak{}Comparison\allowbreak{}Category}} & The arrow category Arrow E, and the comparison category Arrow(Karoubi E). \\
\addlinespace[3pt]
Theorem~\ref{thm:6.27} & \href{https://github.com/iroha1203/AlgebraicArchitectureTheoryV2/blob/719f81f47d410701fd82c2bc88613cc140c59377/research/lean/ResearchLean/AG/RealizationComparisonIdempotents/KaroubiArrowEquivalence.lean\#L336}{\texttt{karoubi\allowbreak{}Arrow\allowbreak{}Equivalence}} & Karoubi(Arrow E)$\simeq$Arrow(Karoubi E) for an arbitrary category E. Constructs the functor, its inverse, and the unit and counit from the idempotents at the ends and the sandwich morphisms. \\
\addlinespace[3pt]
Proposition~\ref{prop:6.28} & \href{https://github.com/iroha1203/AlgebraicArchitectureTheoryV2/blob/719f81f47d410701fd82c2bc88613cc140c59377/research/lean/ResearchLean/AG/RealizationComparisonIdempotents/FunctorNaturality.lean\#L201}{\texttt{karoubi\allowbreak{}Arrow\allowbreak{}Naturality\allowbreak{}Iso}} / \href{https://github.com/iroha1203/AlgebraicArchitectureTheoryV2/blob/719f81f47d410701fd82c2bc88613cc140c59377/research/lean/ResearchLean/AG/RealizationComparisonIdempotents/FunctorNaturality.lean\#L280}{\texttt{karoubi\allowbreak{}Arrow\allowbreak{}Naturality\allowbreak{}Iso\allowbreak{}App\_\allowbreak{}comp}} & The natural isomorphism between the Karoubi and Arrow actions of an arbitrary functor F and the equivalence above, and the coherence with composition. In AAT it is applied to the two-level projection functors. \\
\addlinespace[3pt]
Construction~\ref{cons:6.29} & \href{https://github.com/iroha1203/AlgebraicArchitectureTheoryV2/blob/719f81f47d410701fd82c2bc88613cc140c59377/research/lean/ResearchLean/AG/RealizationComparisonIdempotents/MaximalSubgroupoid.lean\#L33}{\texttt{Reversible\allowbreak{}Representation\allowbreak{}Changes}} / \href{https://github.com/iroha1203/AlgebraicArchitectureTheoryV2/blob/719f81f47d410701fd82c2bc88613cc140c59377/research/lean/ResearchLean/AG/RealizationComparisonIdempotents/MaximalSubgroupoid.lean\#L63}{\texttt{reversible\allowbreak{}Representation\allowbreak{}Hom\allowbreak{}Equiv}} & The morphisms of Core(Arrow(Karoubi E)) represent invertible changes of comparison. Invertibility of the compared morphisms themselves is not assumed. \\
\addlinespace[3pt]
Proposition~\ref{prop:6.31} & \href{https://github.com/iroha1203/AlgebraicArchitectureTheoryV2/blob/719f81f47d410701fd82c2bc88613cc140c59377/research/lean/ResearchLean/AG/RealizationReconstruction/ProtocolIdempotents.lean\#L68}{\texttt{fixed\allowbreak{}Point}} / \href{https://github.com/iroha1203/AlgebraicArchitectureTheoryV2/blob/719f81f47d410701fd82c2bc88613cc140c59377/research/lean/ResearchLean/AG/RealizationReconstruction/ProtocolIdempotents.lean\#L114}{\texttt{fixed\allowbreak{}Point\_\allowbreak{}split\_\allowbreak{}id}} / \href{https://github.com/iroha1203/AlgebraicArchitectureTheoryV2/blob/719f81f47d410701fd82c2bc88613cc140c59377/research/lean/ResearchLean/AG/RealizationReconstruction/ProtocolIdempotents.lean\#L125}{\texttt{fixed\allowbreak{}Point\_\allowbreak{}split\_\allowbreak{}e}} / \href{https://github.com/iroha1203/AlgebraicArchitectureTheoryV2/blob/719f81f47d410701fd82c2bc88613cc140c59377/research/lean/ResearchLean/AG/RealizationReconstruction/ProtocolIdempotents.lean\#L136}{\texttt{protocol\allowbreak{}Realization\_\allowbreak{}is\allowbreak{}Idempotent\allowbreak{}Complete}} & For an idempotent natural transformation of an arbitrary ProtocolRealization, constructs the splitting from the fixed points at the vertices, the action on edges, and the restriction of the observations. The category of realizations is thereby idempotent complete. \\
\bottomrule
\end{xltabular}

\subsection*{Splitting of idempotents for lenses and in the Karoubi category}

\begin{xltabular}{\linewidth}{@{}LLL@{}}
\toprule
Text & Lean declaration & Objects and conditions \\
\midrule
\endhead
\S\ref{sec:6.8} (lenses at a fixed view) & \href{https://github.com/iroha1203/AlgebraicArchitectureTheoryV2/blob/719f81f47d410701fd82c2bc88613cc140c59377/research/lean/ResearchLean/AG/RealizationReconstruction/LensFinitePresentation.lean\#L287}{\texttt{fixed\allowbreak{}Point\allowbreak{}Lens}} / \href{https://github.com/iroha1203/AlgebraicArchitectureTheoryV2/blob/719f81f47d410701fd82c2bc88613cc140c59377/research/lean/ResearchLean/AG/RealizationReconstruction/LensFinitePresentation.lean\#L316}{\texttt{fixed\allowbreak{}Point\_\allowbreak{}split\_\allowbreak{}id}} / \href{https://github.com/iroha1203/AlgebraicArchitectureTheoryV2/blob/719f81f47d410701fd82c2bc88613cc140c59377/research/lean/ResearchLean/AG/RealizationReconstruction/LensFinitePresentation.lean\#L326}{\texttt{fixed\allowbreak{}Point\_\allowbreak{}split\_\allowbreak{}e}} / \href{https://github.com/iroha1203/AlgebraicArchitectureTheoryV2/blob/719f81f47d410701fd82c2bc88613cc140c59377/research/lean/ResearchLean/AG/RealizationReconstruction/LensFinitePresentation.lean\#L335}{\texttt{lens\allowbreak{}Realization\_\allowbreak{}is\allowbreak{}Idempotent\allowbreak{}Complete}} & Splits an idempotent endomorphism of an arbitrary LensRealization by the fixed points of the states. The fixed points are closed under put, and preserve the three laws and the finiteness of the reference fiber. \\
\addlinespace[3pt]
First half of Proposition~\ref{prop:6.11} & \href{https://github.com/leanprover-community/mathlib4/blob/8f9d9cff6bd728b17a24e163c9402775d9e6a365/Mathlib/CategoryTheory/Idempotents/Karoubi.lean\#L210}{\texttt{inst\allowbreak{}Is\allowbreak{}Idempotent\allowbreak{}Complete\allowbreak{}Karoubi}} & For an arbitrary category E, splits an idempotent p on a Karoubi object P by the object (P.X,p.f) and by p.f as both splitting morphisms. Idempotent completeness of E is not assumed. \\
\bottomrule
\end{xltabular}

\section{Chapter 7: Comparison-Preserving Changes and Information}\label{sec:A.7}

\begin{xltabular}{\linewidth}{@{}LLL@{}}
\toprule
Text & Lean declaration & Objects and conditions \\
\midrule
\endhead
Definition~\ref{def:7.2} & \href{https://github.com/iroha1203/AlgebraicArchitectureTheoryV2/blob/719f81f47d410701fd82c2bc88613cc140c59377/research/lean/ResearchLean/AG/ComparisonInformationLoss/KaroubiRestriction.lean\#L179}{\texttt{comparison\allowbreak{}Automorphism\allowbreak{}Subgroup}} / \href{https://github.com/iroha1203/AlgebraicArchitectureTheoryV2/blob/719f81f47d410701fd82c2bc88613cc140c59377/research/lean/ResearchLean/AG/DoctrineFiberProduct/QualifiedComparisonStabilizer.lean\#L33}{\texttt{qualified\allowbreak{}Comparison\allowbreak{}Subgroup}} & The subgroup of commuting pairs of automorphisms for a comparison morphism in an arbitrary category. The restriction to a permitted subgroup is defined by the preimage along the inclusion of products. In AAT we use the version that fixes the base. \\
\addlinespace[3pt]
Proposition~\ref{prop:7.3} & \href{https://github.com/iroha1203/AlgebraicArchitectureTheoryV2/blob/7e68ec6e77ef0249ede6a4c875715cfaf1cb3ee6/research/lean/ResearchLean/AG/ComparisonInformation/F37GeneralComparison.lean\#L28}{\texttt{general\allowbreak{}Comparison\allowbreak{}Subgroup}} / \href{https://github.com/iroha1203/AlgebraicArchitectureTheoryV2/blob/7e68ec6e77ef0249ede6a4c875715cfaf1cb3ee6/research/lean/ResearchLean/AG/ComparisonInformation/F37GeneralComparison.lean\#L66}{\texttt{general\allowbreak{}Comparison\allowbreak{}Arrow\allowbreak{}Mul\allowbreak{}Equiv}} / \href{https://github.com/iroha1203/AlgebraicArchitectureTheoryV2/blob/7e68ec6e77ef0249ede6a4c875715cfaf1cb3ee6/research/lean/ResearchLean/AG/ComparisonInformation/F37GeneralComparison.lean\#L93}{\texttt{general\allowbreak{}Comparison\allowbreak{}Karoubi\allowbreak{}Arrow\allowbreak{}Mul\allowbreak{}Equiv}} / \href{https://github.com/iroha1203/AlgebraicArchitectureTheoryV2/blob/7e68ec6e77ef0249ede6a4c875715cfaf1cb3ee6/research/lean/ResearchLean/AG/ComparisonInformation/F37GeneralComparison.lean\#L148}{\texttt{arrow\allowbreak{}To\allowbreak{}Karoubi\allowbreak{}Arrow\allowbreak{}Aut\allowbreak{}Mul\allowbreak{}Equiv}} & Constructs comparison groups for arbitrary categories and endpoint subgroups. For full endpoint groups, identifies them with automorphisms in the arrow and Karoubi arrow categories, preserving both projections. \\
\addlinespace[3pt]
Theorem~\ref{thm:7.4} (projections and kernels) & \href{https://github.com/iroha1203/AlgebraicArchitectureTheoryV2/blob/7e68ec6e77ef0249ede6a4c875715cfaf1cb3ee6/research/lean/ResearchLean/AG/ComparisonInformation/F37GeneralComparison.lean\#L204}{\texttt{general\allowbreak{}Source\allowbreak{}Kernel\allowbreak{}Mul\allowbreak{}Equiv}} / \href{https://github.com/iroha1203/AlgebraicArchitectureTheoryV2/blob/7e68ec6e77ef0249ede6a4c875715cfaf1cb3ee6/research/lean/ResearchLean/AG/ComparisonInformation/F37GeneralComparison.lean\#L229}{\texttt{general\allowbreak{}Target\allowbreak{}Kernel\allowbreak{}Mul\allowbreak{}Equiv}} / \href{https://github.com/iroha1203/AlgebraicArchitectureTheoryV2/blob/7e68ec6e77ef0249ede6a4c875715cfaf1cb3ee6/research/lean/ResearchLean/AG/ComparisonInformation/F37GeneralComparison.lean\#L264}{\texttt{general\_\allowbreak{}source\_\allowbreak{}range\_\allowbreak{}iff\_\allowbreak{}lift}} / \href{https://github.com/iroha1203/AlgebraicArchitectureTheoryV2/blob/7e68ec6e77ef0249ede6a4c875715cfaf1cb3ee6/research/lean/ResearchLean/AG/ComparisonInformation/F37GeneralComparison.lean\#L275}{\texttt{general\_\allowbreak{}target\_\allowbreak{}range\_\allowbreak{}iff\_\allowbreak{}lift}} & For arbitrary endpoint subgroups, identifies each projection kernel with the opposite stabilizer, and membership in the image with the existence of a following change. \\
\addlinespace[3pt]
Theorem~\ref{thm:7.4} (following changes) & \href{https://github.com/iroha1203/AlgebraicArchitectureTheoryV2/blob/7e68ec6e77ef0249ede6a4c875715cfaf1cb3ee6/research/lean/ResearchLean/AG/ComparisonInformation/F37GeneralComparison.lean\#L315}{\texttt{target\allowbreak{}Lift\_\allowbreak{}exists\allowbreak{}Unique}} / \href{https://github.com/iroha1203/AlgebraicArchitectureTheoryV2/blob/7e68ec6e77ef0249ede6a4c875715cfaf1cb3ee6/research/lean/ResearchLean/AG/ComparisonInformation/F37GeneralComparison.lean\#L349}{\texttt{target\allowbreak{}Lift\allowbreak{}Equiv}} / \href{https://github.com/iroha1203/AlgebraicArchitectureTheoryV2/blob/7e68ec6e77ef0249ede6a4c875715cfaf1cb3ee6/research/lean/ResearchLean/AG/ComparisonInformation/F37GeneralComparison.lean\#L391}{\texttt{source\allowbreak{}Lift\_\allowbreak{}exists\allowbreak{}Unique}} / \href{https://github.com/iroha1203/AlgebraicArchitectureTheoryV2/blob/7e68ec6e77ef0249ede6a4c875715cfaf1cb3ee6/research/lean/ResearchLean/AG/ComparisonInformation/F37GeneralComparison.lean\#L421}{\texttt{source\allowbreak{}Lift\allowbreak{}Equiv}} & Each nonempty fiber on either side is a left torsor under the stabilizer and, after choosing one reference candidate, is uniquely parametrized by group elements. \\
\addlinespace[3pt]
Corollary~\ref{cor:7.5} & \href{https://github.com/iroha1203/AlgebraicArchitectureTheoryV2/blob/7e68ec6e77ef0249ede6a4c875715cfaf1cb3ee6/research/lean/ResearchLean/AG/ComparisonInformation/F37GeneralComparison.lean\#L459}{\texttt{general\allowbreak{}Iso\allowbreak{}Graph\allowbreak{}Mul\allowbreak{}Equiv}} / \href{https://github.com/iroha1203/AlgebraicArchitectureTheoryV2/blob/7e68ec6e77ef0249ede6a4c875715cfaf1cb3ee6/research/lean/ResearchLean/AG/ComparisonInformation/F37GeneralComparison.lean\#L505}{\texttt{general\allowbreak{}Restricted\allowbreak{}Conjugation}} / \href{https://github.com/iroha1203/AlgebraicArchitectureTheoryV2/blob/7e68ec6e77ef0249ede6a4c875715cfaf1cb3ee6/research/lean/ResearchLean/AG/ComparisonInformation/F37GeneralComparison.lean\#L520}{\texttt{general\allowbreak{}Iso\allowbreak{}Target\allowbreak{}Projection\allowbreak{}Mul\allowbreak{}Equiv}} & When conjugation by an invertible comparison matches the endpoint subgroups, the comparison group is its graph and is isomorphic to both endpoint groups. \\
\addlinespace[3pt]
Theorem~\ref{thm:7.6} & \href{https://github.com/iroha1203/AlgebraicArchitectureTheoryV2/blob/719f81f47d410701fd82c2bc88613cc140c59377/research/lean/ResearchLean/AG/DoctrineFiberProduct/QualifiedComparisonGeneratedClassification.lean\#L152}{\texttt{generated\allowbreak{}Qualified\allowbreak{}Comparison\allowbreak{}Relation\_\allowbreak{}diagonal}} / \href{https://github.com/iroha1203/AlgebraicArchitectureTheoryV2/blob/719f81f47d410701fd82c2bc88613cc140c59377/research/lean/ResearchLean/AG/DoctrineFiberProduct/QualifiedComparisonGeneratedClassification.lean\#L177}{\texttt{generated\allowbreak{}Qualified\allowbreak{}Comparison\allowbreak{}Relation\_\allowbreak{}iff\_\allowbreak{}difference\_\allowbreak{}mem}} / \href{https://github.com/iroha1203/AlgebraicArchitectureTheoryV2/blob/719f81f47d410701fd82c2bc88613cc140c59377/research/lean/ResearchLean/AG/DoctrineFiberProduct/QualifiedComparisonGeneratedClassification.lean\#L271}{\texttt{generated\allowbreak{}Qualified\allowbreak{}Comparison\allowbreak{}Relation\_\allowbreak{}iff\_\allowbreak{}exists\_\allowbreak{}kernel\_\allowbreak{}factor}} & Proves, over a fixed compatible generated input, the equivalence between the intertwining of the images along both routes of the same source change and the membership of the difference of the two changes in the pulled stabilizer kernel. \\
\addlinespace[3pt]
Proposition~\ref{prop:7.7} & \href{https://github.com/iroha1203/AlgebraicArchitectureTheoryV2/blob/719f81f47d410701fd82c2bc88613cc140c59377/research/lean/ResearchLean/AG/DoctrineFiberProduct/UpperGeometryCompatibleSourcePresentationNaturalityF0.lean\#L26}{\texttt{Upper\allowbreak{}Geometry\allowbreak{}Compatible\allowbreak{}Source\allowbreak{}Presentation\allowbreak{}Change}} / \href{https://github.com/iroha1203/AlgebraicArchitectureTheoryV2/blob/719f81f47d410701fd82c2bc88613cc140c59377/research/lean/ResearchLean/AG/DoctrineFiberProduct/UpperGeometryCompatibleSourcePresentationNaturalityF3.lean\#L32}{\texttt{generated\allowbreak{}Compatible\allowbreak{}Upper\allowbreak{}Geometry\allowbreak{}Mate\allowbreak{}At\_\allowbreak{}eq\_\allowbreak{}source\allowbreak{}Presentation\_\allowbreak{}conjugation}} / \href{https://github.com/iroha1203/AlgebraicArchitectureTheoryV2/blob/719f81f47d410701fd82c2bc88613cc140c59377/research/lean/ResearchLean/AG/DoctrineFiberProduct/UpperGeometryCompatibleSourcePresentationNaturalityF3.lean\#L94}{\texttt{generated\allowbreak{}Qualified\allowbreak{}Comparison\allowbreak{}Source\allowbreak{}Presentation\allowbreak{}Mul\allowbreak{}Equiv\allowbreak{}At}} & For an isomorphism that preserves the structure of the source presentation, reconstructs the input and obtains the conjugation of the comparison and the isomorphism of comparison-preserving groups. \\
\addlinespace[3pt]
Definition~\ref{def:7.8} & \href{https://github.com/iroha1203/AlgebraicArchitectureTheoryV2/blob/719f81f47d410701fd82c2bc88613cc140c59377/research/lean/ResearchLean/AG/ComparisonInformationLoss/ObservationKernel.lean\#L33}{\texttt{compatible\allowbreak{}Kernel}} & Consists of a group homomorphism $O:Q\to R$ (where $\to^*$ is the Lean notation for group homomorphisms), the comparison subgroup $\Gamma$, and the comap of $\Gamma$ inside $\ker O$. \\
\addlinespace[3pt]
Theorem~\ref{thm:7.9} & \href{https://github.com/iroha1203/AlgebraicArchitectureTheoryV2/blob/719f81f47d410701fd82c2bc88613cc140c59377/research/lean/ResearchLean/AG/ComparisonInformationLoss/ObservationKernel.lean\#L48}{\texttt{exists\_\allowbreak{}observation\_\allowbreak{}predicate\_\allowbreak{}iff\_\allowbreak{}ker\_\allowbreak{}le}} / \href{https://github.com/iroha1203/AlgebraicArchitectureTheoryV2/blob/719f81f47d410701fd82c2bc88613cc140c59377/research/lean/ResearchLean/AG/ComparisonInformationLoss/ObservationKernel.lean\#L89}{\texttt{preimage\_\allowbreak{}image\_\allowbreak{}eq\_\allowbreak{}mul\_\allowbreak{}ker}} & For an arbitrary group homomorphism O and subgroup $\Gamma$, the existence of an observation predicate is equivalent to $\ker O\le\Gamma$, and the saturation is $O^{-1}(O\Gamma)=\Gamma\ker O$. Neither surjectivity of O nor normality of $\Gamma$ is needed. \\
\addlinespace[3pt]
Proposition~\ref{prop:7.10} & \href{https://github.com/iroha1203/AlgebraicArchitectureTheoryV2/blob/719f81f47d410701fd82c2bc88613cc140c59377/research/lean/ResearchLean/AG/ComparisonInformationLoss/ObservationKernel.lean\#L96}{\texttt{observation\_\allowbreak{}fiber\_\allowbreak{}eq\_\allowbreak{}left\allowbreak{}Coset}} / \href{https://github.com/iroha1203/AlgebraicArchitectureTheoryV2/blob/719f81f47d410701fd82c2bc88613cc140c59377/research/lean/ResearchLean/AG/ComparisonInformationLoss/ObservationKernel.lean\#L113}{\texttt{observation\_\allowbreak{}fiber\_\allowbreak{}inter\_\allowbreak{}eq\_\allowbreak{}left\allowbreak{}Coset}} / \href{https://github.com/iroha1203/AlgebraicArchitectureTheoryV2/blob/719f81f47d410701fd82c2bc88613cc140c59377/research/lean/ResearchLean/AG/ComparisonInformationLoss/ObservationKernel.lean\#L143}{\texttt{subsingleton\_\allowbreak{}kernel\_\allowbreak{}quotient\_\allowbreak{}iff\_\allowbreak{}ker\_\allowbreak{}le}} & Expresses the observation fiber and the fiber inside $\Gamma$ as left cosets, and proves the condition for a general (possibly non-normal) coset inside ker to be a subsingleton. \\
\addlinespace[3pt]
Proposition~\ref{prop:7.12} & \href{https://github.com/iroha1203/AlgebraicArchitectureTheoryV2/blob/719f81f47d410701fd82c2bc88613cc140c59377/research/lean/ResearchLean/AG/RealizationComparisonIdempotents/NormalizationComparisonGroup.lean\#L34}{\texttt{functor\allowbreak{}Automorphism\allowbreak{}Hom}} / \href{https://github.com/leanprover-community/mathlib4/blob/8f9d9cff6bd728b17a24e163c9402775d9e6a365/Mathlib/CategoryTheory/Comma/Arrow.lean\#L333}{\texttt{map\allowbreak{}Arrow}} / \href{https://github.com/iroha1203/AlgebraicArchitectureTheoryV2/blob/719f81f47d410701fd82c2bc88613cc140c59377/research/lean/ResearchLean/AG/RealizationComparisonIdempotents/NormalizationComparisonGroup.lean\#L155}{\texttt{normalization\allowbreak{}Endpoint\allowbreak{}Automorphism\_\allowbreak{}preserves\_\allowbreak{}comparison}} & An arbitrary functor maps pairs of automorphisms and commutative squares, and preserves the commutativity of the comparison. It also applies to the core normalization. \\
\addlinespace[3pt]
Construction~\ref{cons:7.13} & \href{https://github.com/iroha1203/AlgebraicArchitectureTheoryV2/blob/719f81f47d410701fd82c2bc88613cc140c59377/research/lean/ResearchLean/AG/ComparisonInformationLoss/KaroubiRestriction.lean\#L129}{\texttt{idempotent\allowbreak{}Restriction\allowbreak{}Hom}} / \href{https://github.com/iroha1203/AlgebraicArchitectureTheoryV2/blob/719f81f47d410701fd82c2bc88613cc140c59377/research/lean/ResearchLean/AG/ComparisonInformationLoss/KaroubiRestriction.lean\#L222}{\texttt{idempotent\allowbreak{}Image\allowbreak{}Comparison}} / \href{https://github.com/iroha1203/AlgebraicArchitectureTheoryV2/blob/719f81f47d410701fd82c2bc88613cc140c59377/research/lean/ResearchLean/AG/ComparisonInformationLoss/KaroubiRestriction.lean\#L333}{\texttt{idempotent\allowbreak{}Endpoint\allowbreak{}Restriction\_\allowbreak{}preserves\_\allowbreak{}comparison}} & From arbitrary E and c, idempotents e and d, and $e\gg c=c\gg d$, constructs the Karoubi restriction group hom of the centralizer Aut and the image comparison, and proves preservation. \\
\addlinespace[3pt]
Theorem~\ref{thm:7.14} & \href{https://github.com/iroha1203/AlgebraicArchitectureTheoryV2/blob/719f81f47d410701fd82c2bc88613cc140c59377/research/lean/ResearchLean/AG/ComparisonInformationLoss/GroupHomRestriction.lean\#L46}{\texttt{comap\_\allowbreak{}eq\_\allowbreak{}iff\_\allowbreak{}ker\_\allowbreak{}le\_\allowbreak{}and\_\allowbreak{}map\_\allowbreak{}eq\_\allowbreak{}inf\_\allowbreak{}range}} / \href{https://github.com/iroha1203/AlgebraicArchitectureTheoryV2/blob/719f81f47d410701fd82c2bc88613cc140c59377/research/lean/ResearchLean/AG/ComparisonInformationLoss/GroupHomRestriction.lean\#L173}{\texttt{restricted\allowbreak{}Fiber\_\allowbreak{}exists\allowbreak{}Unique\_\allowbreak{}smul\_\allowbreak{}eq}} / \href{https://github.com/iroha1203/AlgebraicArchitectureTheoryV2/blob/719f81f47d410701fd82c2bc88613cc140c59377/research/lean/ResearchLean/AG/ComparisonInformationLoss/GroupHomRestriction.lean\#L224}{\texttt{restricted\allowbreak{}Subgroup\allowbreak{}Hom\_\allowbreak{}short\allowbreak{}Exact\_\allowbreak{}iff\_\allowbreak{}map\_\allowbreak{}eq}} & From an arbitrary group hom, source and target subgroups, and the image inclusion, proves that reflection is equivalent to the kernel inclusion together with the image/range condition, that the action on a nonempty fiber is free and transitive, and that a short exact sequence is equivalent to surjectivity. \\
\addlinespace[3pt]
Example~\ref{ex:7.16} & \href{https://github.com/iroha1203/AlgebraicArchitectureTheoryV2/blob/719f81f47d410701fd82c2bc88613cc140c59377/research/lean/ResearchLean/AG/ComparisonInformationLoss/KaroubiRestrictionFiniteWitness.lean\#L220}{\texttt{unequal\allowbreak{}Fiber\allowbreak{}Image\allowbreak{}Swap\allowbreak{}Pair\_\allowbreak{}not\_\allowbreak{}mem\_\allowbreak{}restriction\allowbreak{}Range}} / \href{https://github.com/iroha1203/AlgebraicArchitectureTheoryV2/blob/719f81f47d410701fd82c2bc88613cc140c59377/research/lean/ResearchLean/AG/ComparisonInformationLoss/KaroubiRestrictionFiniteWitness.lean\#L281}{\texttt{unequal\allowbreak{}Fiber\allowbreak{}Image\allowbreak{}Swap\allowbreak{}Pair\_\allowbreak{}has\_\allowbreak{}no\_\allowbreak{}compatible\allowbreak{}Lift}} & For the idempotent map $0\mapsto0$, $1\mapsto1$, $2\mapsto1$ on Fin 3, the permutation that exchanges 0 and 1 in the image does not lift to the centralizer. \\
\addlinespace[3pt]
Construction~\ref{cons:7.17} & \href{https://github.com/iroha1203/AlgebraicArchitectureTheoryV2/blob/719f81f47d410701fd82c2bc88613cc140c59377/research/lean/ResearchLean/AG/FullGeometryNormalization/CanonicalNormalizationObjectSection.lean\#L43}{\texttt{architecture\allowbreak{}Object\allowbreak{}Presentation}} / \href{https://github.com/iroha1203/AlgebraicArchitectureTheoryV2/blob/719f81f47d410701fd82c2bc88613cc140c59377/research/lean/ResearchLean/AG/FullGeometryNormalization/CanonicalNormalizationObjectSection.lean\#L112}{\texttt{canonical\allowbreak{}Normalization\allowbreak{}Section\allowbreak{}Object\allowbreak{}Map}} / \href{https://github.com/iroha1203/AlgebraicArchitectureTheoryV2/blob/719f81f47d410701fd82c2bc88613cc140c59377/research/lean/ResearchLean/AG/FullGeometryNormalization/CanonicalNormalizationObjectSection.lean\#L215}{\texttt{canonical\allowbreak{}Normalization\allowbreak{}Section\allowbreak{}Object\allowbreak{}Map\_\allowbreak{}trans}} & Splits an architecture object into a configuration and a common auxiliary type, and constructs the trivialization that swaps the selected datum with a fixed base datum. Extends the action of Atom isomorphisms across both trivializations, and proves the preservation of selected objects and of identities and composition. \\
\addlinespace[3pt]
Lemma~\ref{lem:7.18} & \href{https://github.com/iroha1203/AlgebraicArchitectureTheoryV2/blob/719f81f47d410701fd82c2bc88613cc140c59377/research/lean/ResearchLean/AG/FullGeometryNormalization/CanonicalNormalizationAutomorphismSection.lean\#L105}{\texttt{canonical\allowbreak{}Normalization\allowbreak{}Automorphism\allowbreak{}Section\allowbreak{}Hom}} / \href{https://github.com/iroha1203/AlgebraicArchitectureTheoryV2/blob/719f81f47d410701fd82c2bc88613cc140c59377/research/lean/ResearchLean/AG/FullGeometryNormalization/CanonicalNormalizationAutomorphismSection.lean\#L127}{\texttt{canonical\allowbreak{}Normalization\allowbreak{}Automorphism\allowbreak{}Section\_\allowbreak{}right\allowbreak{}Inverse}} / \href{https://github.com/iroha1203/AlgebraicArchitectureTheoryV2/blob/719f81f47d410701fd82c2bc88613cc140c59377/research/lean/ResearchLean/AG/FullGeometryNormalization/CanonicalNormalizationGeometrySection.lean\#L87}{\texttt{canonical\allowbreak{}Normalization\allowbreak{}Geometry\allowbreak{}Section}} & Under Ad(G.core), constructs a section from the automorphism group after normalization into Aut G. It preserves the base and the coefficients. \\
\addlinespace[3pt]
Theorem~\ref{thm:7.19} & \href{https://github.com/iroha1203/AlgebraicArchitectureTheoryV2/blob/719f81f47d410701fd82c2bc88613cc140c59377/research/lean/ResearchLean/AG/FullGeometryNormalization/CanonicalNormalizationIsoComparisonSection.lean\#L54}{\texttt{canonical\allowbreak{}Normalization\allowbreak{}Iso\allowbreak{}Comparison\allowbreak{}Section\allowbreak{}Hom}} / \href{https://github.com/iroha1203/AlgebraicArchitectureTheoryV2/blob/719f81f47d410701fd82c2bc88613cc140c59377/research/lean/ResearchLean/AG/FullGeometryNormalization/CanonicalNormalizationIsoComparisonSection.lean\#L128}{\texttt{canonical\allowbreak{}Normalization\allowbreak{}Iso\allowbreak{}Comparison\allowbreak{}Section\_\allowbreak{}right\allowbreak{}Inverse}} & For an isomorphism c of an arbitrary admissible geometry, lifts normalized comparison-preserving pairs to raw comparison-preserving pairs using the source section and conjugation by c. Also proves that it is a right inverse on all pairs and that it preserves the base and the coefficients. \\
\addlinespace[3pt]
Construction~\ref{cons:7.20} & \href{https://github.com/iroha1203/AlgebraicArchitectureTheoryV2/blob/719f81f47d410701fd82c2bc88613cc140c59377/research/lean/ResearchLean/AG/FullGeometryNormalization/AmbientKernelObjectSwap.lean\#L183}{\texttt{ambient\allowbreak{}Kernel\allowbreak{}Object\allowbreak{}Map\_\allowbreak{}ne\_\allowbreak{}id}} / \href{https://github.com/iroha1203/AlgebraicArchitectureTheoryV2/blob/719f81f47d410701fd82c2bc88613cc140c59377/research/lean/ResearchLean/AG/FullGeometryNormalization/AmbientKernelGeometryLift.lean\#L151}{\texttt{ambient\allowbreak{}Kernel\allowbreak{}Geometry\allowbreak{}Aut\_\allowbreak{}ne\_\allowbreak{}one}} / \href{https://github.com/iroha1203/AlgebraicArchitectureTheoryV2/blob/719f81f47d410701fd82c2bc88613cc140c59377/research/lean/ResearchLean/AG/FullGeometryNormalization/AmbientKernelGeometryLift.lean\#L185}{\texttt{geometry\allowbreak{}Normalization\allowbreak{}Functor\_\allowbreak{}map\_\allowbreak{}ambient\allowbreak{}Kernel\allowbreak{}Admissible\allowbreak{}Geometry\allowbreak{}Aut}} & Constructs the non-identity involution that swaps the two decorations other than the selected decoration at each configuration. Extends it from Ad to operation/Law/geometry, and proves that it becomes the identity under normalization. \\
\addlinespace[3pt]
Theorem~\ref{thm:7.21} & \href{https://github.com/iroha1203/AlgebraicArchitectureTheoryV2/blob/7e68ec6e77ef0249ede6a4c875715cfaf1cb3ee6/research/lean/ResearchLean/AG/FullGeometryNormalization/SemanticDerivedCanonicalComparisonExactness.lean\#L144}{\texttt{semantic\allowbreak{}Derived\allowbreak{}Canonical\allowbreak{}Comparison\allowbreak{}Section\allowbreak{}Hom}} / \href{https://github.com/iroha1203/AlgebraicArchitectureTheoryV2/blob/7e68ec6e77ef0249ede6a4c875715cfaf1cb3ee6/research/lean/ResearchLean/AG/FullGeometryNormalization/SemanticDerivedCanonicalComparisonExactness.lean\#L204}{\texttt{semantic\allowbreak{}Derived\allowbreak{}Canonical\allowbreak{}Comparison\_\allowbreak{}short\allowbreak{}Exact}} / \href{https://github.com/iroha1203/AlgebraicArchitectureTheoryV2/blob/7e68ec6e77ef0249ede6a4c875715cfaf1cb3ee6/research/lean/ResearchLean/AG/FullGeometryNormalization/SemanticDerivedCanonicalComparisonExactness.lean\#L390}{\texttt{semantic\allowbreak{}Derived\allowbreak{}Canonical\allowbreak{}Comparison\allowbreak{}Lift\allowbreak{}Fiber\_\allowbreak{}exists\allowbreak{}Unique\_\allowbreak{}smul\_\allowbreak{}eq}} / \href{https://github.com/iroha1203/AlgebraicArchitectureTheoryV2/blob/7e68ec6e77ef0249ede6a4c875715cfaf1cb3ee6/research/lean/ResearchLean/AG/FullGeometryNormalization/SemanticDerivedCanonicalComparisonIdentification.lean\#L26}{\texttt{semantic\allowbreak{}Derived\allowbreak{}Canonical\allowbreak{}Comparison\_\allowbreak{}identification}} / \href{https://github.com/iroha1203/AlgebraicArchitectureTheoryV2/blob/7e68ec6e77ef0249ede6a4c875715cfaf1cb3ee6/research/lean/ResearchLean/AG/FullGeometryNormalization/SemanticDerivedCanonicalComparisonIdentification.lean\#L103}{\texttt{semantic\allowbreak{}Derived\allowbreak{}Canonical\allowbreak{}Bottom\allowbreak{}Comparison\_\allowbreak{}identification}} & For a comparison generated from a general semantic square and admissibility of the source core, gives a section, split short exact sequence, right torsors of fibers, and non-identifiability for both full and base-fixing endpoint groups. \\
\addlinespace[3pt]
Corollary~\ref{cor:7.22} & \href{https://github.com/iroha1203/AlgebraicArchitectureTheoryV2/blob/7e68ec6e77ef0249ede6a4c875715cfaf1cb3ee6/research/lean/ResearchLean/AG/FullGeometryNormalization/SemanticDerivedSelectorGlobalSection.lean\#L845}{\texttt{semantic\allowbreak{}Exact\allowbreak{}Global\allowbreak{}Comparison\allowbreak{}Section\allowbreak{}Hom}} / \href{https://github.com/iroha1203/AlgebraicArchitectureTheoryV2/blob/7e68ec6e77ef0249ede6a4c875715cfaf1cb3ee6/research/lean/ResearchLean/AG/FullGeometryNormalization/SemanticDerivedSelectorGlobalSection.lean\#L859}{\texttt{semantic\allowbreak{}Exact\allowbreak{}Global\allowbreak{}Comparison\allowbreak{}Section\_\allowbreak{}right\allowbreak{}Inverse}} / \href{https://github.com/iroha1203/AlgebraicArchitectureTheoryV2/blob/7e68ec6e77ef0249ede6a4c875715cfaf1cb3ee6/research/lean/ResearchLean/AG/FullGeometryNormalization/SemanticDerivedSelectorGlobalComparison.lean\#L507}{\texttt{semantic\allowbreak{}Exact\allowbreak{}Global\allowbreak{}Restriction\_\allowbreak{}preimage\_\allowbreak{}eq\_\allowbreak{}raw\_\allowbreak{}iff\_\allowbreak{}not\_\allowbreak{}selected}} / \href{https://github.com/iroha1203/AlgebraicArchitectureTheoryV2/blob/7e68ec6e77ef0249ede6a4c875715cfaf1cb3ee6/research/lean/ResearchLean/AG/FullGeometryNormalization/SemanticDerivedSelectorGlobalBottomReflection.lean\#L141}{\texttt{semantic\allowbreak{}Exact\allowbreak{}Global\allowbreak{}Bottom\allowbreak{}Restriction\_\allowbreak{}preimage\_\allowbreak{}eq\_\allowbreak{}raw\_\allowbreak{}iff\_\allowbreak{}not\_\allowbreak{}selected}} & For full endpoint groups commuting with both projectors and their base-fixing subgroups, constructs a section of restriction to the image comparison. Compatibility is reflected exactly when normalization is not selected. \\
\addlinespace[3pt]
Definition~\ref{def:7.23} & \href{https://github.com/iroha1203/AlgebraicArchitectureTheoryV2/blob/719f81f47d410701fd82c2bc88613cc140c59377/research/lean/ResearchLean/AG/RealizationReconstruction/FixedFAllAutomorphismGroup.lean\#L238}{\texttt{Fixed\allowbreak{}FPreserving\allowbreak{}Following\allowbreak{}Pair}} / \href{https://github.com/iroha1203/AlgebraicArchitectureTheoryV2/blob/719f81f47d410701fd82c2bc88613cc140c59377/research/lean/ResearchLean/AG/RealizationReconstruction/FixedFComponentClassification.lean\#L29}{\texttt{fixed\allowbreak{}FDirected\allowbreak{}Edge\allowbreak{}Step}} & For a fixed multigraph, hidden type K, and visible graph automorphism, defines the equivalence of states that preserves the observations and the operations. \\
\addlinespace[3pt]
Theorem~\ref{thm:7.24} & \href{https://github.com/iroha1203/AlgebraicArchitectureTheoryV2/blob/719f81f47d410701fd82c2bc88613cc140c59377/research/lean/ResearchLean/AG/RealizationReconstruction/FixedFComponentClassification.lean\#L145}{\texttt{preserving\allowbreak{}Equiv\allowbreak{}Component\allowbreak{}Permutation\allowbreak{}Families}} & For an arbitrary graph, hidden type, and visible automorphism, the operation-preserving following changes correspond bijectively to the families of hidden permutations, one for each undirected connected component. Finiteness is not assumed. \\
\addlinespace[3pt]
Example~\ref{ex:7.28} & \href{https://github.com/iroha1203/AlgebraicArchitectureTheoryV2/blob/719f81f47d410701fd82c2bc88613cc140c59377/research/lean/ResearchLean/AG/RealizationReconstruction/FixedFFiniteExamples.lean\#L278}{\texttt{pointed\allowbreak{}Lens\allowbreak{}Twist\_\allowbreak{}preserves\_\allowbreak{}get}} / \href{https://github.com/iroha1203/AlgebraicArchitectureTheoryV2/blob/719f81f47d410701fd82c2bc88613cc140c59377/research/lean/ResearchLean/AG/RealizationReconstruction/FixedFFiniteExamples.lean\#L283}{\texttt{pointed\allowbreak{}Lens\allowbreak{}Twist\_\allowbreak{}preserves\_\allowbreak{}section}} / \href{https://github.com/iroha1203/AlgebraicArchitectureTheoryV2/blob/719f81f47d410701fd82c2bc88613cc140c59377/research/lean/ResearchLean/AG/RealizationReconstruction/FixedFFiniteExamples.lean\#L306}{\texttt{pointed\allowbreak{}Lens\allowbreak{}Twist\_\allowbreak{}does\_\allowbreak{}not\_\allowbreak{}preserve\_\allowbreak{}put}} / \href{https://github.com/iroha1203/AlgebraicArchitectureTheoryV2/blob/719f81f47d410701fd82c2bc88613cc140c59377/research/lean/ResearchLean/AG/RealizationReconstruction/FixedFFiniteExamples.lean\#L348}{\texttt{pointed\allowbreak{}Lens\_\allowbreak{}get\_\allowbreak{}section\_\allowbreak{}count}} / \href{https://github.com/iroha1203/AlgebraicArchitectureTheoryV2/blob/719f81f47d410701fd82c2bc88613cc140c59377/research/lean/ResearchLean/AG/RealizationReconstruction/FixedFFiniteExamples.lean\#L356}{\texttt{pointed\allowbreak{}Lens\_\allowbreak{}get\_\allowbreak{}put\_\allowbreak{}section\_\allowbreak{}count}} & An example with Bool view $\times$ Fin 3 hidden, with the visible change fixed to be the identity. It fixes the selected value 0 and exchanges 1 and 2 only on the true side. There are 4 changes that preserve get and the section, and 2 changes that preserve put as well. \\
\addlinespace[3pt]
Proposition~\ref{prop:7.29} & \href{https://github.com/iroha1203/AlgebraicArchitectureTheoryV2/blob/719f81f47d410701fd82c2bc88613cc140c59377/research/lean/ResearchLean/AG/RealizationReconstruction/FixedFProtocolConnection.lean\#L350}{\texttt{equiv\allowbreak{}Preserving\allowbreak{}Following\allowbreak{}Changes}} / \href{https://github.com/iroha1203/AlgebraicArchitectureTheoryV2/blob/719f81f47d410701fd82c2bc88613cc140c59377/research/lean/ResearchLean/AG/RealizationReconstruction/FixedFProtocolGroupConnection.lean\#L245}{\texttt{mul\allowbreak{}Equiv\allowbreak{}Following\allowbreak{}Group}} / \href{https://github.com/iroha1203/AlgebraicArchitectureTheoryV2/blob/719f81f47d410701fd82c2bc88613cc140c59377/research/lean/ResearchLean/AG/RealizationReconstruction/FixedFProtocolConnection.lean\#L524}{\texttt{invertible\allowbreak{}Adapter\allowbreak{}Square\_\allowbreak{}iff\_\allowbreak{}total\allowbreak{}Maps}} & For the identity hidden execution protocol with a finite graph and hidden K, moves in both directions between the original typed-edge squares and observation squares on one side and the following changes on the other, and preserves the group product as well. The adapter square is also equivalent to the original componentwise equations. \\
\bottomrule
\end{xltabular}

\subsection*{Presentation changes, comparison groups, and finite examples}

\begin{xltabular}{\linewidth}{@{}LLL@{}}
\toprule
Text & Lean declaration & Objects and conditions \\
\midrule
\endhead
\S\ref{sec:7.3} (presentation changes) & \href{https://github.com/iroha1203/AlgebraicArchitectureTheoryV2/blob/719f81f47d410701fd82c2bc88613cc140c59377/research/lean/ResearchLean/AG/ComparisonInformationLoss/PresentationTransport.lean\#L89}{\texttt{observation\allowbreak{}Equiv\allowbreak{}At}} / \href{https://github.com/iroha1203/AlgebraicArchitectureTheoryV2/blob/719f81f47d410701fd82c2bc88613cc140c59377/research/lean/ResearchLean/AG/ComparisonInformationLoss/PresentationTransport.lean\#L108}{\texttt{kernel\allowbreak{}Equiv\allowbreak{}At}} / \href{https://github.com/iroha1203/AlgebraicArchitectureTheoryV2/blob/719f81f47d410701fd82c2bc88613cc140c59377/research/lean/ResearchLean/AG/ComparisonInformationLoss/PresentationTransport.lean\#L116}{\texttt{compatible\allowbreak{}Kernel\allowbreak{}Equiv\allowbreak{}At}} / \href{https://github.com/iroha1203/AlgebraicArchitectureTheoryV2/blob/719f81f47d410701fd82c2bc88613cc140c59377/research/lean/ResearchLean/AG/ComparisonInformationLoss/PresentationTransport.lean\#L125}{\texttt{quotient\allowbreak{}Equiv\allowbreak{}At}} / \href{https://github.com/iroha1203/AlgebraicArchitectureTheoryV2/blob/719f81f47d410701fd82c2bc88613cc140c59377/research/lean/ResearchLean/AG/ComparisonInformationLoss/PresentationTransport.lean\#L140}{\texttt{quotient\allowbreak{}Equiv\allowbreak{}At\_\allowbreak{}basepoint}} / \href{https://github.com/iroha1203/AlgebraicArchitectureTheoryV2/blob/719f81f47d410701fd82c2bc88613cc140c59377/research/lean/ResearchLean/AG/ComparisonInformationLoss/PresentationTransport.lean\#L346}{\texttt{observation\allowbreak{}Loss\allowbreak{}Equiv\allowbreak{}Target\allowbreak{}Kernel\_\allowbreak{}naturality}} / \href{https://github.com/iroha1203/AlgebraicArchitectureTheoryV2/blob/719f81f47d410701fd82c2bc88613cc140c59377/research/lean/ResearchLean/AG/ComparisonInformationLoss/ObservationKernel.lean\#L163}{\texttt{subsingleton\_\allowbreak{}kernel\_\allowbreak{}quotient\_\allowbreak{}iff\_\allowbreak{}exists\_\allowbreak{}observation\_\allowbreak{}predicate}} / \href{https://github.com/leanprover-community/mathlib4/blob/8f9d9cff6bd728b17a24e163c9402775d9e6a365/Mathlib/Logic/Equiv/Defs.lean\#L190}{\texttt{subsingleton\_\allowbreak{}congr}} & The presentation change of Proposition~\ref{prop:7.7} commutes with the observation homomorphism, and matches the kernel K, the compatible kernel L, and the pointed cosets. It also preserves the existence of a predicate that determines compatibility from the observation. \\
\addlinespace[3pt]
Example~\ref{ex:7.15} (failure of reflection) & \href{https://github.com/iroha1203/AlgebraicArchitectureTheoryV2/blob/719f81f47d410701fd82c2bc88613cc140c59377/research/lean/ResearchLean/AG/ComparisonInformationLoss/KaroubiRestrictionFiniteWitness.lean\#L39}{\texttt{constant\allowbreak{}Zero\_\allowbreak{}idempotent}} / \href{https://github.com/iroha1203/AlgebraicArchitectureTheoryV2/blob/719f81f47d410701fd82c2bc88613cc140c59377/research/lean/ResearchLean/AG/ComparisonInformationLoss/KaroubiRestrictionFiniteWitness.lean\#L49}{\texttt{swap\allowbreak{}Twelve\_\allowbreak{}centralizes\_\allowbreak{}constant\allowbreak{}Zero}} / \href{https://github.com/iroha1203/AlgebraicArchitectureTheoryV2/blob/719f81f47d410701fd82c2bc88613cc140c59377/research/lean/ResearchLean/AG/ComparisonInformationLoss/KaroubiRestrictionFiniteWitness.lean\#L68}{\texttt{constant\allowbreak{}Zero\allowbreak{}Centralizing\allowbreak{}Pair\_\allowbreak{}raw\_\allowbreak{}mismatch}} / \href{https://github.com/iroha1203/AlgebraicArchitectureTheoryV2/blob/719f81f47d410701fd82c2bc88613cc140c59377/research/lean/ResearchLean/AG/ComparisonInformationLoss/KaroubiRestrictionFiniteWitness.lean\#L85}{\texttt{constant\allowbreak{}Zero\_\allowbreak{}restriction\_\allowbreak{}mem\_\allowbreak{}Gamma\allowbreak{}Image}} / \href{https://github.com/iroha1203/AlgebraicArchitectureTheoryV2/blob/719f81f47d410701fd82c2bc88613cc140c59377/research/lean/ResearchLean/AG/ComparisonInformationLoss/KaroubiRestrictionFiniteWitness.lean\#L114}{\texttt{constant\allowbreak{}Zero\_\allowbreak{}reflection\_\allowbreak{}fails}} & Uses the constant map at 0 on Fin 3 and the exchange of 1 and 2. The original comparison square is not commutative, but it becomes commutative after restriction. \\
\addlinespace[3pt]
Theorem~\ref{thm:7.25} and \eqref{eq:7.40} & \href{https://github.com/iroha1203/AlgebraicArchitectureTheoryV2/blob/719f81f47d410701fd82c2bc88613cc140c59377/research/lean/ResearchLean/AG/RealizationReconstruction/FixedFAllAutomorphismGroup.lean\#L336}{\texttt{mul\_\allowbreak{}fiber\allowbreak{}Perm}} & The reindexing formula for fiber permutations under the group product of following pairs. It can be restricted to an arbitrary permitted subgroup H, and finiteness is not assumed. \\
\addlinespace[3pt]
Theorem~\ref{thm:7.25} and \eqref{eq:7.41} & \href{https://github.com/iroha1203/AlgebraicArchitectureTheoryV2/blob/719f81f47d410701fd82c2bc88613cc140c59377/research/lean/ResearchLean/AG/RealizationReconstruction/FixedFRestrictedAutomorphismGroup.lean\#L49}{\texttt{canonical\allowbreak{}Section}} / \href{https://github.com/iroha1203/AlgebraicArchitectureTheoryV2/blob/719f81f47d410701fd82c2bc88613cc140c59377/research/lean/ResearchLean/AG/RealizationReconstruction/FixedFRestrictedAutomorphismGroup.lean\#L64}{\texttt{projection\_\allowbreak{}section}} / \href{https://github.com/iroha1203/AlgebraicArchitectureTheoryV2/blob/719f81f47d410701fd82c2bc88613cc140c59377/research/lean/ResearchLean/AG/RealizationReconstruction/FixedFRestrictedKernelIdentification.lean\#L70}{\texttt{component\allowbreak{}Kernel\allowbreak{}Hom}} / \href{https://github.com/iroha1203/AlgebraicArchitectureTheoryV2/blob/719f81f47d410701fd82c2bc88613cc140c59377/research/lean/ResearchLean/AG/RealizationReconstruction/FixedFRestrictedKernelIdentification.lean\#L147}{\texttt{range\_\allowbreak{}component\allowbreak{}Kernel\allowbreak{}Hom\_\allowbreak{}eq\_\allowbreak{}ker\_\allowbreak{}projection}} / \href{https://github.com/iroha1203/AlgebraicArchitectureTheoryV2/blob/719f81f47d410701fd82c2bc88613cc140c59377/research/lean/ResearchLean/AG/RealizationReconstruction/FixedFSplitExactSequenceAndTorsor.lean\#L143}{\texttt{is\allowbreak{}Group\allowbreak{}Short\allowbreak{}Exact}} / \href{https://github.com/iroha1203/AlgebraicArchitectureTheoryV2/blob/719f81f47d410701fd82c2bc88613cc140c59377/research/lean/ResearchLean/AG/RealizationReconstruction/FixedFSplitExactSequenceAndTorsor.lean\#L157}{\texttt{canonical\allowbreak{}Section\_\allowbreak{}right\allowbreak{}Inverse}} & Takes H to be an arbitrary subgroup of the group of visible graph automorphisms. The image of the component group agrees with the kernel of the projection, and one obtains a split short exact sequence with a canonical section. \\
\addlinespace[3pt]
Theorem~\ref{thm:7.25} (fibers of the projection) & \href{https://github.com/iroha1203/AlgebraicArchitectureTheoryV2/blob/719f81f47d410701fd82c2bc88613cc140c59377/research/lean/ResearchLean/AG/RealizationReconstruction/FixedFSplitExactSequenceAndTorsor.lean\#L57}{\texttt{projection\allowbreak{}Fiber\allowbreak{}Mul\allowbreak{}Action}} / \href{https://github.com/iroha1203/AlgebraicArchitectureTheoryV2/blob/719f81f47d410701fd82c2bc88613cc140c59377/research/lean/ResearchLean/AG/RealizationReconstruction/FixedFSplitExactSequenceAndTorsor.lean\#L128}{\texttt{projection\allowbreak{}Fiber\_\allowbreak{}exists\allowbreak{}Unique\_\allowbreak{}smul\_\allowbreak{}eq}} / \href{https://github.com/iroha1203/AlgebraicArchitectureTheoryV2/blob/719f81f47d410701fd82c2bc88613cc140c59377/research/lean/ResearchLean/AG/RealizationReconstruction/FixedFSplitExactSequenceAndTorsor.lean\#L275}{\texttt{component\allowbreak{}Group\allowbreak{}Equiv\allowbreak{}Projection\allowbreak{}Fiber}} & The free and transitive right action of the kernel of the projection, and the bijection between the component group M\_Q and each fiber given by the canonical section. The identification of the component group with the kernel is by \eqref{eq:7.41}. \\
\addlinespace[3pt]
Theorem~\ref{thm:7.25} and \eqref{eq:7.42} & \href{https://github.com/iroha1203/AlgebraicArchitectureTheoryV2/blob/719f81f47d410701fd82c2bc88613cc140c59377/research/lean/ResearchLean/AG/RealizationReconstruction/FixedFFiberCardinality.lean\#L89}{\texttt{nat\allowbreak{}Card\_\allowbreak{}projection\allowbreak{}Fiber}} & If the vertex set and the hidden type K are finite, gives the number of elements of the fiber over a fixed visible change. Finiteness of H or of the edge set is not needed. \\
\addlinespace[3pt]
Theorem~\ref{thm:7.25}, \eqref{eq:7.43} & \href{https://github.com/iroha1203/AlgebraicArchitectureTheoryV2/blob/7e68ec6e77ef0249ede6a4c875715cfaf1cb3ee6/research/lean/ResearchLean/AG/RealizationReconstruction/FixedFSemidirectProduct.lean\#L48}{\texttt{component\allowbreak{}Action}} / \href{https://github.com/iroha1203/AlgebraicArchitectureTheoryV2/blob/7e68ec6e77ef0249ede6a4c875715cfaf1cb3ee6/research/lean/ResearchLean/AG/RealizationReconstruction/FixedFSemidirectProduct.lean\#L209}{\texttt{actual\allowbreak{}Equiv\allowbreak{}Semidirect}} & For any visible subgroup and hidden type, gives a group isomorphism with the semidirect product for the reindexing action on connected components. Families are indexed by destination components. \\
\addlinespace[3pt]
Corollary~\ref{cor:7.26} (number of elements of the fiber) & \href{https://github.com/iroha1203/AlgebraicArchitectureTheoryV2/blob/719f81f47d410701fd82c2bc88613cc140c59377/research/lean/ResearchLean/AG/RealizationReconstruction/FixedFFiberCardinality.lean\#L53}{\texttt{pointed\allowbreak{}Permutation\allowbreak{}Equiv\allowbreak{}Complement}} / \href{https://github.com/iroha1203/AlgebraicArchitectureTheoryV2/blob/719f81f47d410701fd82c2bc88613cc140c59377/research/lean/ResearchLean/AG/RealizationReconstruction/FixedFFiberCardinality.lean\#L61}{\texttt{nat\allowbreak{}Card\_\allowbreak{}pointed\allowbreak{}Permutation}} / \href{https://github.com/iroha1203/AlgebraicArchitectureTheoryV2/blob/719f81f47d410701fd82c2bc88613cc140c59377/research/lean/ResearchLean/AG/RealizationReconstruction/FixedFFiberCardinality.lean\#L116}{\texttt{nat\allowbreak{}Card\_\allowbreak{}pointed\allowbreak{}Projection\allowbreak{}Fiber}} & The bijection between the permutations of K that fix the selected value k0 and the permutations of its complement. If K is finite, one obtains a count given by a factorial and the corresponding number of elements of the fiber. \\
\addlinespace[3pt]
Corollary~\ref{cor:7.26} (semidirect product) & \href{https://github.com/iroha1203/AlgebraicArchitectureTheoryV2/blob/7e68ec6e77ef0249ede6a4c875715cfaf1cb3ee6/research/lean/ResearchLean/AG/RealizationReconstruction/FixedFSemidirectProduct.lean\#L234}{\texttt{pointed\allowbreak{}Component\allowbreak{}Action}} / \href{https://github.com/iroha1203/AlgebraicArchitectureTheoryV2/blob/7e68ec6e77ef0249ede6a4c875715cfaf1cb3ee6/research/lean/ResearchLean/AG/RealizationReconstruction/FixedFSemidirectProduct.lean\#L356}{\texttt{pointed\allowbreak{}Actual\allowbreak{}Equiv\allowbreak{}Semidirect}} & Identifies the actual changes preserving the selected section with the semidirect product formed from permutations fixing the designated point on each component. \\
\addlinespace[3pt]
Proposition~\ref{prop:7.27} & \href{https://github.com/iroha1203/AlgebraicArchitectureTheoryV2/blob/719f81f47d410701fd82c2bc88613cc140c59377/research/lean/ResearchLean/AG/RealizationReconstruction/FixedFLensConnection.lean\#L90}{\texttt{equiv\allowbreak{}Preserving\allowbreak{}Following\allowbreak{}Changes}} / \href{https://github.com/iroha1203/AlgebraicArchitectureTheoryV2/blob/719f81f47d410701fd82c2bc88613cc140c59377/research/lean/ResearchLean/AG/RealizationReconstruction/FixedFLensConnection.lean\#L136}{\texttt{normal\allowbreak{}Form}} / \href{https://github.com/iroha1203/AlgebraicArchitectureTheoryV2/blob/719f81f47d410701fd82c2bc88613cc140c59377/research/lean/ResearchLean/AG/RealizationReconstruction/FixedFLensConnection.lean\#L179}{\texttt{equiv\allowbreak{}Hidden\allowbreak{}Permutations}} / \href{https://github.com/iroha1203/AlgebraicArchitectureTheoryV2/blob/719f81f47d410701fd82c2bc88613cc140c59377/research/lean/ResearchLean/AG/RealizationReconstruction/FixedFLensConnection.lean\#L198}{\texttt{nat\allowbreak{}Card\_\allowbreak{}product\allowbreak{}Lens\allowbreak{}Changes}} / \href{https://github.com/iroha1203/AlgebraicArchitectureTheoryV2/blob/719f81f47d410701fd82c2bc88613cc140c59377/research/lean/ResearchLean/AG/RealizationReconstruction/FixedFLensConnection.lean\#L220}{\texttt{preserves\allowbreak{}Section\_\allowbreak{}iff}} & Correspondence of operation-preserving changes, hidden permutations, counts, and section preservation for product lenses. Transfers this through the product decomposition of Proposition~\ref{prop:1.33} and \eqref{eq:1.20} to general total lenses, and applies Theorem~\ref{thm:7.25} to the connected update graph. \\
\addlinespace[3pt]
Example~\ref{ex:7.30} (two workers) & \href{https://github.com/iroha1203/AlgebraicArchitectureTheoryV2/blob/719f81f47d410701fd82c2bc88613cc140c59377/research/lean/ResearchLean/AG/RealizationReconstruction/FixedFFiniteExamples.lean\#L489}{\texttt{protocol\allowbreak{}Component\allowbreak{}Equiv\allowbreak{}Bool}} / \href{https://github.com/iroha1203/AlgebraicArchitectureTheoryV2/blob/719f81f47d410701fd82c2bc88613cc140c59377/research/lean/ResearchLean/AG/RealizationReconstruction/FixedFFiniteExamples.lean\#L46}{\texttt{nat\allowbreak{}Card\_\allowbreak{}following\allowbreak{}State\allowbreak{}Change}} / \href{https://github.com/iroha1203/AlgebraicArchitectureTheoryV2/blob/719f81f47d410701fd82c2bc88613cc140c59377/research/lean/ResearchLean/AG/RealizationReconstruction/FixedFFiniteExamples.lean\#L384}{\texttt{nat\allowbreak{}Card\_\allowbreak{}protocol\allowbreak{}Vertex}} / \href{https://github.com/iroha1203/AlgebraicArchitectureTheoryV2/blob/719f81f47d410701fd82c2bc88613cc140c59377/research/lean/ResearchLean/AG/RealizationReconstruction/FixedFFiniteExamples.lean\#L515}{\texttt{protocol\_\allowbreak{}observation\_\allowbreak{}count\_\allowbreak{}identity}} / \href{https://github.com/iroha1203/AlgebraicArchitectureTheoryV2/blob/719f81f47d410701fd82c2bc88613cc140c59377/research/lean/ResearchLean/AG/RealizationReconstruction/FixedFFiniteExamples.lean\#L523}{\texttt{protocol\_\allowbreak{}operation\_\allowbreak{}count\_\allowbreak{}identity}} / \href{https://github.com/iroha1203/AlgebraicArchitectureTheoryV2/blob/719f81f47d410701fd82c2bc88613cc140c59377/research/lean/ResearchLean/AG/RealizationReconstruction/FixedFFiniteExamples.lean\#L531}{\texttt{protocol\_\allowbreak{}operation\_\allowbreak{}count\_\allowbreak{}session\allowbreak{}Swap}} / \href{https://github.com/iroha1203/AlgebraicArchitectureTheoryV2/blob/719f81f47d410701fd82c2bc88613cc140c59377/research/lean/ResearchLean/AG/RealizationReconstruction/FixedFFiniteExamples.lean\#L427}{\texttt{protocol\allowbreak{}Session\allowbreak{}Swap\allowbreak{}Lift\_\allowbreak{}fiber\allowbreak{}Perm}} / \href{https://github.com/iroha1203/AlgebraicArchitectureTheoryV2/blob/719f81f47d410701fd82c2bc88613cc140c59377/research/lean/ResearchLean/AG/RealizationReconstruction/FixedFFiniteExamples.lean\#L435}{\texttt{protocol\allowbreak{}Session\allowbreak{}Swap\_\allowbreak{}operation\allowbreak{}Map\_\allowbreak{}false}} / \href{https://github.com/iroha1203/AlgebraicArchitectureTheoryV2/blob/719f81f47d410701fd82c2bc88613cc140c59377/research/lean/ResearchLean/AG/RealizationReconstruction/FixedFFiniteExamples.lean\#L450}{\texttt{protocol\allowbreak{}Session\allowbreak{}Swap\_\allowbreak{}operation\allowbreak{}Map\_\allowbreak{}true}} & For the graph of two workers, gives the number of elements of the fibers over the identity change and over the session swap, and the action on edges of the lift. Taking the visible group to be the full automorphism group of the graph, with Bool states and 4 vertices, for both the identity change and the session swap there are 16 candidates that follow the readout of the control points, and the fiber of the changes that preserve the operations as well has 4 elements. \\
\addlinespace[3pt]
Example~\ref{ex:7.30} (the full change group) & \href{https://github.com/iroha1203/AlgebraicArchitectureTheoryV2/blob/7e68ec6e77ef0249ede6a4c875715cfaf1cb3ee6/research/lean/ResearchLean/AG/RealizationReconstruction/FixedFSemidirectProduct.lean\#L672}{\texttt{actual\allowbreak{}Equiv\allowbreak{}Worker\allowbreak{}Semidirect}} / \href{https://github.com/iroha1203/AlgebraicArchitectureTheoryV2/blob/7e68ec6e77ef0249ede6a4c875715cfaf1cb3ee6/research/lean/ResearchLean/AG/RealizationReconstruction/FixedFSemidirectProduct.lean\#L679}{\texttt{nat\allowbreak{}Card\_\allowbreak{}worker\allowbreak{}Change\allowbreak{}Group}} & For the visible group of order two consisting of identity and worker exchange, constructs the semidirect product whose exchange swaps the two hidden-permutation factors. The full change group has order eight. \\
\addlinespace[3pt]
\S\ref{sec:7.8} (fixed view and fixed schema) & \href{https://github.com/iroha1203/AlgebraicArchitectureTheoryV2/blob/719f81f47d410701fd82c2bc88613cc140c59377/research/lean/ResearchLean/AG/RealizationReconstruction/CSAATFullyFaithfulComparisonTransport.lean\#L25}{\texttt{fully\allowbreak{}Faithful\allowbreak{}Endpoint\allowbreak{}Aut\allowbreak{}Mul\allowbreak{}Equiv}} / \href{https://github.com/iroha1203/AlgebraicArchitectureTheoryV2/blob/719f81f47d410701fd82c2bc88613cc140c59377/research/lean/ResearchLean/AG/RealizationReconstruction/CSAATFullyFaithfulComparisonTransport.lean\#L93}{\texttt{generated\allowbreak{}Arrow\allowbreak{}Comparison\allowbreak{}Mul\allowbreak{}Equiv\allowbreak{}Of\allowbreak{}Fully\allowbreak{}Faithful}} / \href{https://github.com/iroha1203/AlgebraicArchitectureTheoryV2/blob/719f81f47d410701fd82c2bc88613cc140c59377/research/lean/ResearchLean/AG/RealizationReconstruction/CSAATFullyFaithfulComparisonTransport.lean\#L137}{\texttt{lens\allowbreak{}AATIndependent\allowbreak{}Package\allowbreak{}Comparison\allowbreak{}Mul\allowbreak{}Equiv}} / \href{https://github.com/iroha1203/AlgebraicArchitectureTheoryV2/blob/719f81f47d410701fd82c2bc88613cc140c59377/research/lean/ResearchLean/AG/RealizationReconstruction/CSAATFullyFaithfulComparisonTransport.lean\#L150}{\texttt{protocol\allowbreak{}AATIndependent\allowbreak{}Package\allowbreak{}Comparison\allowbreak{}Mul\allowbreak{}Equiv}} / \href{https://github.com/iroha1203/AlgebraicArchitectureTheoryV2/blob/719f81f47d410701fd82c2bc88613cc140c59377/research/lean/ResearchLean/AG/RealizationReconstruction/CSAATRestrictionKernelFiberTransport.lean\#L162}{\texttt{generated\allowbreak{}Arrow\allowbreak{}Comparison\allowbreak{}Source\allowbreak{}Kernel\allowbreak{}Mul\allowbreak{}Equiv}} / \href{https://github.com/iroha1203/AlgebraicArchitectureTheoryV2/blob/719f81f47d410701fd82c2bc88613cc140c59377/research/lean/ResearchLean/AG/RealizationReconstruction/CSAATRestrictionKernelFiberTransport.lean\#L175}{\texttt{generated\allowbreak{}Arrow\allowbreak{}Comparison\allowbreak{}Source\allowbreak{}Fiber\allowbreak{}Equiv}} / \href{https://github.com/iroha1203/AlgebraicArchitectureTheoryV2/blob/719f81f47d410701fd82c2bc88613cc140c59377/research/lean/ResearchLean/AG/RealizationReconstruction/CSAATRestrictionKernelFiberTransport.lean\#L193}{\texttt{generated\allowbreak{}Arrow\allowbreak{}Comparison\allowbreak{}Source\allowbreak{}Fiber\allowbreak{}Equiv\_\allowbreak{}smul}} & From the full faithfulness of each semantic functor, identifies the comparison groups of CS and of typed AAT, and matches the kernel of the source projection, each fiber, and the right action. Treats the full automorphism group of typed objects. \\
\bottomrule
\end{xltabular}

\section{Chapter 8: Presentation and Local Reconstruction}\label{sec:A.8}

\begin{xltabular}{\linewidth}{@{}LLL@{}}
\toprule
Text & Lean declaration & Objects and conditions \\
\midrule
\endhead
Definition~\ref{def:8.1} & \href{https://github.com/iroha1203/AlgebraicArchitectureTheoryV2/blob/719f81f47d410701fd82c2bc88613cc140c59377/research/lean/ResearchLean/AG/DoctrineFiberProduct/Schema.lean\#L47}{\texttt{Atom\allowbreak{}Predicate\allowbreak{}Code}} / \href{https://github.com/iroha1203/AlgebraicArchitectureTheoryV2/blob/719f81f47d410701fd82c2bc88613cc140c59377/research/lean/ResearchLean/AG/DoctrineFiberProduct/Schema.lean\#L65}{\texttt{eval}} / \href{https://github.com/iroha1203/AlgebraicArchitectureTheoryV2/blob/719f81f47d410701fd82c2bc88613cc140c59377/research/lean/ResearchLean/AG/FiniteDecoderRepresentability/OnePointCode.lean\#L42}{\texttt{atom\allowbreak{}Predicate\allowbreak{}Code\allowbreak{}To\allowbreak{}Continuous\allowbreak{}Map}} & Evaluates a code from a Bool default and a finite exception set. The value at the point at infinity of the OnePoint of the discrete Atoms corresponds to the default. \\
\addlinespace[3pt]
Proposition~\ref{prop:8.2} & \href{https://github.com/iroha1203/AlgebraicArchitectureTheoryV2/blob/719f81f47d410701fd82c2bc88613cc140c59377/research/lean/ResearchLean/AG/FiniteDecoderRepresentability/OnePointCode.lean\#L140}{\texttt{atom\allowbreak{}Predicate\allowbreak{}Code\allowbreak{}Continuous\allowbreak{}Map\allowbreak{}Equiv}} / \href{https://github.com/iroha1203/AlgebraicArchitectureTheoryV2/blob/719f81f47d410701fd82c2bc88613cc140c59377/research/lean/ResearchLean/AG/FiniteDecoderRepresentability/CodeFibers.lean\#L74}{\texttt{atom\allowbreak{}Predicate\allowbreak{}Code\_\allowbreak{}eval\_\allowbreak{}injective\_\allowbreak{}of\_\allowbreak{}infinite}} / \href{https://github.com/iroha1203/AlgebraicArchitectureTheoryV2/blob/719f81f47d410701fd82c2bc88613cc140c59377/research/lean/ResearchLean/AG/FiniteDecoderRepresentability/CodeFibers.lean\#L188}{\texttt{finite\allowbreak{}Predicate\allowbreak{}Code\allowbreak{}Fiber\allowbreak{}Equiv}} & Fixing the discrete Atoms and Bool, the bijection between raw codes and continuous maps out of OnePoint. Proves that evaluation is injective for Infinite Atom, and that for Finite Atom the code fiber of an arbitrary predicate is equal to Bool. \\
\addlinespace[3pt]
Proposition~\ref{prop:8.3} & \href{https://github.com/iroha1203/AlgebraicArchitectureTheoryV2/blob/719f81f47d410701fd82c2bc88613cc140c59377/research/lean/ResearchLean/AG/FiniteDecoderRepresentability/FixedArrowClassification.lean\#L244}{\texttt{exists\_\allowbreak{}fixed\allowbreak{}Presentation\_\allowbreak{}decode\_\allowbreak{}iff}} / \href{https://github.com/iroha1203/AlgebraicArchitectureTheoryV2/blob/719f81f47d410701fd82c2bc88613cc140c59377/research/lean/ResearchLean/AG/FiniteDecoderRepresentability/FixedArrowClassification.lean\#L294}{\texttt{exists\_\allowbreak{}finite\allowbreak{}Code\allowbreak{}Cart\allowbreak{}Hom\_\allowbreak{}map\_\allowbreak{}iff}} & For a fixed source and target FiniteInstanceCode and a semantic hom, the existence of an exact code presentation is equivalent to the finiteness of the support of the Atom permutation together with the preservation of the default after normalization. Definition~\ref{def:5.14} is read with equality of codes. \\
\addlinespace[3pt]
Corollary~\ref{cor:8.5} & \href{https://github.com/iroha1203/AlgebraicArchitectureTheoryV2/blob/719f81f47d410701fd82c2bc88613cc140c59377/research/lean/ResearchLean/AG/FiniteDecoderRepresentability/FiniteCodeNormalization.lean\#L302}{\texttt{finite\allowbreak{}Code\allowbreak{}Normalization\allowbreak{}Functor}} / \href{https://github.com/iroha1203/AlgebraicArchitectureTheoryV2/blob/719f81f47d410701fd82c2bc88613cc140c59377/research/lean/ResearchLean/AG/FiniteDecoderRepresentability/FiniteFullSubcategory.lean\#L129}{\texttt{false\allowbreak{}Default\allowbreak{}Finite\allowbreak{}Code\allowbreak{}Realization\_\allowbreak{}full}} / \href{https://github.com/iroha1203/AlgebraicArchitectureTheoryV2/blob/719f81f47d410701fd82c2bc88613cc140c59377/research/lean/ResearchLean/AG/FiniteDecoderRepresentability/FiniteNormalizationRealizationIso.lean\#L226}{\texttt{finite\allowbreak{}Normalization\allowbreak{}Realization\allowbreak{}Iso}} & A functor that normalizes the default to false over finite Atoms while preserving evaluation. On the full subcategory of false defaults the decoder is fully faithful, and it has a natural isomorphism to the original decoder. \\
\addlinespace[3pt]
Definition~\ref{def:8.6} & \href{https://github.com/iroha1203/AlgebraicArchitectureTheoryV2/blob/719f81f47d410701fd82c2bc88613cc140c59377/research/lean/ResearchLean/AG/LocalSemanticReconstruction/IndependentFiniteFragments.lean\#L27}{\texttt{Fragment}} / \href{https://github.com/iroha1203/AlgebraicArchitectureTheoryV2/blob/719f81f47d410701fd82c2bc88613cc140c59377/research/lean/ResearchLean/AG/LocalSemanticReconstruction/IndependentFiniteFragments.lean\#L41}{\texttt{Compatible}} & Defines dependent tables over finite subsets and compatible families that agree under restriction, for an arbitrary dependent value type V\_q. What is finite is the number of queries, not the value type. \\
\addlinespace[3pt]
Lemma~\ref{lem:8.7} & \href{https://github.com/iroha1203/AlgebraicArchitectureTheoryV2/blob/719f81f47d410701fd82c2bc88613cc140c59377/research/lean/ResearchLean/AG/LocalSemanticReconstruction/IndependentFiniteFragments.lean\#L55}{\texttt{glue\_\allowbreak{}fragments}} / \href{https://github.com/iroha1203/AlgebraicArchitectureTheoryV2/blob/719f81f47d410701fd82c2bc88613cc140c59377/research/lean/ResearchLean/AG/LocalSemanticReconstruction/IndependentFiniteFragments.lean\#L58}{\texttt{fragments\_\allowbreak{}glue}} & The gluing from values on singletons and the restriction to each finite subset are mutually inverse. \\
\addlinespace[3pt]
Definition~\ref{def:8.8} & \href{https://github.com/iroha1203/AlgebraicArchitectureTheoryV2/blob/719f81f47d410701fd82c2bc88613cc140c59377/research/lean/ResearchLean/AG/LocalSemanticReconstruction/IndependentCarrierGraphReadings.lean\#L35}{\texttt{Is\allowbreak{}Typed}} / \href{https://github.com/iroha1203/AlgebraicArchitectureTheoryV2/blob/719f81f47d410701fd82c2bc88613cc140c59377/research/lean/ResearchLean/AG/LocalSemanticReconstruction/IndependentCarrierGraphReadings.lean\#L39}{\texttt{Is\allowbreak{}Total}} / \href{https://github.com/iroha1203/AlgebraicArchitectureTheoryV2/blob/719f81f47d410701fd82c2bc88613cc140c59377/research/lean/ResearchLean/AG/LocalSemanticReconstruction/IndependentCarrierGraphReadings.lean\#L43}{\texttt{Is\allowbreak{}Lawful}} & The definition of a Bool graph that is false on carrier pairs with wrong type references and requires a unique true output in each row on the selected pairs. \\
\addlinespace[3pt]
Lemma~\ref{lem:8.9} & \href{https://github.com/iroha1203/AlgebraicArchitectureTheoryV2/blob/719f81f47d410701fd82c2bc88613cc140c59377/research/lean/ResearchLean/AG/LocalSemanticReconstruction/IndependentCarrierGraphReadings.lean\#L103}{\texttt{function\allowbreak{}Equiv}} / \href{https://github.com/iroha1203/AlgebraicArchitectureTheoryV2/blob/719f81f47d410701fd82c2bc88613cc140c59377/research/lean/ResearchLean/AG/LocalSemanticReconstruction/IndependentCarrierGraphReadings.lean\#L117}{\texttt{assemble\_\allowbreak{}read}} / \href{https://github.com/iroha1203/AlgebraicArchitectureTheoryV2/blob/719f81f47d410701fd82c2bc88613cc140c59377/research/lean/ResearchLean/AG/LocalSemanticReconstruction/IndependentCarrierGraphReadings.lean\#L121}{\texttt{read\_\allowbreak{}assemble}} / \href{https://github.com/iroha1203/AlgebraicArchitectureTheoryV2/blob/719f81f47d410701fd82c2bc88613cc140c59377/research/lean/ResearchLean/AG/LocalSemanticReconstruction/IndependentCarrierGraphReadings.lean\#L274}{\texttt{read\_\allowbreak{}compose}} & The bijection between arbitrary functions and typed total functional Bool tables. The relational composition of tables by the existence of an intermediate value corresponds to ordinary composition of functions. \\
\addlinespace[3pt]
Example~\ref{ex:8.10} & \href{https://github.com/iroha1203/AlgebraicArchitectureTheoryV2/blob/719f81f47d410701fd82c2bc88613cc140c59377/research/lean/ResearchLean/AG/LocalSemanticReconstruction/IndependentCarrierGraphReadings.lean\#L155}{\texttt{false\_\allowbreak{}table\_\allowbreak{}rejected}} / \href{https://github.com/iroha1203/AlgebraicArchitectureTheoryV2/blob/719f81f47d410701fd82c2bc88613cc140c59377/research/lean/ResearchLean/AG/LocalSemanticReconstruction/IndependentCarrierGraphReadings.lean\#L149}{\texttt{duplicate\_\allowbreak{}outputs\_\allowbreak{}rejected}} / \href{https://github.com/iroha1203/AlgebraicArchitectureTheoryV2/blob/719f81f47d410701fd82c2bc88613cc140c59377/research/lean/ResearchLean/AG/LocalSemanticReconstruction/IndependentFiniteFragments.lean\#L51}{\texttt{fragments\_\allowbreak{}compatible}} & An all-false table over a nonempty source is not total, and a table with two distinct true outputs for one input is not functional. These tables also have compatible families of finite restrictions. \\
\addlinespace[3pt]
Definition~\ref{def:8.11} & \href{https://github.com/iroha1203/AlgebraicArchitectureTheoryV2/blob/719f81f47d410701fd82c2bc88613cc140c59377/research/lean/ResearchLean/AG/LocalSemanticReconstruction/IndependentFiniteLawFormula.lean\#L31}{\texttt{Bool\allowbreak{}Formula}} / \href{https://github.com/iroha1203/AlgebraicArchitectureTheoryV2/blob/719f81f47d410701fd82c2bc88613cc140c59377/research/lean/ResearchLean/AG/LocalSemanticReconstruction/IndependentFiniteLawFormula.lean\#L55}{\texttt{support}} / \href{https://github.com/iroha1203/AlgebraicArchitectureTheoryV2/blob/719f81f47d410701fd82c2bc88613cc140c59377/research/lean/ResearchLean/AG/LocalSemanticReconstruction/IndependentFiniteLawFormula.lean\#L68}{\texttt{evaluate\_\allowbreak{}iff\_\allowbreak{}of\_\allowbreak{}support}} & The syntax built from finite Bool cells and equations of types and values, its finite support, and the dependence of evaluation on the support. Since the quantified CarrierRows.Instances expresses totality, no finite check of all the conditions is claimed. \\
\addlinespace[3pt]
Theorem~\ref{thm:8.12} & \href{https://github.com/iroha1203/AlgebraicArchitectureTheoryV2/blob/719f81f47d410701fd82c2bc88613cc140c59377/research/lean/ResearchLean/AG/LocalSemanticReconstruction/LocalReconstructionEquivalence.lean\#L114}{\texttt{Reconstruction\allowbreak{}Data}} / \href{https://github.com/iroha1203/AlgebraicArchitectureTheoryV2/blob/719f81f47d410701fd82c2bc88613cc140c59377/research/lean/ResearchLean/AG/LocalSemanticReconstruction/LocalReconstructionEquivalence.lean\#L155}{\texttt{equivalence}} / \href{https://github.com/iroha1203/AlgebraicArchitectureTheoryV2/blob/719f81f47d410701fd82c2bc88613cc140c59377/research/lean/ResearchLean/AG/LocalSemanticReconstruction/LocalReconstructionEquivalence.lean\#L167}{\texttt{exists\allowbreak{}Unique\_\allowbreak{}preimage}} & For an arbitrary functor F, obtains fullness, faithfulness, and essential surjectivity, and an equivalence of categories, from the three conditions of separation of Homs, agreement with the map after assembly, and the assembly isomorphism of objects. \\
\addlinespace[3pt]
Construction~\ref{cons:8.13} & \href{https://github.com/iroha1203/AlgebraicArchitectureTheoryV2/blob/719f81f47d410701fd82c2bc88613cc140c59377/research/lean/ResearchLean/AG/LocalSemanticReconstruction/IndependentAATPrimitiveReconstruction.lean\#L55}{\texttt{Parameter}} / \href{https://github.com/iroha1203/AlgebraicArchitectureTheoryV2/blob/719f81f47d410701fd82c2bc88613cc140c59377/research/lean/ResearchLean/AG/LocalSemanticReconstruction/IndependentAATPrimitiveReconstruction.lean\#L61}{\texttt{Native\allowbreak{}Category}} / \href{https://github.com/iroha1203/AlgebraicArchitectureTheoryV2/blob/719f81f47d410701fd82c2bc88613cc140c59377/research/lean/ResearchLean/AG/LocalSemanticReconstruction/IndependentAATPrimitiveReconstruction.lean\#L72}{\texttt{Local\allowbreak{}Category}} & Fixes the carrier of Atoms and the representative and explicit variants of morphisms, and selects the category of realizations for each. \\
\addlinespace[3pt]
Definition~\ref{def:8.14} & \href{https://github.com/iroha1203/AlgebraicArchitectureTheoryV2/blob/719f81f47d410701fd82c2bc88613cc140c59377/research/lean/ResearchLean/AG/RealizationReconstruction/CSAATExplicitExactGeometryHom.lean\#L25}{\texttt{Explicit\allowbreak{}Exact\allowbreak{}Geometry\allowbreak{}Hom}} / \href{https://github.com/iroha1203/AlgebraicArchitectureTheoryV2/blob/719f81f47d410701fd82c2bc88613cc140c59377/research/lean/ResearchLean/AG/LocalSemanticReconstruction/IndependentGeometryHomPrimitiveDeclaration.lean\#L100}{\texttt{Query}} & The representative variant uses raw\_eq of GeomReadHom, and the explicit variant holds the corresponding maps of raw, context, and realization. A common HomQuery reads the primitive components of each variant. \\
\addlinespace[3pt]
Definition~\ref{def:8.15} & \href{https://github.com/iroha1203/AlgebraicArchitectureTheoryV2/blob/719f81f47d410701fd82c2bc88613cc140c59377/research/lean/ResearchLean/AG/LocalSemanticReconstruction/IndependentAATPrimitiveReconstruction.lean\#L1114}{\texttt{Object\allowbreak{}Table\allowbreak{}Laws}} / \href{https://github.com/iroha1203/AlgebraicArchitectureTheoryV2/blob/719f81f47d410701fd82c2bc88613cc140c59377/research/lean/ResearchLean/AG/LocalSemanticReconstruction/IndependentAATPrimitiveReconstruction.lean\#L1176}{\texttt{Geometry\allowbreak{}Hom\allowbreak{}Table\allowbreak{}Certificate}} / \href{https://github.com/iroha1203/AlgebraicArchitectureTheoryV2/blob/719f81f47d410701fd82c2bc88613cc140c59377/research/lean/ResearchLean/AG/LocalSemanticReconstruction/IndependentGeometryHomInvariantWitnesses.lean\#L108}{\texttt{Presentation}} / \href{https://github.com/iroha1203/AlgebraicArchitectureTheoryV2/blob/719f81f47d410701fd82c2bc88613cc140c59377/research/lean/ResearchLean/AG/LocalSemanticReconstruction/IndependentGeometryHomInvariantQuotient.lean\#L29}{\texttt{Local}} / \href{https://github.com/iroha1203/AlgebraicArchitectureTheoryV2/blob/719f81f47d410701fd82c2bc88613cc140c59377/research/lean/ResearchLean/AG/LocalSemanticReconstruction/IndependentAATPrimitiveReconstruction.lean\#L1329}{\texttt{Lawful\allowbreak{}Hom\allowbreak{}Family}} & Objects are defined by typed primitive tables, laws at each point, and conditions on finite families. A presentation of a Hom consists of compatible finite Bool tables for the preserved components and the auxiliary components together with row laws, and the quotient is taken by the agreement of the preserved components. The choice of representatives for the auxiliary tables is fixed, but no conclusion about global morphisms or transport is assumed. \\
\addlinespace[3pt]
Lemma~\ref{lem:8.16} & \href{https://github.com/iroha1203/AlgebraicArchitectureTheoryV2/blob/719f81f47d410701fd82c2bc88613cc140c59377/research/lean/ResearchLean/AG/LocalSemanticReconstruction/IndependentGeometryCategoryReconstruction.lean\#L168}{\texttt{representative\allowbreak{}Local\allowbreak{}Category}} / \href{https://github.com/iroha1203/AlgebraicArchitectureTheoryV2/blob/719f81f47d410701fd82c2bc88613cc140c59377/research/lean/ResearchLean/AG/LocalSemanticReconstruction/IndependentGeometryCategoryReconstruction.lean\#L189}{\texttt{explicit\allowbreak{}Local\allowbreak{}Category}} / \href{https://github.com/iroha1203/AlgebraicArchitectureTheoryV2/blob/719f81f47d410701fd82c2bc88613cc140c59377/research/lean/ResearchLean/AG/LocalSemanticReconstruction/IndependentGeometryCategoryReconstruction.lean\#L459}{\texttt{representative\allowbreak{}Reading\allowbreak{}Functor}} / \href{https://github.com/iroha1203/AlgebraicArchitectureTheoryV2/blob/719f81f47d410701fd82c2bc88613cc140c59377/research/lean/ResearchLean/AG/LocalSemanticReconstruction/IndependentGeometryCategoryReconstruction.lean\#L475}{\texttt{explicit\allowbreak{}Reading\allowbreak{}Functor}} & Defines the identity and compose of local graphs directly, and constructs the category by proving that they preserve the point laws and the quotient. The identity and compose of the reading are also proved. \\
\addlinespace[3pt]
Lemma~\ref{lem:8.17} & \href{https://github.com/iroha1203/AlgebraicArchitectureTheoryV2/blob/719f81f47d410701fd82c2bc88613cc140c59377/research/lean/ResearchLean/AG/LocalSemanticReconstruction/IndependentGeometryPrimitiveAssembly.lean\#L347}{\texttt{object\allowbreak{}Equiv}} / \href{https://github.com/iroha1203/AlgebraicArchitectureTheoryV2/blob/719f81f47d410701fd82c2bc88613cc140c59377/research/lean/ResearchLean/AG/LocalSemanticReconstruction/IndependentGeometryPrimitiveAssembly.lean\#L361}{\texttt{read\_\allowbreak{}assemble}} / \href{https://github.com/iroha1203/AlgebraicArchitectureTheoryV2/blob/719f81f47d410701fd82c2bc88613cc140c59377/research/lean/ResearchLean/AG/LocalSemanticReconstruction/IndependentGeometryHomInvariantWitnesses.lean\#L126}{\texttt{transported\_\allowbreak{}of\_\allowbreak{}row}} / \href{https://github.com/iroha1203/AlgebraicArchitectureTheoryV2/blob/719f81f47d410701fd82c2bc88613cc140c59377/research/lean/ResearchLean/AG/LocalSemanticReconstruction/IndependentGeometryCategoryReconstruction.lean\#L617}{\texttt{representative\_\allowbreak{}read\_\allowbreak{}assemble}} / \href{https://github.com/iroha1203/AlgebraicArchitectureTheoryV2/blob/719f81f47d410701fd82c2bc88613cc140c59377/research/lean/ResearchLean/AG/LocalSemanticReconstruction/IndependentGeometryCategoryReconstruction.lean\#L637}{\texttt{explicit\_\allowbreak{}read\_\allowbreak{}assemble}} / \href{https://github.com/iroha1203/AlgebraicArchitectureTheoryV2/blob/719f81f47d410701fd82c2bc88613cc140c59377/research/lean/ResearchLean/AG/LocalSemanticReconstruction/IndependentAATPrimitiveReconstruction.lean\#L1884}{\texttt{native\allowbreak{}Hom\_\allowbreak{}eq\_\allowbreak{}of\_\allowbreak{}common\allowbreak{}Family\_\allowbreak{}eq}} & The object Equiv from primitive tables $\to$ dependent stages $\to$ full geometry. Derives Invariant.TransportedAlong from the row law of the auxiliary inverse graph. Proves that read and assemble are mutually inverse on all Homs of both variants, and the separation by a common finite family. \\
\addlinespace[3pt]
Theorem~\ref{thm:8.18} & \href{https://github.com/iroha1203/AlgebraicArchitectureTheoryV2/blob/719f81f47d410701fd82c2bc88613cc140c59377/research/lean/ResearchLean/AG/LocalSemanticReconstruction/IndependentAATPrimitiveReconstruction.lean\#L1908}{\texttt{reconstruction\allowbreak{}Data}} / \href{https://github.com/iroha1203/AlgebraicArchitectureTheoryV2/blob/719f81f47d410701fd82c2bc88613cc140c59377/research/lean/ResearchLean/AG/LocalSemanticReconstruction/IndependentAATPrimitiveReconstruction.lean\#L1953}{\texttt{equivalence}} / \href{https://github.com/iroha1203/AlgebraicArchitectureTheoryV2/blob/719f81f47d410701fd82c2bc88613cc140c59377/research/lean/ResearchLean/AG/LocalSemanticReconstruction/IndependentAATPrimitiveReconstruction.lean\#L1958}{\texttt{equivalence\_\allowbreak{}functor}} / \href{https://github.com/iroha1203/AlgebraicArchitectureTheoryV2/blob/719f81f47d410701fd82c2bc88613cc140c59377/research/lean/ResearchLean/AG/LocalSemanticReconstruction/IndependentAATPrimitiveReconstruction.lean\#L1963}{\texttt{hom\allowbreak{}Equiv}} & The equivalence of the local graph category and the category of realizations for a fixed carrier and a fixed variant of morphisms. The forward functor is the primitive reading, and it reconstructs all Homs, including non-invertible morphisms. \\
\addlinespace[3pt]
Corollary~\ref{cor:8.19} & \href{https://github.com/iroha1203/AlgebraicArchitectureTheoryV2/blob/719f81f47d410701fd82c2bc88613cc140c59377/research/lean/ResearchLean/AG/LocalSemanticReconstruction/IndependentGeometryCategoryReconstruction.lean\#L490}{\texttt{representative\allowbreak{}Assembly\allowbreak{}Functor}} / \href{https://github.com/iroha1203/AlgebraicArchitectureTheoryV2/blob/719f81f47d410701fd82c2bc88613cc140c59377/research/lean/ResearchLean/AG/LocalSemanticReconstruction/IndependentGeometryCategoryReconstruction.lean\#L503}{\texttt{explicit\allowbreak{}Assembly\allowbreak{}Functor}} / \href{https://github.com/iroha1203/AlgebraicArchitectureTheoryV2/blob/719f81f47d410701fd82c2bc88613cc140c59377/research/lean/ResearchLean/AG/LocalSemanticReconstruction/IndependentAATPrimitiveReconstruction.lean\#L1843}{\texttt{local\allowbreak{}Hom\allowbreak{}Family\_\allowbreak{}read\_\allowbreak{}assemble}} / \href{https://github.com/leanprover-community/mathlib4/blob/8f9d9cff6bd728b17a24e163c9402775d9e6a365/Mathlib/CategoryTheory/Functor/ReflectsIso/Basic.lean\#L47}{\texttt{is\allowbreak{}Iso\_\allowbreak{}iff\_\allowbreak{}of\_\allowbreak{}reflects\_\allowbreak{}iso}} & The assembly functors of both variants recover the primitive values and reflect isomorphisms. \\
\addlinespace[3pt]
Construction~\ref{cons:8.22} & \href{https://github.com/iroha1203/AlgebraicArchitectureTheoryV2/blob/719f81f47d410701fd82c2bc88613cc140c59377/research/lean/ResearchLean/AG/LocalSemanticReconstruction/IndependentAATPrimitiveReconstruction.lean\#L55}{\texttt{Parameter}} / \href{https://github.com/iroha1203/AlgebraicArchitectureTheoryV2/blob/719f81f47d410701fd82c2bc88613cc140c59377/research/lean/ResearchLean/AG/LocalSemanticReconstruction/IndependentAATPrimitiveReconstruction.lean\#L61}{\texttt{Native\allowbreak{}Category}} / \href{https://github.com/iroha1203/AlgebraicArchitectureTheoryV2/blob/719f81f47d410701fd82c2bc88613cc140c59377/research/lean/ResearchLean/AG/LocalSemanticReconstruction/IndependentLensPrimitiveReconstruction.lean\#L301}{\texttt{Object}} (lens) / \href{https://github.com/iroha1203/AlgebraicArchitectureTheoryV2/blob/719f81f47d410701fd82c2bc88613cc140c59377/research/lean/ResearchLean/AG/LocalSemanticReconstruction/IndependentProtocolPrimitiveReconstruction.lean\#L449}{\texttt{Object}} (protocol) & Takes the category of all semantic homs from a fixed View and reference, or from a ProtocolSchema and observation. Local objects hold carrier refs, graph rows, primitive laws, and a finite cover, and all homs are given by the conditions that preserve get/put or the generating edges and the observations. \\
\addlinespace[3pt]
Lemma~\ref{lem:8.23} & \href{https://github.com/iroha1203/AlgebraicArchitectureTheoryV2/blob/719f81f47d410701fd82c2bc88613cc140c59377/research/lean/ResearchLean/AG/LocalSemanticReconstruction/IndependentLensPrimitiveReconstruction.lean\#L1370}{\texttt{reconstruction\allowbreak{}Data}} / \href{https://github.com/iroha1203/AlgebraicArchitectureTheoryV2/blob/719f81f47d410701fd82c2bc88613cc140c59377/research/lean/ResearchLean/AG/LocalSemanticReconstruction/IndependentLensPrimitiveReconstruction.lean\#L1378}{\texttt{assemble\_\allowbreak{}read}} / \href{https://github.com/iroha1203/AlgebraicArchitectureTheoryV2/blob/719f81f47d410701fd82c2bc88613cc140c59377/research/lean/ResearchLean/AG/LocalSemanticReconstruction/IndependentLensPrimitiveReconstruction.lean\#L1387}{\texttt{read\_\allowbreak{}assemble}} & Assembles get and put from the row conditions of the primitive lens tables, and obtains the three laws and the finiteness of the reference fiber. For the state maps that preserve get and put as well, the readout and the reconstruction are mutually inverse. \\
\addlinespace[3pt]
Lemma~\ref{lem:8.24} & \href{https://github.com/iroha1203/AlgebraicArchitectureTheoryV2/blob/719f81f47d410701fd82c2bc88613cc140c59377/research/lean/ResearchLean/AG/LocalSemanticReconstruction/IndependentProtocolPrimitiveReconstruction.lean\#L1671}{\texttt{reconstruction\allowbreak{}Data}} / \href{https://github.com/iroha1203/AlgebraicArchitectureTheoryV2/blob/719f81f47d410701fd82c2bc88613cc140c59377/research/lean/ResearchLean/AG/LocalSemanticReconstruction/IndependentProtocolPrimitiveReconstruction.lean\#L1678}{\texttt{assemble\_\allowbreak{}read}} / \href{https://github.com/iroha1203/AlgebraicArchitectureTheoryV2/blob/719f81f47d410701fd82c2bc88613cc140c59377/research/lean/ResearchLean/AG/LocalSemanticReconstruction/IndependentProtocolPrimitiveReconstruction.lean\#L1686}{\texttt{read\_\allowbreak{}assemble}} / \href{https://github.com/iroha1203/AlgebraicArchitectureTheoryV2/blob/719f81f47d410701fd82c2bc88613cc140c59377/research/lean/ResearchLean/AG/RealizationReconstruction/ProtocolReconstruction.lean\#L60}{\texttt{generator\_\allowbreak{}path\_\allowbreak{}naturality}} & Reconstructs all the objects and morphisms from the primitive tables of generating edges and observations and from a finite cover of the states. The naturality of the generating edges extends to all paths, and descends to the quotient by the declared relations. \\
\addlinespace[3pt]
Corollary~\ref{cor:8.25} & \href{https://github.com/iroha1203/AlgebraicArchitectureTheoryV2/blob/719f81f47d410701fd82c2bc88613cc140c59377/research/lean/ResearchLean/AG/LocalSemanticReconstruction/IndependentAATPrimitiveReconstruction.lean\#L1953}{\texttt{equivalence}} / \href{https://github.com/iroha1203/AlgebraicArchitectureTheoryV2/blob/719f81f47d410701fd82c2bc88613cc140c59377/research/lean/ResearchLean/AG/LocalSemanticReconstruction/IndependentAATPrimitiveReconstruction.lean\#L1958}{\texttt{equivalence\_\allowbreak{}functor}} / \href{https://github.com/iroha1203/AlgebraicArchitectureTheoryV2/blob/719f81f47d410701fd82c2bc88613cc140c59377/research/lean/ResearchLean/AG/LocalSemanticReconstruction/IndependentAATPrimitiveReconstruction.lean\#L2658}{\texttt{lens\allowbreak{}Semantic\allowbreak{}Primitive\allowbreak{}Equivalence}} / \href{https://github.com/iroha1203/AlgebraicArchitectureTheoryV2/blob/719f81f47d410701fd82c2bc88613cc140c59377/research/lean/ResearchLean/AG/LocalSemanticReconstruction/IndependentAATPrimitiveReconstruction.lean\#L2667}{\texttt{protocol\allowbreak{}Semantic\allowbreak{}Primitive\allowbreak{}Equivalence}} & The equivalences of categories for lenses and protocols. They treat arbitrary semantic morphisms, preserve identities, composition, and primitive evaluation, and reflect isomorphisms. \\
\addlinespace[3pt]
Proposition~\ref{prop:8.26} & \href{https://github.com/iroha1203/AlgebraicArchitectureTheoryV2/blob/719f81f47d410701fd82c2bc88613cc140c59377/research/lean/ResearchLean/AG/RealizationReconstruction/LensSemantics.lean\#L145}{\texttt{ext}} / \href{https://github.com/iroha1203/AlgebraicArchitectureTheoryV2/blob/719f81f47d410701fd82c2bc88613cc140c59377/research/lean/ResearchLean/AG/RealizationReconstruction/LensSemantics.lean\#L170}{\texttt{res\_\allowbreak{}ext}} / \href{https://github.com/iroha1203/AlgebraicArchitectureTheoryV2/blob/719f81f47d410701fd82c2bc88613cc140c59377/research/lean/ResearchLean/AG/RealizationReconstruction/LensSemantics.lean\#L191}{\texttt{ext\_\allowbreak{}res}} & Between arbitrary lenses with the same View and the same reference view, extends a function on the reference fiber to all states by put. It preserves get and put, and restriction and extension are mutually inverse. Finiteness of View and invertibility of the maps are not needed. \\
\addlinespace[3pt]
Proposition~\ref{prop:8.28} & \href{https://github.com/iroha1203/AlgebraicArchitectureTheoryV2/blob/719f81f47d410701fd82c2bc88613cc140c59377/research/lean/ResearchLean/AG/RealizationReconstruction/ProtocolReconstruction.lean\#L36}{\texttt{Generator\allowbreak{}Map}} / \href{https://github.com/iroha1203/AlgebraicArchitectureTheoryV2/blob/719f81f47d410701fd82c2bc88613cc140c59377/research/lean/ResearchLean/AG/RealizationReconstruction/ProtocolReconstruction.lean\#L60}{\texttt{generator\_\allowbreak{}path\_\allowbreak{}naturality}} / \href{https://github.com/iroha1203/AlgebraicArchitectureTheoryV2/blob/719f81f47d410701fd82c2bc88613cc140c59377/research/lean/ResearchLean/AG/RealizationReconstruction/ProtocolReconstruction.lean\#L118}{\texttt{hom\allowbreak{}Equiv\allowbreak{}Generator\allowbreak{}Map}} & Constructs the natural transformation for all paths from a function at each vertex, the commutativity of the named generating edges, and the preservation of observations. The restriction to the components and the reconstruction are mutually inverse. \\
\addlinespace[3pt]
Corollary~\ref{cor:8.30} & \href{https://github.com/iroha1203/AlgebraicArchitectureTheoryV2/blob/719f81f47d410701fd82c2bc88613cc140c59377/research/lean/ResearchLean/AG/RealizationReconstruction/LensFinitePresentation.lean\#L237}{\texttt{lens\allowbreak{}Presentation\allowbreak{}Equivalence}} / \href{https://github.com/iroha1203/AlgebraicArchitectureTheoryV2/blob/719f81f47d410701fd82c2bc88613cc140c59377/research/lean/ResearchLean/AG/RealizationReconstruction/ProtocolFinitePresentation.lean\#L597}{\texttt{presentation\allowbreak{}Equivalence}} / \href{https://github.com/iroha1203/AlgebraicArchitectureTheoryV2/blob/719f81f47d410701fd82c2bc88613cc140c59377/research/lean/ResearchLean/AG/LocalSemanticReconstruction/IndependentAATPrimitiveReconstruction.lean\#L2677}{\texttt{lens\allowbreak{}Semantic\allowbreak{}Primitive\allowbreak{}Fiber\allowbreak{}Comparison\allowbreak{}Iso}} / \href{https://github.com/iroha1203/AlgebraicArchitectureTheoryV2/blob/719f81f47d410701fd82c2bc88613cc140c59377/research/lean/ResearchLean/AG/LocalSemanticReconstruction/IndependentAATPrimitiveReconstruction.lean\#L2691}{\texttt{protocol\allowbreak{}Semantic\allowbreak{}Primitive\allowbreak{}Observed\allowbreak{}Comparison\allowbreak{}Iso}} & Obtains the fullness, faithfulness, and essential surjectivity of the finite-syntax decoder from the enumerations of the finite fibers and of the vertex set. Constructs the composite with the equivalence of categories of primitive presentations, and the natural isomorphism with the readouts of fibers and observations. \\
\addlinespace[3pt]
Proposition~\ref{prop:8.31} & \href{https://github.com/iroha1203/AlgebraicArchitectureTheoryV2/blob/719f81f47d410701fd82c2bc88613cc140c59377/research/lean/ResearchLean/AG/RealizationReconstruction/CSKaroubiReconstruction.lean\#L40}{\texttt{lens\allowbreak{}Karoubi\allowbreak{}Reconstruction\allowbreak{}Equivalence}} / \href{https://github.com/iroha1203/AlgebraicArchitectureTheoryV2/blob/719f81f47d410701fd82c2bc88613cc140c59377/research/lean/ResearchLean/AG/RealizationReconstruction/CSKaroubiReconstruction.lean\#L77}{\texttt{protocol\allowbreak{}Karoubi\allowbreak{}Reconstruction\allowbreak{}Equivalence}} / \href{https://github.com/iroha1203/AlgebraicArchitectureTheoryV2/blob/719f81f47d410701fd82c2bc88613cc140c59377/research/lean/ResearchLean/AG/LocalSemanticReconstruction/IndependentAATPrimitiveReconstruction.lean\#L2816}{\texttt{lens\allowbreak{}Karoubi\allowbreak{}Arrow\allowbreak{}Equivalence}} / \href{https://github.com/iroha1203/AlgebraicArchitectureTheoryV2/blob/719f81f47d410701fd82c2bc88613cc140c59377/research/lean/ResearchLean/AG/LocalSemanticReconstruction/IndependentAATPrimitiveReconstruction.lean\#L2825}{\texttt{protocol\allowbreak{}Karoubi\allowbreak{}Arrow\allowbreak{}Equivalence}} / \href{https://github.com/iroha1203/AlgebraicArchitectureTheoryV2/blob/719f81f47d410701fd82c2bc88613cc140c59377/research/lean/ResearchLean/AG/RealizationComparisonIdempotents/FunctorNaturality.lean\#L201}{\texttt{karoubi\allowbreak{}Arrow\allowbreak{}Naturality\allowbreak{}Iso}} & The splitting of idempotents in the lens and protocol categories of realizations, and the equivalences of Karoubi categories and arrow categories from the full faithfulness of the decoder and generation by retracts. The correspondence is by restriction to the decoder and natural isomorphisms, and includes non-invertible morphisms. \\
\addlinespace[3pt]
Proposition~\ref{prop:8.32} & \href{https://github.com/iroha1203/AlgebraicArchitectureTheoryV2/blob/719f81f47d410701fd82c2bc88613cc140c59377/research/lean/ResearchLean/AG/FiniteDecoderRepresentability/NatSubsetSwaps.lean\#L212}{\texttt{nat\allowbreak{}Subset\allowbreak{}Semantic\allowbreak{}Aut\_\allowbreak{}injective}} / \href{https://github.com/iroha1203/AlgebraicArchitectureTheoryV2/blob/719f81f47d410701fd82c2bc88613cc140c59377/research/lean/ResearchLean/AG/FiniteDecoderRepresentability/CountableSyntaxObstruction.lean\#L55}{\texttt{countable\allowbreak{}Decoder\_\allowbreak{}not\_\allowbreak{}surjective}} / \href{https://github.com/iroha1203/AlgebraicArchitectureTheoryV2/blob/719f81f47d410701fd82c2bc88613cc140c59377/research/lean/ResearchLean/AG/FiniteDecoderRepresentability/CountableSyntaxObstruction.lean\#L63}{\texttt{finite\allowbreak{}List\allowbreak{}Decoder\_\allowbreak{}not\_\allowbreak{}surjective}} & With Nat Atoms, a singleton Source, and the fixed code of the full extraction, sends arbitrary Nat subsets injectively to the semantic automorphisms given by adjacent pair swaps. Proves that there is no surjection from an arbitrary Countable syntax onto that Aut, and the application to finite-list syntax. \\
\bottomrule
\end{xltabular}

\subsection*{Covers, finite presentations, and infinitely indexed families}

\begin{xltabular}{\linewidth}{@{}LLL@{}}
\toprule
Text & Lean declaration & Objects and conditions \\
\midrule
\endhead
\S\ref{sec:8.1} (finite coverage) & \href{https://github.com/iroha1203/AlgebraicArchitectureTheoryV2/blob/719f81f47d410701fd82c2bc88613cc140c59377/research/lean/ResearchLean/AG/FiniteDecoderRepresentability/CoverageTopology.lean\#L113}{\texttt{nonempty\_\allowbreak{}anchored\allowbreak{}Coverage\allowbreak{}Witness\_\allowbreak{}iff\_\allowbreak{}finite\_\allowbreak{}continuous\allowbreak{}Extensions}} / \href{https://github.com/iroha1203/AlgebraicArchitectureTheoryV2/blob/719f81f47d410701fd82c2bc88613cc140c59377/research/lean/ResearchLean/AG/FiniteDecoderRepresentability/DiscreteCompactness.lean\#L33}{\texttt{finite\_\allowbreak{}iff\_\allowbreak{}compact\allowbreak{}Space\_\allowbreak{}bot}} / \href{https://github.com/iroha1203/AlgebraicArchitectureTheoryV2/blob/719f81f47d410701fd82c2bc88613cc140c59377/research/lean/ResearchLean/AG/FiniteDecoderRepresentability/DiscreteCompactness.lean\#L50}{\texttt{nonempty\_\allowbreak{}anchored\allowbreak{}Coverage\allowbreak{}Witness\_\allowbreak{}iff\_\allowbreak{}compact\_\allowbreak{}continuous\allowbreak{}Extensions}} & For an arbitrary CartSemanticInput, the finiteness of the Sources at both ends and the continuous extension of all the extractions of the target to OnePoint $\to$ Bool are equivalent to the existence of an anchored coverage. For the discrete topology on the source, finiteness and compactness are equivalent. \\
\addlinespace[3pt]
\S\ref{sec:8.8} (fixed bit width and finite types) & \href{https://github.com/leanprover-community/mathlib4/blob/8f9d9cff6bd728b17a24e163c9402775d9e6a365/Mathlib/Data/Fintype/BigOperators.lean\#L199}{\texttt{card\_\allowbreak{}fun}} / \href{https://github.com/leanprover-community/mathlib4/blob/8f9d9cff6bd728b17a24e163c9402775d9e6a365/Mathlib/Data/Fintype/Card.lean\#L172}{\texttt{card\_\allowbreak{}bool}} / \href{https://github.com/leanprover-community/mathlib4/blob/8f9d9cff6bd728b17a24e163c9402775d9e6a365/Mathlib/Data/Fintype/Card.lean\#L488}{\texttt{card\_\allowbreak{}fin}} / \href{https://github.com/leanprover-community/mathlib4/blob/8f9d9cff6bd728b17a24e163c9402775d9e6a365/Mathlib/Data/Fintype/Card.lean\#L229}{\texttt{card\_\allowbreak{}le\_\allowbreak{}of\_\allowbreak{}injective}} / \href{https://github.com/leanprover-community/mathlib4/blob/8f9d9cff6bd728b17a24e163c9402775d9e6a365/Mathlib/Data/Fintype/BigOperators.lean\#L204}{\texttt{card\_\allowbreak{}vector}} / \href{https://github.com/leanprover-community/mathlib4/blob/8f9d9cff6bd728b17a24e163c9402775d9e6a365/Mathlib/Data/Set/Finite/List.lean\#L34}{\texttt{finite\_\allowbreak{}length\_\allowbreak{}le}} & If b bits are represented by Fin b $\to$ Bool, the number of states is $2^b$, which is an upper bound for the number of states that can be injectively encoded into it. If a finite alphabet and an upper bound on the length are fixed, the set of all strings is also finite. Integers of fixed bit width are also treated as finite bit states. \\
\addlinespace[3pt]
\S\ref{sec:8.8} (the whole table for a finite index set) & \href{https://github.com/iroha1203/AlgebraicArchitectureTheoryV2/blob/719f81f47d410701fd82c2bc88613cc140c59377/research/lean/ResearchLean/AG/LocalSemanticReconstruction/IndependentFiniteFragments.lean\#L58}{\texttt{fragments\_\allowbreak{}glue}} / \href{https://github.com/iroha1203/AlgebraicArchitectureTheoryV2/blob/719f81f47d410701fd82c2bc88613cc140c59377/research/lean/ResearchLean/AG/LocalSemanticReconstruction/IndependentFiniteFragments.lean\#L55}{\texttt{glue\_\allowbreak{}fragments}} & If the index set $\Lambda$ is finite, the gluing of the finite fragments recovers the table over all of $\Lambda$. \\
\addlinespace[3pt]
\S\ref{sec:8.8} (finite universal decision) & \href{https://github.com/leanprover-community/mathlib4/blob/8f9d9cff6bd728b17a24e163c9402775d9e6a365/Mathlib/Data/Fintype/Defs.lean\#L208}{\texttt{decidable\allowbreak{}Forall\allowbreak{}Fintype}} & If the index of the conditions is a Fintype and each condition has a DecidablePred, the universal predicate can be decided by finite enumeration. \\
\addlinespace[3pt]
Example~\ref{ex:8.21} (generated comparisons) & \href{https://github.com/iroha1203/AlgebraicArchitectureTheoryV2/blob/7e68ec6e77ef0249ede6a4c875715cfaf1cb3ee6/research/lean/ResearchLean/AG/LocalSemanticReconstruction/SemanticGeneratedComparisonReconstruction.lean\#L261}{\texttt{semantic\allowbreak{}Comparison\allowbreak{}Original\_\allowbreak{}reconstruction\allowbreak{}Iso}} / \href{https://github.com/iroha1203/AlgebraicArchitectureTheoryV2/blob/7e68ec6e77ef0249ede6a4c875715cfaf1cb3ee6/research/lean/ResearchLean/AG/LocalSemanticReconstruction/SemanticGeneratedComparisonReconstruction.lean\#L277}{\texttt{semantic\allowbreak{}Comparison\allowbreak{}Direct\_\allowbreak{}reconstruction\allowbreak{}Iso}} / \href{https://github.com/iroha1203/AlgebraicArchitectureTheoryV2/blob/7e68ec6e77ef0249ede6a4c875715cfaf1cb3ee6/research/lean/ResearchLean/AG/LocalSemanticReconstruction/SemanticGeneratedComparisonReconstruction.lean\#L300}{\texttt{semantic\allowbreak{}Comparison\allowbreak{}Via\allowbreak{}Base\_\allowbreak{}reconstruction\allowbreak{}Iso}} / \href{https://github.com/iroha1203/AlgebraicArchitectureTheoryV2/blob/7e68ec6e77ef0249ede6a4c875715cfaf1cb3ee6/research/lean/ResearchLean/AG/LocalSemanticReconstruction/SemanticGeneratedComparisonReconstruction.lean\#L49}{\texttt{semantic\allowbreak{}Comparison\allowbreak{}Hom\_\allowbreak{}assemble\_\allowbreak{}read}} / \href{https://github.com/iroha1203/AlgebraicArchitectureTheoryV2/blob/7e68ec6e77ef0249ede6a4c875715cfaf1cb3ee6/research/lean/ResearchLean/AG/LocalSemanticReconstruction/SemanticGeneratedComparisonReconstruction.lean\#L676}{\texttt{semantic\allowbreak{}Comparison\allowbreak{}Generated\allowbreak{}Square\_\allowbreak{}iff}} & Reconstructs the original geometry and both routes generated from the same general semantic square, all morphisms, endpoint automorphisms, and commutative squares from primitive tables in the representative variant. \\
\addlinespace[3pt]
Example~\ref{ex:8.21} (generated morphisms and the three-axis example) & \href{https://github.com/iroha1203/AlgebraicArchitectureTheoryV2/blob/7e68ec6e77ef0249ede6a4c875715cfaf1cb3ee6/research/lean/ResearchLean/AG/LocalSemanticReconstruction/SemanticGeneratedComparisonReconstruction.lean\#L418}{\texttt{semantic\allowbreak{}Comparison\allowbreak{}Alpha\_\allowbreak{}eq\_\allowbreak{}G118\allowbreak{}Five\allowbreak{}Factor}} / \href{https://github.com/iroha1203/AlgebraicArchitectureTheoryV2/blob/7e68ec6e77ef0249ede6a4c875715cfaf1cb3ee6/research/lean/ResearchLean/AG/LocalSemanticReconstruction/SemanticGeneratedComparisonReconstruction.lean\#L508}{\texttt{semantic\allowbreak{}Comparison\allowbreak{}Beta\_\allowbreak{}factor}} / \href{https://github.com/iroha1203/AlgebraicArchitectureTheoryV2/blob/7e68ec6e77ef0249ede6a4c875715cfaf1cb3ee6/research/lean/ResearchLean/AG/LocalSemanticReconstruction/SemanticGeneratedComparisonReconstruction.lean\#L604}{\texttt{semantic\allowbreak{}Comparison\_\allowbreak{}read\_\allowbreak{}projector\_\allowbreak{}laws}} / \href{https://github.com/iroha1203/AlgebraicArchitectureTheoryV2/blob/7e68ec6e77ef0249ede6a4c875715cfaf1cb3ee6/research/lean/ResearchLean/AG/LocalSemanticReconstruction/IndependentAATPrimitiveReconstruction.lean\#L3139}{\texttt{finite\allowbreak{}Axis\allowbreak{}Fold\allowbreak{}Bar\allowbreak{}Beta\allowbreak{}Native\allowbreak{}Hom\_\allowbreak{}factor}} / \href{https://github.com/iroha1203/AlgebraicArchitectureTheoryV2/blob/7e68ec6e77ef0249ede6a4c875715cfaf1cb3ee6/research/lean/ResearchLean/AG/LocalSemanticReconstruction/IndependentAATPrimitiveReconstruction.lean\#L3251}{\texttt{finite\allowbreak{}Axis\allowbreak{}Fold\allowbreak{}Identity\allowbreak{}Cochain\_\allowbreak{}bar\allowbreak{}Beta\_\allowbreak{}eq\_\allowbreak{}bar\allowbreak{}Alpha\_\allowbreak{}common}} & Recovers the five-factor comparison, generated comparison, and both projectors as the same morphisms. The three-axis example treats normalization selected by a nonidentity defect and identity selected by the identity cochain. \\
\addlinespace[3pt]
\S\ref{sec:8.9} (swapping adjacent pairs) & \href{https://github.com/iroha1203/AlgebraicArchitectureTheoryV2/blob/719f81f47d410701fd82c2bc88613cc140c59377/research/lean/ResearchLean/AG/FiniteDecoderRepresentability/NatAdjacentSwap.lean\#L103}{\texttt{nat\allowbreak{}Adjacent\allowbreak{}Swap\_\allowbreak{}even}} / \href{https://github.com/iroha1203/AlgebraicArchitectureTheoryV2/blob/719f81f47d410701fd82c2bc88613cc140c59377/research/lean/ResearchLean/AG/FiniteDecoderRepresentability/NatAdjacentSwap.lean\#L109}{\texttt{nat\allowbreak{}Adjacent\allowbreak{}Swap\_\allowbreak{}odd}} / \href{https://github.com/iroha1203/AlgebraicArchitectureTheoryV2/blob/719f81f47d410701fd82c2bc88613cc140c59377/research/lean/ResearchLean/AG/FiniteDecoderRepresentability/NatAdjacentSwap.lean\#L197}{\texttt{nat\allowbreak{}Adjacent\allowbreak{}Swap\_\allowbreak{}support\_\allowbreak{}infinite}} / \href{https://github.com/iroha1203/AlgebraicArchitectureTheoryV2/blob/719f81f47d410701fd82c2bc88613cc140c59377/research/lean/ResearchLean/AG/FiniteDecoderRepresentability/NatAdjacentSwap.lean\#L203}{\texttt{not\_\allowbreak{}exists\_\allowbreak{}nat\allowbreak{}Swap\allowbreak{}Code\allowbreak{}Hom}} & The map that exchanges all the adjacent pairs of the natural numbers is determined by a finite arithmetic rule on parity, but its support is infinite and it cannot be presented as a morphism with a fixed code. \\
\addlinespace[3pt]
Proposition~\ref{prop:8.33} (the group of abstract Bool families) & \href{https://github.com/iroha1203/AlgebraicArchitectureTheoryV2/blob/719f81f47d410701fd82c2bc88613cc140c59377/research/lean/ResearchLean/AG/LocalSemanticReconstruction/TagChangeFiniteGroupReconstruction.lean\#L134}{\texttt{global\allowbreak{}Tag\allowbreak{}Change\allowbreak{}Add\allowbreak{}Equiv\allowbreak{}Coherent\allowbreak{}Family}} / \href{https://github.com/iroha1203/AlgebraicArchitectureTheoryV2/blob/719f81f47d410701fd82c2bc88613cc140c59377/research/lean/ResearchLean/AG/LocalSemanticReconstruction/TagChangeFiniteGroupReconstruction.lean\#L140}{\texttt{global\allowbreak{}Tag\allowbreak{}Change\allowbreak{}Add\allowbreak{}Equiv\allowbreak{}Coherent\allowbreak{}Family\_\allowbreak{}apply}} / \href{https://github.com/iroha1203/AlgebraicArchitectureTheoryV2/blob/719f81f47d410701fd82c2bc88613cc140c59377/research/lean/ResearchLean/AG/LocalSemanticReconstruction/TagChangeFiniteGroupReconstruction.lean\#L146}{\texttt{global\allowbreak{}Tag\allowbreak{}Change\allowbreak{}Add\allowbreak{}Equiv\allowbreak{}Coherent\allowbreak{}Family\_\allowbreak{}symm\_\allowbreak{}apply}} & The isomorphism of additive groups between the global Bool families and the families of all the compatible finite fragments. The objects are abstract families, and the identification with the automorphisms of an operation package is not included. \\
\addlinespace[3pt]
Proposition~\ref{prop:8.33} (finite readouts) & \href{https://github.com/iroha1203/AlgebraicArchitectureTheoryV2/blob/719f81f47d410701fd82c2bc88613cc140c59377/research/lean/ResearchLean/AG/LocalSemanticReconstruction/TagChangeFiniteReconstruction.lean\#L188}{\texttt{finite\_\allowbreak{}reading\_\allowbreak{}not\_\allowbreak{}separating}} & For an infinite set $\Omega$ and a finite subset S, a Bool family that is true at one point outside S has the same readout on S as the zero family. Hence finite readouts do not separate these abstract Bool families. \\
\bottomrule
\end{xltabular}

\chapter{Repository and Lean Build}\label{app:B}

The proof sources are publicly available in the \href{https://github.com/iroha1203/AlgebraicArchitectureTheoryV2}{AlgebraicArchitectureTheoryV2 repository}.

With Git and elan installed, run the following to build the main body (\texttt{Formal}).
This uses the Lean version specified by \texttt{lean-toolchain} and the dependency libraries fixed by \texttt{lake-manifest.json}.

{\small\begin{verbatim}
git clone https://github.com/iroha1203/AlgebraicArchitectureTheoryV2.git
cd AlgebraicArchitectureTheoryV2
git checkout --detach 719f81f47d410701fd82c2bc88613cc140c59377
lake exe cache get
lake build +Formal.AG
\end{verbatim}}

Next, run the following in \texttt{research/lean}.
The targets below, together with the dependency modules that are built automatically, contain the Research-side proof sources that Appendix~\ref{app:A} cites at version \texttt{719f81f47d41}.

{\small\begin{verbatim}
cd research/lean
lake build \
  ResearchLean.AG.AtomFoundation.RefinementSupply \
  ResearchLean.AG.CanonicalResolution.NegativeWitness \
  ResearchLean.AG.ComparisonInformationLoss.KaroubiRestrictionFiniteWitness \
  ResearchLean.AG.ComparisonInformationLoss.PresentationTransport \
  ResearchLean.AG.DiagnosticConservativity.ObstructionExactness \
  ResearchLean.AG.DiagnosticConservativity.OrbitExactness \
  ResearchLean.AG.DoctrineFiberProduct.ConfigurationDescent \
  ResearchLean.AG.DoctrineFiberProduct.IndexedRawFamilyClassification \
  ResearchLean.AG.DoctrineFiberProduct.InternalNormalizationSplitNoGo \
  ResearchLean.AG.FiniteDecoderRepresentability.CountableSyntaxObstruction \
  ResearchLean.AG.LocalSemanticReconstruction.IndependentAATPrimitiveReconstruction \
  ResearchLean.AG.ObstructionDiagnosticBridge.SelectedFiniteObstructionExamples \
  ResearchLean.AG.RealizationComparisonIdempotents.MaximalSubgroupoid \
  ResearchLean.AG.RealizationReconstruction.CSAATProtocolAdapterSquares \
  ResearchLean.AG.RealizationReconstruction.CSAATRestrictionKernelFiberTransport \
  ResearchLean.AG.RealizationReconstruction.FixedFLensGroupConnection \
  ResearchLean.AG.RealizationReconstruction.FixedFProtocolGroupConnection \
  ResearchLean.AG.StructuralCover.GeneratedH1Vanishing \
  ResearchLean.AG.UniformInvariance.AtlasPositioning \
  ResearchLean.AG.UniformInvariance.ConditionCAllAFiring \
  ResearchLean.AG.UniformInvariance.GLocalV1T3T6Uniformity
\end{verbatim}}

For general transport, lenses, base change, and comparison groups, run the following in the same \texttt{research/lean} directory.
The declarations cited at this version in Appendix~\ref{app:A} are all contained in these targets and their dependency modules.

{\small\begin{verbatim}
git checkout --detach 7e68ec6e77ef0249ede6a4c875715cfaf1cb3ee6
lake build \
  ResearchLean.AG.ComparisonInformation.F37GeneralComparison \
  ResearchLean.AG.CrossStageCoherence.ArbitraryStrongLiftPseudofunctor \
  ResearchLean.AG.CrossStageCoherence.CompositeQualification \
  ResearchLean.AG.DoctrineFiberProduct.ArbitrarySemanticCleavageCoherence \
  ResearchLean.AG.FullGeometryNormalization.\
SemanticDerivedCanonicalComparisonIdentification \
  ResearchLean.AG.FullGeometryNormalization.SemanticDerivedDiagnosticEndpointBridge \
  ResearchLean.AG.FullGeometryNormalization.\
SemanticDerivedSelectorGlobalBottomReflection \
  ResearchLean.AG.FullGeometryNormalization.SemanticExactBarBetaKaroubiProjection \
  ResearchLean.AG.LocalSemanticReconstruction.\
SemanticGeneratedComparisonReconstruction \
  ResearchLean.AG.RealizationReconstruction.FixedFSemidirectProduct \
  ResearchLean.AG.RealizationReconstruction.GeneralRelativeLensComparison \
  ResearchLean.AG.TransportCoherence.ArbitraryObstruction
\end{verbatim}}

\chapter{Disclosure of AI Use}\label{app:C}

This research was led by the author. The author formulated the questions, set the
direction of the theory, decided which results to accept, and carried out the final
check of the manuscript; the author bears responsibility for the content.

AI was used for constructing the theory and proving theorems. The formalization of
theorems in Lean and the writing of the proofs relied mainly on Codex (OpenAI), and
the examination of the design of the theory and the drafting of the manuscript relied
mainly on Claude (Anthropic). For the independent review of the
manuscript and the proofs, several AI systems were used, including these two.

The correspondence between the propositions in the text and the Lean declarations is
given in Appendix~\ref{app:A}; the fixed version of the proof sources and the build
procedure are given in Appendix~\ref{app:B}.
The tools used, their purposes, the scope covered, and the verification methods are
recorded in the repository given in Appendix~\ref{app:B}, in the file
\texttt{outreach/\allowbreak{}paper/\allowbreak{}rising-sea/\allowbreak{}ai-use.md}.

\end{document}